\documentclass{aa}  

\usepackage{graphicx}
\usepackage{txfonts}
\usepackage{natbib}
\bibpunct[]{(}{)}{;}{a}{}{,}
\begin{document}
   \title{Complex organic molecules in the interstellar medium:\\
   IRAM 30~m line survey of Sagittarius~B2(N) and (M)
   \thanks{Based on observations carried out with the IRAM 30~m telescope.
   IRAM is supported by INSU/CNRS (France), MPG (Germany), and IGN (Spain).}$^,$
   \thanks{Figures \ref{f:diag_radtrans} to \ref{f:survey_b2m_1mm} and
   Tables \ref{t:modparam_lmh_c2h3cn} to \ref{t:modparam_b2m_so} are only
   available in electronic form at http://www.aanda.org. 
   The observed and synthetic 3~mm spectra of Sgr~B2(N) and (M) are available 
   in electronic form in both FITS and ASCII formats at the CDS via 
   anonymous ftp to cdsarc.u-strasbg.fr (130.79.128.5) or via 
   http://cdsweb.u-strasbg.fr/cgi-bin/qcat?J/A+A/. They are also available in 
   CLASS format on request to the first author. 
   The list of identifications corresponding to the blue labels in 
   Figs.~\ref{f:survey_lmh_3mm} to \ref{f:survey_b2m_1mm} where the labels are 
   often too crowded to be easily readable are available in electronic form
   at the CDS via anonymous ftp to cdsarc.u-strasbg.fr (130.79.128.5) or 
   via http://cdsweb.u-strasbg.fr/cgi-bin/qcat?J/A+A/.}
   }
   \titlerunning{Single-dish line survey of Sagittarius~B2(N) and (M) at 3~mm}
   \author{A. Belloche\inst{1}
          \and
          H.~S.~P. M\"uller\inst{1,2}
          \and
          K.~M. Menten\inst{1}
          \and
          P. Schilke\inst{1,2}
          \and
          C. Comito\inst{1}
          }

   \institute{Max-Planck-Institut f\"ur Radioastronomie, Auf dem H\"ugel 69,
              53121 Bonn, Germany\\
              \email{belloche@mpifr-bonn.mpg.de}
         \and
             I. Physikalisches Institut, Universit{\"a}t zu K{\"o}ln, 
             Z{\"u}lpicher Str. 77, 50937 K{\"o}ln, Germany
             }

   \date{Received 14 January 2013; accepted 11 July 2013}

 
  \abstract
   {The discovery of amino acids in meteorites fallen to Earth and the 
   detection of glycine, the simplest of them, in samples returned from a 
   comet to Earth strongly suggest that the chemistry of the interstellar 
   medium is capable of producing such complex 
   organic molecules and that they may be widespread in our Galaxy.
   }
   {Our goal is to investigate the degree of chemical complexity that can be
   reached in the interstellar medium, in particular in dense star-forming 
   regions.
   }
   {We performed an unbiased, spectral line survey toward Sgr~B2(N) and (M),
   two regions where high-mass stars are formed, with the IRAM 30~m telescope 
   in the 3~mm
   atmospheric transmission window. Partial surveys at 2 and 1.3~mm were
   performed in parallel. The spectra were analyzed with a simple radiative 
   transfer model that assumes local thermodynamic equilibrium but takes
   optical depth effects into account.
   }
   {About 3675 and 945 spectral lines with a peak signal-to-noise ratio higher 
   than 4 are detected at 3~mm toward Sgr~B2(N) and (M), i.e. about 102 and 26 
   lines per GHz, respectively. This represents an increase by about a 
   factor of two over previous surveys of Sgr~B2. About 70\% and 47\% 
   of the lines detected toward Sgr~B2(N) and (M) are identified and assigned 
   to 56 and 46 distinct molecules as well as to 66 and 54 less abundant 
   isotopologues of these molecules, respectively. In 
   addition, we report the detection of transitions from 59 and 24 catalog 
   entries corresponding to vibrationally or torsionally excited states of 
   some of these molecules, respectively, up to a vibration energy of
   1400~cm$^{-1}$ (2000~K). Excitation temperatures and column
   densities were derived for each species but should be used with 
   caution. The rotation temperatures of the detected complex 
   molecules typically range from $\sim$~50 to 200~K. Among the detected 
   molecules, aminoacetonitrile, \textit{n-}propyl cyanide, and ethyl formate
   were reported for the first time in space based on this survey, as were 
   five rare isotopologues of vinyl cyanide, cyanoacetylene, and hydrogen 
   cyanide. We also report the detection of transitions from within 
   twelve new vibrationally or torsionally excited states of known molecules.
   Absorption features produced by diffuse clouds along the line of 
   sight are detected in transitions with low rotation quantum numbers of many 
   simple molecules and are modeled with $\sim$~30--40 velocity components 
   with typical linewidths of $\sim$~3--5~km~s$^{-1}$.
   }
   {Although the large number of unidentified lines may still allow
   future identification of new molecules, we expect most of these lines to 
   belong to 
   vibrationally or torsionally excited states or to rare isotopologues of 
   known molecules for which spectroscopic predictions are currently missing.
   Significant progress in extending the inventory of complex organic molecules
   in Sgr~B2(N) and deriving tighter constraints on their location, origin, 
   and abundance is expected in the near future thanks to an ongoing spectral 
   line survey at 3~mm with ALMA in its cycles 0 and 1. The present 
   single-dish survey will serve as a solid basis for the line identification
   and analysis of such an interferometric survey.
   }


   \maketitle
%

\section{Introduction}
\label{s:intro}

In the field of astrochemistry, organic molecules are considered to be
complex if they contain six atoms or more \citep[][]{Herbst09}. The search for 
such complex organic molecules in the interstellar medium is in part motivated 
by the discovery of more than 80 distinct amino acids in meteorites fallen to 
Earth. The isotopic composition and racemic distribution of these amino acids,
in particular, suggest that they, or at least 
their direct precursors, have an extraterrestrial origin \citep[see, e.g.,][, 
and references therein]{Ehrenfreund01,Bernstein02,Elsila07}. More recently, 
the detection of glycine (NH$_2$CH$_2$COOH), the simplest amino acid, has even 
been claimed in samples from comet 81P/Wild~2 returned to Earth by NASA's
Stardust spacecraft \citep[][]{Elsila09}. Even if it is unclear whether the 
meteoritic amino acids and the cometary glycine represent pristine 
\textit{interstellar} molecules, these discoveries strongly suggest that 
interstellar chemistry is capable of producing such complex molecules and that
they could be widespread in our Galaxy. 

A direct detection of amino acids and other similarly complex molecules 
in the interstellar medium is challenging because of their large partition 
functions and the related weakness of their molecular line emission. The most 
efficient and least ambiguous way to search for such complex molecules is to 
perform unbiased spectral line surveys of suitable astronomical sources over a 
broad frequency range, as was realized three decades ago already 
\citep[see, e.g., the spectral survey of][]{Cummins86}. The reliability of a
new detection is significantly improved by checking the presence of a 
molecule as a whole instead of relying on only detecting a few specific 
transitions 
\citep[see, e.g.,][ for the disputed case of interstellar glycine]{Snyder05}. 
Such an approach also allows the contribution of all known molecules to be
modeled at once and reduces the risks of misassignment.

The best hunting ground for complex organic molecules has turned out to be the 
Sagittarius~B2 molecular cloud -- hereafter \object{Sgr~B2}. It is the most 
massive star-forming region in our Galaxy, at a projected distance of 
$\sim\,$100~pc from the Galactic center, which is located at $7.9 \pm 0.8$~kpc 
from the Sun \citep{Reid09}. Sgr~B2 has a diameter of 34~pc 
\citep[][, after correction for the distance]{Scoville75} and contains two 
main sites of star formation, \object{Sgr~B2(N)} and \object{Sgr~B2(M)}, which 
are separated by $\sim\,$48$\arcsec$ ($\sim\,$1.8~pc in projection). 
Their luminosities are $8.6 \times 10^5$ and $6.5 \times 10^6$~$L_\odot$ 
\citep[][, after correction for the distance]{Goldsmith92}, and their masses 
are 3--$10 \times 10^4$ and 1.5--$4.1 \times 10^4$~$M_\odot$, respectively, 
which are small fractions of the total mass of the cloud, 
5--$10 \times 10^6$~$M_\odot$ \citep[][]{Lis90}. Both Sgr~B2(N) and (M) contain 
a number of ultracompact \ion{H}{ii} regions 
\citep[see, e.g.,][]{Gaume90,Mehringer93,Gaume95}. The star formation rate of 
Sgr~B2 is about 0.04~$M_\odot$~yr$^{-1}$, which qualifies Sgr~B2 as a 
ministarburst (see Appendix~\ref{a:sfr}).

Most complex organic molecules reported so far, such as 
acetic acid \citep[CH$_3$COOH,][]{Mehringer97}, 
glycolaldehyde \citep[CH$_2$(OH)CHO,][]{Hollis00}, 
acetamide \citep[CH$_3$CONH$_2$,][]{Hollis06}, 
aminoacetonitrile \citep[NH$_2$CH$_2$CN,][]{Belloche08}, 
or ethyl formate \citep[C$_2$H$_5$OCHO,][]{Belloche09}, 
were found for the first time in space toward Sgr~B2(N). This source 
contains two hot cores separated by $5\arcsec$ ($\sim\,$0.2~pc in 
projection) in the north--south direction, the less prominent one 
being located north of the most prominent one. The positions 
of these hot cores as obtained with the IRAM Plateau de Bure 
interferometer are 
($\alpha_{\rm J2000}$=17$^{\rm h}$47$^{\rm m}$19$\fs$886$\pm$0$\fs$005,
$\delta_{\rm J2000}$=$-28^\circ$22$\arcmin$18.4$\arcsec$$\pm$0.1$\arcsec$)
and ($\alpha_{\rm J2000}$=17$^{\rm h}$47$^{\rm m}$19$\fs$88$\pm$0$\fs$01,
$\delta_{\rm J2000}$=$-28^\circ$22$\arcmin$13.5$\arcsec$$\pm$0.2$\arcsec$)
\citep[][, see also the SMA map of \citealp{Qin11}]{Belloche08}. Their 
molecular emissions have similar linewidths ($\sim 7$~km~s$^{-1}$), but 
significantly different systemic velocities of 64 and $\sim 73$~km~s$^{-1}$, 
respectively, which allows their emissions to be separated even with single-dish
telescopes that do not resolve them spatially. \textit{Their immense hydrogen 
column densities} of $> 10^{25}$~cm$^{-2}$ on scales of a few arcseconds 
\citep[][]{Belloche08,Qin11} \textit{boost weak lines from rare species into 
detectability.}

Here, we report the analysis of a deep, unbiased, spectral line survey 
obtained at 3~mm toward Sgr~B2(N) and (M) with the IRAM 30~m telescope. The 
main goal of the survey was to search for new complex organic molecules in the
gas phase of the interstellar medium. This has been partially 
addressed in a number of earlier publications 
\citep[][]{Belloche08,Mueller08,Belloche09,Bruenken10,Braakman10,Mollendal12,Ordu12}. 
This article now presents the complete analysis of the full 
survey, the main goal being to provide a reliable identification of 
the detected lines. Section~\ref{s:obs} summarizes the observational details. 
The procedure to model the line survey is described in Sect.~\ref{s:modeling}. 
The results are presented in Sect.~\ref{s:results} and are discussed in 
Sect.~\ref{s:discussion}. Section~\ref{s:conclusions} summarizes our
conclusions.


\section{Observations and data reduction}
\label{s:obs}

\subsection{Observations}
\label{ss:obs}

We carried out millimeter line observations with the IRAM 30~m telescope on
Pico Veleta, Spain, in 2004 January (project 217-03), 2004 September 
(project 102-04), and 2005 January (project D18-04). Four SIS 
heterodyne receivers were used simultaneously, two in the 3~mm window connected
to the autocorrelation spectrometer VESPA and two in the 1.3~mm window with
filter banks as backends. A few selected frequency ranges were also observed
with one SIS receiver at 2~mm in 2004 January. The channel spacing and
bandwidth were 0.313 and 420~MHz for each receiver at 3 and 2~mm,
and 1 and 512~MHz for each receiver at 1.3~mm, respectively. The 
channel spacing corresponds to 0.8--1.2, 0.5--0.7, and 1.1--1.5~km~s$^{-1}$ in 
velocity scale at 3, 2, and 1~mm, respectively.
The observations were done in single-sideband mode with sideband rejection 
gains of \hbox{$\sim 1$--3~$\%$} at 3~mm, $\sim 5$--7~$\%$ at 2~mm, and 
$\sim 5$--8~$\%$ at 1.3~mm. The half-power beamwidth can be computed with 
the equation \hbox{$HPBW(\arcsec) = \frac{2460}{\nu(\rm{GHz})}$}. The 
forward efficiencies, $F_{\rm eff}$, were 0.95 at 3~mm, 0.93 at 2~mm, and 
0.91 at 1.3~mm, 
respectively. The main-beam efficiencies were computed using the Ruze function
$B_{\rm eff} = 1.2 \epsilon \,\, e^{-(4 \pi R \sigma / \lambda)^2}$, with
$\epsilon = 0.69$, $R \sigma = 0.07$, and $\lambda$ the wavelength in 
mm\footnote{See the system summary of the IRAM 30m telescope as of September 
2007 on http://www.iram.es/IRAMES/telescope/
\hbox{telescopeSummary/telescope\_summary.html}.}. 
The system temperatures ranged from 96 to 600 K at 3~mm, from 220 to 720 K at 
2~mm (except at 176 GHz where they ranged from 2400 to 3000 K), and from 280 
to 1200~K at 1.3~mm. The telescope pointing was checked every $\sim 1.5$~h on 
Mercury, Mars, 1757$-$240, or G10.62-0.38, and was found to be
accurate to 2$\arcsec$--3$\arcsec$ (rms). The telescope focus was optimized on 
Mercury, Jupiter, Mars, or G34.3+0.2 every $\sim 1.5$--3~h.
The observations were performed toward both sources
Sgr~B2(N) ($\alpha_{\rm J2000}$=17$^{\rm h}$47$^{\rm m}$20$\fs$0,
$\delta_{\rm J2000}$=$-28^\circ$22$\arcmin$19.0$\arcsec$, 
$V_{\rm lsr}$ = 64 km~s$^{-1}$) and Sgr~B2(M)
($\alpha_{\rm J2000}$=17$^{\rm h}$47$^{\rm m}$20$\fs$4, 
$\delta_{\rm J2000}$=$-28^\circ$23$\arcmin$07.0$\arcsec$,
\hbox{$V_{\rm lsr}$ = 62 km~s$^{-1}$)} in
position-switching mode with a reference position offset by 
($\Delta\alpha$,$\Delta\delta$)=($-752\arcsec$,$+342\arcsec$) with respect
to the former. The emission toward this reference position was found to be
weak: $T_{\rm a}^\star$($^{12}$CO~1$-$0) $\sim$ 2 K,
$T_{\rm a}^\star$(CS~2$-$1) $\la$ 0.05 K,
$T_{\rm a}^\star$($^{12}$CO~2$-$1) $\sim$ 1.5 K,
$T_{\rm a}^\star$($^{13}$CO~2$-$1) $\la$ 0.1 K,
and it is negligible for higher excitation lines and/or complex species.

We observed the full 3~mm window between 79.990 and 115.985~GHz toward both 
sources. The step between two adjacent tuning frequencies was 395~MHz, which 
yielded an overlap of 50~MHz. Half of the integration time at each tuning 
frequency was spent with the backend shifted by 50 MHz.
At 2~mm, we observed Sgr~B2(M) at only six selected frequencies (136.250, 
145.300, 154.500, 163.500, 172.650, and 176.200~GHz) and Sgr~B2(N) at
8 frequencies (the same ones plus 147.600 and 166.400~GHz). At 
1.3~mm, we covered the frequency ranges 201.805 to 204.615~GHz and 205.025 to 
217.750~GHz, plus a number of selected spots at higher frequency between 
219.285 and 267.175~GHz, 26 for Sgr~B2(N) and 19 for Sgr~B2(M). The total 
observing time from 2004 January to 2005 January was on the order of 100~h.

\subsection{Data reduction}
\label{ss:reduction}

The data were reduced with the Fortran 77 version of the CLASS software, which 
was part of the GILDAS software 
package\footnote{See http://www.iram.fr/IRAMFR/GILDAS.} until 2010. 
The autocorrelator 
VESPA produces artificial spikes with a width of 3--5 channels at the junction 
between subbands (typically two or three spikes per spectrum). We used the 
redundancy provided by our observing strategy of shifting the backend for half 
of the integration time (see Sect.~\ref{ss:obs}) to get rid of these artefacts 
at 3~mm without producing any hole in the frequency coverage of the final 
average spectra. In each 3~mm spectrum, we removed five channels at each of 
the six junctions between subbands that were possibly affected by this 
phenomenon. The artificial spikes at 2~mm were removed in the same way, but 
without redundancy, thus producing some holes in each frequency band.

For each individual spectrum, we removed a zeroth-order (constant) 
baseline by selecting a group of channels which seemed to be free of emission 
or absorption. However, many spectra are full of lines and we may have 
overestimated the level of the baseline in many cases, especially at 1.3~mm 
where the confusion limit was often reached.

The spectra obtained toward Sgr~B2(N) and (M) have a very high density of
spectral lines (see Sect.~\ref{ss:overview}), which makes a direct measurement 
of the rms noise level difficult to perform in many cases. Instead, we 
estimate the rms noise level with the radiometer formula: 
\begin{equation}
\sigma = 2 T_{\rm sys}/\sqrt{B t_{\rm int}}\,\,
\end{equation} 
with $T_{\rm sys}$ the system temperature, $B$ the \textit{effective} 
spectral resolution, and $t_{\rm int}$ the sum of the on- and off-source 
integration times. The spectral
resolution of the autocorrelator VESPA is worse than the channel spacing by
a factor of about 1.2 (G. Paubert, \textit{priv. comm.}). We verified on the
spectra taken toward the reference position, which consist mostly of 
noise apart from the narrow frequency range over which weak emission of CO is 
detected (see Sect.~\ref{ss:obs}), that the rms noise level estimated with the 
radiometer formula assuming this factor 1.2 is consistent with the measured 
noise level. As far as the 1~MHz filter banks are concerned, the measured rms 
noise levels toward the reference position are consistent with the radiometer 
formula when we use a factor $\sim 1.1$. Table~\ref{t:rms} 
lists for Sgr~B2(N) and (M) the rms noise levels estimated with the radiometer
formula as described above. For both sources, the median noise levels at 3 and 
1.3~mm are 18 and 53~mK, respectively, in $T_{\rm a}^*$ scale. The median 
noise level at 2~mm is 27~mK for Sgr~B2(N) and 51~mK for Sgr~B2(M).

\begin{table}
 \caption{
 Estimated rms noise level achieved toward Sgr~B2(N) and (M).
 }
 \label{t:rms}
 \centering
 \begin{tabular}{@{\extracolsep{-1ex}}cccccc}
 \hline\hline
  \multicolumn{1}{c}{} & \multicolumn{2}{c}{Sgr~B2(N)} & & \multicolumn{2}{c}{Sgr~B2(M)} \\ 
 \cline{2-3} \cline{5-6}
 \noalign{\smallskip}
  \multicolumn{1}{c}{Frequency range} & \multicolumn{5}{c}{Estimated rms noise level (mK)} \\ 
  \multicolumn{1}{c}{(GHz)} & \multicolumn{1}{c}{range} & \multicolumn{1}{c}{median} & & \multicolumn{1}{c}{range} & \multicolumn{1}{c}{median} \\ 
 \hline
$ 80.0- 89.9$ & $ 10- 25$ &  13 &  & $ 12- 22$ &  15 \\ 
$ 89.9-100.2$ & $ 10- 23$ &  17 &  & $ 14- 42$ &  18 \\ 
$100.2-114.6$ & $ 12- 35$ &  22 &  & $ 14- 34$ &  22 \\ 
$114.6-116.0$ & $ 42- 56$ &  43 &  & $ 43- 57$ &  45 \\ 
\hline 
$136.0-147.9$ & $ 17- 21$ &  20 &  & $ 24- 26$ &  25 \\ 
$154.2-154.8$ & $ 74$ &  74 &  & $ 75$ &  75 \\ 
$163.2-172.9$ & $ 25- 42$ &  28 &  & $ 47- 55$ &  51 \\ 
$176.0-176.5$ & $227$ & 227 &  & $289$ & 289 \\ 
\hline 
$201.8-204.6$ & $ 66-220$ &  91 &  & $ 72-118$ &  85 \\ 
$205.0-217.8$ & $ 19-169$ &  44 &  & $ 18- 91$ &  44 \\ 
$219.2-228.0$ & $ 47-155$ &  54 &  & $ 42- 69$ &  52 \\ 
$230.9-236.6$ & $ 10-100$ &  22 &  & $  9-123$ & 115 \\ 
$240.7-241.3$ & $114$ & 114 &  & $169$ & 169 \\ 
$243.9-247.5$ & $ 21- 49$ &  32 &  & $ 36- 40$ &  38 \\ 
$251.9-254.5$ & $ 16- 21$ &  19 &  & $ 14- 19$ &  17 \\ 
$255.2-259.3$ & $107-784$ & 551 &  & $ 98-339$ & 263 \\ 
$260.0-267.2$ & $ 37-200$ &  49 &  & $ 50- 85$ &  84 \\ 
 \hline
 \end{tabular}
 \tablefoot{
 The rms noise levels are given in $T_a^\star$ scale and are estimated from the radiometer formula, assuming a spectral resolution of 1.2 times the channel spacing below 180~GHz and 1.1 above. Some frequency ranges listed in the first column above 116~GHz include holes that were not covered by the observations.
 }
 \end{table}

\section{Modeling of the line survey}
\label{s:modeling}

Given the high line density of the Sgr~B2(N) and (M) spectra (see 
Sect.~\ref{ss:overview}), the assignment of a line to a given molecule can
be trusted only if all lines emitted by this molecule in our frequency coverage
are detected with the right intensity predicted by a model (see below) and no
predicted line is missing in the observed spectrum. This approach of modeling 
the full spectrum at once helps reducing the risks of misassignment and
allows to deal with lines that overlap in frequency.

\subsection{Radiative transfer with XCLASS}
\label{ss:xclass}

We used the XCLASS software\footnote{See 
http://www.astro.uni-koeln.de/projects/schilke/XCLASS. Details about the 
software can also be found in \citet{Schilke99}, \citet{Comito05},
\citet{Hieret05}, and \citet{Zernickel12}.} to model both the emission and 
absorption lines. The method used to compute the radiative transfer is 
described in the next paragraphs and illustrated in Fig.~\ref{f:diag_radtrans} 
(online material). The molecular spectroscopic parameters are taken from our 
line catalog which contains all entries from the catalog of the Cologne 
Database for Molecular Spectroscopy 
\citep[CDMS\footnote{See http://www.cdms.de.}, see][]{Mueller01,Mueller05} and 
most entries from the molecular spectroscopic database of the Jet Propulsion 
Laboratory 
\citep[JPL\footnote{See http://spec.jpl.nasa.gov.}, see][]{Pickett98}, as well 
as additional ``private'' entries. In the version of XCLASS used for this 
work, a global linear fit to the tabulated values of the partition function 
between 9.375 and 300~K is performed in log-log space for each species. The 
parameters of this fit are used to interpolate or extrapolate the partition 
functions at any temperature. The extrapolation at temperatures below 9.375~K, 
which concerns the species seen in absorption, is in some cases significantly 
wrong. In the following, we will mention when the column densities were 
corrected a posteriori for the inaccuracy of the linear extrapolation.

The modeling was done species per species, a species being one molecule, or only
one particular vibrationally or torsionally excited state of a molecule.
The rare isotopologues of each main species were also modeled separately, but we
tried to stay close to the isotopic ratios listed in Table~\ref{t:iso}.
For each species, XCLASS assumes that the level populations are described by a
single excitation temperature, which we refer to as the rotation temperature,
$T_{\rm rot}$. This assumption is valid in two cases: at high density
where collisions are frequent enough for the local thermodynamic equilibrium 
(LTE) approximation to be valid, and at very low density, where 
$T_{\rm rot}$ equals the temperature of the cosmic microwave background, 
$T_{\rm CMB} = 2.73$~K. 

\begin{table*}[t]
\caption{Velocity ranges and isotopic ratios of the Galactic emission and 
absorption components along the line of sight of Sgr~B2.}
\label{t:iso}
\centering
\begin{tabular}{cccccl}
\hline\hline
\noalign{\smallskip}
\multicolumn{1}{c}{Velocity range\tablefootmark{a}} & \multicolumn{1}{c}{$^{12}$C/$^{13}$C\tablefootmark{b}} & \multicolumn{1}{c}{$^{16}$O/$^{18}$O\tablefootmark{c}} & \multicolumn{1}{c}{$^{18}$O/$^{17}$O\tablefootmark{d}} & \multicolumn{1}{c}{$^{14}$N/$^{15}$N\tablefootmark{e}} & \multicolumn{1}{c}{Location\tablefootmark{a}} \\
\multicolumn{1}{c}{(km~s$^{-1}$)} & & & & & \\
\multicolumn{1}{c}{(1)} & \multicolumn{1}{c}{(2)} & \multicolumn{1}{c}{(3)} & \multicolumn{1}{c}{(4)} & \multicolumn{1}{c}{(5)} & \multicolumn{1}{c}{(6)} \\ 
\hline
\noalign{\smallskip}
$\sim 62$      & 20 & 250 & 2.88 & 300 & Galactic Center, Sgr~B2 \\
39 -- 25       & 40 & 327 & 4.16 & 350 & Scutum/Crux arm \\
22 -- 12       & 60 & 560 & 4.16 & 450 & Sagittarius arm \\
8 -- $-9$      & 20 & 250 & 2.88 & 300 & Galactic Center \\
$-13$ -- $-50$ & 40 & 327 & 4.16 & 350 & 3-kpc arm \\
$< -53$        & 20 & 250 & 2.88 & 300 & Galactic Center \\
\hline
\end{tabular}
\tablefoot{The following isotopic ratios were assumed to be constant: 
$^{32}$S/$^{34}$S = 22\tablefootmark{c}, 
$^{34}$S/$^{33}$S = 5\tablefootmark{f},
$^{28}$SiO/$^{29}$SiO = 20\tablefootmark{g},
$^{29}$SiO/$^{30}$SiO = 1.5\tablefootmark{g}.
References:
\tablefoottext{a}{\citet{Menten11}} and references therein,
\tablefoottext{b}{\citet{Milam05},}
\tablefoottext{c}{\citet{Wilson94},}
\tablefoottext{d}{\citet{Wouterloot08},}
\tablefoottext{e}{\citet{Dahmen95},}
\tablefoottext{f}{\citet{Nummelin00},}
\tablefoottext{g}{\citet{Penzias81}.}
}
\end{table*}

For each species $m$, several components can be modeled at once. For each 
component $c$, the fixed parameters are the diameter of the telescope, $D$, and 
the spectral resolution, and the free parameters are: source size, 
$\theta_{m,c}$, temperature, $T_{\rm rot}^{m,c}$, column density, 
$N_{\rm tot}^{m,c}$, velocity linewidth, ${\Delta \varv}^{m,c}$ ($FWHM$), 
velocity offset, $\varv_{\rm off}^{m,c}$, with respect to the systemic velocity 
of the source, and a flag indicating if it belongs
to the ``emission'' or ``absorption'' group. 
The velocity components in the ``emission'' group are supposed to be 
non-interacting, i.e. the intensities add up linearly. This approximation is 
valid for two distinct, non-overlapping sources smaller than the beam of the 
telescope, but it is a priori less good for, e.g., a source that 
consists of a hot, compact region surrounded by a cold, extended envelope 
or for two angularly-overlapping sources with spectrally-overlapping 
optically-thick emissions. The radiative transfer is performed in the 
following way.
The full spectrum of the ``emission'' group is computed in a first step. In 
a second step, this spectrum is taken as a background for the calculation of
the contribution of the components in the ``absorption'' group (see 
Fig.~\ref{f:diag_radtrans}).

The set of equations used to compute the spectrum of the ``emission'' group is 
the following:

\begin{equation}
\label{e:radtrans_em}
T_{\rm emg}(f) = \sum_m \sum_c \eta_f^{m,c} 
\left( J_{f}(T^{m,c}_{\rm rot}) - J_{f}(T_{\rm bg}) \right) 
\left(1-e^{-\tau_f^{m,c}}\right)\,,
\end{equation}
with
\begin{equation}
J_f(T) = \frac{hf/k}{e^{hf/kT}-1}\,,
\end{equation}
\begin{equation}
\eta_f^{m,c} = \frac{{\theta_{m,c}}^2}{{\Theta_f}^2+{\theta_{m,c}}^2}\,,
\end{equation}
with
\begin{equation}
\Theta_f = 1.22\,\frac{c_{\rm l}}{fD}\,,
\end{equation}
and
\begin{equation}
\tau_f^{m,c} = \sum_t \tau_f^{m,c,t}\,,
\end{equation}
where
\begin{equation}
\tau_f^{m,c,t} = \frac{{c_{\rm l}}^2}{8\pi f^2}\,g_u^t\,A_{ul}^t\,
\frac{N_{\rm tot}^{m,c}}{Q_m(T_{\rm ex}^{m,c})}\,
\left(e^{-E_l^t/kT_{\rm rot}^{m,c}} - e^{-E_u^t/kT_{\rm rot}^{m,c}}\right)\,
\phi_f^{m,c}
\end{equation}
and
\begin{equation}
\phi_f^{m,c} = \frac{1}{\sqrt{2\pi}\,\sigma_f}\,e^{-\frac{(f-f_t-f_{\rm off}^{m,c})^2}{2{\sigma_f}^2}}\,,
\end{equation}
with
\begin{equation}
f_{\rm off}^{m,c} = -\frac{\varv_{off}^{m,c}}{c_{\rm l}}\,f_t
\end{equation}
and
\begin{equation}
\sigma_f = \frac{1}{\sqrt{8\ln2}}\, \frac{{\Delta \varv}^{m,c}}{c_{\rm l}} \left(f_t+f_{\rm off}^{m,c}\right)\,,
\end{equation}
with $f$ the frequency, $c_{\rm l}$ the speed of light, 
$m$ the index on the species, 
$Q_m$ the partition function of species $m$, 
$c$ the index on the components in the ``emission'' group,
$t$ the index on the transitions of each species, 
$f_t$ the transition rest frequency, 
$g_u^t$ the upper state degeneracy, 
$A_{ul}^t$ the Einstein coefficient for spontaneous emission, 
and $E_l^t$ and $E_u^t$ the lower and upper state energy levels, respectively.

The spectrum of the ``absorption'' group is computed with the following 
equation, $c$ being now the index on the components in the ``absorption'' 
group:

\begin{eqnarray}
\lefteqn{T_{\rm abg}(f) = \sum_m \sum_c} \nonumber \\
& & \:\:\:\:\: T_{c-1}(f)\,e^{-\tau_f^{m,c}} +\,
\eta_f^{m,c} \left( J_{f}(T^{m,c}_{\rm rot}) - J_{f}(T_{\rm bg}) \right) 
\left(1-e^{-\tau_f^{m,c}}\right),
\end{eqnarray}
where $T_{c-1}$ is initialized to the spectrum of the ``emission'' group and
then contains in addition the contribution of each component of the 
``absorption'' group up to index $c-1$. 

\onlfig{
\begin{figure}
\centerline{\resizebox{0.7\hsize}{!}{\includegraphics[angle=0]{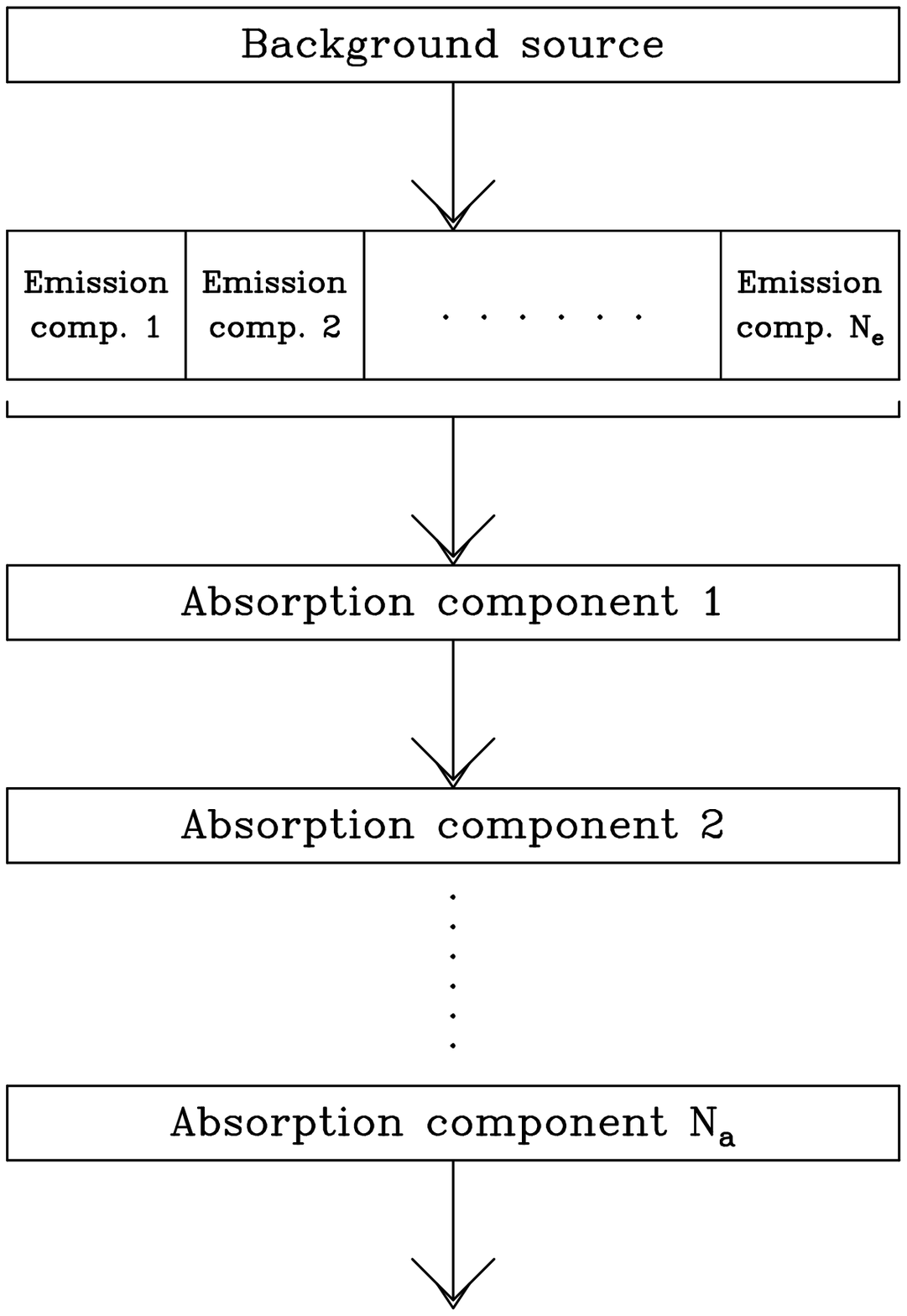}}}
\caption{Diagram illustrating how the radiative transfer of 
$N_{\mathrm e}$ emission components and $N_{\mathrm a}$ absorption components is 
computed with XCLASS (see Sect.~\ref{ss:xclass}).
}
\label{f:diag_radtrans}
\end{figure}
}

\subsection{Background temperature and continuum emission}
\label{ss:tbg}

The effective background temperature, $T_{\rm bg}$, was estimated from the 
absorption lines that looked saturated, assuming a beam filling factor of 
unity. Its values for each source are 
listed in Table~\ref{t:tbg}. It is only an effective temperature related to 
the beam-averaged continuum emission, and its values may not be physically 
meaningful especially if the continuum emission is optically thin, which is 
very likely at scales comparable to the beam.

\begin{table}
\caption{Effective background temperatures used to model the spectra of 
Sgr~B2(N) and (M).}
\label{t:tbg}
\centering
\begin{tabular}{ccc}
\hline\hline
\multicolumn{1}{c}{Atmospheric window} & \multicolumn{1}{c}{Sgr~B2(N)} &\multicolumn{1}{c}{Sgr~B2(M)} \\
\hline
3~mm &  5.2 & 5.9 \\
2~mm &  6.5 & 7.0 \\
1.3~mm & 10.0 & 8.5 \\
\hline
\end{tabular}
\tablefoot{The background temperatures, $T_{\rm bg}$, are given in K.}
\end{table}

For simplicity reasons and because we do not know the structure of the 
continuum emission at small scales very well, we use the background 
temperature $T_{\rm bg}$ derived from the saturated absorption lines to 
compute the spectrum of the ``emission'' group (equation~\ref{e:radtrans_em}). 
This assumption is, strictly speaking, not correct, especially for the 
components that are much more compact than the beam because the continuum 
emission is certainly not uniform within the single-dish beam. To evaluate the 
error introduced by this, we compare the continuum and molecular emissions in 
Table~\ref{t:contflux}. We compiled continuum measurements (Cols.~2 and 3) 
made at different frequencies (Col.~1) and small scales (Col.~4) from the 
literature. We compute $\Omega B(T_{\rm rot})$, the product of the solid 
angle over which the continuum emission was measured (Col.~5) and the Planck 
function at typical rotation temperatures of the molecular transitions 
(Col.~6). The ratio of the flux density of the continuum emission to 
$\Omega B(T_{\rm rot})$ is given in Col.~7. It is equal to the ratio 
$R = J_f(T_{\rm bg})/J_f(T_{\rm rot})$ (see 
equation~\ref{e:radtrans_em}), provided the continuum emission 
is a true background, i.e. the particles producing the continuum emission are 
not mixed with the molecules. Under this assumption, we see that the error
done in equation~(\ref{e:radtrans_em}) is on the order of 10--20\% at 3~mm for
a source size of $\sim 3\arcsec$ for the main hot core of Sgr~B2(N). This 
error will propagate to the derived column density in case of optically thin 
emission or to the derived rotation temperature/beam filling factor 
in case of optically thick emission. The error is less than 10\% for sizes
$\ga 7\arcsec$ at 3~mm. For Sgr~B2(M), the error can be as high as 30--60\% 
for a source size of $\sim 4\arcsec$ at 3~mm\footnote{We note, however, that 
the continuum measurements performed at 85 and 91~GHz toward Sgr~B2(M) do not 
seem to be consistent with each other.}. At 227~GHz, the 
error is on the order of 15--30\% for a source size of $\sim 4\arcsec$ in 
Sgr~B2(N), and somewhat smaller (9--18\%) in Sgr~B2(M).

However, the assumption that the continuum emission is a true background for 
the molecular emission is not correct. It is probably a good approximation for 
the free-free continuum emission which is concentrated in the ultra-compact 
\ion{H}{ii} regions of size $\sim 0.1\arcsec$ \citep[][]{dePree98} and 
contributes to more than 50\% of the continuum flux density of Sgr~B2(M) at 
3~mm on scales of 10--20$\arcsec$ \citep[][]{Kuan96}. But it is certainly not 
a good approximation for the thermal dust emission because we expect the dust 
particles and the molecules in the gas phase to be well mixed together. In 
this case, equation~(\ref{e:radtrans_em}) is not fully correct: 
attenuation of the molecular emission by dust should also be taken
into account. This can be done in XCLASS by multiplying 
equation~\ref{e:radtrans_em} by a factor $e^{-\tau_{\rm dust}}$ 
\citep[see equation 1 of][]{Zernickel12}. It is, however, not completely 
correct either because it assumes that all the absorbing dust is in front of 
the gas while dust and gas are likely well mixed. Given 
that the 2 and 1.3~mm spectra are also severely affected by the uncertainty on
the baseline level due to line confusion, it is difficult to break the 
degeneracy between dust attenuation and overestimate of the baseline level, 
and we did not use this option of dust attenuation.

At 1.3~mm, the continuum emission is dominated by the thermal dust emission. 
Column~7 of Table~\ref{t:contflux} is then equivalent to the dust optical 
depth. At 3~mm, it is only an upper limit, because the continuum flux density 
includes a (significant) contribution from the free-free emission. This dust 
attenuation effect should thus be small at 3~mm, but significantly more 
important at 1.3~mm, especially for Sgr~B2(N) with maybe 30--40\% of 
attenuation for a source size of 3$\arcsec$.

\begin{table}
 \caption{
 Properties of the continuum emission toward Sgr~B2(N) and (M).
}
\label{t:contflux}
 \centering
 \begin{tabular}{@{\extracolsep{-0.6ex}}rcccccc}
 \hline\hline
\multicolumn{1}{c}{$f$\tablefootmark{a}} & \multicolumn{1}{c}{$F_f$\tablefootmark{b}} & \multicolumn{1}{c}{$S_f$\tablefootmark{c}} & \multicolumn{1}{c}{Size\tablefootmark{d}} & \multicolumn{1}{c}{$\Omega$\tablefootmark{e}} & \multicolumn{1}{c}{$T$\tablefootmark{f}} & \multicolumn{1}{c}{$S_f/\Omega B_f$\tablefootmark{g}} \\ 
 \multicolumn{1}{c}{\scriptsize (GHz)} & \multicolumn{1}{c}{\scriptsize (Jy/beam)} & \multicolumn{1}{c}{\scriptsize (Jy)} & \multicolumn{1}{c}{\scriptsize ($\arcsec \times \arcsec$)} & \multicolumn{1}{c}{\scriptsize ($10^{-10}$ sr)} & \multicolumn{1}{c}{\scriptsize (K)} & \multicolumn{1}{c}{} \\ 
 \multicolumn{1}{c}{(1)} & \multicolumn{1}{c}{(2)} & \multicolumn{1}{c}{(3)} & \multicolumn{1}{c}{(4)} & \multicolumn{1}{c}{(5)} & \multicolumn{1}{c}{(6)} & \multicolumn{1}{c}{(7)} \\ 
\hline
\noalign{\smallskip}\multicolumn{2}{l}{Sgr~B2(N)} & & & & & \\ 
\noalign{\smallskip}  85  & -- &  0.90 & $3.2 \times 2.8$ &  2.1&  200 & 0.10 \\ 
 & & & & &  150 & 0.13 \\ 
 & & & & &  100 & 0.20 \\ 
  91  &  2.0 & -- & $12.2 \times 4.4$ & 14.3&  200 & 0.03 \\ 
 & & & & &  150 & 0.04 \\ 
 & & & & &  100 & 0.06 \\ 
 227  & 20.0 & -- & $4.5 \times 3.7$ &  4.4&  200 & 0.15 \\ 
 & & & & &  150 & 0.20 \\ 
 & & & & &  100 & 0.30 \\ 
\hline
\noalign{\smallskip}\multicolumn{2}{l}{Sgr~B2(M)} & & & & & \\ 
\noalign{\smallskip}  85  & -- &  4.38 & $4.8 \times 3.2$ &  3.6&  200 & 0.28 \\ 
 & & & & &  150 & 0.37 \\ 
 & & & & &  100 & 0.56 \\ 
  91  &  3.0 & -- & $10.5 \times 4.9$ & 13.7&  200 & 0.04 \\ 
 & & & & &  150 & 0.06 \\ 
 & & & & &  100 & 0.09 \\ 
 153  &  3.5 & -- & $6.5 \times 3.0$ &  5.2&  200 & 0.05 \\ 
 & & & & &  150 & 0.06 \\ 
 & & & & &  100 & 0.10 \\ 
 227  & 12.0 & -- & $4.5 \times 3.7$ &  4.4&  200 & 0.09 \\ 
 & & & & &  150 & 0.12 \\ 
 & & & & &  100 & 0.18 \\ 
\hline
 \end{tabular}
 \tablefoot{
\tablefoottext{a}{Frequency.}
 \tablefoottext{b}{Peak flux density from \citet{deVicente00} at 91~GHz, \citet{Furuya03} at 153~GHz, \citet{Lis93} at 227~GHz.}
 \tablefoottext{c}{Flux density from \citet{Liu99} at 85~GHz.}
 \tablefoottext{d}{Beam size or size of the box over which the flux density was integrated.}
\tablefoottext{e}{Solid angle of the beam or integration box.}
 \tablefoottext{f}{Assumed temperature for the Planck function.}
 \tablefoottext{g}{Ratio of the measured continuum flux density to the product of the solid angle $\Omega$ and the Planck function $B_f(T)$.}
}
 \end{table}

\subsection{Optimization}
\label{ss:optimization}

The input parameters ($\theta_{m,c}$, $T_{\rm rot}^{m,c}$, 
$N_{\rm tot}^{m,c}$, ${\Delta \varv}^{m,c}$, $\varv_{\rm off}^{m,c}$) were 
varied until a good fit to the data was obtained for each species. The whole 
spectrum including all the identified species was then computed at once, and 
the parameters for each species were adjusted again when necessary. The 
quality of the fit was checked by eye for each species over the whole 
frequency coverage of the 3~mm line survey. 

We limited the fit optimization to the 3~mm range because the synthetic 
spectrum systematically overestimates by a significant amount (factor two or 
even more) the molecular emission of the compact components
at 2 and 1.3~mm. This could be the 
combined result of at least four effects, some of them already mentioned 
before. First, the confusion limit is reached in most parts of the 1.3~mm 
spectrum, which hides the true continuum level (baseline subtraction) and 
implies that the level of the baseline must have been significantly 
overestimated. Second, the distance between the two hot cores of Sgr~B2(N), 
$5\arcsec$, is about half the size of the beam at 1.3~mm, which means that the 
observed emission of the northern hot core is attenuated by nearly a factor 
of two. The model does not take this into account because the 3 and 1.3~mm 
ranges are modeled with the same parameters. Finally, the last two effects are 
related to the radiative transfer. As detailed in Sect.~\ref{ss:tbg}, the 
contribution of the ``background'' term in equation~(\ref{e:radtrans_em}) was 
likely underestimated and the attenuation by the dust was not taken into 
account, both effects being more pronounced at 1.3~mm than at 3~mm. Because of 
these limitations, we used the information provided by the 2 and 1.3~mm ranges 
only in few cases (extended components).

\section{Results}
\label{s:results}

\subsection{Overview of the detected lines and molecules}
\label{ss:overview}

The 3~mm spectra of Sgr~B2(N) and (M) contain, respectively, about 3675 and 
945 lines detected with a main-beam peak temperature higher than 90~mK, about 
four times the median noise level ($4\sigma$). These numbers correspond to an 
average line density of 102 and 26 lines per GHz for Sgr~B2(N) and (M), 
respectively. For a higher threshold of 0.3~K ($\sim 13\sigma$), we found 1377 
and 193 lines toward Sgr~B2(N) and (M), i.e. an average line density of 38 and 
5.4 lines per GHz, respectively. 

The counting mentioned in the previous paragraph was performed by eye. Many 
transitions toward Sgr~B2(N), which consists of two hot cores separated by 
about 5$\arcsec$, have a double-peaked profile because they are emitted by both 
hot cores with a difference in line-of-sight velocity of 9--10~km~s$^{-1}$
(see Sect.~\ref{s:intro}). 
Distinguishing between a true double-peaked line profile and the chance 
juxtaposition of two independent lines is possible only once the modeling has 
been performed. The counting mentioned above being purely observational, such 
transitions with a ``true'' double-peaked profile were counted twice. 

So far, we have identified 56 molecules and 66 of their isotopologues toward 
Sgr~B2(N). In addition, transitions from 59 catalog entries corresponding to 
vibrationally or torsionally excited states of some of the identified 
molecules or their isotopologues were also identified. For Sgr~B2(M), the 
corresponding numbers are 46, 54, and 24, respectively.

The spectra observed toward Sgr~B2(N) in the 3, 2, and 1.3~mm atmospheric 
windows are shown in Figs.~\ref{f:survey_lmh_3mm}, \ref{f:survey_lmh_2mm}, and
\ref{f:survey_lmh_1mm}, respectively (online material). The corresponding 
spectra toward Sgr~B2(M) are shown in Figs.~\ref{f:survey_b2m_3mm}, 
\ref{f:survey_b2m_2mm}, and \ref{f:survey_b2m_1mm}, respectively (online
material). For each source, the best-fit 
synthetic model produced with XCLASS is overlaid in green, and the relevant 
molecular lines are marked with the name of the emitting or absorbing molecule 
in blue. Recombination lines cannot be modeled with XCLASS but some of them 
are clearly detected toward Sgr~B2(N) (several H\,$\alpha$, H\,$\beta$, and 
H\,$\gamma$ lines) and Sgr~B2(M) (several H\,$\alpha$, H\,$\beta$, 
H\,$\gamma$, and He\,$\alpha$ lines). They are labeled in pink in all figures.
Finally, because of the high dynamic range of our spectra ($\sim 2000$), 
strong lines from the image band can contaminate the signal band. We address 
this issue only for the 3~mm window. With a typical sideband rejection gain of 
$\sim 1$--3~$\%$ (see Sect.~\ref{ss:obs}), we estimate that only lines with a 
peak temperature higher than $\sim 2$~K in $T_{\rm mb}$ scale can 
significantly contaminate the spectra. We find 120 and 116 locations in the 
full 3~mm spectra of Sgr~B2(N) and (M), respectively, where contamination by
a strong line from the image band may occur, depending on the exact sideband 
rejection gain which is not precisely known. These locations are marked with 
red labels in Figs.~\ref{f:survey_lmh_3mm} to \ref{f:survey_b2m_1mm}.
Among these, an inspection by eye reveals only a few cases 
where the contamination is obvious: at 81.246 and 94.254~GHz by CH$_3$OH (image 
band frequencies: 84.521 and 97.583~GHz), at 112.111 and 112.124~GHz by CO 
(115.275 and 115.263~GHz) for Sgr~B2(N); at 81.246 GHz by CH$_3$OH 
(84.521~GHz), at 94.642 and 94.651~GHz by CS (97.984 and 97.975~GHz), at 
112.111 and 112.123~GHz by CO (115.275 and 115.263~GHz) for Sgr~B2(M).

\onlfig{\clearpage
\begin{figure*}
\centerline{\resizebox{1.0\hsize}{!}{\includegraphics[angle=270]{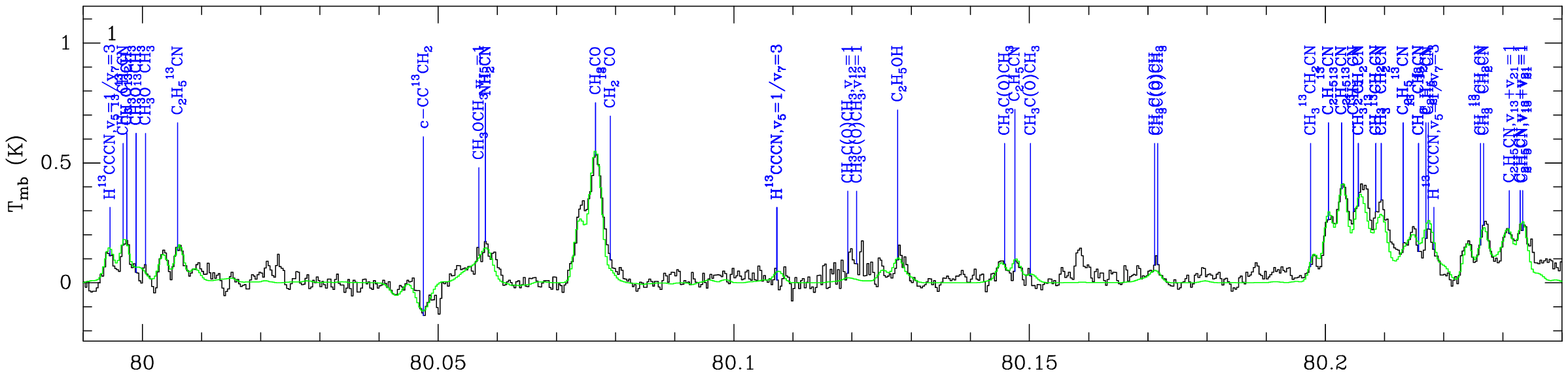}}}
\vspace*{1ex}\centerline{\resizebox{1.0\hsize}{!}{\includegraphics[angle=270]{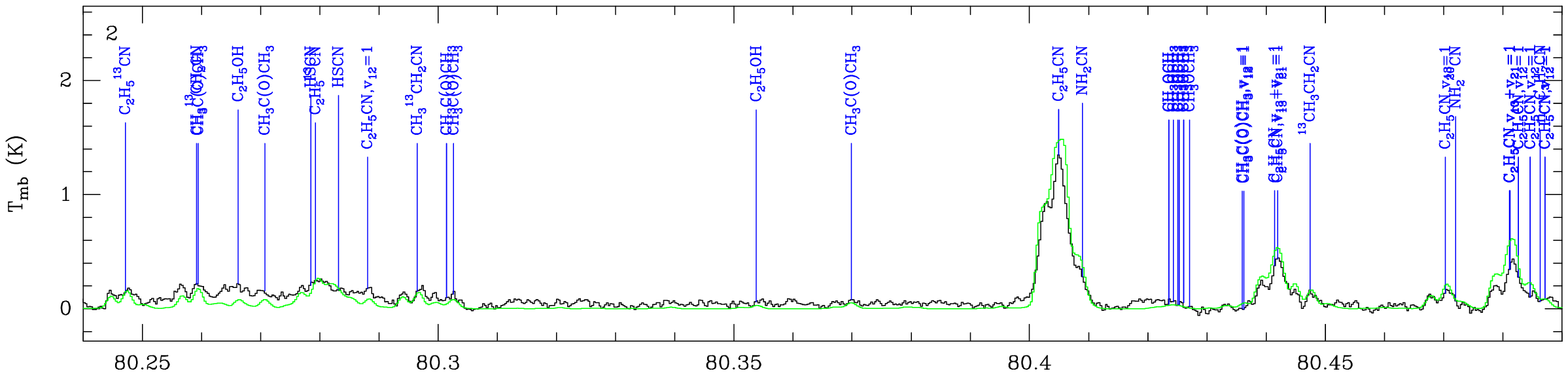}}}
\vspace*{1ex}\centerline{\resizebox{1.0\hsize}{!}{\includegraphics[angle=270]{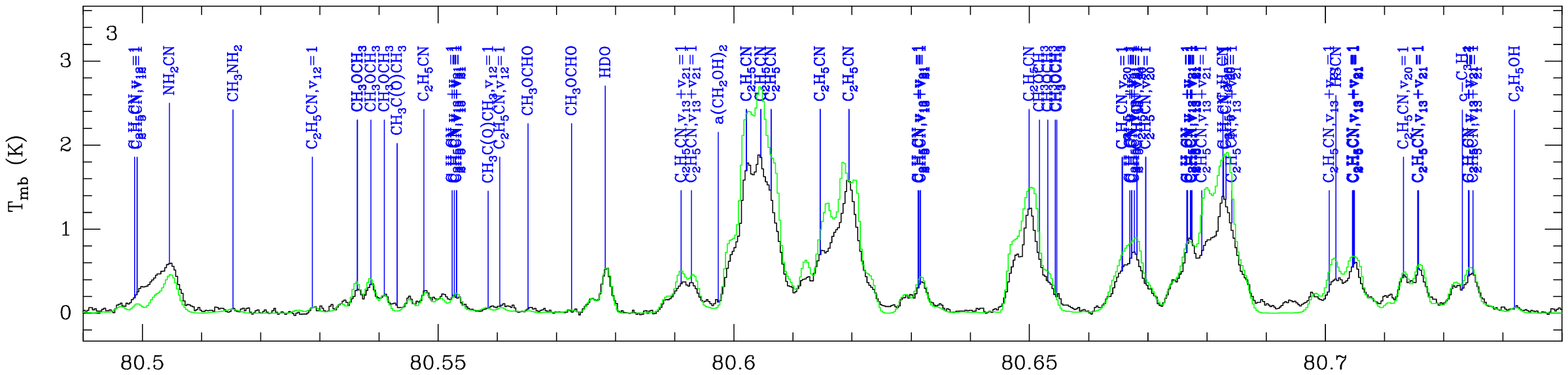}}}
\vspace*{1ex}\centerline{\resizebox{1.0\hsize}{!}{\includegraphics[angle=270]{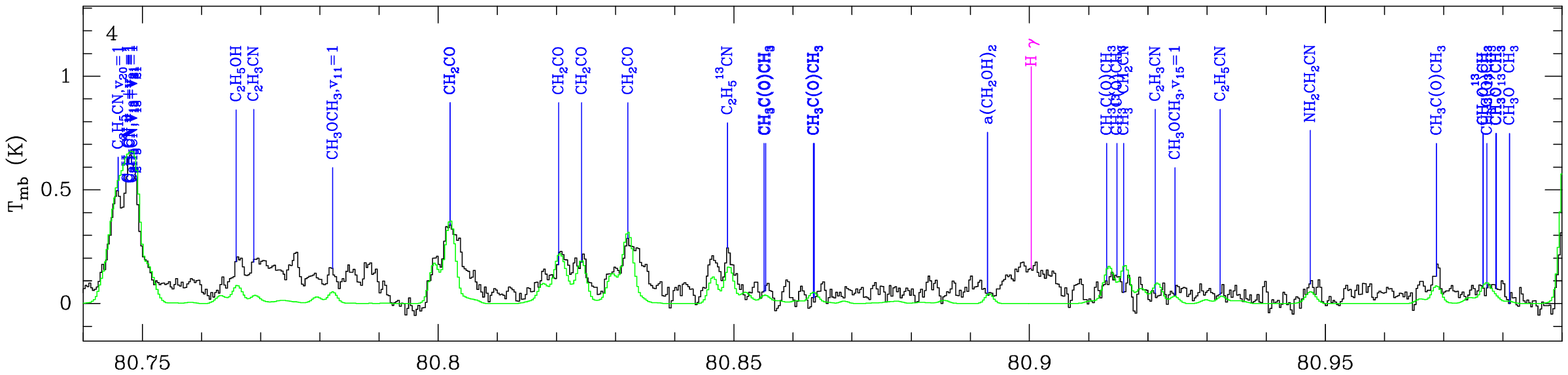}}}
\vspace*{1ex}\centerline{\resizebox{1.0\hsize}{!}{\includegraphics[angle=270]{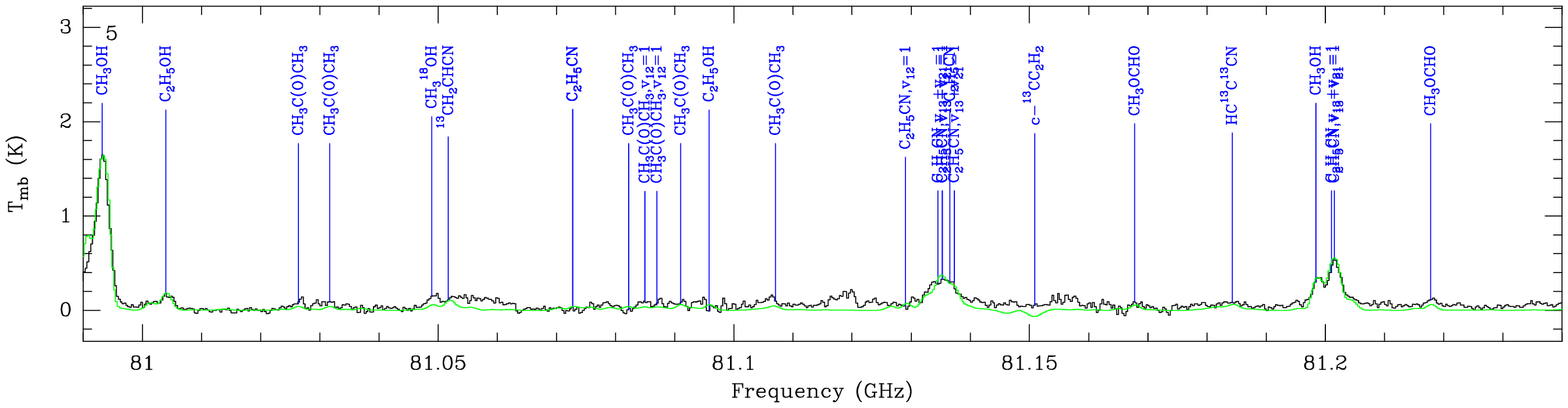}}}
\caption{
Spectrum obtained toward Sgr~B2(N)
in the 3~mm window
with the IRAM~30\,m telescope in main-beam temperature scale. The synthetic model is overlaid in green and its relevant lines are labeled in blue.
The frequencies of the hydrogen recombination lines are indicated with a pink label.
The position of the lines with a peak temperature higher than 2~K in the image band and possibly contaminating the spectrum are marked with a red label indicating their rest frequency and their peak temperature in K in the image band.
}
\label{f:survey_lmh_3mm}
\end{figure*}
 \clearpage
\begin{figure*}
\addtocounter{figure}{-1}
\centerline{\resizebox{1.0\hsize}{!}{\includegraphics[angle=270]{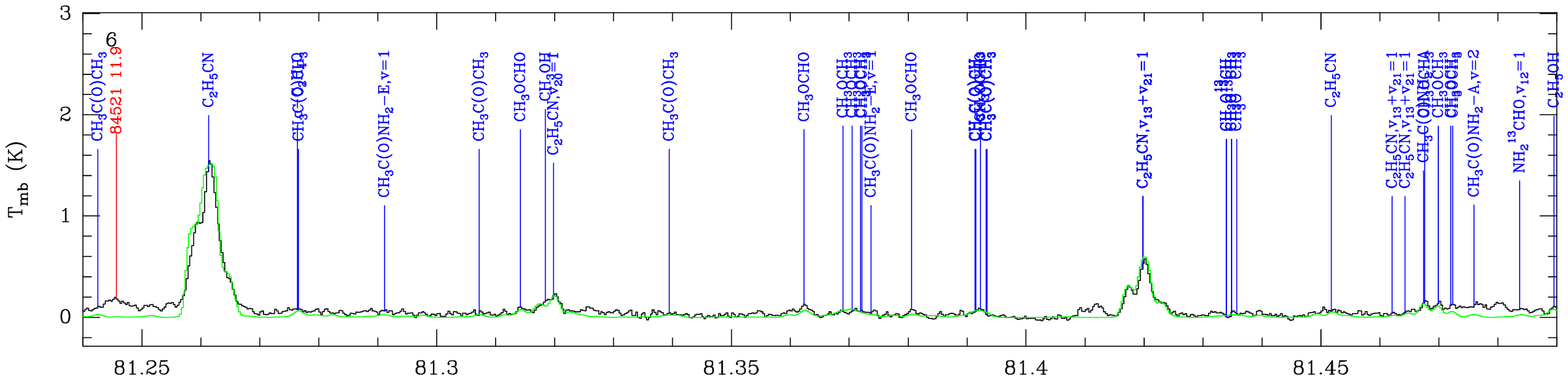}}}
\vspace*{1ex}\centerline{\resizebox{1.0\hsize}{!}{\includegraphics[angle=270]{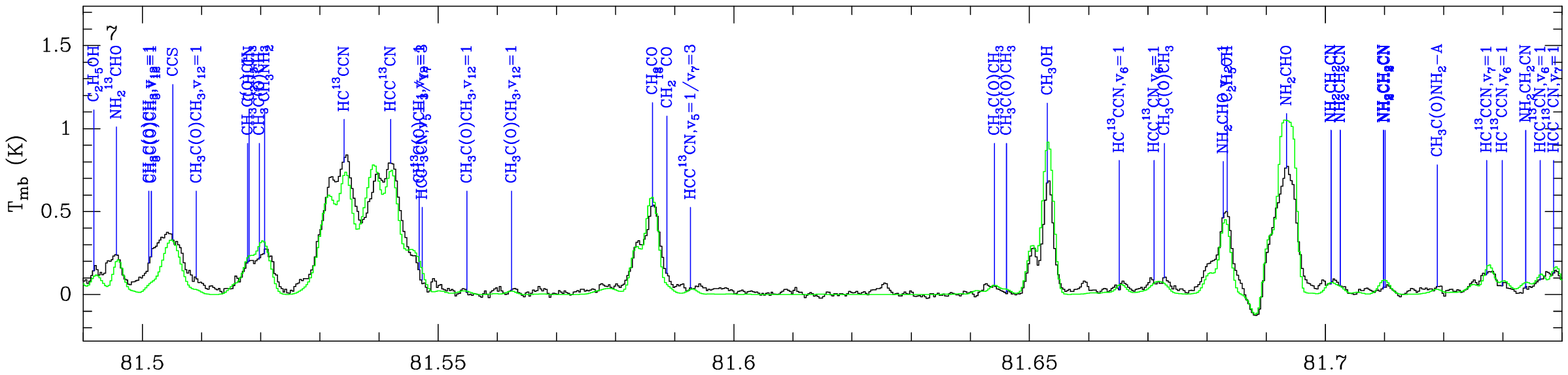}}}
\vspace*{1ex}\centerline{\resizebox{1.0\hsize}{!}{\includegraphics[angle=270]{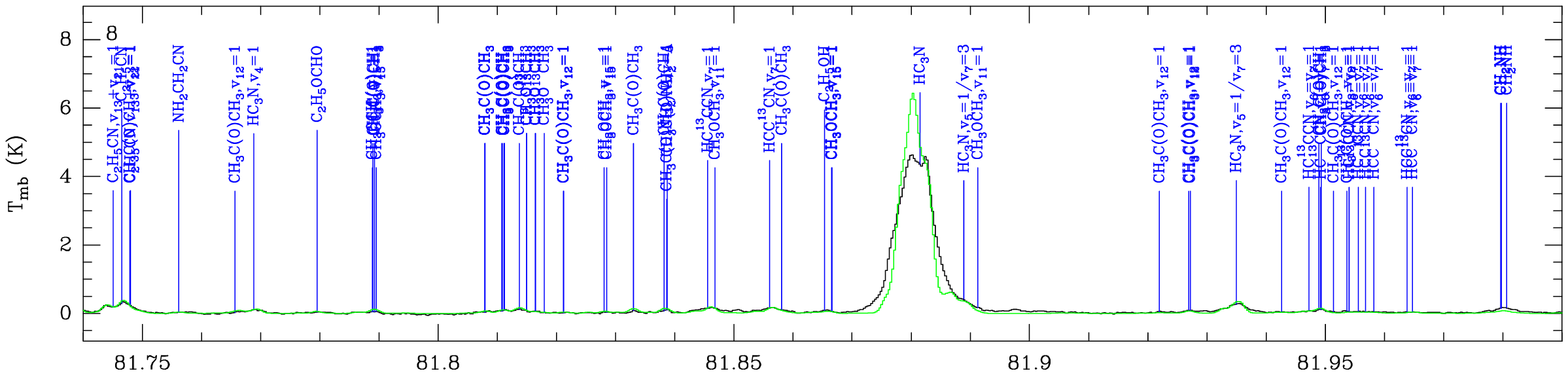}}}
\vspace*{1ex}\centerline{\resizebox{1.0\hsize}{!}{\includegraphics[angle=270]{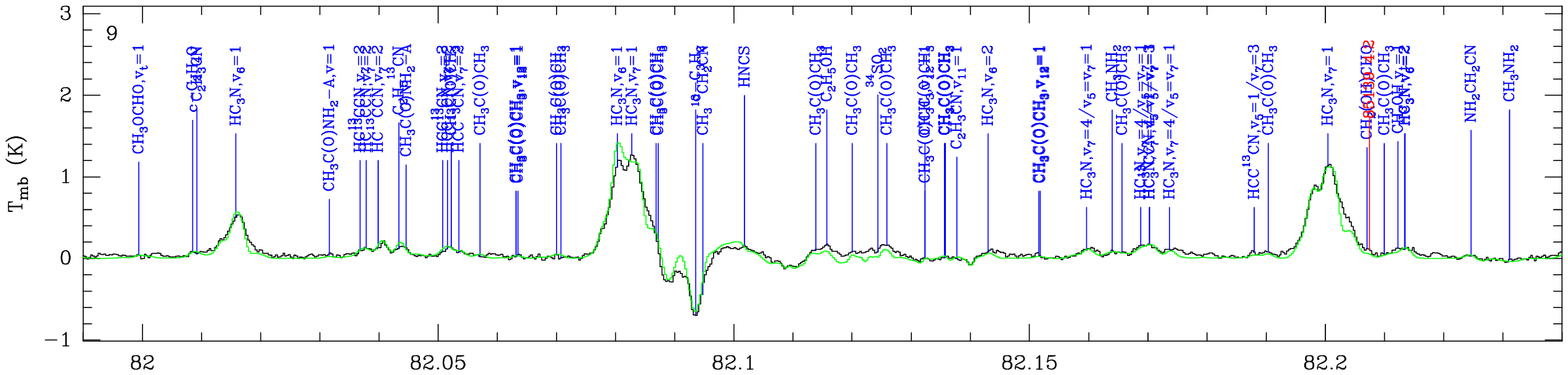}}}
\vspace*{1ex}\centerline{\resizebox{1.0\hsize}{!}{\includegraphics[angle=270]{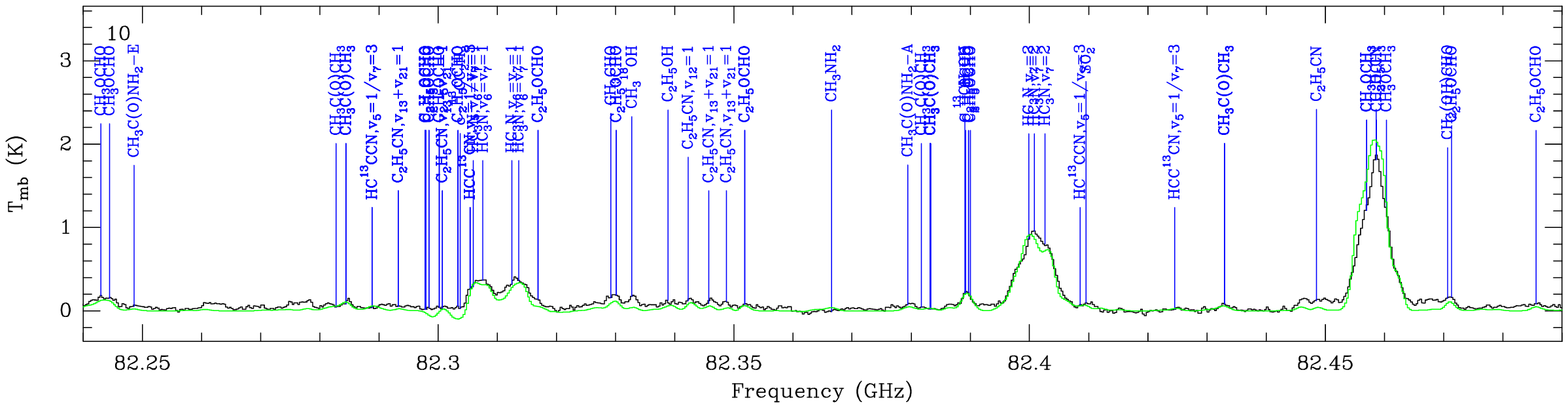}}}
\caption{
continued.
}
\end{figure*}
 \clearpage
\begin{figure*}
\addtocounter{figure}{-1}
\centerline{\resizebox{1.0\hsize}{!}{\includegraphics[angle=270]{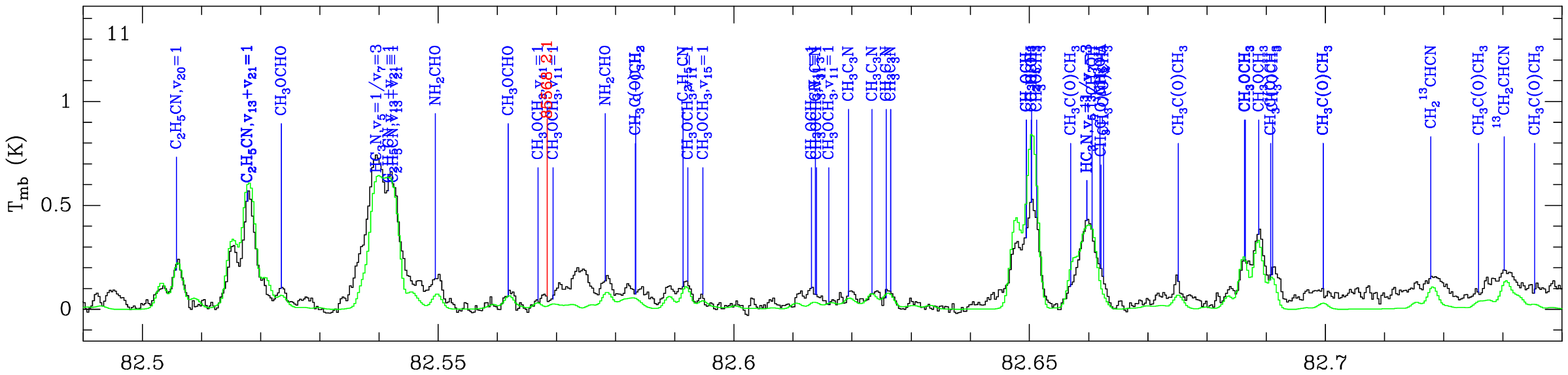}}}
\vspace*{1ex}\centerline{\resizebox{1.0\hsize}{!}{\includegraphics[angle=270]{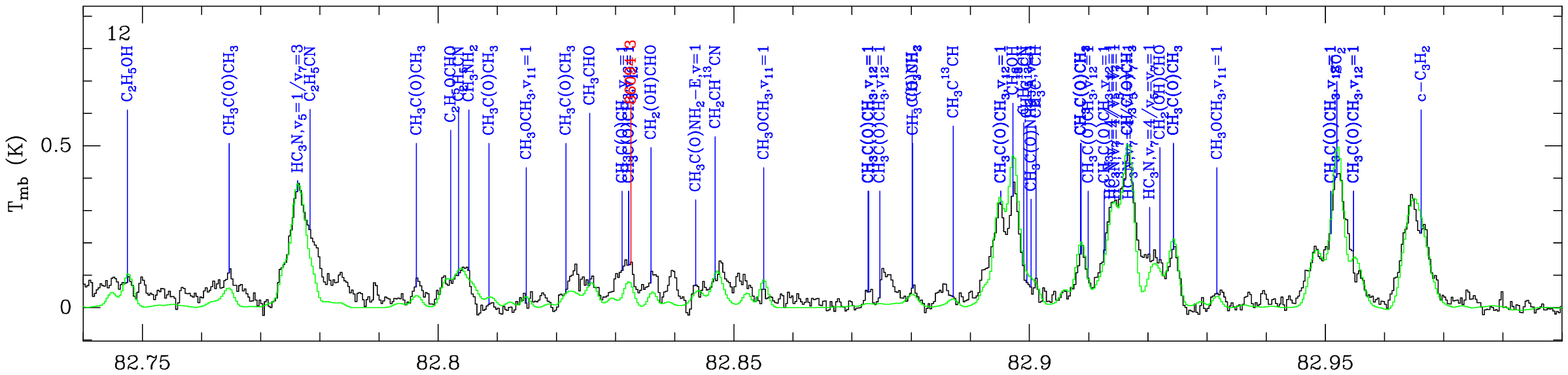}}}
\vspace*{1ex}\centerline{\resizebox{1.0\hsize}{!}{\includegraphics[angle=270]{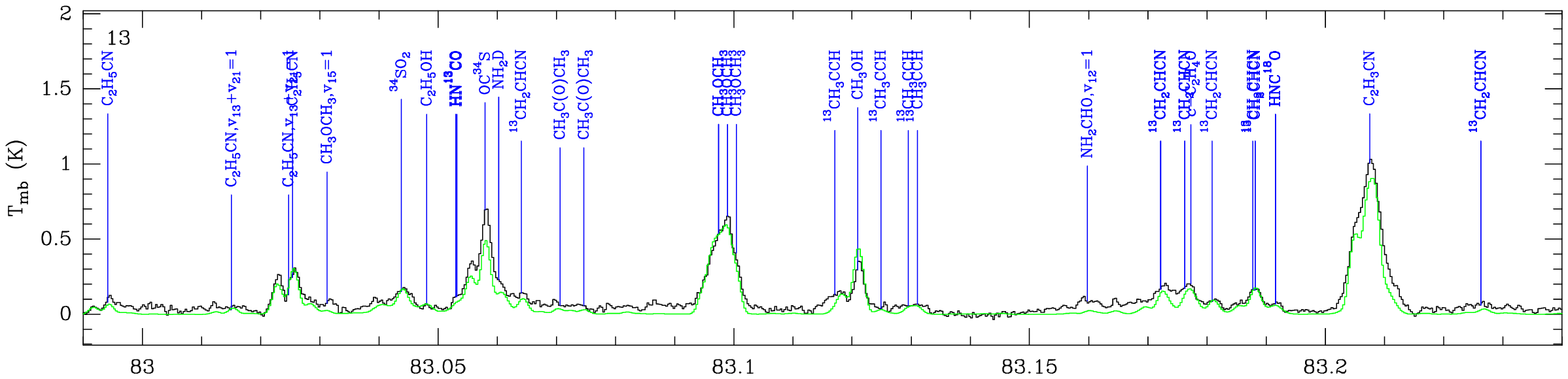}}}
\vspace*{1ex}\centerline{\resizebox{1.0\hsize}{!}{\includegraphics[angle=270]{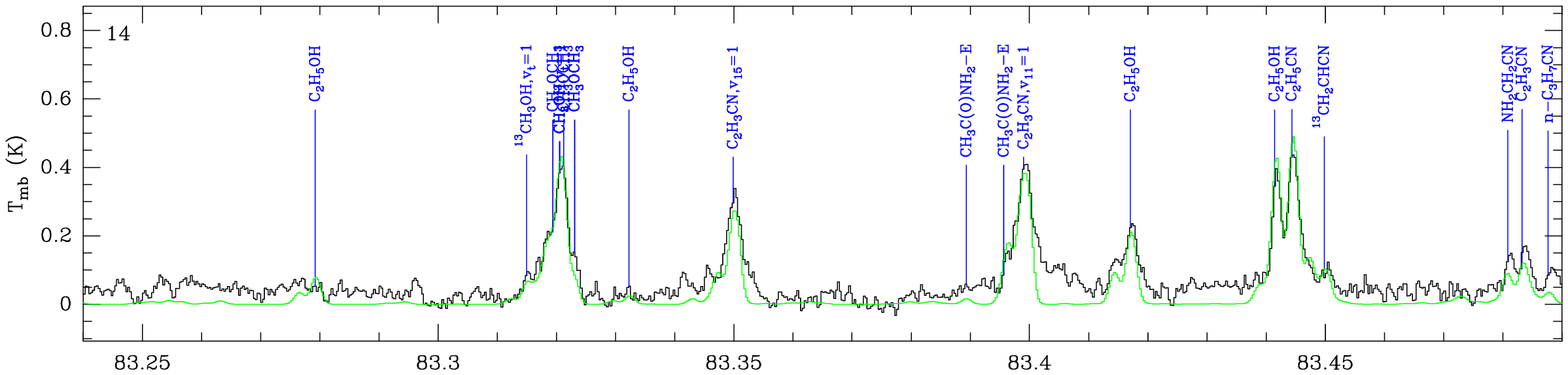}}}
\vspace*{1ex}\centerline{\resizebox{1.0\hsize}{!}{\includegraphics[angle=270]{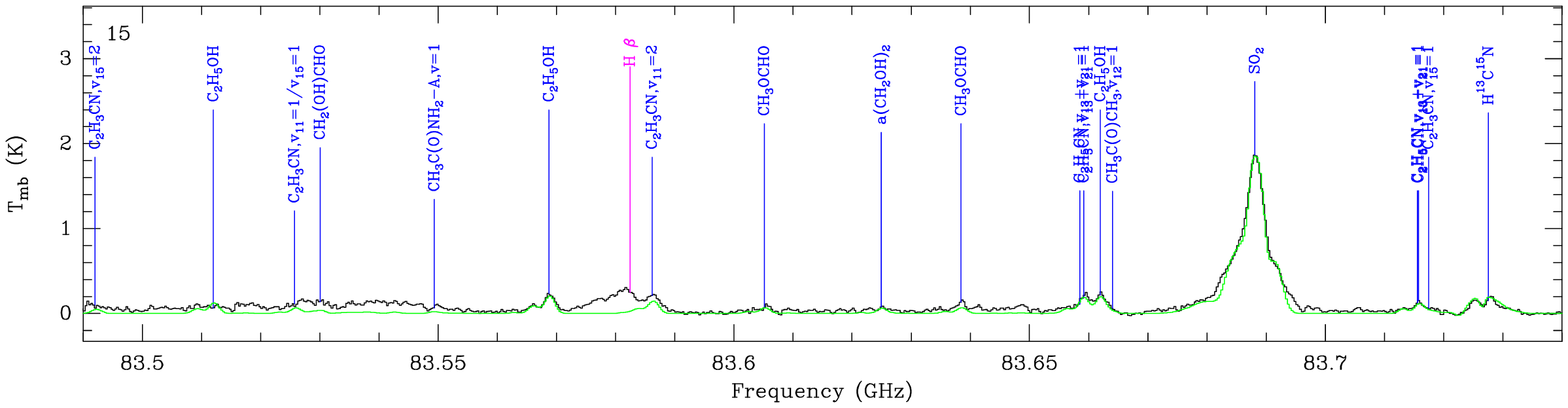}}}
\caption{
continued.
}
\end{figure*}
 \clearpage
\begin{figure*}
\addtocounter{figure}{-1}
\centerline{\resizebox{1.0\hsize}{!}{\includegraphics[angle=270]{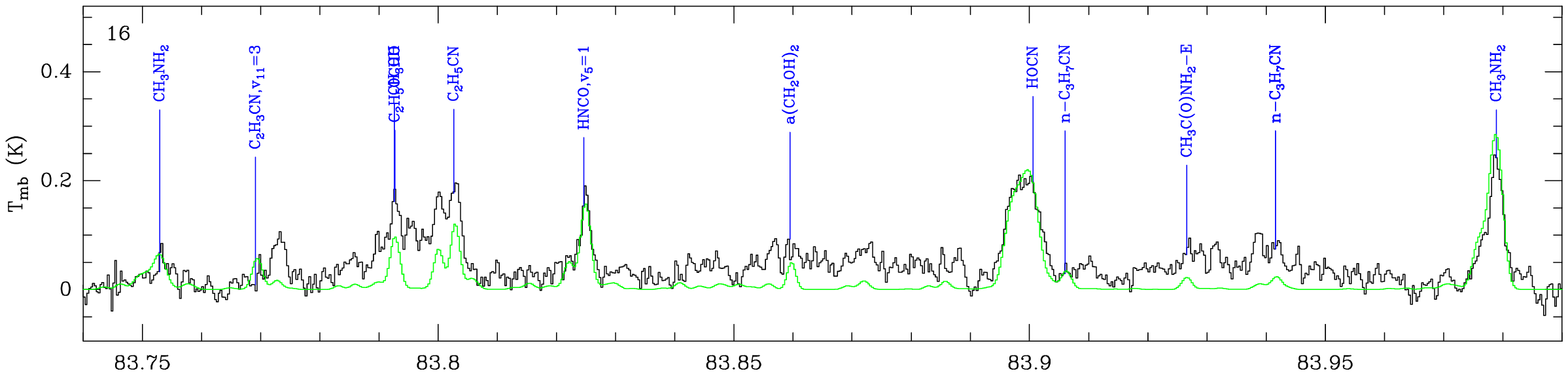}}}
\vspace*{1ex}\centerline{\resizebox{1.0\hsize}{!}{\includegraphics[angle=270]{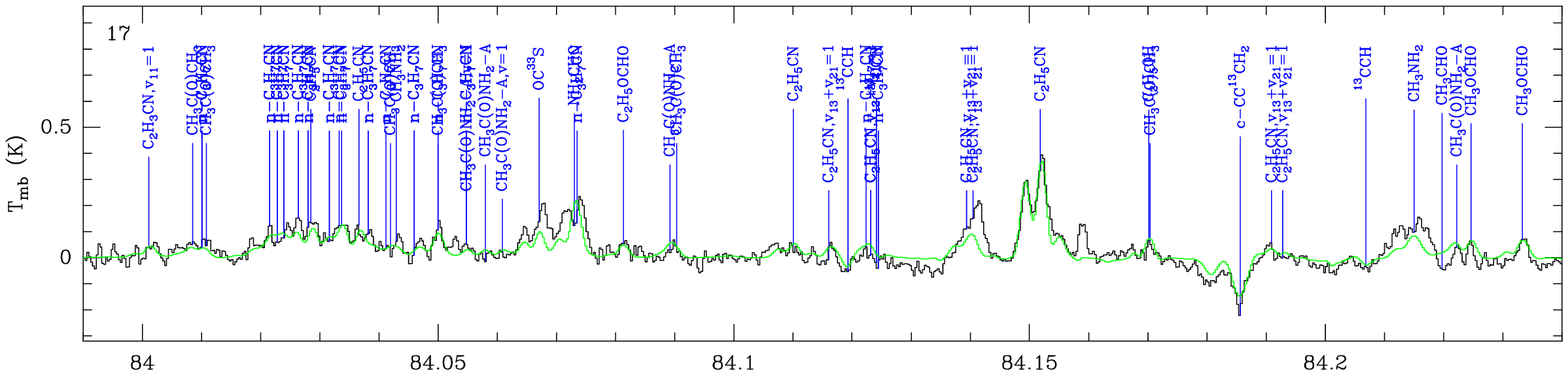}}}
\vspace*{1ex}\centerline{\resizebox{1.0\hsize}{!}{\includegraphics[angle=270]{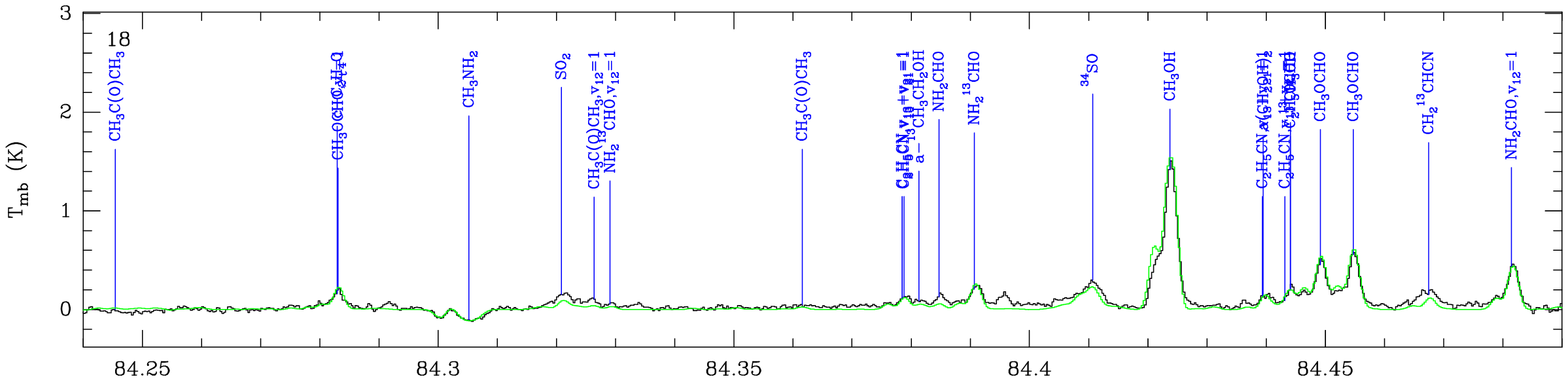}}}
\vspace*{1ex}\centerline{\resizebox{1.0\hsize}{!}{\includegraphics[angle=270]{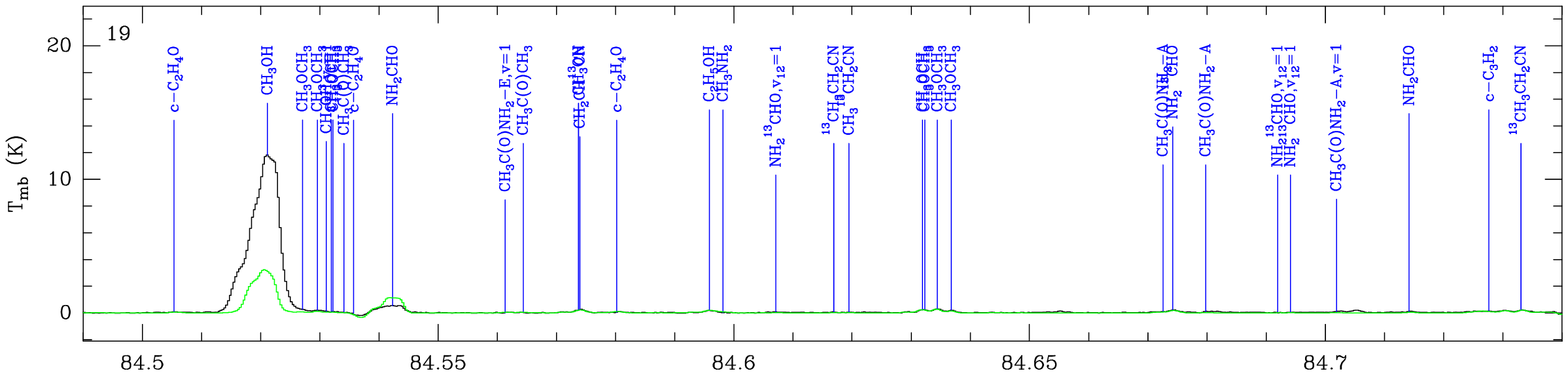}}}
\vspace*{1ex}\centerline{\resizebox{1.0\hsize}{!}{\includegraphics[angle=270]{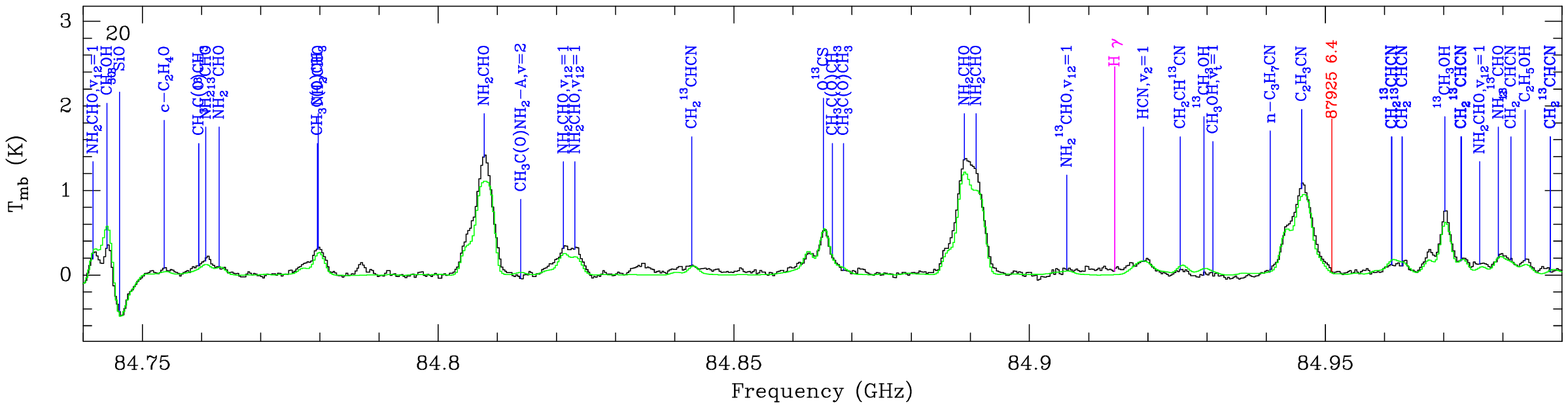}}}
\caption{
continued.
}
\end{figure*}
 \clearpage
\begin{figure*}
\addtocounter{figure}{-1}
\centerline{\resizebox{1.0\hsize}{!}{\includegraphics[angle=270]{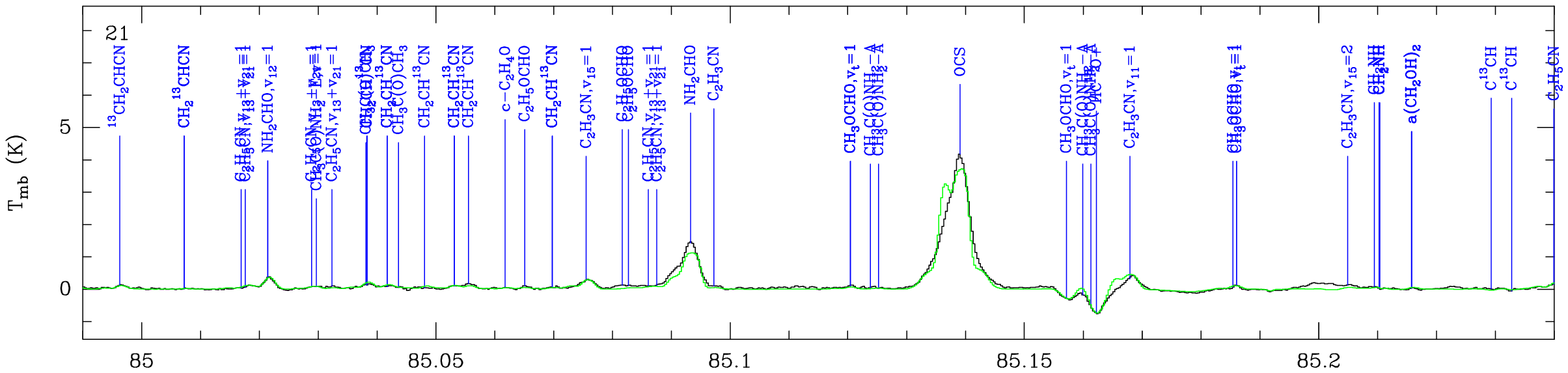}}}
\vspace*{1ex}\centerline{\resizebox{1.0\hsize}{!}{\includegraphics[angle=270]{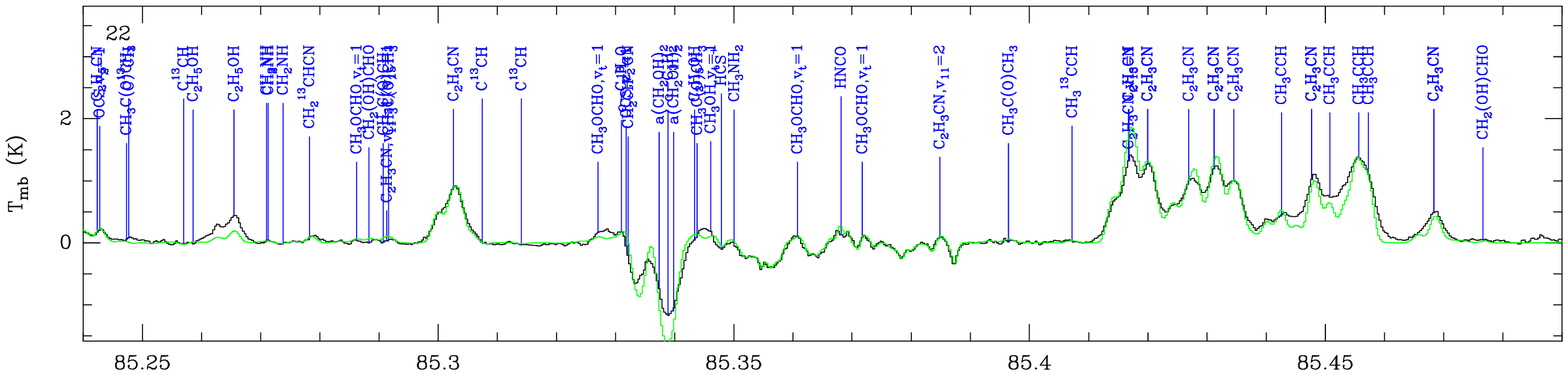}}}
\vspace*{1ex}\centerline{\resizebox{1.0\hsize}{!}{\includegraphics[angle=270]{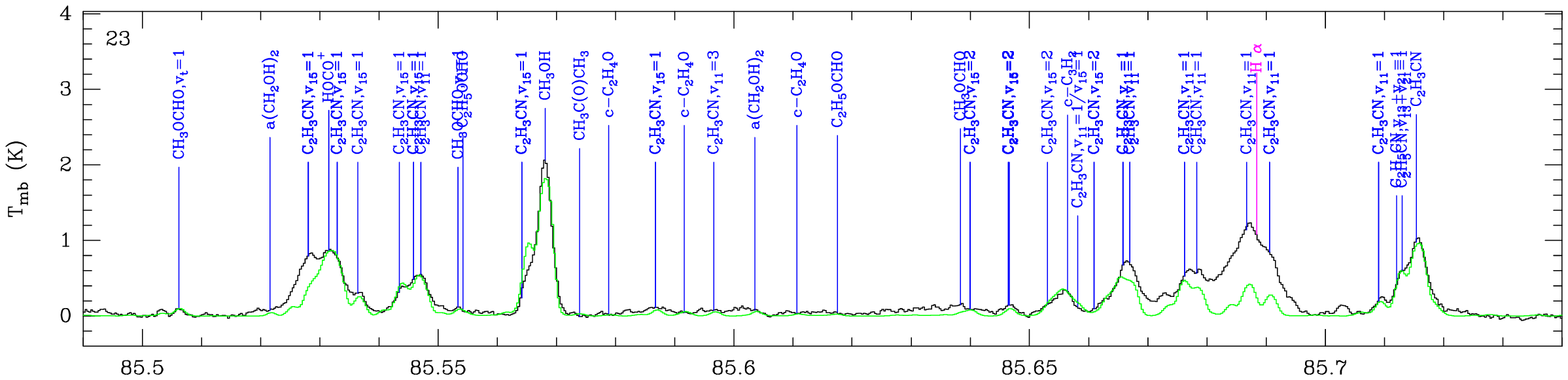}}}
\vspace*{1ex}\centerline{\resizebox{1.0\hsize}{!}{\includegraphics[angle=270]{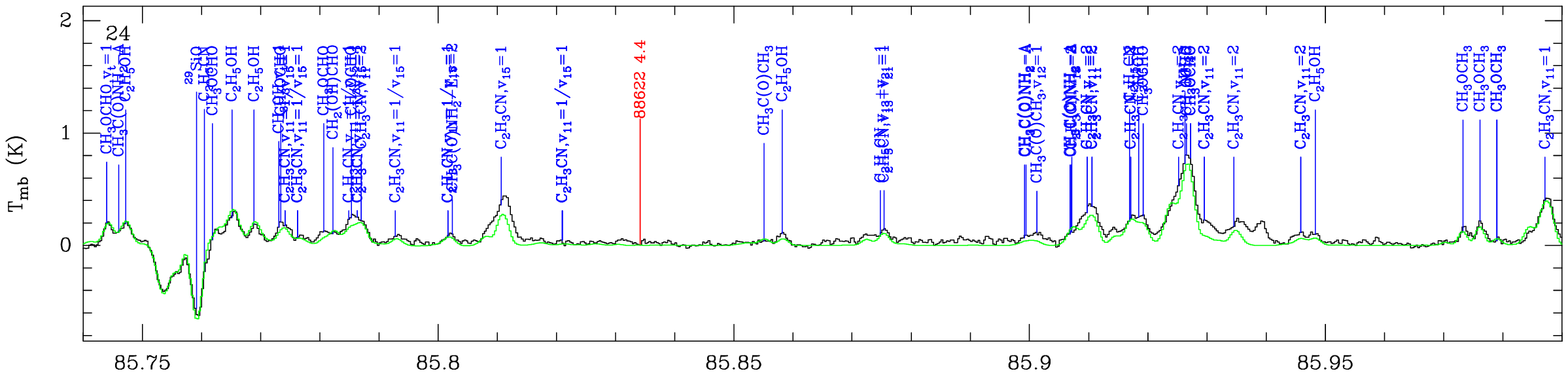}}}
\vspace*{1ex}\centerline{\resizebox{1.0\hsize}{!}{\includegraphics[angle=270]{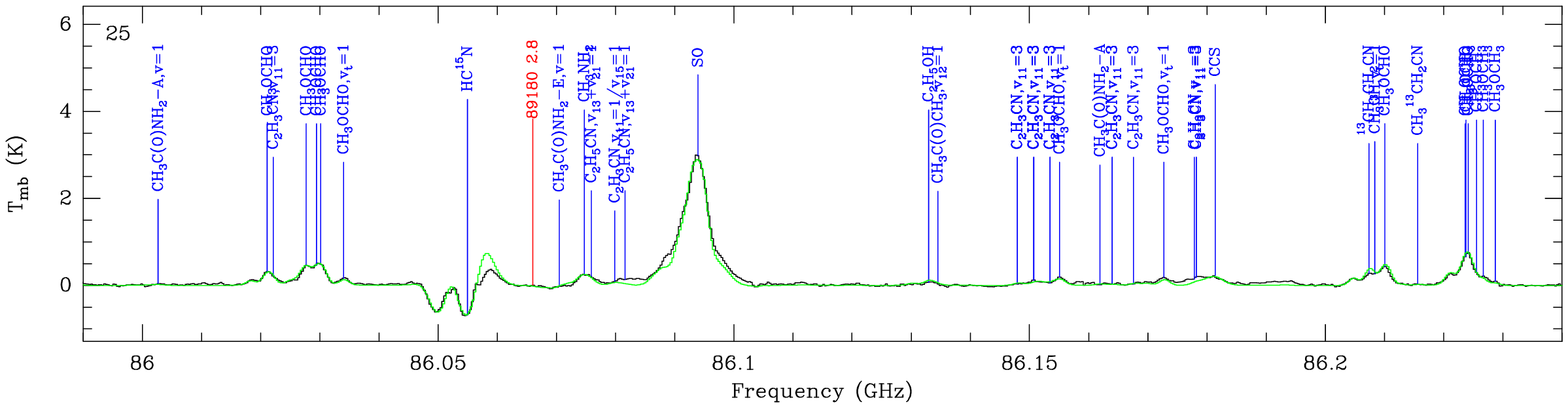}}}
\caption{
continued.
}
\end{figure*}
 \clearpage
\begin{figure*}
\addtocounter{figure}{-1}
\centerline{\resizebox{1.0\hsize}{!}{\includegraphics[angle=270]{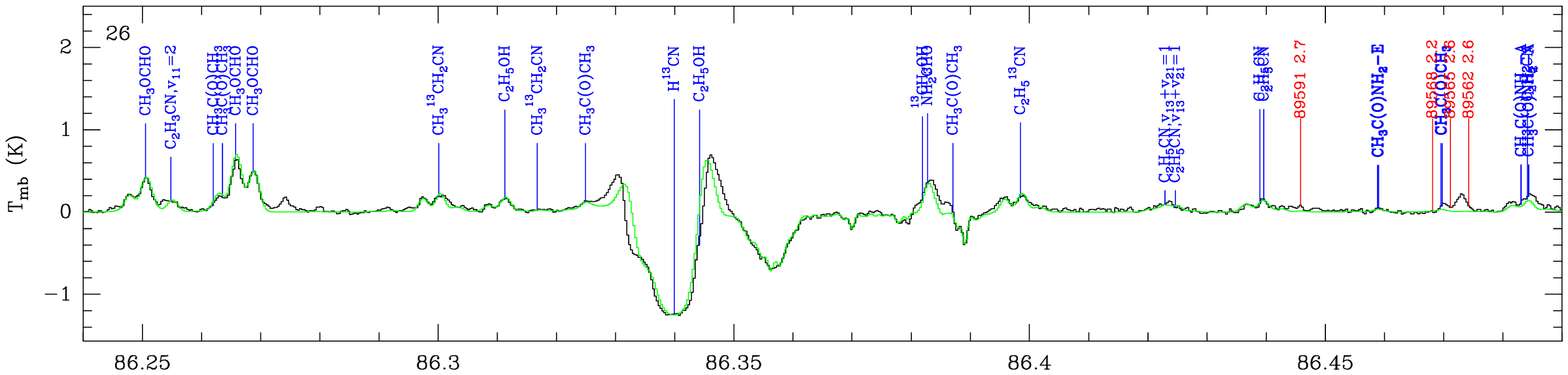}}}
\vspace*{1ex}\centerline{\resizebox{1.0\hsize}{!}{\includegraphics[angle=270]{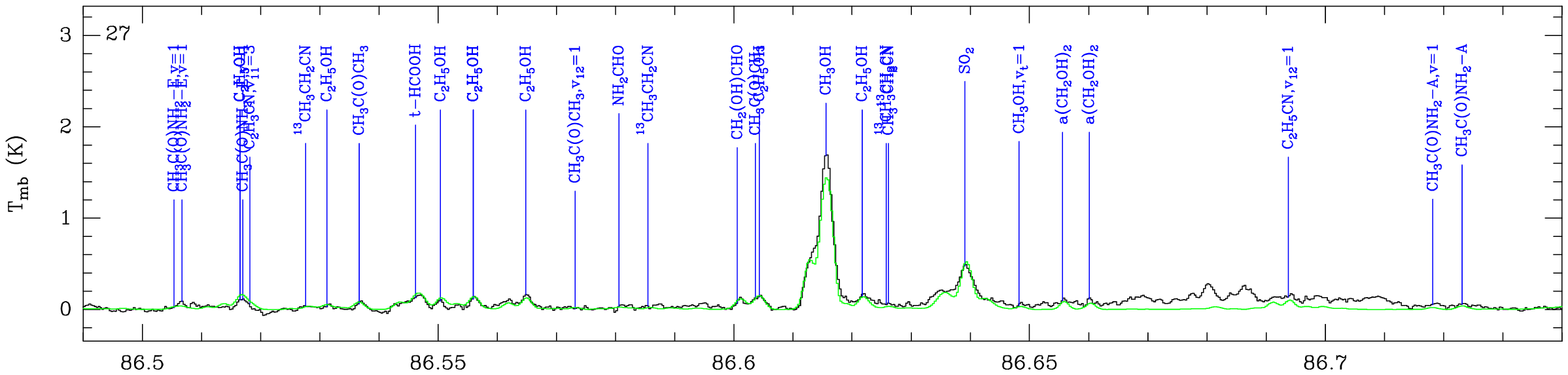}}}
\vspace*{1ex}\centerline{\resizebox{1.0\hsize}{!}{\includegraphics[angle=270]{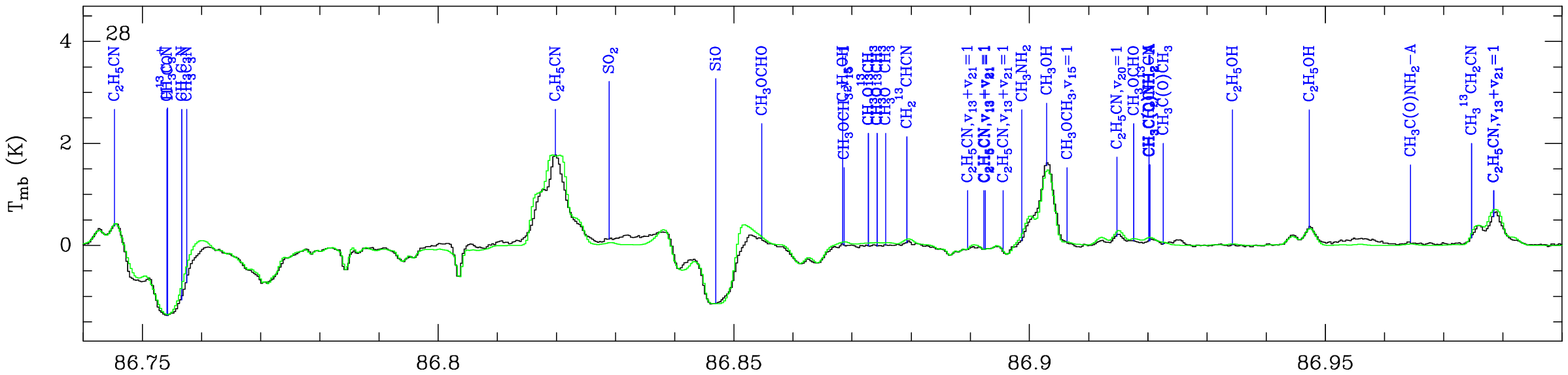}}}
\vspace*{1ex}\centerline{\resizebox{1.0\hsize}{!}{\includegraphics[angle=270]{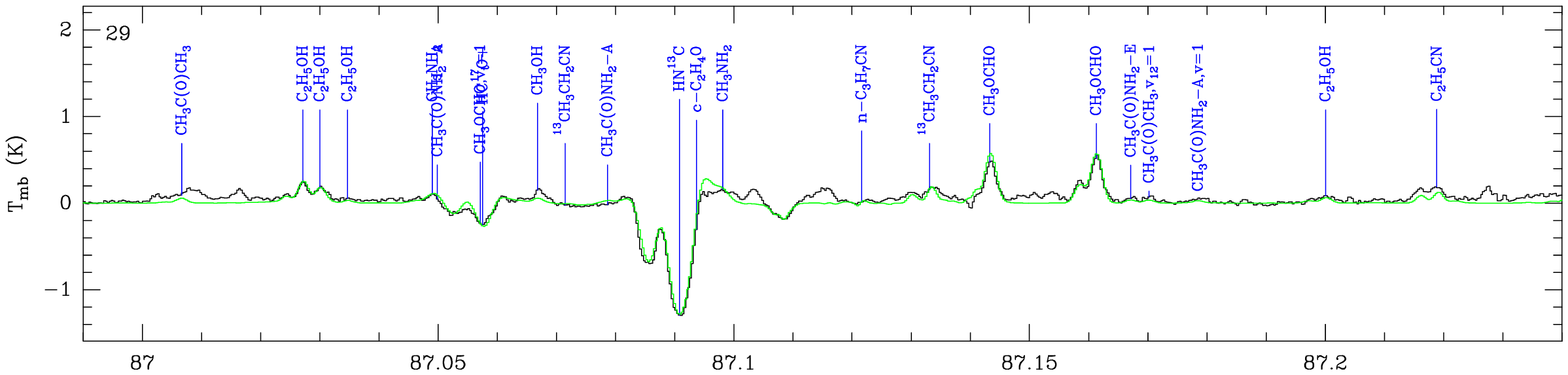}}}
\vspace*{1ex}\centerline{\resizebox{1.0\hsize}{!}{\includegraphics[angle=270]{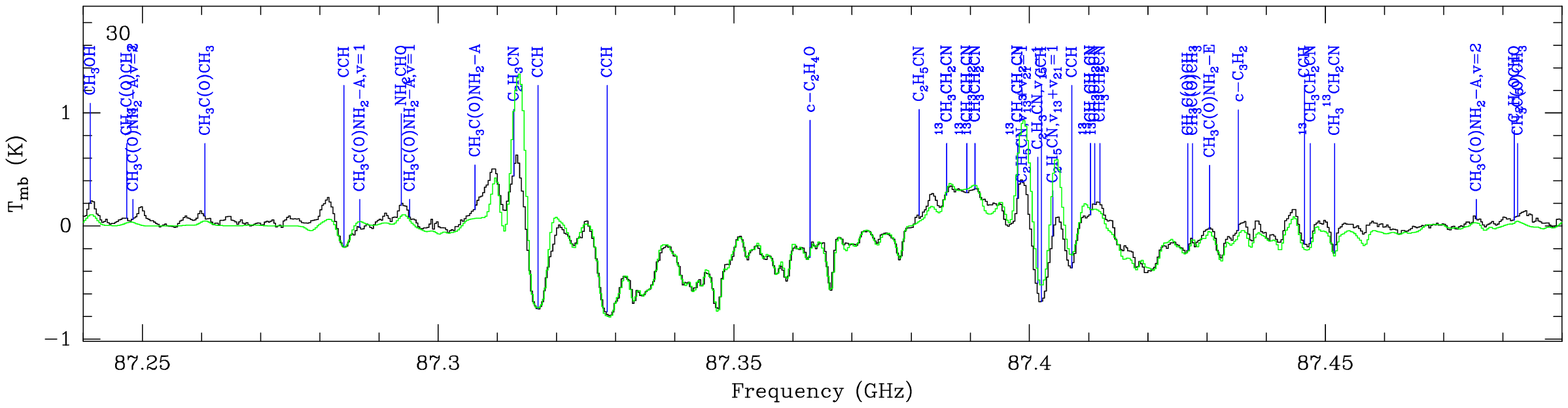}}}
\caption{
continued.
}
\end{figure*}
 \clearpage
\begin{figure*}
\addtocounter{figure}{-1}
\centerline{\resizebox{1.0\hsize}{!}{\includegraphics[angle=270]{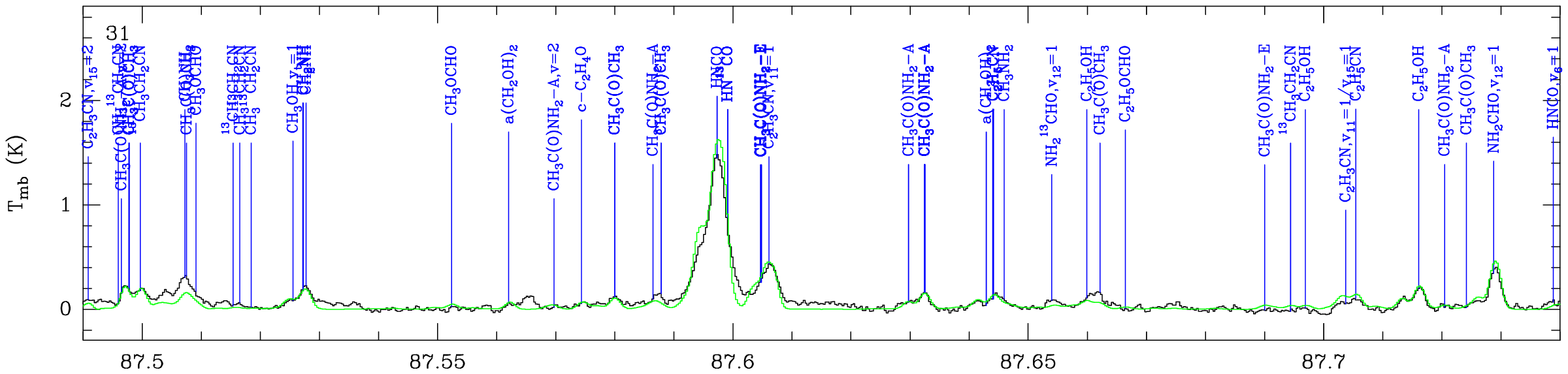}}}
\vspace*{1ex}\centerline{\resizebox{1.0\hsize}{!}{\includegraphics[angle=270]{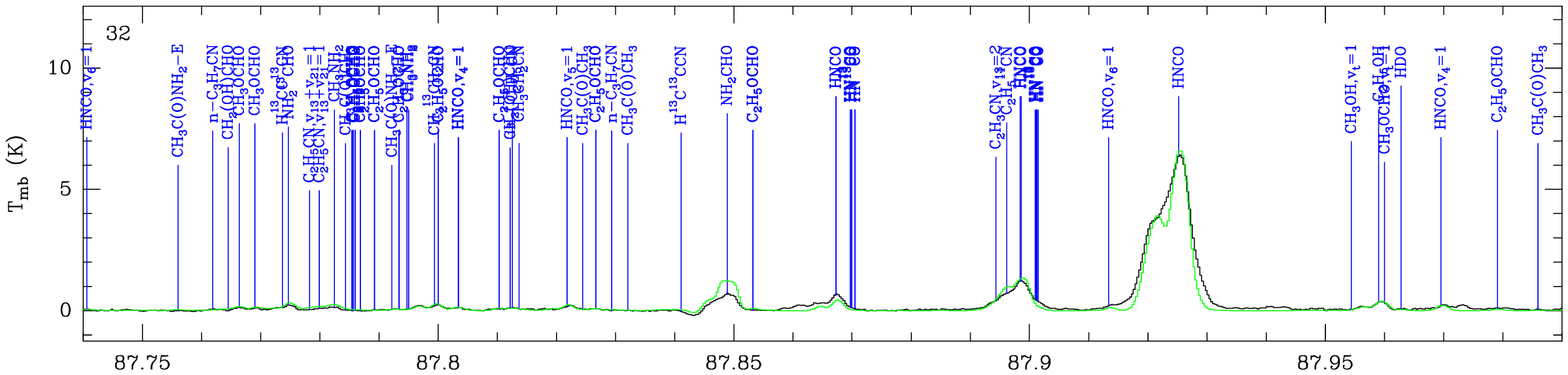}}}
\vspace*{1ex}\centerline{\resizebox{1.0\hsize}{!}{\includegraphics[angle=270]{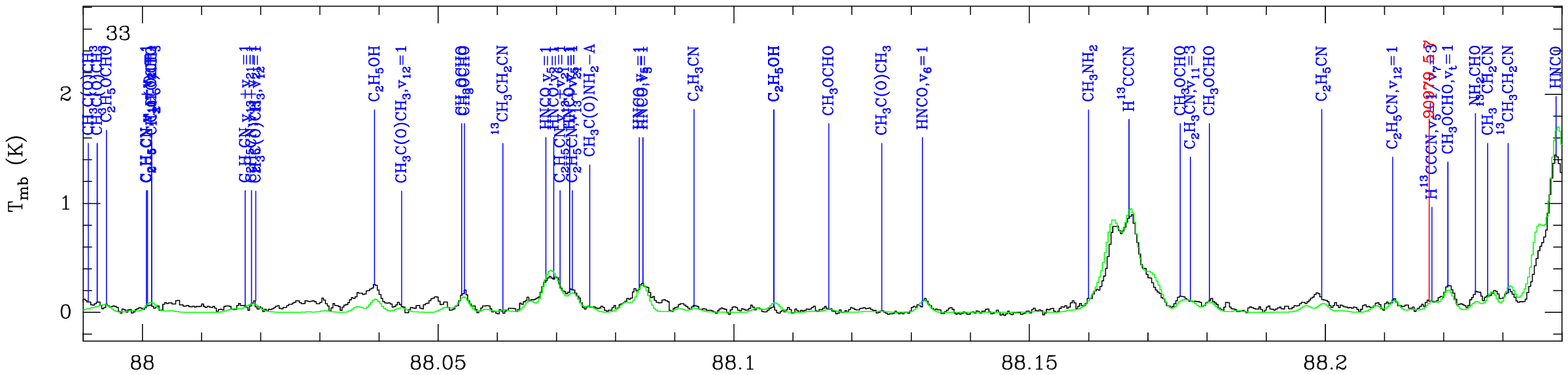}}}
\vspace*{1ex}\centerline{\resizebox{1.0\hsize}{!}{\includegraphics[angle=270]{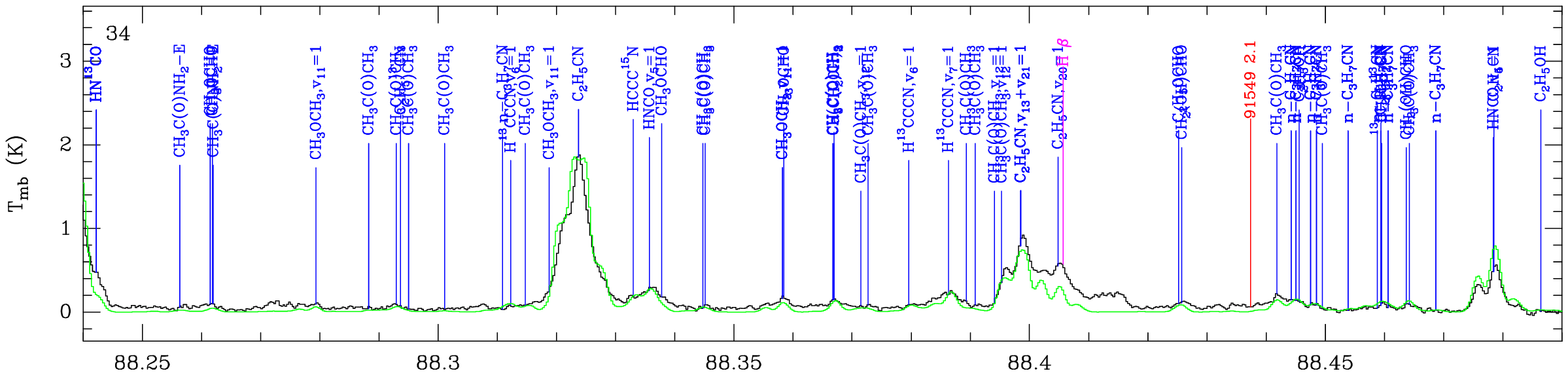}}}
\vspace*{1ex}\centerline{\resizebox{1.0\hsize}{!}{\includegraphics[angle=270]{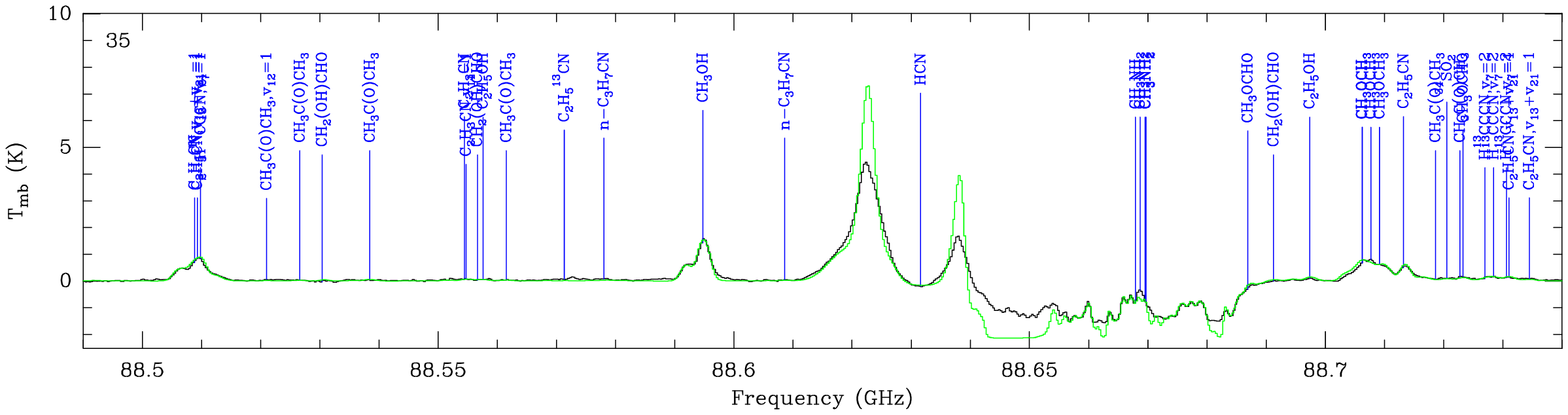}}}
\caption{
continued.
}
\end{figure*}
 \clearpage
\begin{figure*}
\addtocounter{figure}{-1}
\centerline{\resizebox{1.0\hsize}{!}{\includegraphics[angle=270]{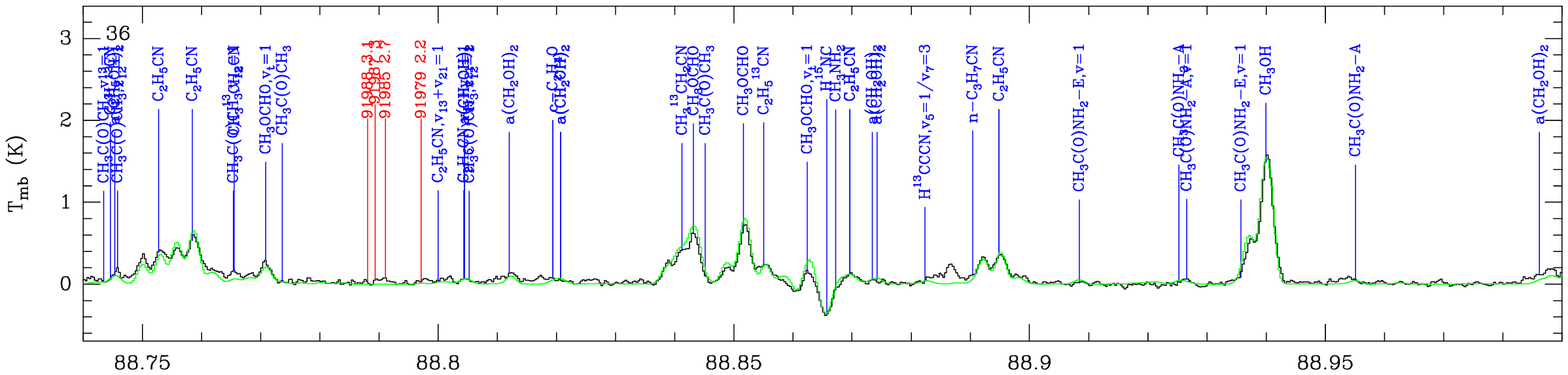}}}
\vspace*{1ex}\centerline{\resizebox{1.0\hsize}{!}{\includegraphics[angle=270]{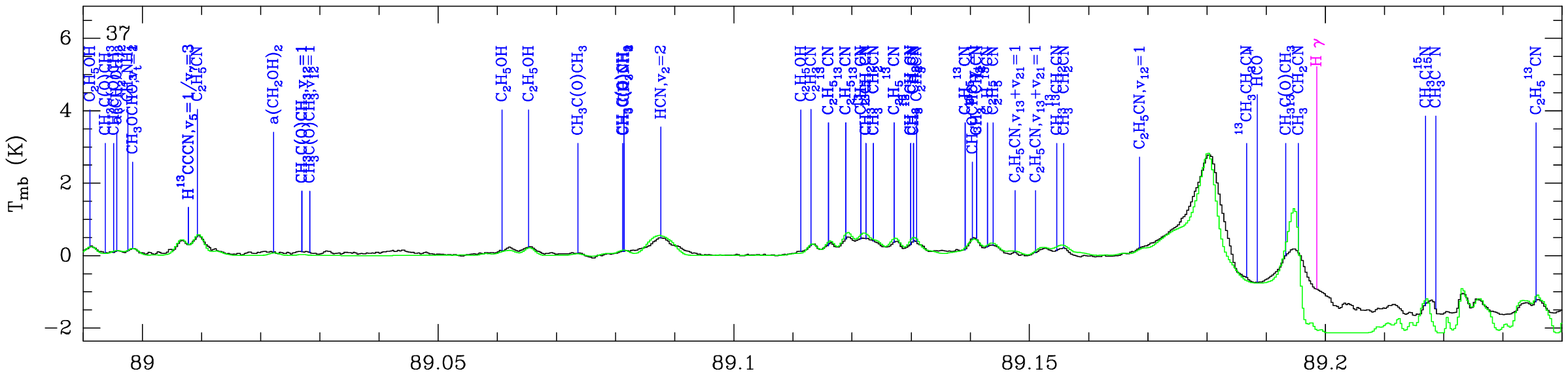}}}
\vspace*{1ex}\centerline{\resizebox{1.0\hsize}{!}{\includegraphics[angle=270]{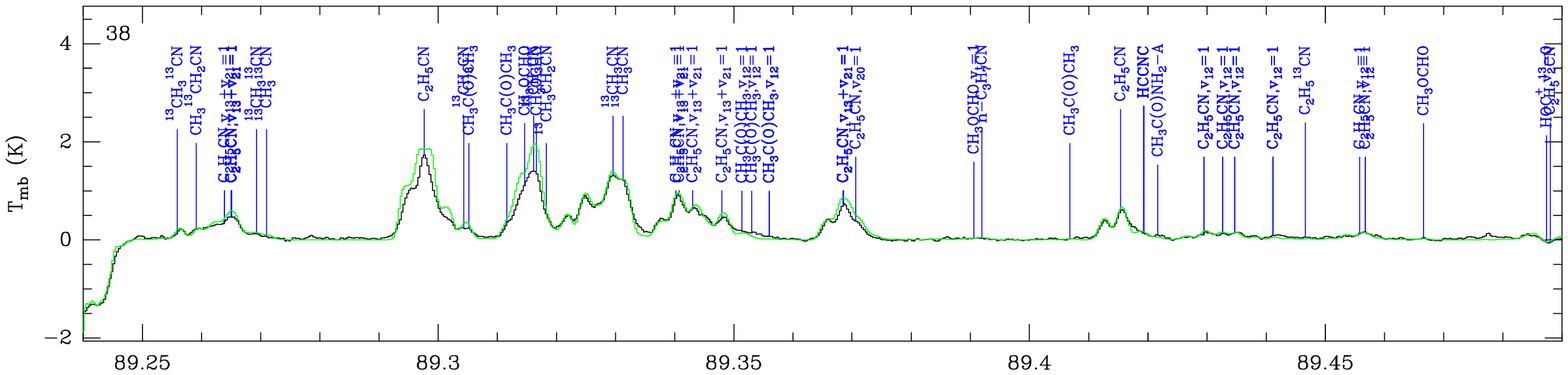}}}
\vspace*{1ex}\centerline{\resizebox{1.0\hsize}{!}{\includegraphics[angle=270]{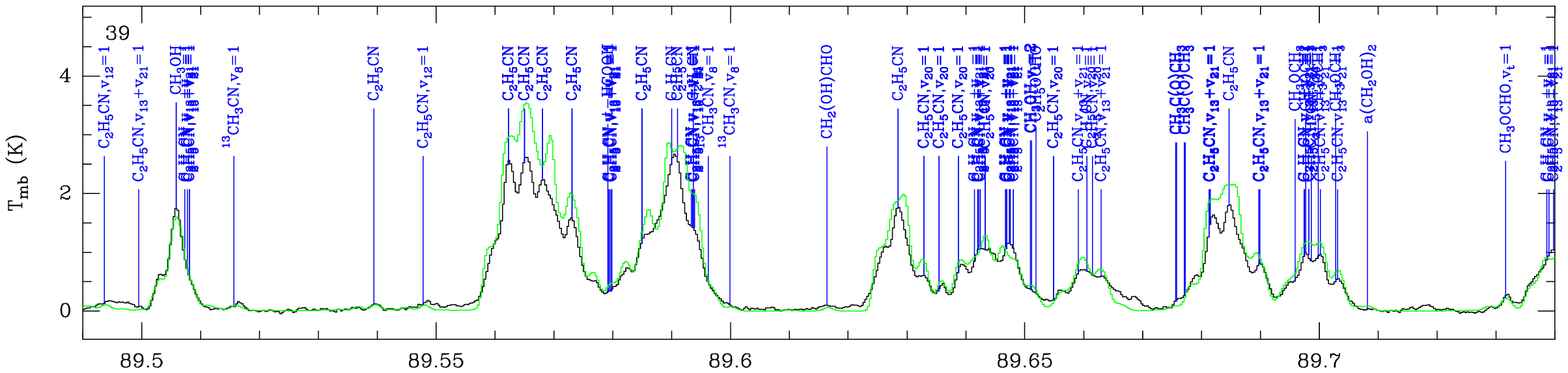}}}
\vspace*{1ex}\centerline{\resizebox{1.0\hsize}{!}{\includegraphics[angle=270]{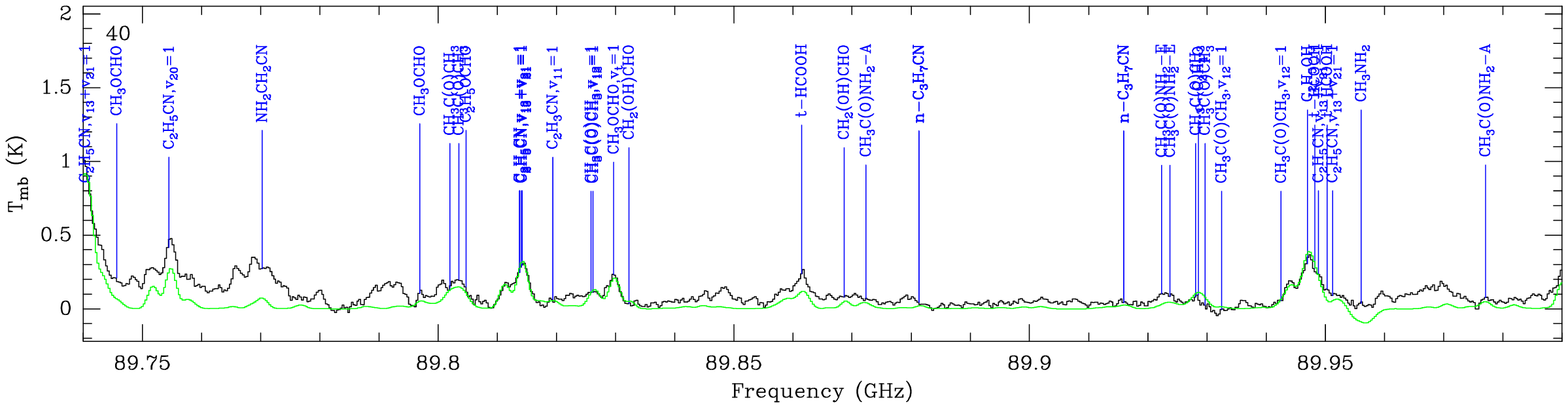}}}
\caption{
continued.
}
\end{figure*}
 \clearpage
\begin{figure*}
\addtocounter{figure}{-1}
\centerline{\resizebox{1.0\hsize}{!}{\includegraphics[angle=270]{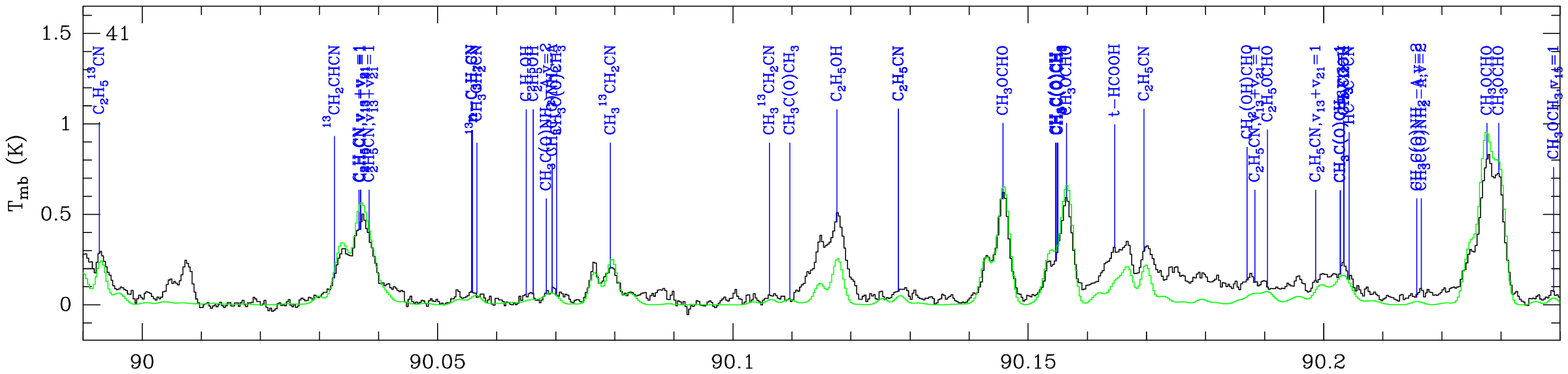}}}
\vspace*{1ex}\centerline{\resizebox{1.0\hsize}{!}{\includegraphics[angle=270]{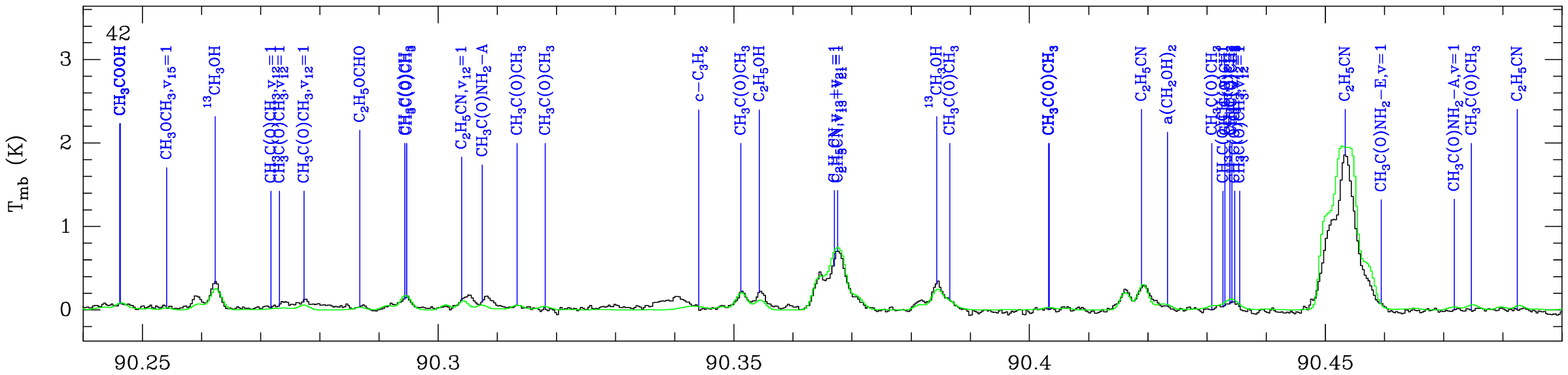}}}
\vspace*{1ex}\centerline{\resizebox{1.0\hsize}{!}{\includegraphics[angle=270]{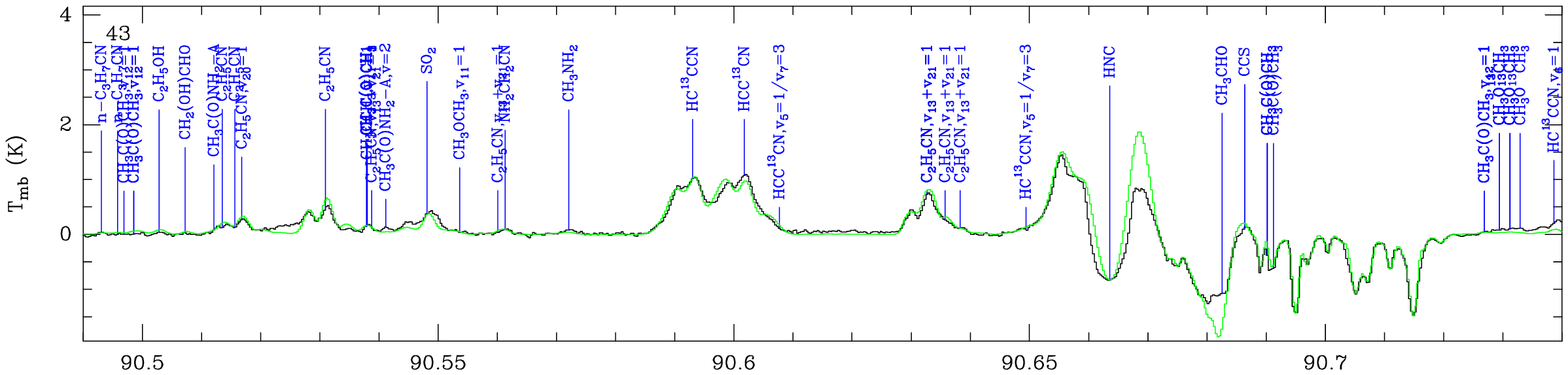}}}
\vspace*{1ex}\centerline{\resizebox{1.0\hsize}{!}{\includegraphics[angle=270]{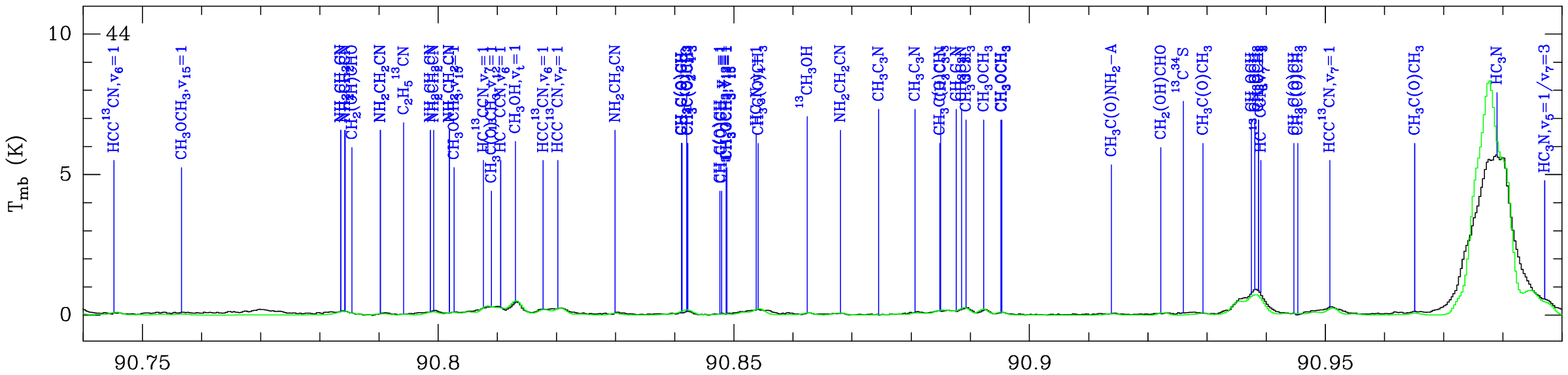}}}
\vspace*{1ex}\centerline{\resizebox{1.0\hsize}{!}{\includegraphics[angle=270]{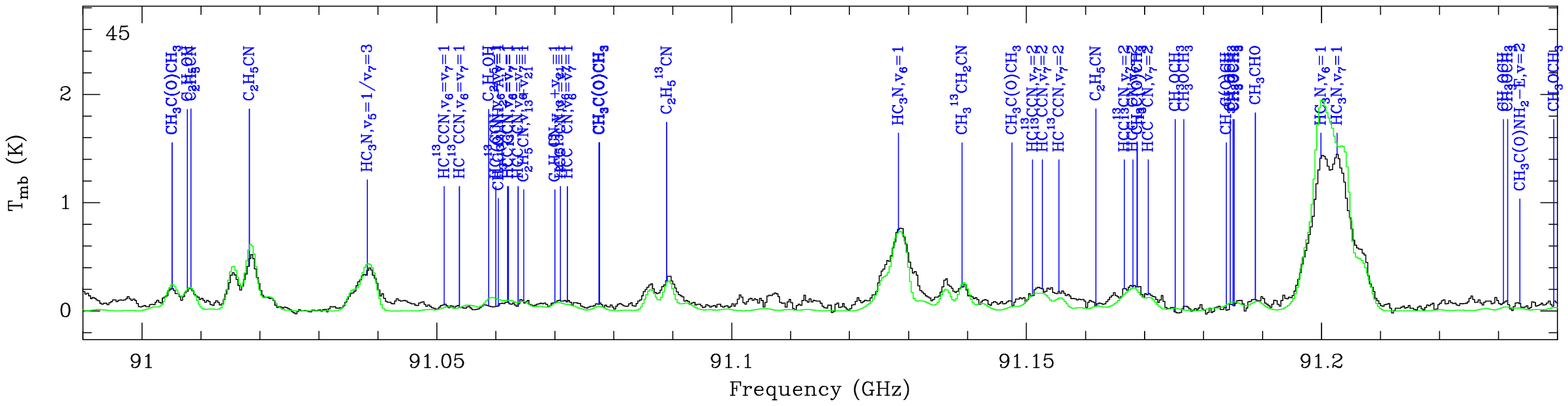}}}
\caption{
continued.
}
\end{figure*}
 \clearpage
\begin{figure*}
\addtocounter{figure}{-1}
\centerline{\resizebox{1.0\hsize}{!}{\includegraphics[angle=270]{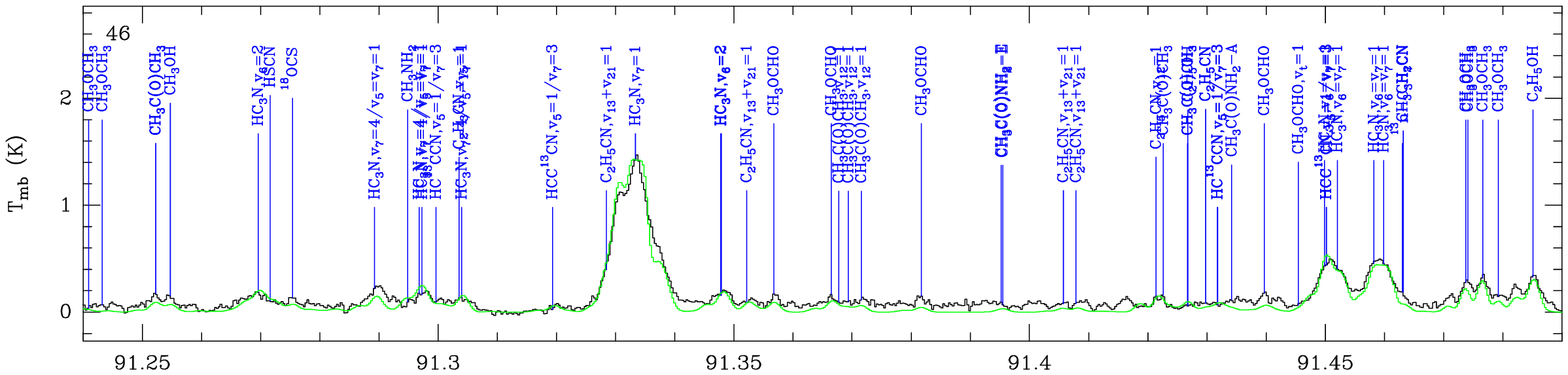}}}
\vspace*{1ex}\centerline{\resizebox{1.0\hsize}{!}{\includegraphics[angle=270]{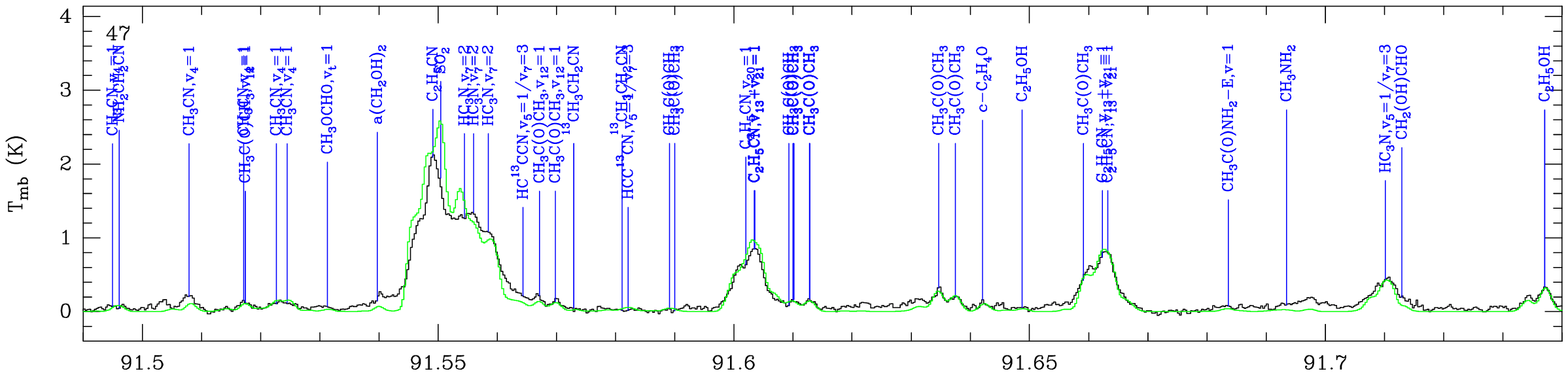}}}
\vspace*{1ex}\centerline{\resizebox{1.0\hsize}{!}{\includegraphics[angle=270]{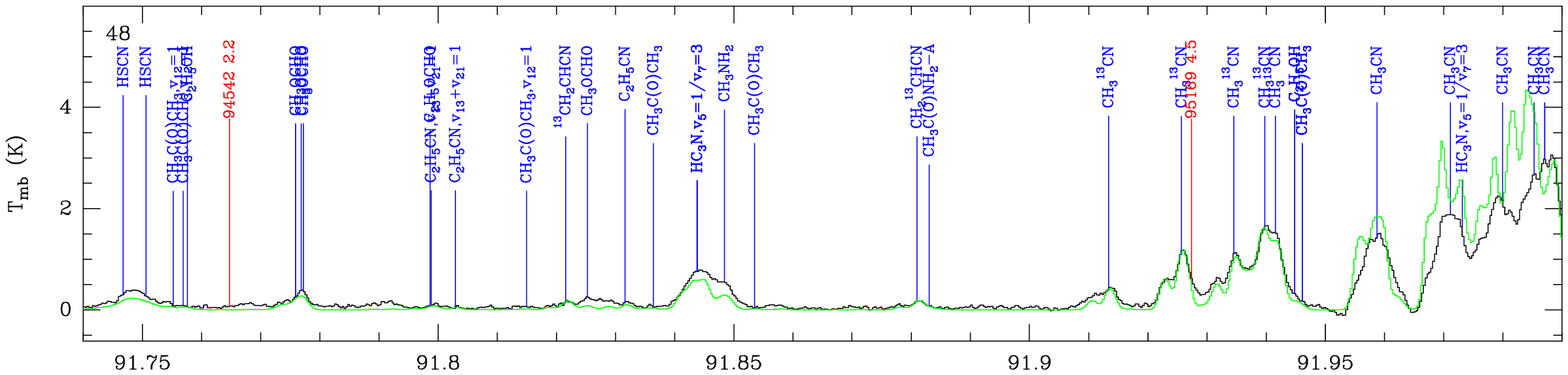}}}
\vspace*{1ex}\centerline{\resizebox{1.0\hsize}{!}{\includegraphics[angle=270]{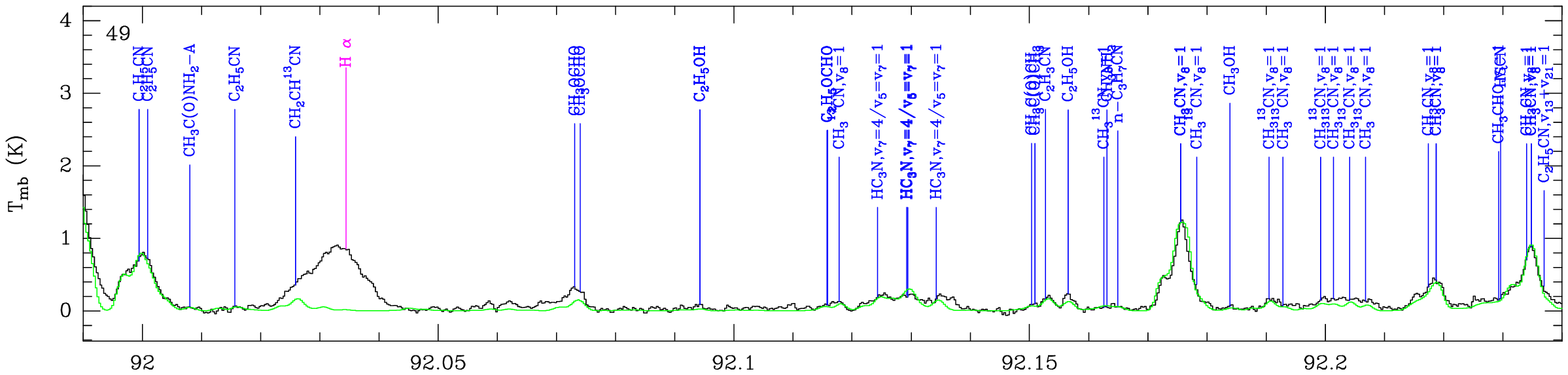}}}
\vspace*{1ex}\centerline{\resizebox{1.0\hsize}{!}{\includegraphics[angle=270]{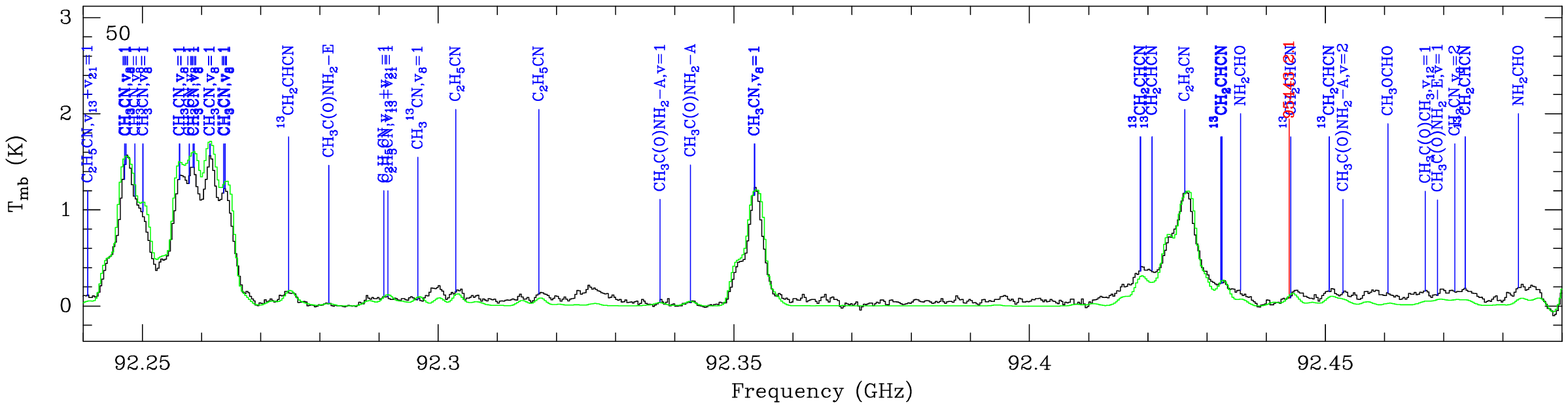}}}
\caption{
continued.
}
\end{figure*}
 \clearpage
\begin{figure*}
\addtocounter{figure}{-1}
\centerline{\resizebox{1.0\hsize}{!}{\includegraphics[angle=270]{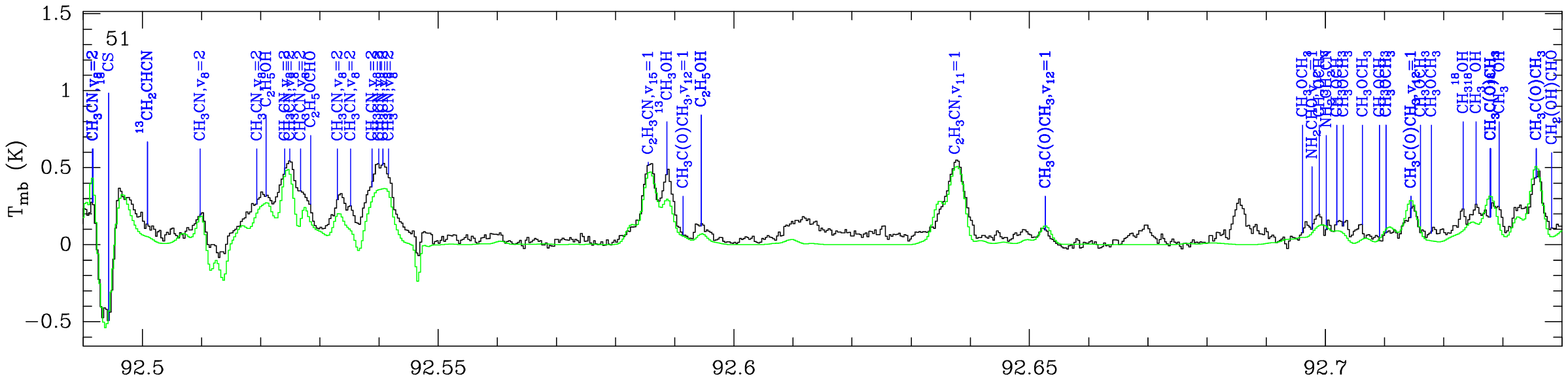}}}
\vspace*{1ex}\centerline{\resizebox{1.0\hsize}{!}{\includegraphics[angle=270]{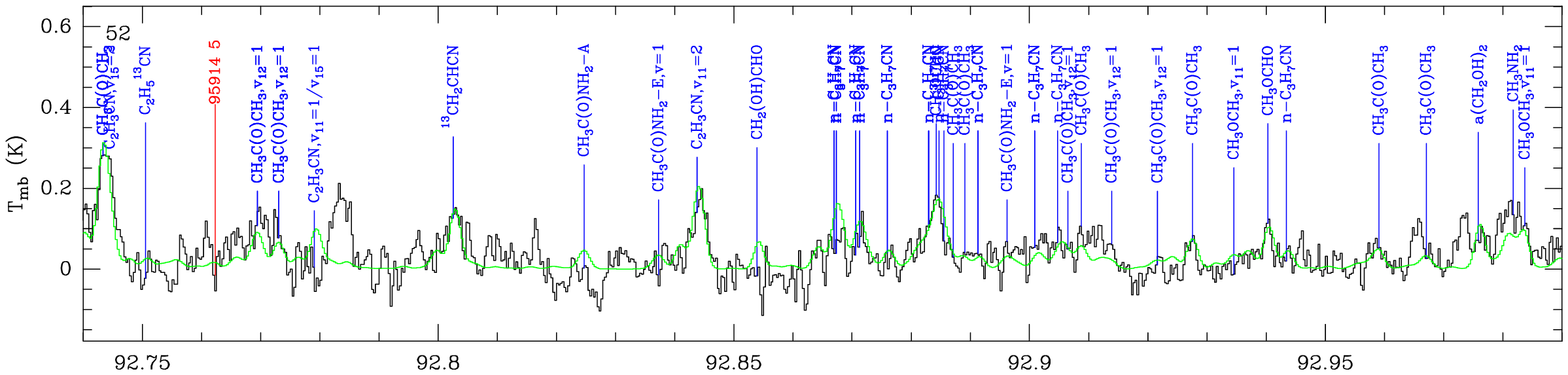}}}
\vspace*{1ex}\centerline{\resizebox{1.0\hsize}{!}{\includegraphics[angle=270]{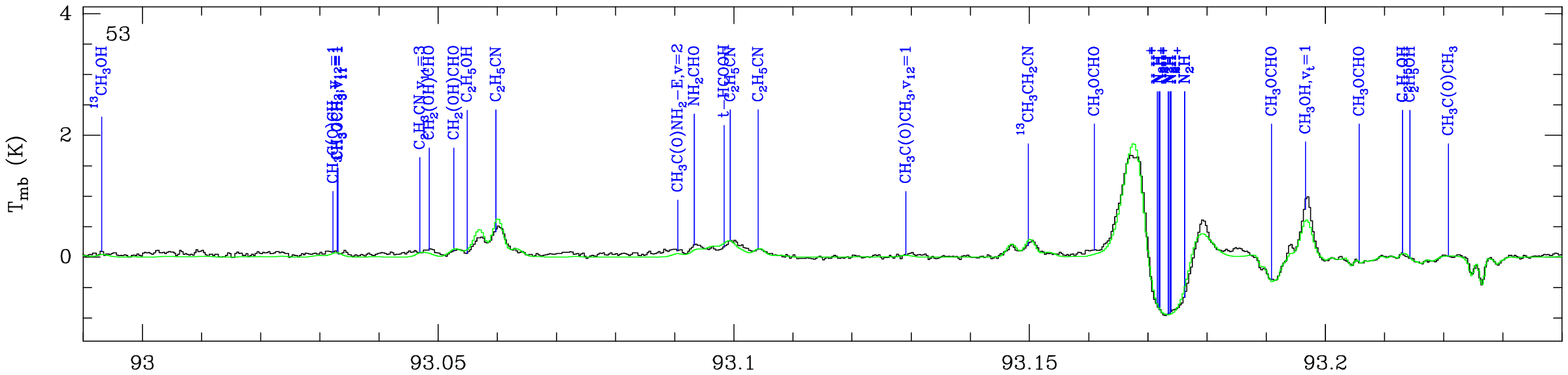}}}
\vspace*{1ex}\centerline{\resizebox{1.0\hsize}{!}{\includegraphics[angle=270]{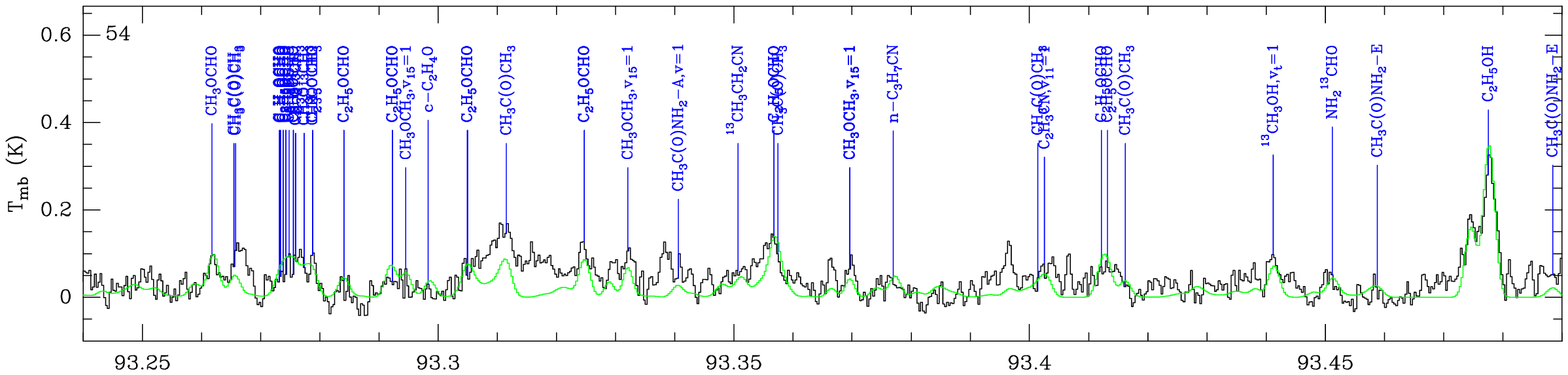}}}
\vspace*{1ex}\centerline{\resizebox{1.0\hsize}{!}{\includegraphics[angle=270]{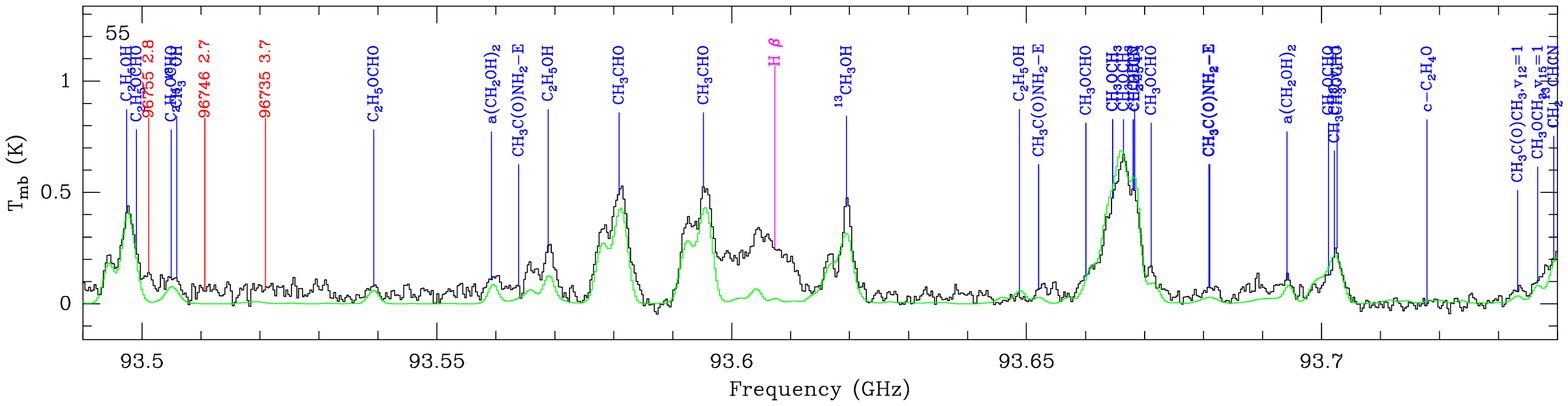}}}
\caption{
continued.
}
\end{figure*}
 \clearpage
\begin{figure*}
\addtocounter{figure}{-1}
\centerline{\resizebox{1.0\hsize}{!}{\includegraphics[angle=270]{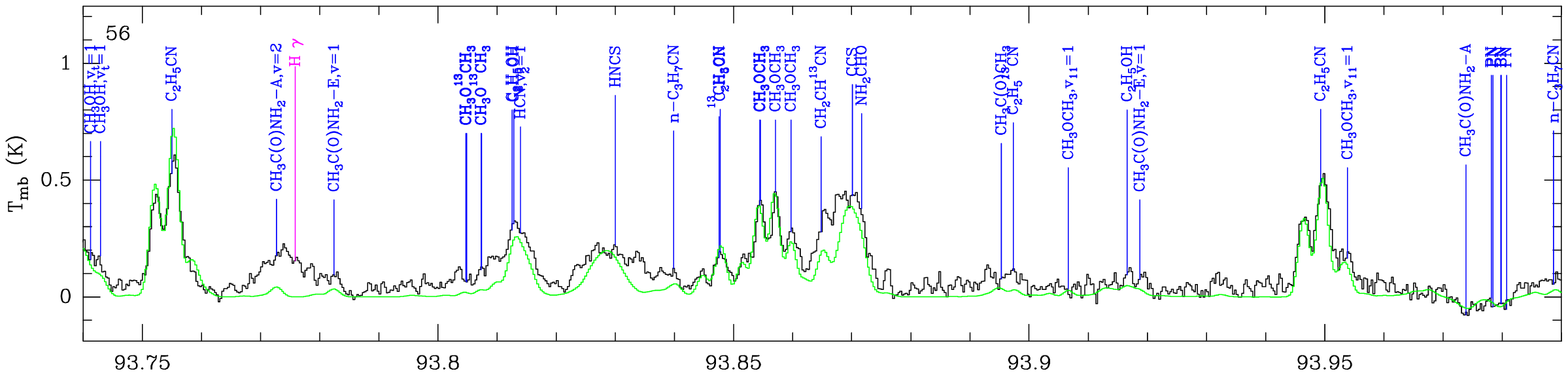}}}
\vspace*{1ex}\centerline{\resizebox{1.0\hsize}{!}{\includegraphics[angle=270]{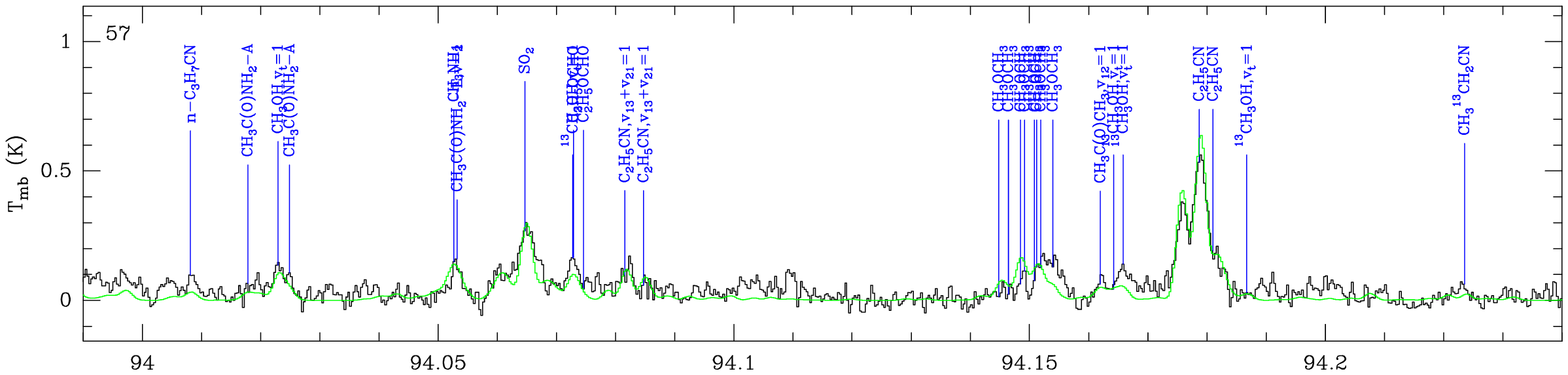}}}
\vspace*{1ex}\centerline{\resizebox{1.0\hsize}{!}{\includegraphics[angle=270]{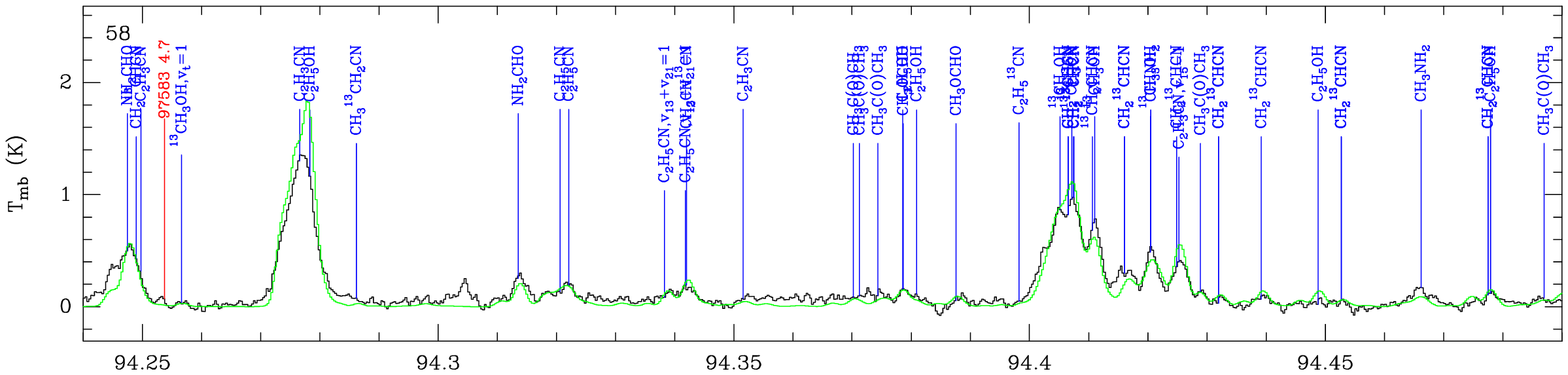}}}
\vspace*{1ex}\centerline{\resizebox{1.0\hsize}{!}{\includegraphics[angle=270]{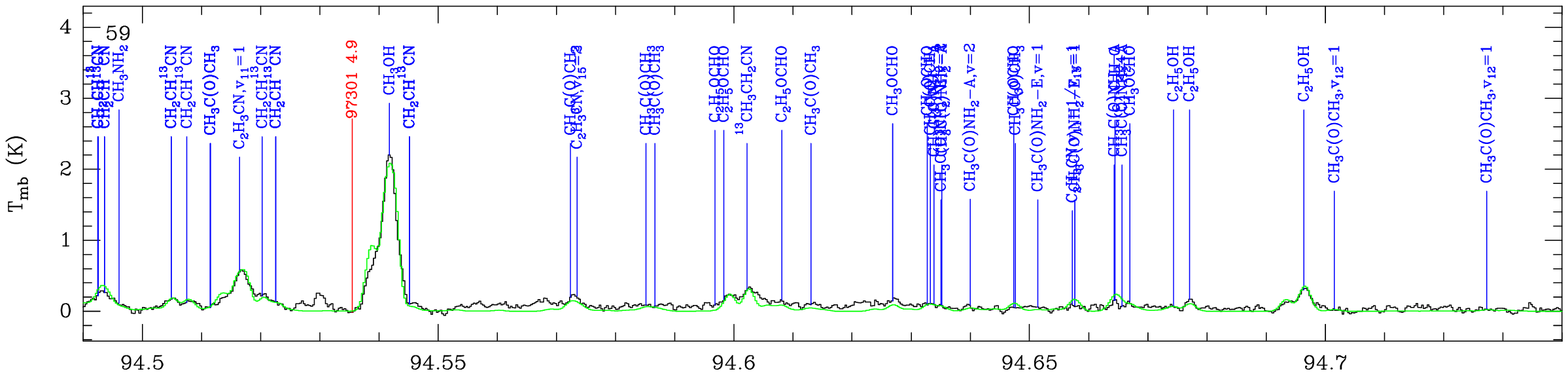}}}
\vspace*{1ex}\centerline{\resizebox{1.0\hsize}{!}{\includegraphics[angle=270]{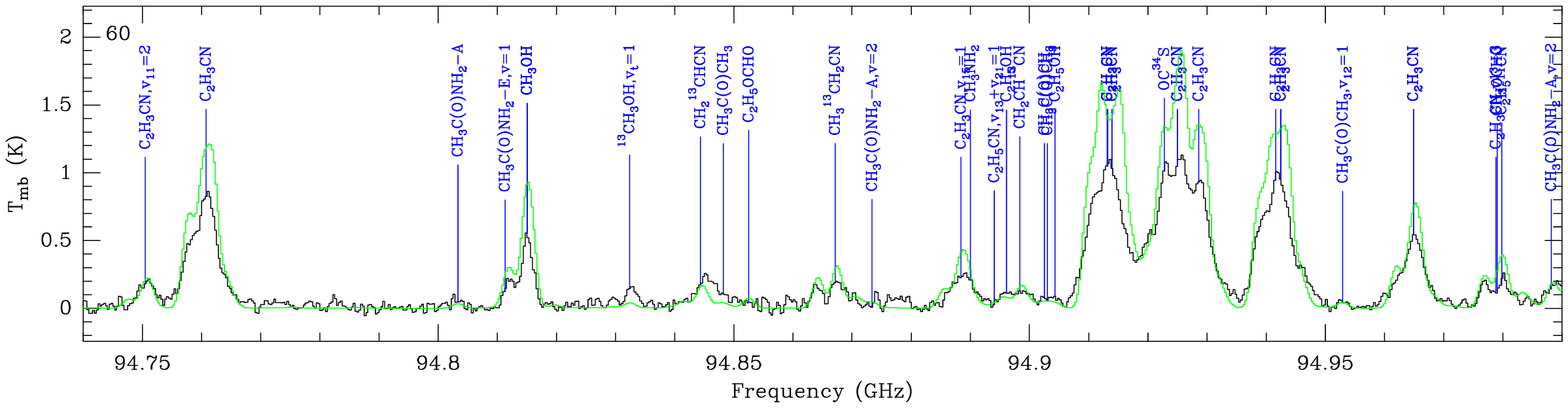}}}
\caption{
continued.
}
\end{figure*}
 \clearpage
\begin{figure*}
\addtocounter{figure}{-1}
\centerline{\resizebox{1.0\hsize}{!}{\includegraphics[angle=270]{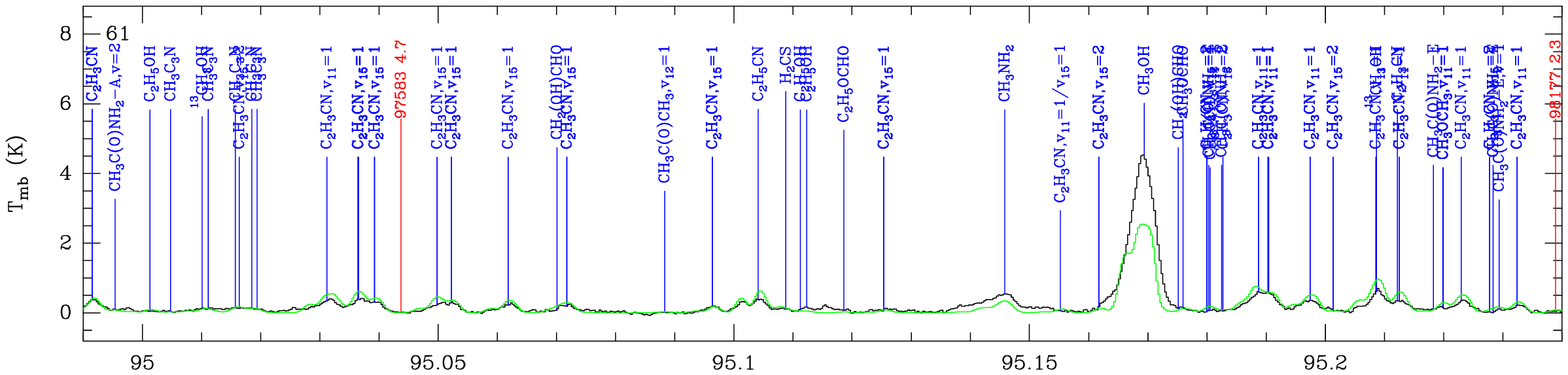}}}
\vspace*{1ex}\centerline{\resizebox{1.0\hsize}{!}{\includegraphics[angle=270]{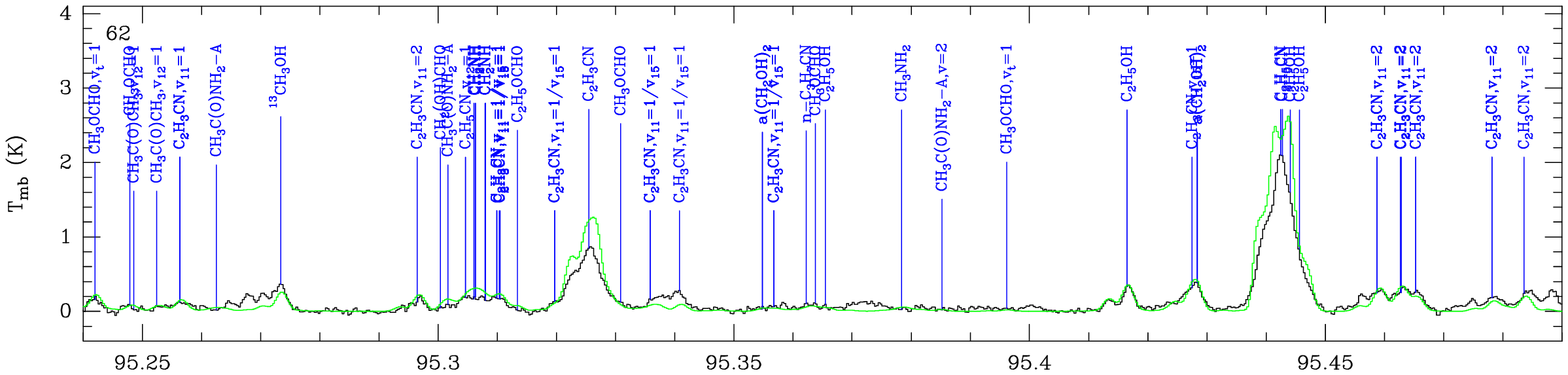}}}
\vspace*{1ex}\centerline{\resizebox{1.0\hsize}{!}{\includegraphics[angle=270]{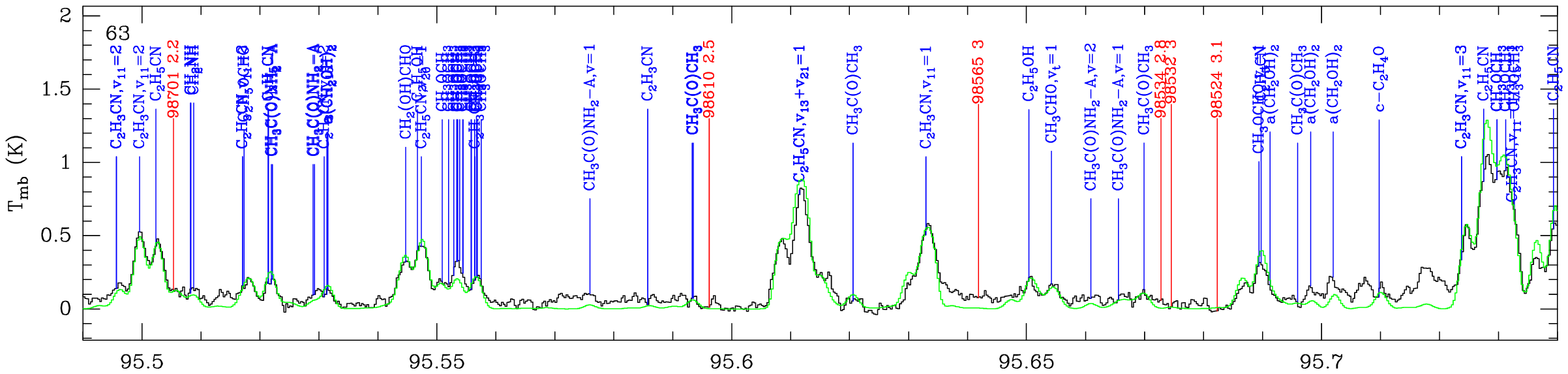}}}
\vspace*{1ex}\centerline{\resizebox{1.0\hsize}{!}{\includegraphics[angle=270]{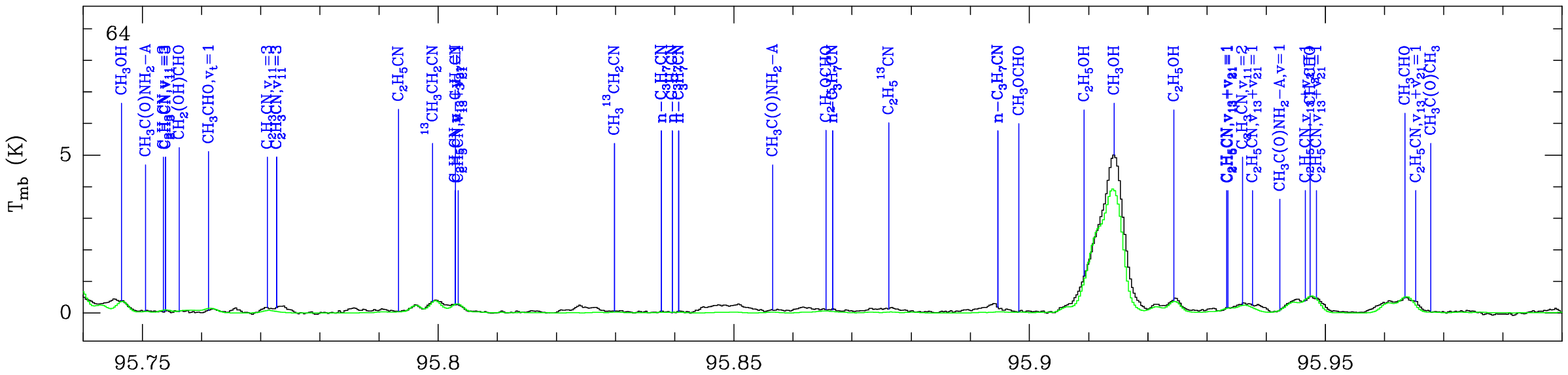}}}
\vspace*{1ex}\centerline{\resizebox{1.0\hsize}{!}{\includegraphics[angle=270]{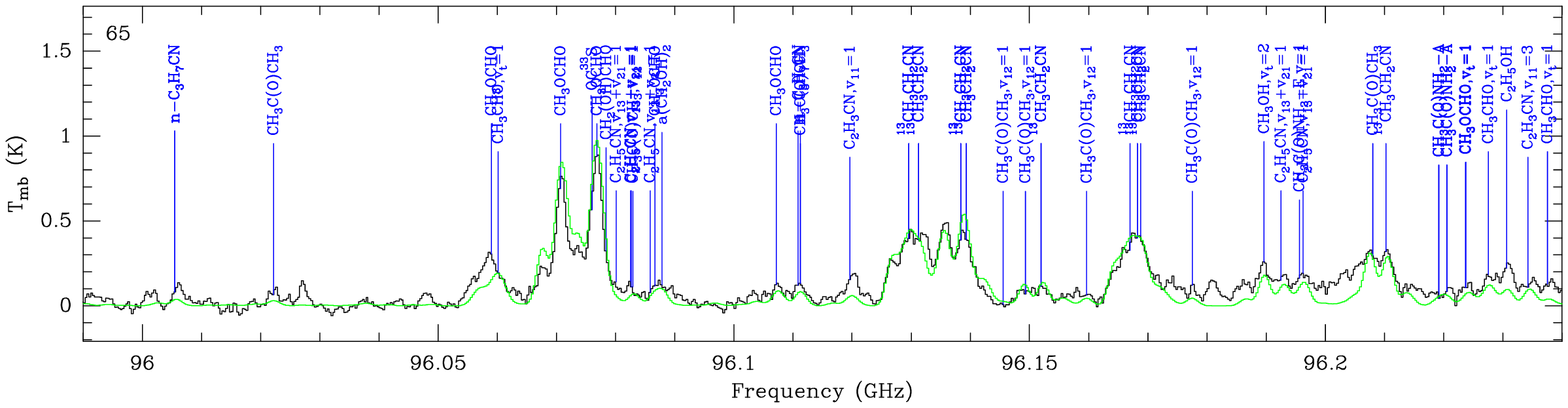}}}
\caption{
continued.
}
\end{figure*}
 \clearpage
\begin{figure*}
\addtocounter{figure}{-1}
\centerline{\resizebox{1.0\hsize}{!}{\includegraphics[angle=270]{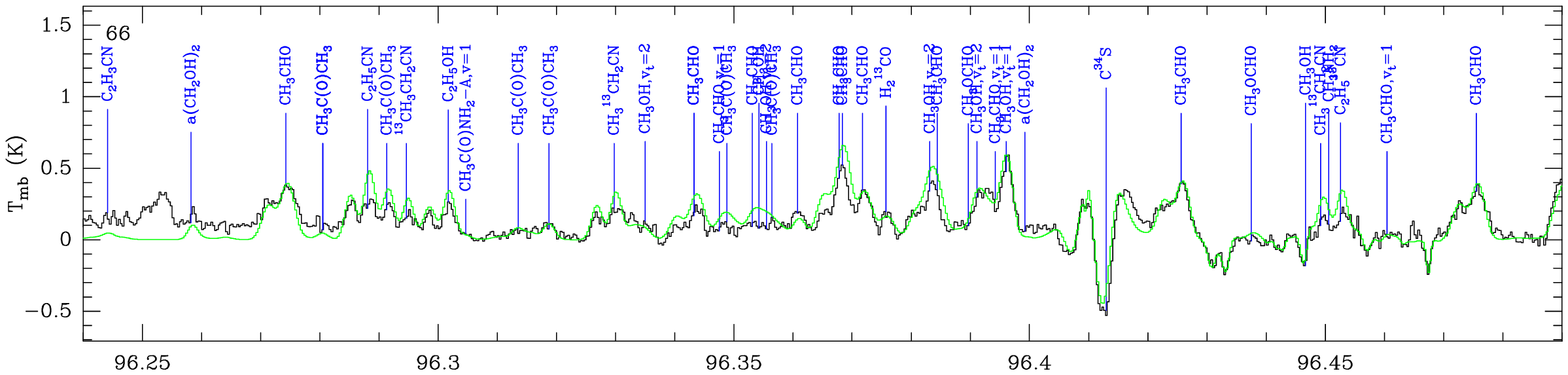}}}
\vspace*{1ex}\centerline{\resizebox{1.0\hsize}{!}{\includegraphics[angle=270]{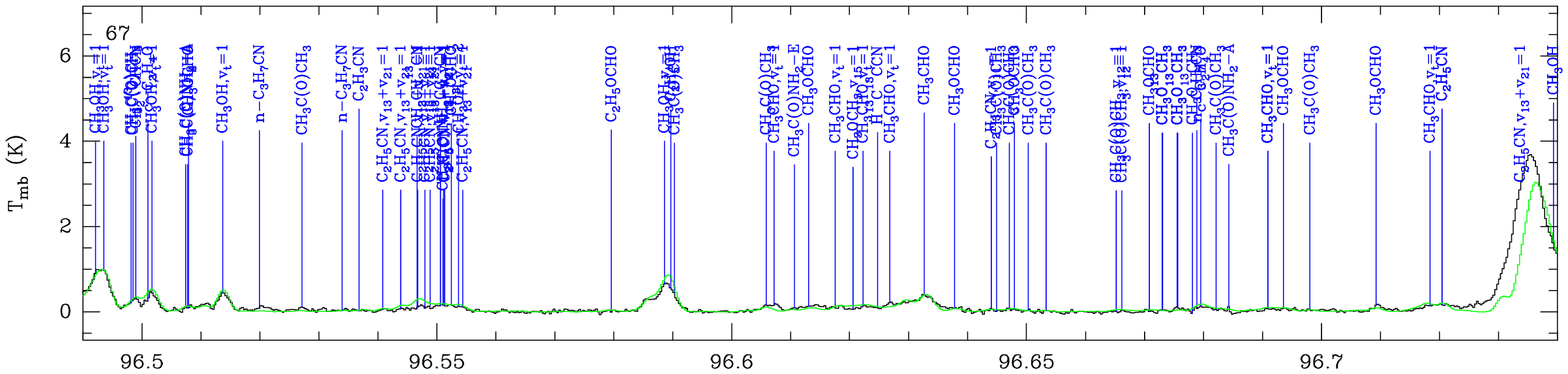}}}
\vspace*{1ex}\centerline{\resizebox{1.0\hsize}{!}{\includegraphics[angle=270]{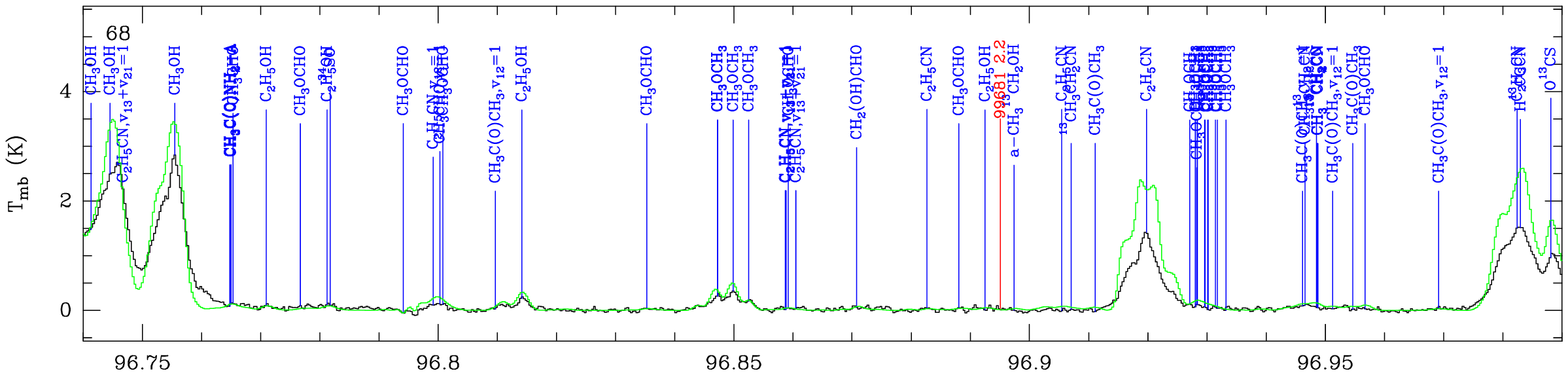}}}
\vspace*{1ex}\centerline{\resizebox{1.0\hsize}{!}{\includegraphics[angle=270]{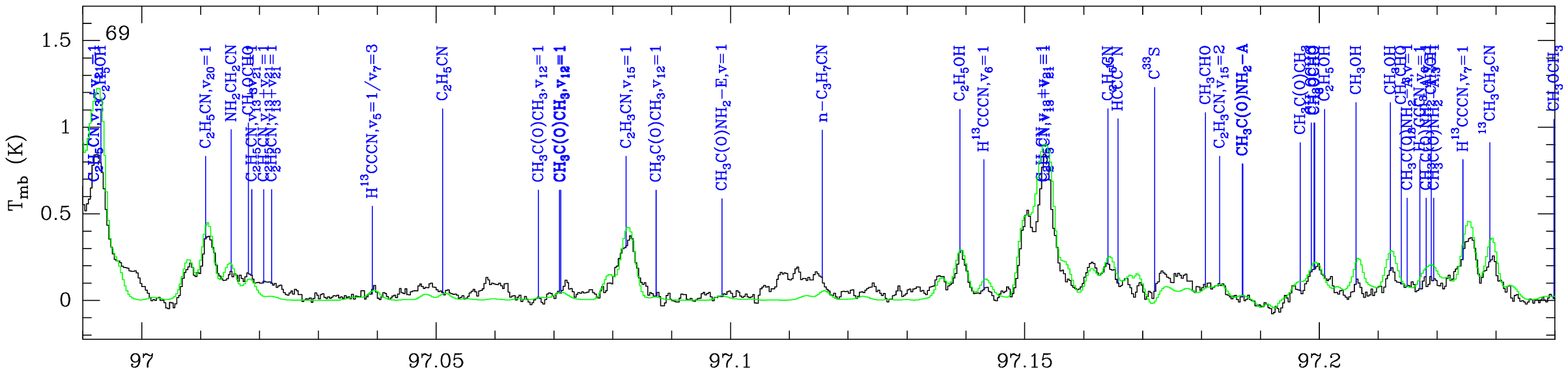}}}
\vspace*{1ex}\centerline{\resizebox{1.0\hsize}{!}{\includegraphics[angle=270]{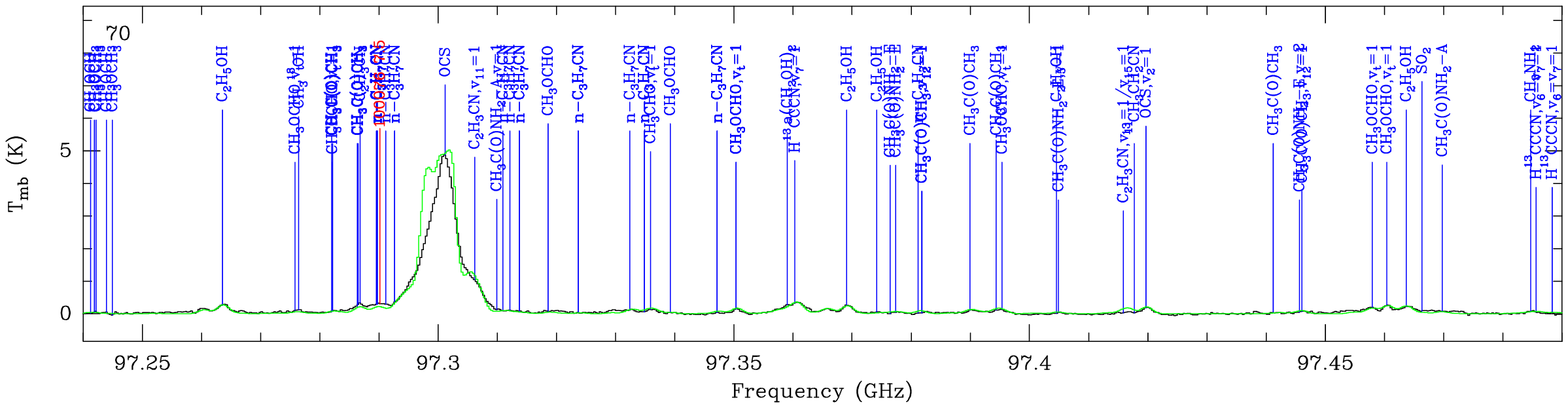}}}
\caption{
continued.
}
\end{figure*}
 \clearpage
\begin{figure*}
\addtocounter{figure}{-1}
\centerline{\resizebox{1.0\hsize}{!}{\includegraphics[angle=270]{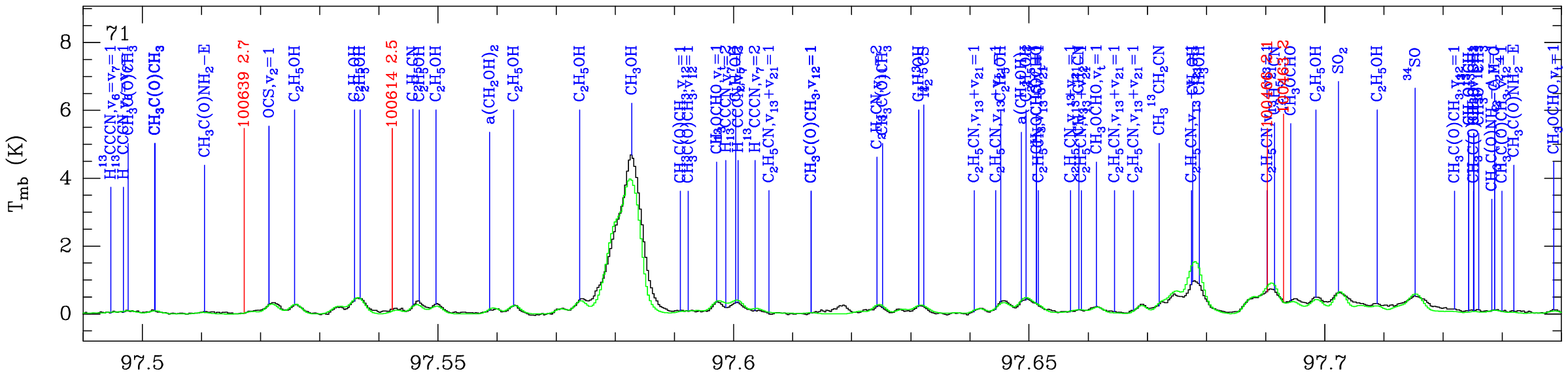}}}
\vspace*{1ex}\centerline{\resizebox{1.0\hsize}{!}{\includegraphics[angle=270]{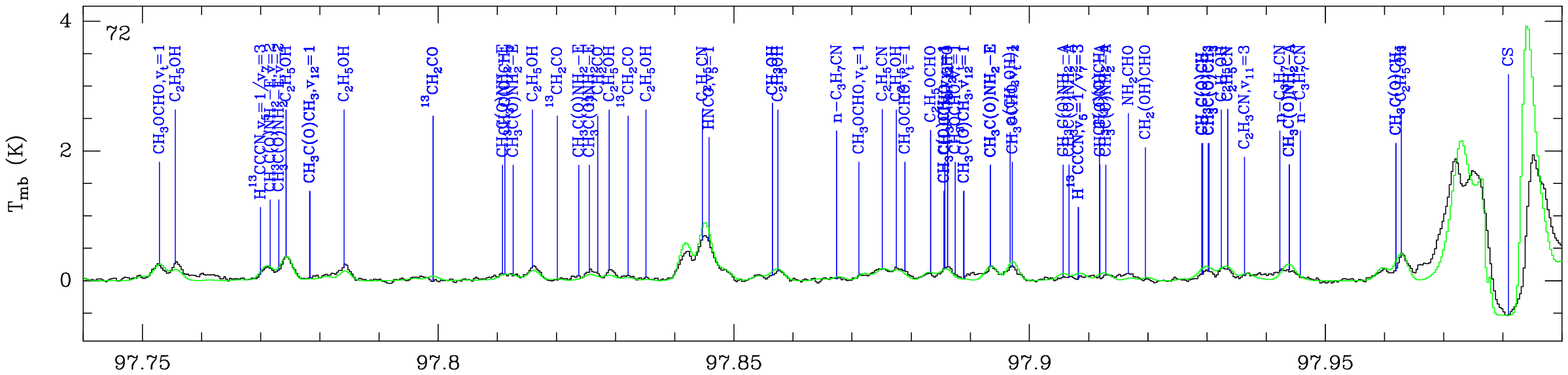}}}
\vspace*{1ex}\centerline{\resizebox{1.0\hsize}{!}{\includegraphics[angle=270]{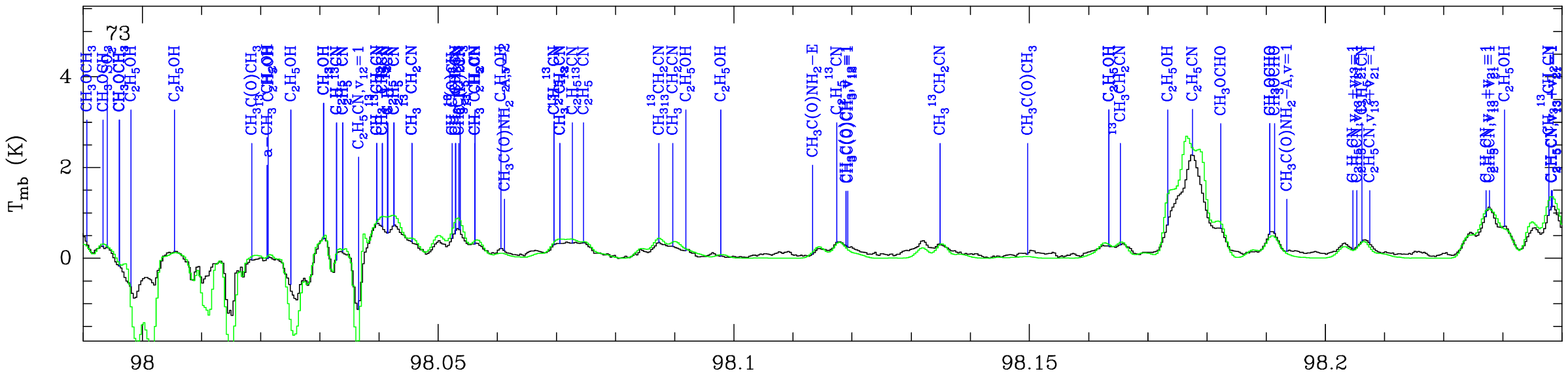}}}
\vspace*{1ex}\centerline{\resizebox{1.0\hsize}{!}{\includegraphics[angle=270]{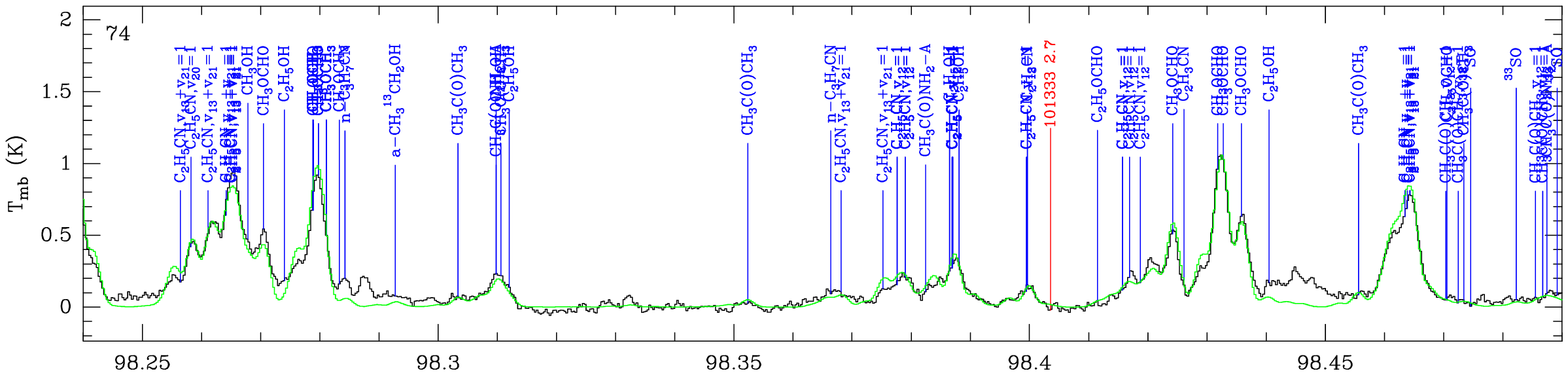}}}
\vspace*{1ex}\centerline{\resizebox{1.0\hsize}{!}{\includegraphics[angle=270]{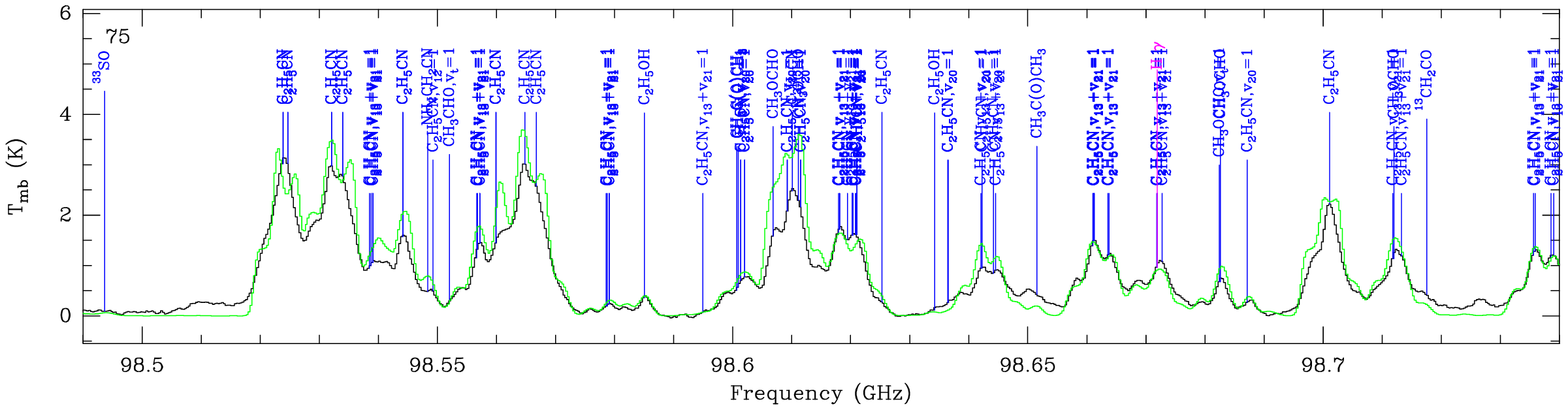}}}
\caption{
continued.
}
\end{figure*}
 \clearpage
\begin{figure*}
\addtocounter{figure}{-1}
\centerline{\resizebox{1.0\hsize}{!}{\includegraphics[angle=270]{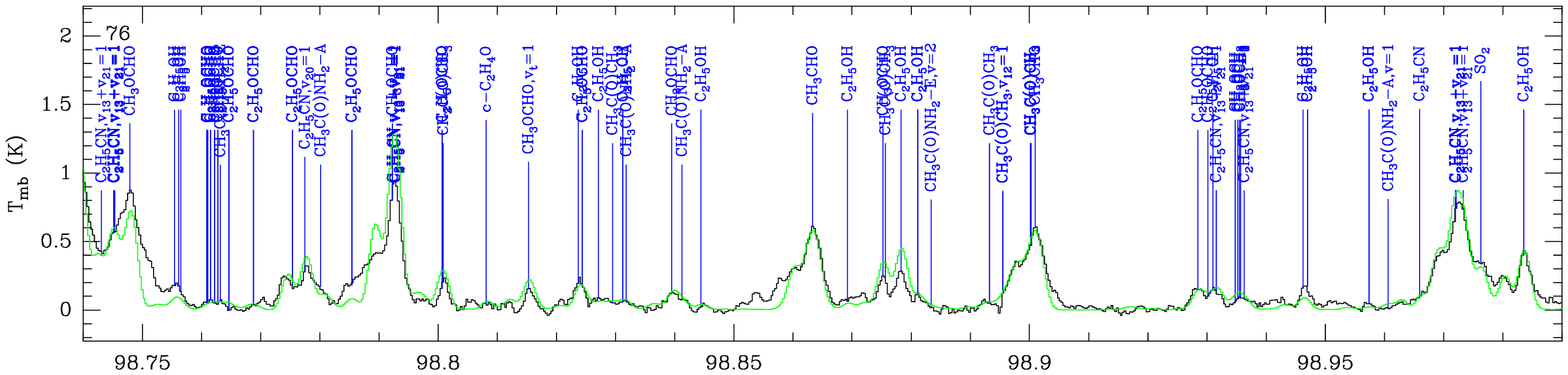}}}
\vspace*{1ex}\centerline{\resizebox{1.0\hsize}{!}{\includegraphics[angle=270]{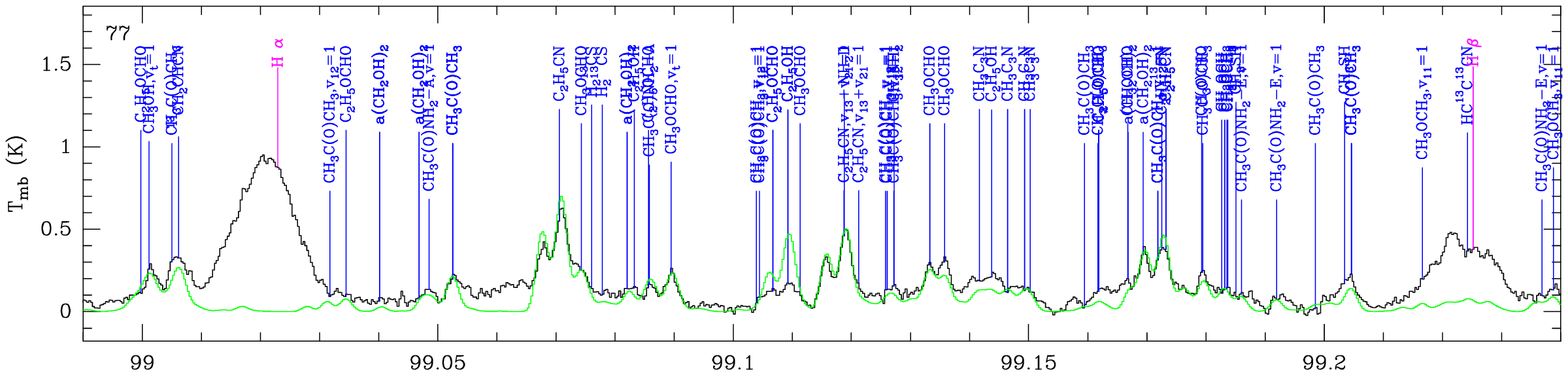}}}
\vspace*{1ex}\centerline{\resizebox{1.0\hsize}{!}{\includegraphics[angle=270]{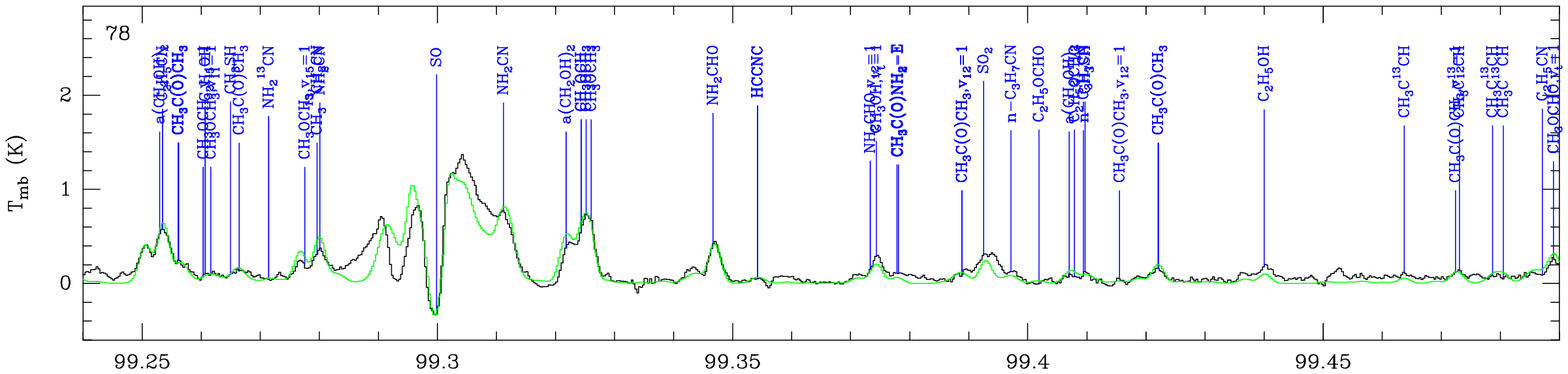}}}
\vspace*{1ex}\centerline{\resizebox{1.0\hsize}{!}{\includegraphics[angle=270]{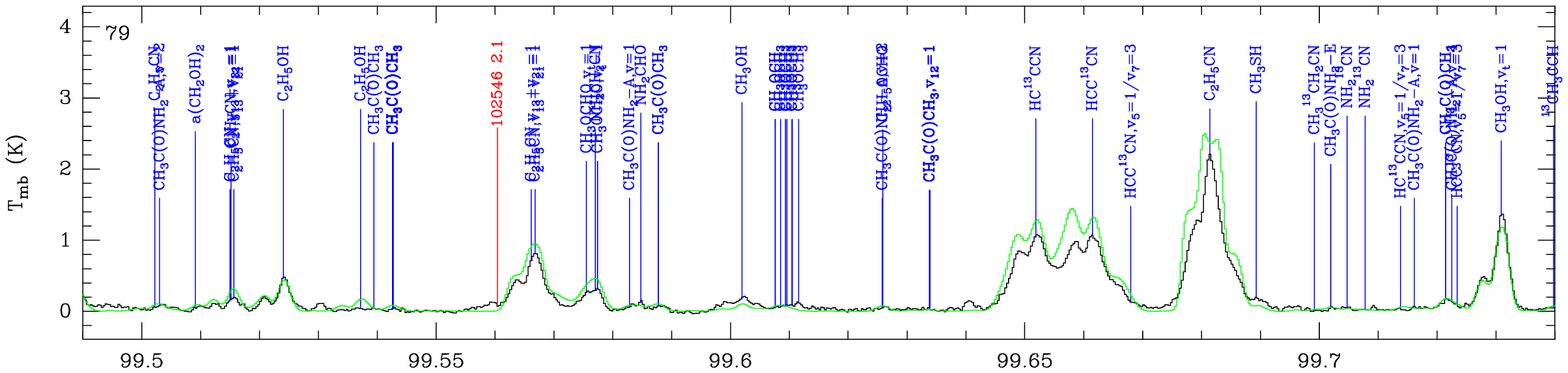}}}
\vspace*{1ex}\centerline{\resizebox{1.0\hsize}{!}{\includegraphics[angle=270]{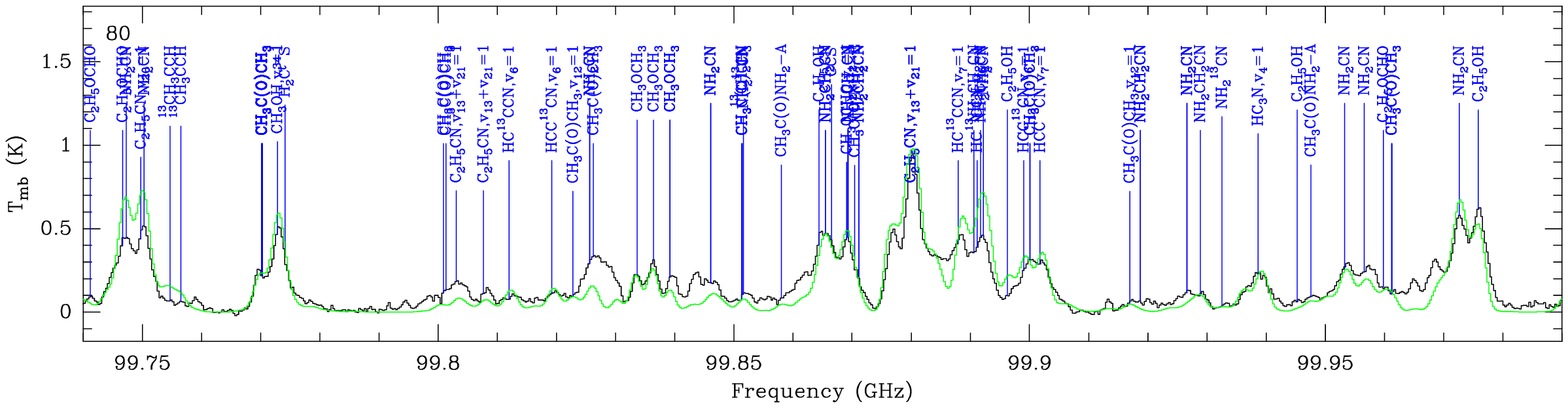}}}
\caption{
continued.
}
\end{figure*}
 \clearpage
\begin{figure*}
\addtocounter{figure}{-1}
\centerline{\resizebox{1.0\hsize}{!}{\includegraphics[angle=270]{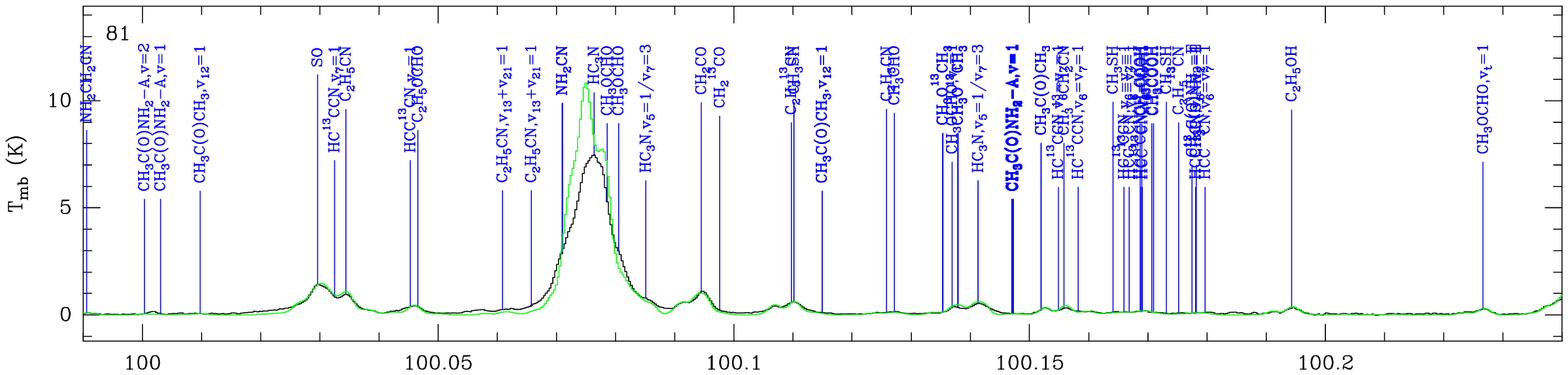}}}
\vspace*{1ex}\centerline{\resizebox{1.0\hsize}{!}{\includegraphics[angle=270]{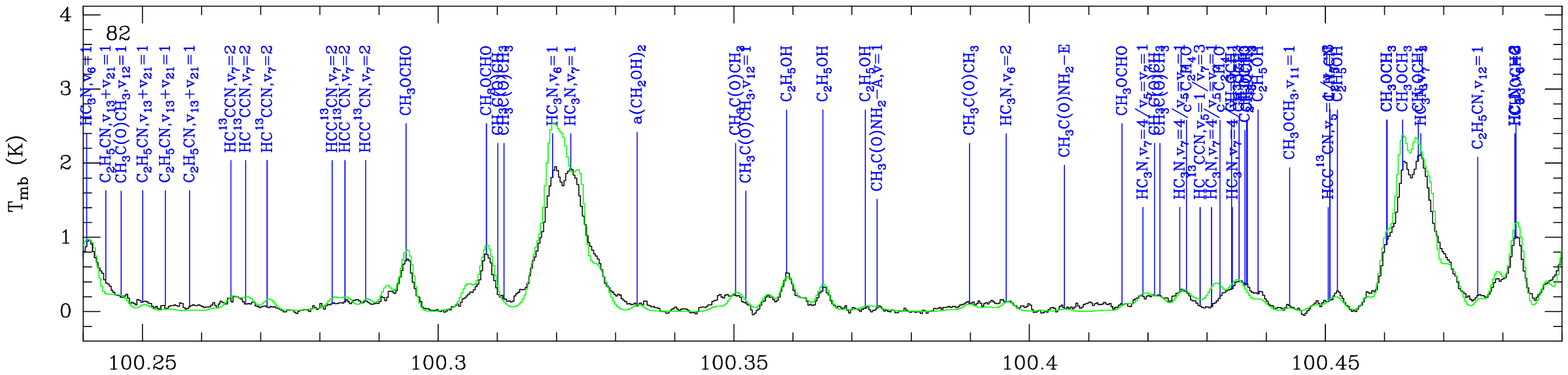}}}
\vspace*{1ex}\centerline{\resizebox{1.0\hsize}{!}{\includegraphics[angle=270]{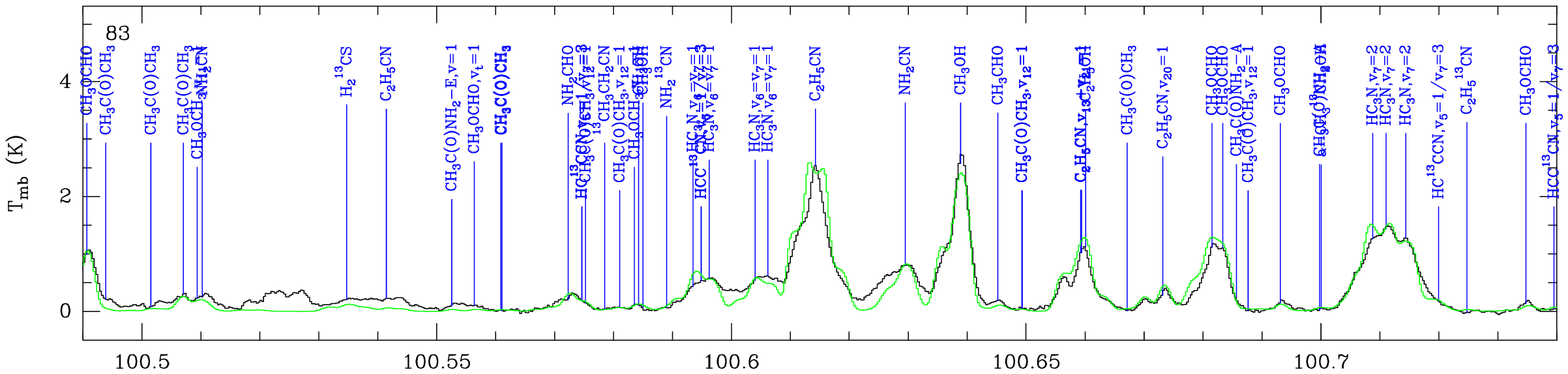}}}
\vspace*{1ex}\centerline{\resizebox{1.0\hsize}{!}{\includegraphics[angle=270]{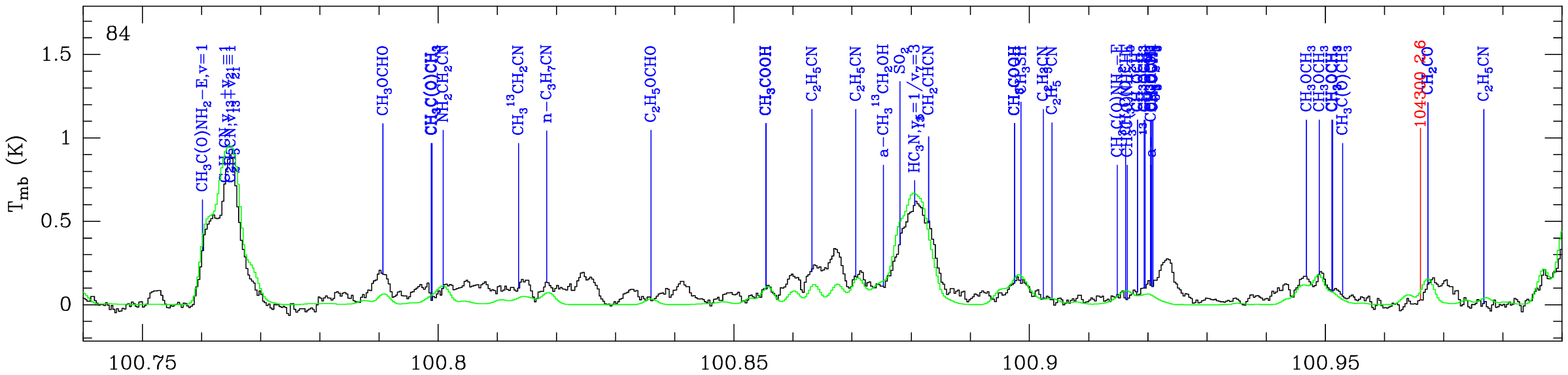}}}
\vspace*{1ex}\centerline{\resizebox{1.0\hsize}{!}{\includegraphics[angle=270]{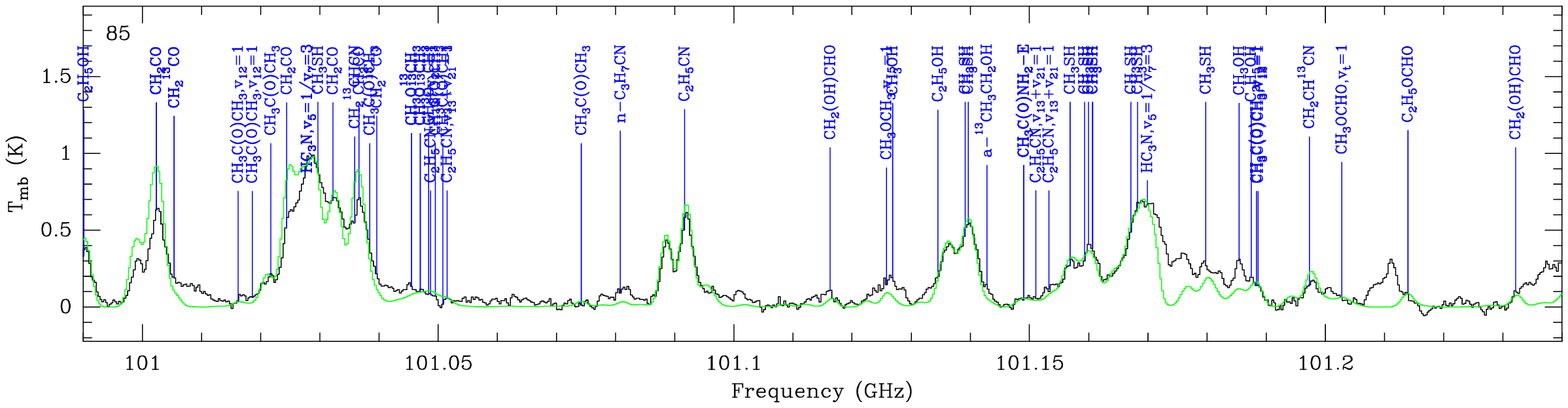}}}
\caption{
continued.
}
\end{figure*}
 \clearpage
\begin{figure*}
\addtocounter{figure}{-1}
\centerline{\resizebox{1.0\hsize}{!}{\includegraphics[angle=270]{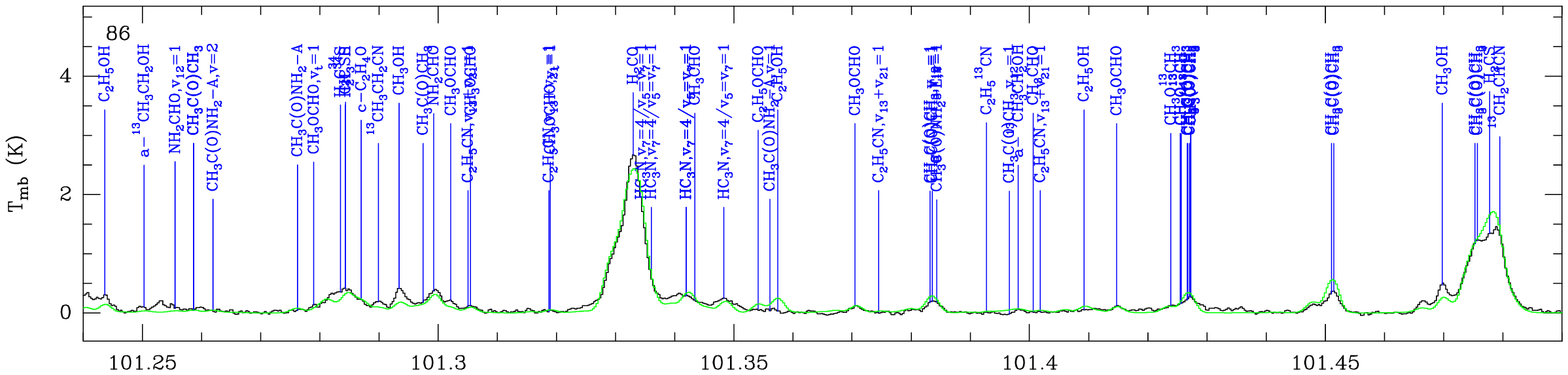}}}
\vspace*{1ex}\centerline{\resizebox{1.0\hsize}{!}{\includegraphics[angle=270]{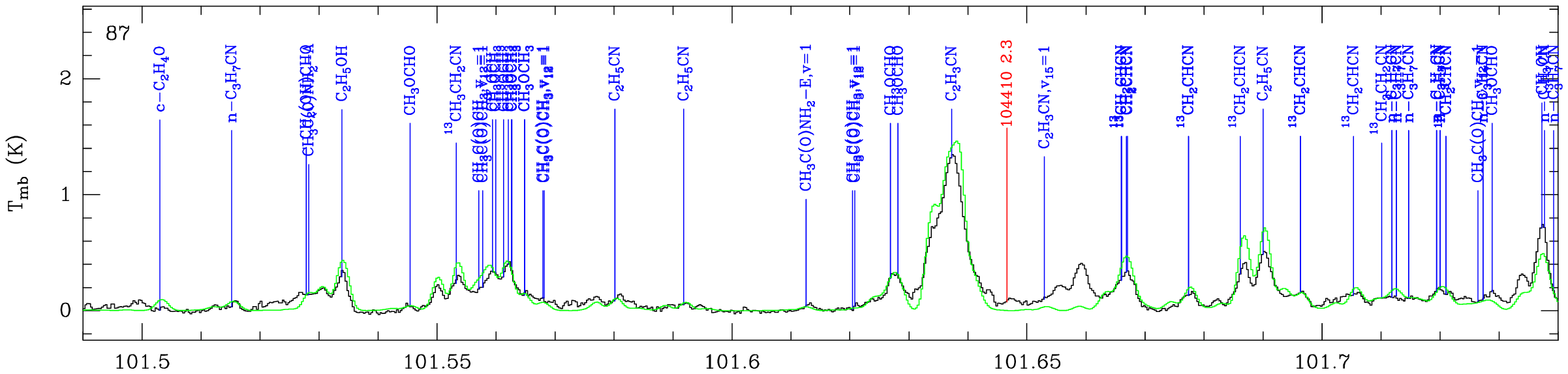}}}
\vspace*{1ex}\centerline{\resizebox{1.0\hsize}{!}{\includegraphics[angle=270]{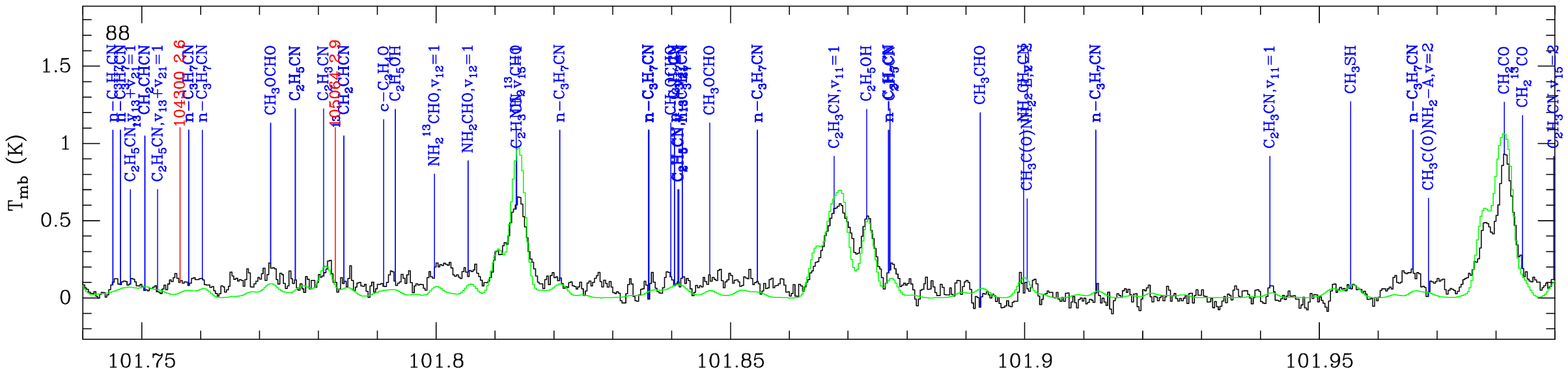}}}
\vspace*{1ex}\centerline{\resizebox{1.0\hsize}{!}{\includegraphics[angle=270]{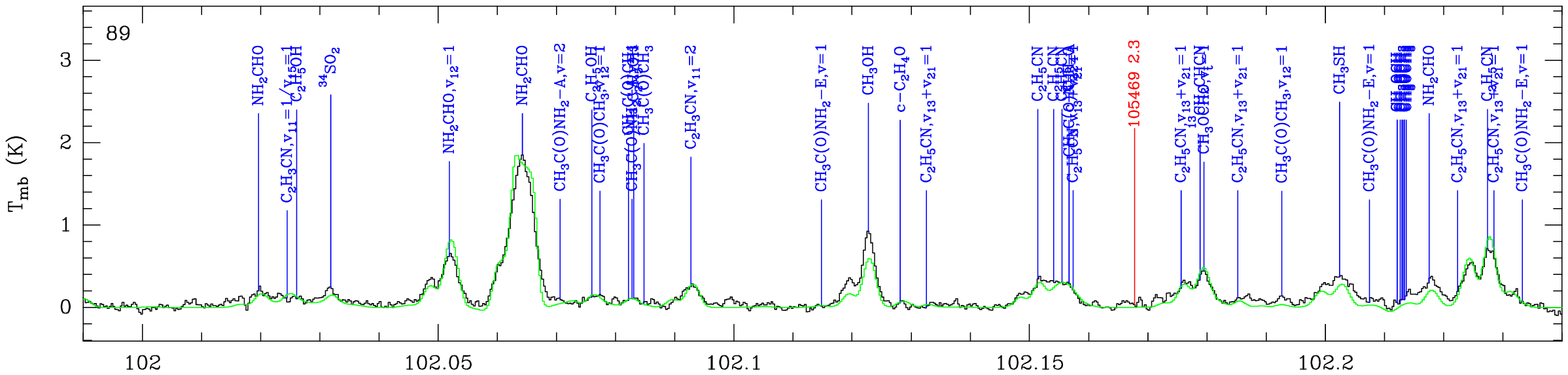}}}
\vspace*{1ex}\centerline{\resizebox{1.0\hsize}{!}{\includegraphics[angle=270]{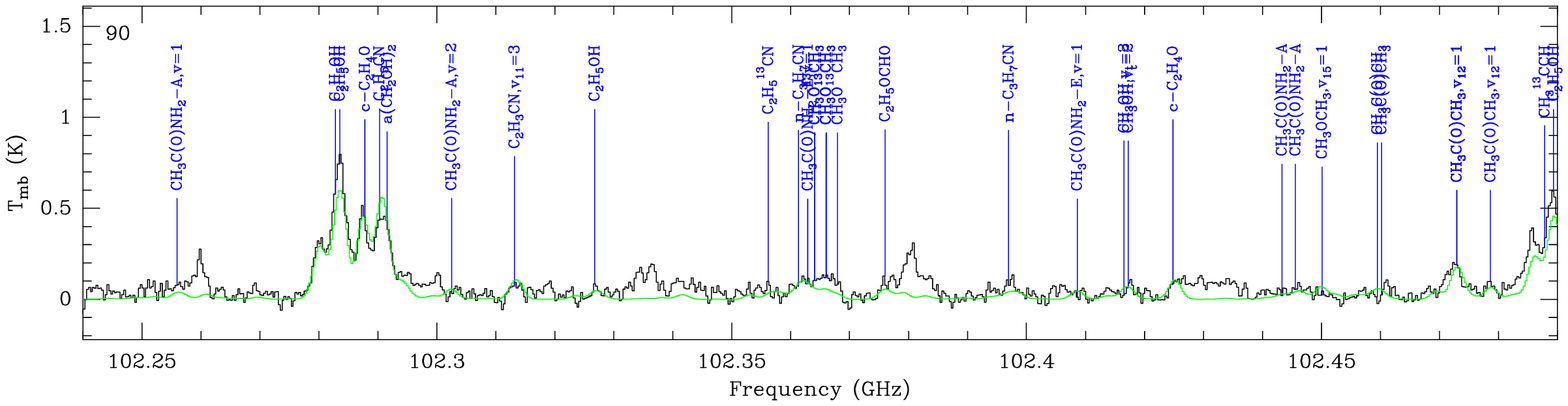}}}
\caption{
continued.
}
\end{figure*}
 \clearpage
\begin{figure*}
\addtocounter{figure}{-1}
\centerline{\resizebox{1.0\hsize}{!}{\includegraphics[angle=270]{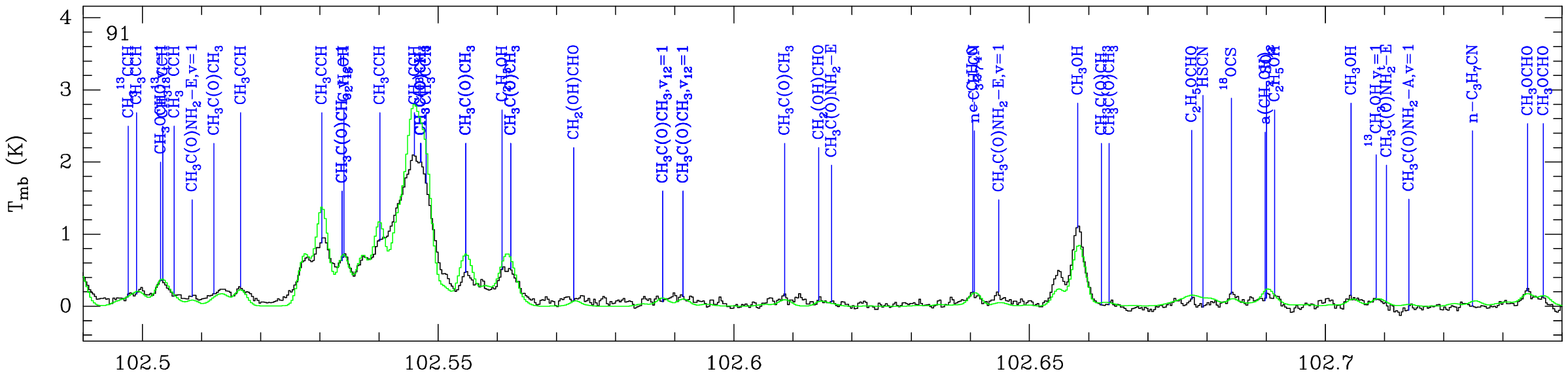}}}
\vspace*{1ex}\centerline{\resizebox{1.0\hsize}{!}{\includegraphics[angle=270]{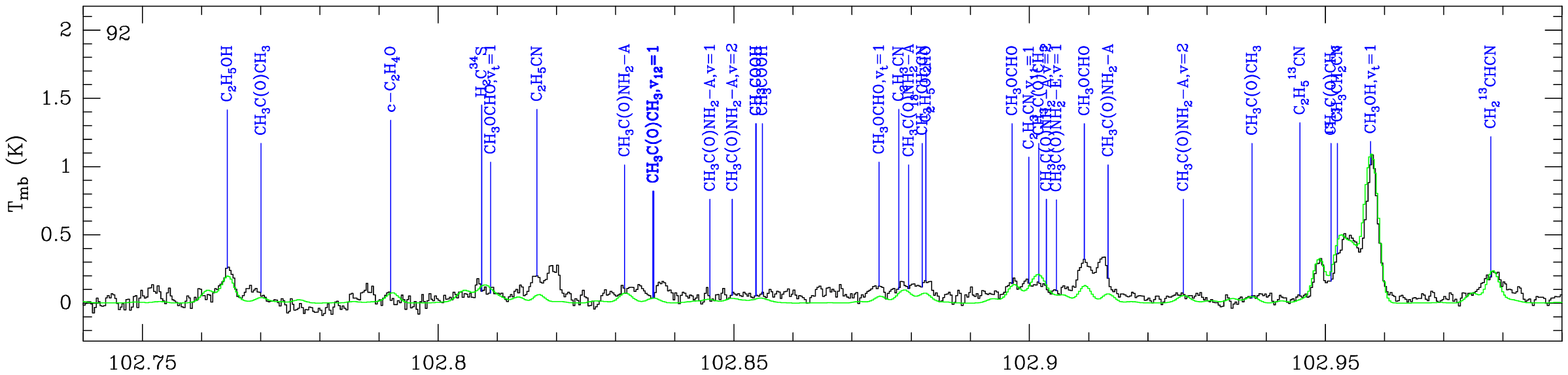}}}
\vspace*{1ex}\centerline{\resizebox{1.0\hsize}{!}{\includegraphics[angle=270]{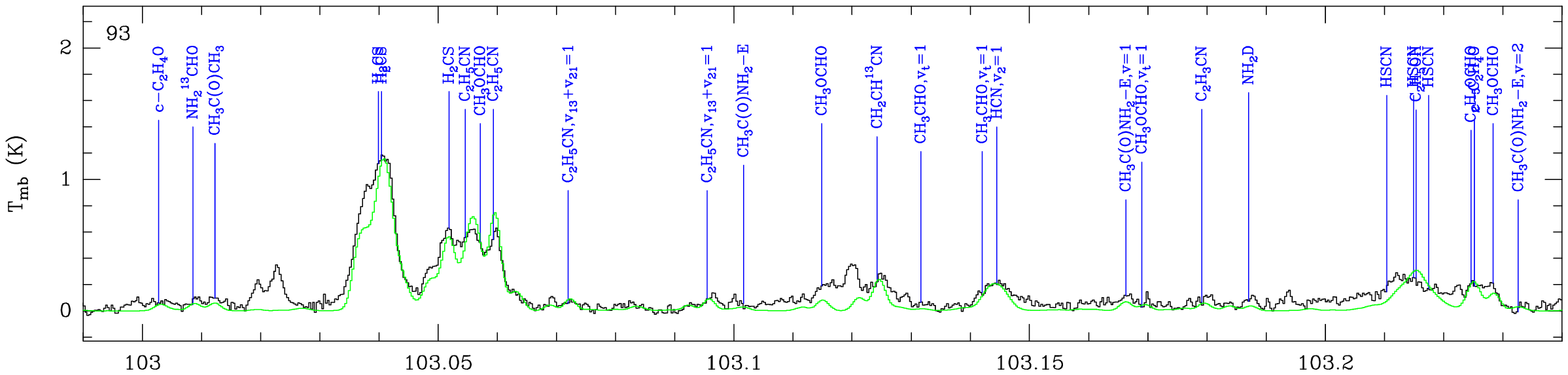}}}
\vspace*{1ex}\centerline{\resizebox{1.0\hsize}{!}{\includegraphics[angle=270]{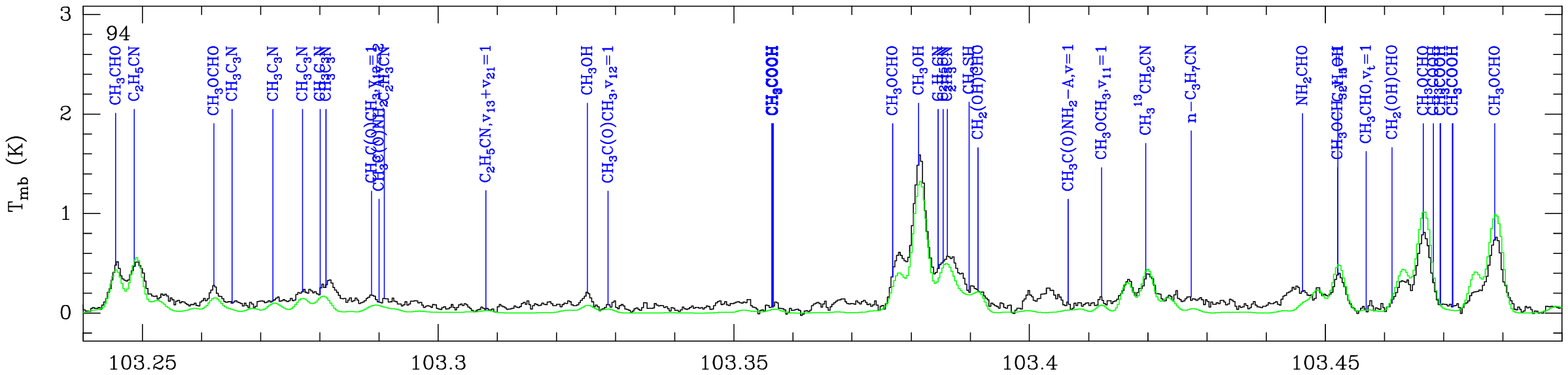}}}
\vspace*{1ex}\centerline{\resizebox{1.0\hsize}{!}{\includegraphics[angle=270]{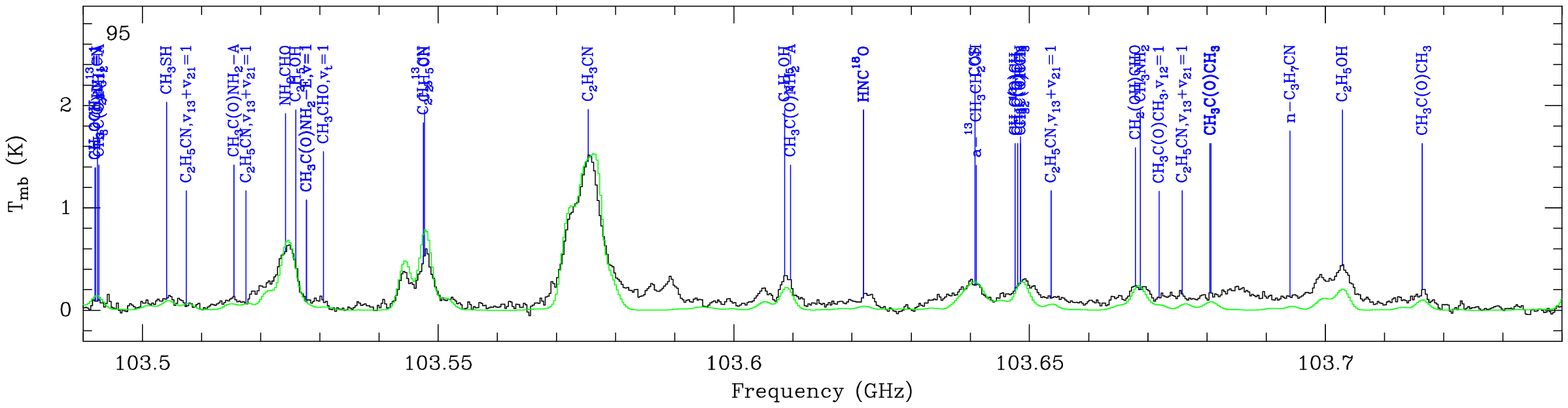}}}
\caption{
continued.
}
\end{figure*}
 \clearpage
\begin{figure*}
\addtocounter{figure}{-1}
\centerline{\resizebox{1.0\hsize}{!}{\includegraphics[angle=270]{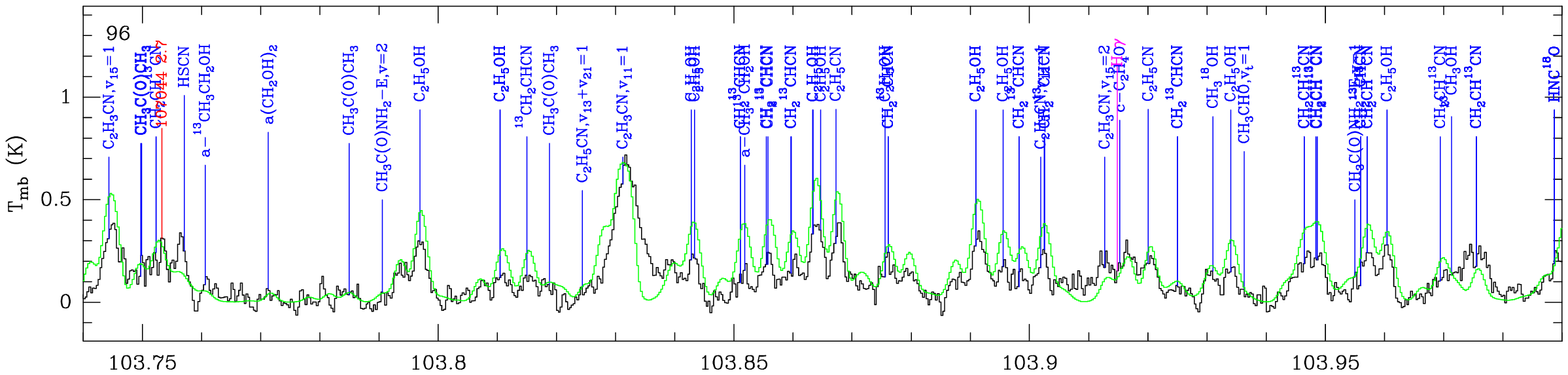}}}
\vspace*{1ex}\centerline{\resizebox{1.0\hsize}{!}{\includegraphics[angle=270]{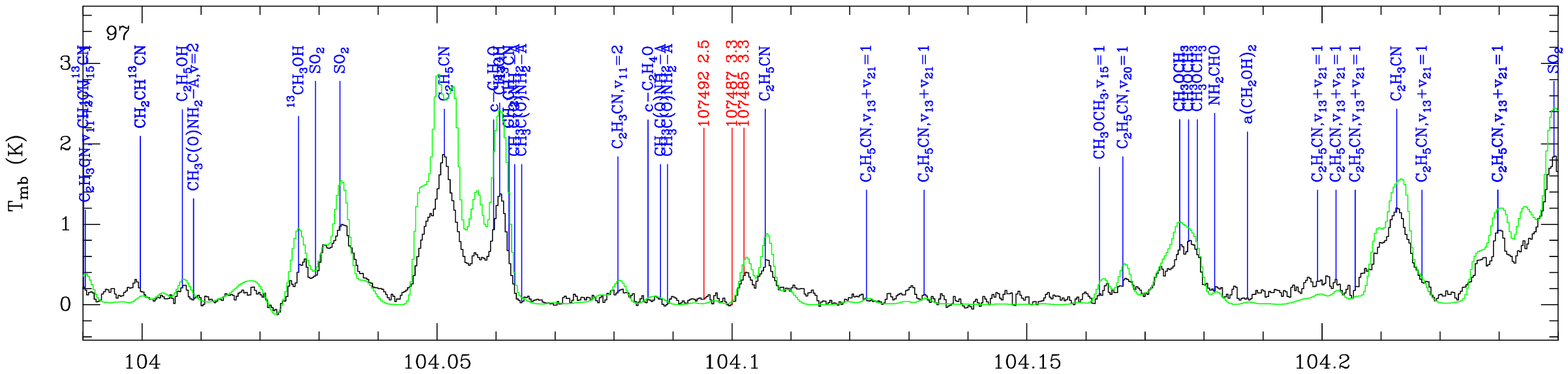}}}
\vspace*{1ex}\centerline{\resizebox{1.0\hsize}{!}{\includegraphics[angle=270]{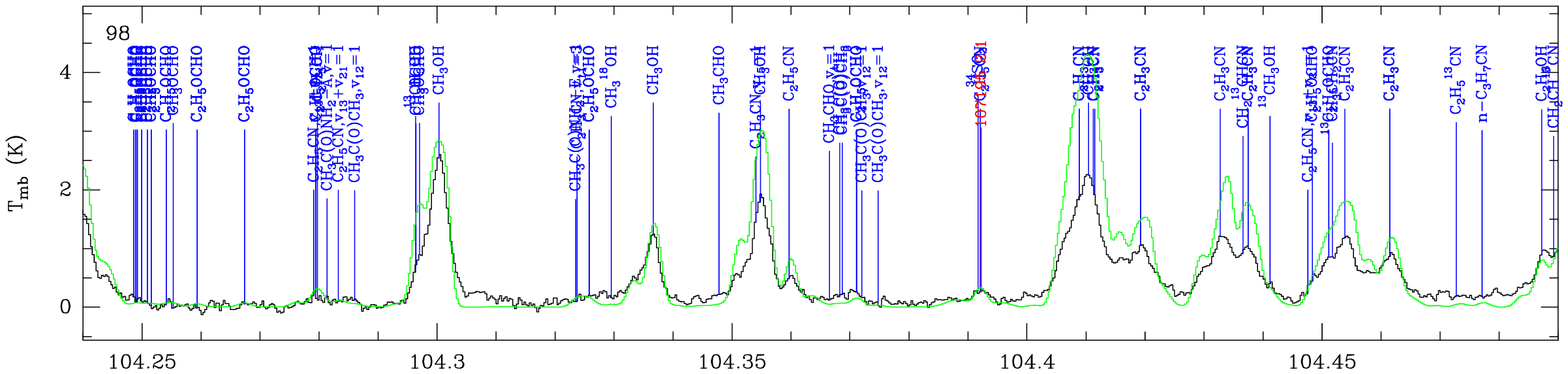}}}
\vspace*{1ex}\centerline{\resizebox{1.0\hsize}{!}{\includegraphics[angle=270]{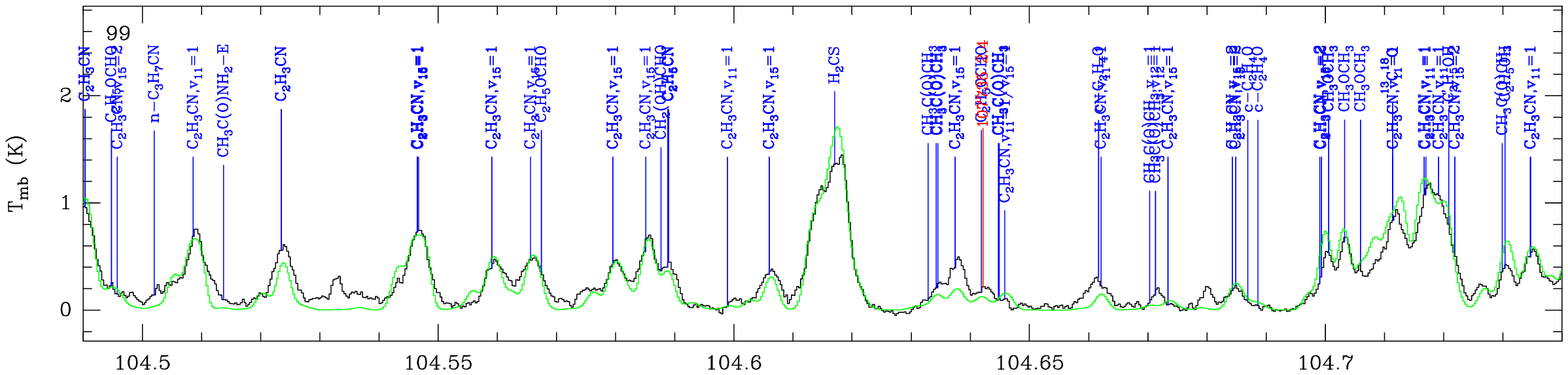}}}
\vspace*{1ex}\centerline{\resizebox{1.0\hsize}{!}{\includegraphics[angle=270]{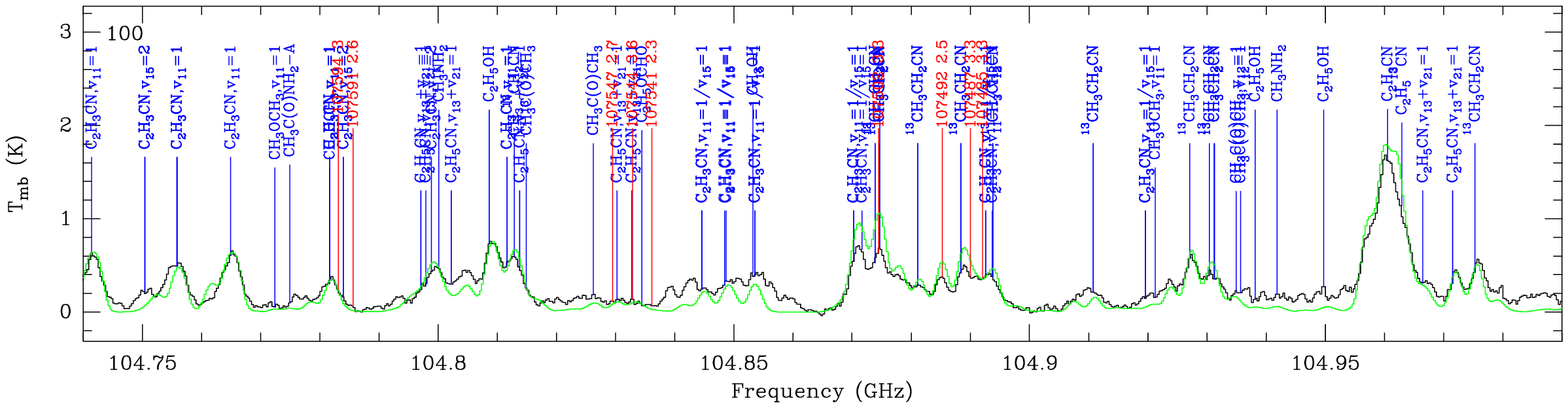}}}
\caption{
continued.
}
\end{figure*}
 \clearpage
\begin{figure*}
\addtocounter{figure}{-1}
\centerline{\resizebox{1.0\hsize}{!}{\includegraphics[angle=270]{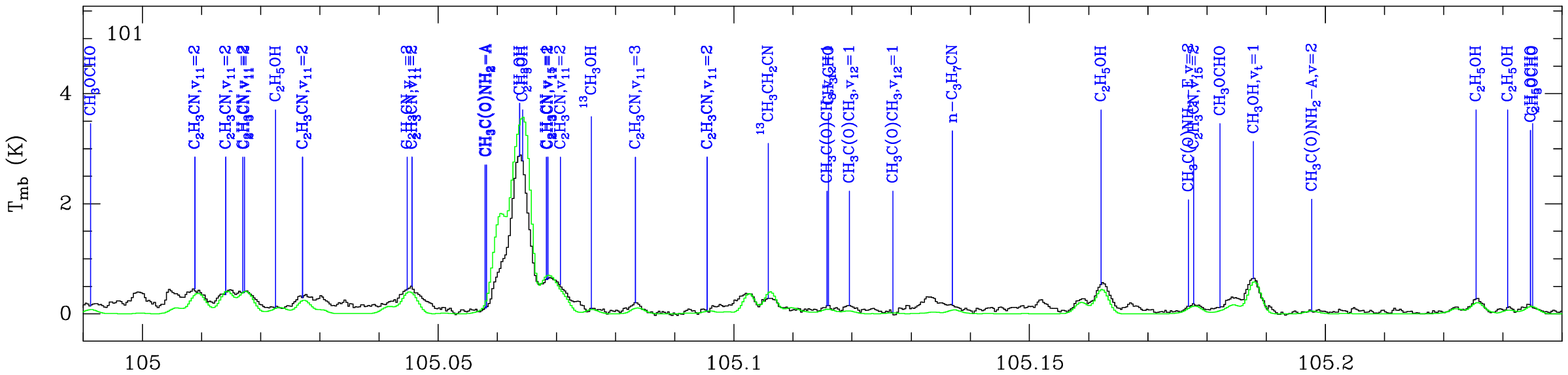}}}
\vspace*{1ex}\centerline{\resizebox{1.0\hsize}{!}{\includegraphics[angle=270]{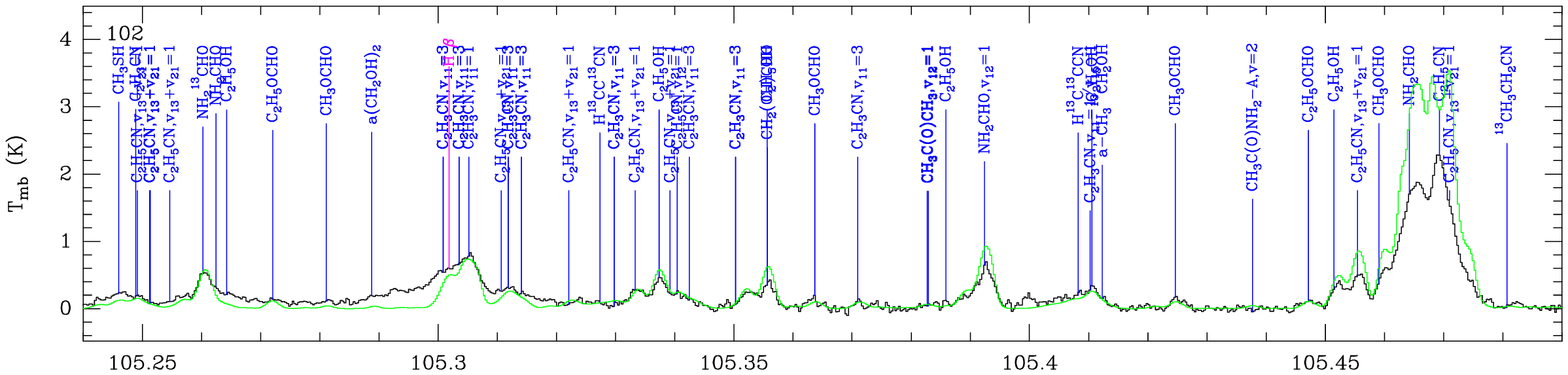}}}
\vspace*{1ex}\centerline{\resizebox{1.0\hsize}{!}{\includegraphics[angle=270]{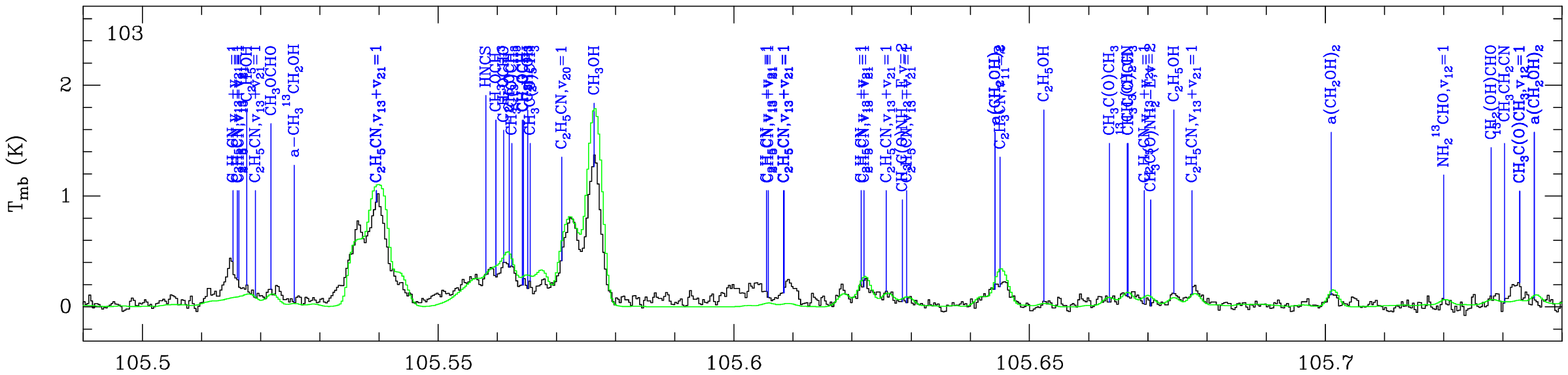}}}
\vspace*{1ex}\centerline{\resizebox{1.0\hsize}{!}{\includegraphics[angle=270]{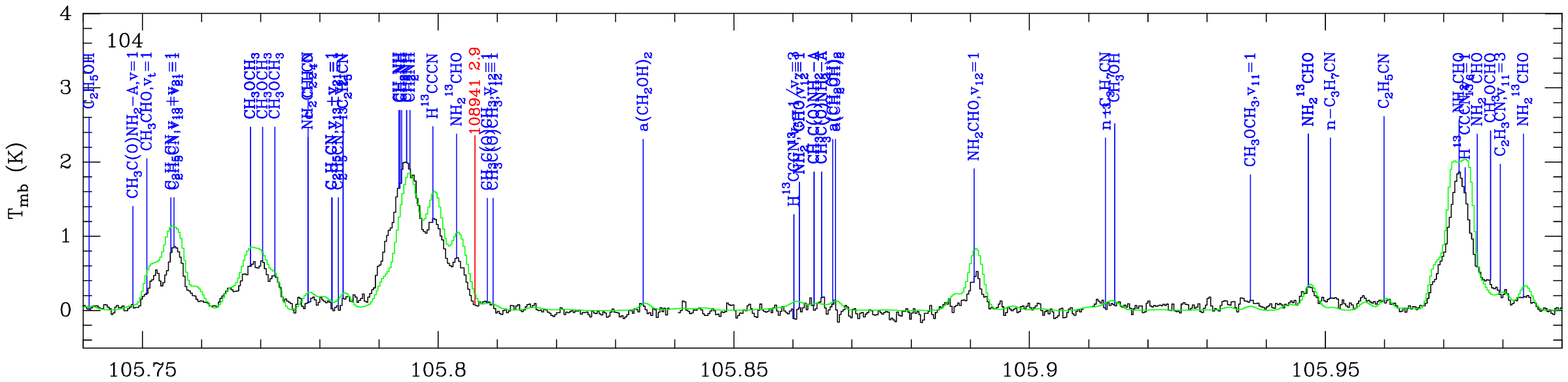}}}
\vspace*{1ex}\centerline{\resizebox{1.0\hsize}{!}{\includegraphics[angle=270]{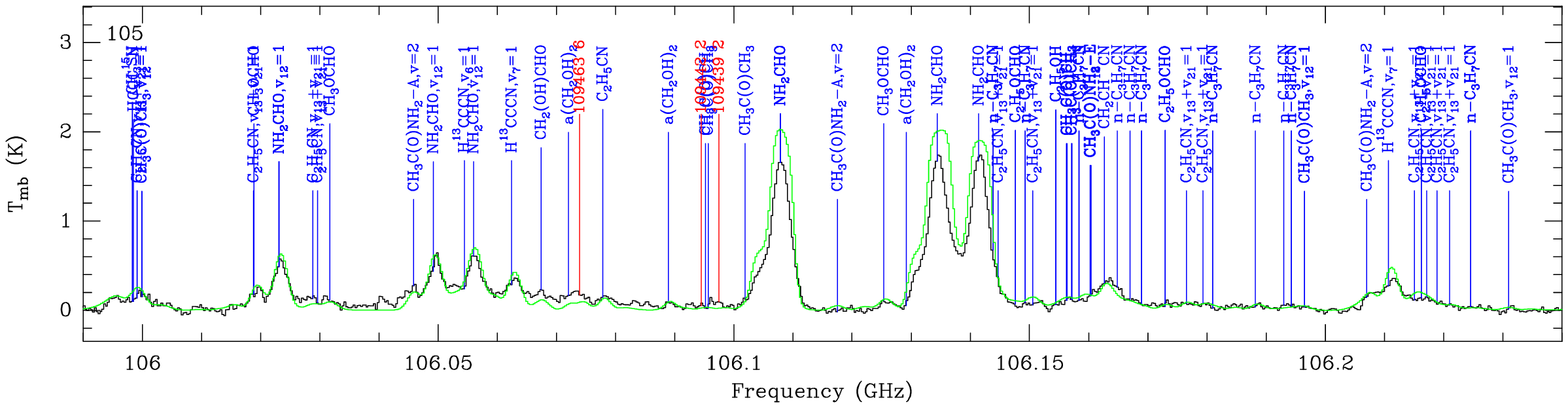}}}
\caption{
continued.
}
\end{figure*}
 \clearpage
\begin{figure*}
\addtocounter{figure}{-1}
\centerline{\resizebox{1.0\hsize}{!}{\includegraphics[angle=270]{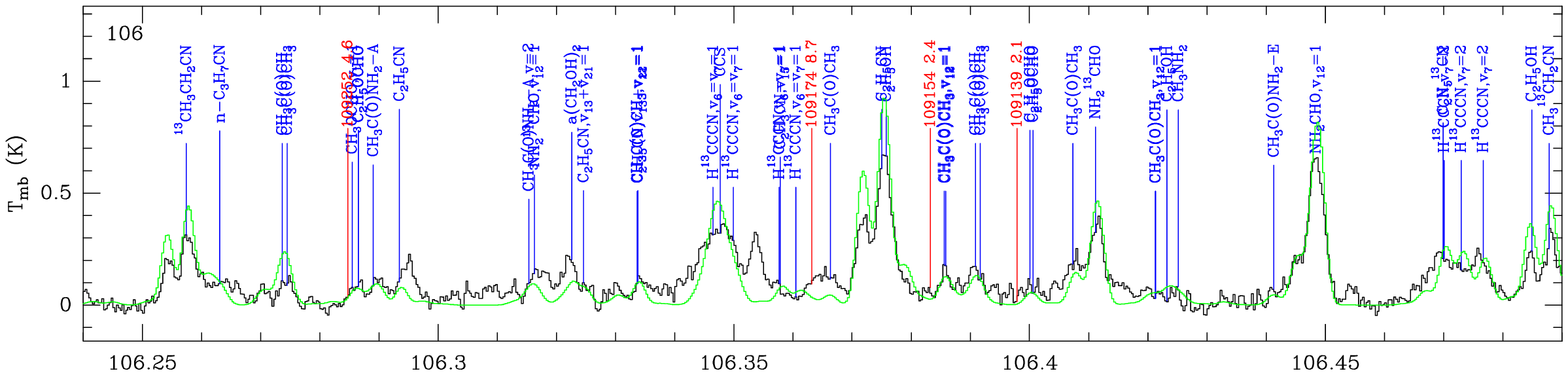}}}
\vspace*{1ex}\centerline{\resizebox{1.0\hsize}{!}{\includegraphics[angle=270]{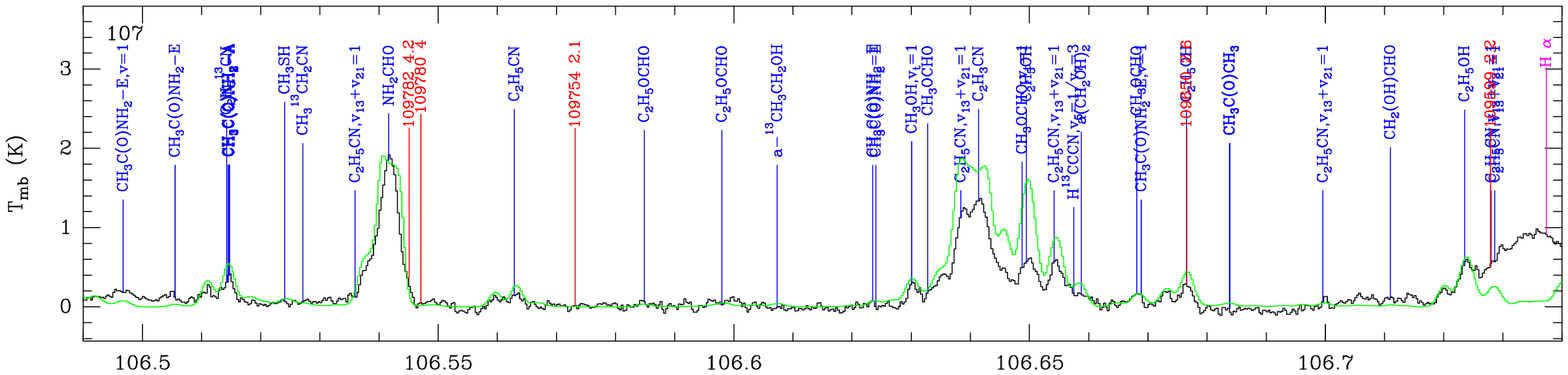}}}
\vspace*{1ex}\centerline{\resizebox{1.0\hsize}{!}{\includegraphics[angle=270]{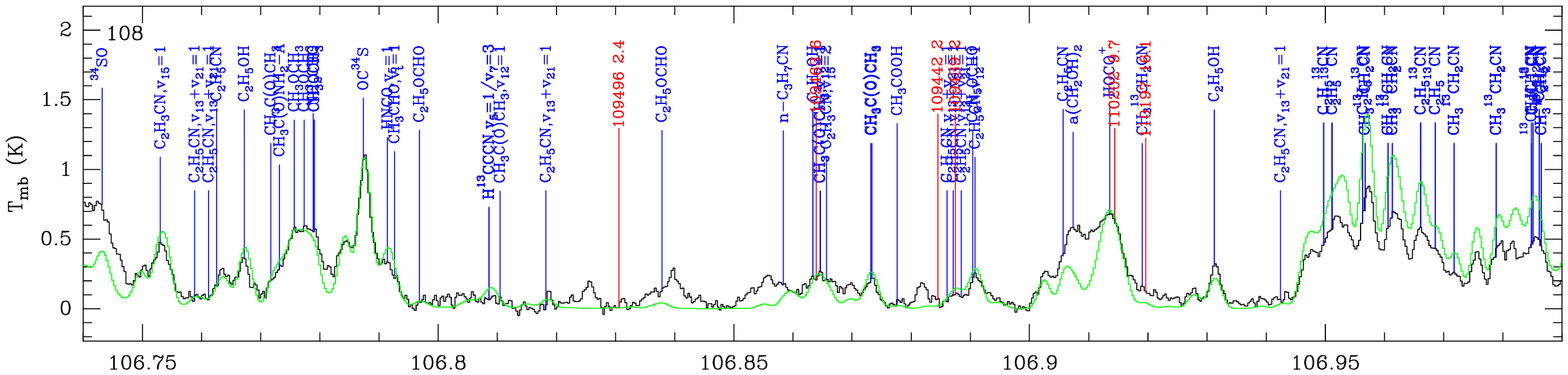}}}
\vspace*{1ex}\centerline{\resizebox{1.0\hsize}{!}{\includegraphics[angle=270]{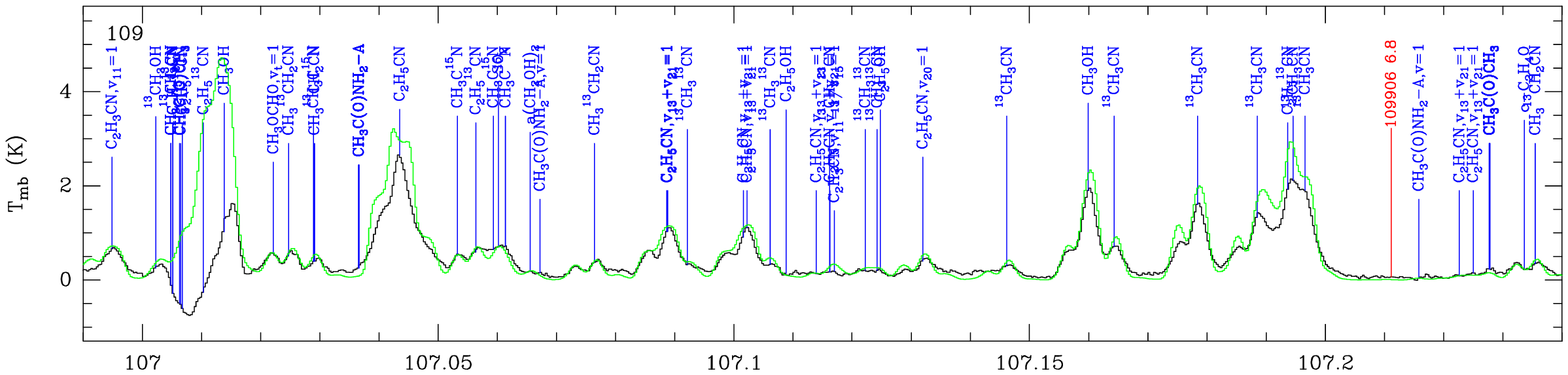}}}
\vspace*{1ex}\centerline{\resizebox{1.0\hsize}{!}{\includegraphics[angle=270]{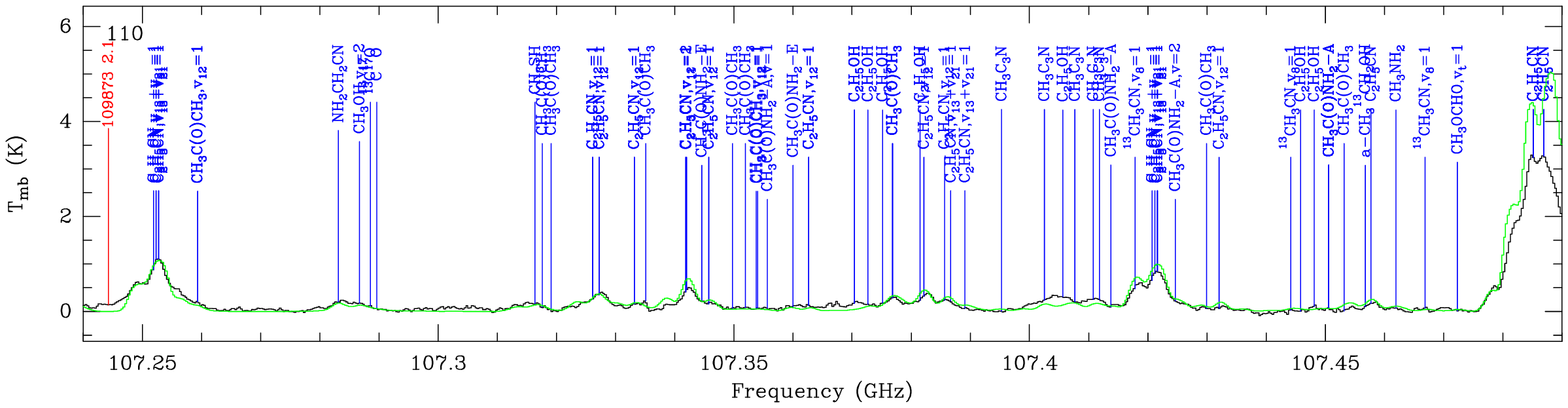}}}
\caption{
continued.
}
\end{figure*}
 \clearpage
\begin{figure*}
\addtocounter{figure}{-1}
\centerline{\resizebox{1.0\hsize}{!}{\includegraphics[angle=270]{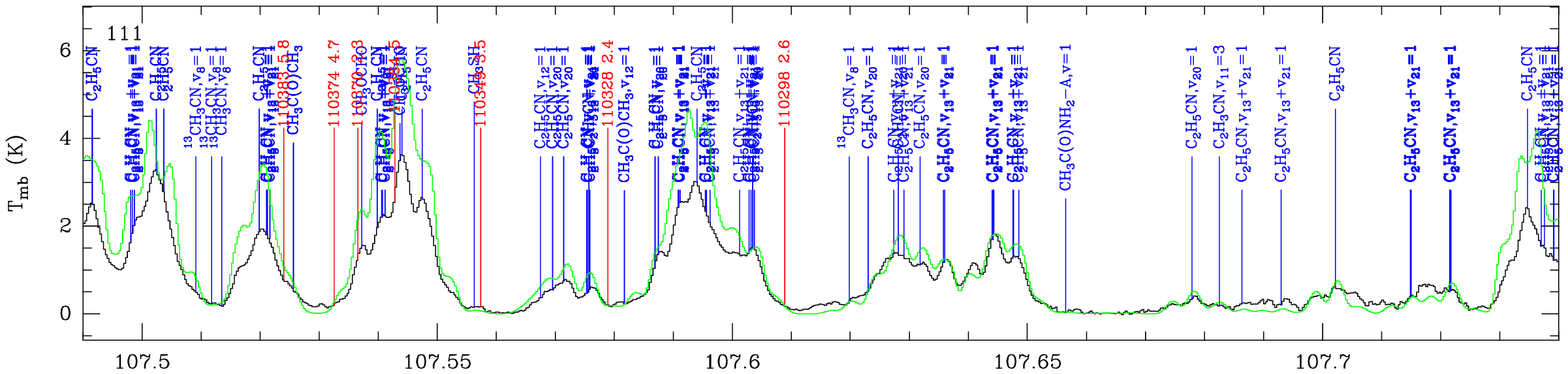}}}
\vspace*{1ex}\centerline{\resizebox{1.0\hsize}{!}{\includegraphics[angle=270]{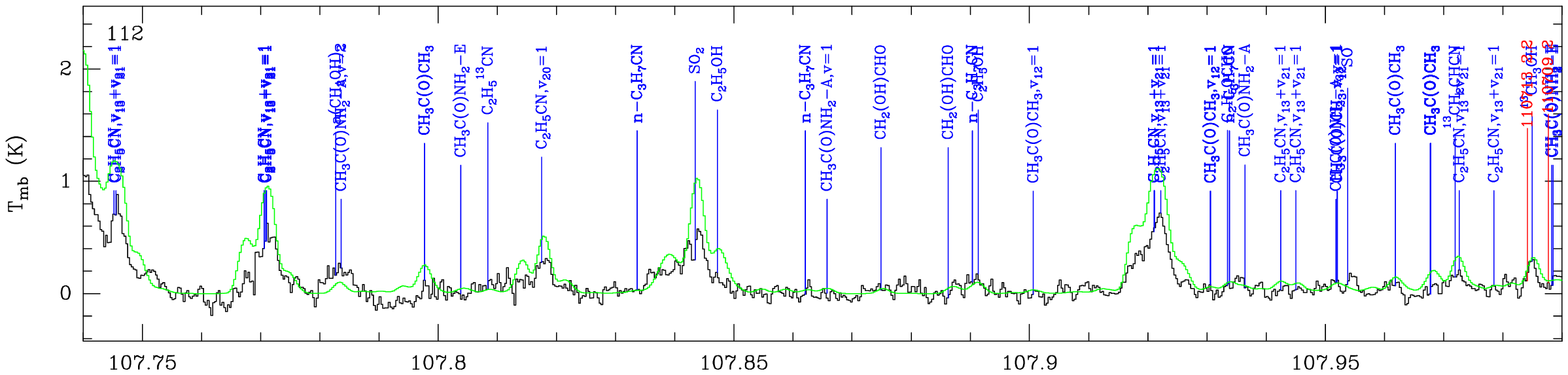}}}
\vspace*{1ex}\centerline{\resizebox{1.0\hsize}{!}{\includegraphics[angle=270]{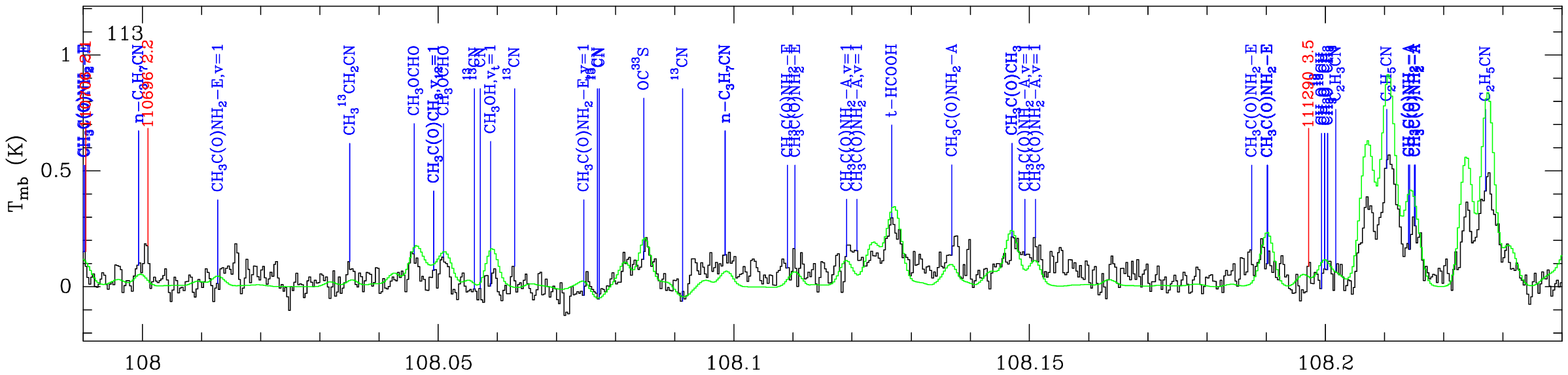}}}
\vspace*{1ex}\centerline{\resizebox{1.0\hsize}{!}{\includegraphics[angle=270]{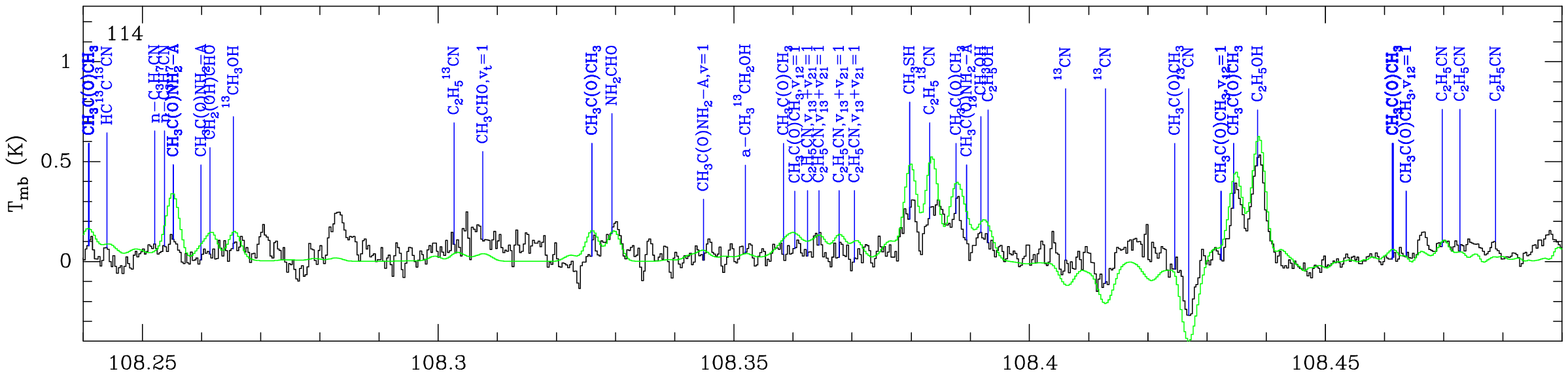}}}
\vspace*{1ex}\centerline{\resizebox{1.0\hsize}{!}{\includegraphics[angle=270]{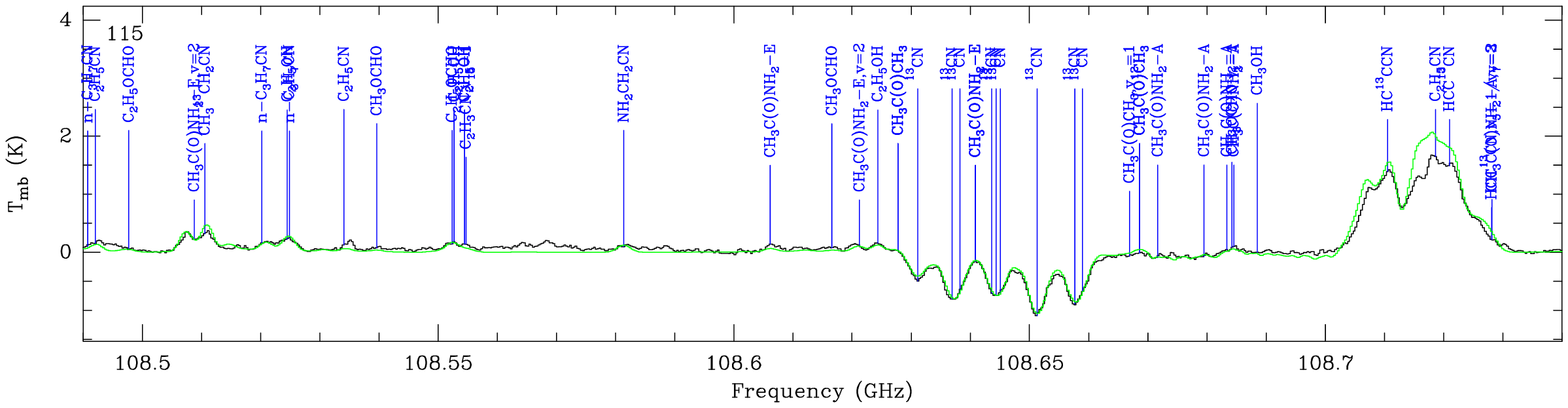}}}
\caption{
continued.
}
\end{figure*}
 \clearpage
\begin{figure*}
\addtocounter{figure}{-1}
\centerline{\resizebox{1.0\hsize}{!}{\includegraphics[angle=270]{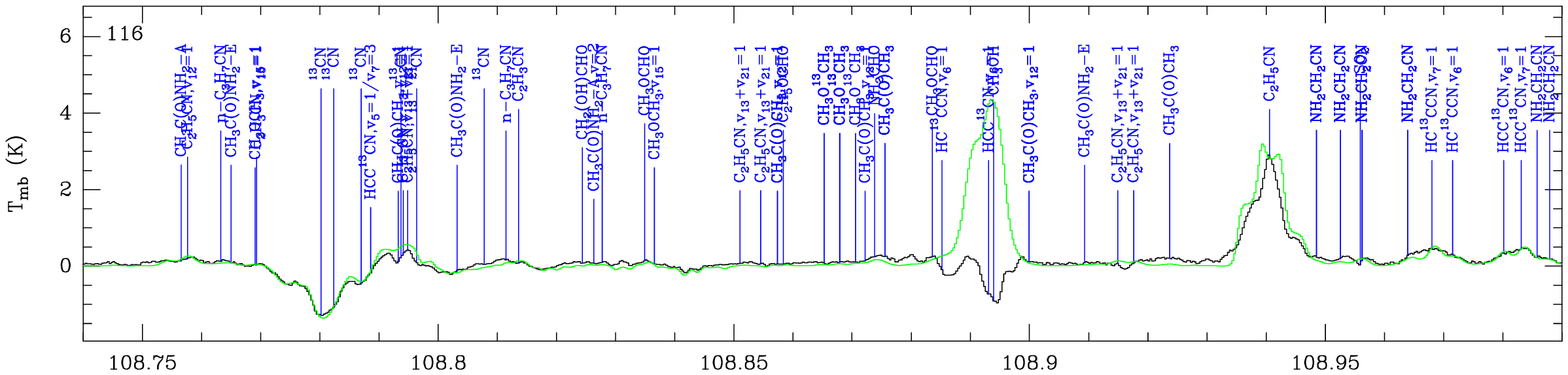}}}
\vspace*{1ex}\centerline{\resizebox{1.0\hsize}{!}{\includegraphics[angle=270]{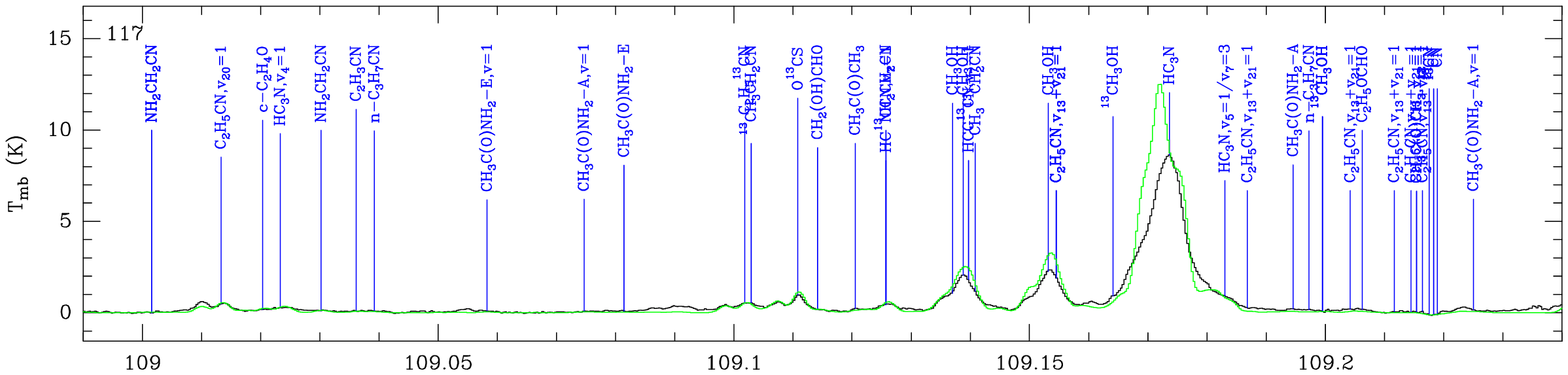}}}
\vspace*{1ex}\centerline{\resizebox{1.0\hsize}{!}{\includegraphics[angle=270]{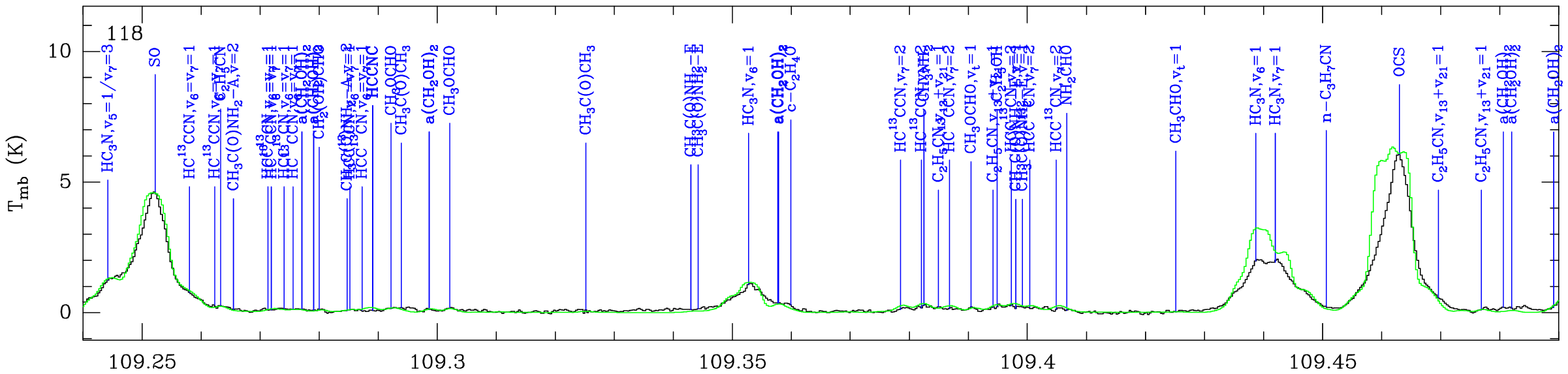}}}
\vspace*{1ex}\centerline{\resizebox{1.0\hsize}{!}{\includegraphics[angle=270]{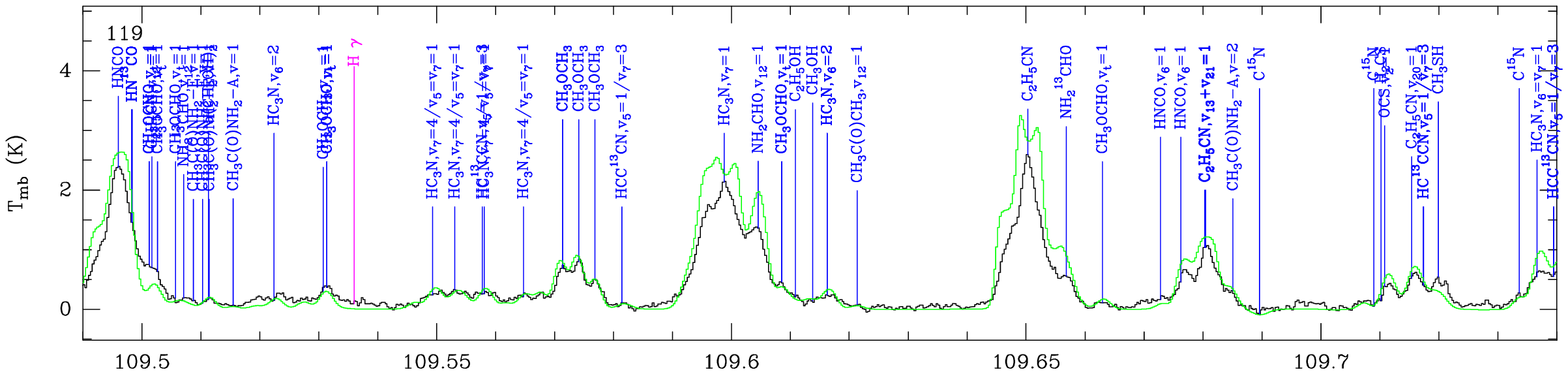}}}
\vspace*{1ex}\centerline{\resizebox{1.0\hsize}{!}{\includegraphics[angle=270]{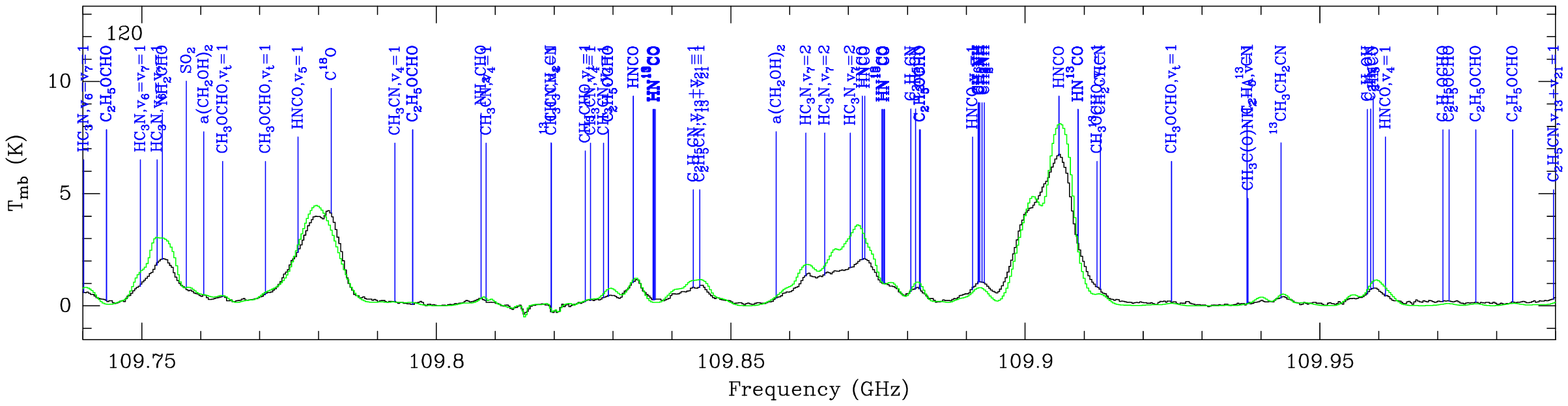}}}
\caption{
continued.
}
\end{figure*}
 \clearpage
\begin{figure*}
\addtocounter{figure}{-1}
\centerline{\resizebox{1.0\hsize}{!}{\includegraphics[angle=270]{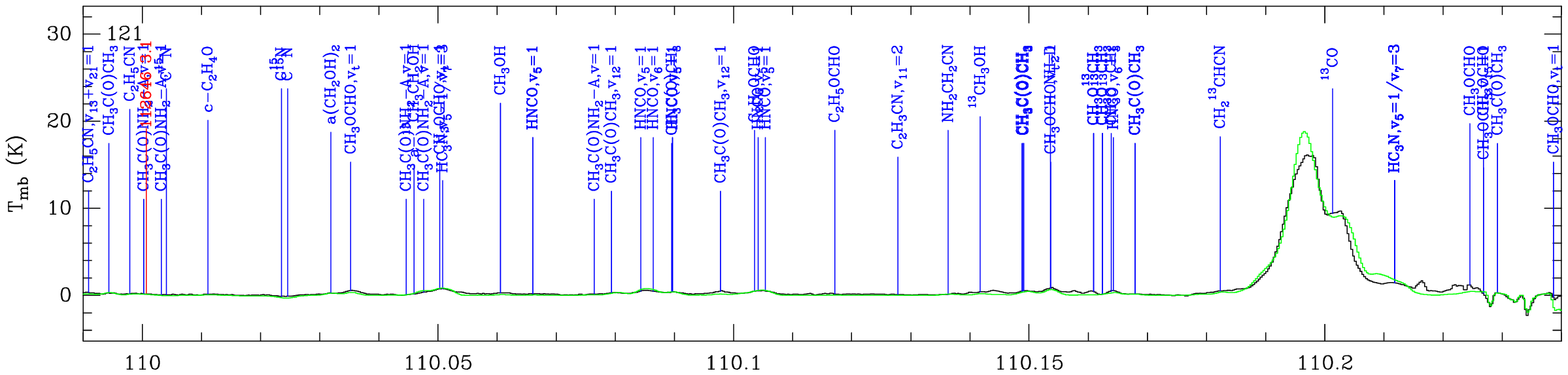}}}
\vspace*{1ex}\centerline{\resizebox{1.0\hsize}{!}{\includegraphics[angle=270]{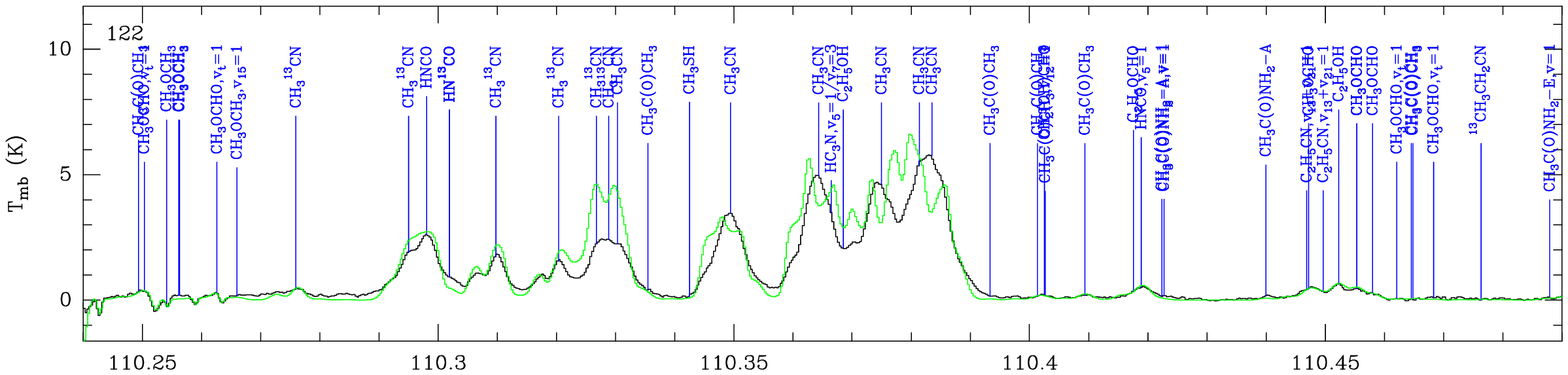}}}
\vspace*{1ex}\centerline{\resizebox{1.0\hsize}{!}{\includegraphics[angle=270]{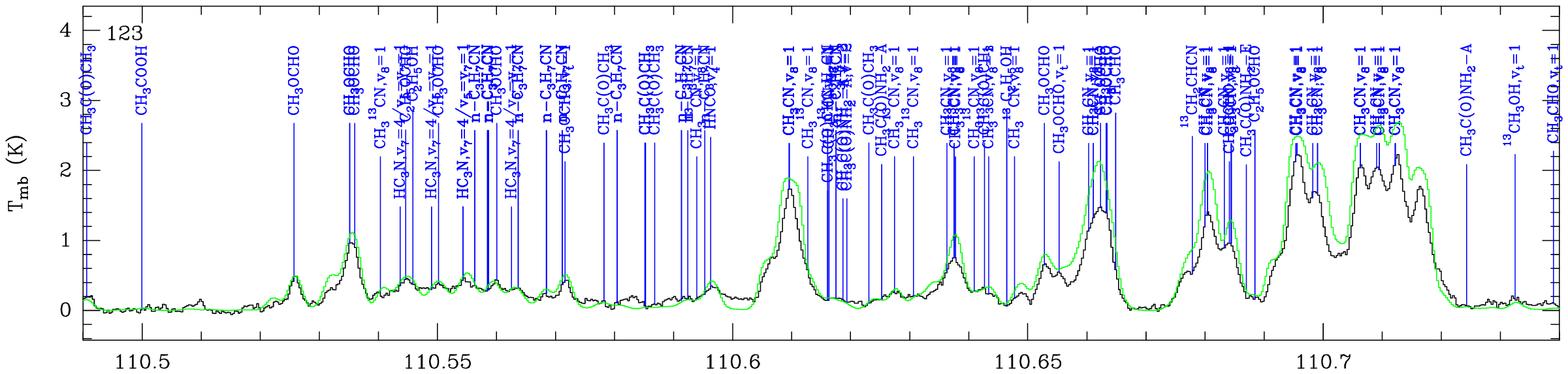}}}
\vspace*{1ex}\centerline{\resizebox{1.0\hsize}{!}{\includegraphics[angle=270]{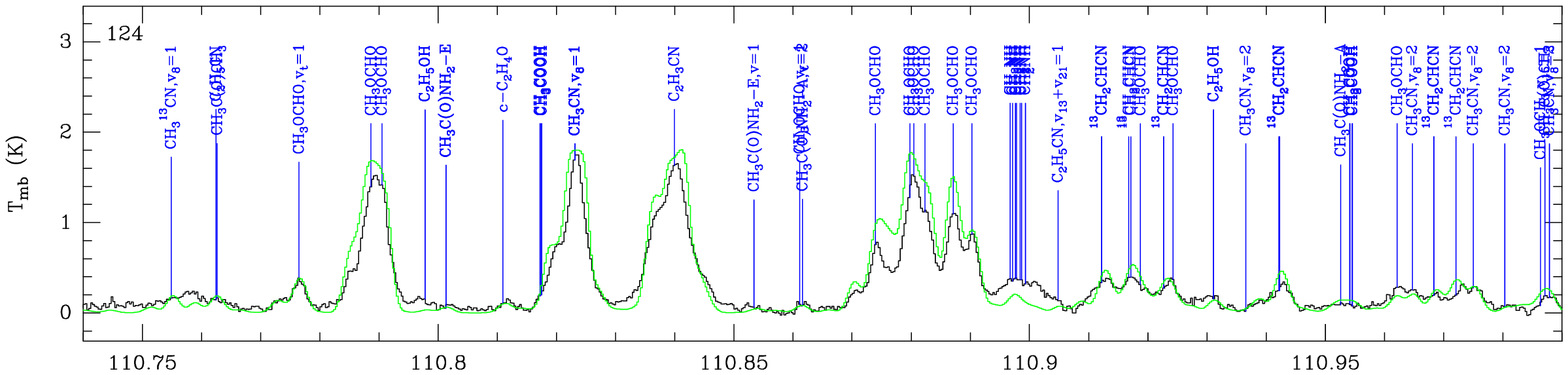}}}
\vspace*{1ex}\centerline{\resizebox{1.0\hsize}{!}{\includegraphics[angle=270]{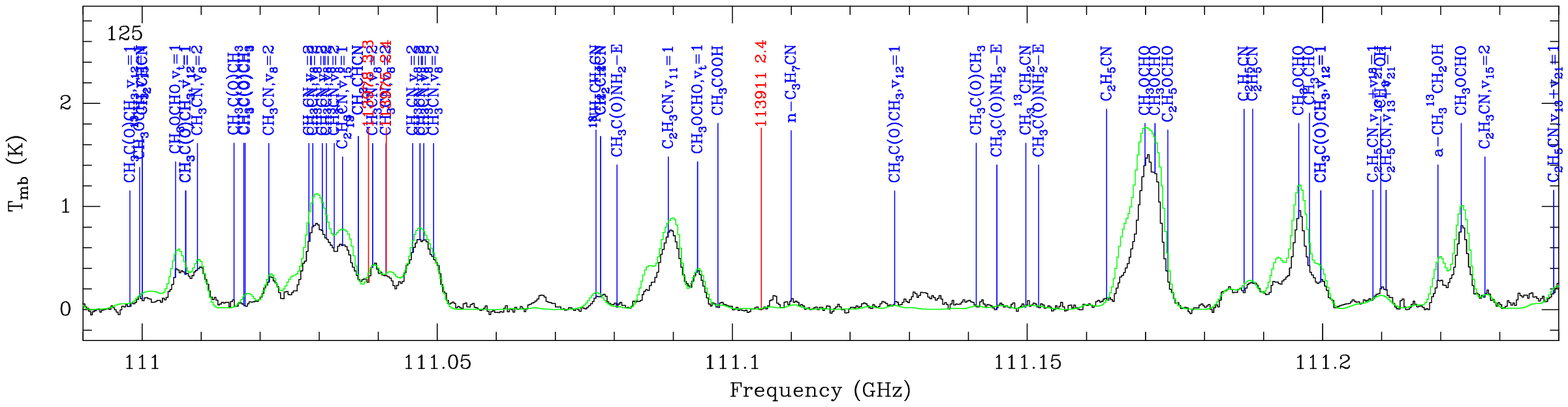}}}
\caption{
continued.
}
\end{figure*}
 \clearpage
\begin{figure*}
\addtocounter{figure}{-1}
\centerline{\resizebox{1.0\hsize}{!}{\includegraphics[angle=270]{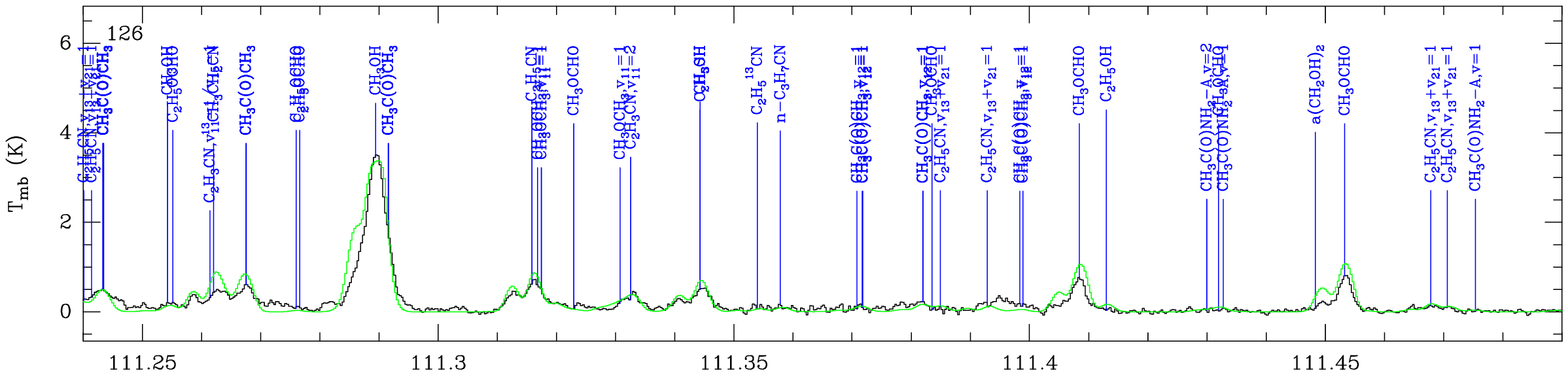}}}
\vspace*{1ex}\centerline{\resizebox{1.0\hsize}{!}{\includegraphics[angle=270]{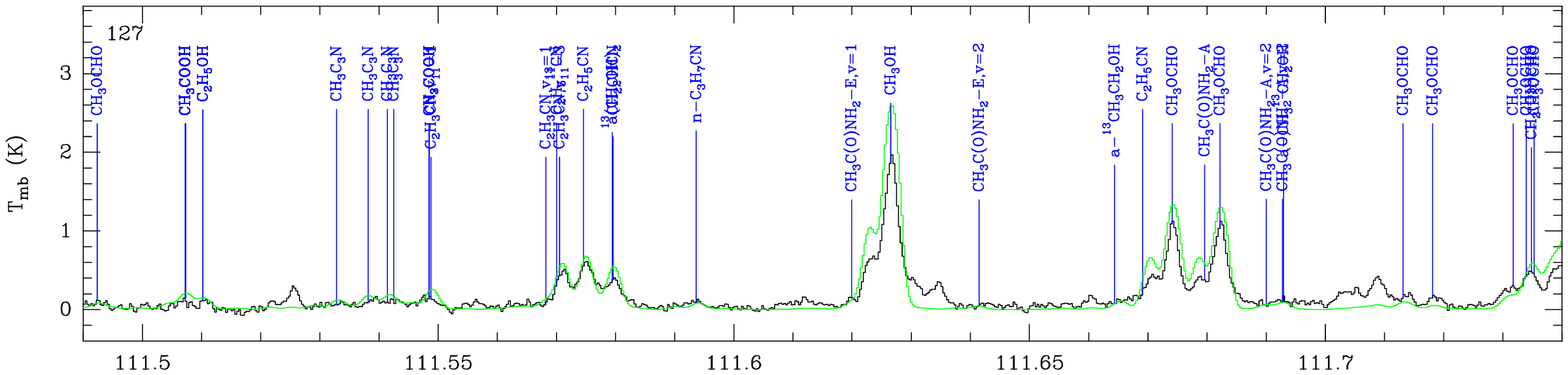}}}
\vspace*{1ex}\centerline{\resizebox{1.0\hsize}{!}{\includegraphics[angle=270]{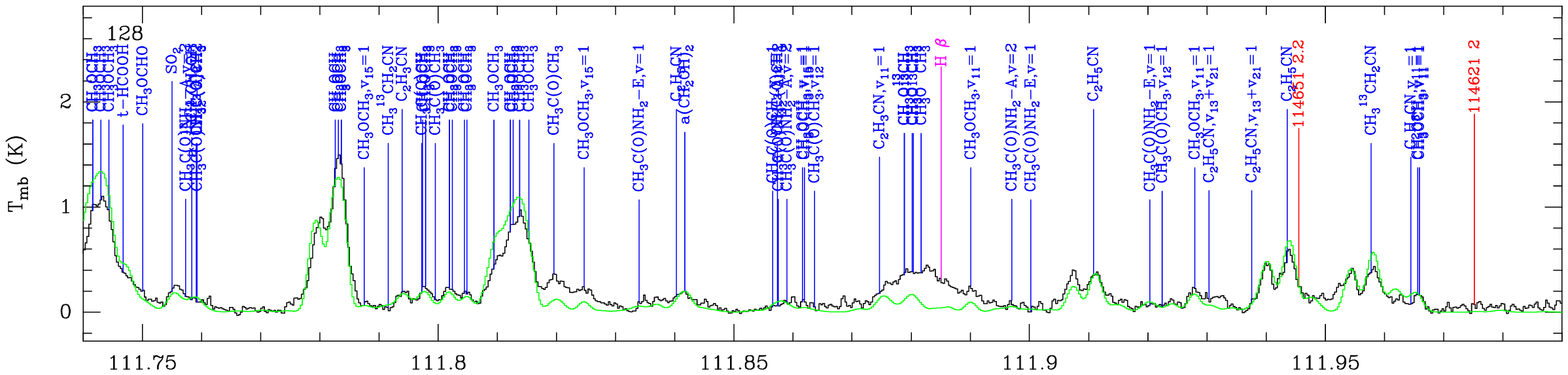}}}
\vspace*{1ex}\centerline{\resizebox{1.0\hsize}{!}{\includegraphics[angle=270]{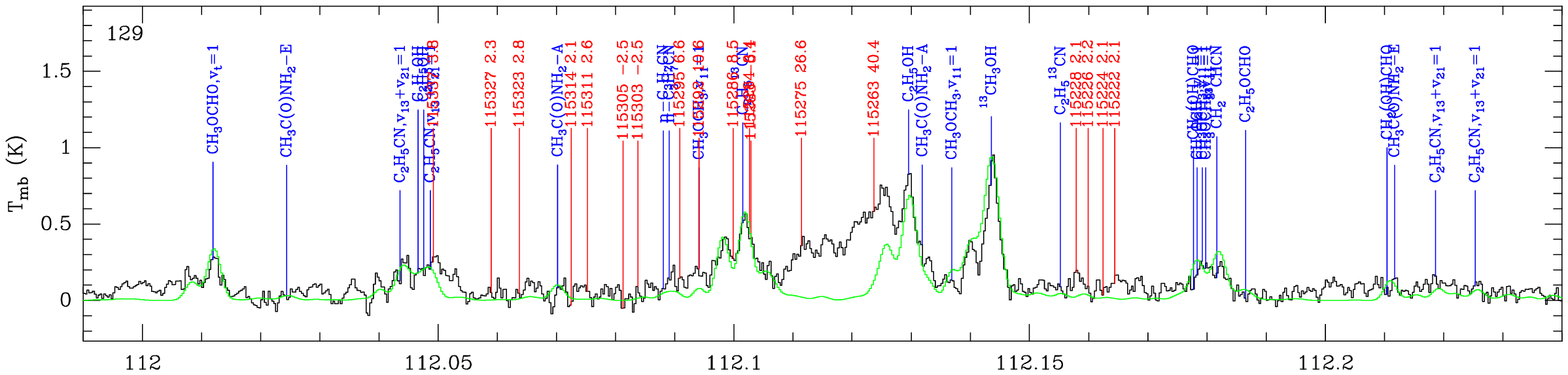}}}
\vspace*{1ex}\centerline{\resizebox{1.0\hsize}{!}{\includegraphics[angle=270]{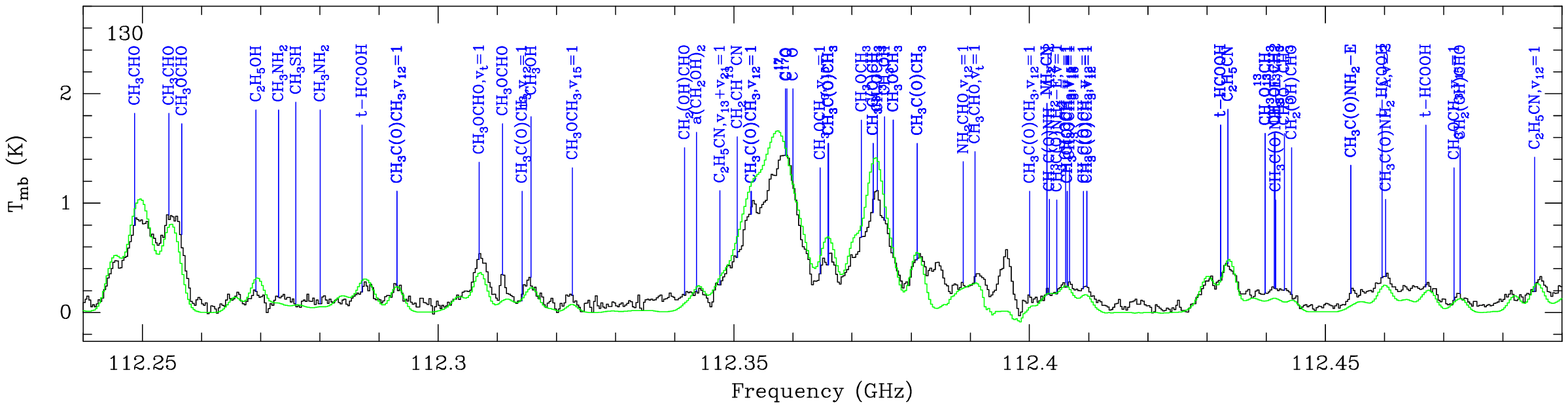}}}
\caption{
continued.
}
\end{figure*}
 \clearpage
\begin{figure*}
\addtocounter{figure}{-1}
\centerline{\resizebox{1.0\hsize}{!}{\includegraphics[angle=270]{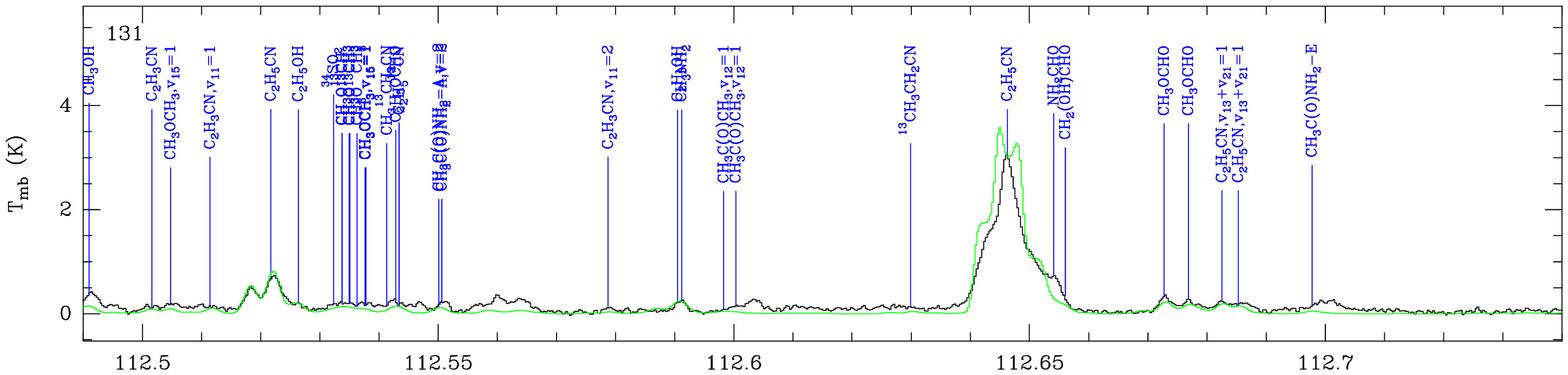}}}
\vspace*{1ex}\centerline{\resizebox{1.0\hsize}{!}{\includegraphics[angle=270]{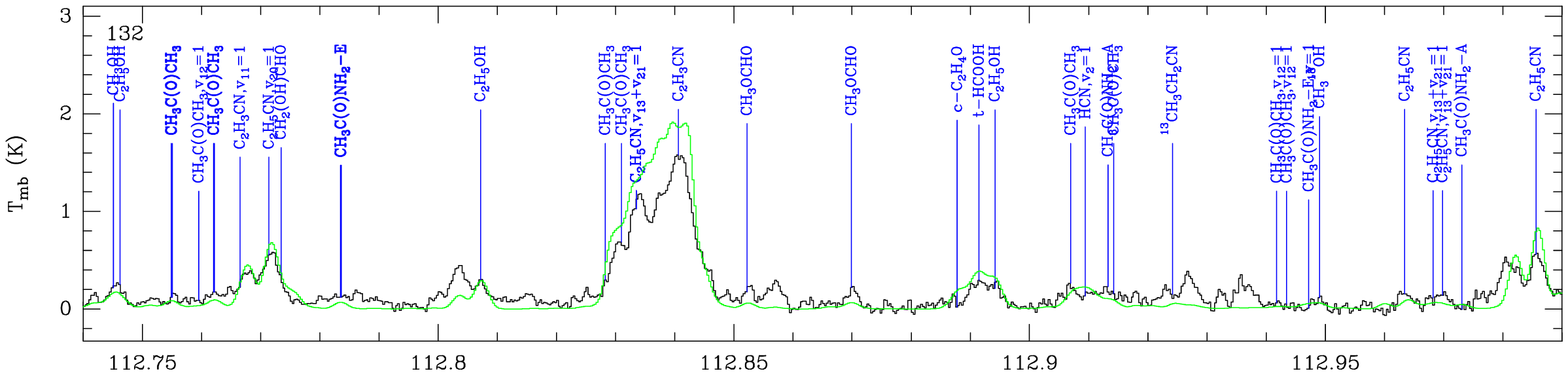}}}
\vspace*{1ex}\centerline{\resizebox{1.0\hsize}{!}{\includegraphics[angle=270]{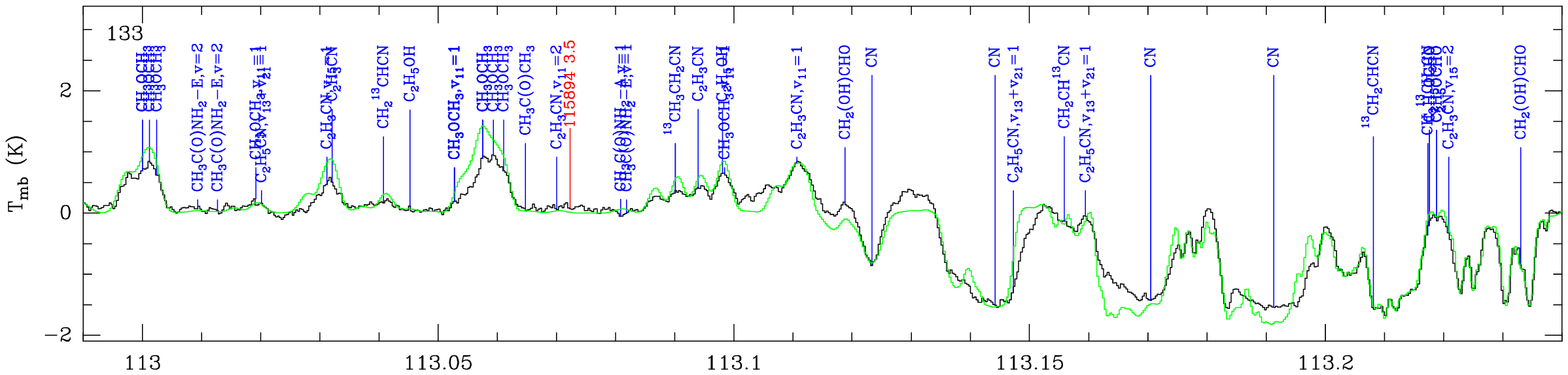}}}
\vspace*{1ex}\centerline{\resizebox{1.0\hsize}{!}{\includegraphics[angle=270]{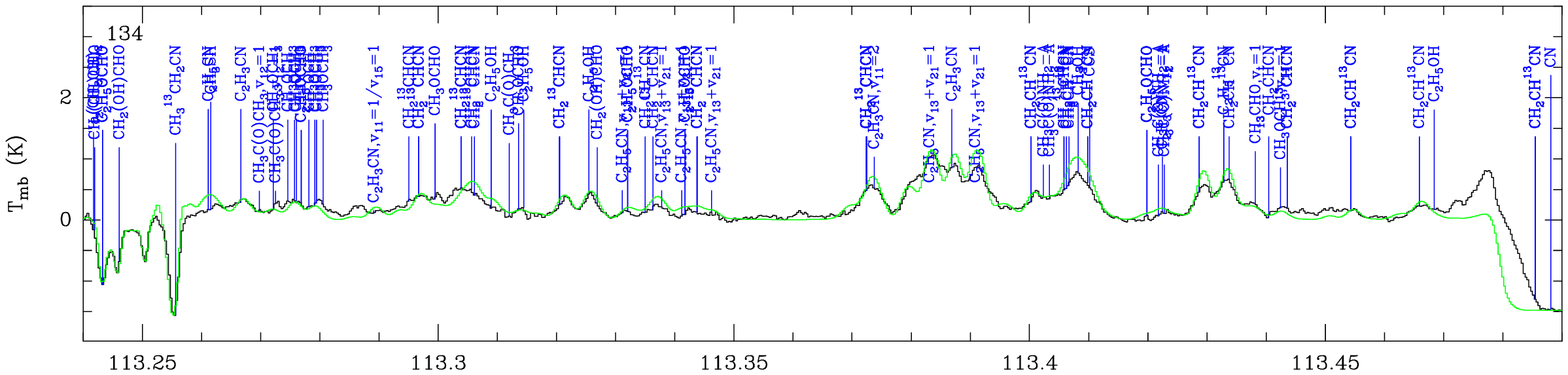}}}
\vspace*{1ex}\centerline{\resizebox{1.0\hsize}{!}{\includegraphics[angle=270]{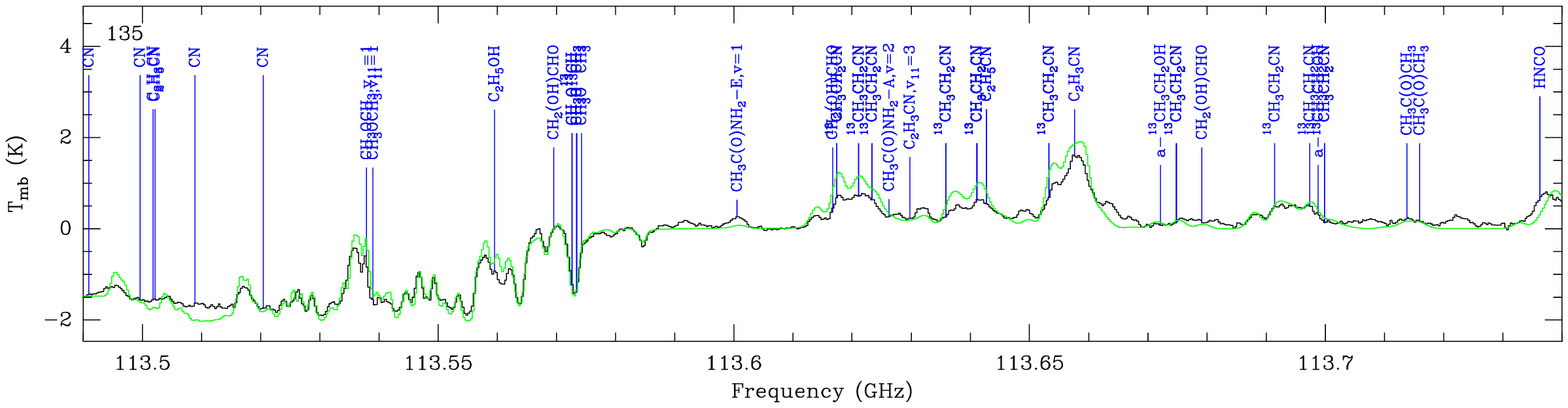}}}
\caption{
continued.
}
\end{figure*}
 \clearpage
\begin{figure*}
\addtocounter{figure}{-1}
\centerline{\resizebox{1.0\hsize}{!}{\includegraphics[angle=270]{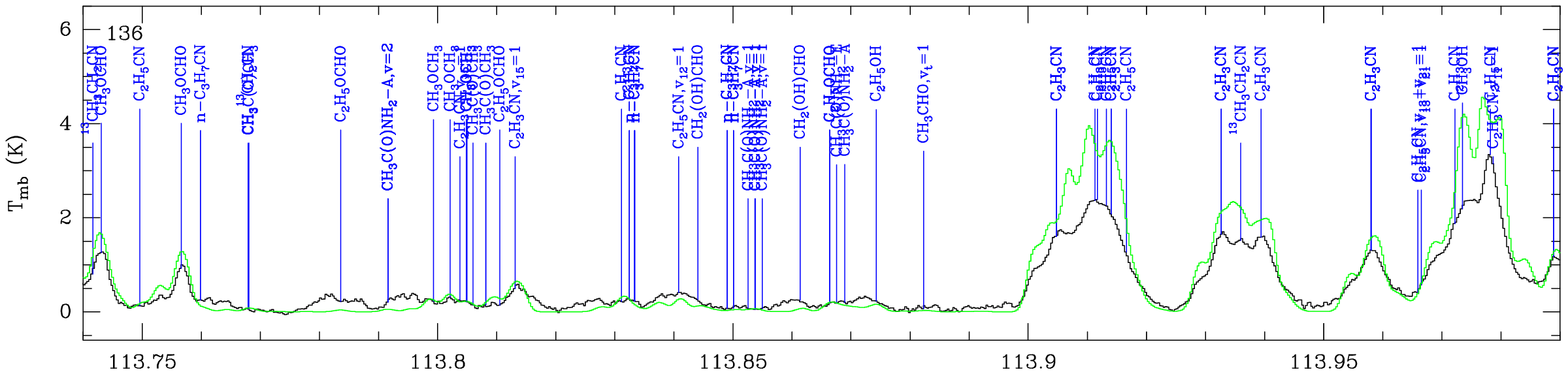}}}
\vspace*{1ex}\centerline{\resizebox{1.0\hsize}{!}{\includegraphics[angle=270]{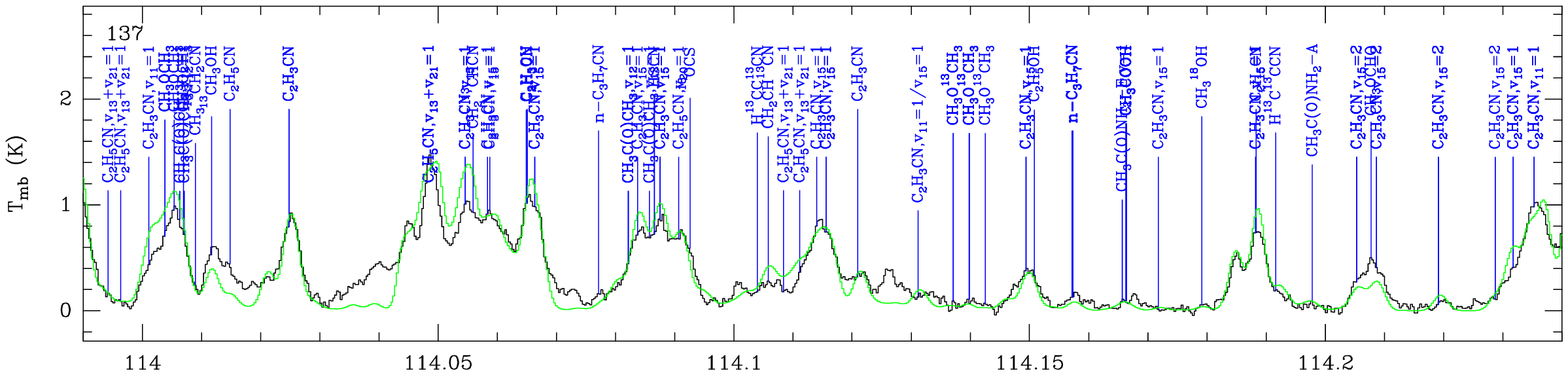}}}
\vspace*{1ex}\centerline{\resizebox{1.0\hsize}{!}{\includegraphics[angle=270]{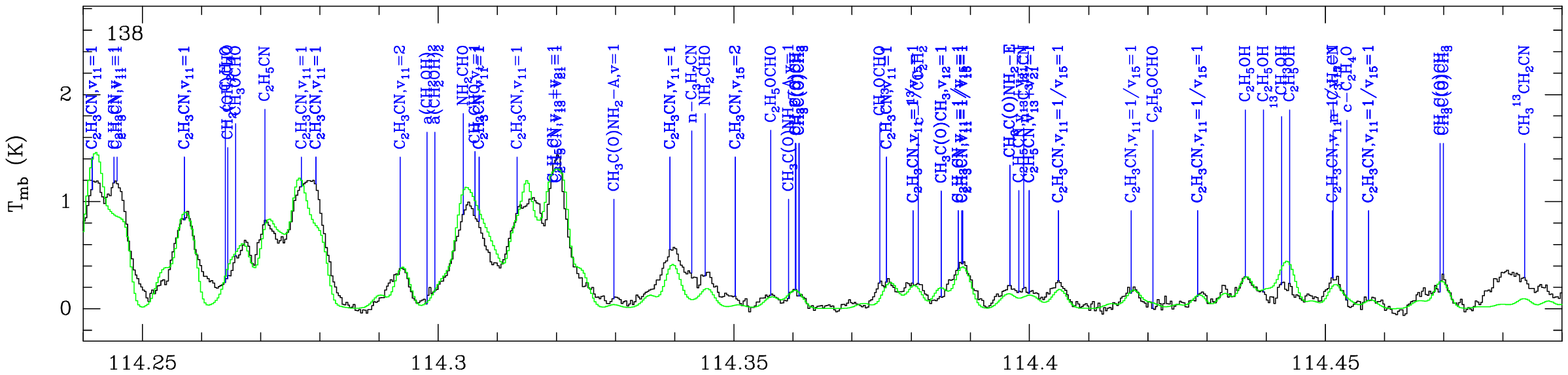}}}
\vspace*{1ex}\centerline{\resizebox{1.0\hsize}{!}{\includegraphics[angle=270]{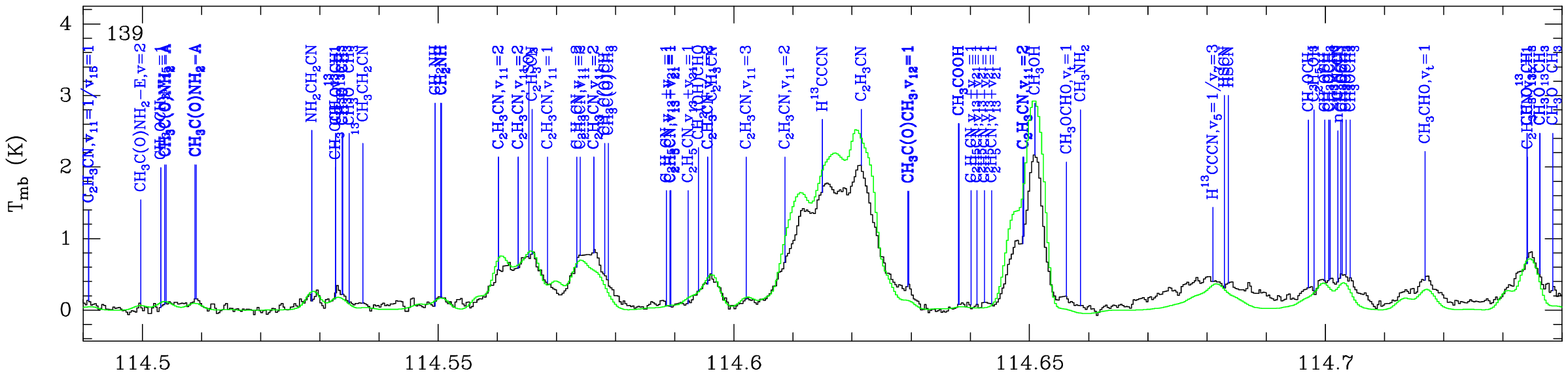}}}
\vspace*{1ex}\centerline{\resizebox{1.0\hsize}{!}{\includegraphics[angle=270]{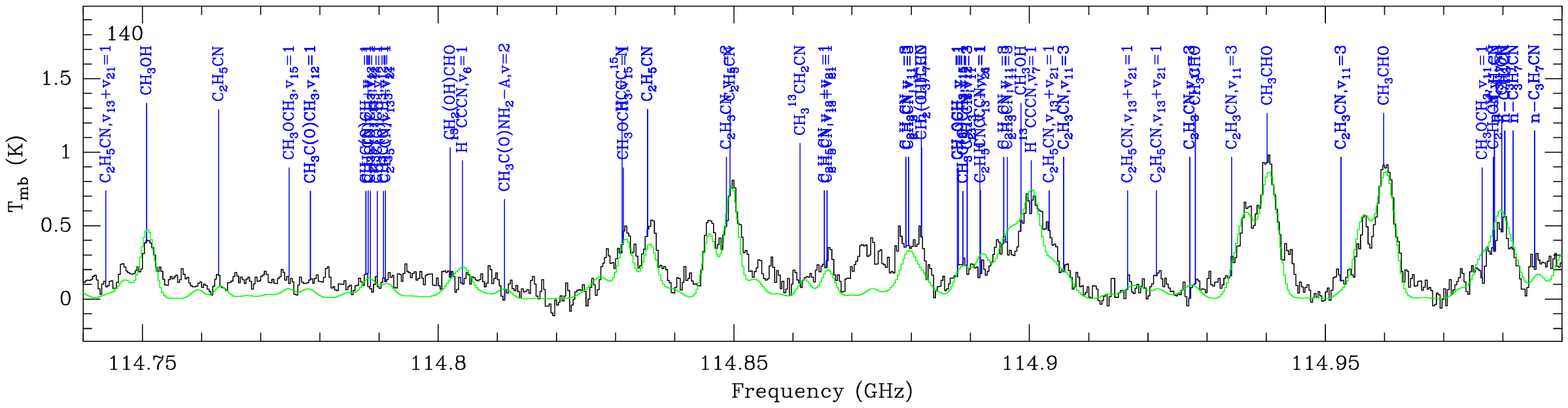}}}
\caption{
continued.
}
\end{figure*}
 \clearpage
\begin{figure*}
\addtocounter{figure}{-1}
\centerline{\resizebox{1.0\hsize}{!}{\includegraphics[angle=270]{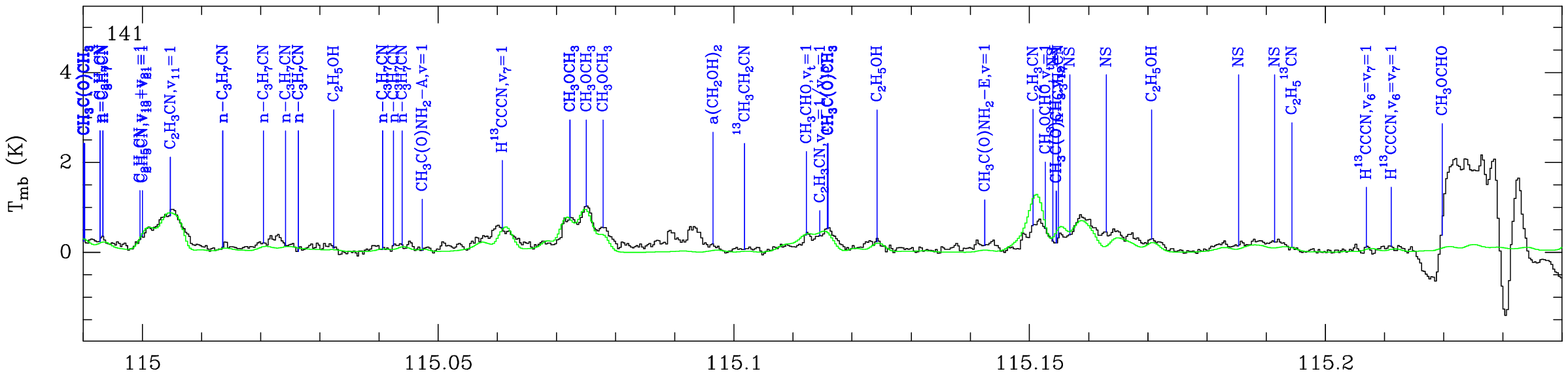}}}
\vspace*{1ex}\centerline{\resizebox{1.0\hsize}{!}{\includegraphics[angle=270]{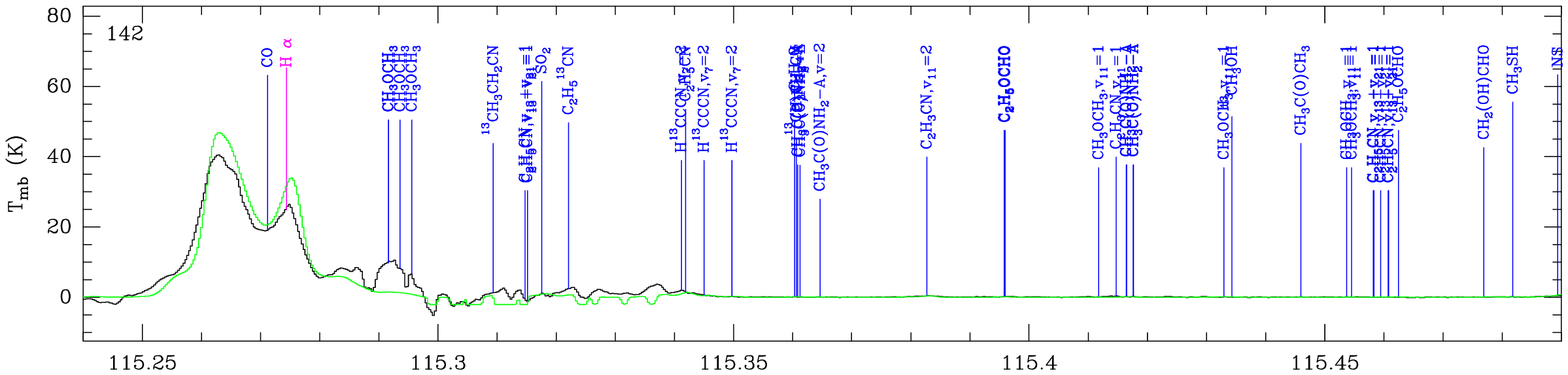}}}
\vspace*{1ex}\centerline{\resizebox{1.0\hsize}{!}{\includegraphics[angle=270]{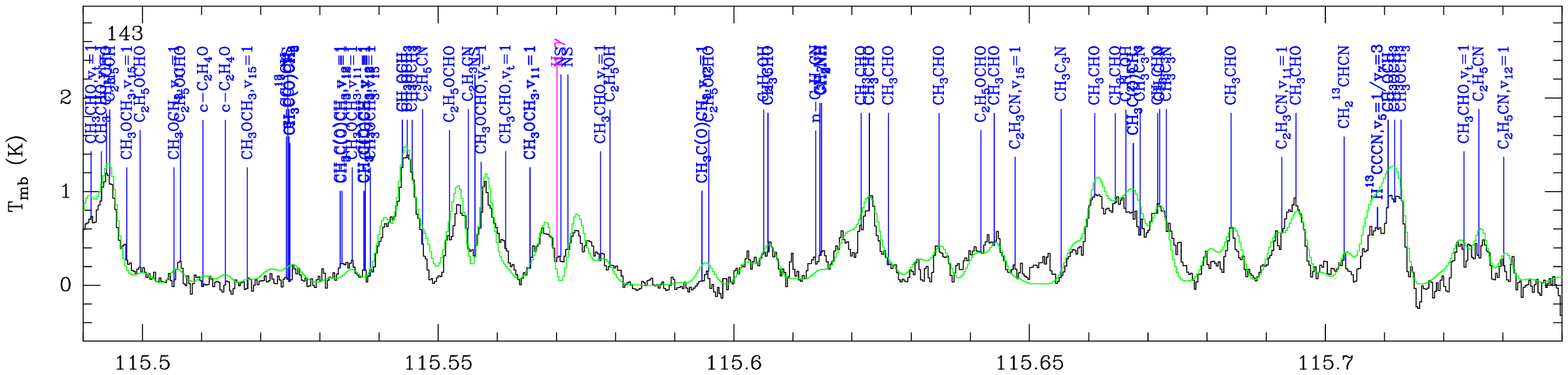}}}
\vspace*{1ex}\centerline{\resizebox{1.0\hsize}{!}{\includegraphics[angle=270]{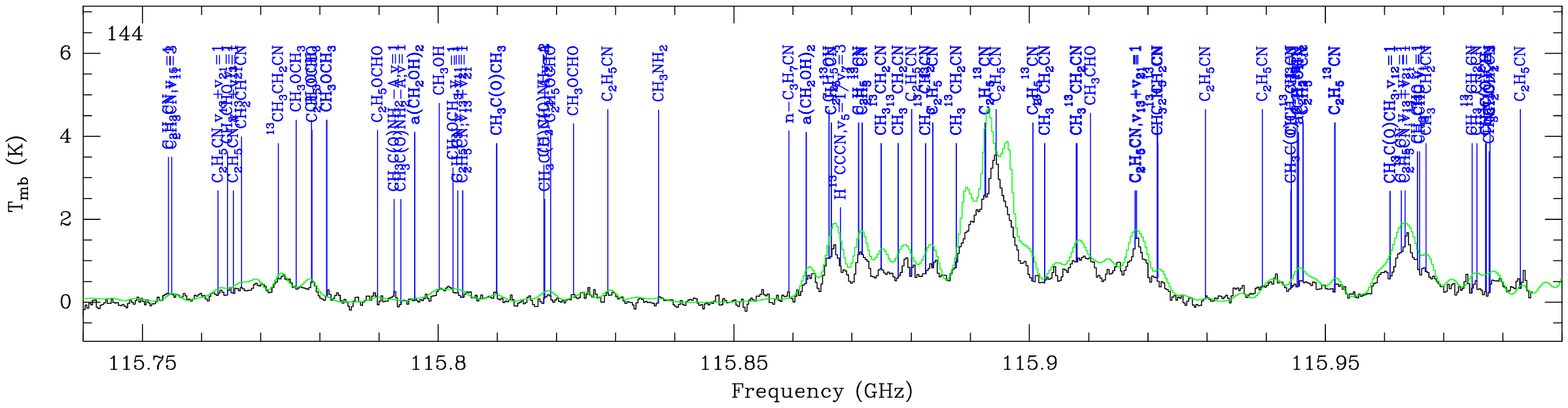}}}
\caption{
continued.
}
\end{figure*}
 \clearpage

}
\onlfig{\clearpage
\begin{figure*}
\centerline{\resizebox{1.0\hsize}{!}{\includegraphics[angle=270]{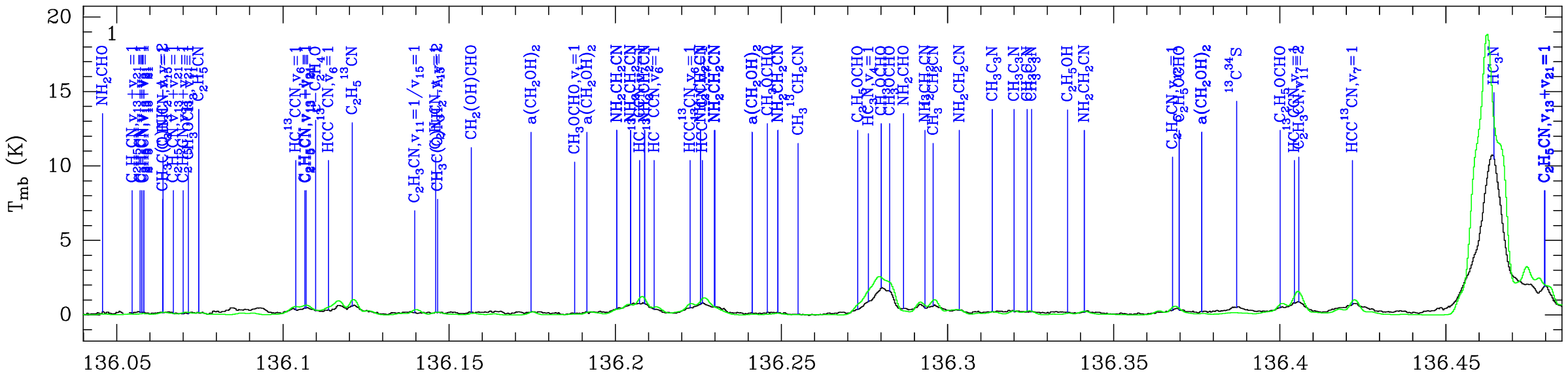}}}
\vspace*{1ex}\centerline{\resizebox{1.0\hsize}{!}{\includegraphics[angle=270]{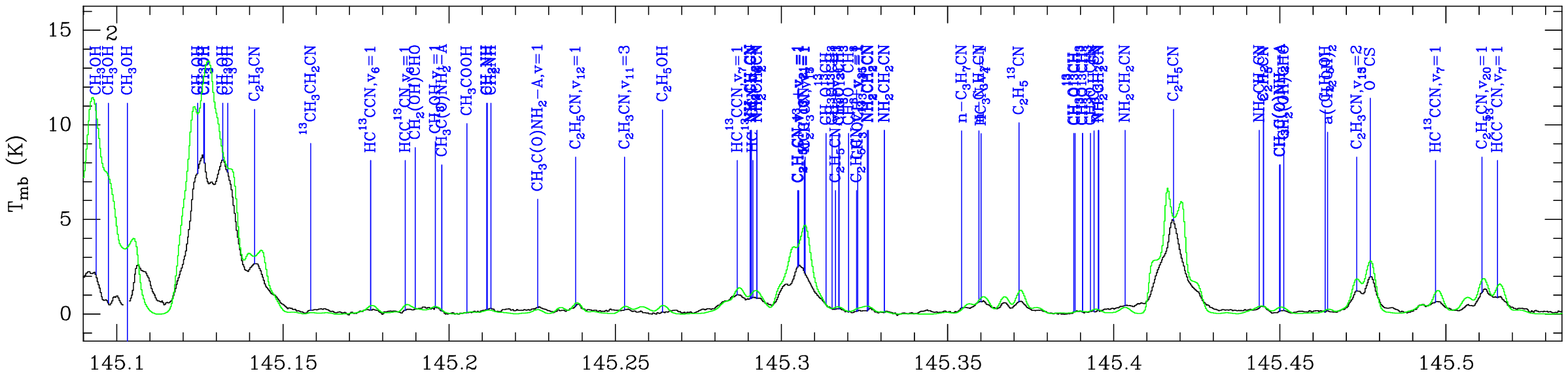}}}
\vspace*{1ex}\centerline{\resizebox{1.0\hsize}{!}{\includegraphics[angle=270]{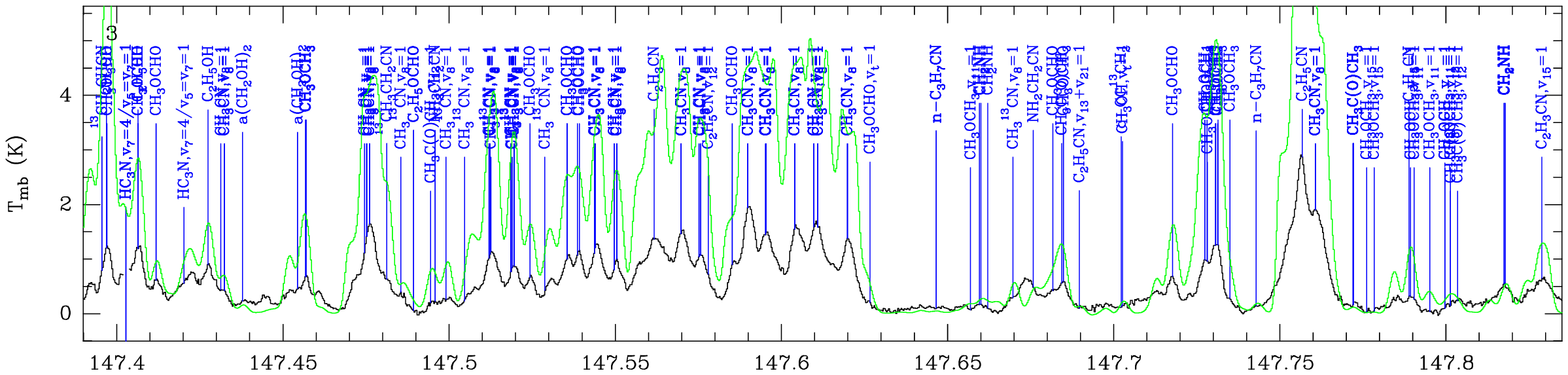}}}
\vspace*{1ex}\centerline{\resizebox{1.0\hsize}{!}{\includegraphics[angle=270]{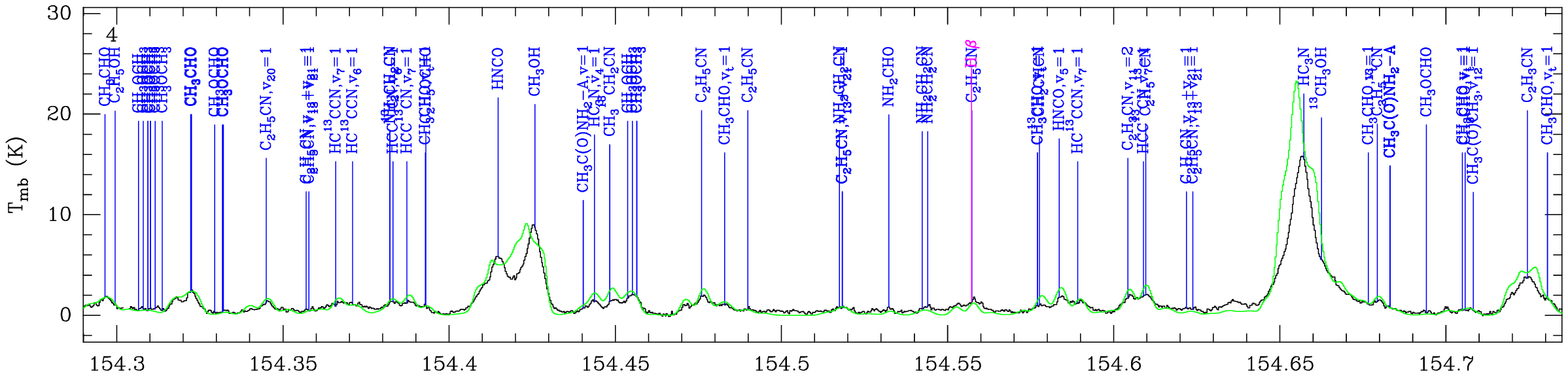}}}
\vspace*{1ex}\centerline{\resizebox{1.0\hsize}{!}{\includegraphics[angle=270]{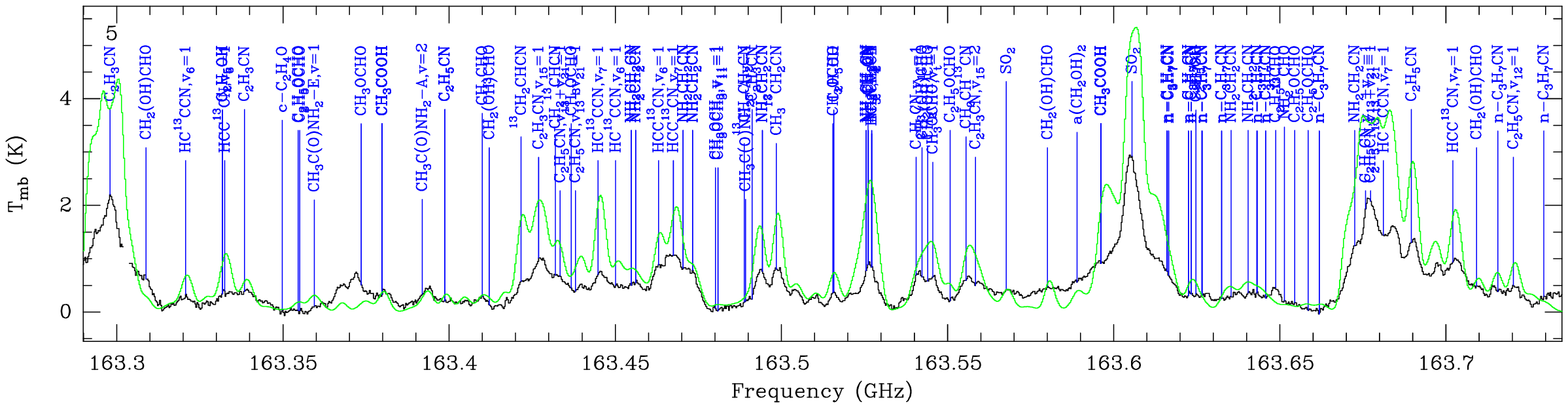}}}
\caption{
Spectrum obtained toward Sgr~B2(N)
in the 2~mm window
with the IRAM~30\,m telescope in main-beam temperature scale. The synthetic model is overlaid in green and its relevant lines are labeled in blue.
The frequencies of the hydrogen recombination lines are indicated with a pink label.
}
\label{f:survey_lmh_2mm}
\end{figure*}
 \clearpage
\begin{figure*}
\addtocounter{figure}{-1}
\centerline{\resizebox{1.0\hsize}{!}{\includegraphics[angle=270]{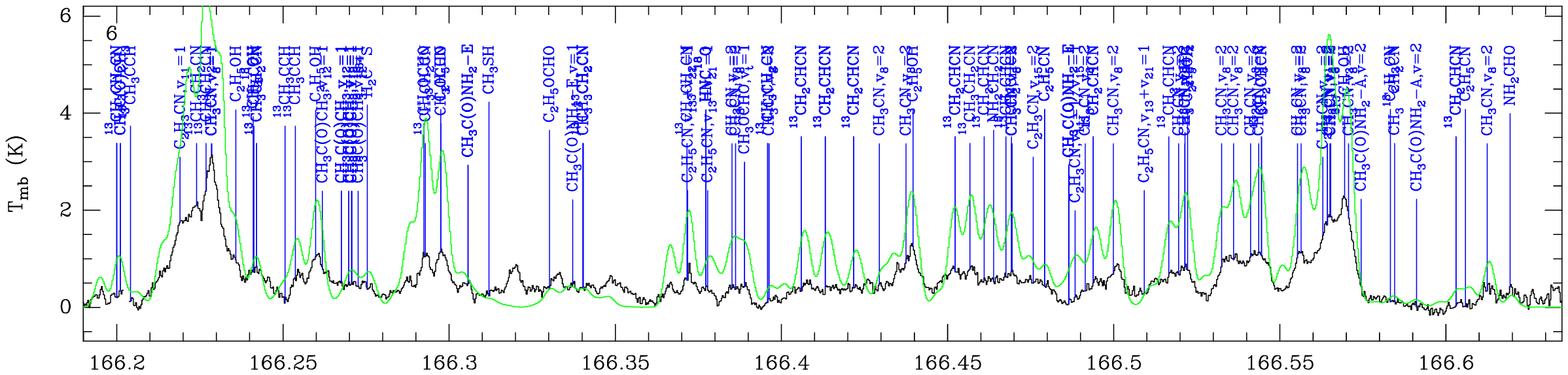}}}
\vspace*{1ex}\centerline{\resizebox{1.0\hsize}{!}{\includegraphics[angle=270]{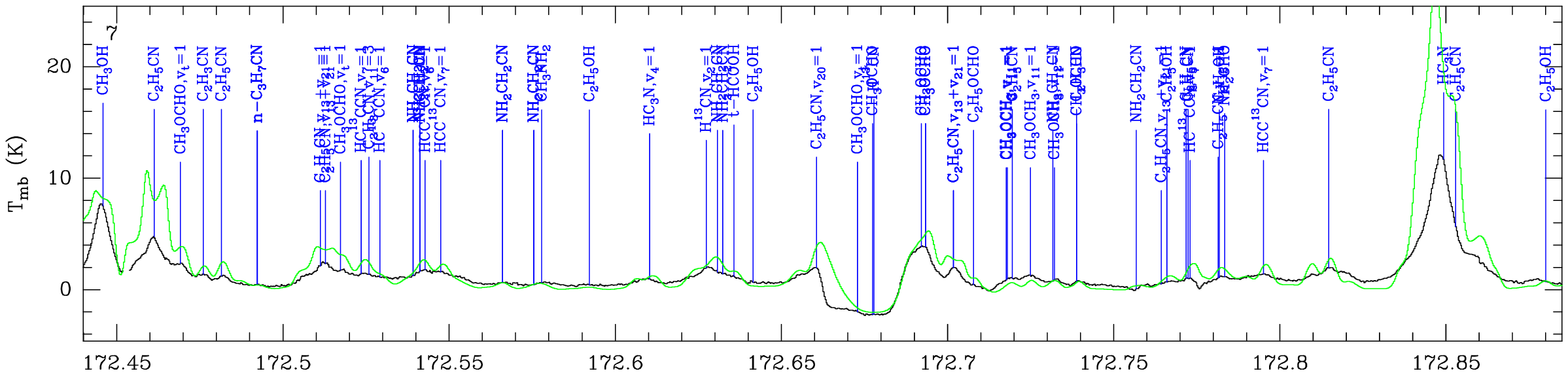}}}
\vspace*{1ex}\centerline{\resizebox{1.0\hsize}{!}{\includegraphics[angle=270]{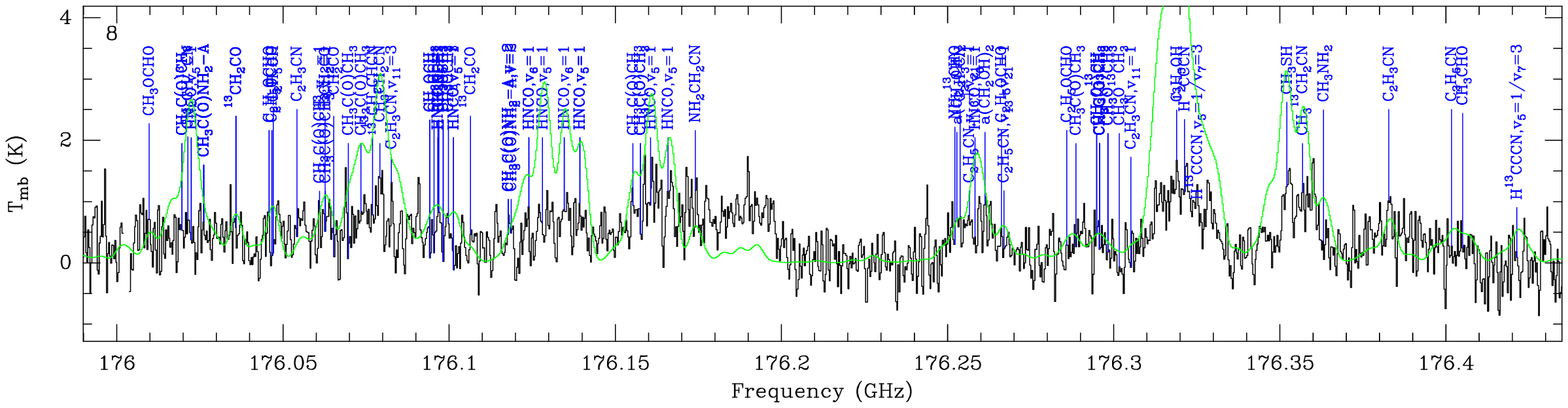}}}
\caption{
continued.
}
\end{figure*}
 \clearpage

}
\onlfig{\clearpage
\begin{figure*}
\centerline{\resizebox{1.0\hsize}{!}{\includegraphics[angle=270]{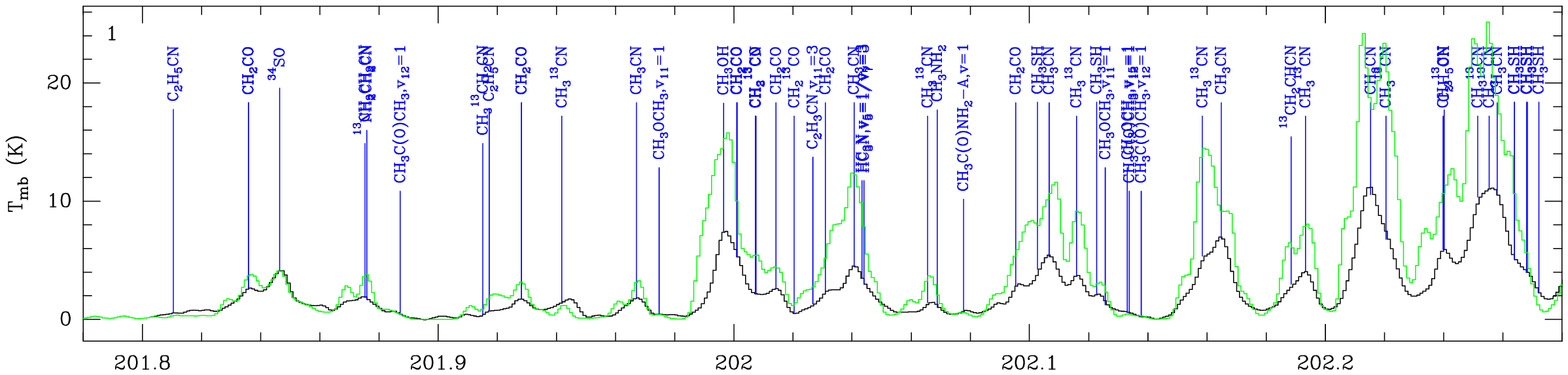}}}
\vspace*{1ex}\centerline{\resizebox{1.0\hsize}{!}{\includegraphics[angle=270]{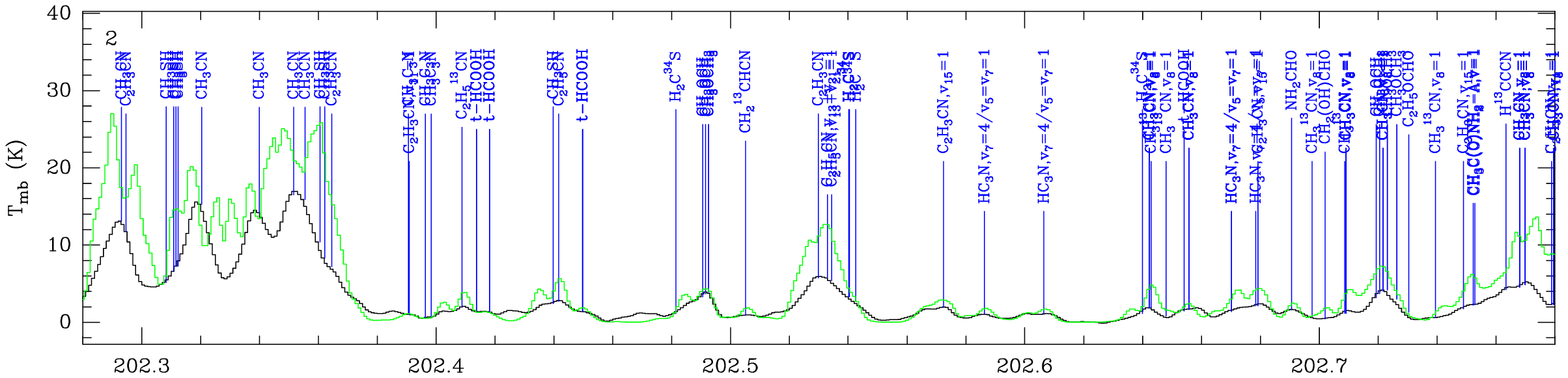}}}
\vspace*{1ex}\centerline{\resizebox{1.0\hsize}{!}{\includegraphics[angle=270]{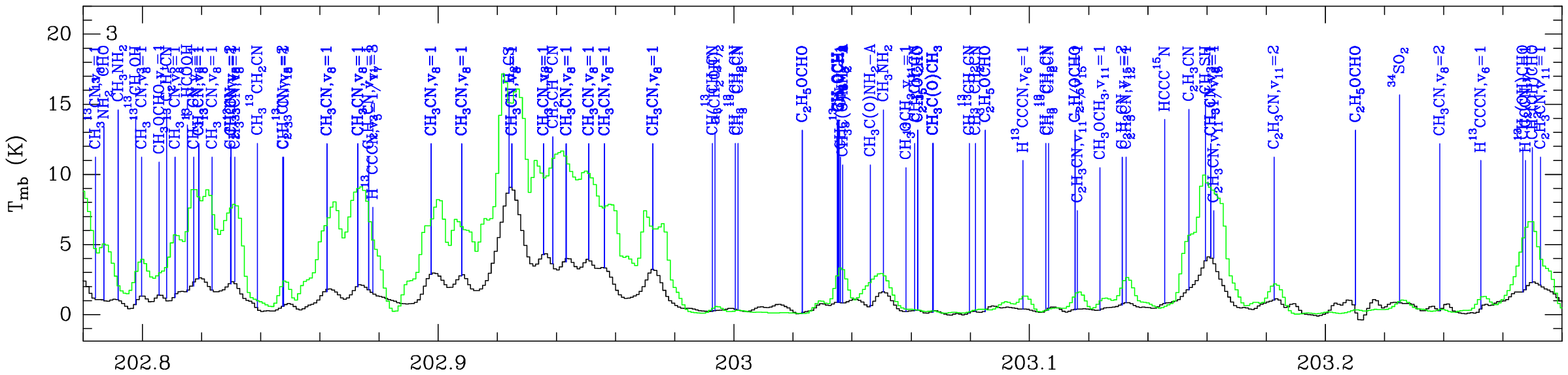}}}
\vspace*{1ex}\centerline{\resizebox{1.0\hsize}{!}{\includegraphics[angle=270]{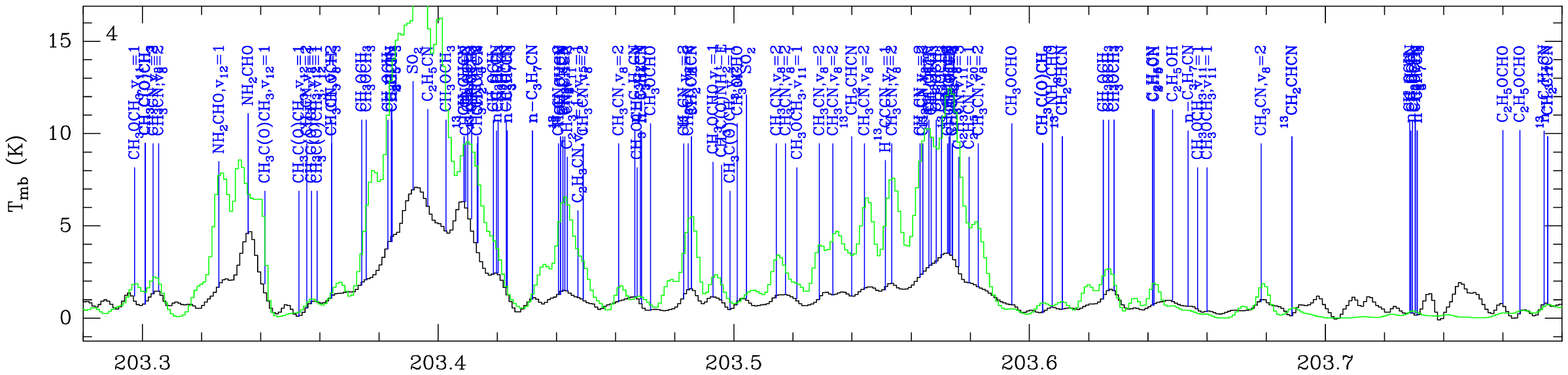}}}
\vspace*{1ex}\centerline{\resizebox{1.0\hsize}{!}{\includegraphics[angle=270]{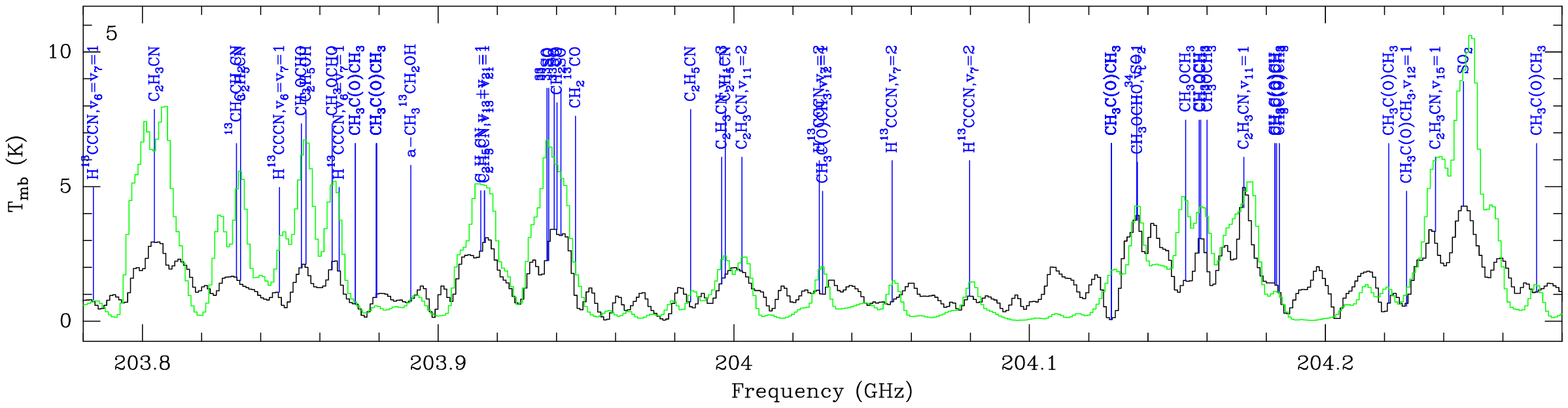}}}
\caption{
Spectrum obtained toward Sgr~B2(N)
in the 1~mm window
with the IRAM~30\,m telescope in main-beam temperature scale. The synthetic model is overlaid in green and its relevant lines are labeled in blue.
The frequencies of the hydrogen recombination lines are indicated with a pink label.
}
\label{f:survey_lmh_1mm}
\end{figure*}
 \clearpage
\begin{figure*}
\addtocounter{figure}{-1}
\centerline{\resizebox{1.0\hsize}{!}{\includegraphics[angle=270]{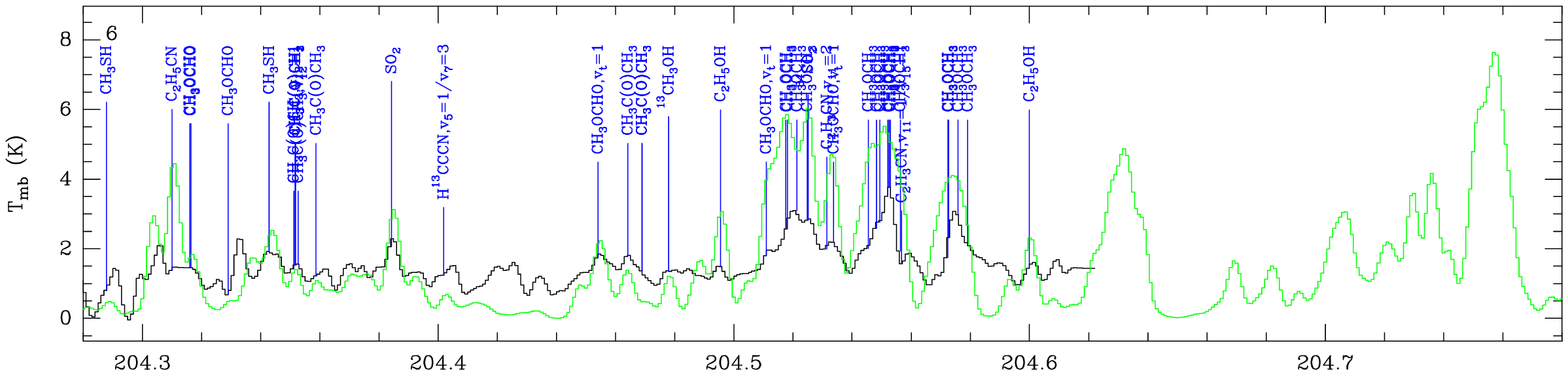}}}
\vspace*{1ex}\centerline{\resizebox{1.0\hsize}{!}{\includegraphics[angle=270]{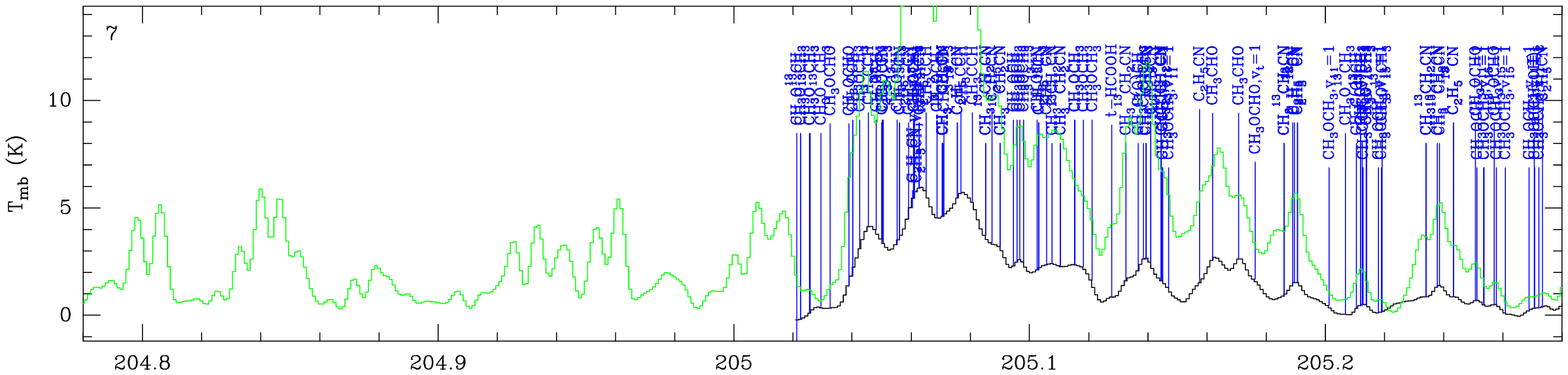}}}
\vspace*{1ex}\centerline{\resizebox{1.0\hsize}{!}{\includegraphics[angle=270]{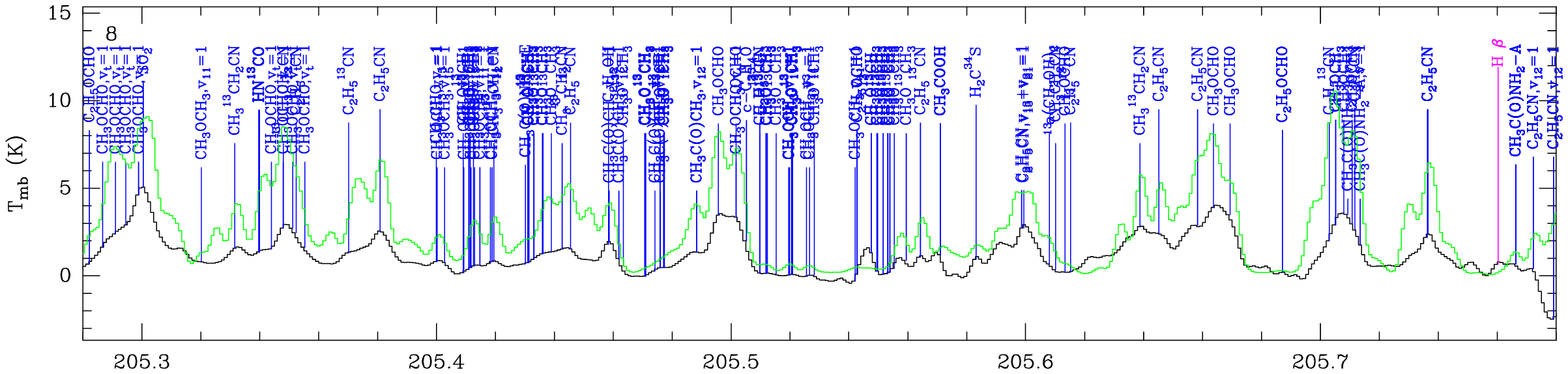}}}
\vspace*{1ex}\centerline{\resizebox{1.0\hsize}{!}{\includegraphics[angle=270]{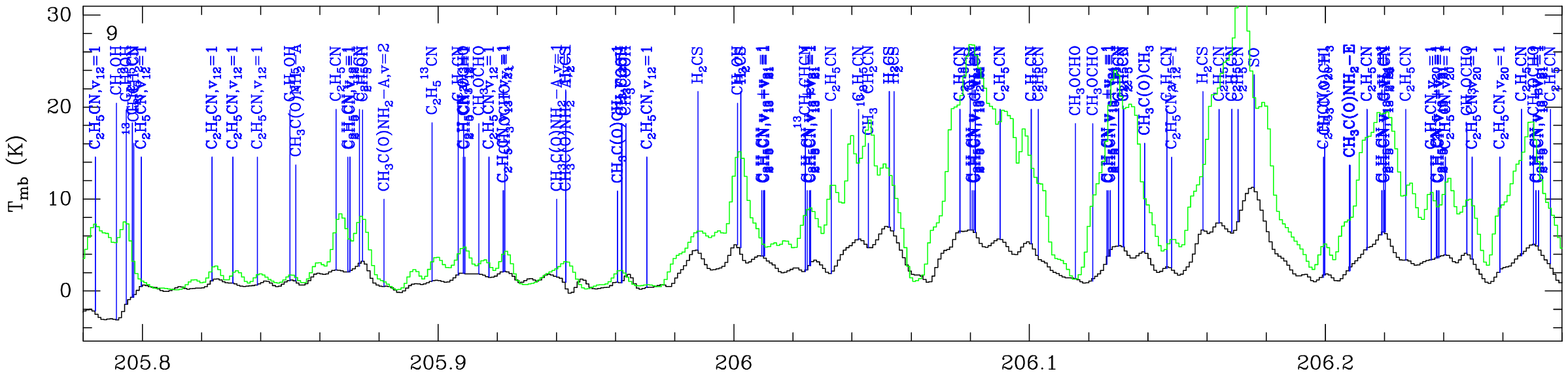}}}
\vspace*{1ex}\centerline{\resizebox{1.0\hsize}{!}{\includegraphics[angle=270]{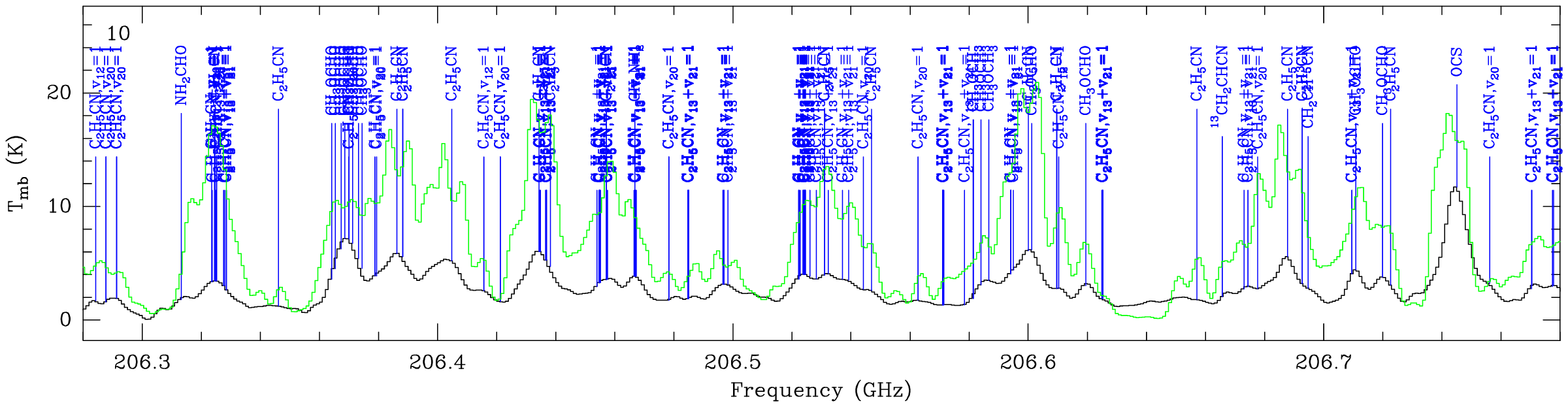}}}
\caption{
continued.
}
\end{figure*}
 \clearpage
\begin{figure*}
\addtocounter{figure}{-1}
\centerline{\resizebox{1.0\hsize}{!}{\includegraphics[angle=270]{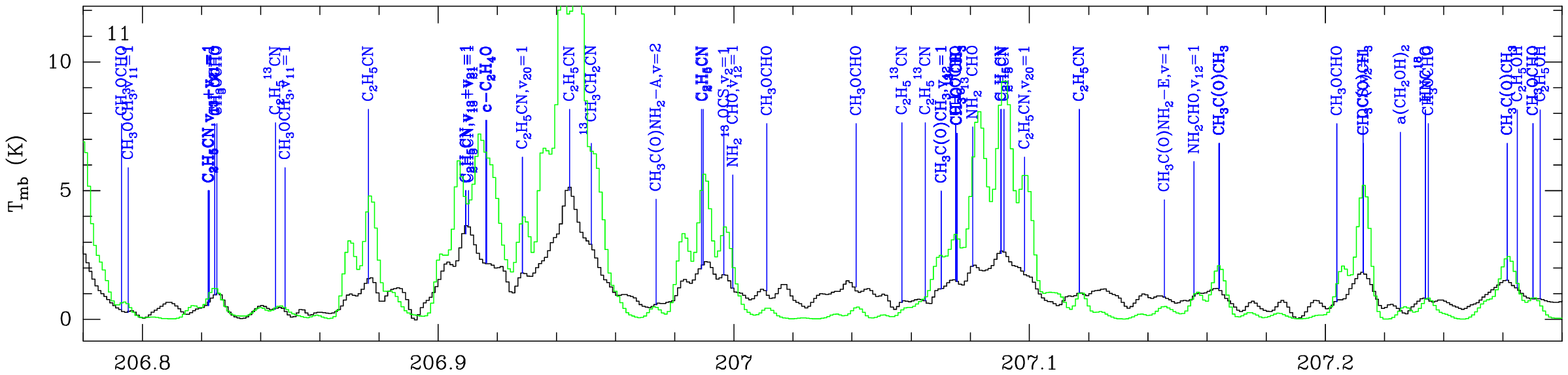}}}
\vspace*{1ex}\centerline{\resizebox{1.0\hsize}{!}{\includegraphics[angle=270]{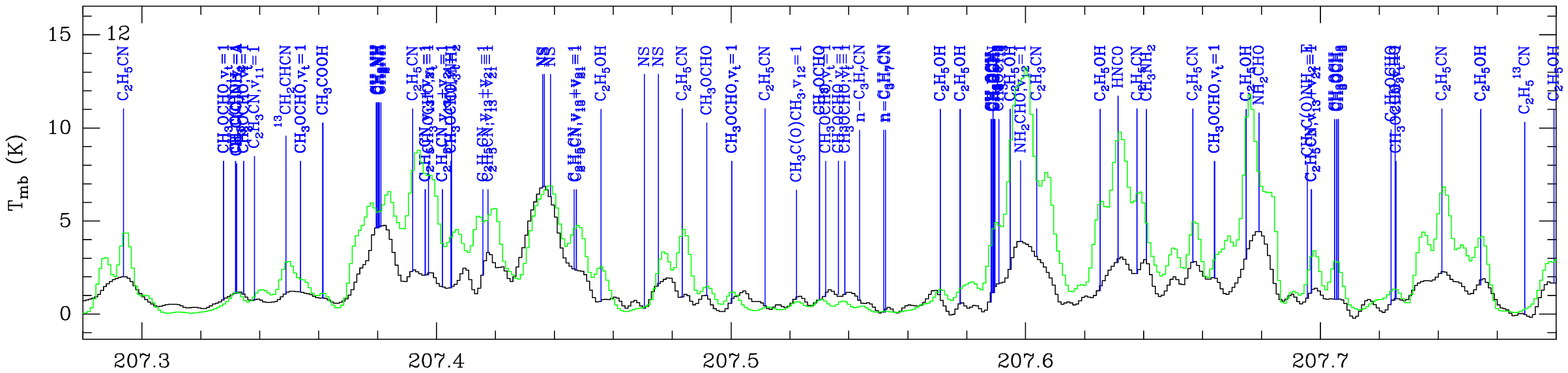}}}
\vspace*{1ex}\centerline{\resizebox{1.0\hsize}{!}{\includegraphics[angle=270]{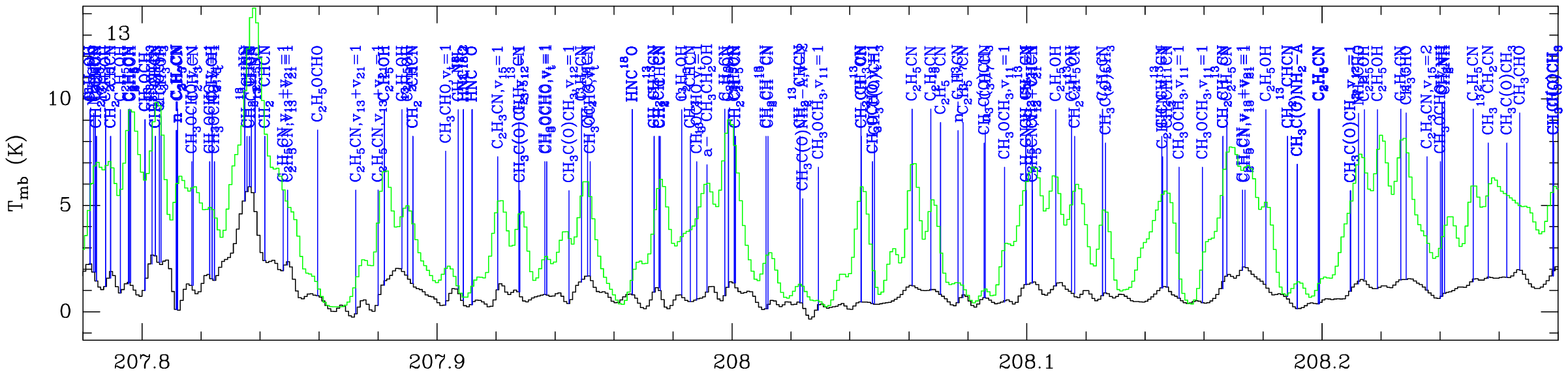}}}
\vspace*{1ex}\centerline{\resizebox{1.0\hsize}{!}{\includegraphics[angle=270]{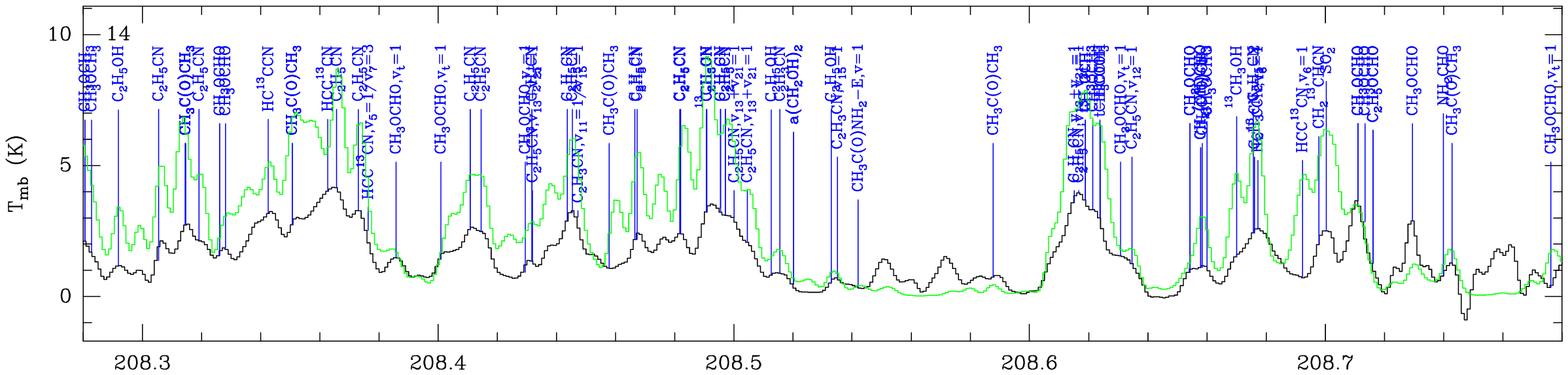}}}
\vspace*{1ex}\centerline{\resizebox{1.0\hsize}{!}{\includegraphics[angle=270]{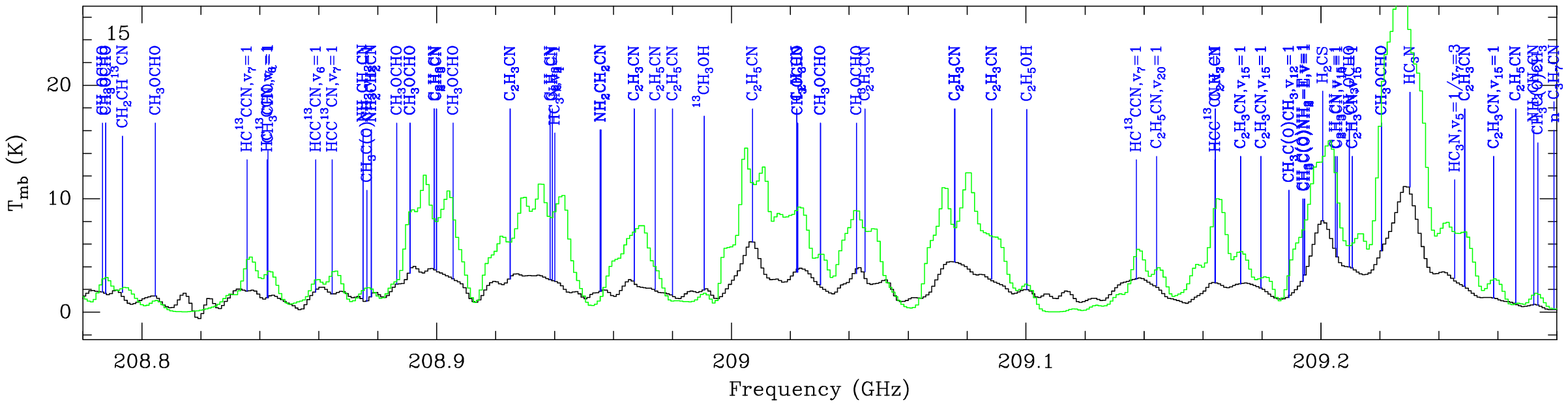}}}
\caption{
continued.
}
\end{figure*}
 \clearpage
\begin{figure*}
\addtocounter{figure}{-1}
\centerline{\resizebox{1.0\hsize}{!}{\includegraphics[angle=270]{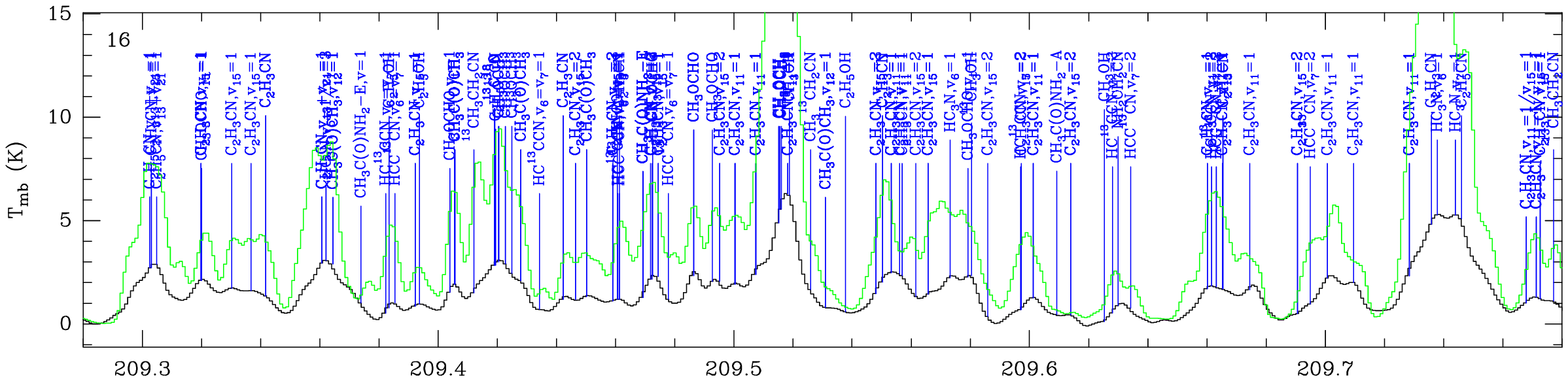}}}
\vspace*{1ex}\centerline{\resizebox{1.0\hsize}{!}{\includegraphics[angle=270]{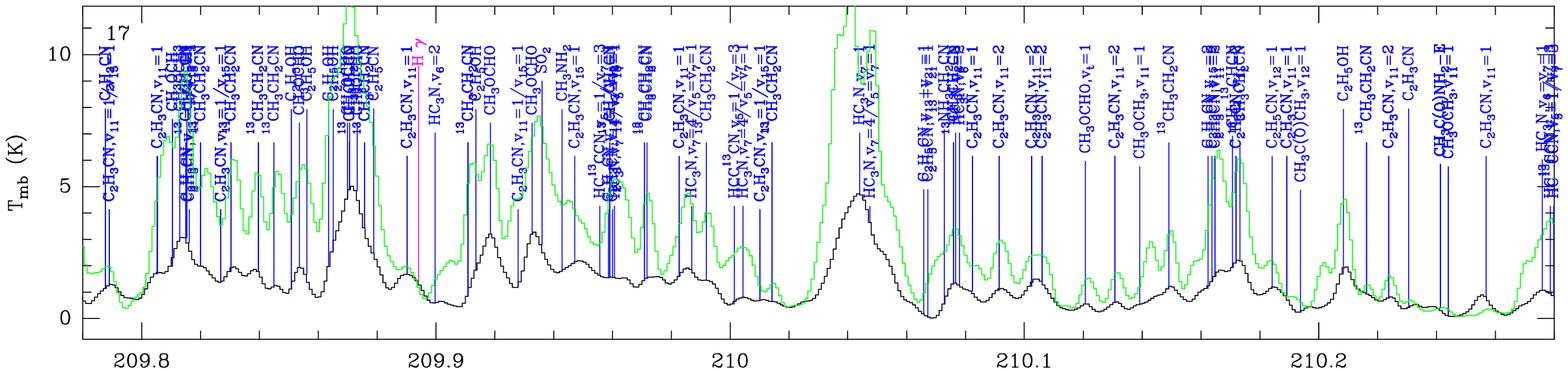}}}
\vspace*{1ex}\centerline{\resizebox{1.0\hsize}{!}{\includegraphics[angle=270]{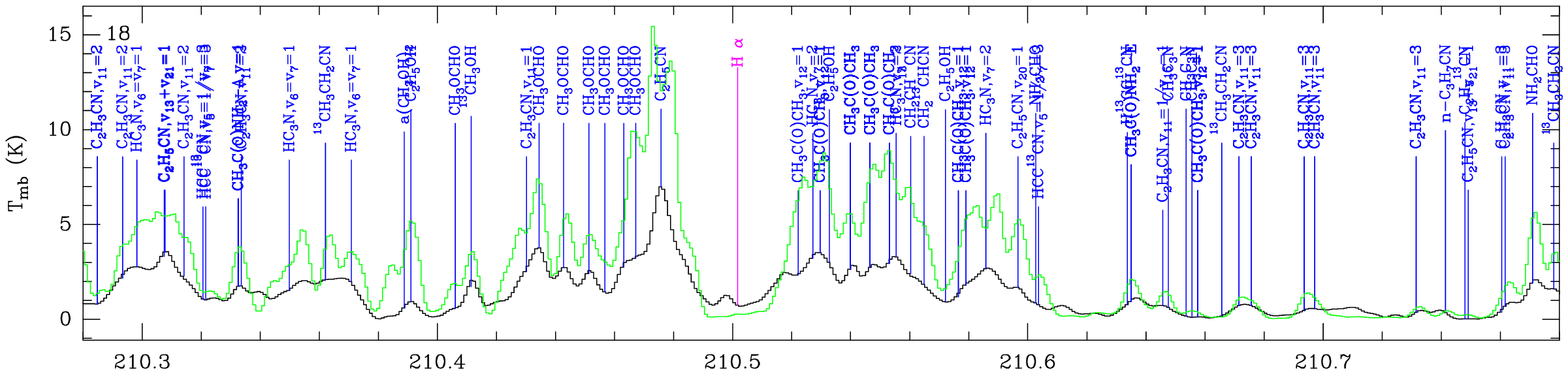}}}
\vspace*{1ex}\centerline{\resizebox{1.0\hsize}{!}{\includegraphics[angle=270]{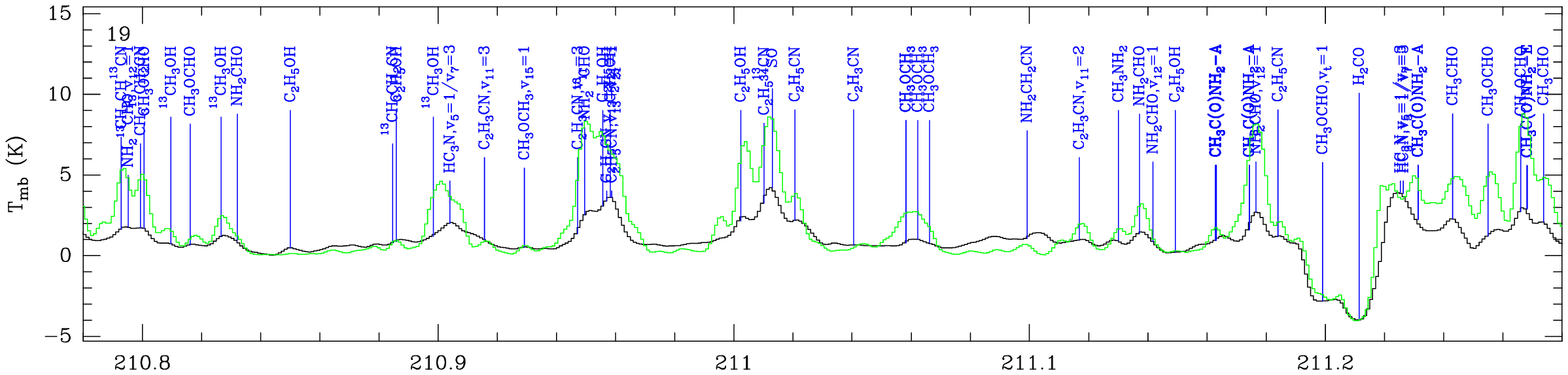}}}
\vspace*{1ex}\centerline{\resizebox{1.0\hsize}{!}{\includegraphics[angle=270]{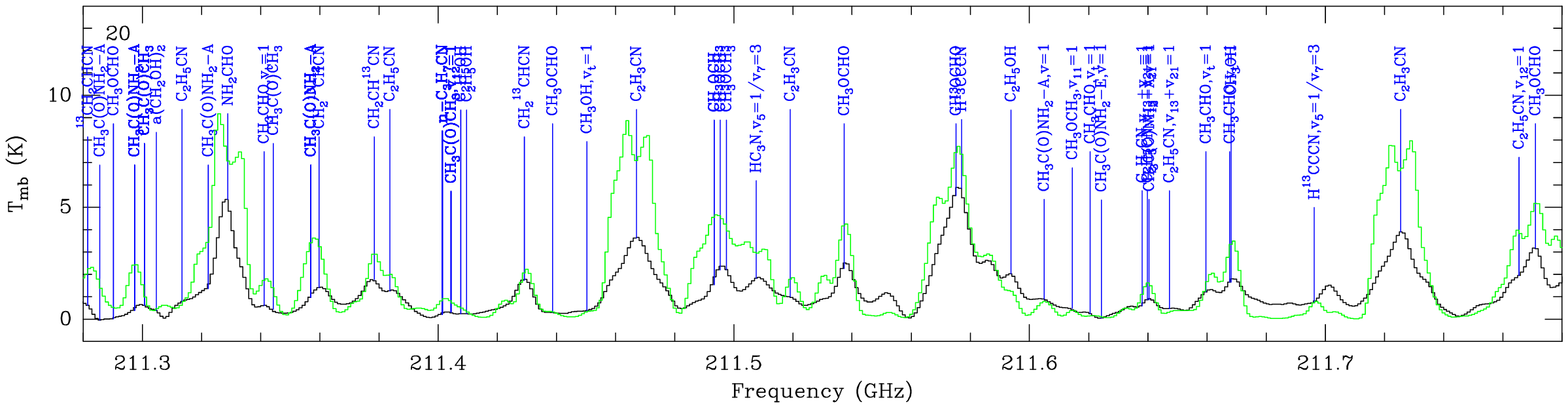}}}
\caption{
continued.
}
\end{figure*}
 \clearpage
\begin{figure*}
\addtocounter{figure}{-1}
\centerline{\resizebox{1.0\hsize}{!}{\includegraphics[angle=270]{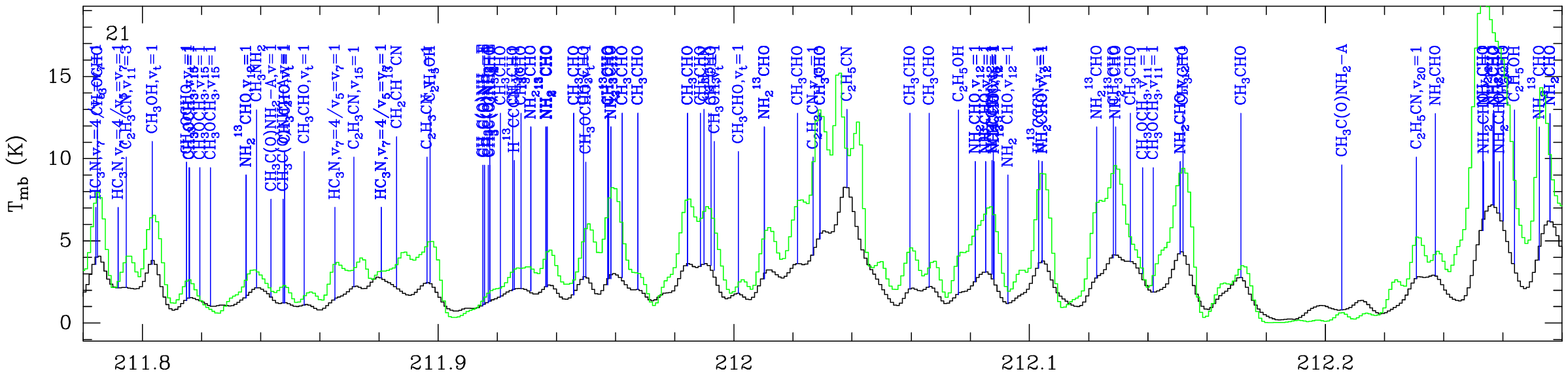}}}
\vspace*{1ex}\centerline{\resizebox{1.0\hsize}{!}{\includegraphics[angle=270]{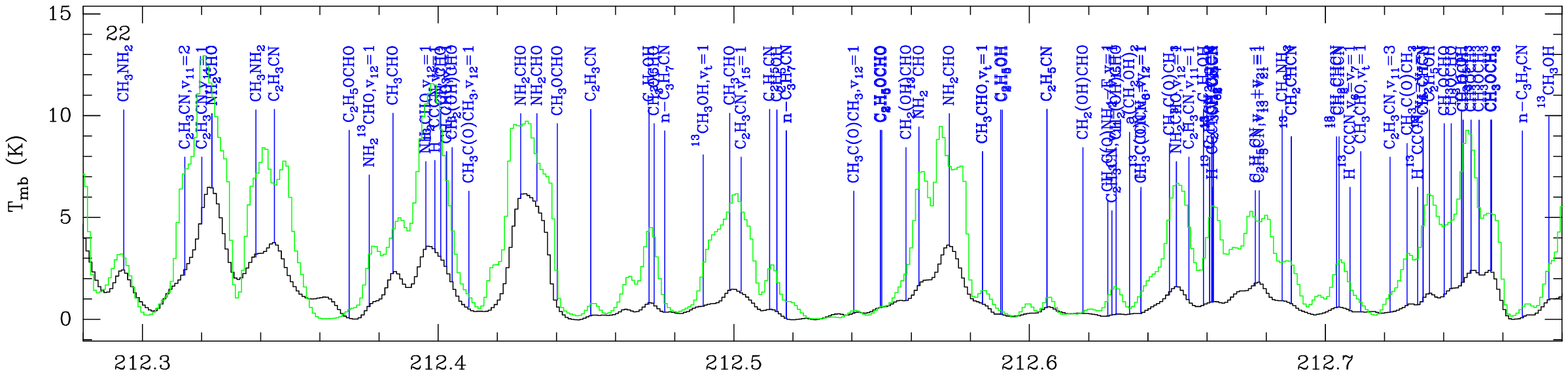}}}
\vspace*{1ex}\centerline{\resizebox{1.0\hsize}{!}{\includegraphics[angle=270]{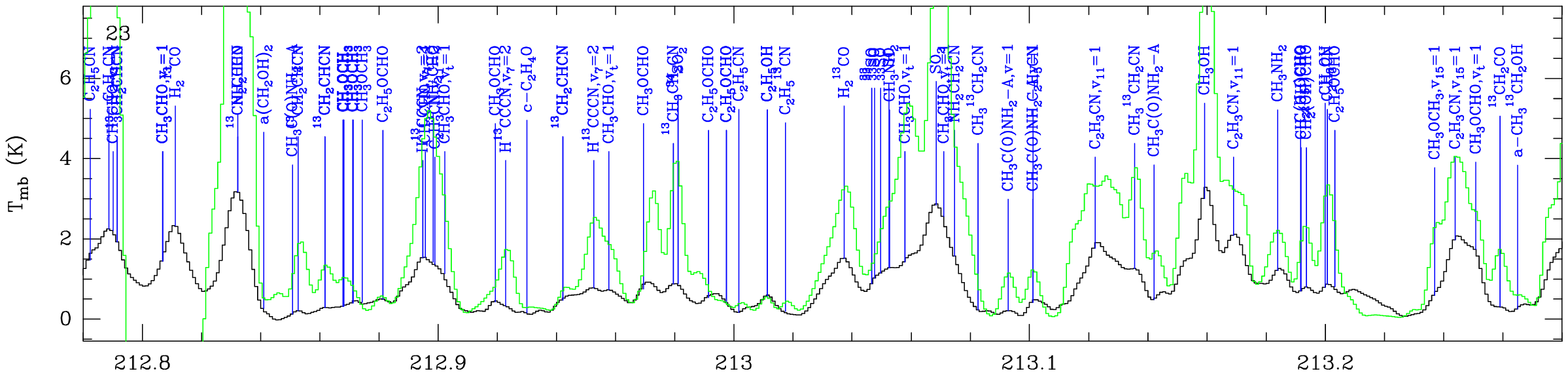}}}
\vspace*{1ex}\centerline{\resizebox{1.0\hsize}{!}{\includegraphics[angle=270]{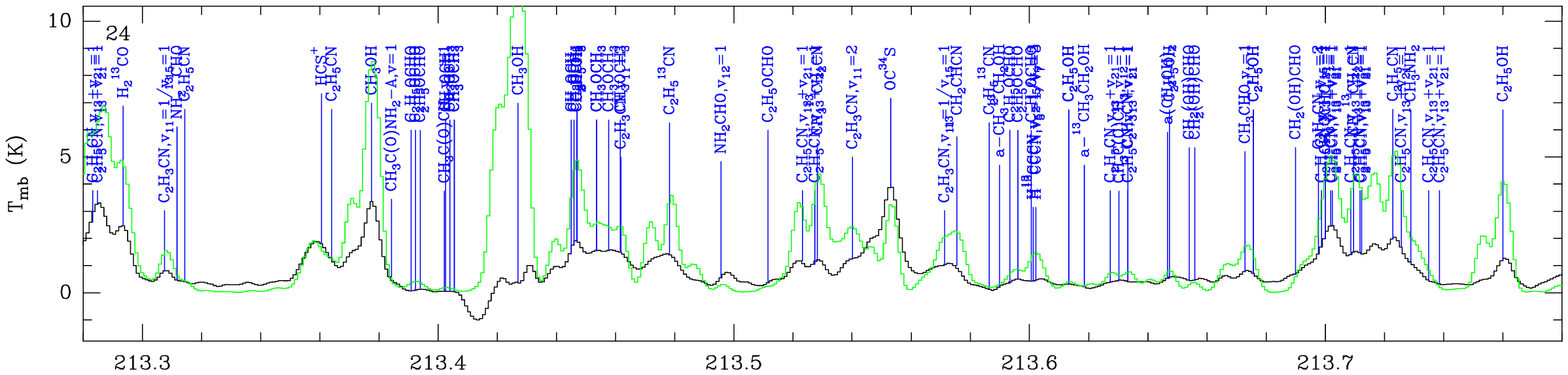}}}
\vspace*{1ex}\centerline{\resizebox{1.0\hsize}{!}{\includegraphics[angle=270]{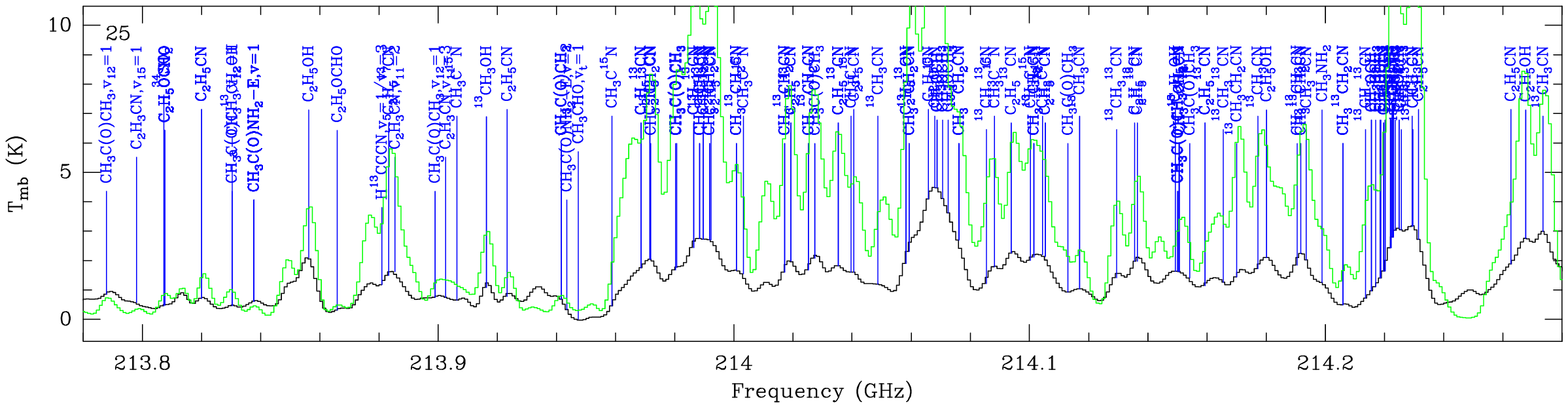}}}
\caption{
continued.
}
\end{figure*}
 \clearpage
\begin{figure*}
\addtocounter{figure}{-1}
\centerline{\resizebox{1.0\hsize}{!}{\includegraphics[angle=270]{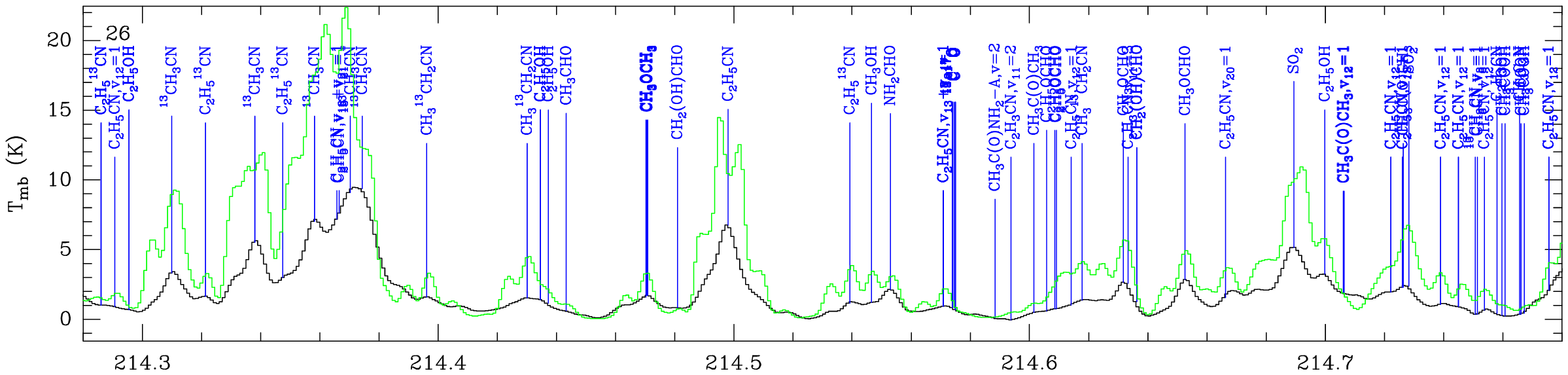}}}
\vspace*{1ex}\centerline{\resizebox{1.0\hsize}{!}{\includegraphics[angle=270]{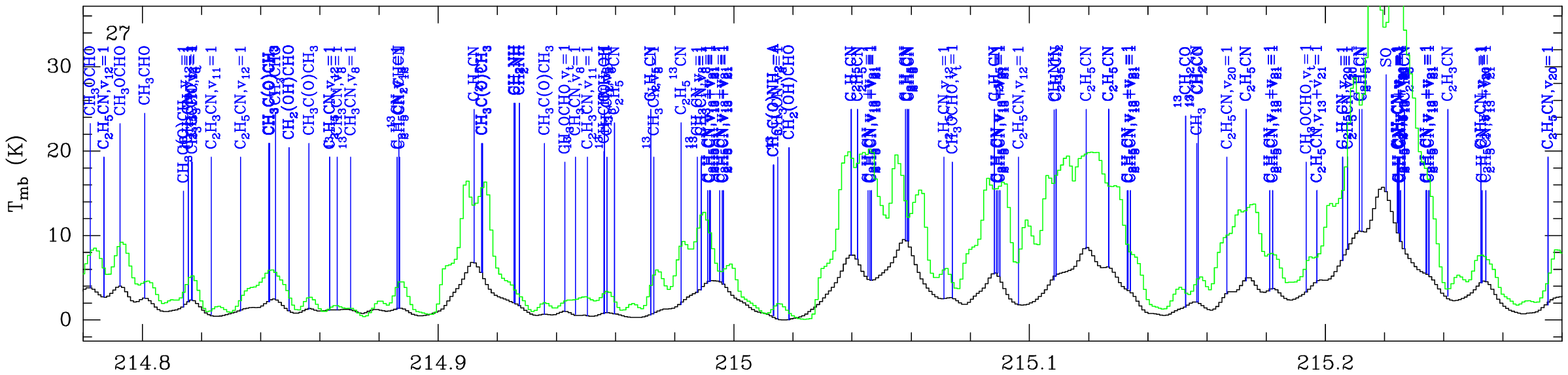}}}
\vspace*{1ex}\centerline{\resizebox{1.0\hsize}{!}{\includegraphics[angle=270]{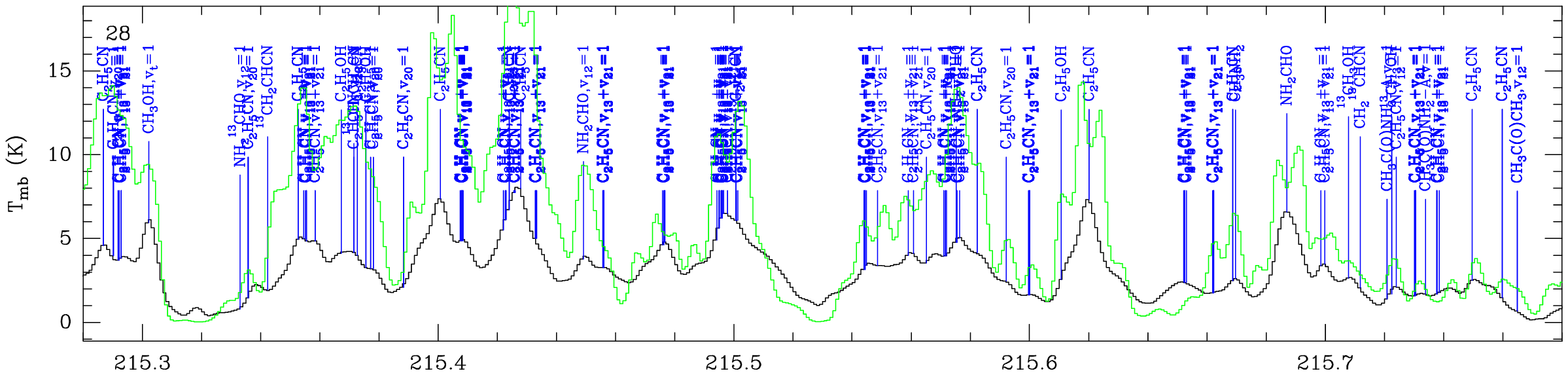}}}
\vspace*{1ex}\centerline{\resizebox{1.0\hsize}{!}{\includegraphics[angle=270]{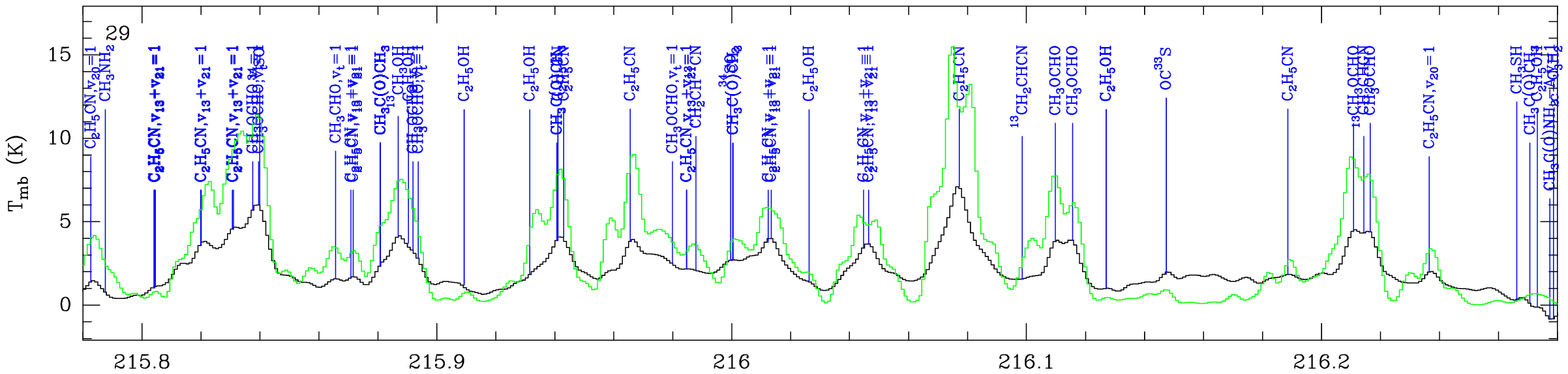}}}
\vspace*{1ex}\centerline{\resizebox{1.0\hsize}{!}{\includegraphics[angle=270]{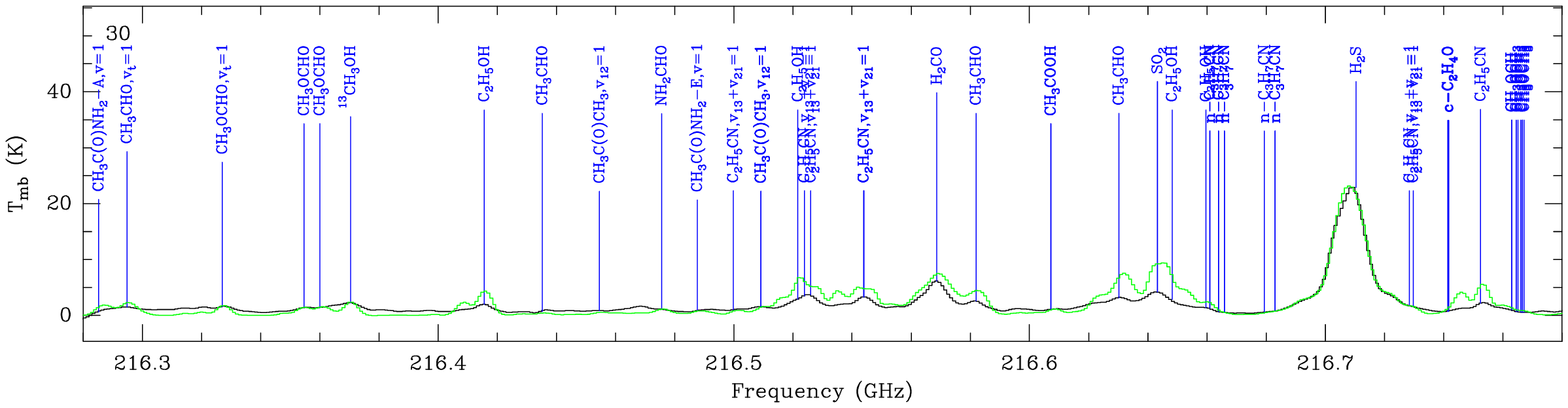}}}
\caption{
continued.
}
\end{figure*}
 \clearpage
\begin{figure*}
\addtocounter{figure}{-1}
\centerline{\resizebox{1.0\hsize}{!}{\includegraphics[angle=270]{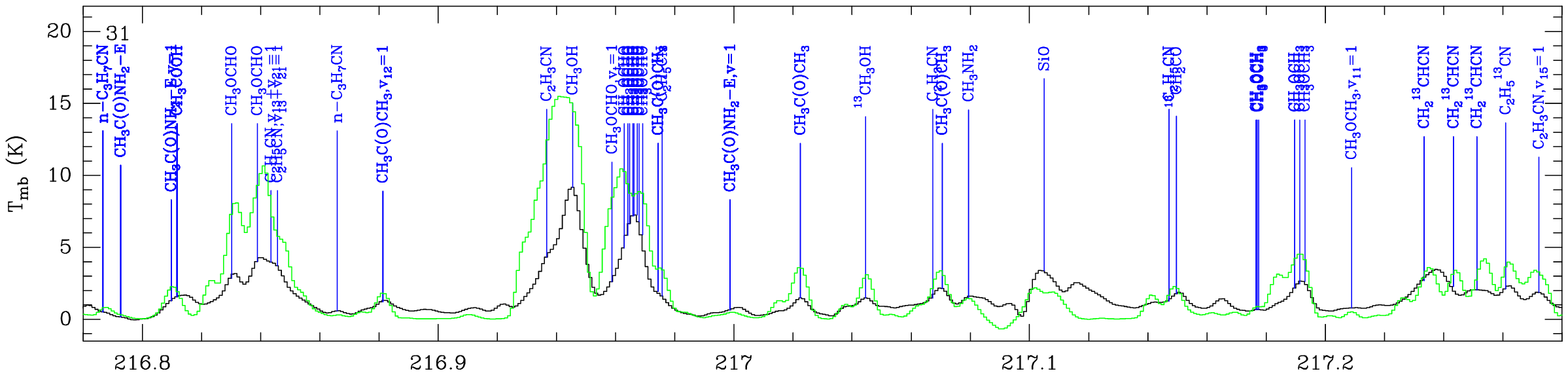}}}
\vspace*{1ex}\centerline{\resizebox{1.0\hsize}{!}{\includegraphics[angle=270]{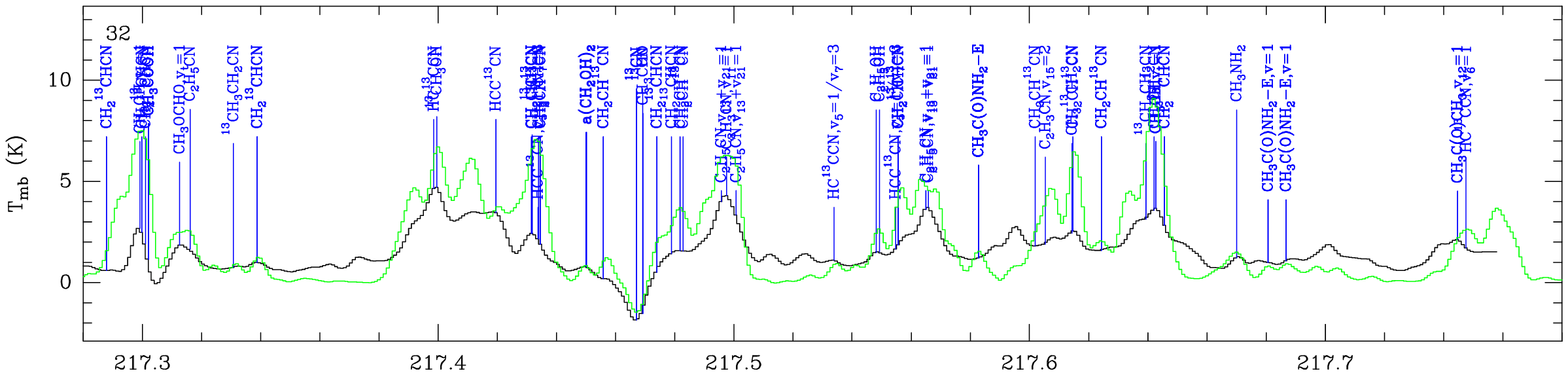}}}
\vspace*{1ex}\centerline{\resizebox{1.0\hsize}{!}{\includegraphics[angle=270]{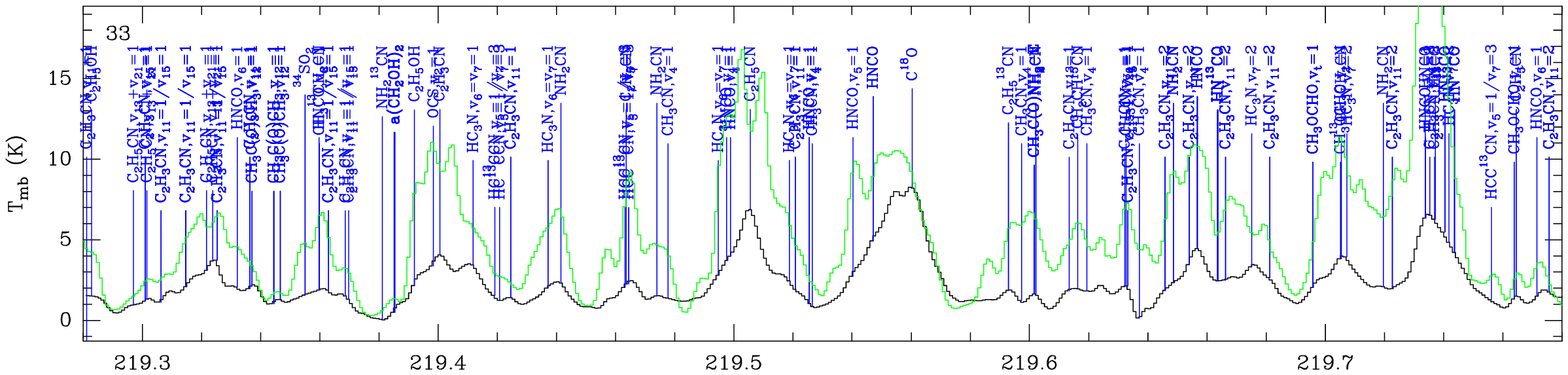}}}
\vspace*{1ex}\centerline{\resizebox{1.0\hsize}{!}{\includegraphics[angle=270]{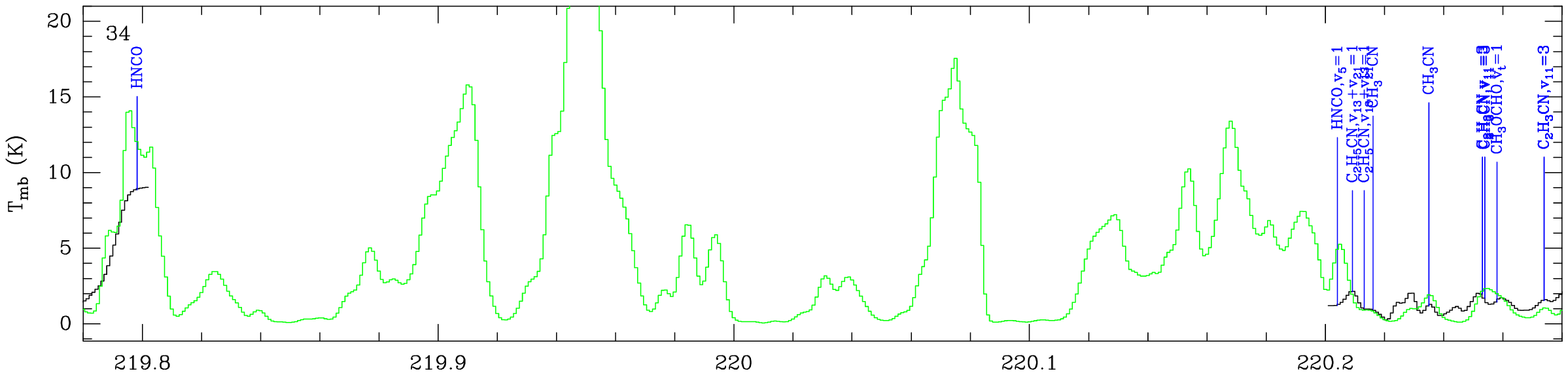}}}
\vspace*{1ex}\centerline{\resizebox{1.0\hsize}{!}{\includegraphics[angle=270]{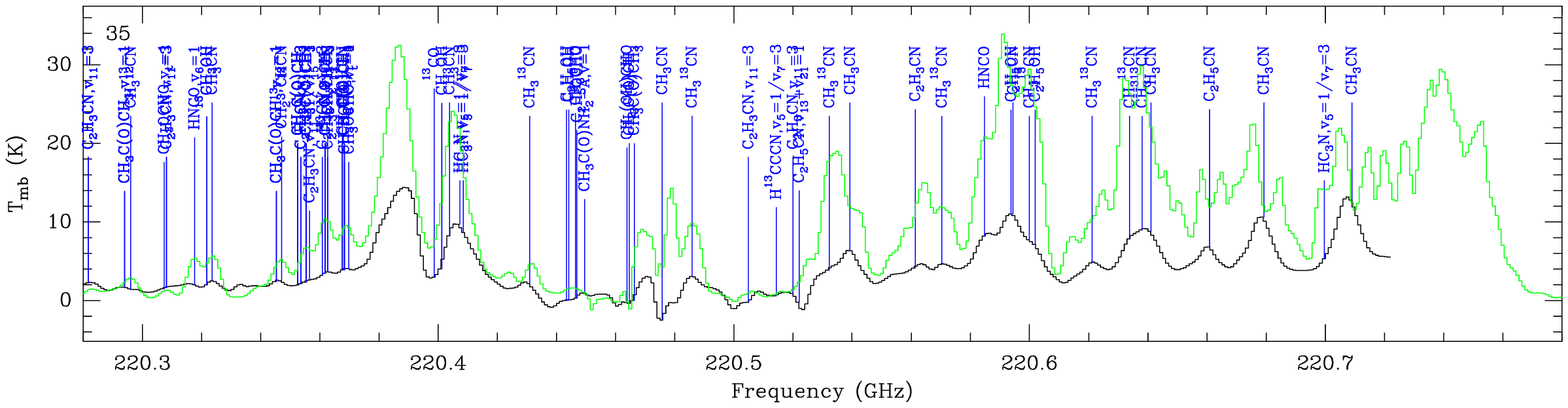}}}
\caption{
continued.
}
\end{figure*}
 \clearpage
\begin{figure*}
\addtocounter{figure}{-1}
\centerline{\resizebox{1.0\hsize}{!}{\includegraphics[angle=270]{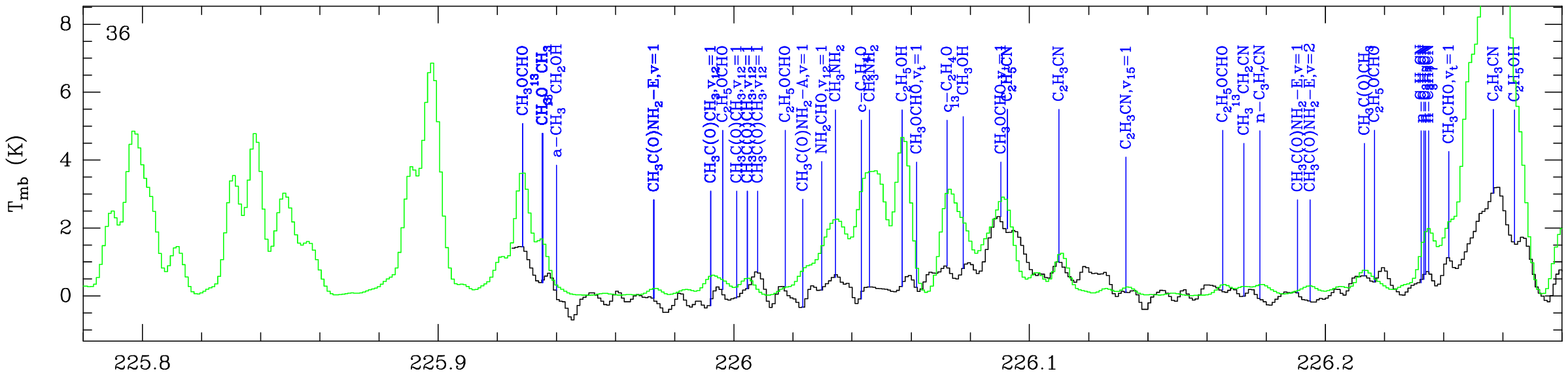}}}
\vspace*{1ex}\centerline{\resizebox{1.0\hsize}{!}{\includegraphics[angle=270]{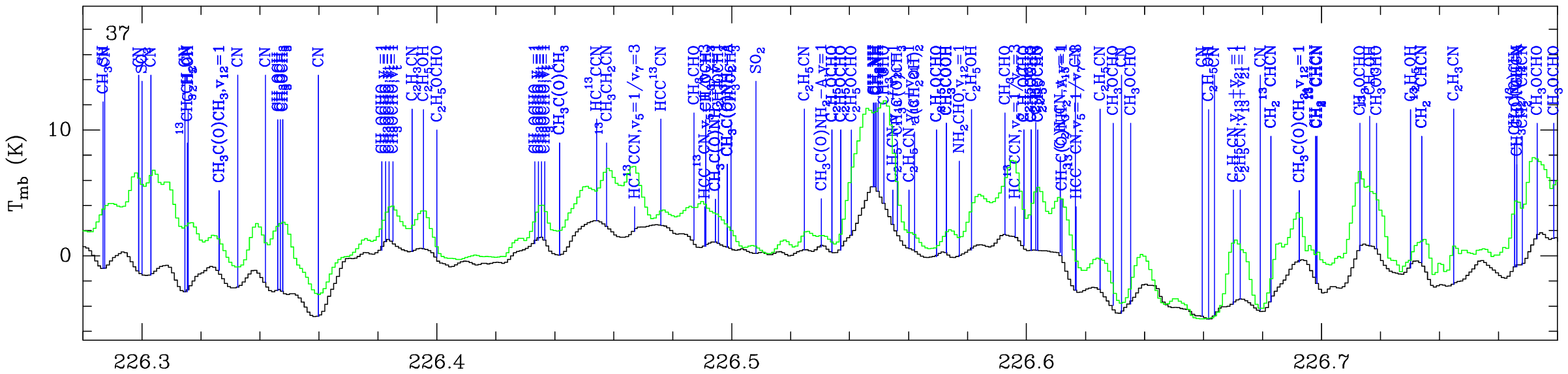}}}
\vspace*{1ex}\centerline{\resizebox{1.0\hsize}{!}{\includegraphics[angle=270]{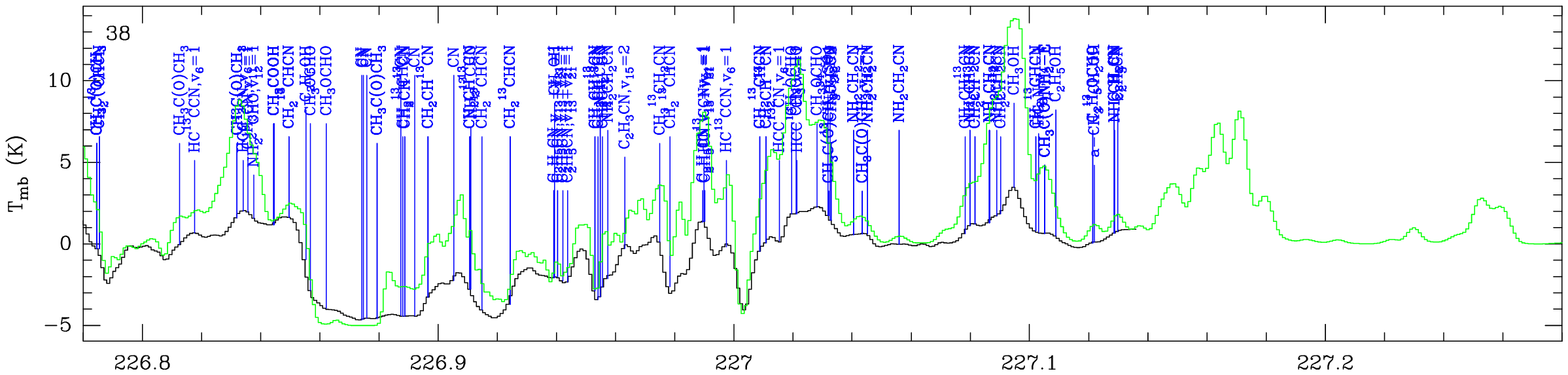}}}
\vspace*{1ex}\centerline{\resizebox{1.0\hsize}{!}{\includegraphics[angle=270]{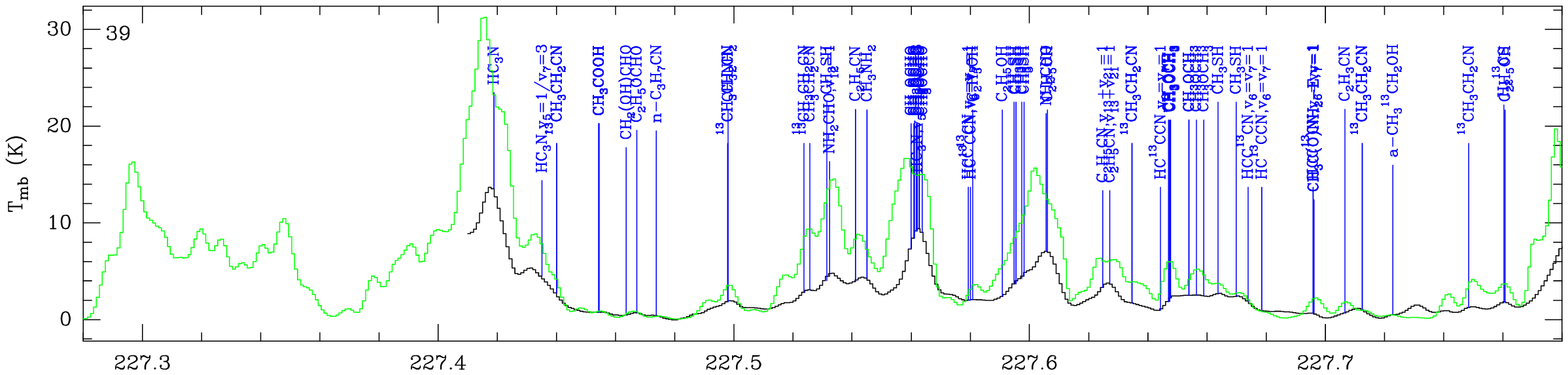}}}
\vspace*{1ex}\centerline{\resizebox{1.0\hsize}{!}{\includegraphics[angle=270]{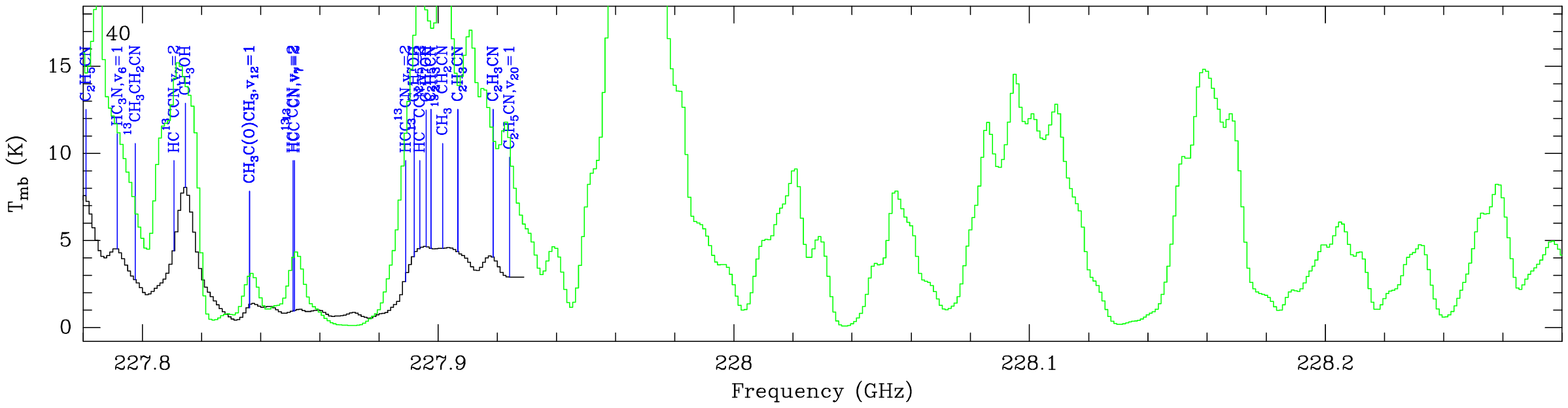}}}
\caption{
continued.
}
\end{figure*}
 \clearpage
\begin{figure*}
\addtocounter{figure}{-1}
\centerline{\resizebox{1.0\hsize}{!}{\includegraphics[angle=270]{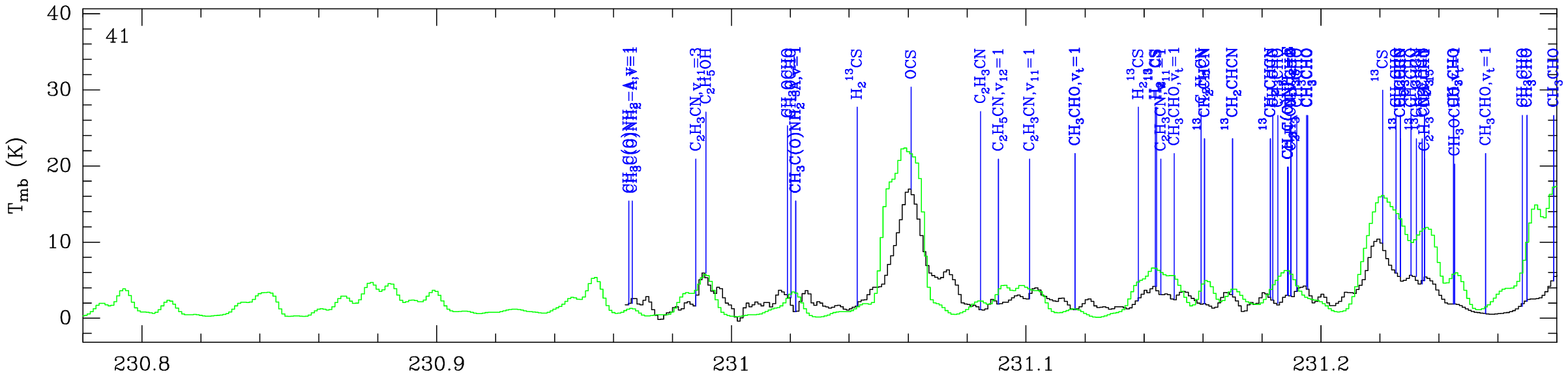}}}
\vspace*{1ex}\centerline{\resizebox{1.0\hsize}{!}{\includegraphics[angle=270]{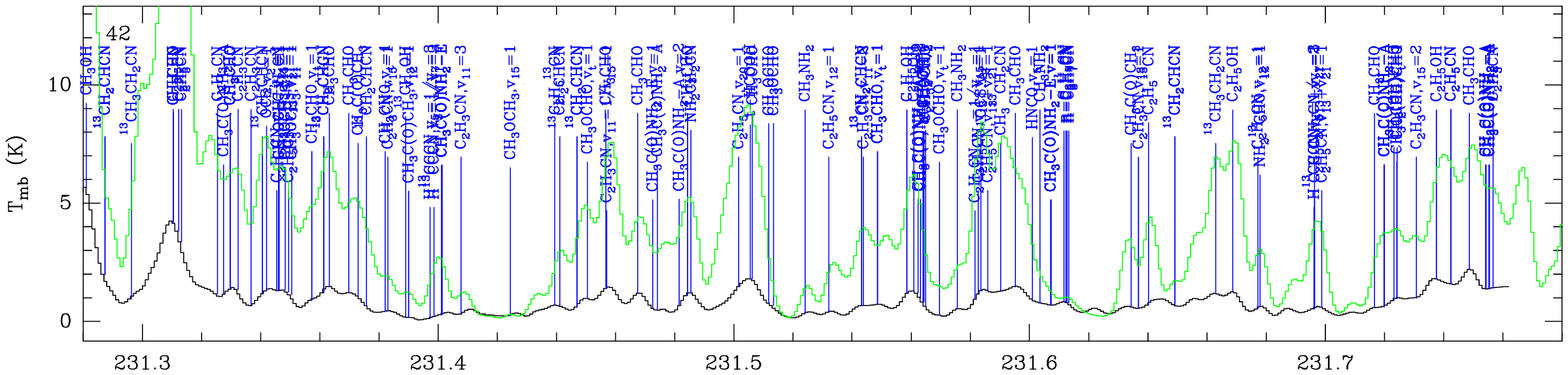}}}
\vspace*{1ex}\centerline{\resizebox{1.0\hsize}{!}{\includegraphics[angle=270]{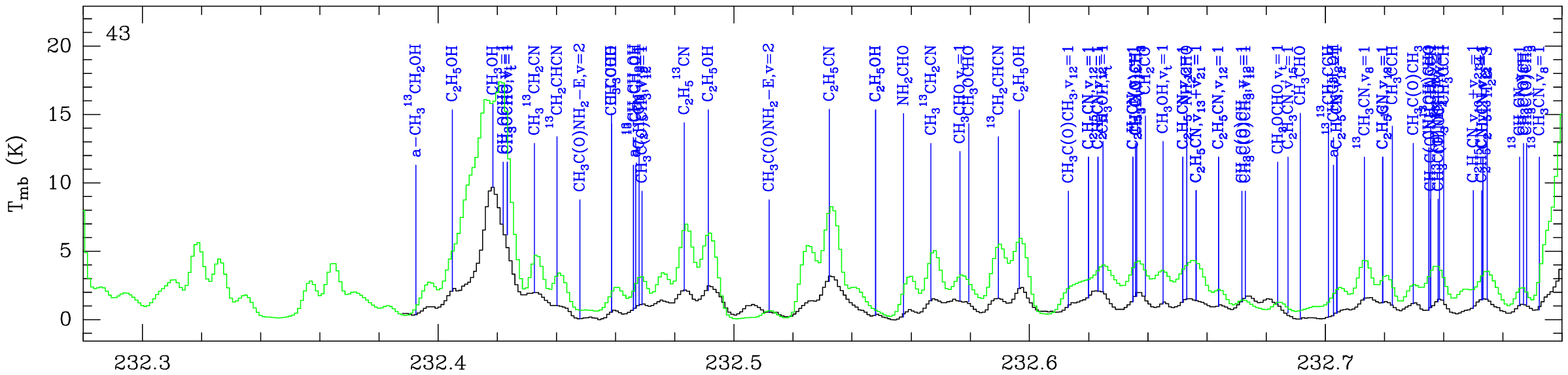}}}
\vspace*{1ex}\centerline{\resizebox{1.0\hsize}{!}{\includegraphics[angle=270]{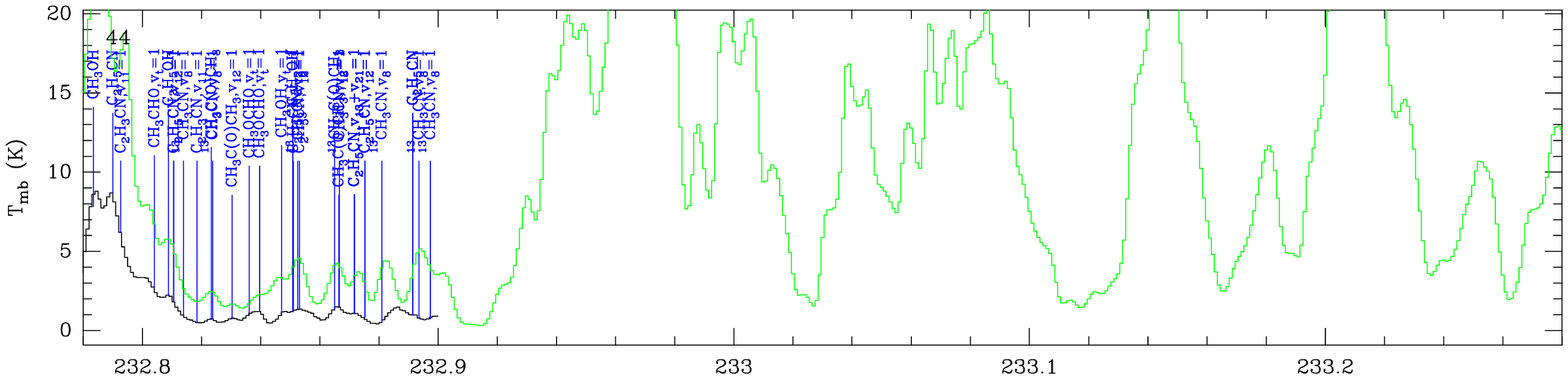}}}
\vspace*{1ex}\centerline{\resizebox{1.0\hsize}{!}{\includegraphics[angle=270]{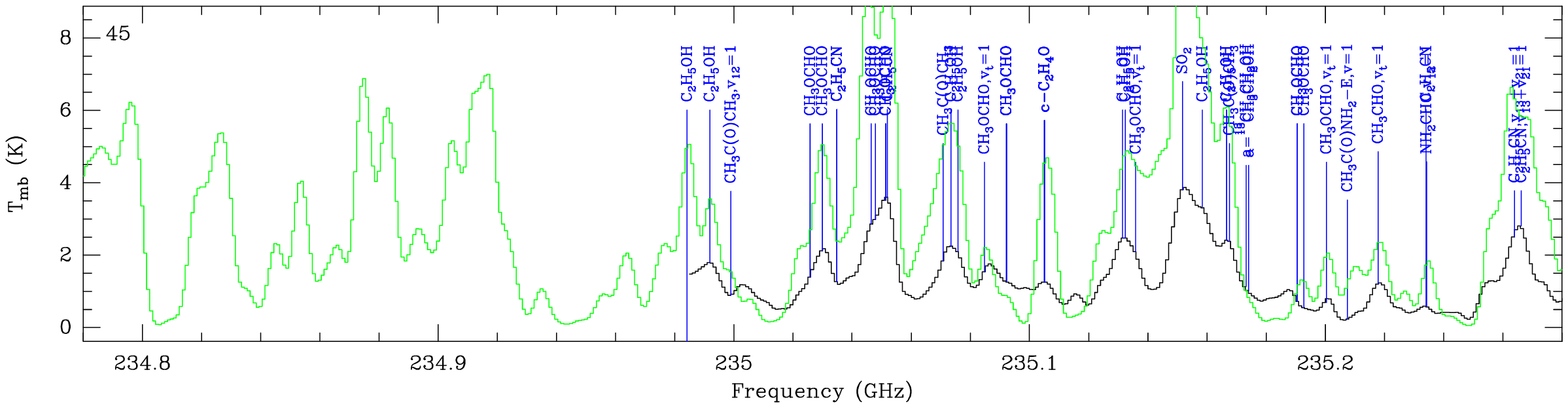}}}
\caption{
continued.
}
\end{figure*}
 \clearpage
\begin{figure*}
\addtocounter{figure}{-1}
\centerline{\resizebox{1.0\hsize}{!}{\includegraphics[angle=270]{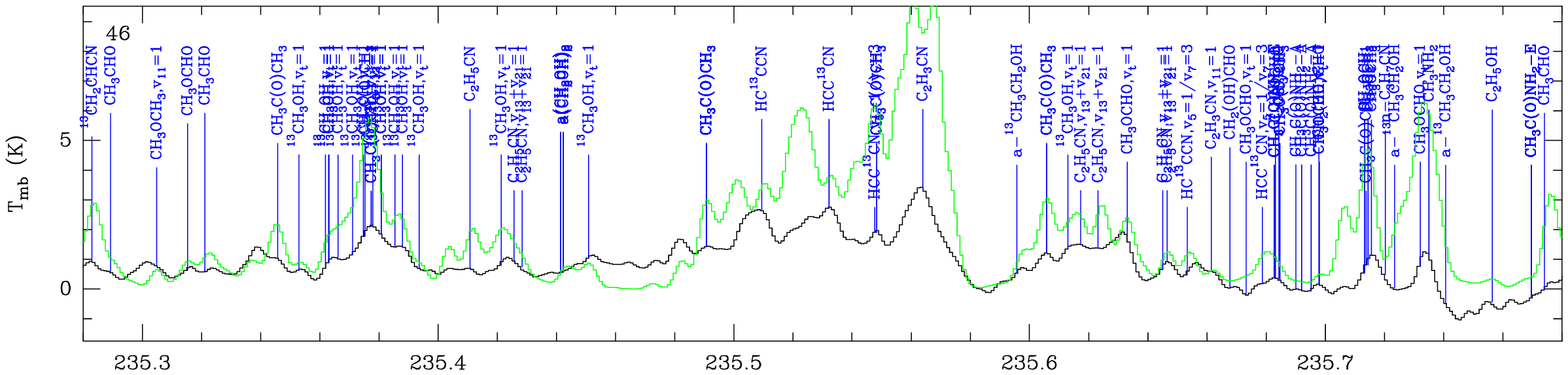}}}
\vspace*{1ex}\centerline{\resizebox{1.0\hsize}{!}{\includegraphics[angle=270]{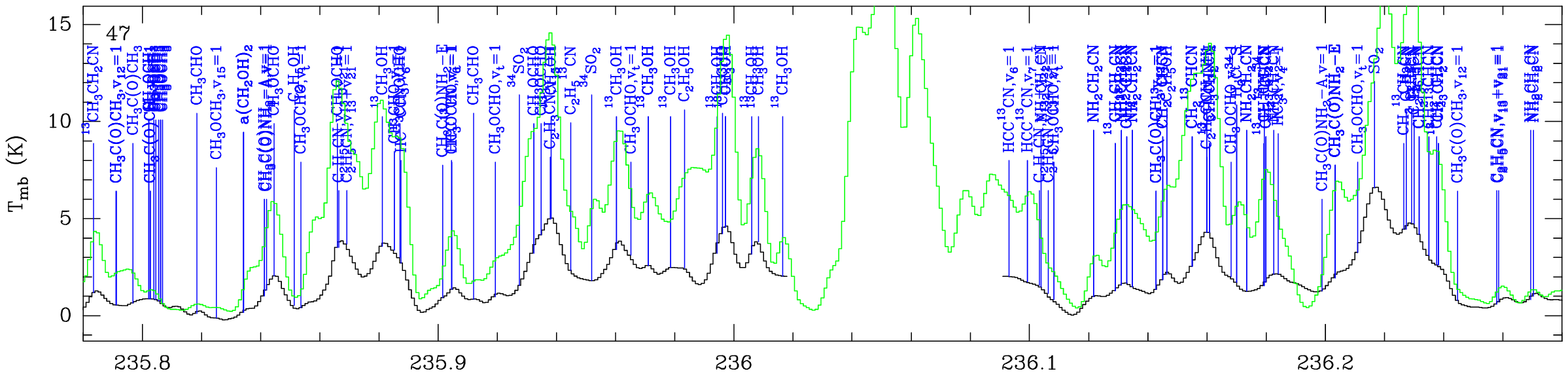}}}
\vspace*{1ex}\centerline{\resizebox{1.0\hsize}{!}{\includegraphics[angle=270]{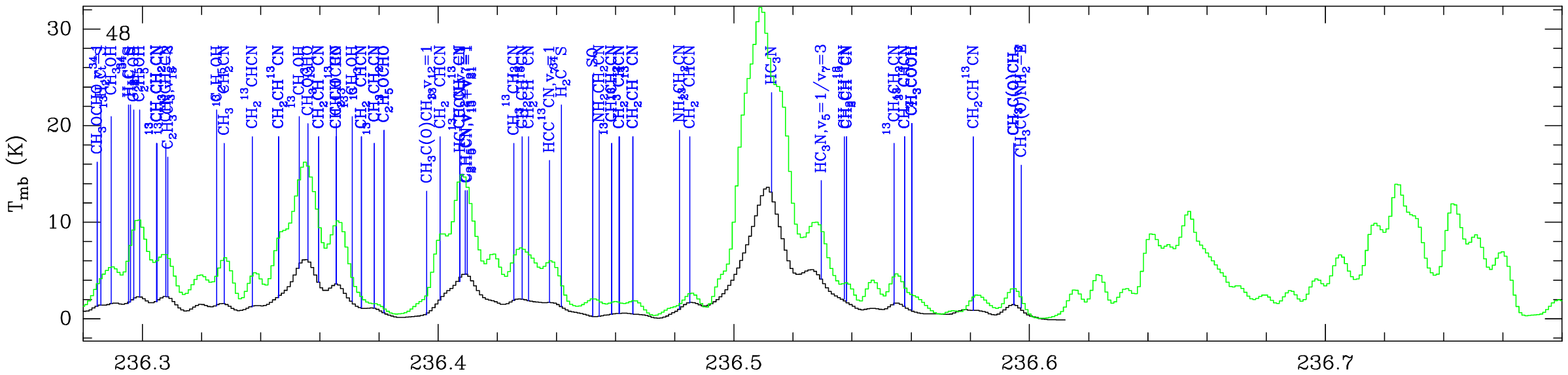}}}
\vspace*{1ex}\centerline{\resizebox{1.0\hsize}{!}{\includegraphics[angle=270]{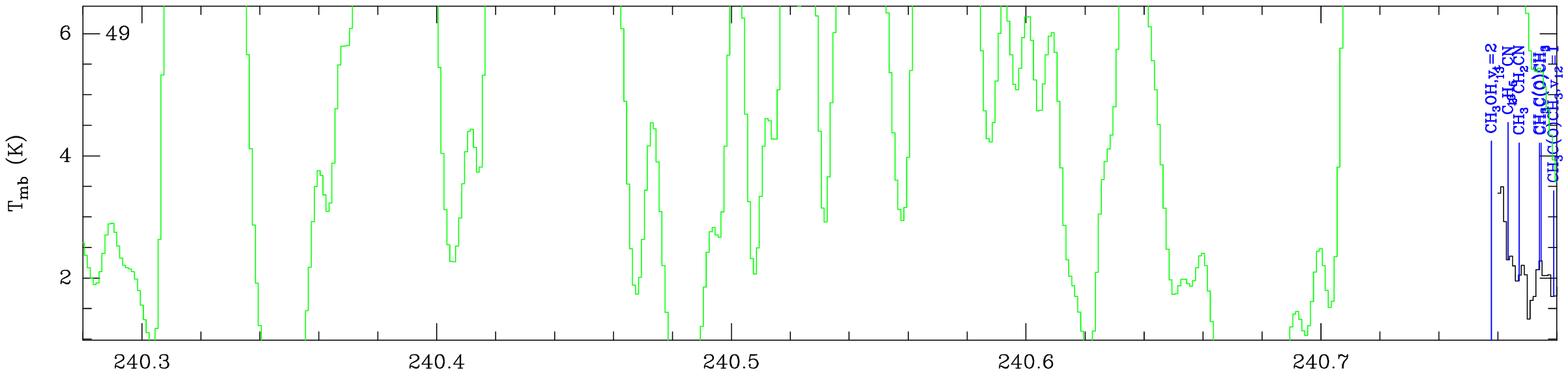}}}
\vspace*{1ex}\centerline{\resizebox{1.0\hsize}{!}{\includegraphics[angle=270]{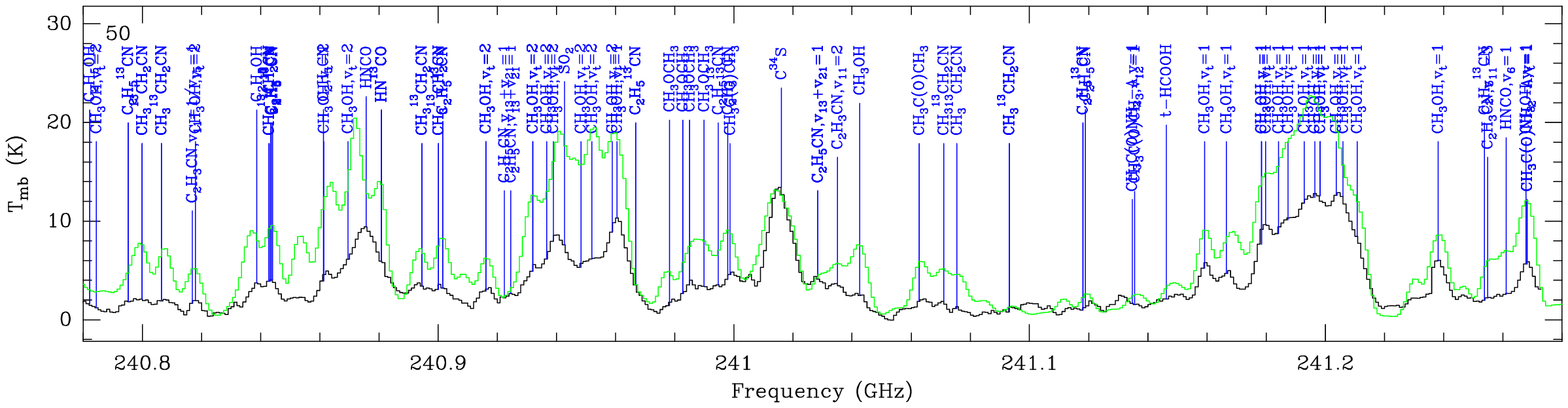}}}
\caption{
continued.
}
\end{figure*}
 \clearpage
\begin{figure*}
\addtocounter{figure}{-1}
\centerline{\resizebox{1.0\hsize}{!}{\includegraphics[angle=270]{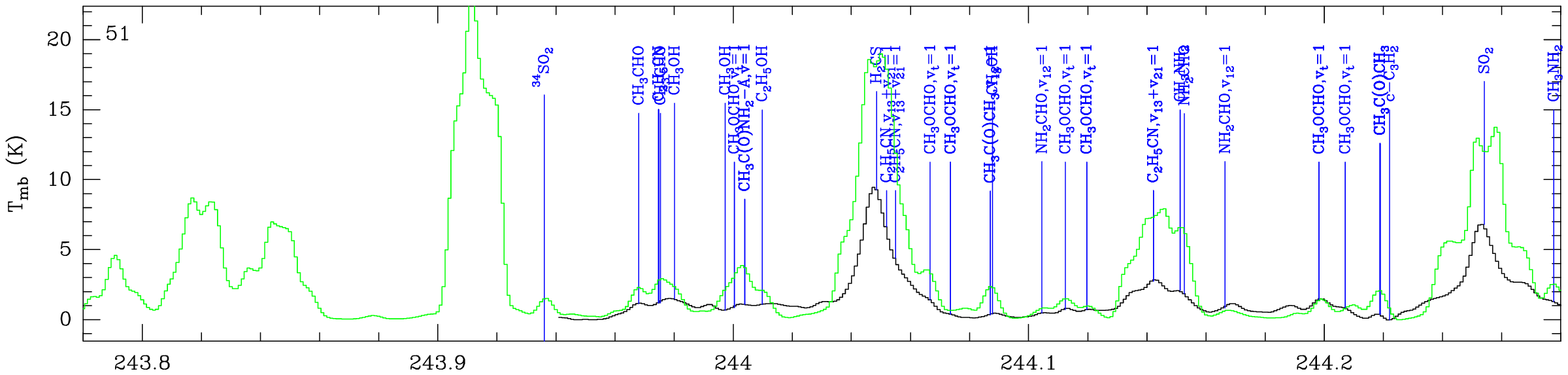}}}
\vspace*{1ex}\centerline{\resizebox{1.0\hsize}{!}{\includegraphics[angle=270]{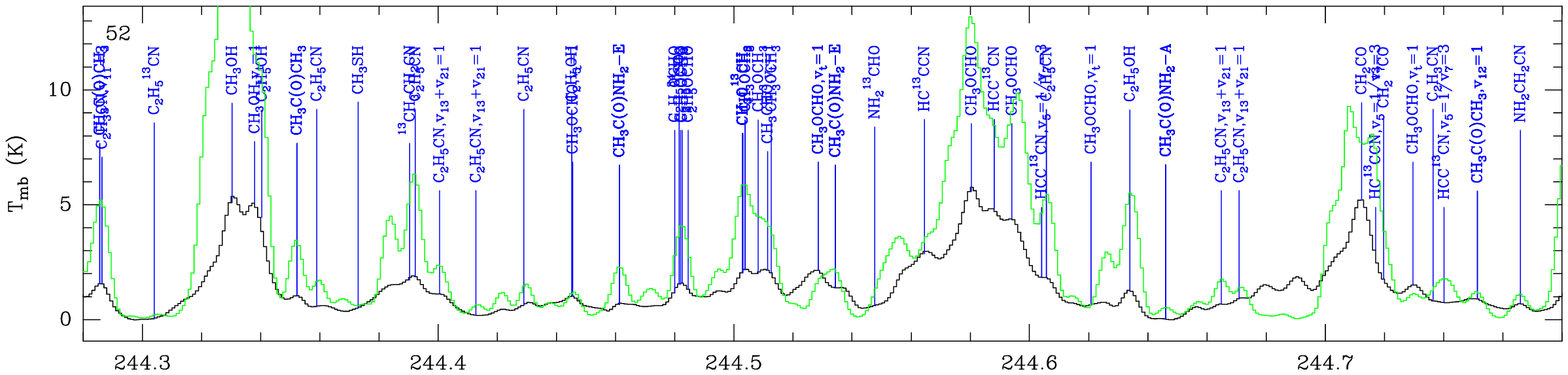}}}
\vspace*{1ex}\centerline{\resizebox{1.0\hsize}{!}{\includegraphics[angle=270]{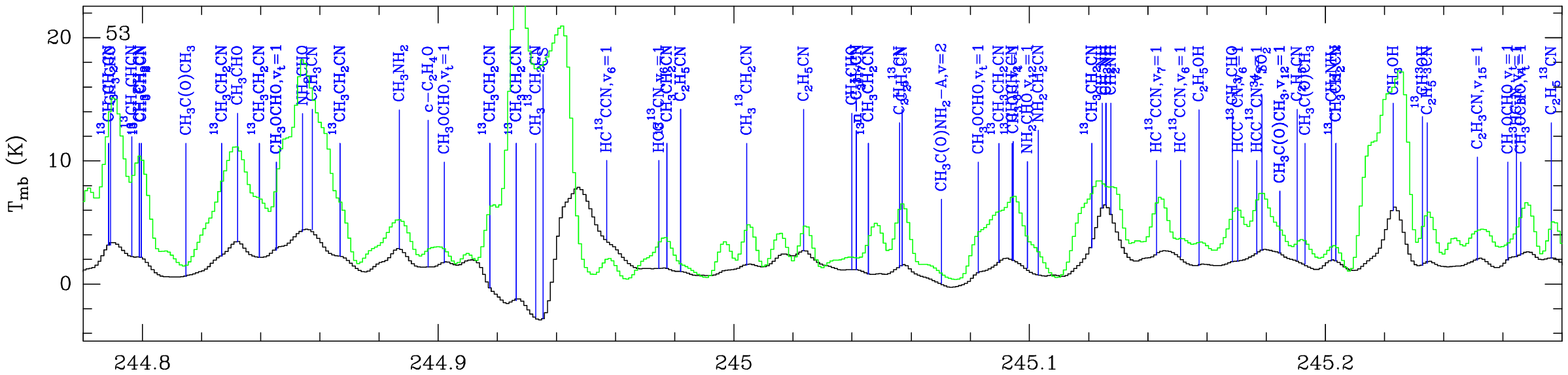}}}
\vspace*{1ex}\centerline{\resizebox{1.0\hsize}{!}{\includegraphics[angle=270]{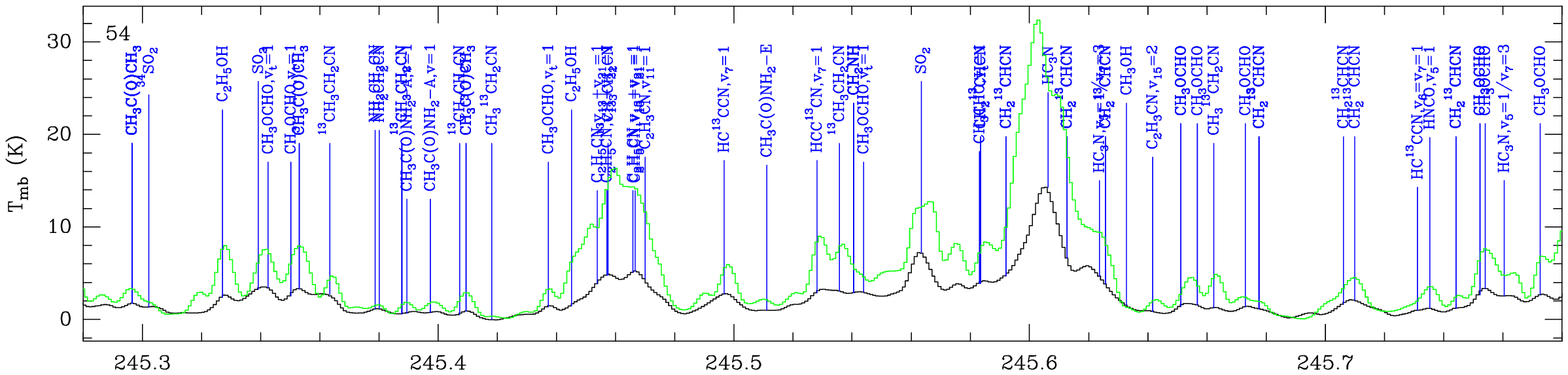}}}
\vspace*{1ex}\centerline{\resizebox{1.0\hsize}{!}{\includegraphics[angle=270]{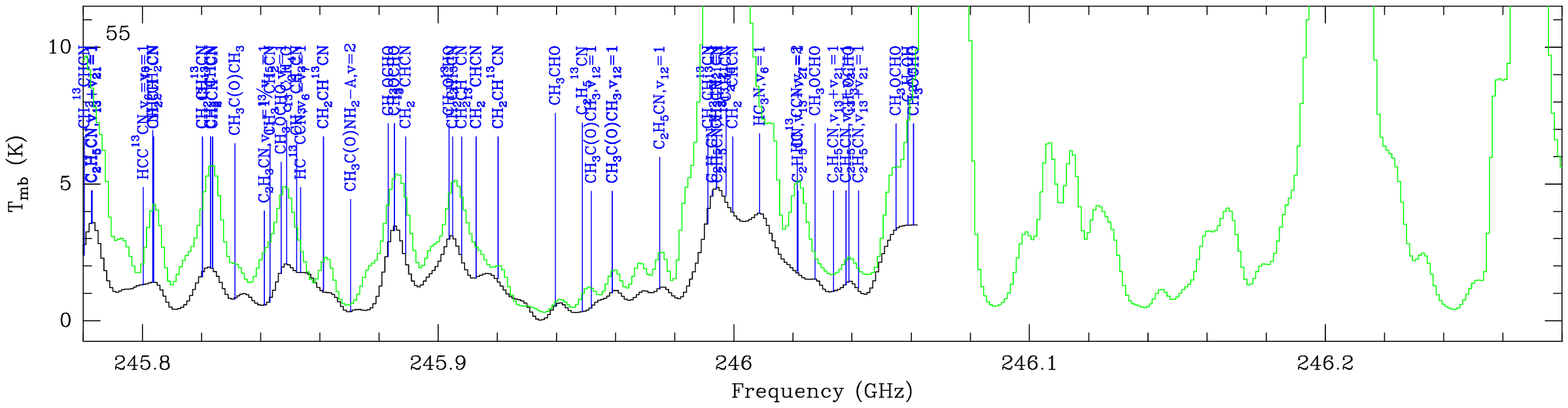}}}
\caption{
continued.
}
\end{figure*}
 \clearpage
\begin{figure*}
\addtocounter{figure}{-1}
\centerline{\resizebox{1.0\hsize}{!}{\includegraphics[angle=270]{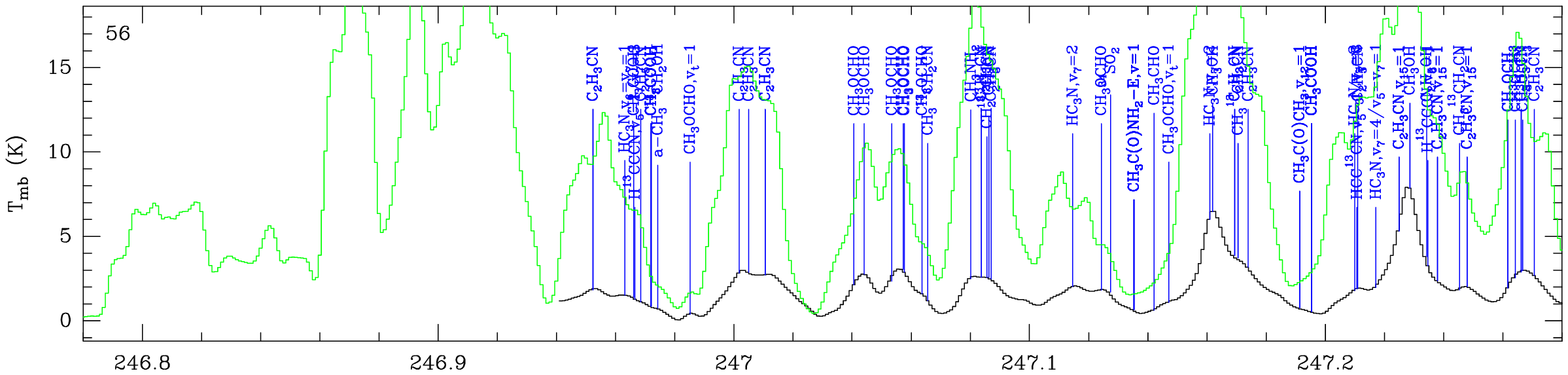}}}
\vspace*{1ex}\centerline{\resizebox{1.0\hsize}{!}{\includegraphics[angle=270]{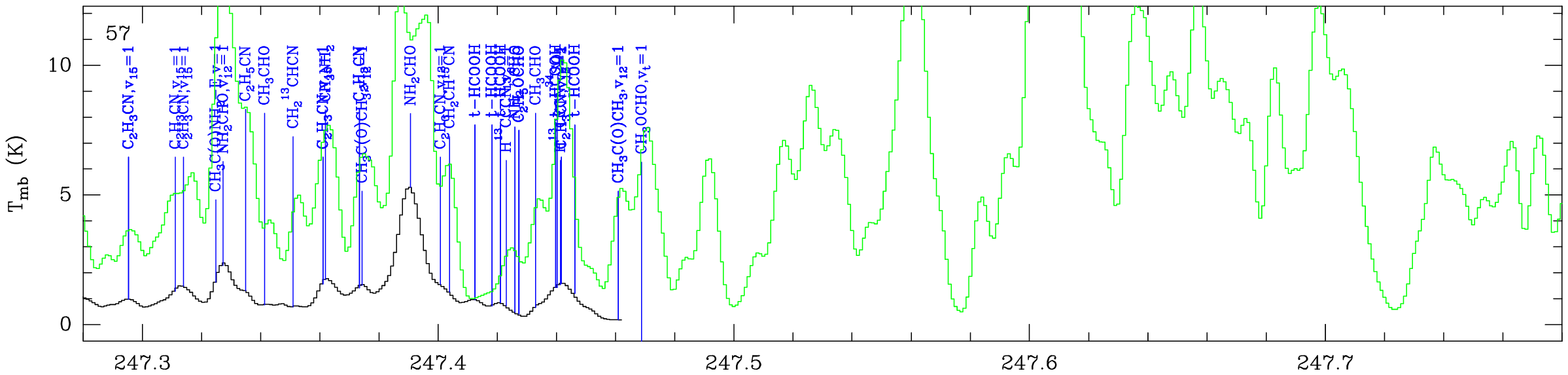}}}
\vspace*{1ex}\centerline{\resizebox{1.0\hsize}{!}{\includegraphics[angle=270]{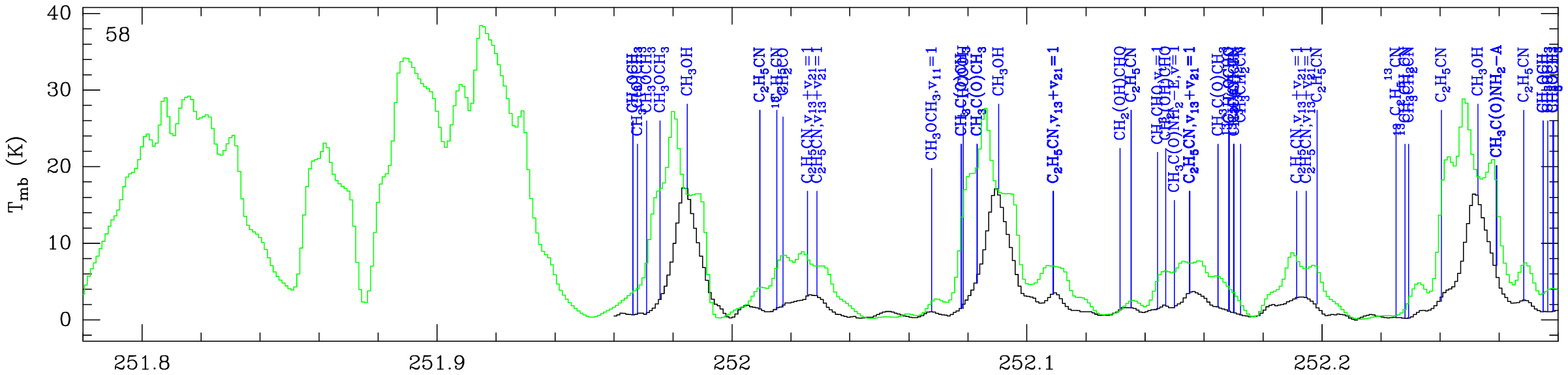}}}
\vspace*{1ex}\centerline{\resizebox{1.0\hsize}{!}{\includegraphics[angle=270]{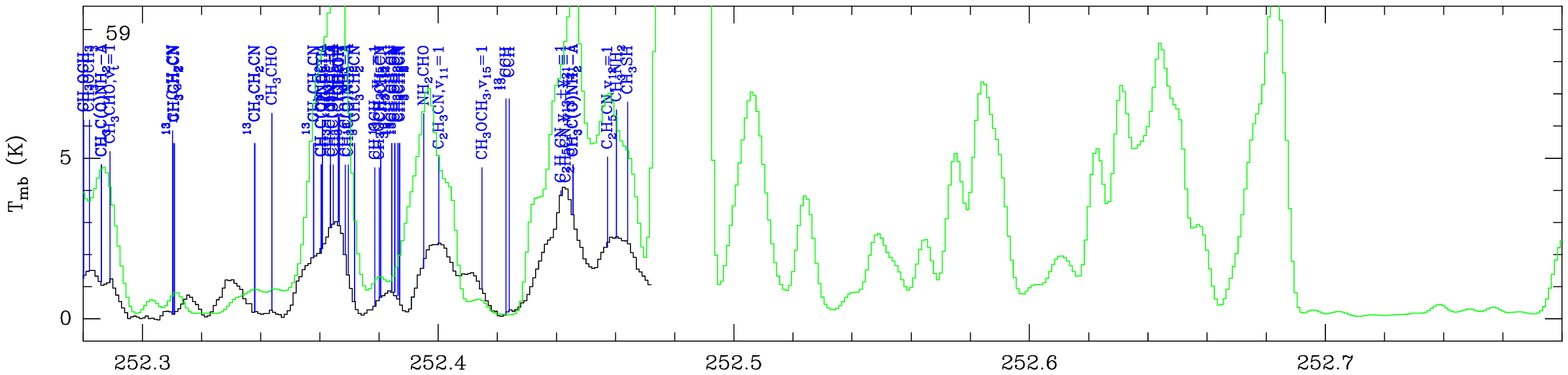}}}
\vspace*{1ex}\centerline{\resizebox{1.0\hsize}{!}{\includegraphics[angle=270]{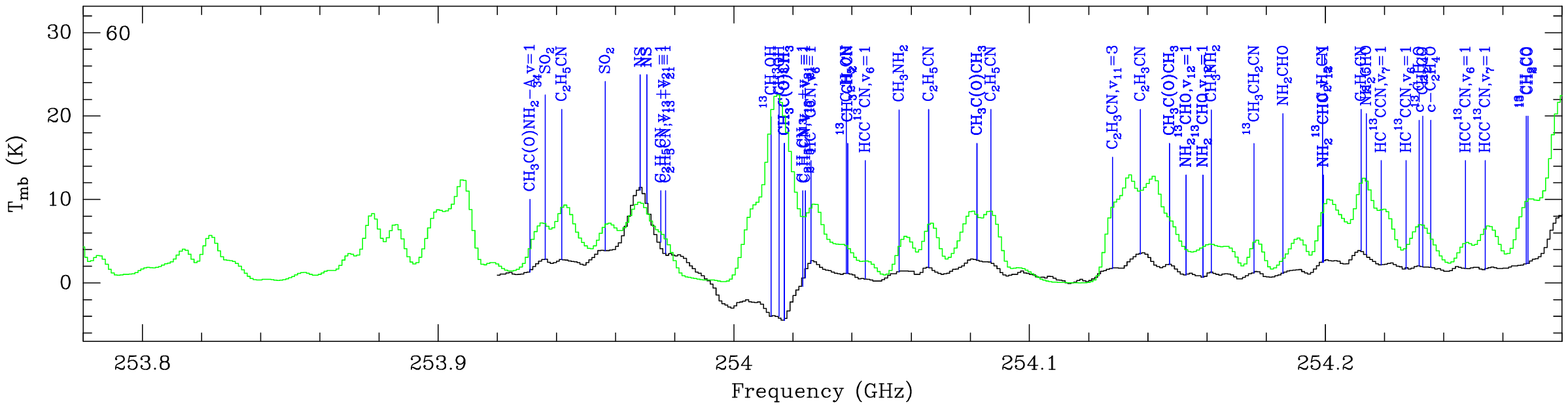}}}
\caption{
continued.
}
\end{figure*}
 \clearpage
\begin{figure*}
\addtocounter{figure}{-1}
\centerline{\resizebox{1.0\hsize}{!}{\includegraphics[angle=270]{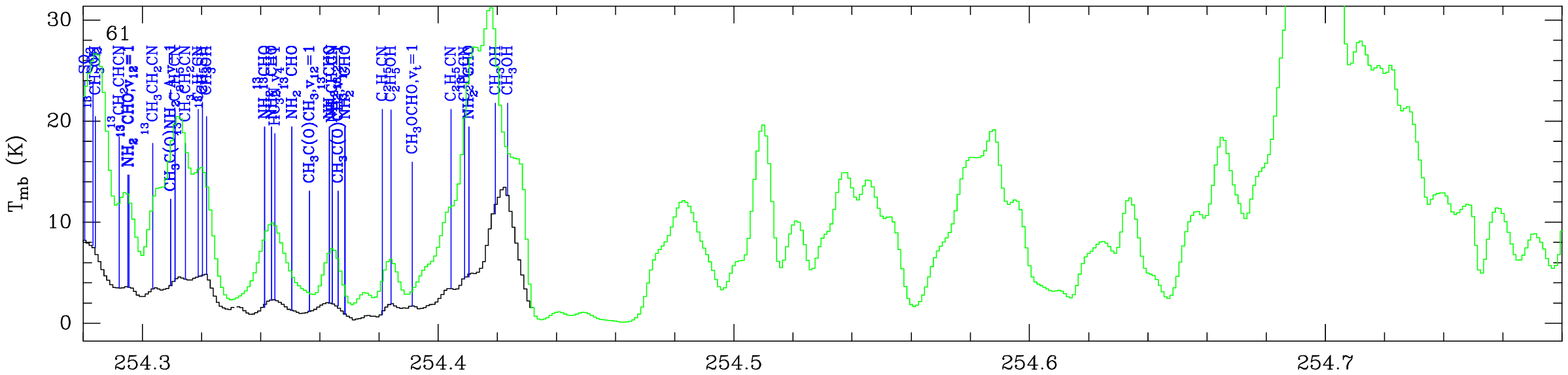}}}
\vspace*{1ex}\centerline{\resizebox{1.0\hsize}{!}{\includegraphics[angle=270]{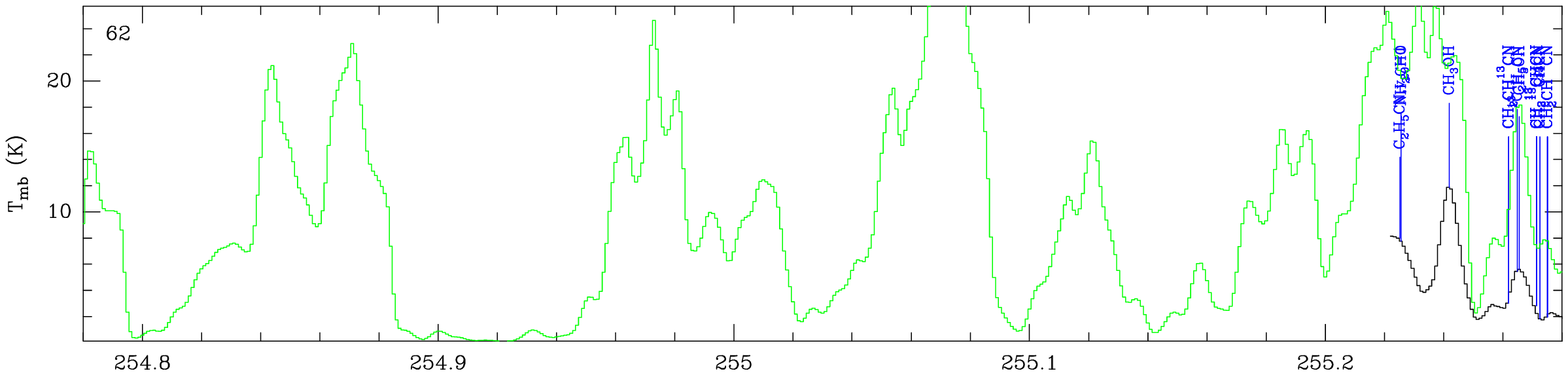}}}
\vspace*{1ex}\centerline{\resizebox{1.0\hsize}{!}{\includegraphics[angle=270]{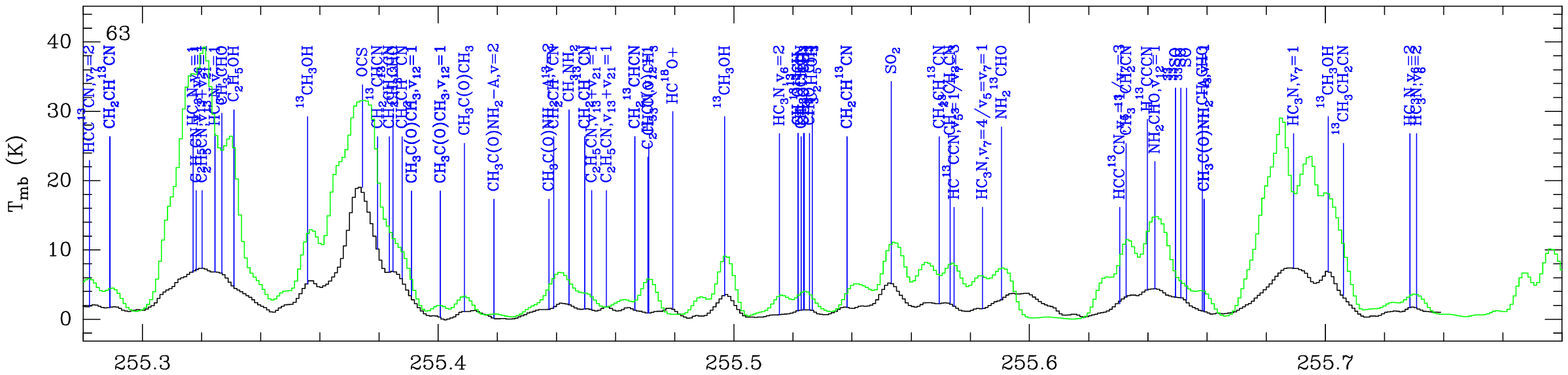}}}
\vspace*{1ex}\centerline{\resizebox{1.0\hsize}{!}{\includegraphics[angle=270]{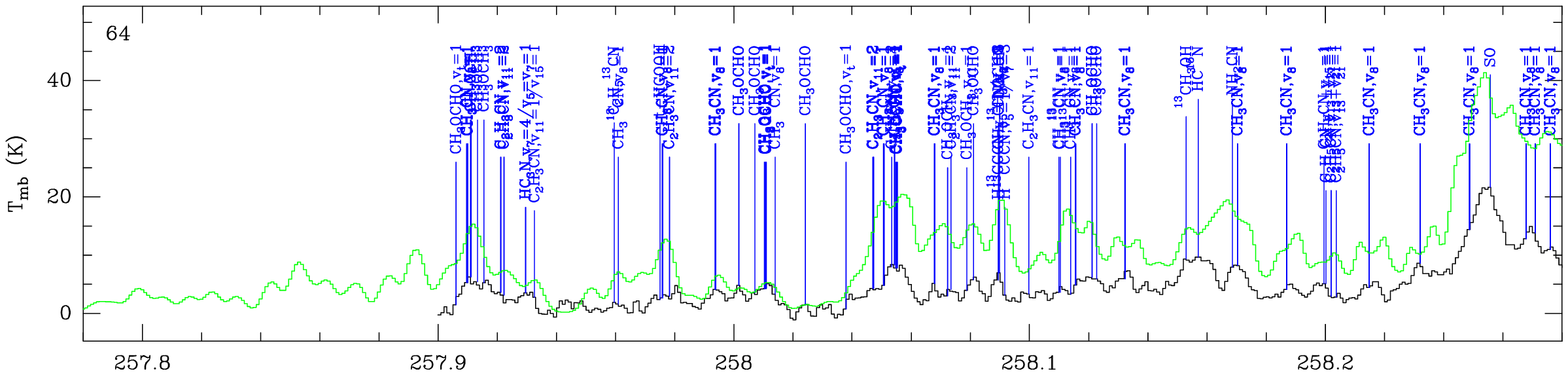}}}
\vspace*{1ex}\centerline{\resizebox{1.0\hsize}{!}{\includegraphics[angle=270]{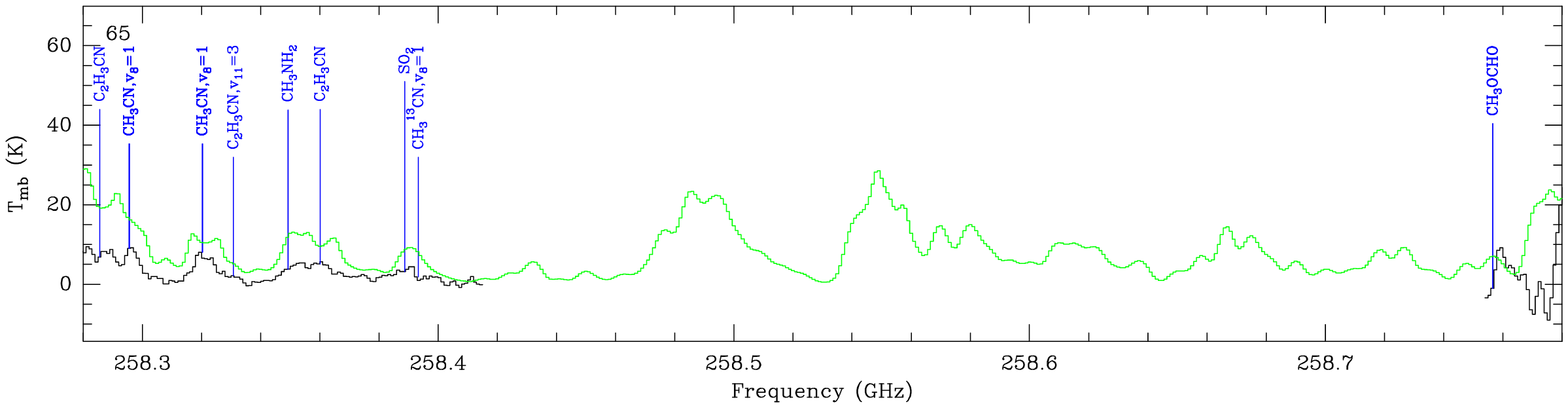}}}
\caption{
continued.
}
\end{figure*}
 \clearpage
\begin{figure*}
\addtocounter{figure}{-1}
\centerline{\resizebox{1.0\hsize}{!}{\includegraphics[angle=270]{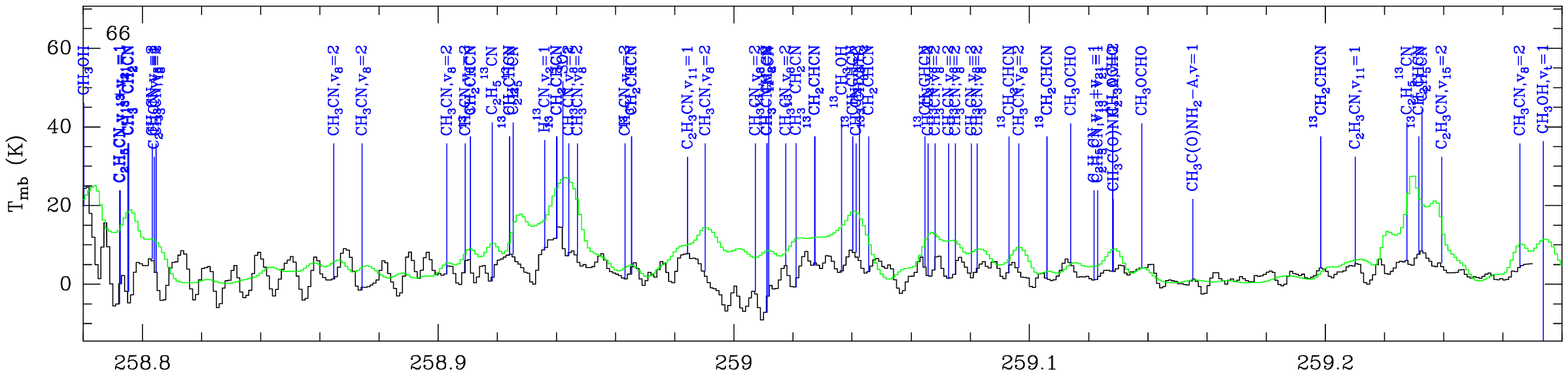}}}
\vspace*{1ex}\centerline{\resizebox{1.0\hsize}{!}{\includegraphics[angle=270]{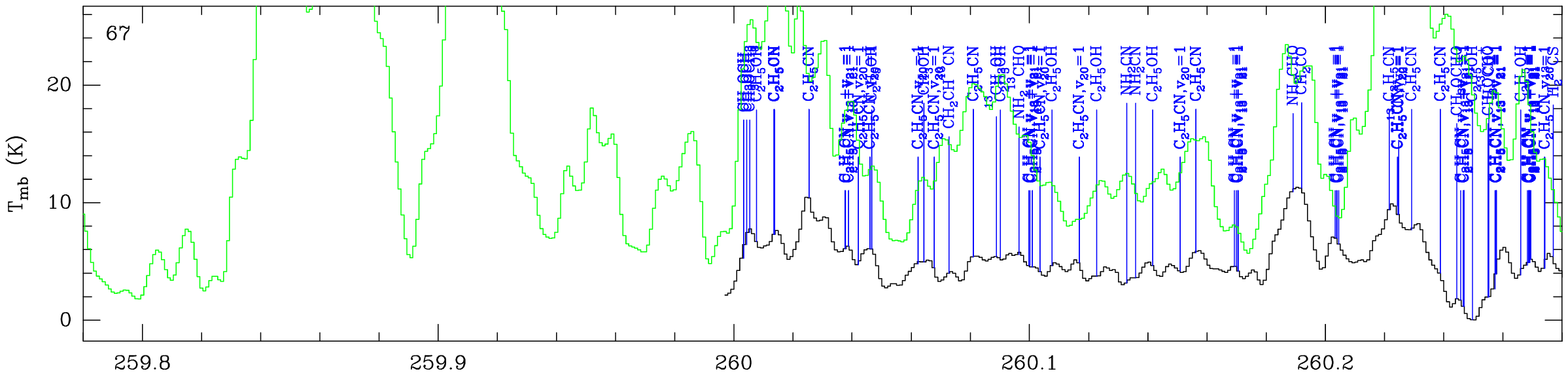}}}
\vspace*{1ex}\centerline{\resizebox{1.0\hsize}{!}{\includegraphics[angle=270]{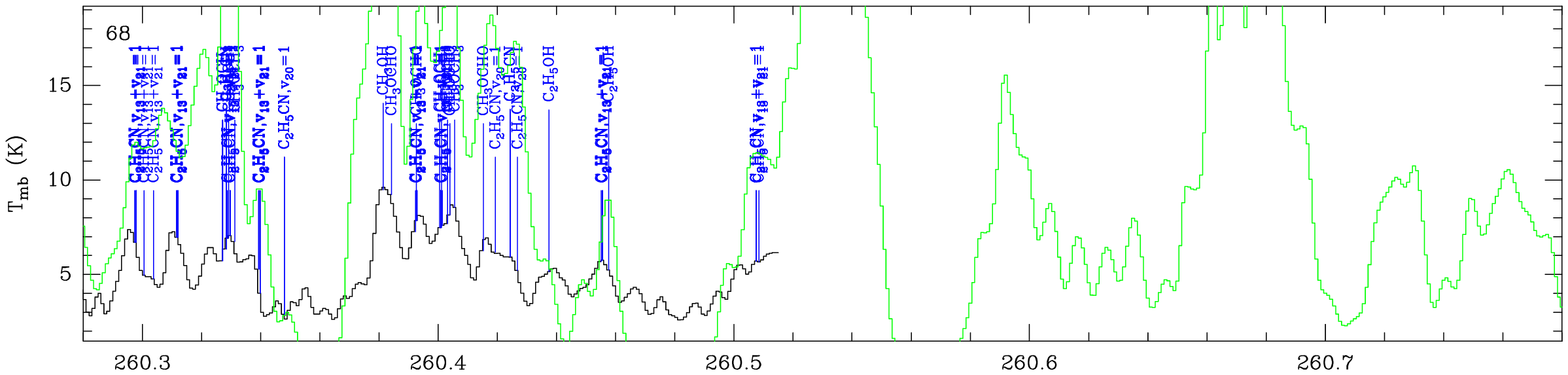}}}
\vspace*{1ex}\centerline{\resizebox{1.0\hsize}{!}{\includegraphics[angle=270]{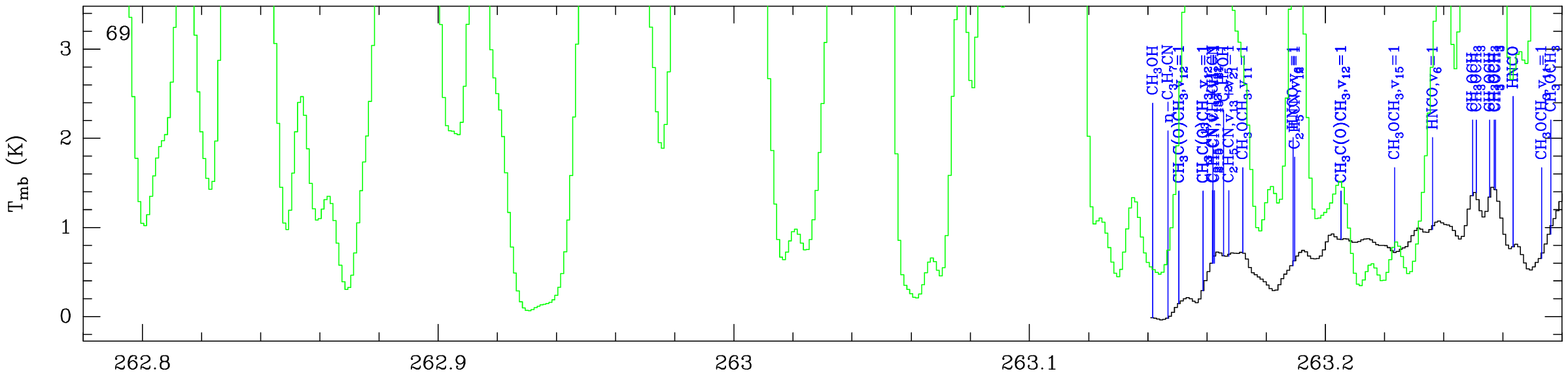}}}
\vspace*{1ex}\centerline{\resizebox{1.0\hsize}{!}{\includegraphics[angle=270]{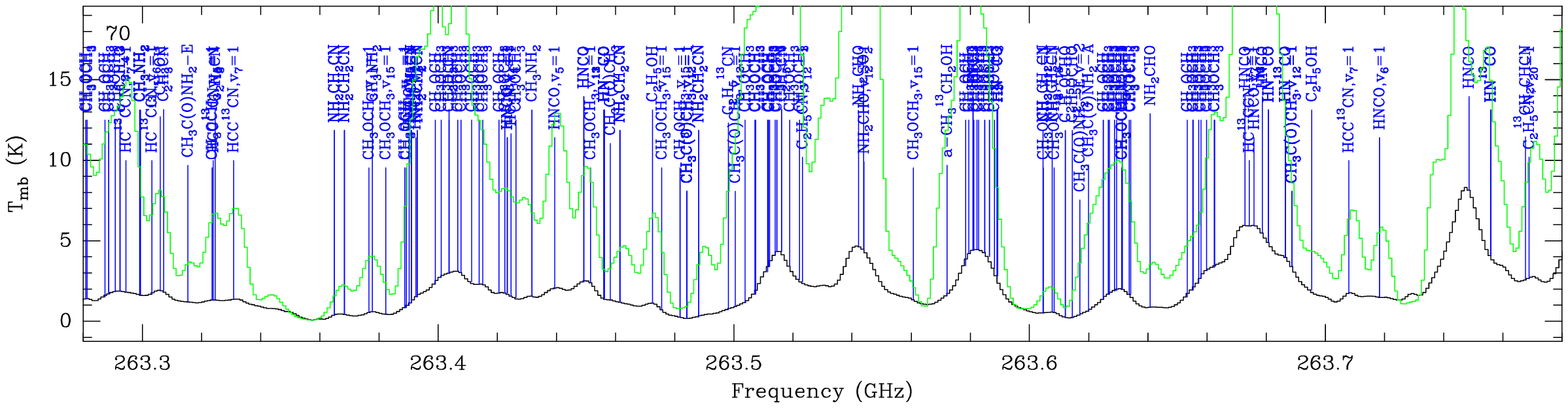}}}
\caption{
continued.
}
\end{figure*}
 \clearpage
\begin{figure*}
\addtocounter{figure}{-1}
\centerline{\resizebox{1.0\hsize}{!}{\includegraphics[angle=270]{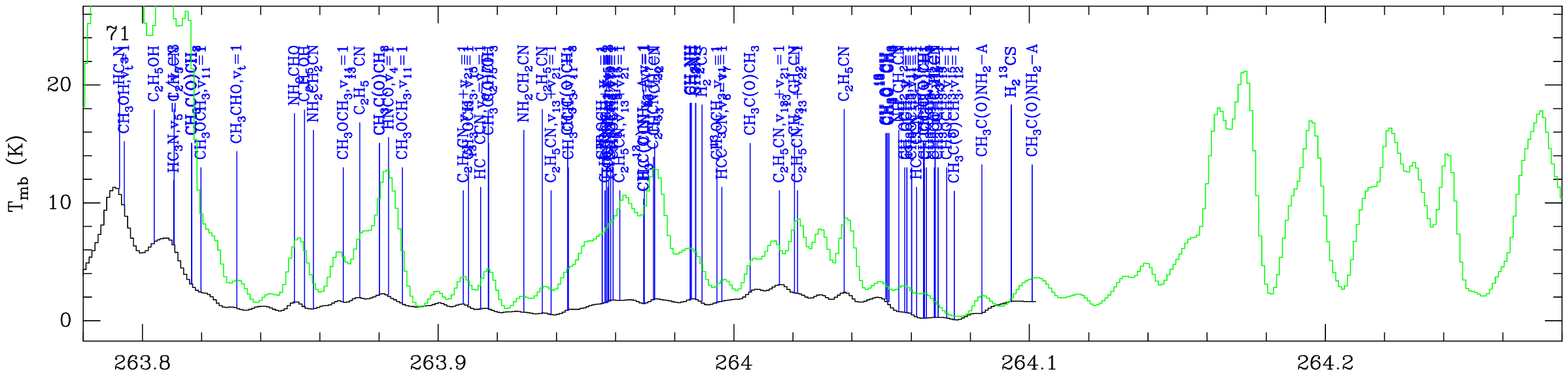}}}
\vspace*{1ex}\centerline{\resizebox{1.0\hsize}{!}{\includegraphics[angle=270]{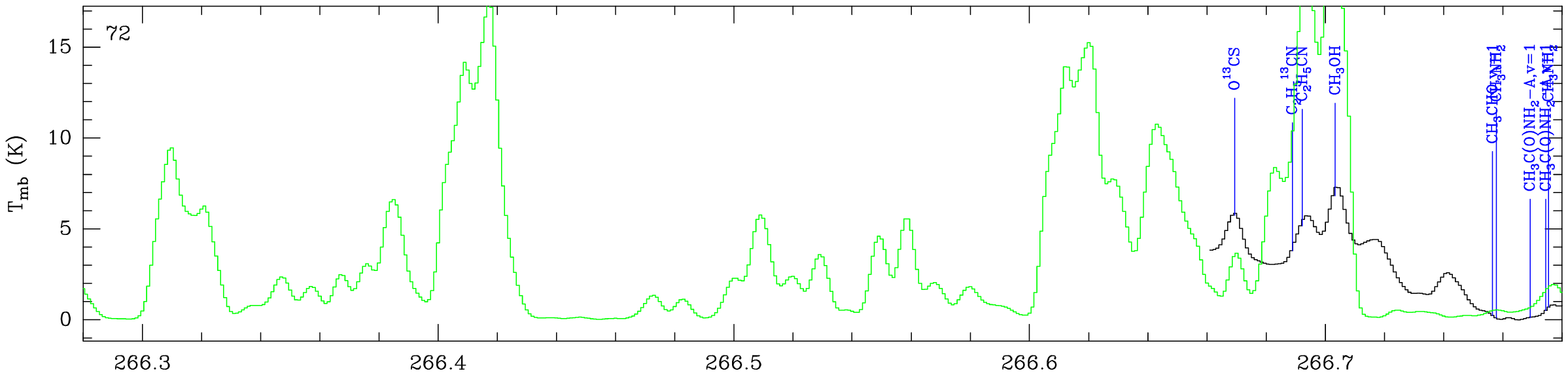}}}
\vspace*{1ex}\centerline{\resizebox{1.0\hsize}{!}{\includegraphics[angle=270]{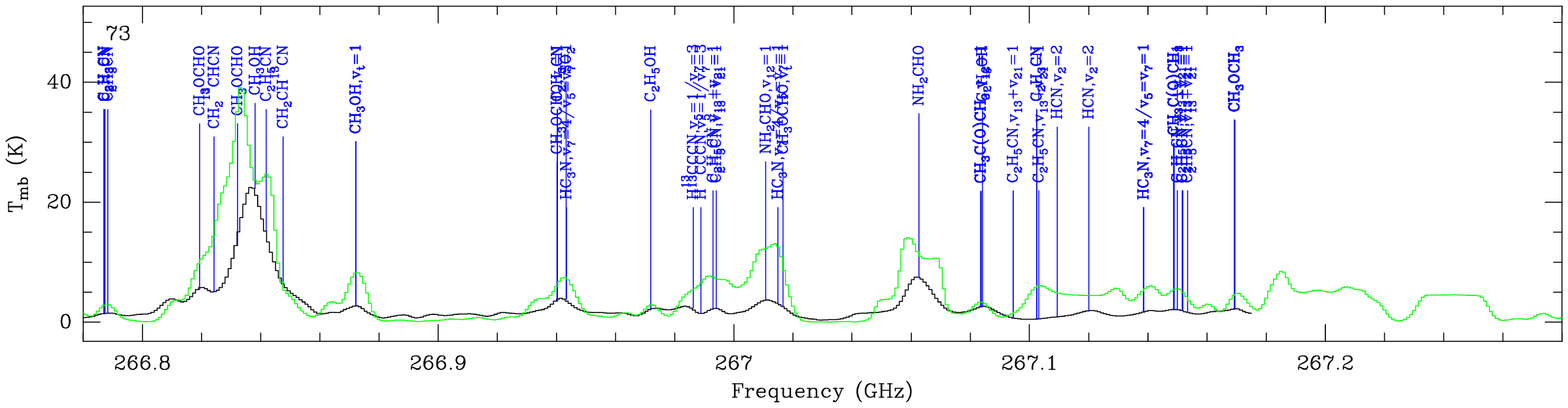}}}
\caption{
continued.
}
\end{figure*}
 \clearpage

}
\onlfig{\clearpage
\begin{figure*}
\centerline{\resizebox{1.0\hsize}{!}{\includegraphics[angle=270]{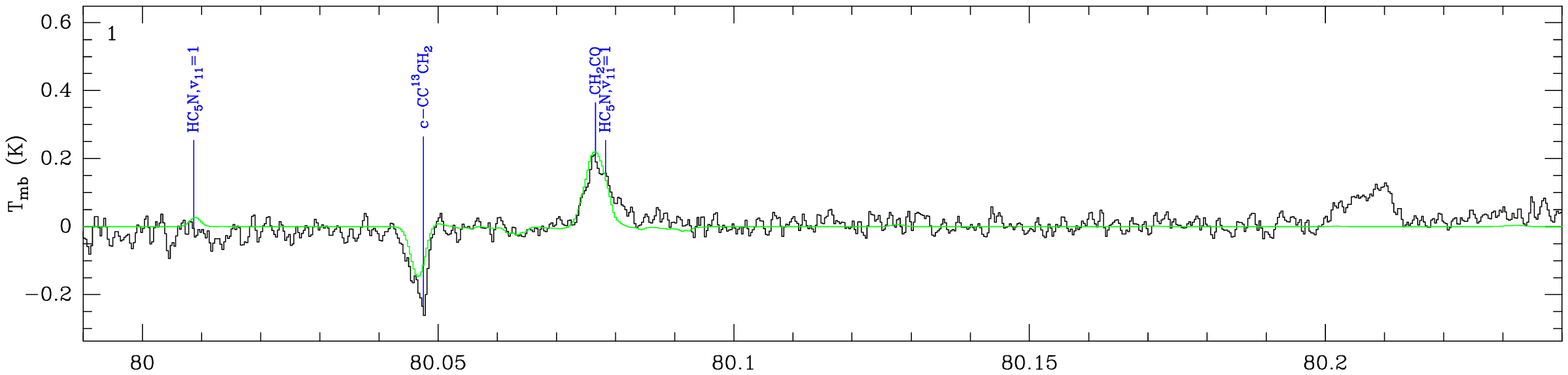}}}
\vspace*{1ex}\centerline{\resizebox{1.0\hsize}{!}{\includegraphics[angle=270]{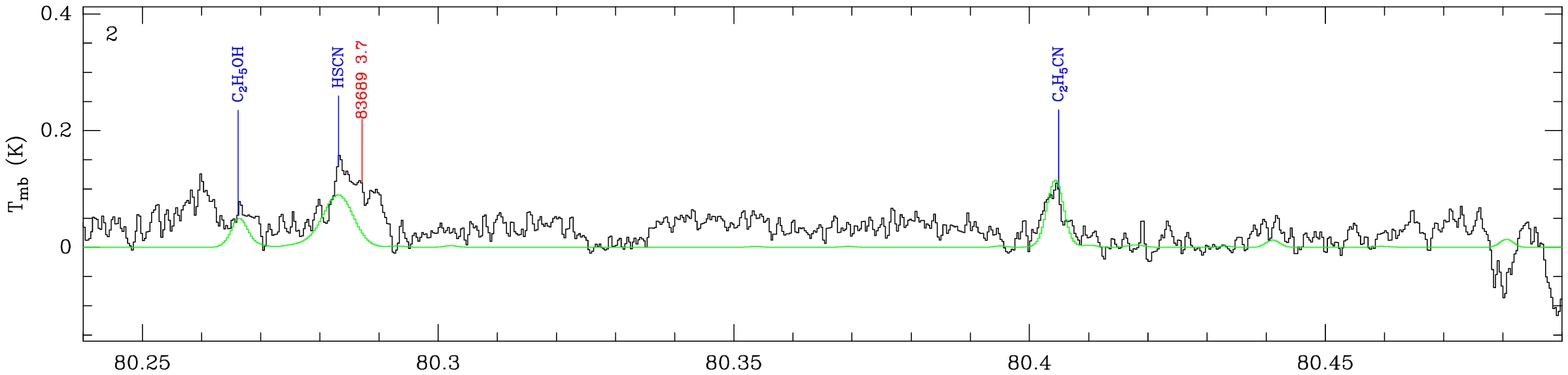}}}
\vspace*{1ex}\centerline{\resizebox{1.0\hsize}{!}{\includegraphics[angle=270]{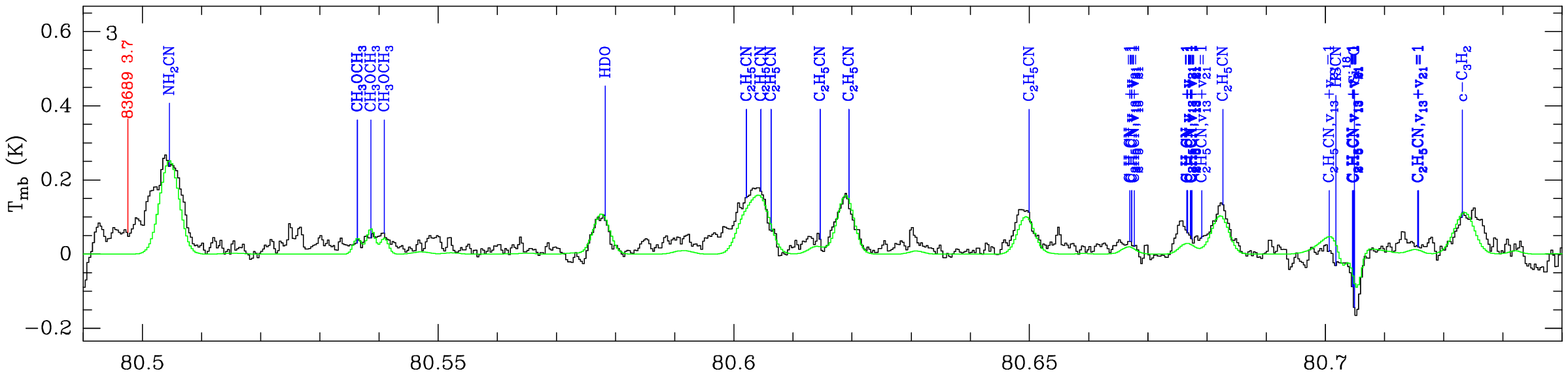}}}
\vspace*{1ex}\centerline{\resizebox{1.0\hsize}{!}{\includegraphics[angle=270]{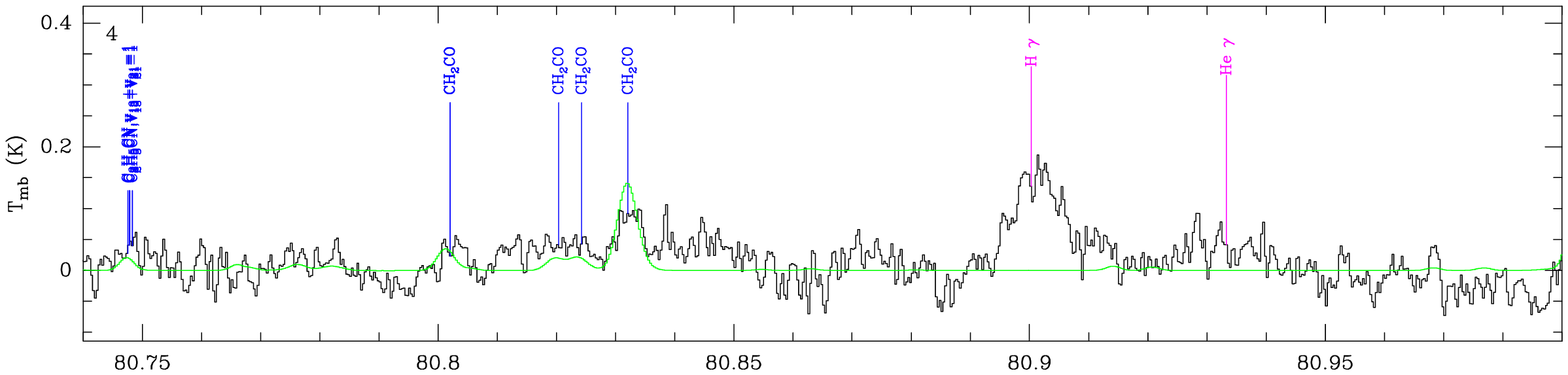}}}
\vspace*{1ex}\centerline{\resizebox{1.0\hsize}{!}{\includegraphics[angle=270]{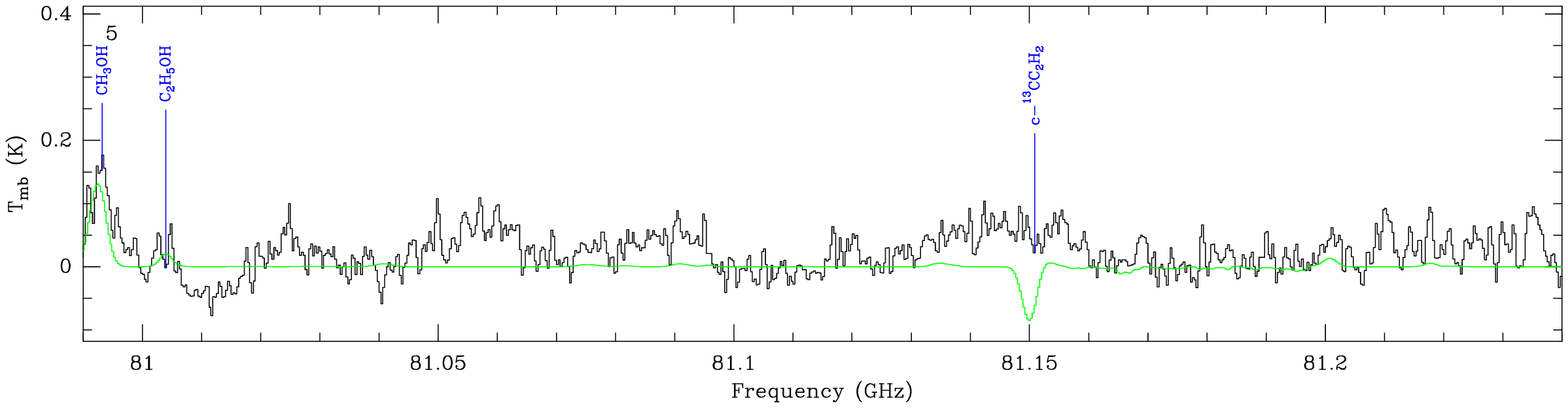}}}
\caption{
Spectrum obtained toward Sgr~B2(M)
in the 3~mm window
with the IRAM~30\,m telescope in main-beam temperature scale. The synthetic model is overlaid in green and its relevant lines are labeled in blue.
The frequencies of the hydrogen and helium recombination lines are indicated with a pink label.
The position of the lines with a peak temperature higher than 2~K in the image band and possibly contaminating the spectrum are marked with a red label indicating their rest frequency and their peak temperature in K in the image band.
}
\label{f:survey_b2m_3mm}
\end{figure*}
 \clearpage
\begin{figure*}
\addtocounter{figure}{-1}
\centerline{\resizebox{1.0\hsize}{!}{\includegraphics[angle=270]{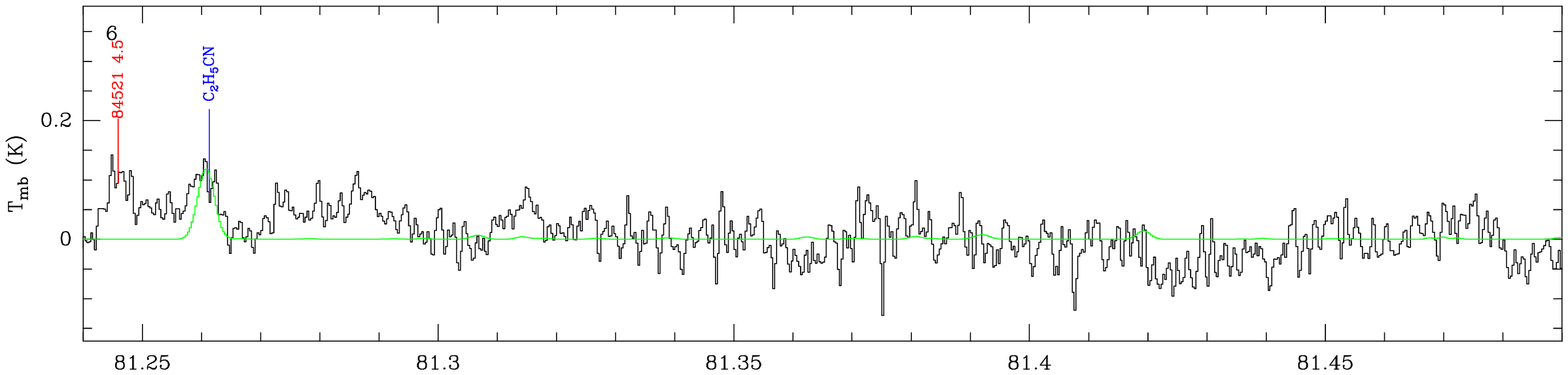}}}
\vspace*{1ex}\centerline{\resizebox{1.0\hsize}{!}{\includegraphics[angle=270]{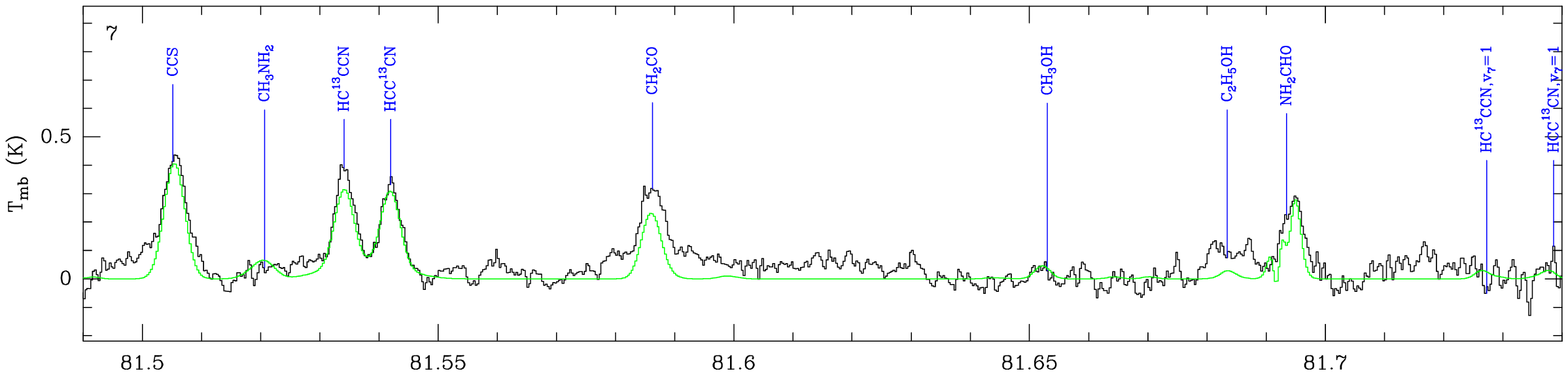}}}
\vspace*{1ex}\centerline{\resizebox{1.0\hsize}{!}{\includegraphics[angle=270]{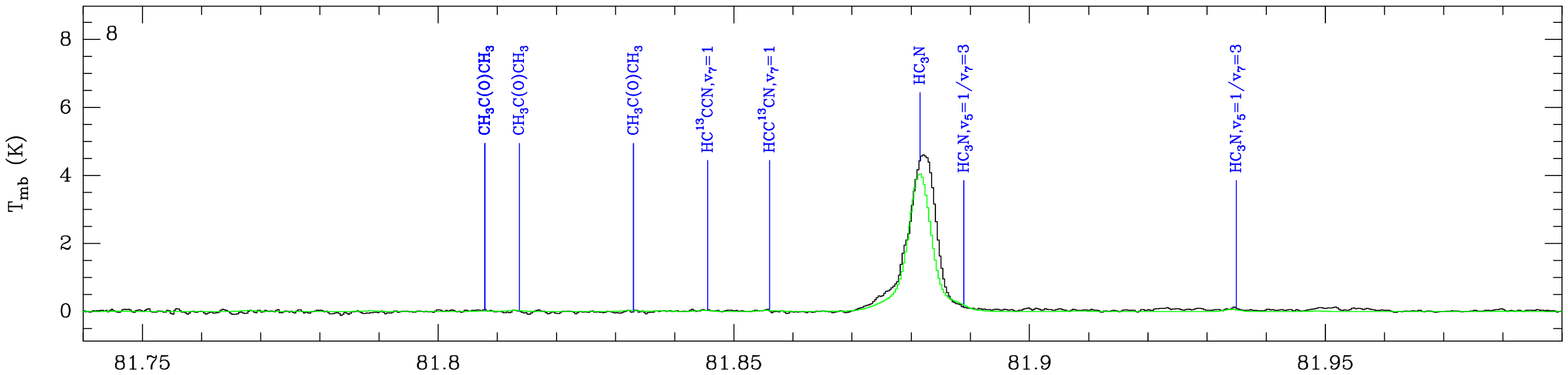}}}
\vspace*{1ex}\centerline{\resizebox{1.0\hsize}{!}{\includegraphics[angle=270]{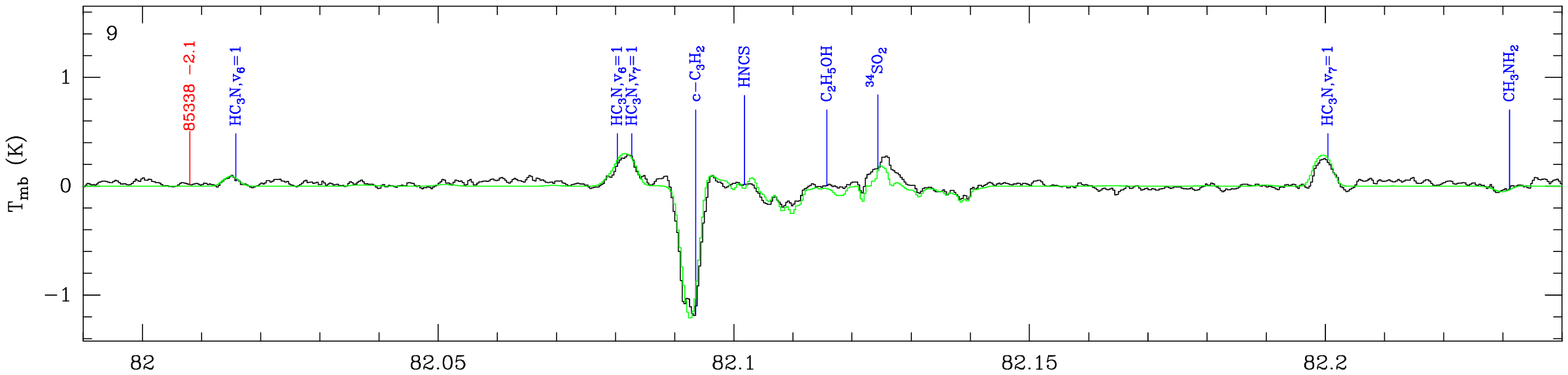}}}
\vspace*{1ex}\centerline{\resizebox{1.0\hsize}{!}{\includegraphics[angle=270]{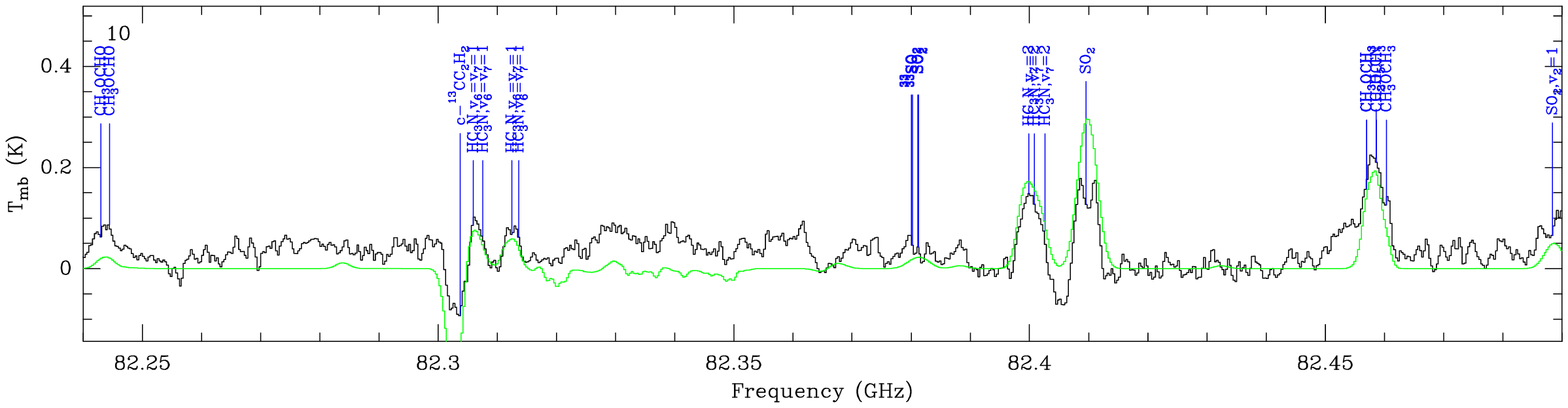}}}
\caption{
continued.
}
\end{figure*}
 \clearpage
\begin{figure*}
\addtocounter{figure}{-1}
\centerline{\resizebox{1.0\hsize}{!}{\includegraphics[angle=270]{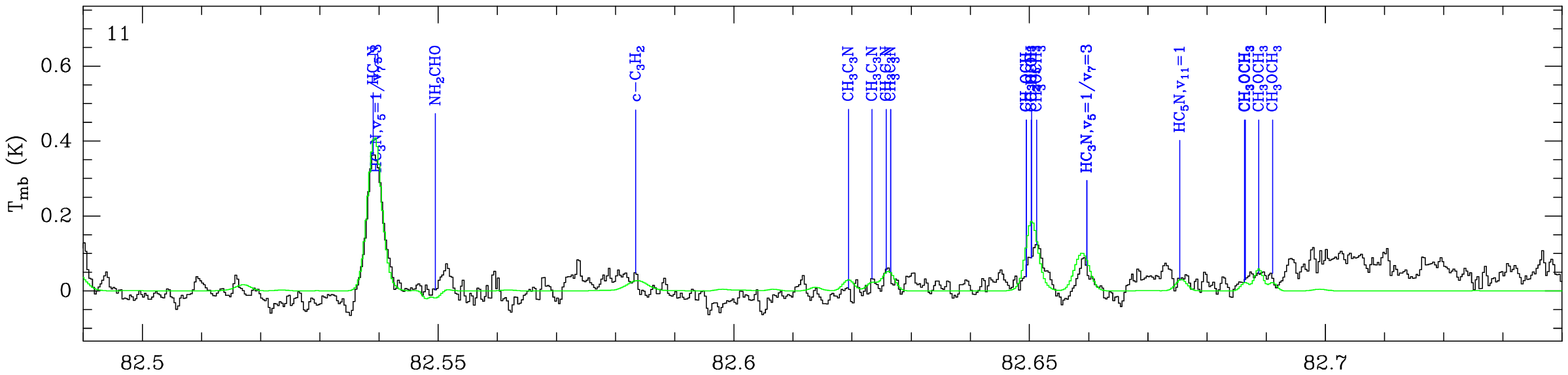}}}
\vspace*{1ex}\centerline{\resizebox{1.0\hsize}{!}{\includegraphics[angle=270]{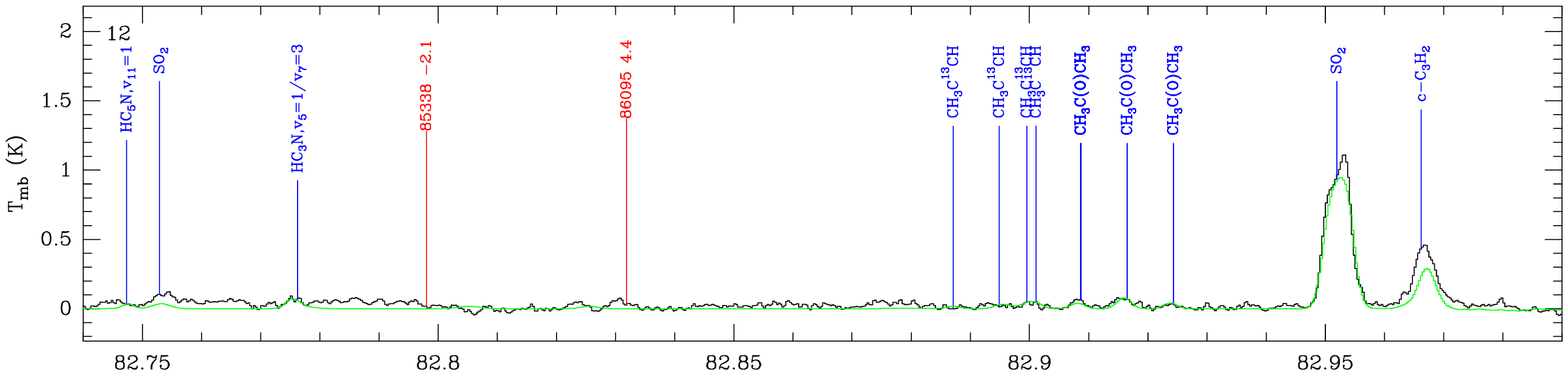}}}
\vspace*{1ex}\centerline{\resizebox{1.0\hsize}{!}{\includegraphics[angle=270]{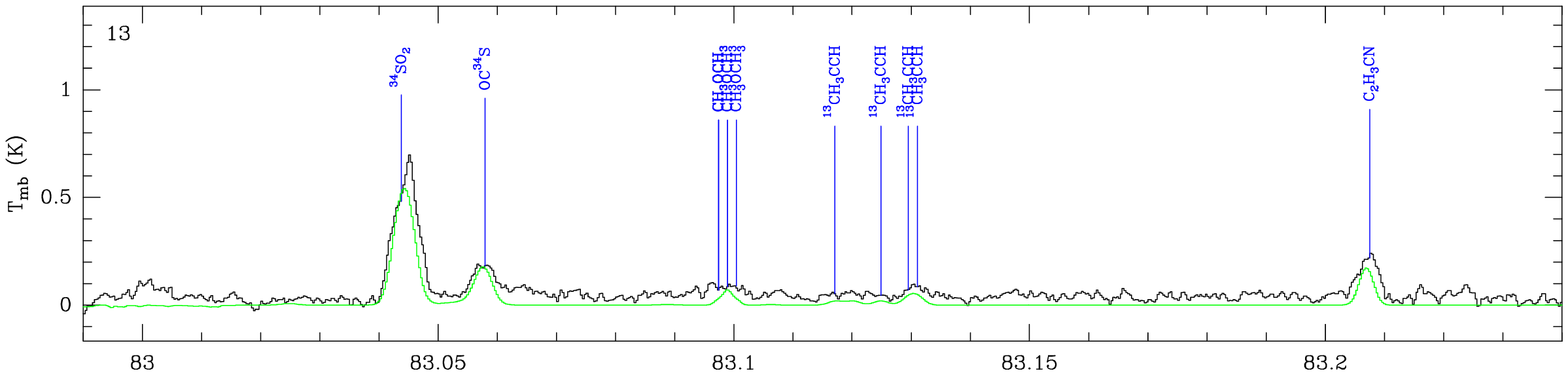}}}
\vspace*{1ex}\centerline{\resizebox{1.0\hsize}{!}{\includegraphics[angle=270]{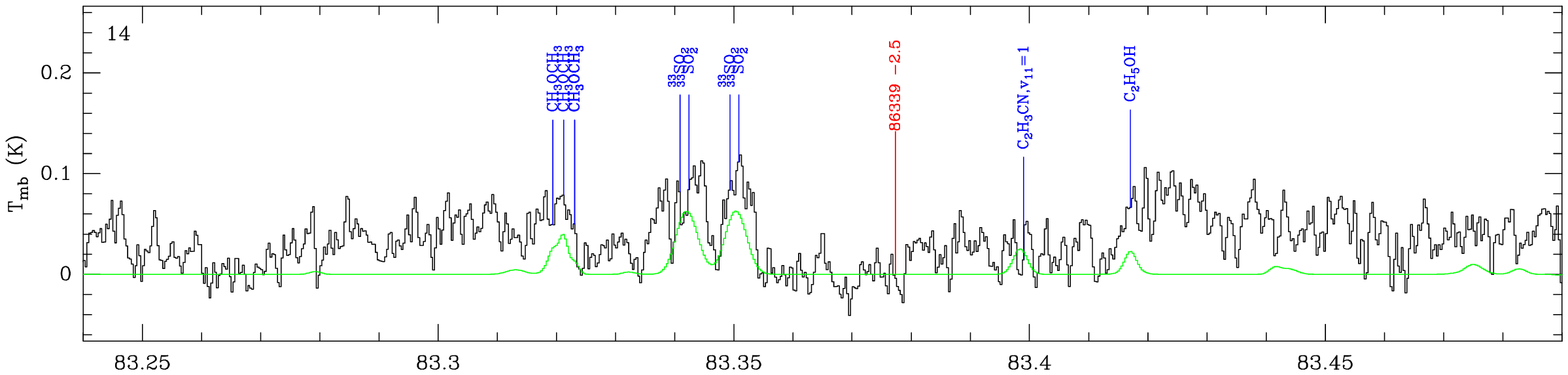}}}
\vspace*{1ex}\centerline{\resizebox{1.0\hsize}{!}{\includegraphics[angle=270]{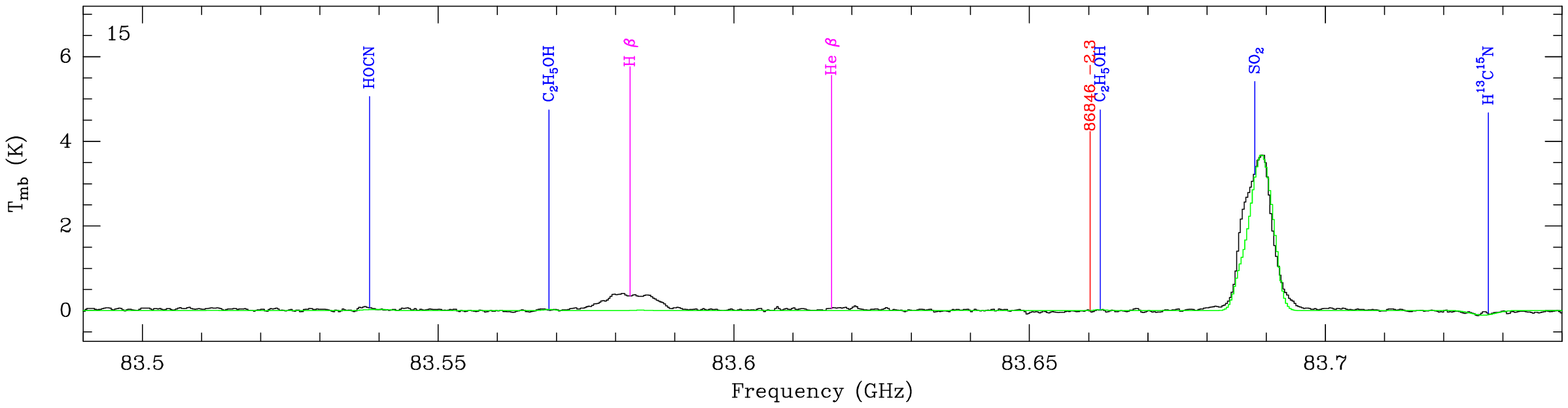}}}
\caption{
continued.
}
\end{figure*}
 \clearpage
\begin{figure*}
\addtocounter{figure}{-1}
\centerline{\resizebox{1.0\hsize}{!}{\includegraphics[angle=270]{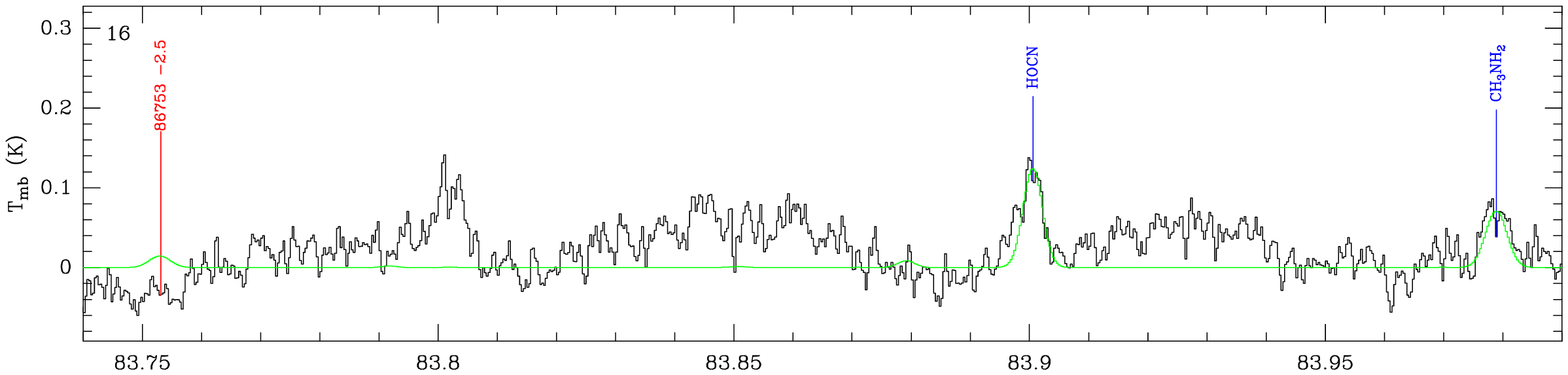}}}
\vspace*{1ex}\centerline{\resizebox{1.0\hsize}{!}{\includegraphics[angle=270]{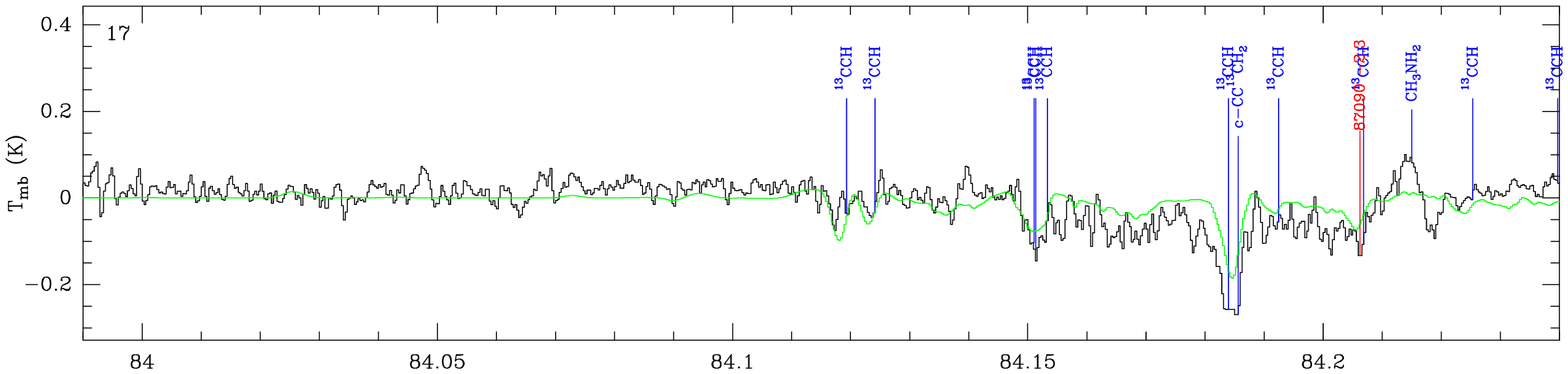}}}
\vspace*{1ex}\centerline{\resizebox{1.0\hsize}{!}{\includegraphics[angle=270]{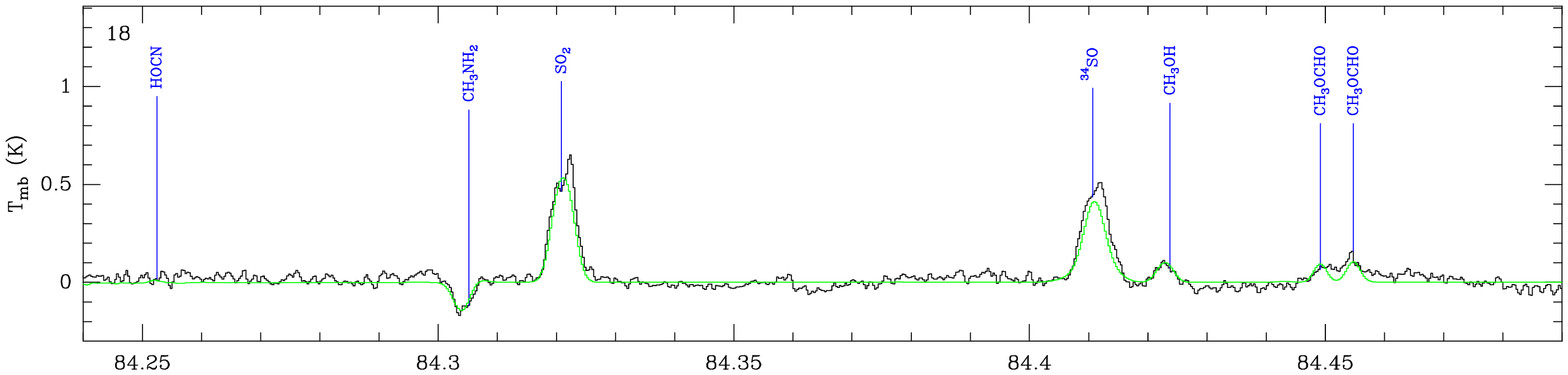}}}
\vspace*{1ex}\centerline{\resizebox{1.0\hsize}{!}{\includegraphics[angle=270]{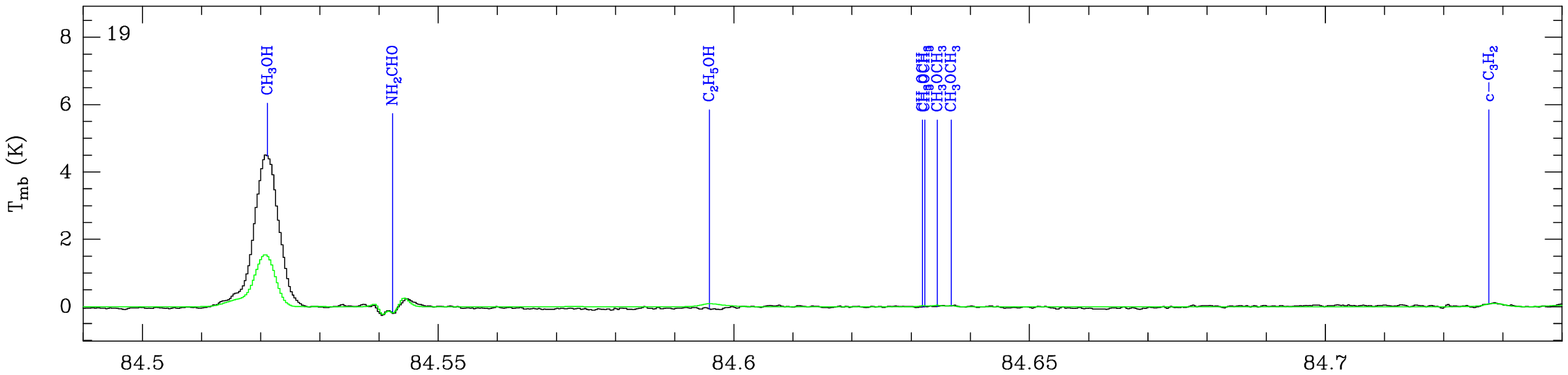}}}
\vspace*{1ex}\centerline{\resizebox{1.0\hsize}{!}{\includegraphics[angle=270]{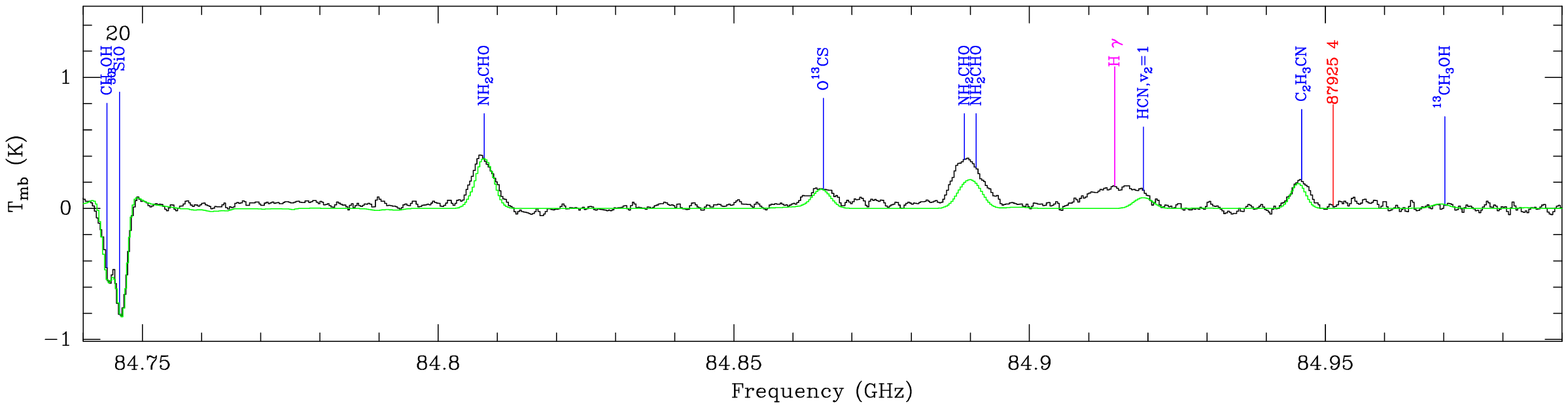}}}
\caption{
continued.
}
\end{figure*}
 \clearpage
\begin{figure*}
\addtocounter{figure}{-1}
\centerline{\resizebox{1.0\hsize}{!}{\includegraphics[angle=270]{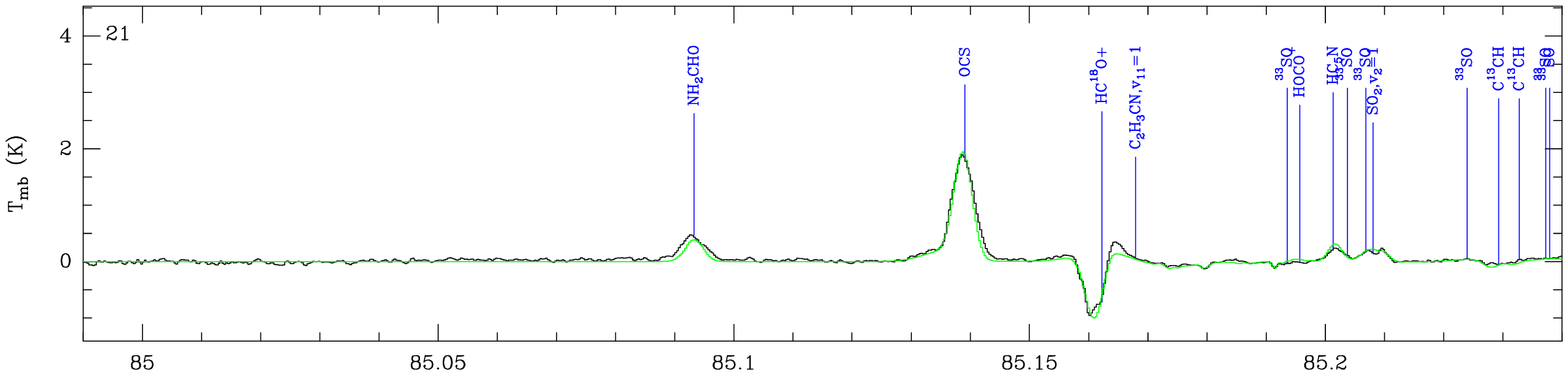}}}
\vspace*{1ex}\centerline{\resizebox{1.0\hsize}{!}{\includegraphics[angle=270]{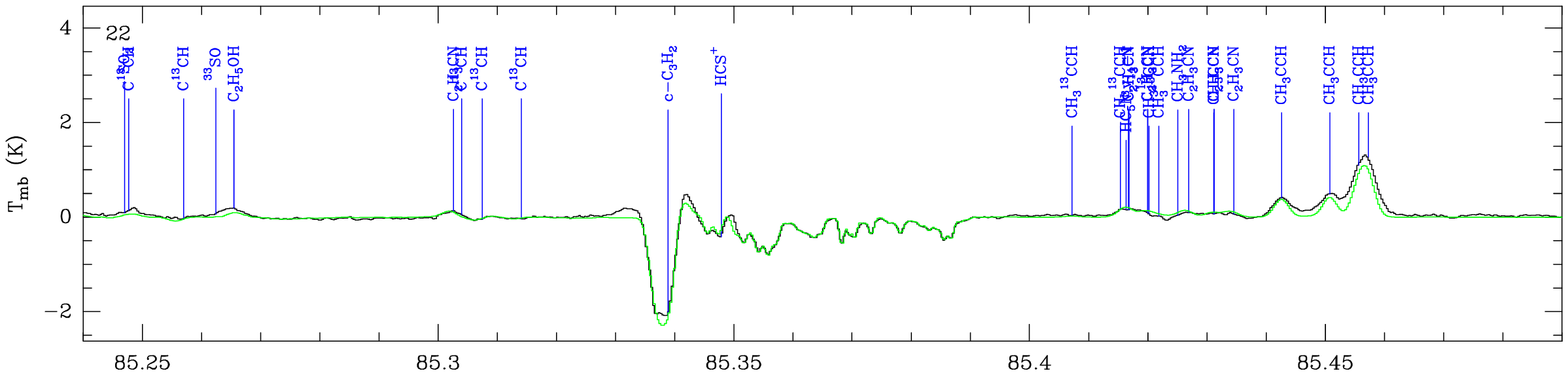}}}
\vspace*{1ex}\centerline{\resizebox{1.0\hsize}{!}{\includegraphics[angle=270]{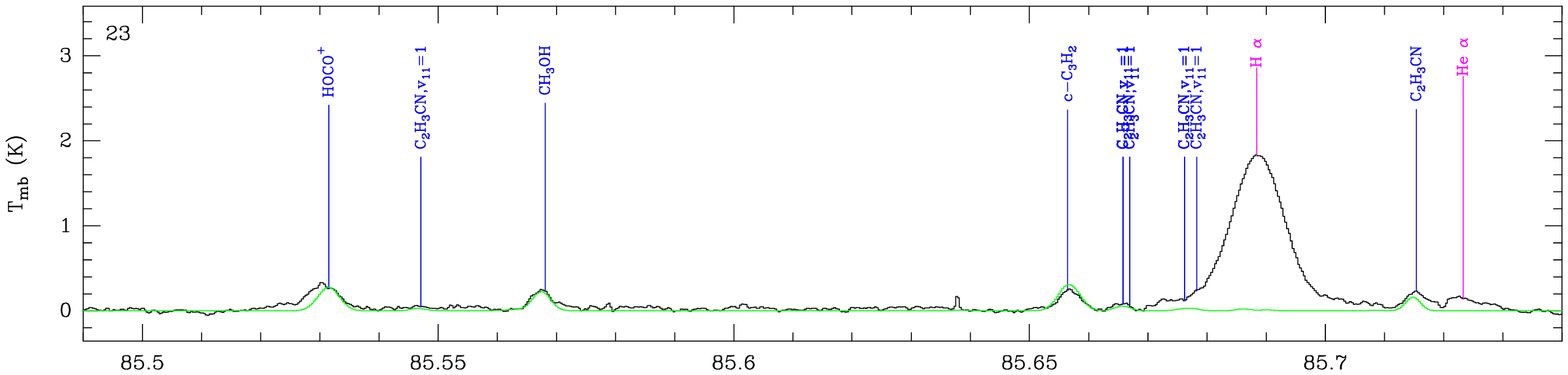}}}
\vspace*{1ex}\centerline{\resizebox{1.0\hsize}{!}{\includegraphics[angle=270]{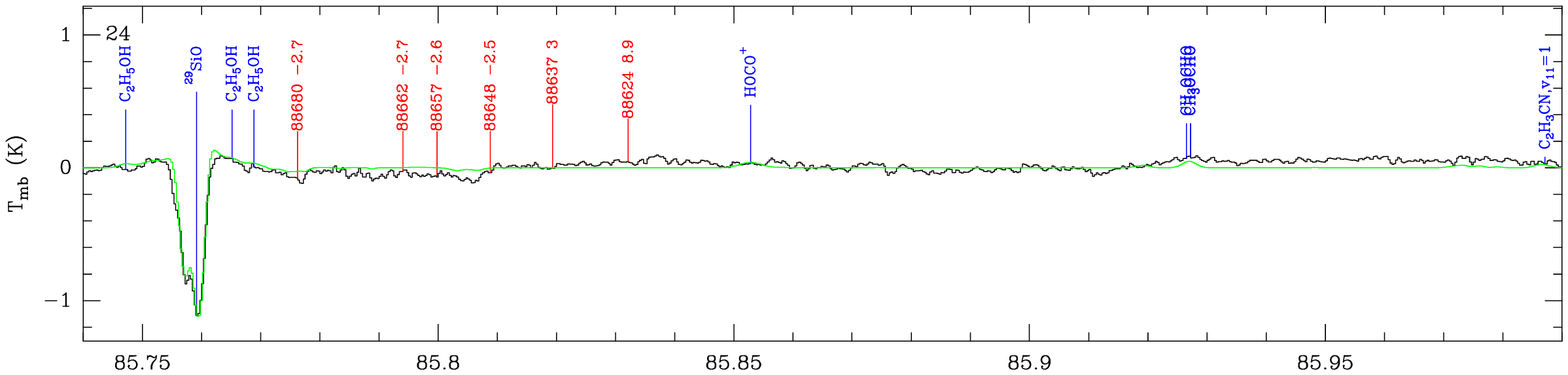}}}
\vspace*{1ex}\centerline{\resizebox{1.0\hsize}{!}{\includegraphics[angle=270]{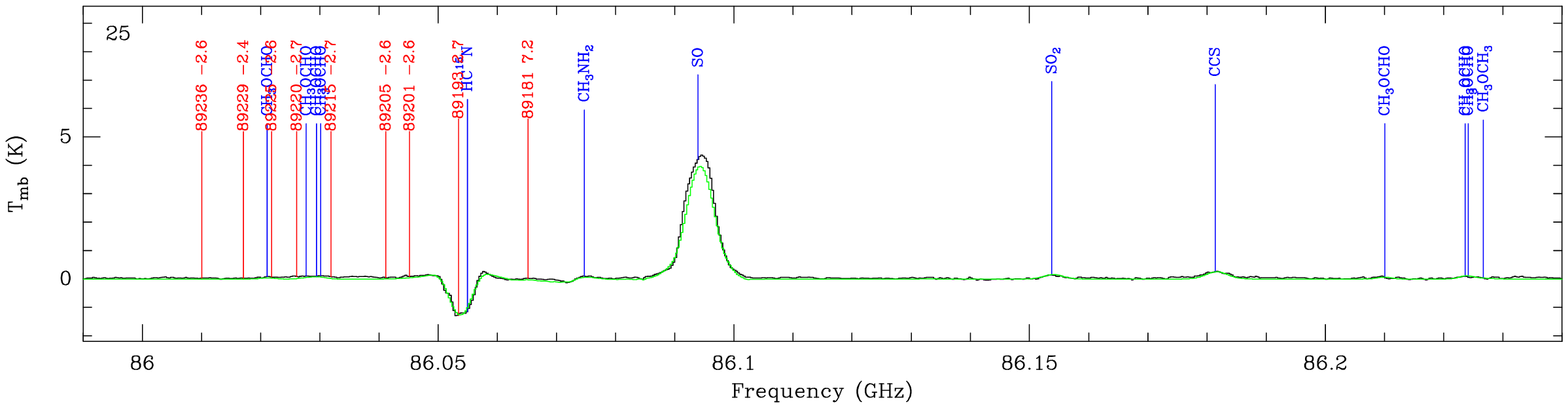}}}
\caption{
continued.
}
\end{figure*}
 \clearpage
\begin{figure*}
\addtocounter{figure}{-1}
\centerline{\resizebox{1.0\hsize}{!}{\includegraphics[angle=270]{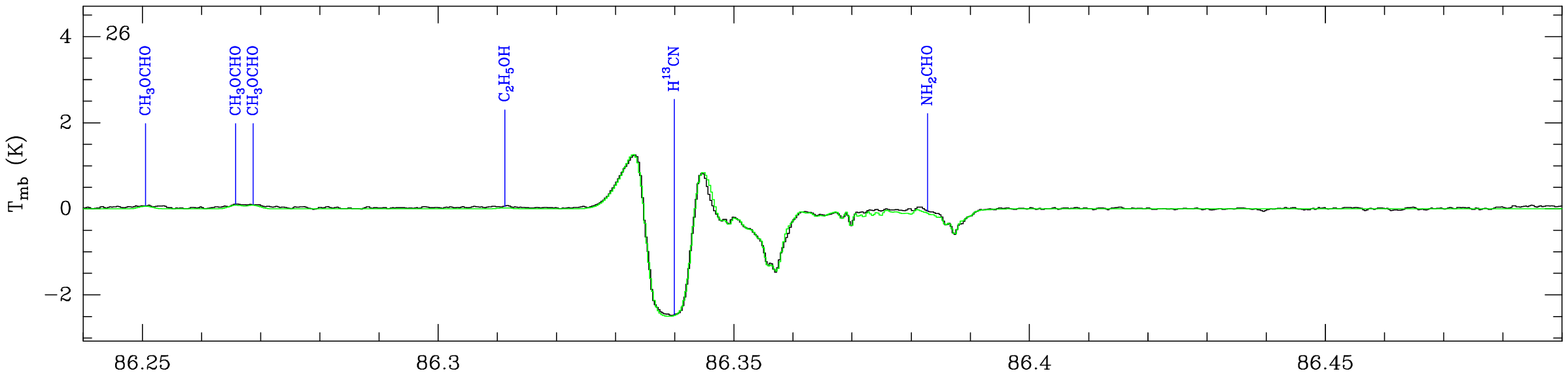}}}
\vspace*{1ex}\centerline{\resizebox{1.0\hsize}{!}{\includegraphics[angle=270]{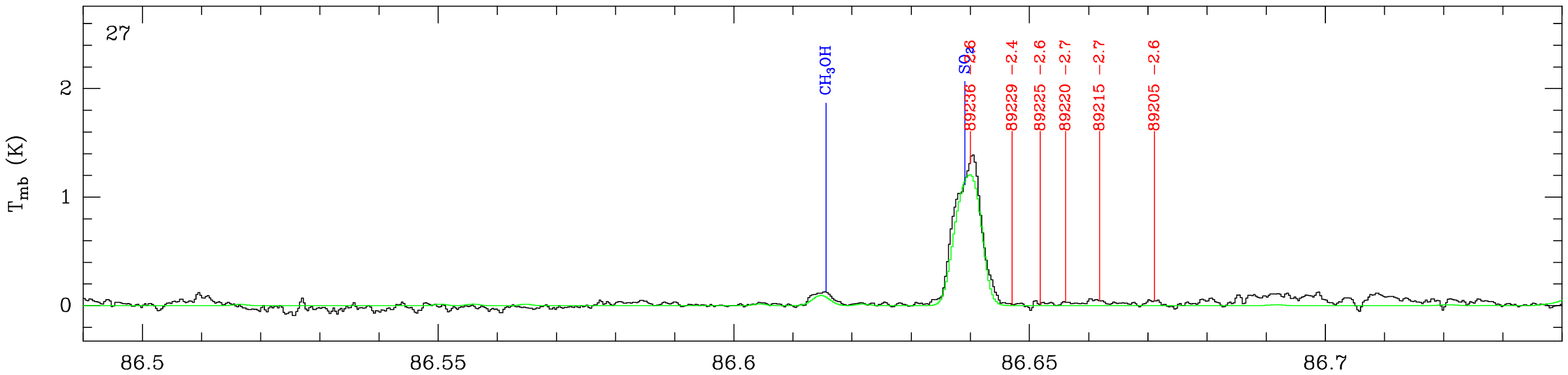}}}
\vspace*{1ex}\centerline{\resizebox{1.0\hsize}{!}{\includegraphics[angle=270]{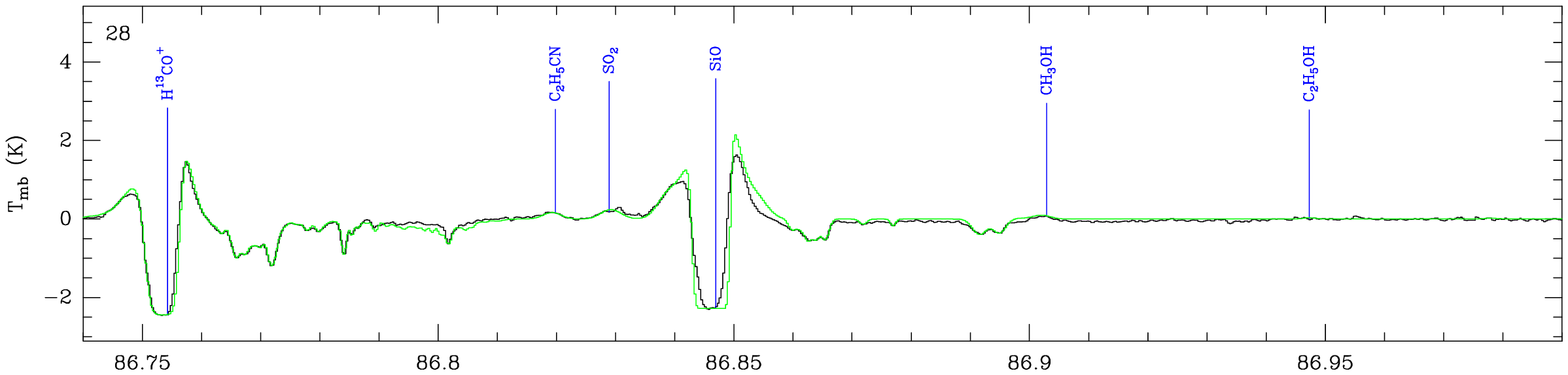}}}
\vspace*{1ex}\centerline{\resizebox{1.0\hsize}{!}{\includegraphics[angle=270]{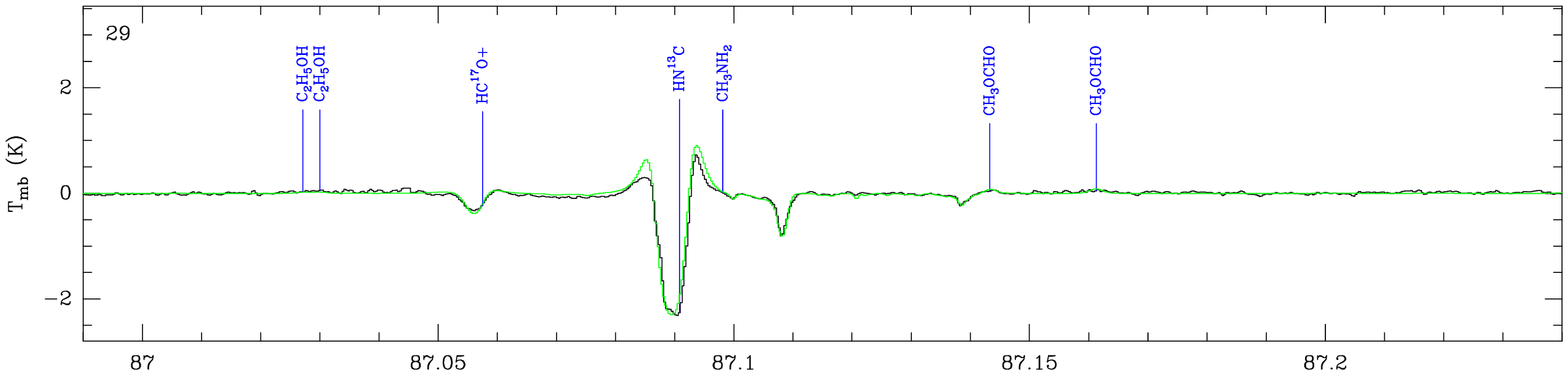}}}
\vspace*{1ex}\centerline{\resizebox{1.0\hsize}{!}{\includegraphics[angle=270]{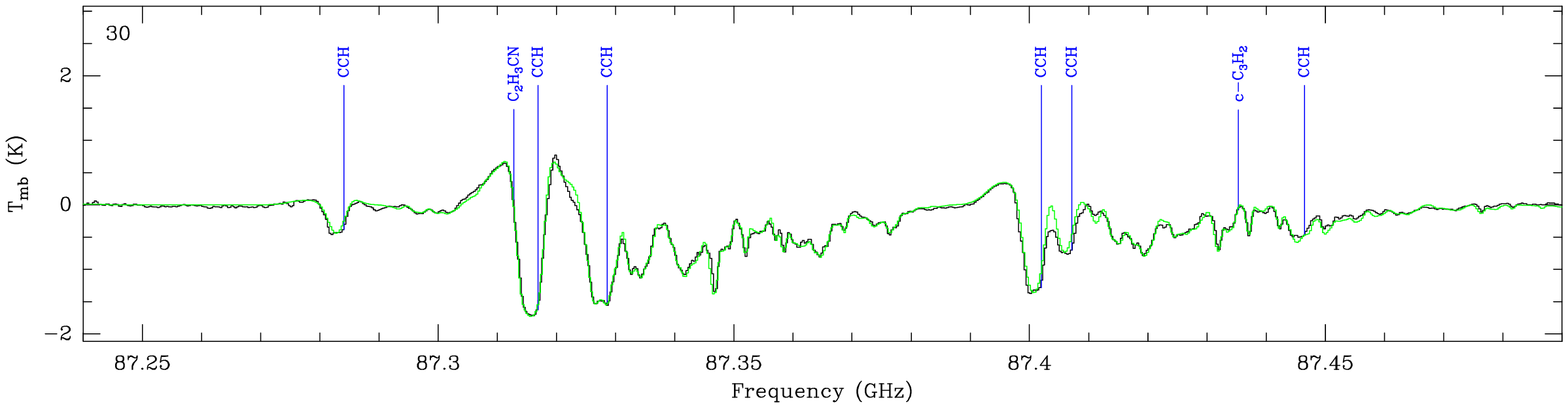}}}
\caption{
continued.
}
\end{figure*}
 \clearpage
\begin{figure*}
\addtocounter{figure}{-1}
\centerline{\resizebox{1.0\hsize}{!}{\includegraphics[angle=270]{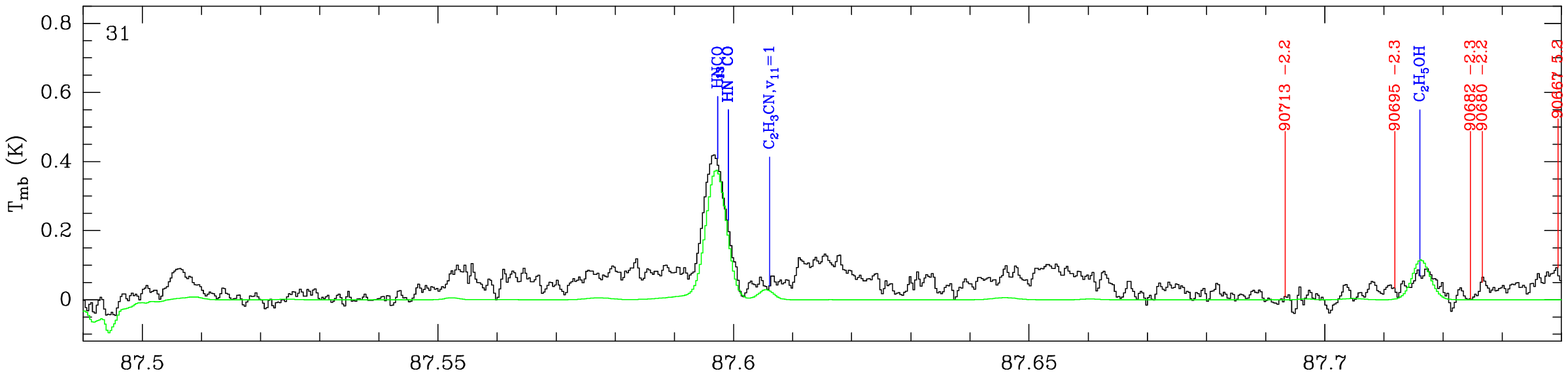}}}
\vspace*{1ex}\centerline{\resizebox{1.0\hsize}{!}{\includegraphics[angle=270]{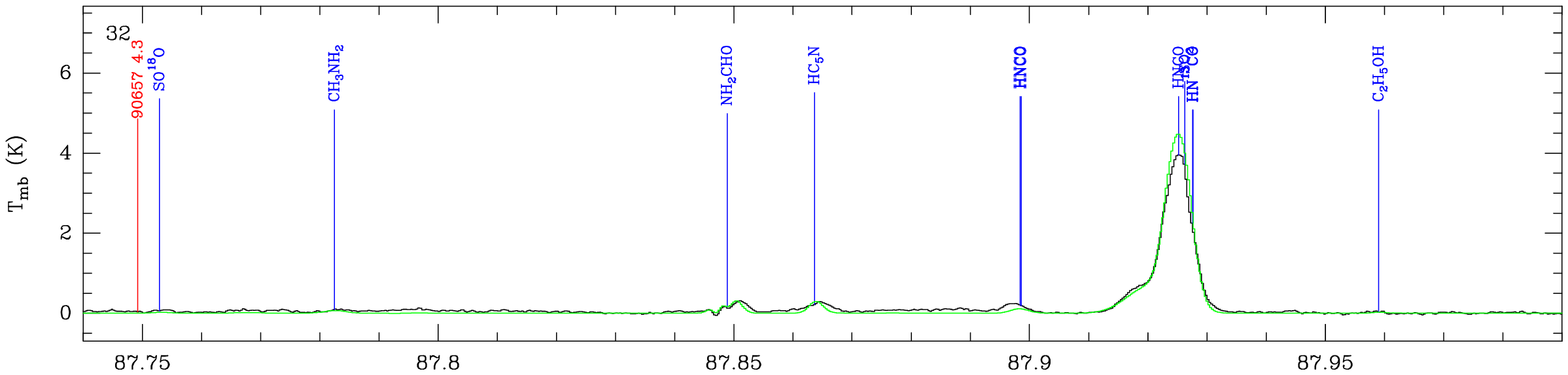}}}
\vspace*{1ex}\centerline{\resizebox{1.0\hsize}{!}{\includegraphics[angle=270]{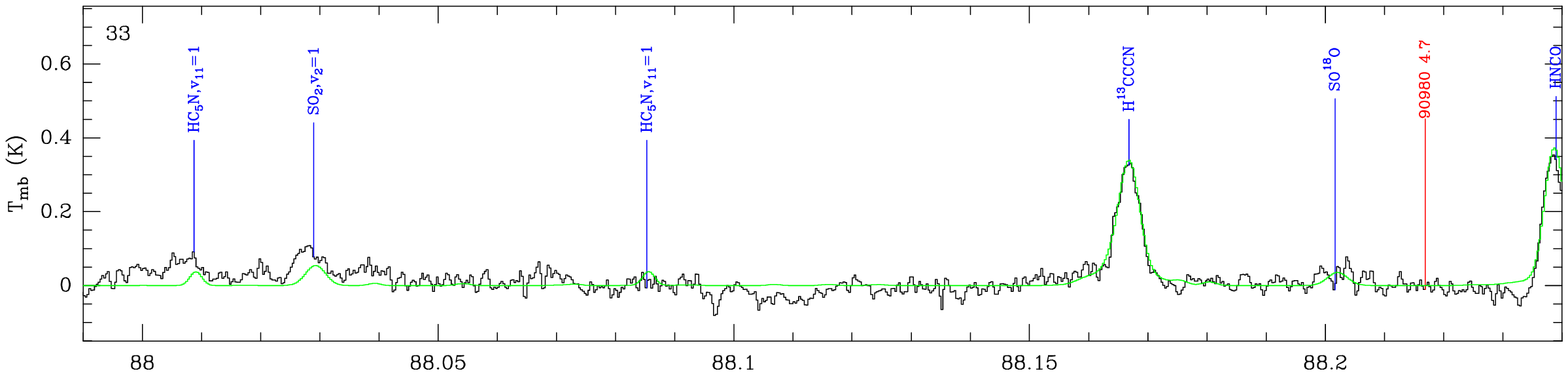}}}
\vspace*{1ex}\centerline{\resizebox{1.0\hsize}{!}{\includegraphics[angle=270]{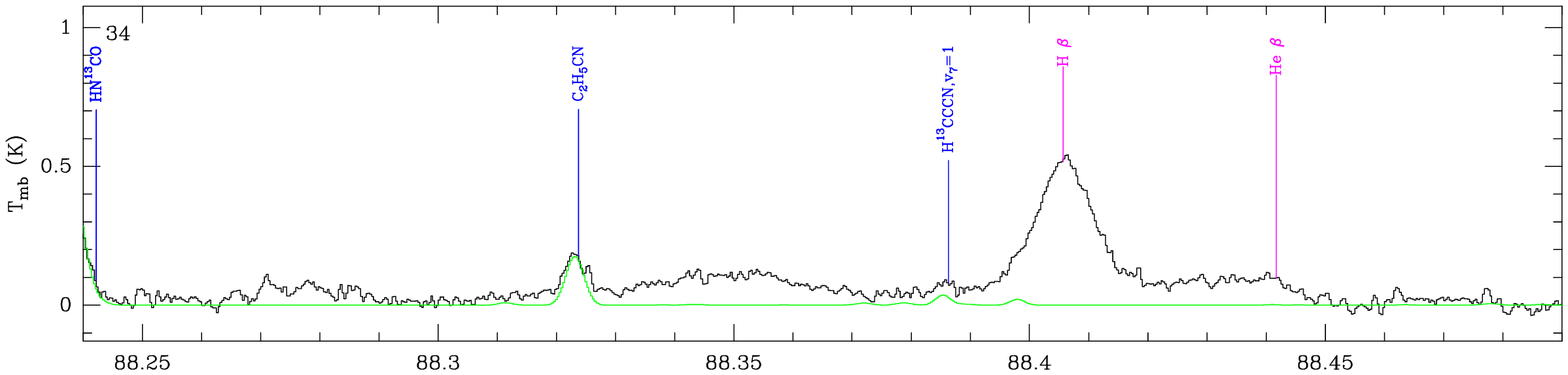}}}
\vspace*{1ex}\centerline{\resizebox{1.0\hsize}{!}{\includegraphics[angle=270]{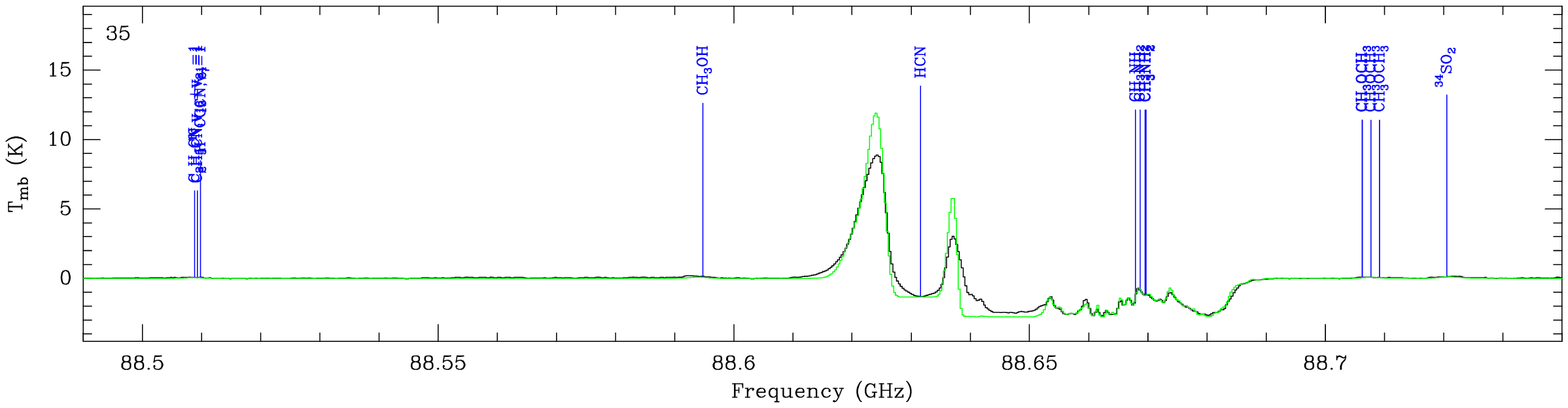}}}
\caption{
continued.
}
\end{figure*}
 \clearpage
\begin{figure*}
\addtocounter{figure}{-1}
\centerline{\resizebox{1.0\hsize}{!}{\includegraphics[angle=270]{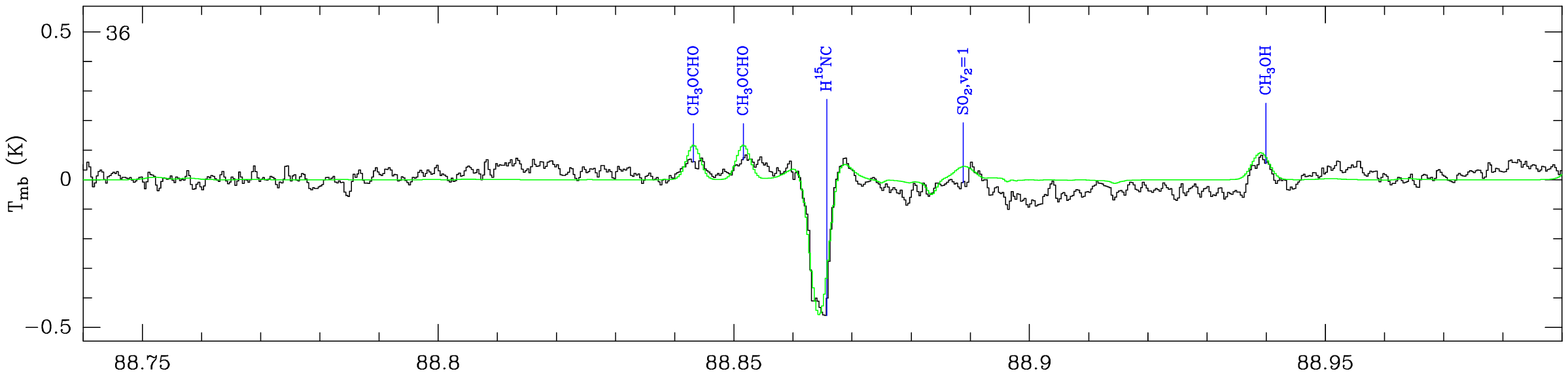}}}
\vspace*{1ex}\centerline{\resizebox{1.0\hsize}{!}{\includegraphics[angle=270]{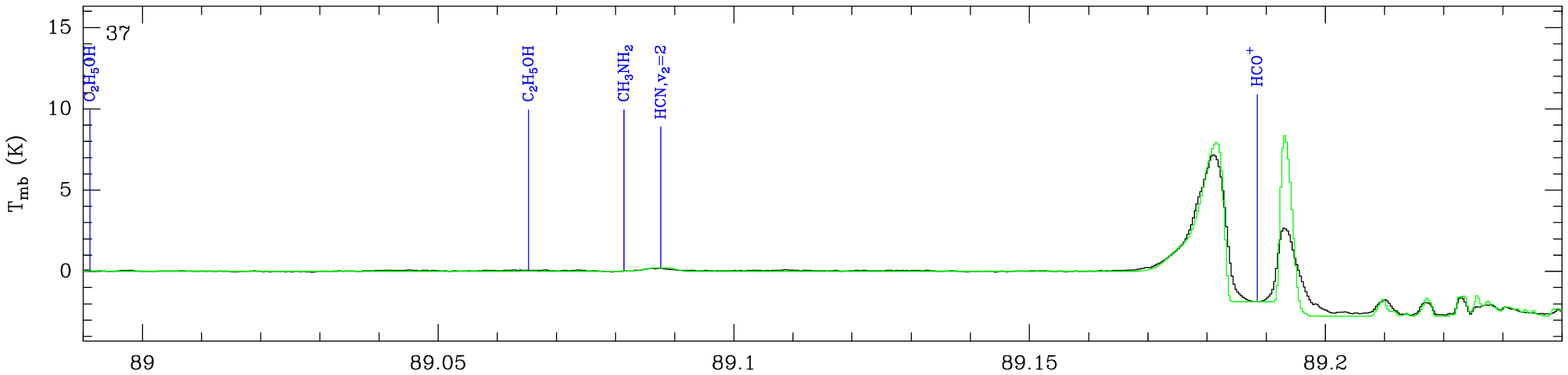}}}
\vspace*{1ex}\centerline{\resizebox{1.0\hsize}{!}{\includegraphics[angle=270]{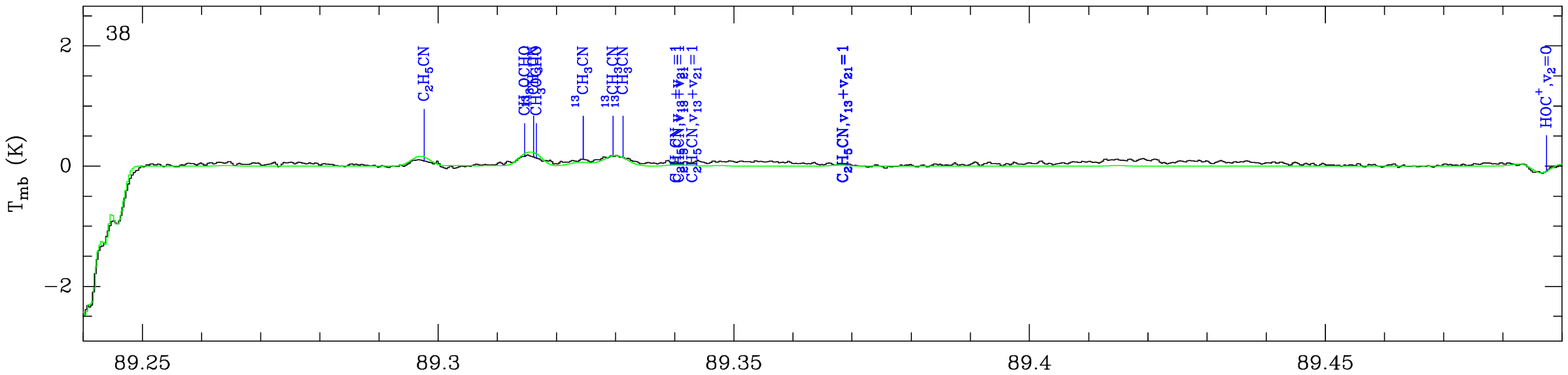}}}
\vspace*{1ex}\centerline{\resizebox{1.0\hsize}{!}{\includegraphics[angle=270]{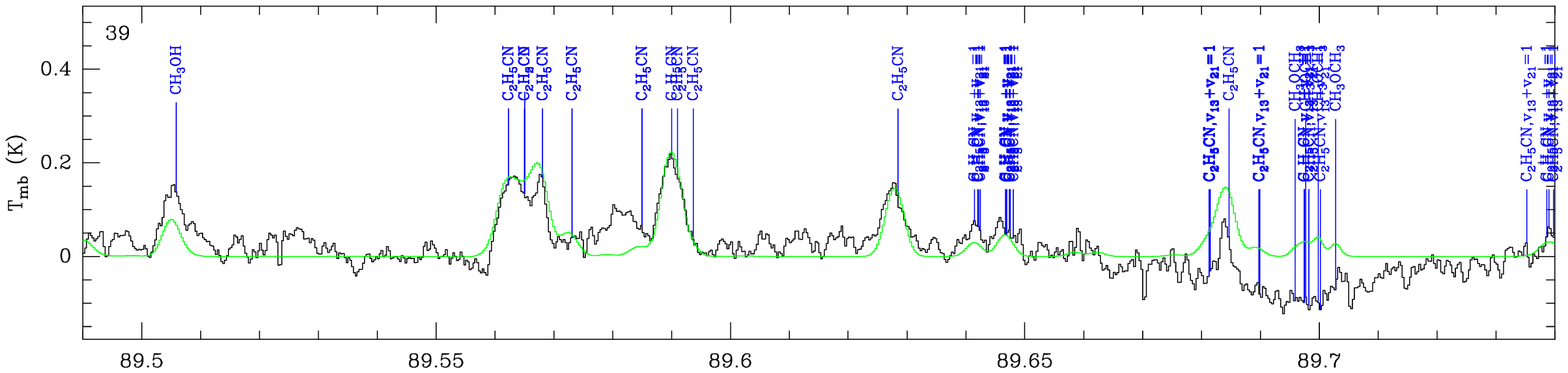}}}
\vspace*{1ex}\centerline{\resizebox{1.0\hsize}{!}{\includegraphics[angle=270]{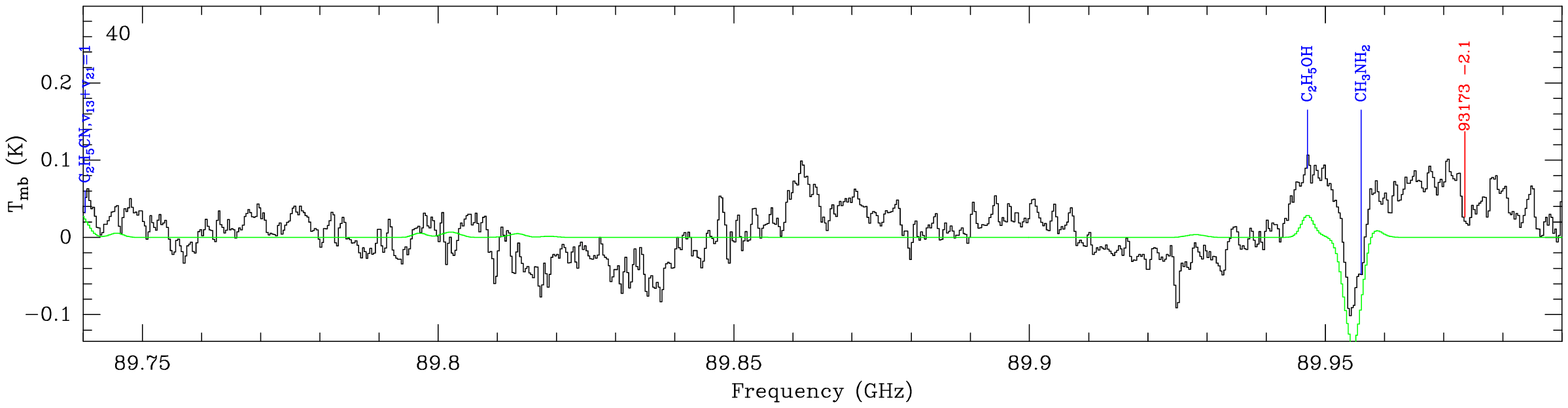}}}
\caption{
continued.
}
\end{figure*}
 \clearpage
\begin{figure*}
\addtocounter{figure}{-1}
\centerline{\resizebox{1.0\hsize}{!}{\includegraphics[angle=270]{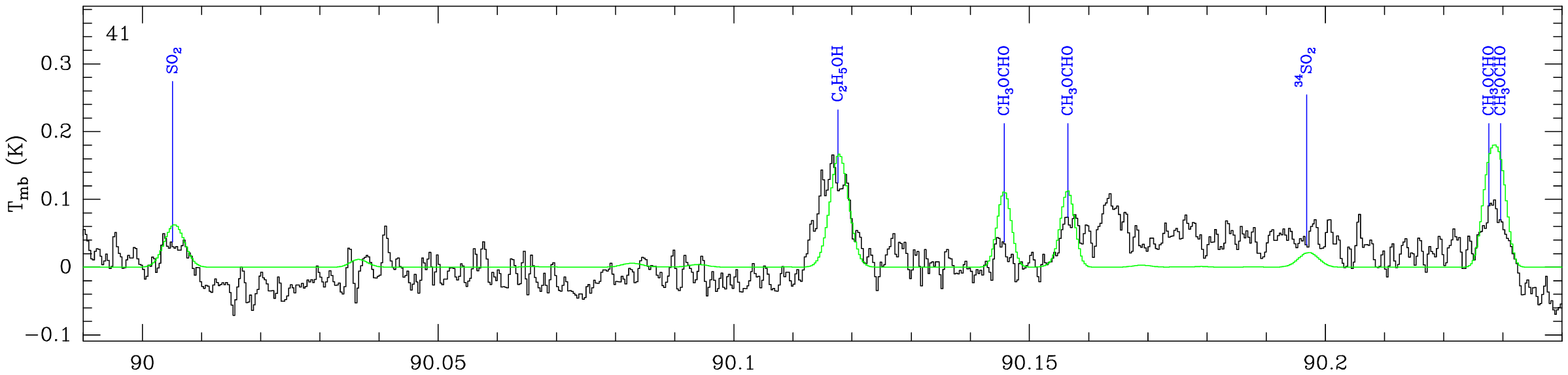}}}
\vspace*{1ex}\centerline{\resizebox{1.0\hsize}{!}{\includegraphics[angle=270]{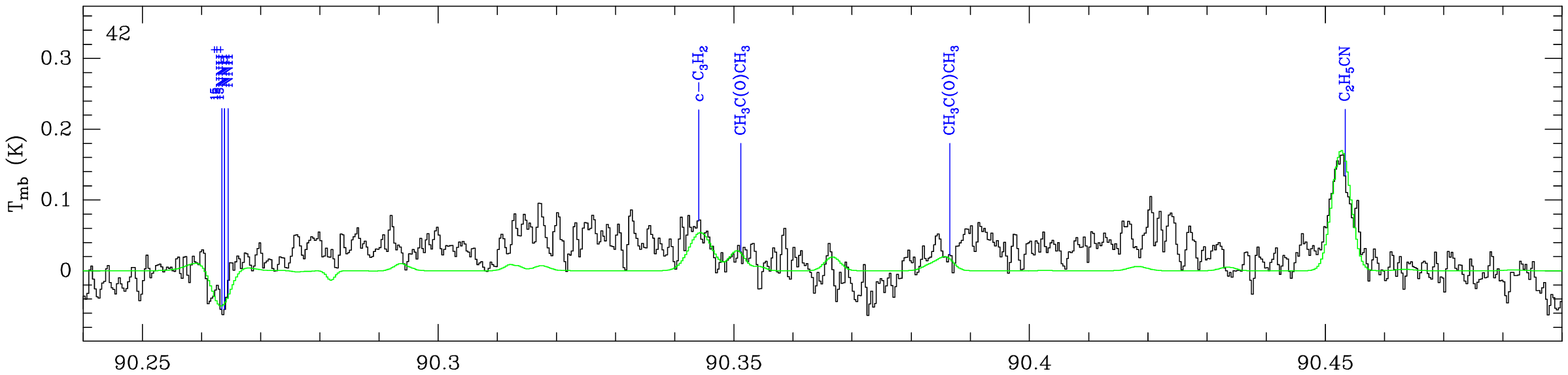}}}
\vspace*{1ex}\centerline{\resizebox{1.0\hsize}{!}{\includegraphics[angle=270]{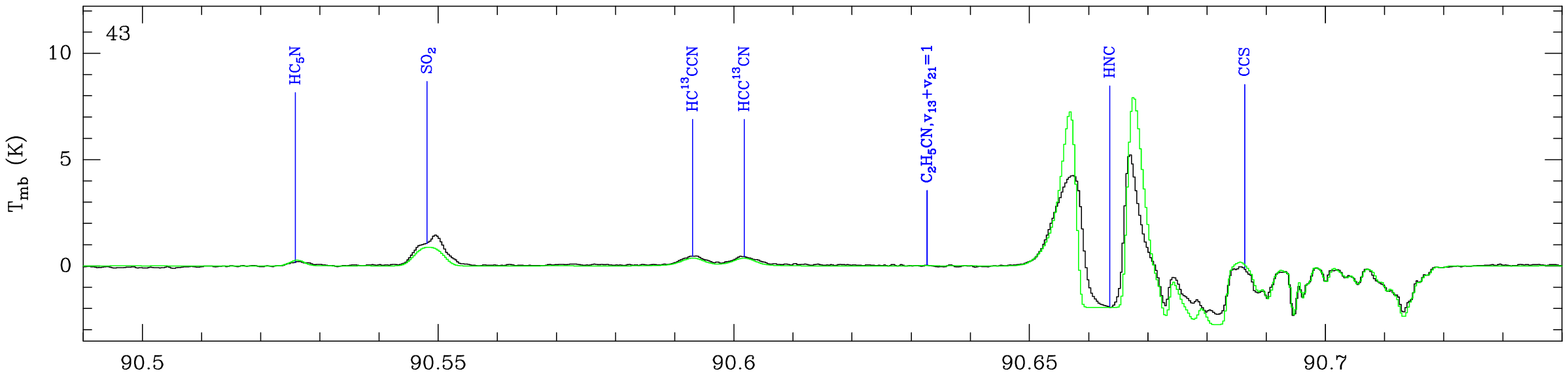}}}
\vspace*{1ex}\centerline{\resizebox{1.0\hsize}{!}{\includegraphics[angle=270]{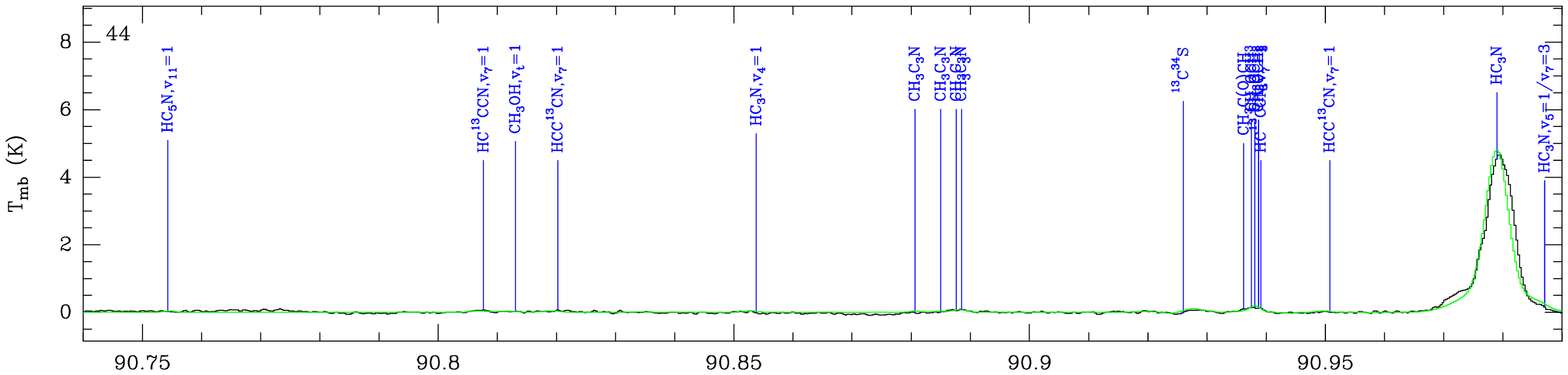}}}
\vspace*{1ex}\centerline{\resizebox{1.0\hsize}{!}{\includegraphics[angle=270]{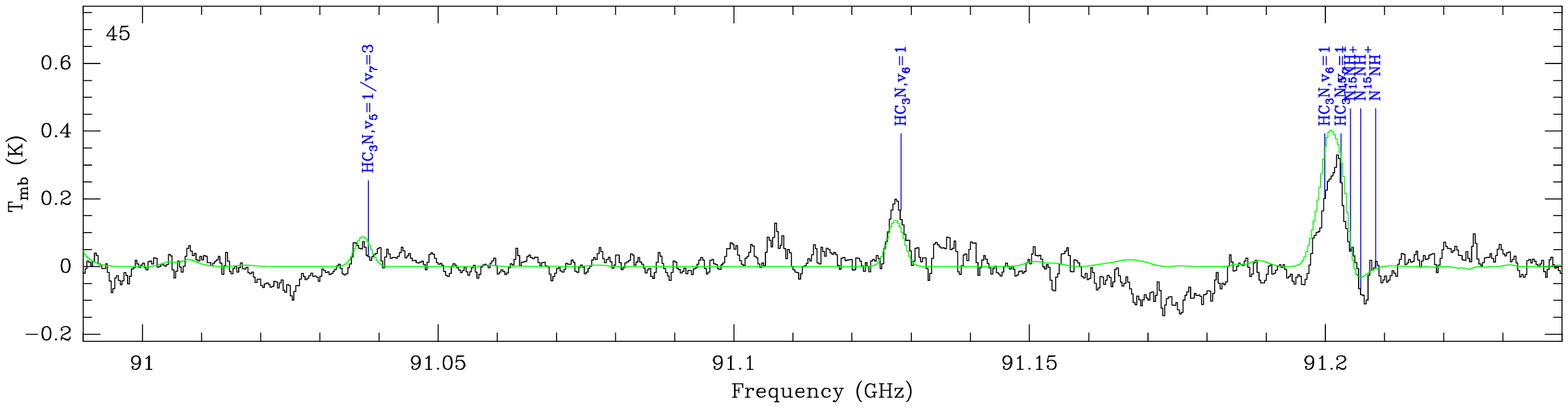}}}
\caption{
continued.
}
\end{figure*}
 \clearpage
\begin{figure*}
\addtocounter{figure}{-1}
\centerline{\resizebox{1.0\hsize}{!}{\includegraphics[angle=270]{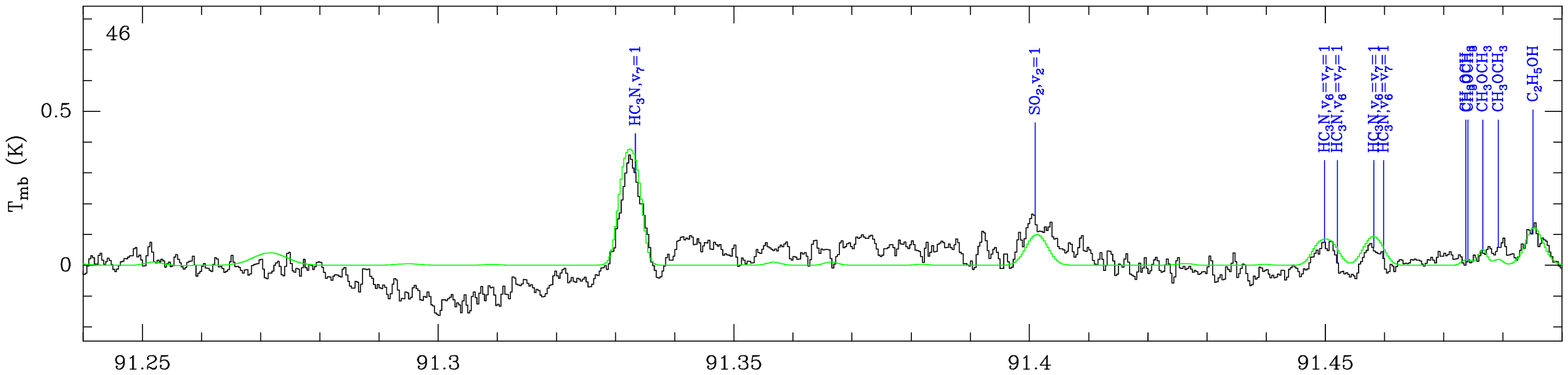}}}
\vspace*{1ex}\centerline{\resizebox{1.0\hsize}{!}{\includegraphics[angle=270]{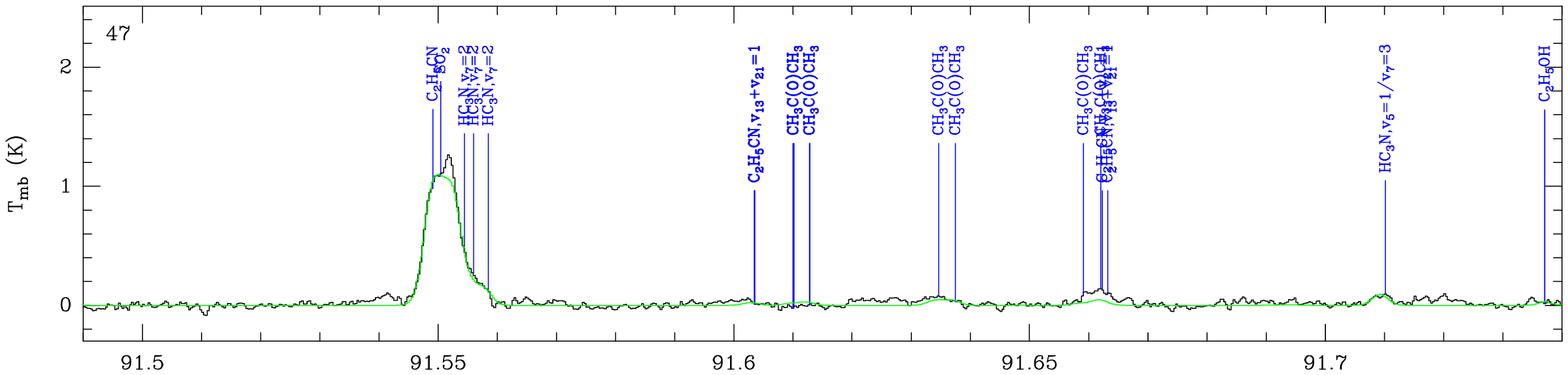}}}
\vspace*{1ex}\centerline{\resizebox{1.0\hsize}{!}{\includegraphics[angle=270]{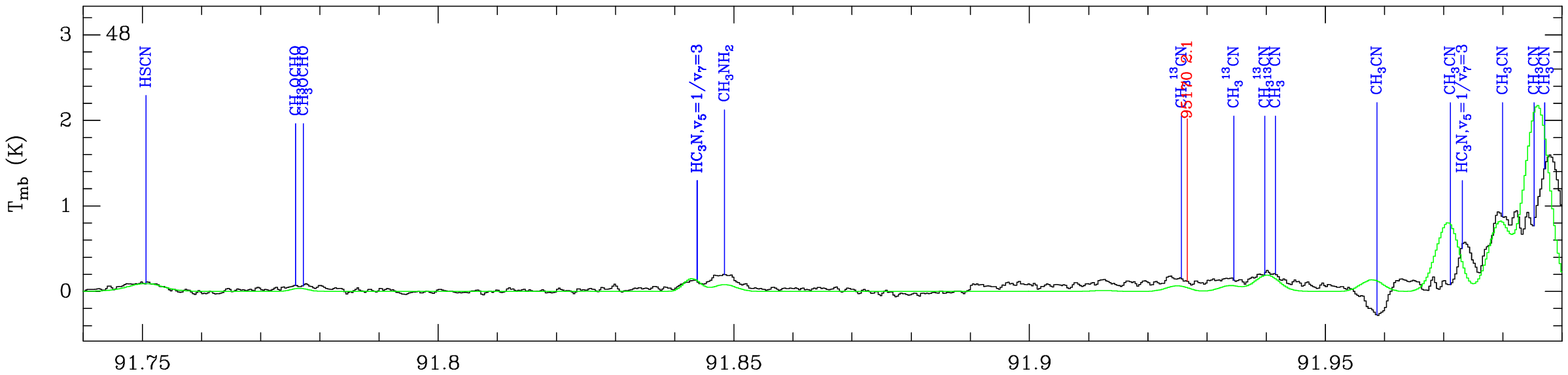}}}
\vspace*{1ex}\centerline{\resizebox{1.0\hsize}{!}{\includegraphics[angle=270]{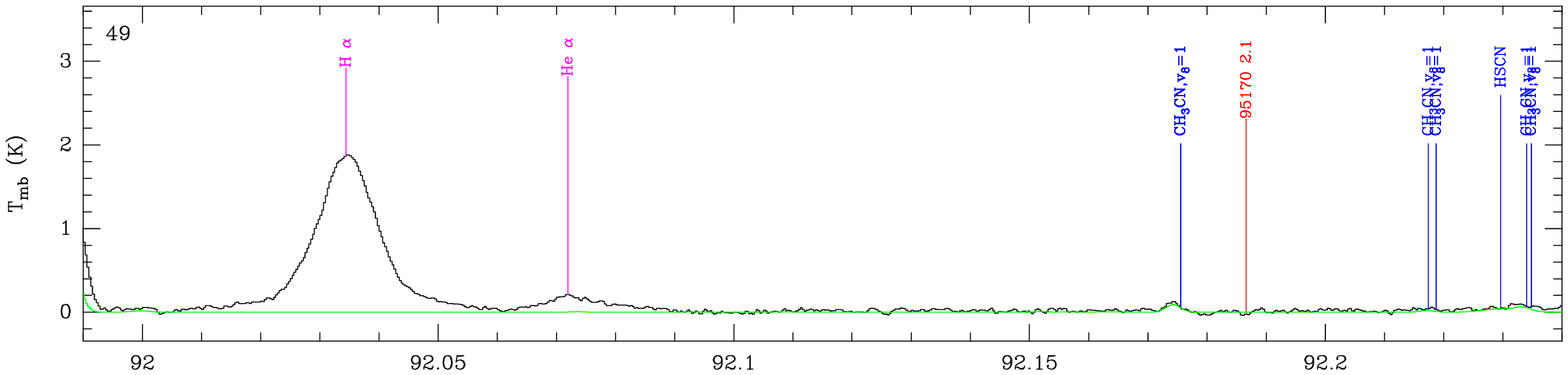}}}
\vspace*{1ex}\centerline{\resizebox{1.0\hsize}{!}{\includegraphics[angle=270]{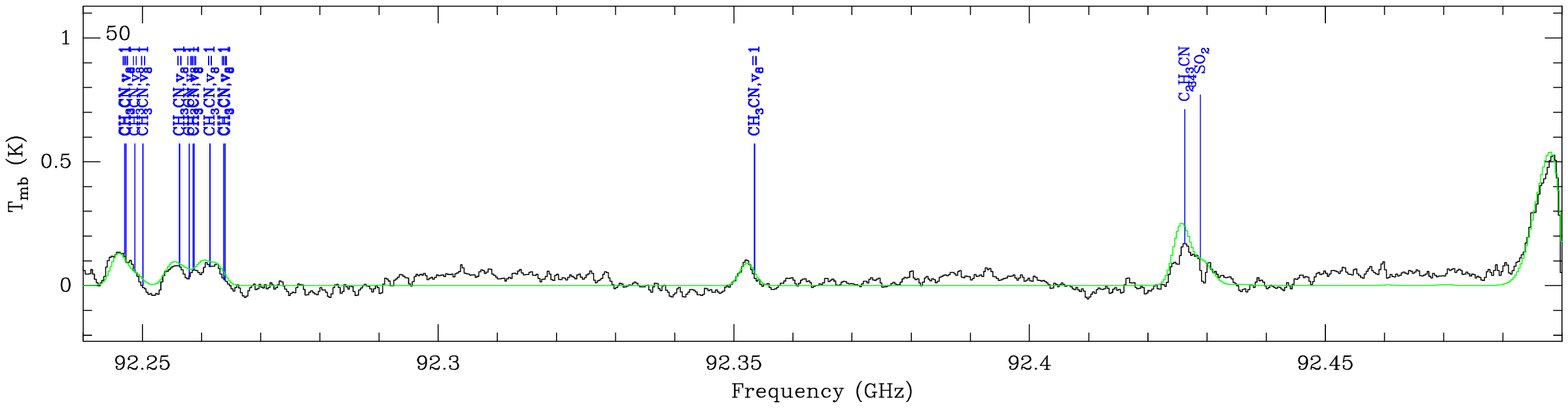}}}
\caption{
continued.
}
\end{figure*}
 \clearpage
\begin{figure*}
\addtocounter{figure}{-1}
\centerline{\resizebox{1.0\hsize}{!}{\includegraphics[angle=270]{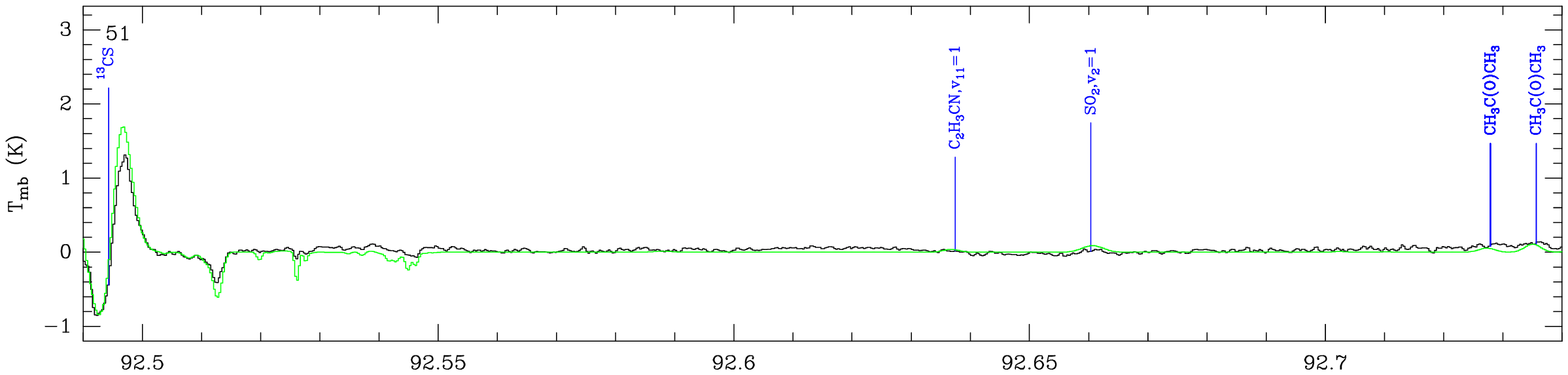}}}
\vspace*{1ex}\centerline{\resizebox{1.0\hsize}{!}{\includegraphics[angle=270]{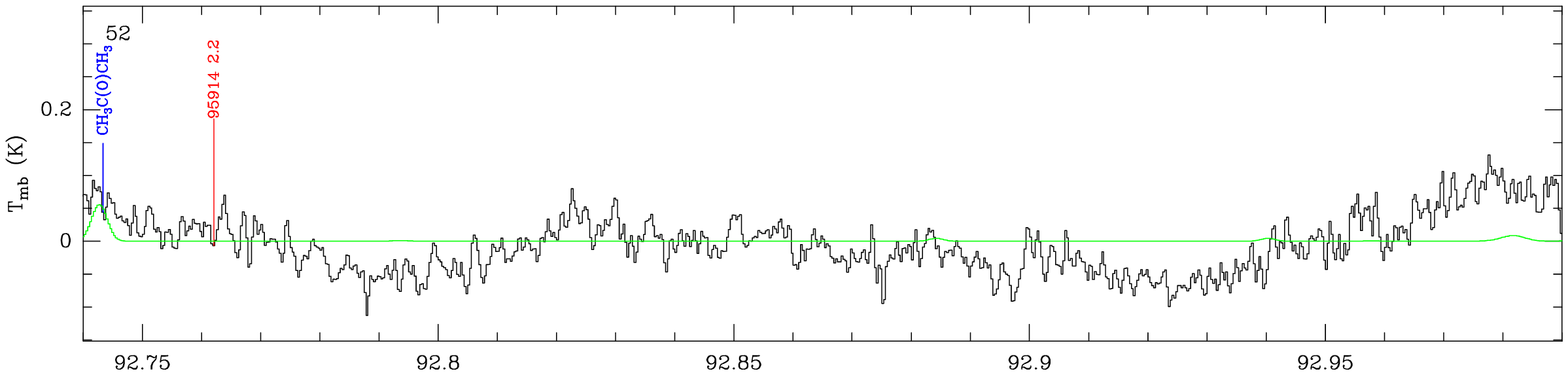}}}
\vspace*{1ex}\centerline{\resizebox{1.0\hsize}{!}{\includegraphics[angle=270]{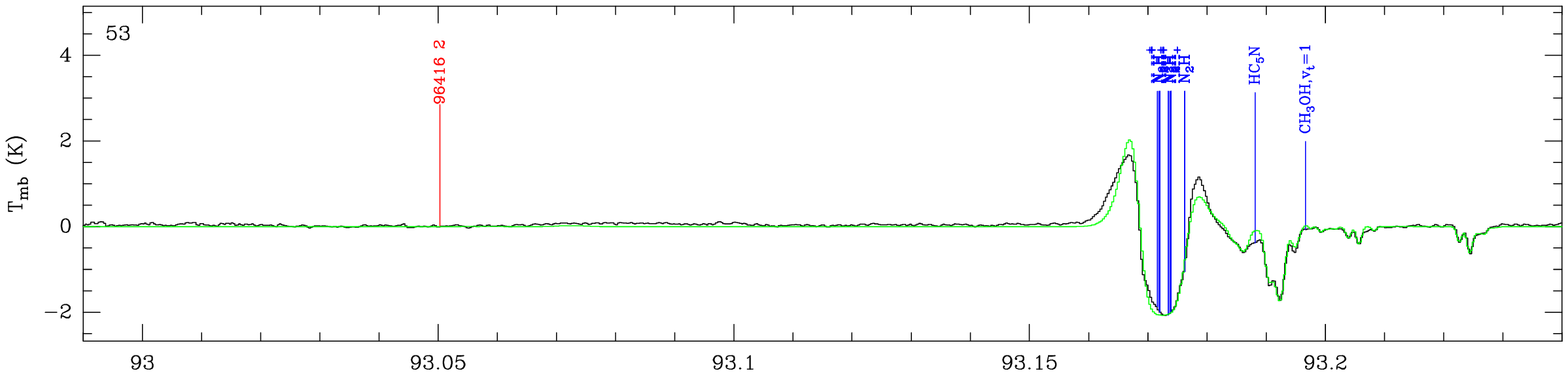}}}
\vspace*{1ex}\centerline{\resizebox{1.0\hsize}{!}{\includegraphics[angle=270]{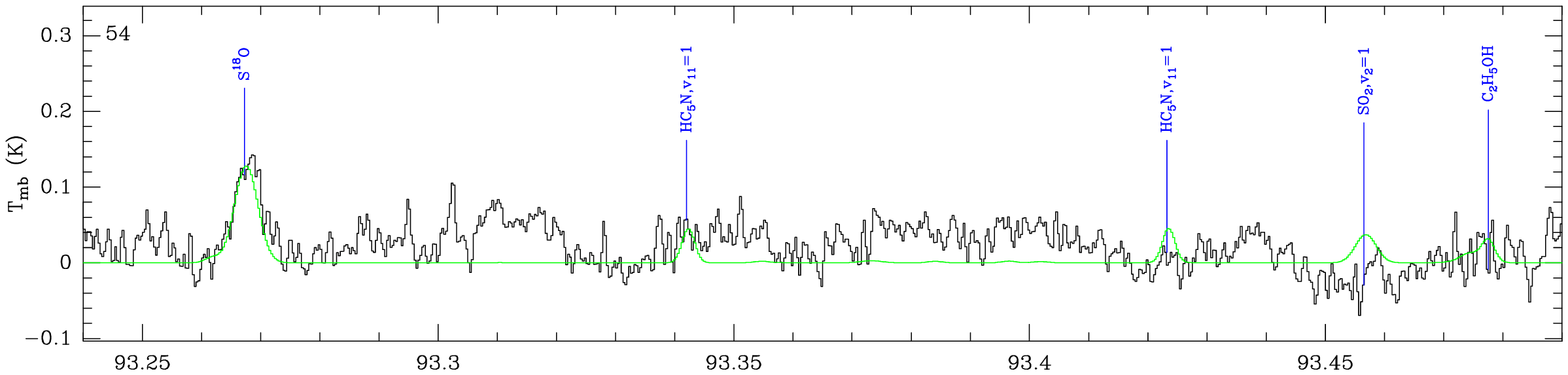}}}
\vspace*{1ex}\centerline{\resizebox{1.0\hsize}{!}{\includegraphics[angle=270]{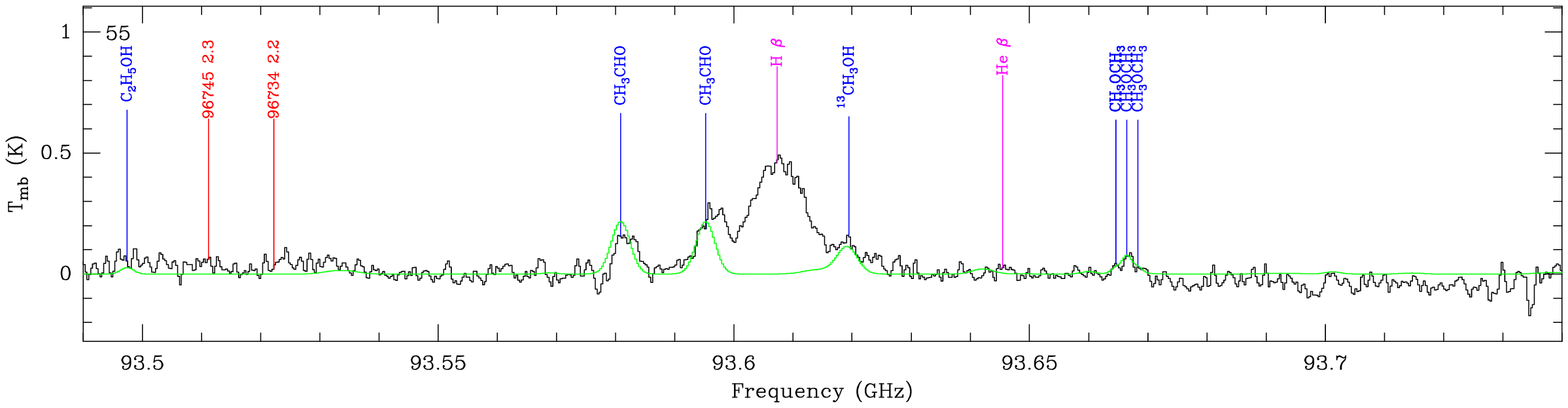}}}
\caption{
continued.
}
\end{figure*}
 \clearpage
\begin{figure*}
\addtocounter{figure}{-1}
\centerline{\resizebox{1.0\hsize}{!}{\includegraphics[angle=270]{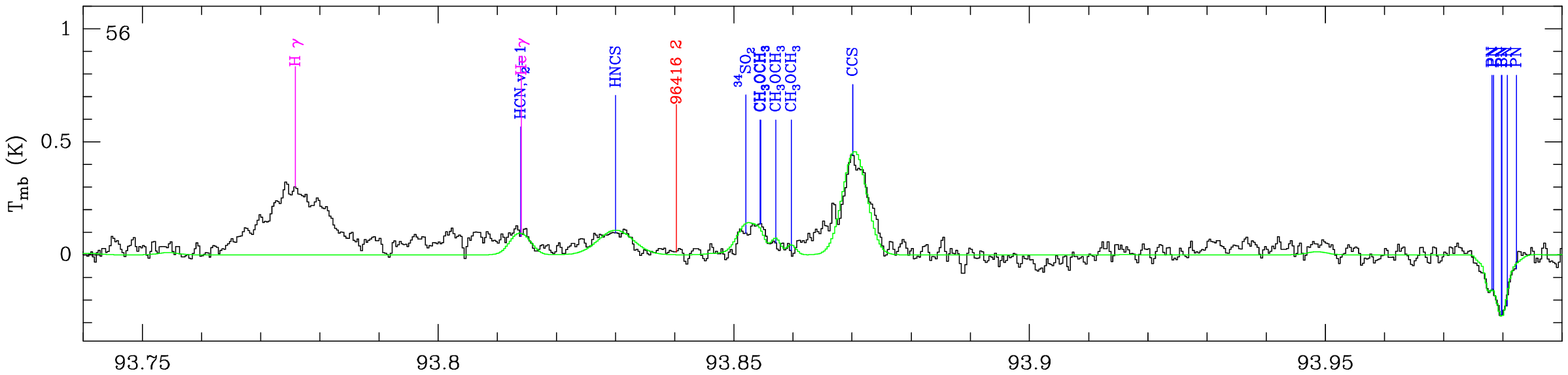}}}
\vspace*{1ex}\centerline{\resizebox{1.0\hsize}{!}{\includegraphics[angle=270]{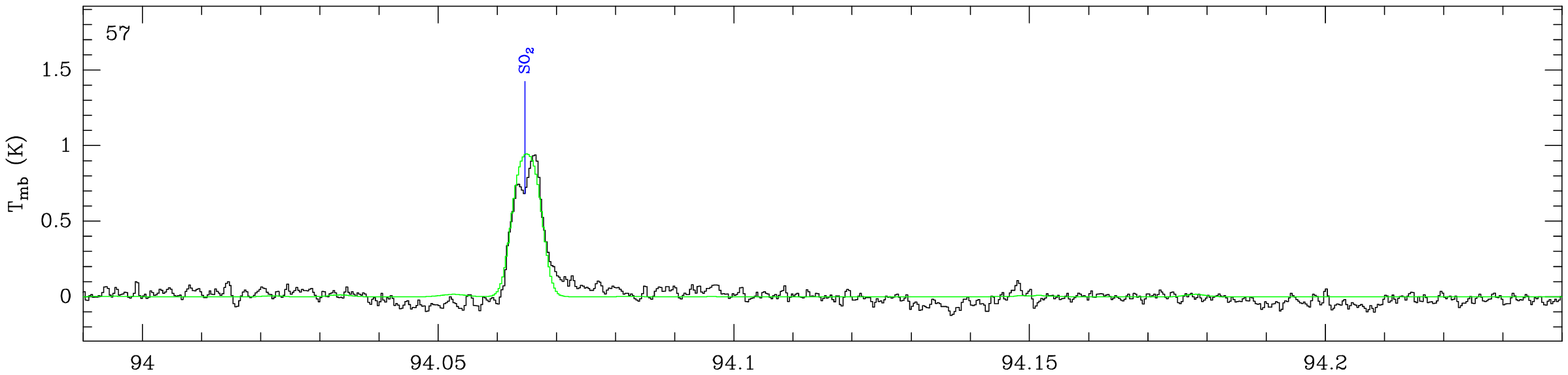}}}
\vspace*{1ex}\centerline{\resizebox{1.0\hsize}{!}{\includegraphics[angle=270]{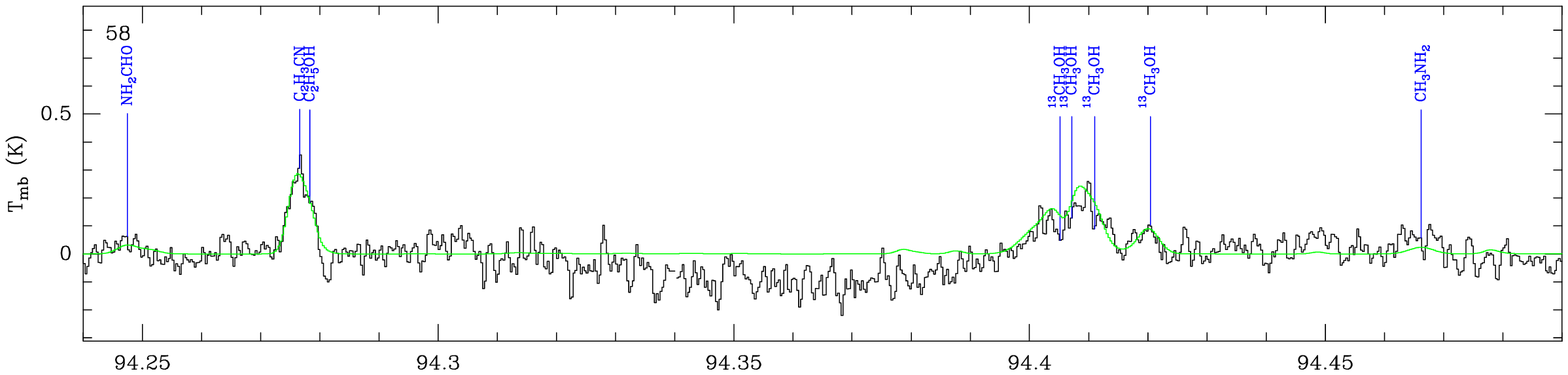}}}
\vspace*{1ex}\centerline{\resizebox{1.0\hsize}{!}{\includegraphics[angle=270]{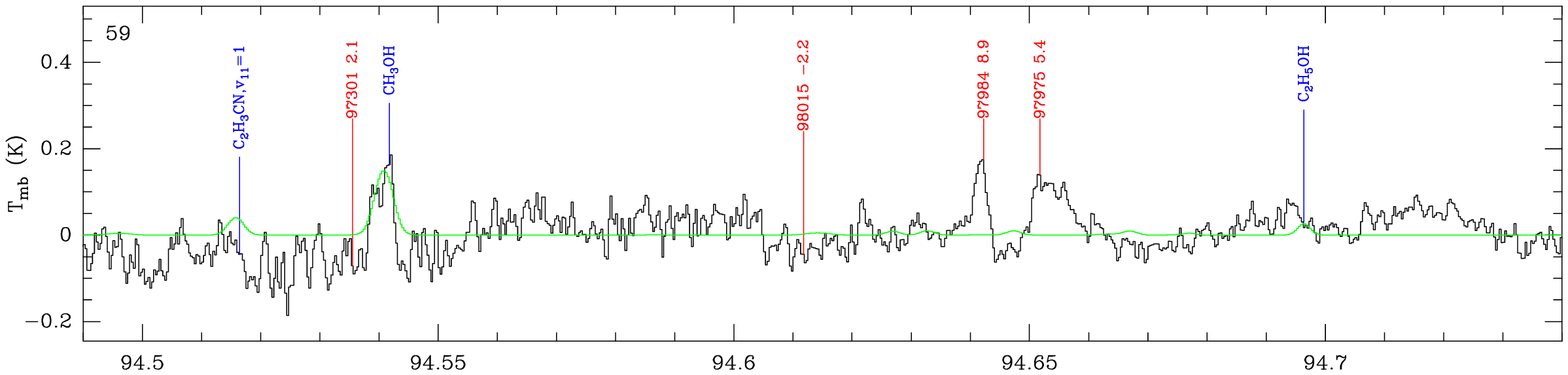}}}
\vspace*{1ex}\centerline{\resizebox{1.0\hsize}{!}{\includegraphics[angle=270]{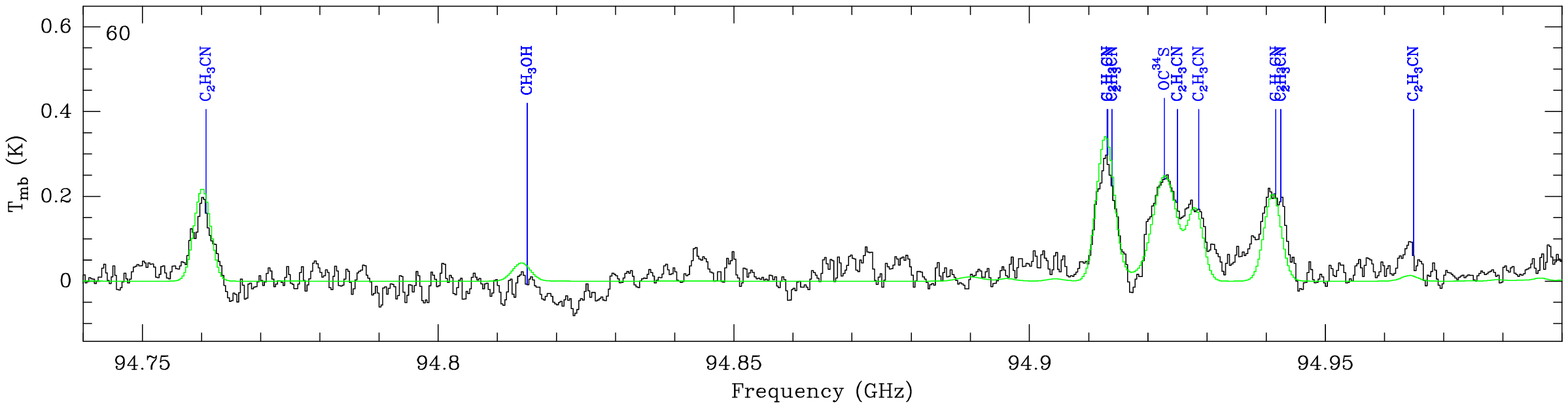}}}
\caption{
continued.
}
\end{figure*}
 \clearpage
\begin{figure*}
\addtocounter{figure}{-1}
\centerline{\resizebox{1.0\hsize}{!}{\includegraphics[angle=270]{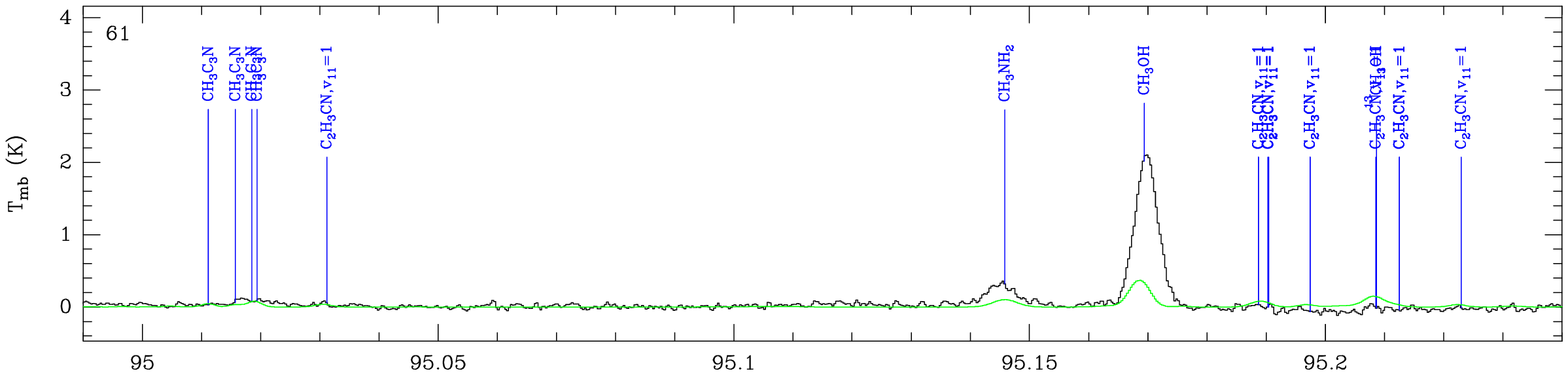}}}
\vspace*{1ex}\centerline{\resizebox{1.0\hsize}{!}{\includegraphics[angle=270]{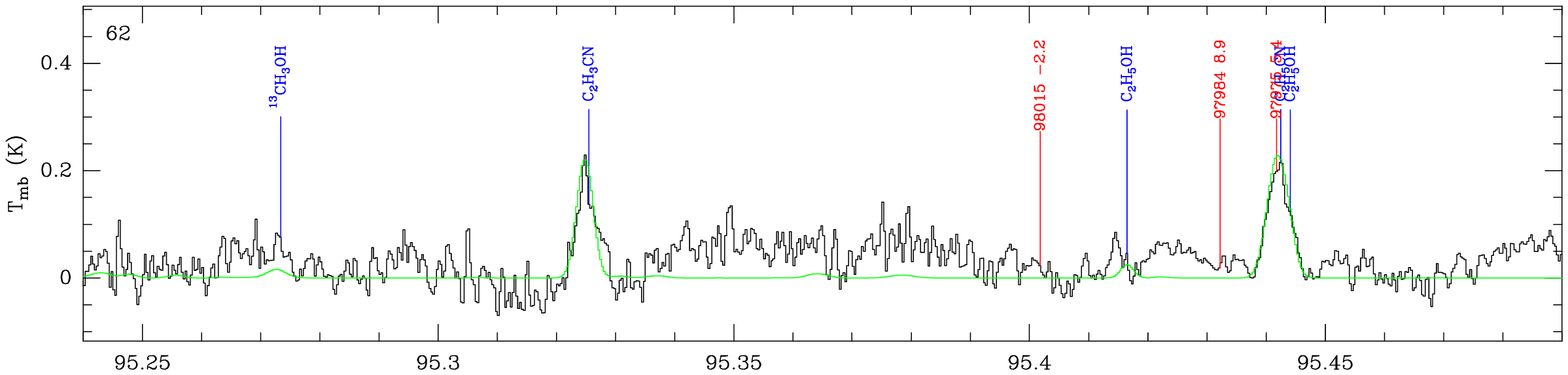}}}
\vspace*{1ex}\centerline{\resizebox{1.0\hsize}{!}{\includegraphics[angle=270]{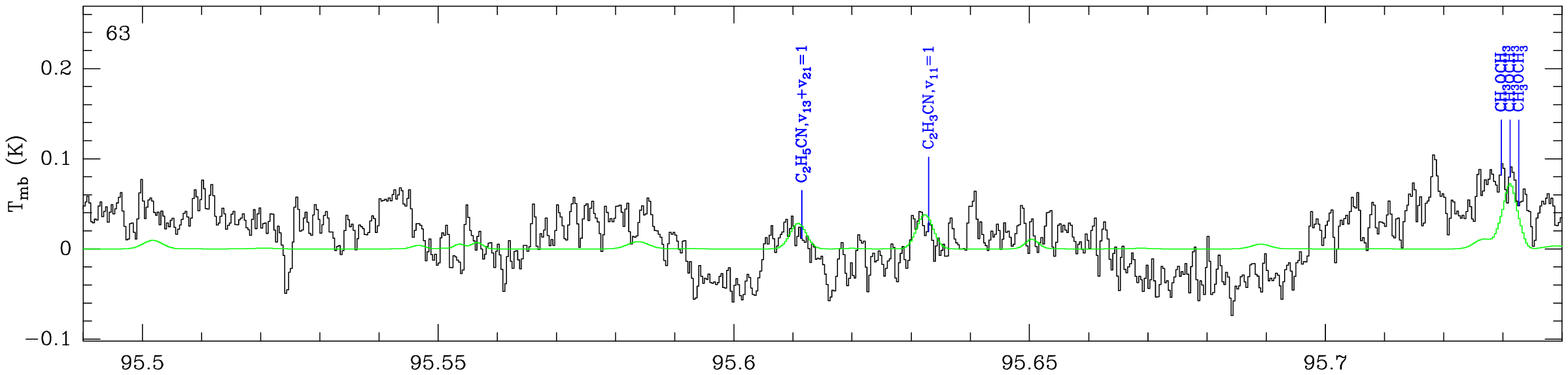}}}
\vspace*{1ex}\centerline{\resizebox{1.0\hsize}{!}{\includegraphics[angle=270]{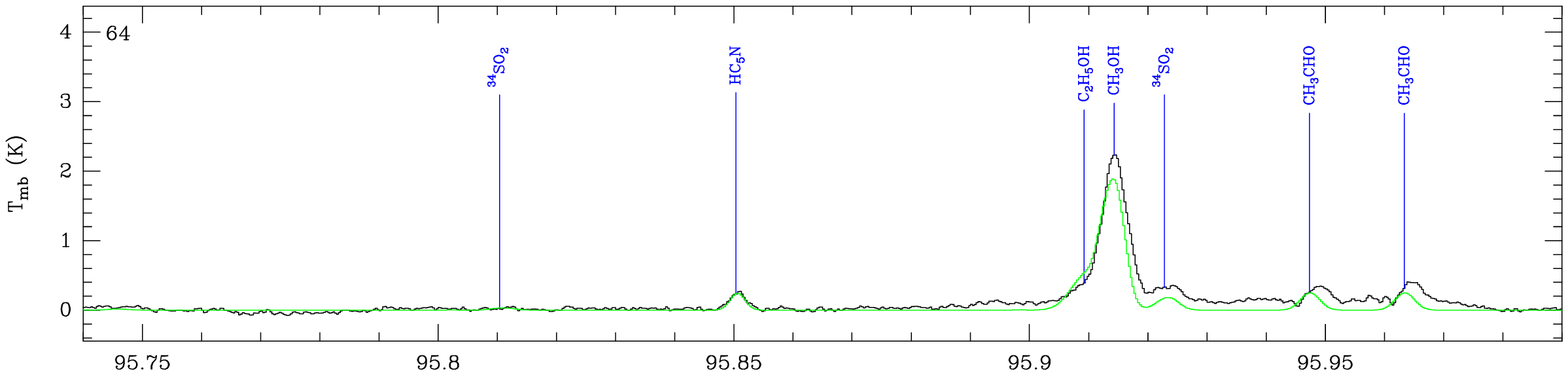}}}
\vspace*{1ex}\centerline{\resizebox{1.0\hsize}{!}{\includegraphics[angle=270]{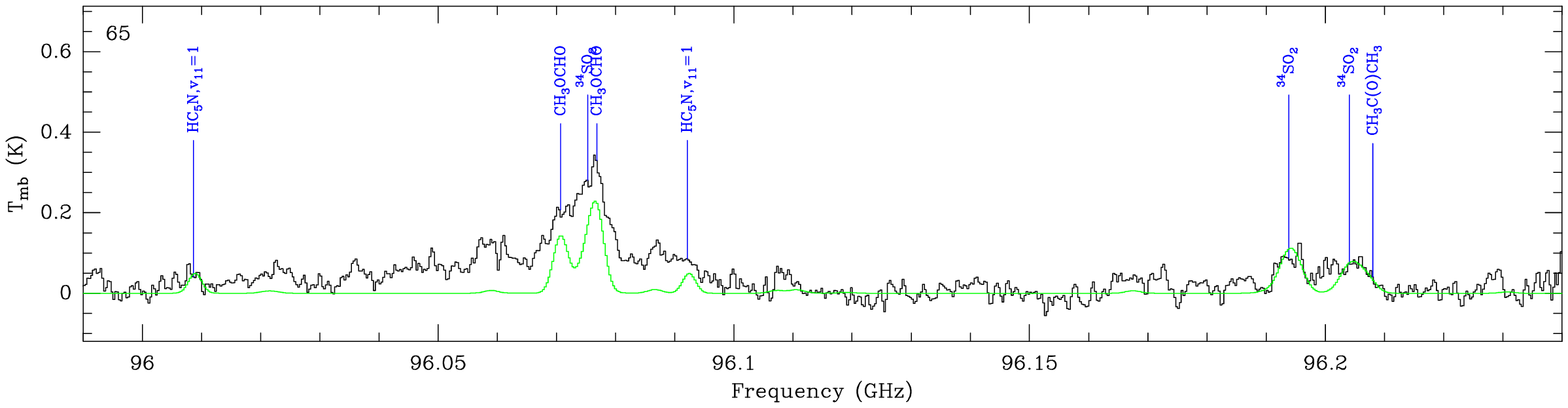}}}
\caption{
continued.
}
\end{figure*}
 \clearpage
\begin{figure*}
\addtocounter{figure}{-1}
\centerline{\resizebox{1.0\hsize}{!}{\includegraphics[angle=270]{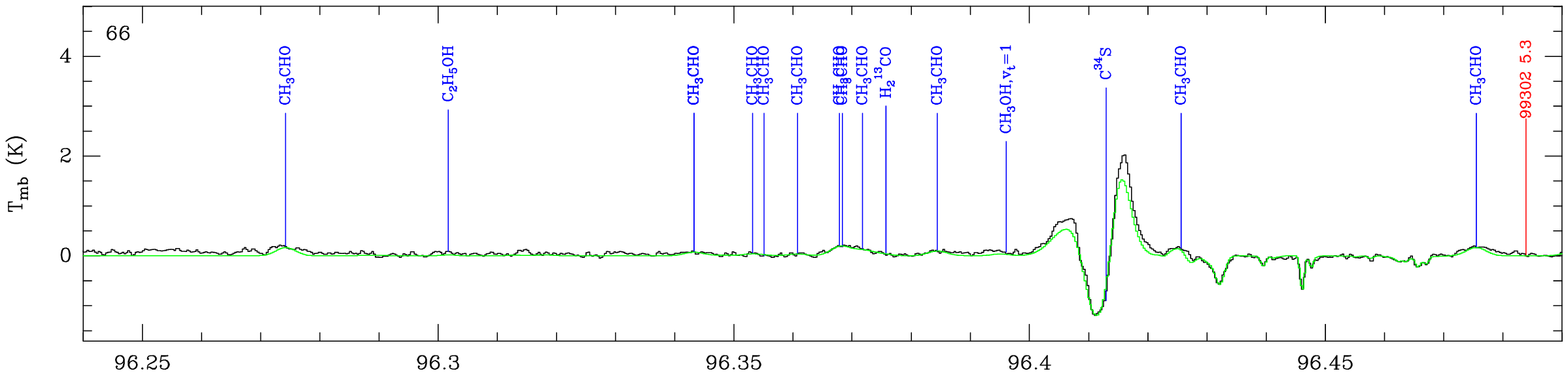}}}
\vspace*{1ex}\centerline{\resizebox{1.0\hsize}{!}{\includegraphics[angle=270]{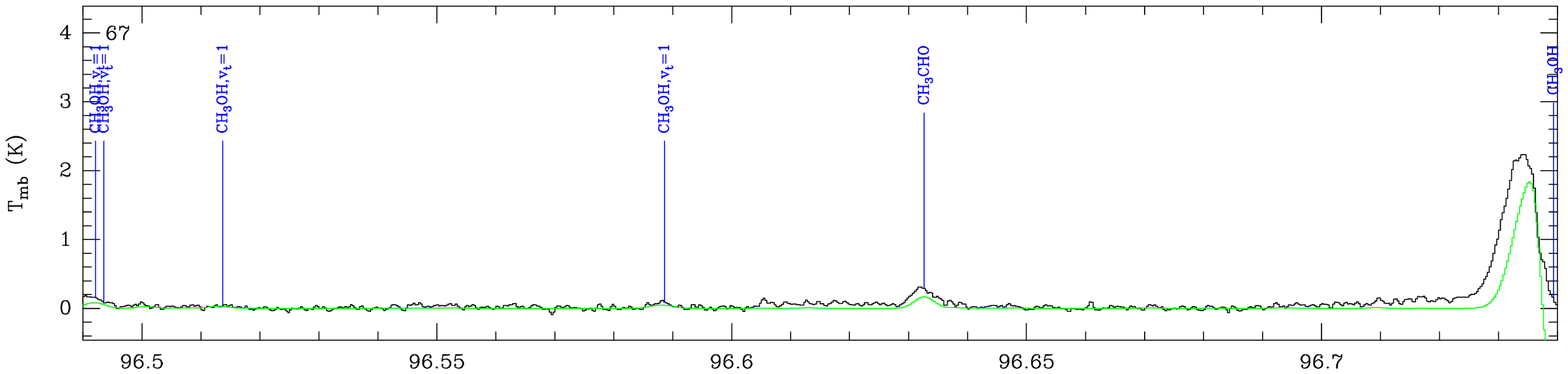}}}
\vspace*{1ex}\centerline{\resizebox{1.0\hsize}{!}{\includegraphics[angle=270]{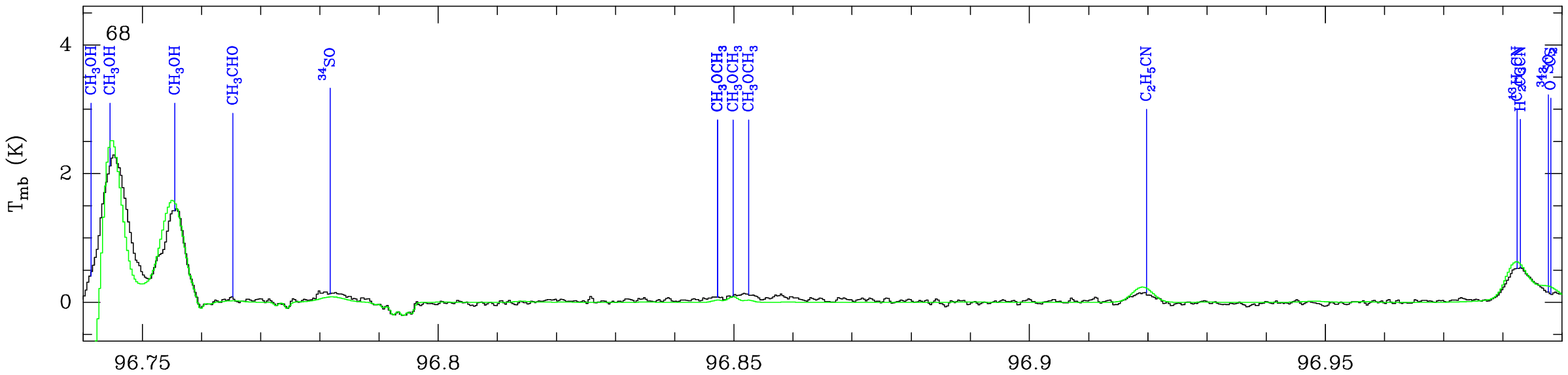}}}
\vspace*{1ex}\centerline{\resizebox{1.0\hsize}{!}{\includegraphics[angle=270]{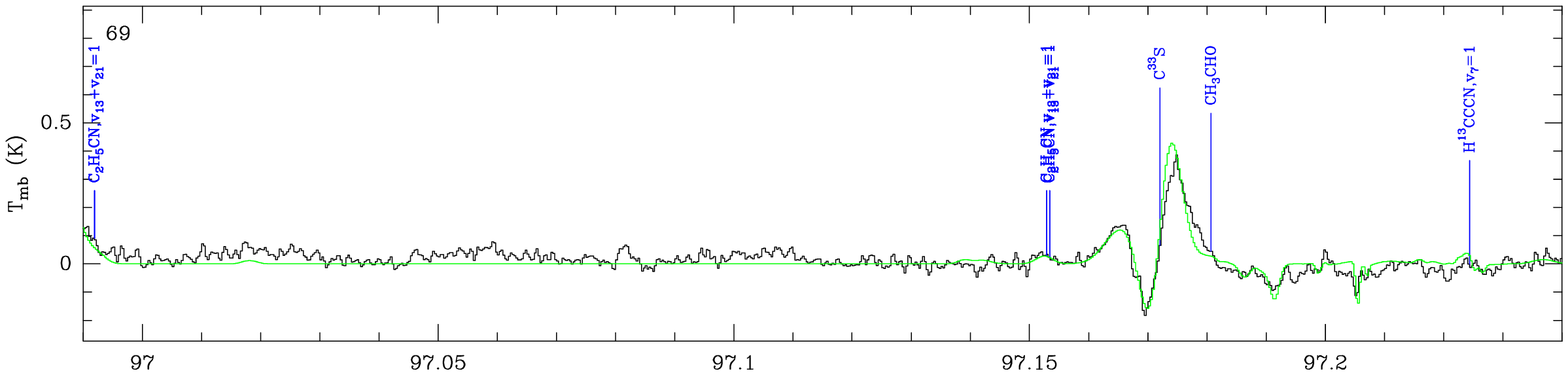}}}
\vspace*{1ex}\centerline{\resizebox{1.0\hsize}{!}{\includegraphics[angle=270]{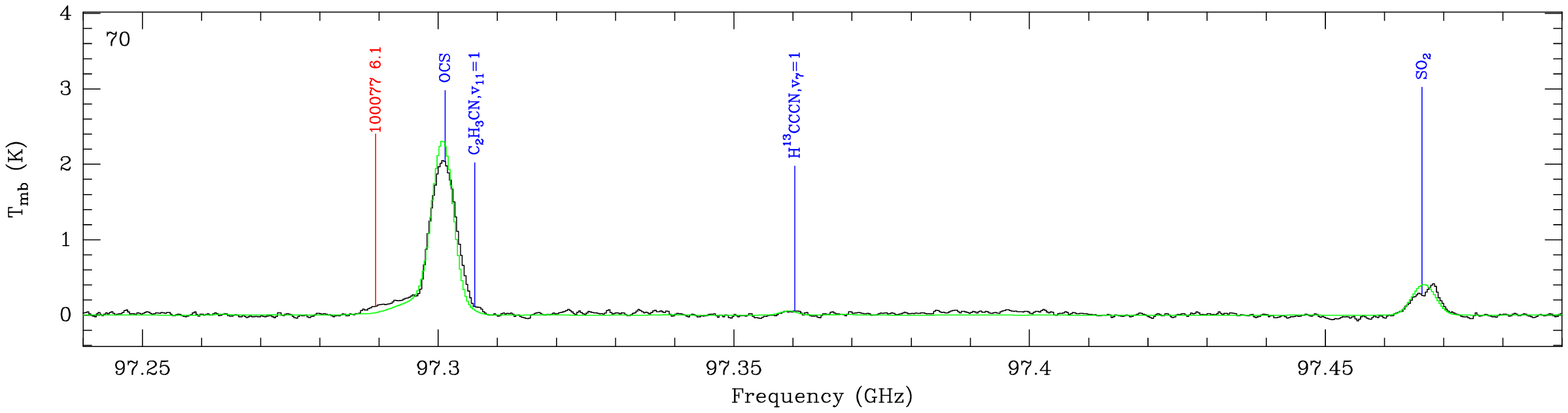}}}
\caption{
continued.
}
\end{figure*}
 \clearpage
\begin{figure*}
\addtocounter{figure}{-1}
\centerline{\resizebox{1.0\hsize}{!}{\includegraphics[angle=270]{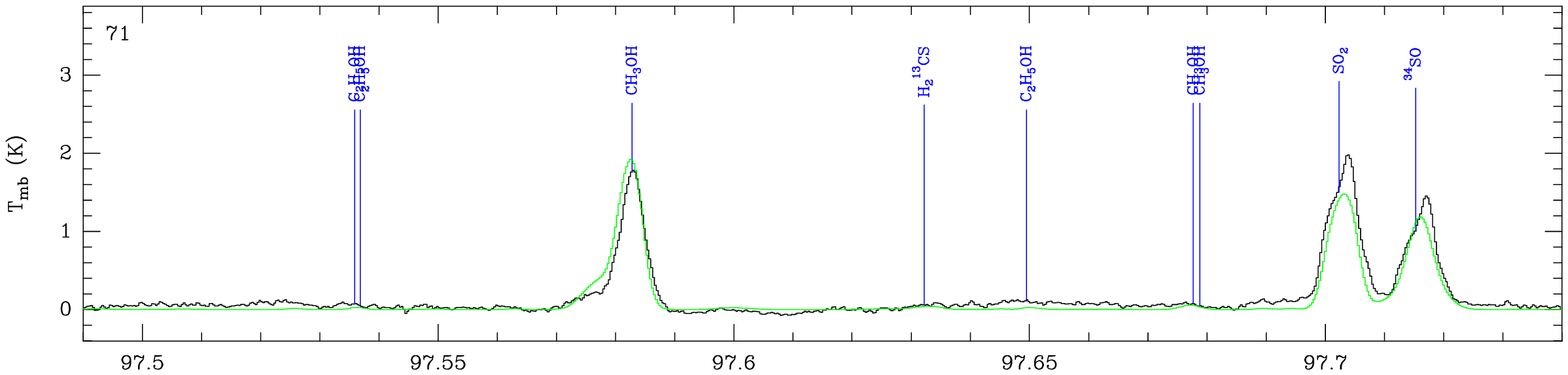}}}
\vspace*{1ex}\centerline{\resizebox{1.0\hsize}{!}{\includegraphics[angle=270]{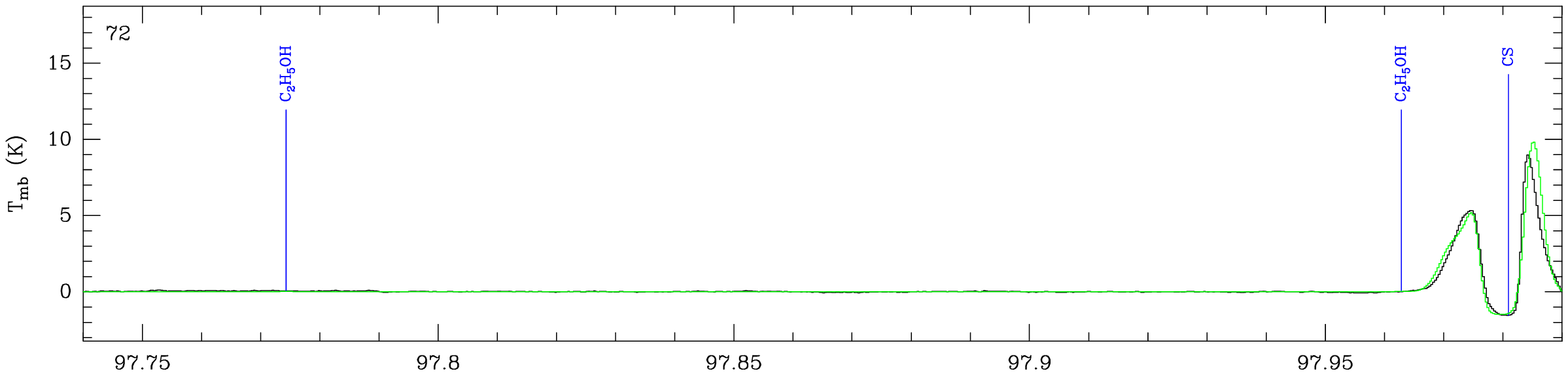}}}
\vspace*{1ex}\centerline{\resizebox{1.0\hsize}{!}{\includegraphics[angle=270]{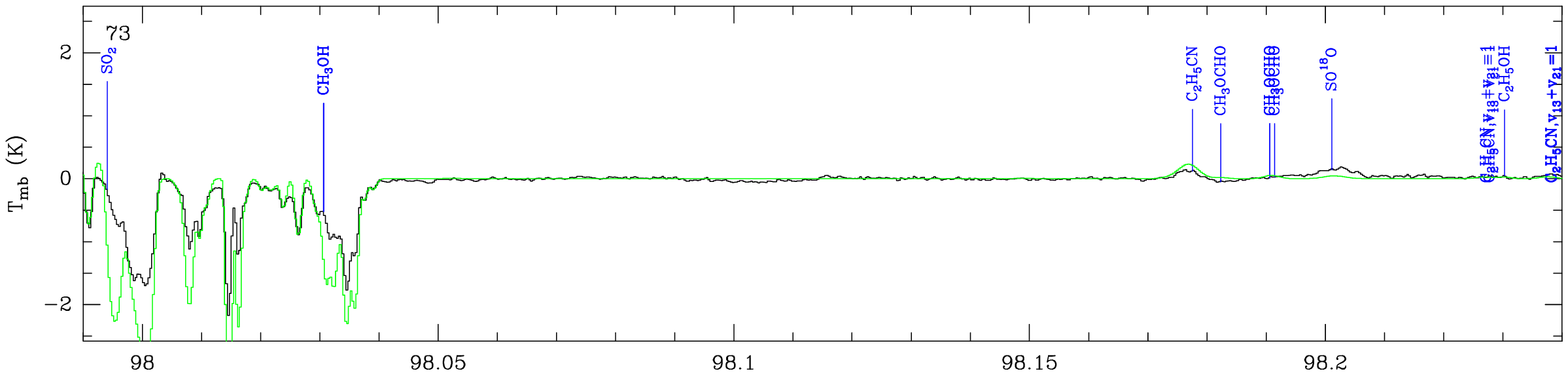}}}
\vspace*{1ex}\centerline{\resizebox{1.0\hsize}{!}{\includegraphics[angle=270]{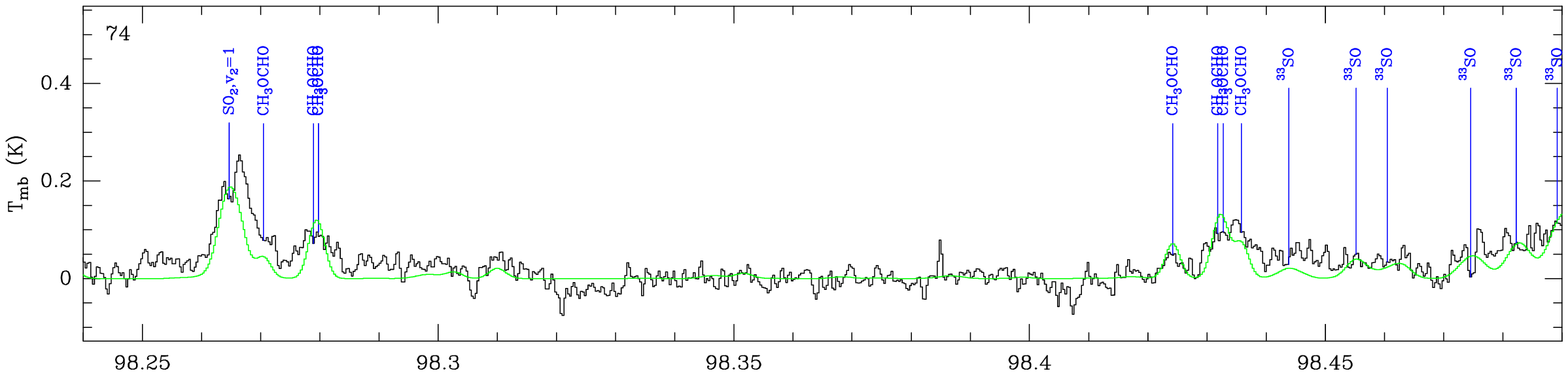}}}
\vspace*{1ex}\centerline{\resizebox{1.0\hsize}{!}{\includegraphics[angle=270]{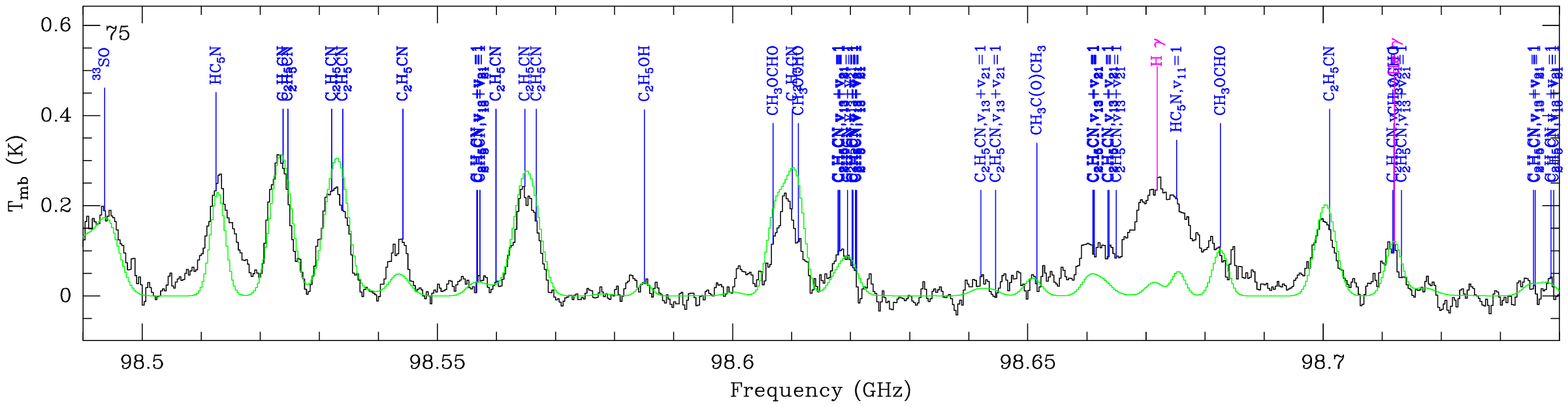}}}
\caption{
continued.
}
\end{figure*}
 \clearpage
\begin{figure*}
\addtocounter{figure}{-1}
\centerline{\resizebox{1.0\hsize}{!}{\includegraphics[angle=270]{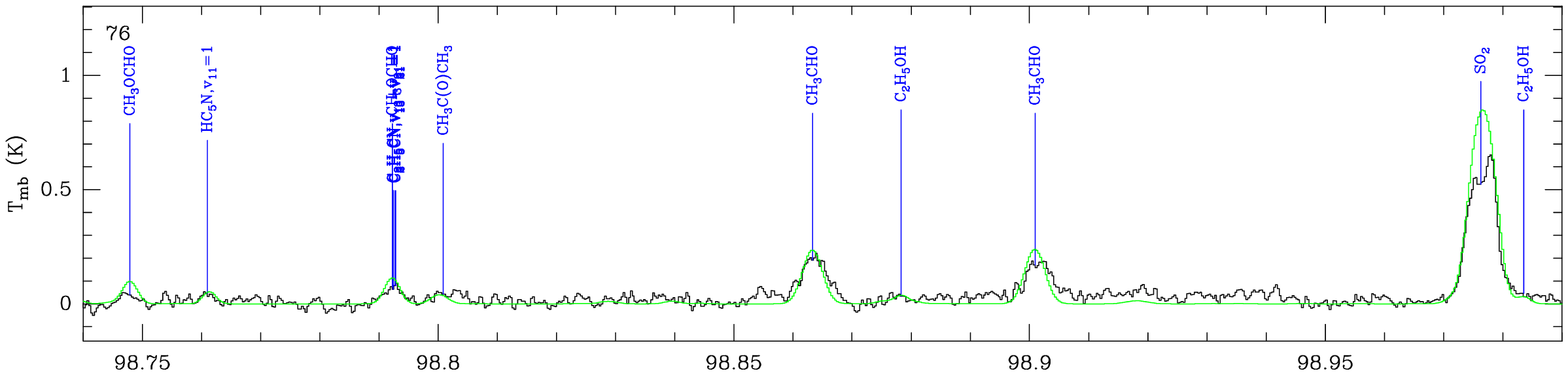}}}
\vspace*{1ex}\centerline{\resizebox{1.0\hsize}{!}{\includegraphics[angle=270]{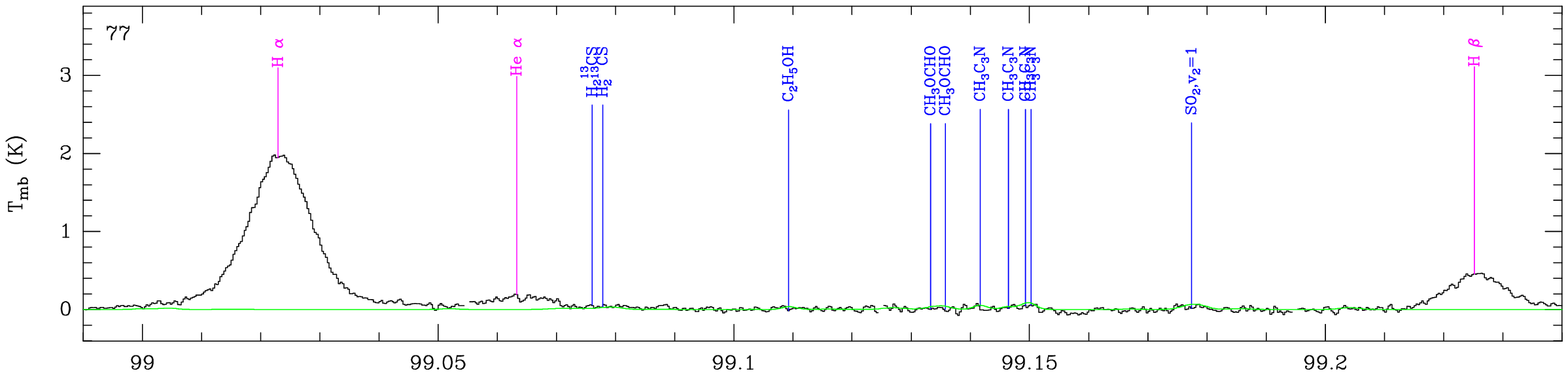}}}
\vspace*{1ex}\centerline{\resizebox{1.0\hsize}{!}{\includegraphics[angle=270]{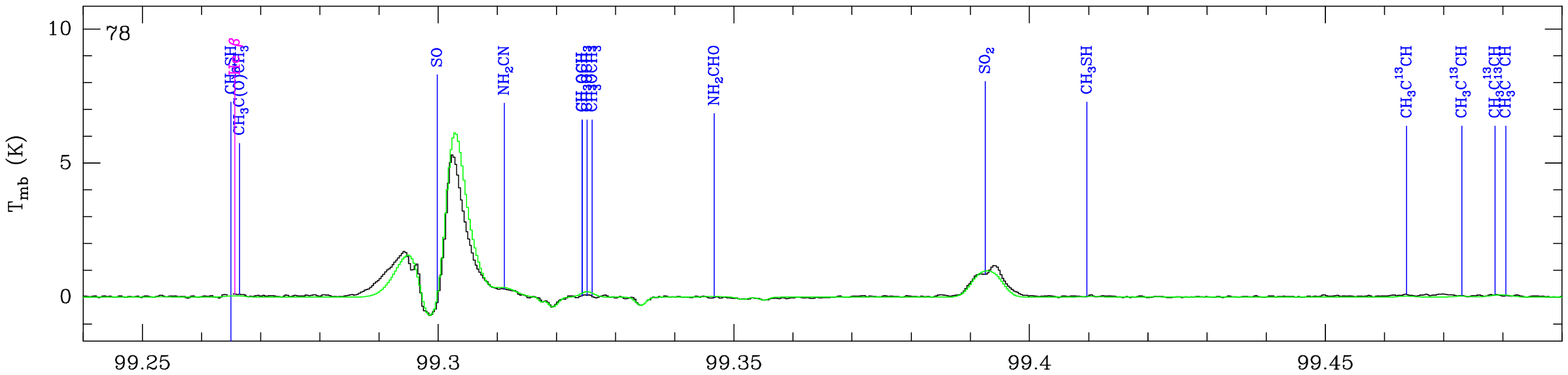}}}
\vspace*{1ex}\centerline{\resizebox{1.0\hsize}{!}{\includegraphics[angle=270]{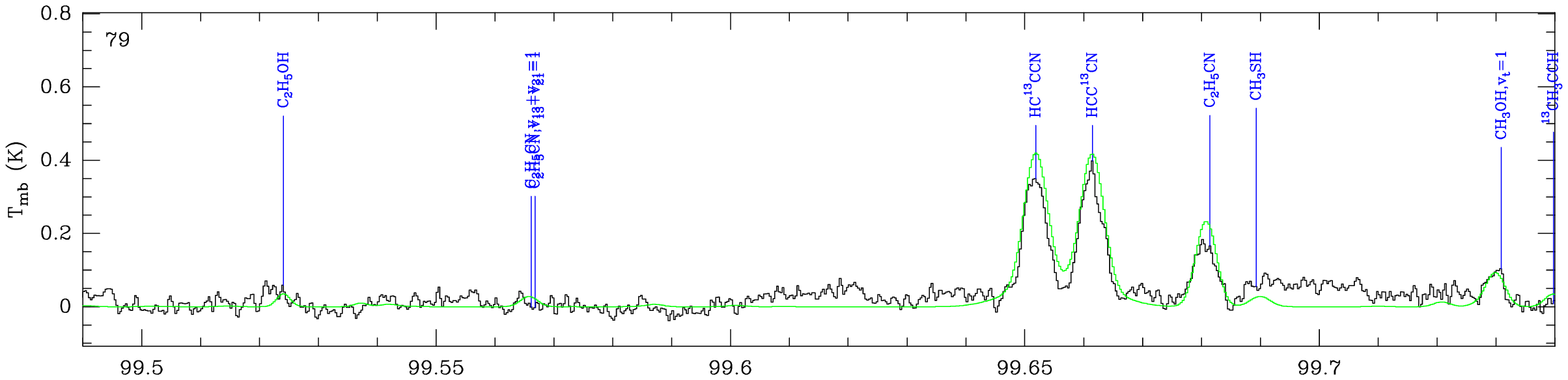}}}
\vspace*{1ex}\centerline{\resizebox{1.0\hsize}{!}{\includegraphics[angle=270]{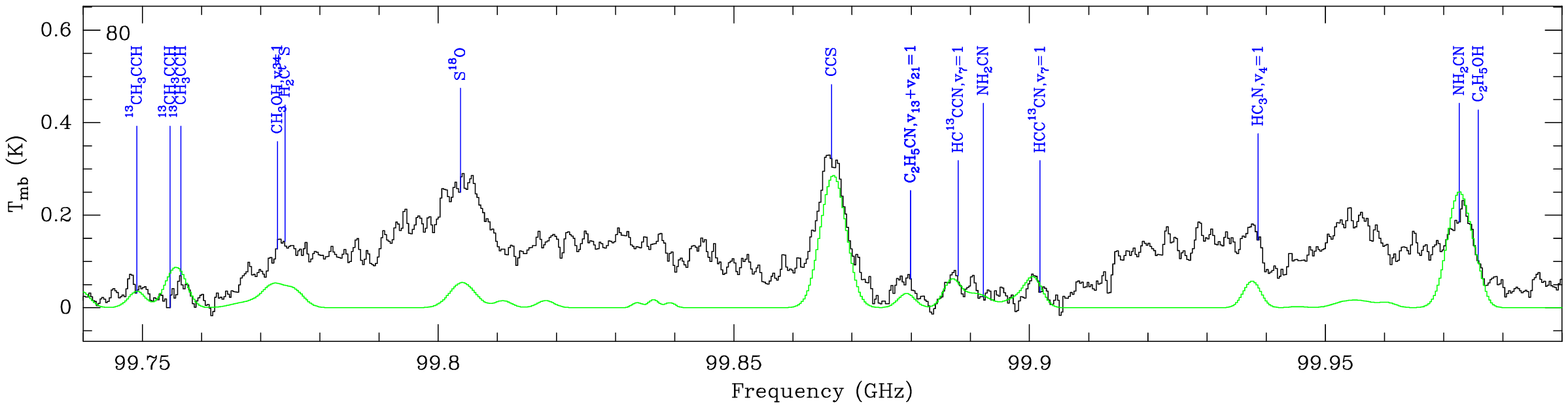}}}
\caption{
continued.
}
\end{figure*}
 \clearpage
\begin{figure*}
\addtocounter{figure}{-1}
\centerline{\resizebox{1.0\hsize}{!}{\includegraphics[angle=270]{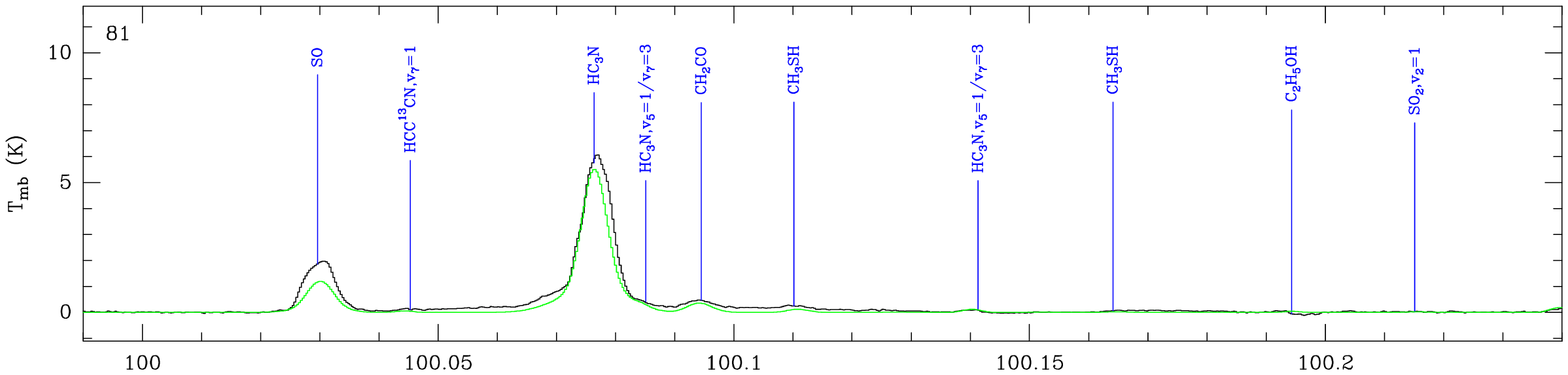}}}
\vspace*{1ex}\centerline{\resizebox{1.0\hsize}{!}{\includegraphics[angle=270]{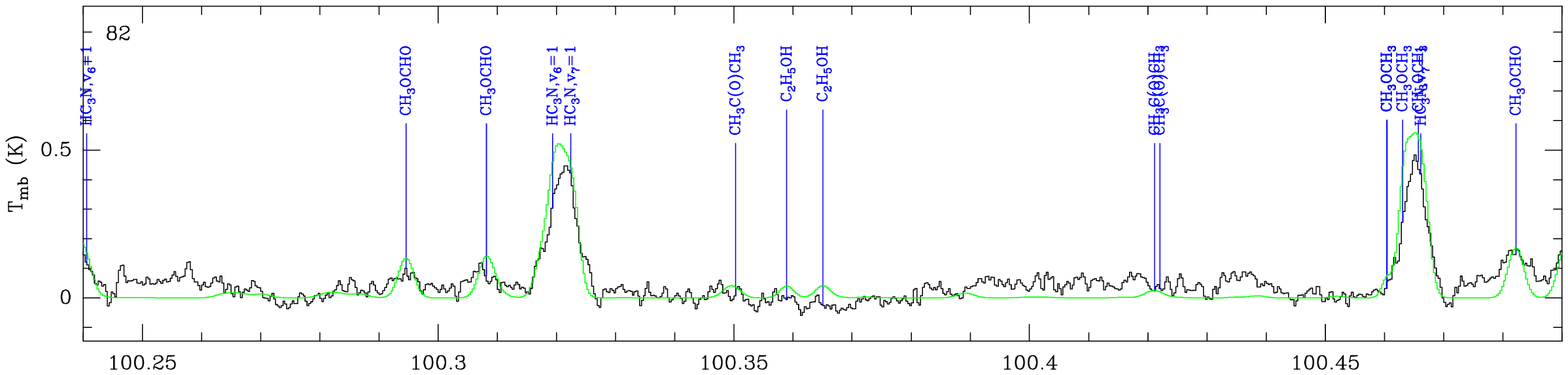}}}
\vspace*{1ex}\centerline{\resizebox{1.0\hsize}{!}{\includegraphics[angle=270]{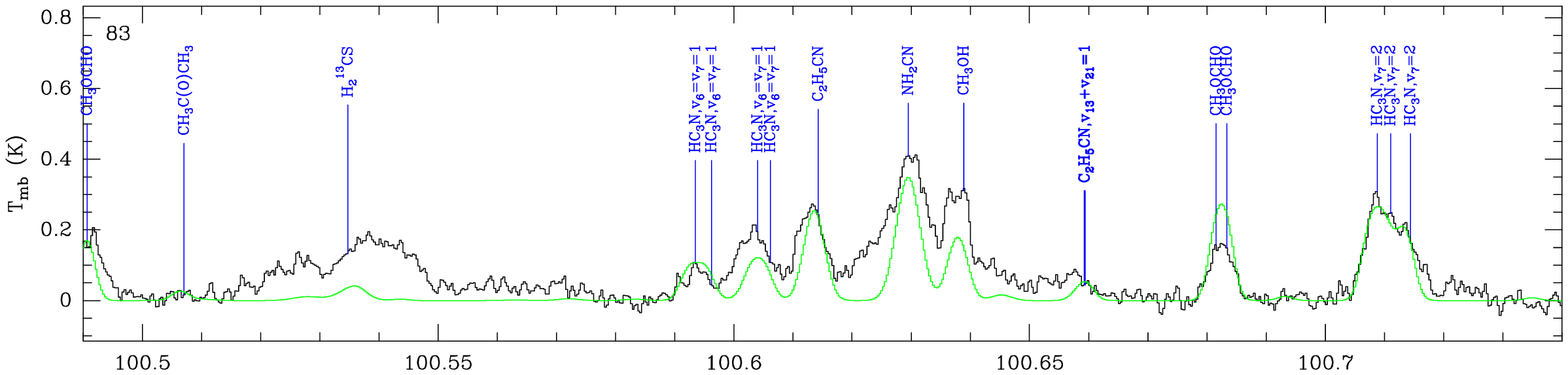}}}
\vspace*{1ex}\centerline{\resizebox{1.0\hsize}{!}{\includegraphics[angle=270]{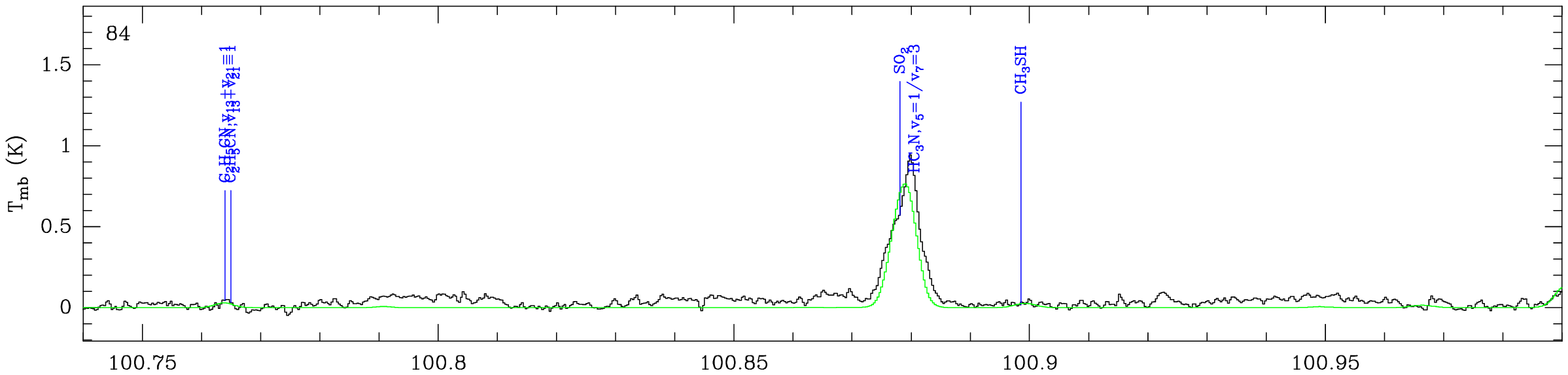}}}
\vspace*{1ex}\centerline{\resizebox{1.0\hsize}{!}{\includegraphics[angle=270]{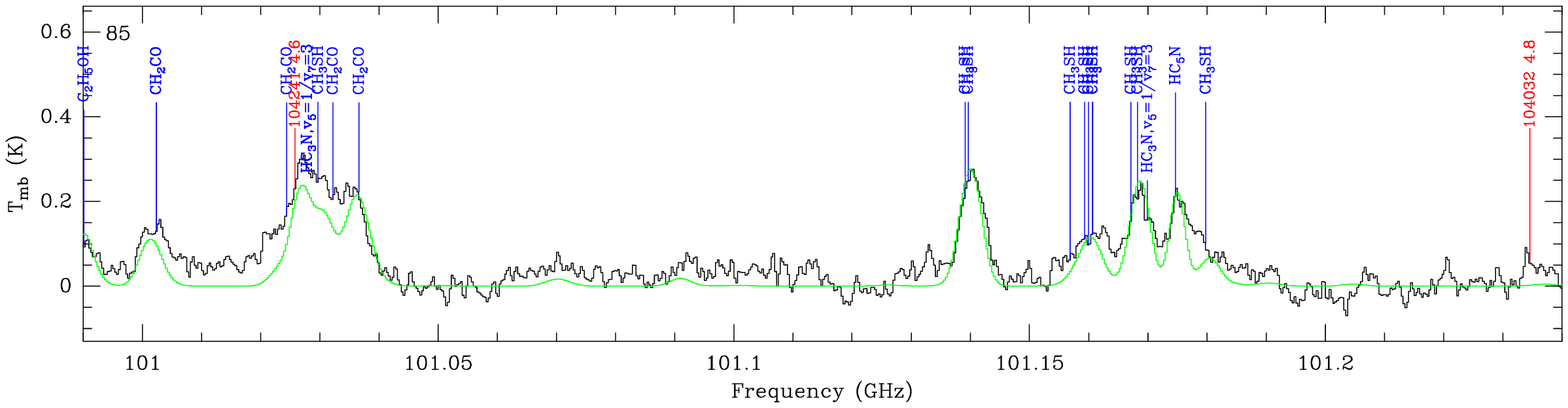}}}
\caption{
continued.
}
\end{figure*}
 \clearpage
\begin{figure*}
\addtocounter{figure}{-1}
\centerline{\resizebox{1.0\hsize}{!}{\includegraphics[angle=270]{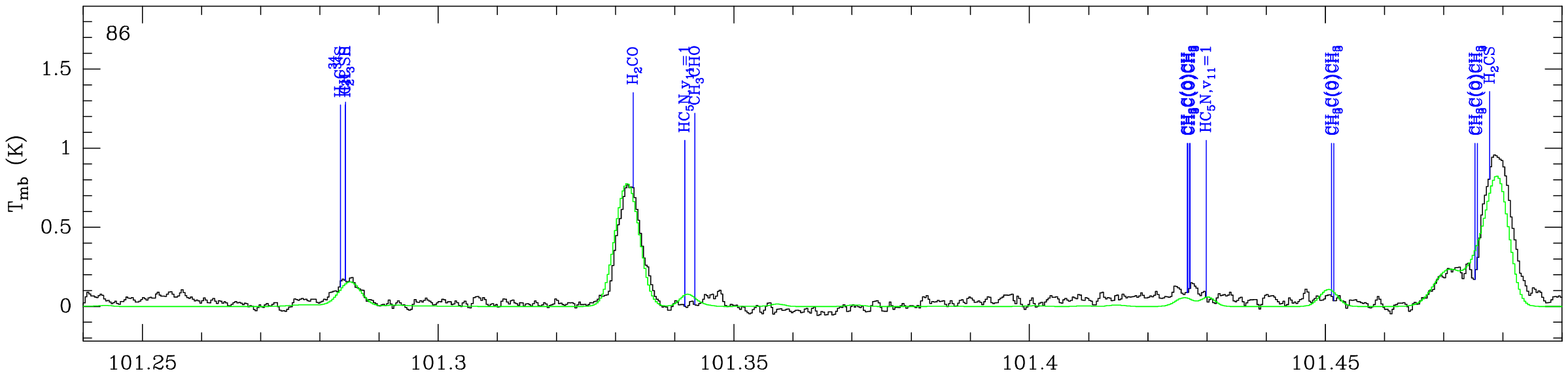}}}
\vspace*{1ex}\centerline{\resizebox{1.0\hsize}{!}{\includegraphics[angle=270]{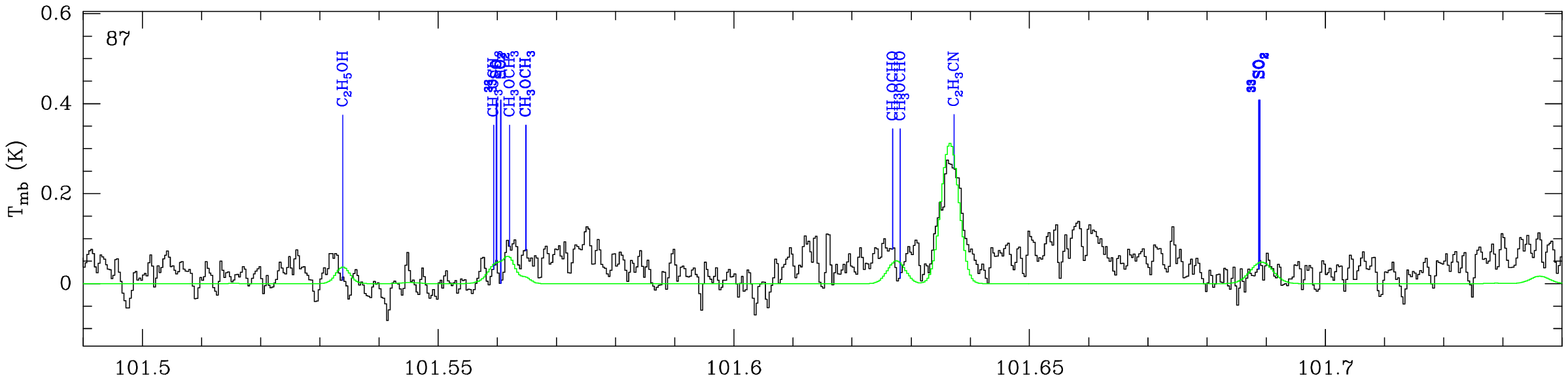}}}
\vspace*{1ex}\centerline{\resizebox{1.0\hsize}{!}{\includegraphics[angle=270]{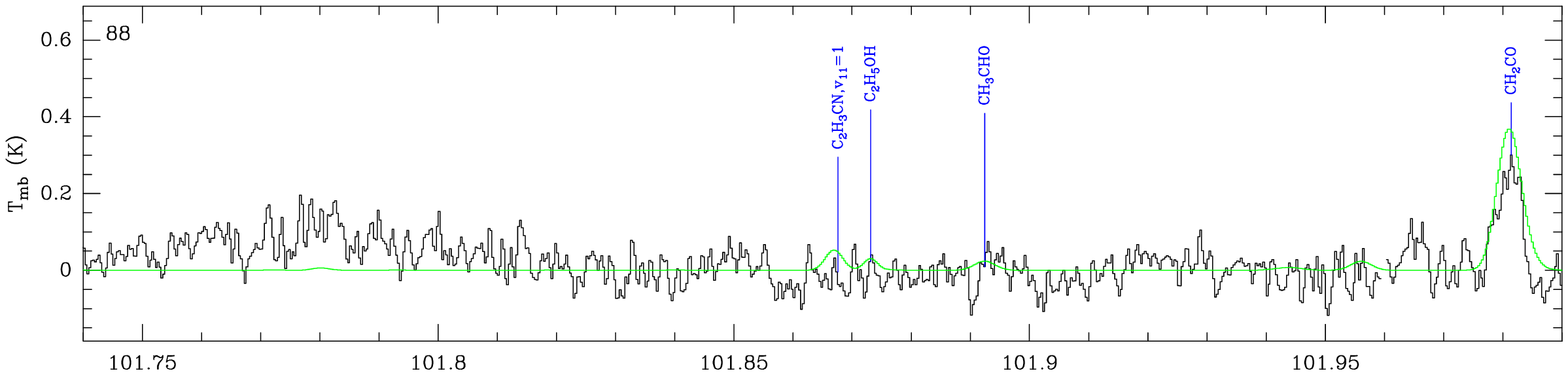}}}
\vspace*{1ex}\centerline{\resizebox{1.0\hsize}{!}{\includegraphics[angle=270]{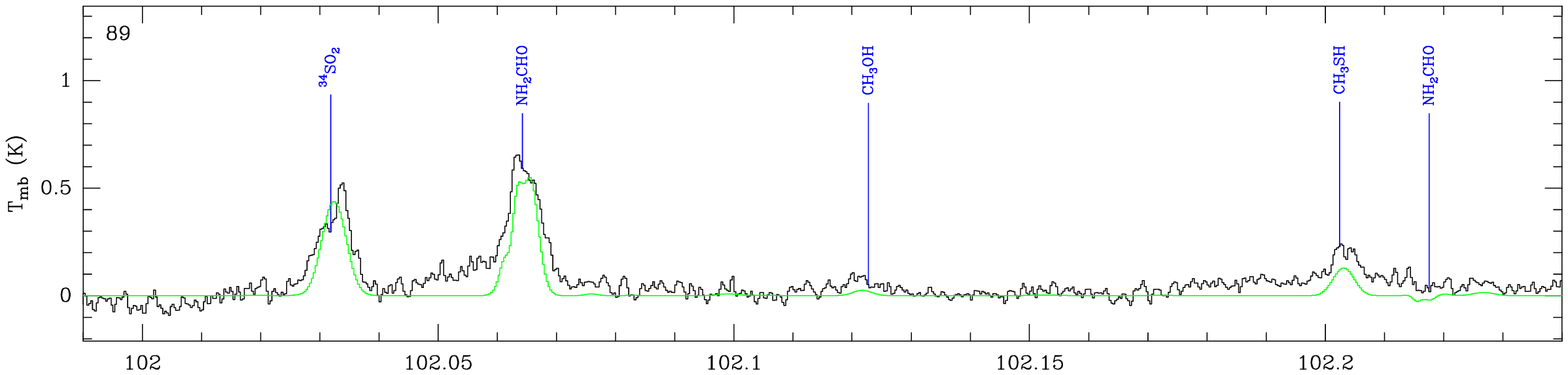}}}
\vspace*{1ex}\centerline{\resizebox{1.0\hsize}{!}{\includegraphics[angle=270]{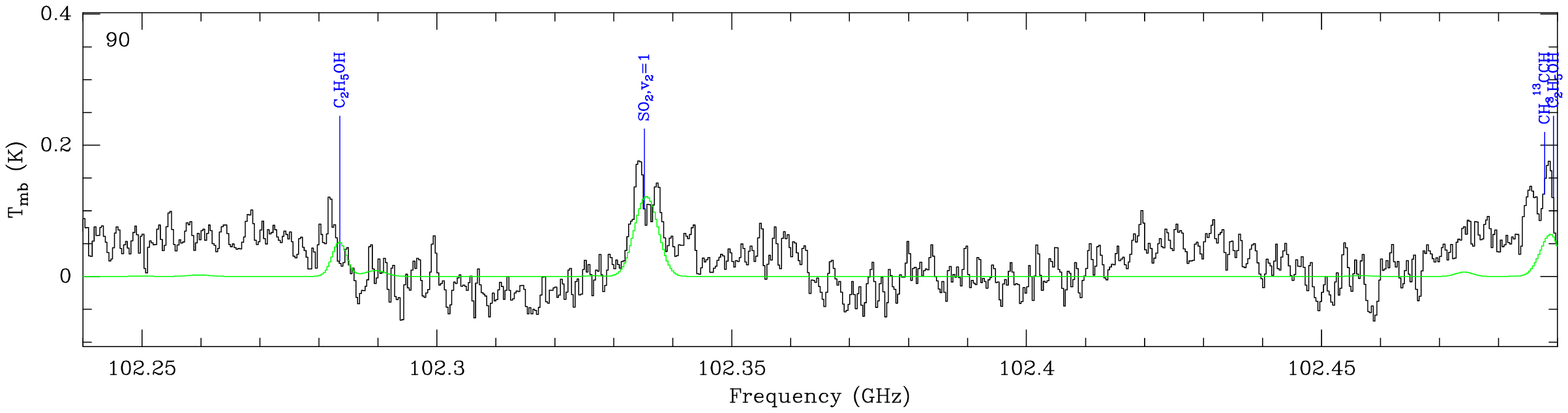}}}
\caption{
continued.
}
\end{figure*}
 \clearpage
\begin{figure*}
\addtocounter{figure}{-1}
\centerline{\resizebox{1.0\hsize}{!}{\includegraphics[angle=270]{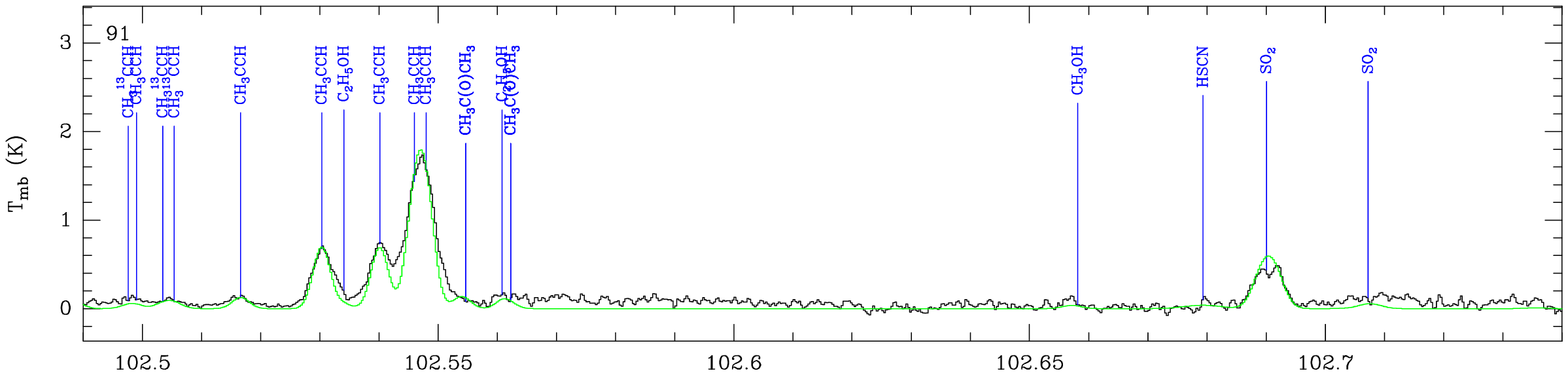}}}
\vspace*{1ex}\centerline{\resizebox{1.0\hsize}{!}{\includegraphics[angle=270]{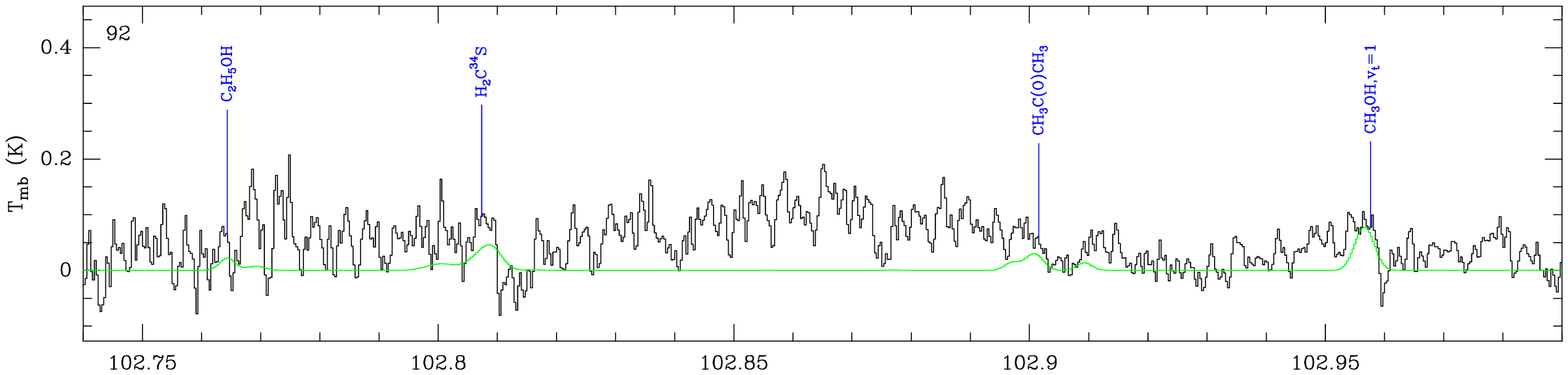}}}
\vspace*{1ex}\centerline{\resizebox{1.0\hsize}{!}{\includegraphics[angle=270]{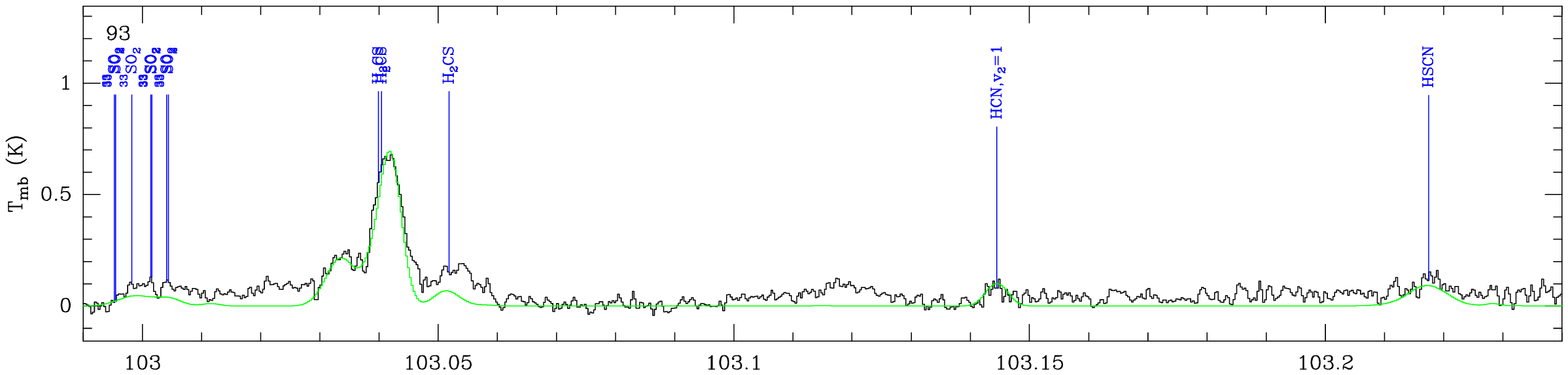}}}
\vspace*{1ex}\centerline{\resizebox{1.0\hsize}{!}{\includegraphics[angle=270]{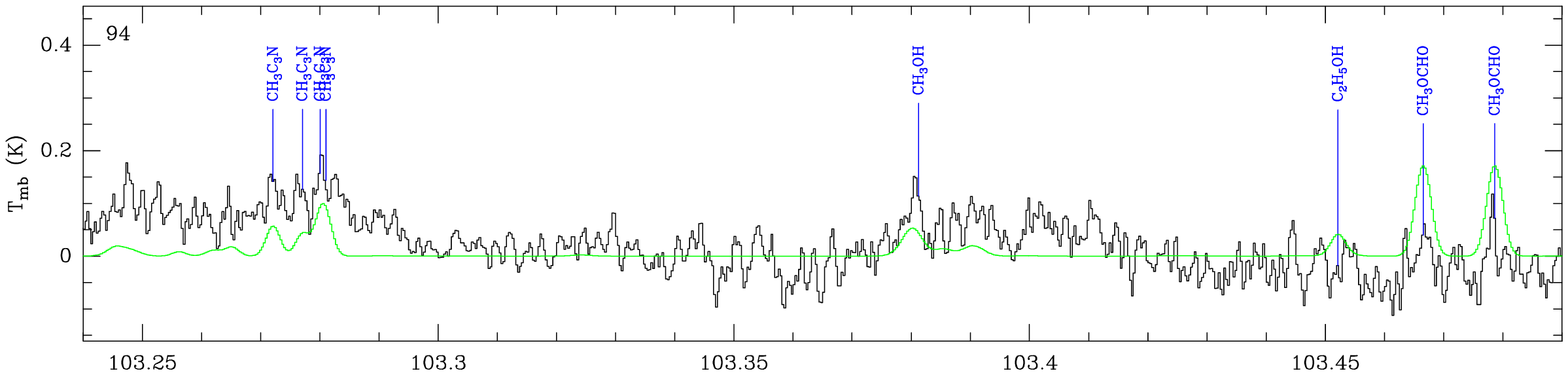}}}
\vspace*{1ex}\centerline{\resizebox{1.0\hsize}{!}{\includegraphics[angle=270]{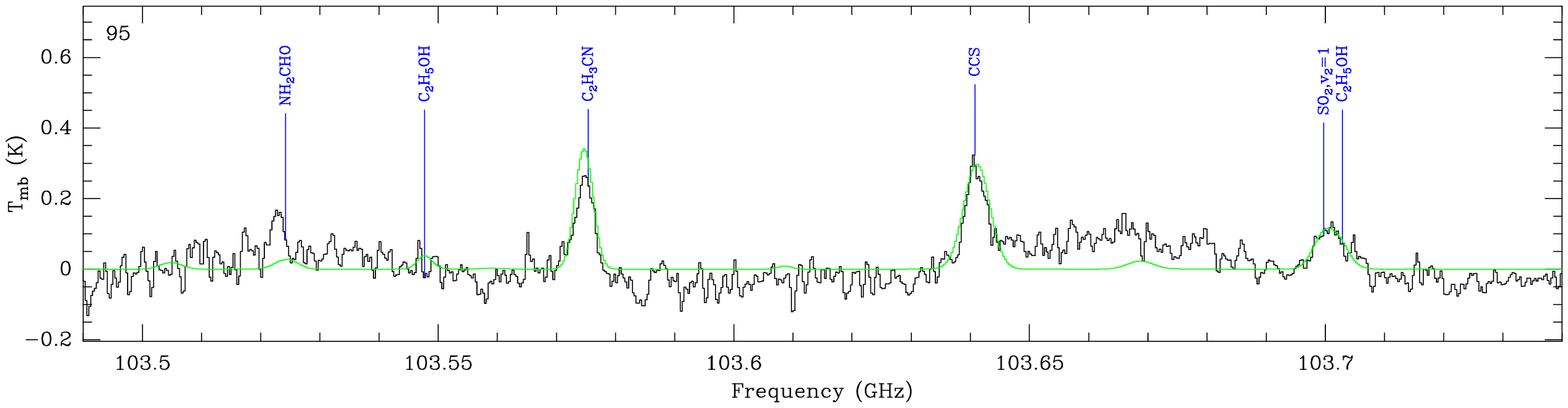}}}
\caption{
continued.
}
\end{figure*}
 \clearpage
\begin{figure*}
\addtocounter{figure}{-1}
\centerline{\resizebox{1.0\hsize}{!}{\includegraphics[angle=270]{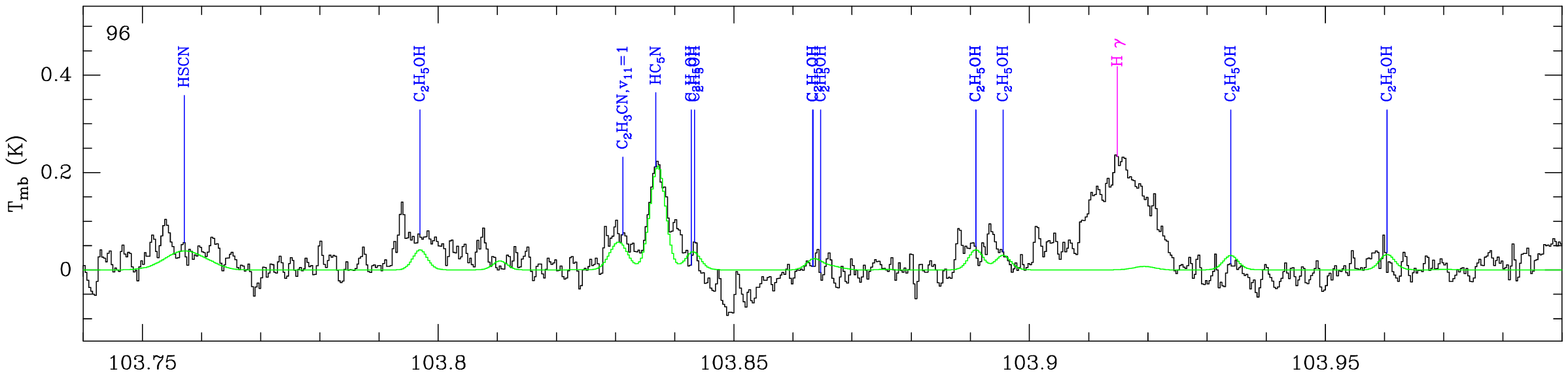}}}
\vspace*{1ex}\centerline{\resizebox{1.0\hsize}{!}{\includegraphics[angle=270]{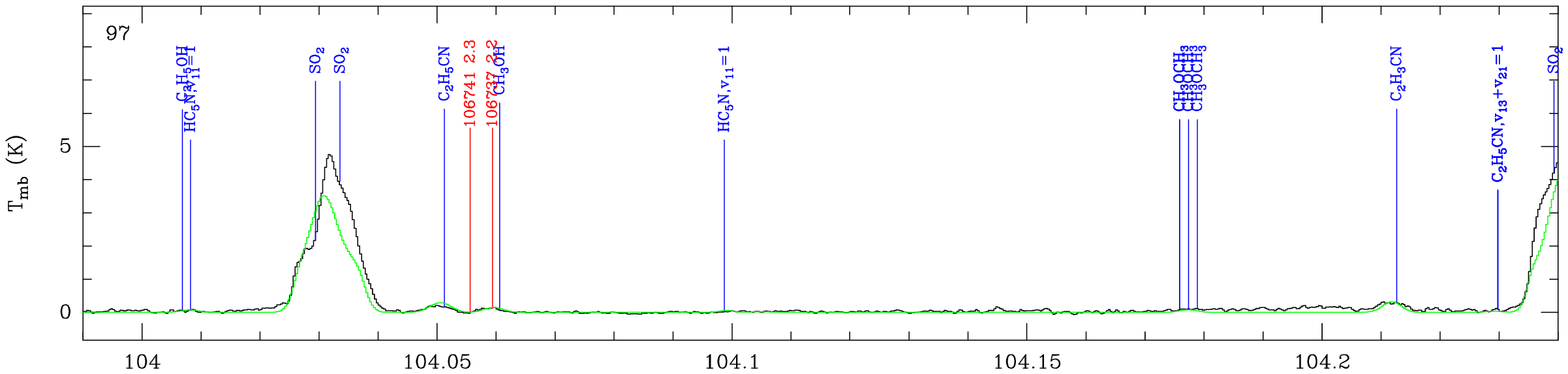}}}
\vspace*{1ex}\centerline{\resizebox{1.0\hsize}{!}{\includegraphics[angle=270]{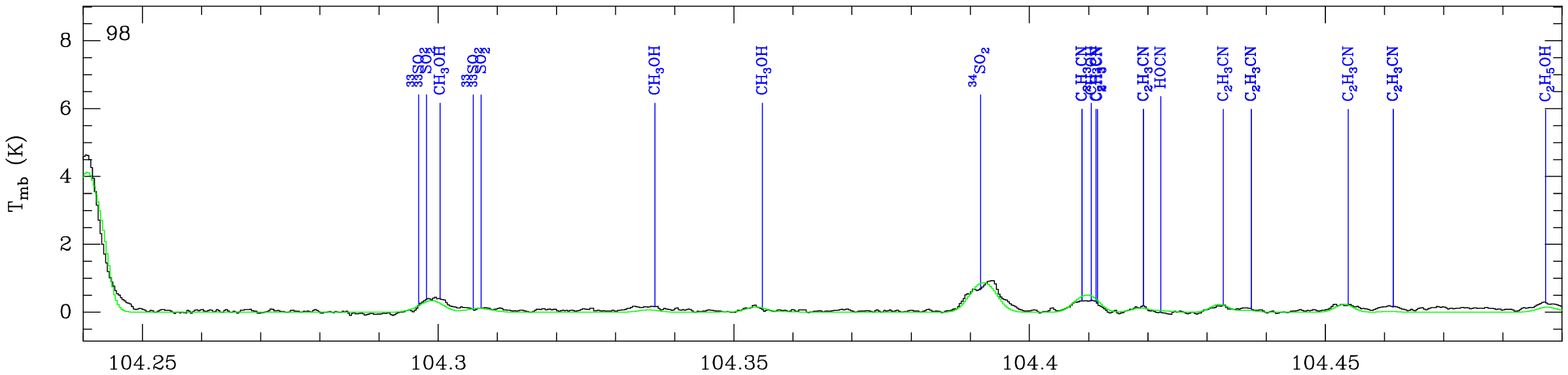}}}
\vspace*{1ex}\centerline{\resizebox{1.0\hsize}{!}{\includegraphics[angle=270]{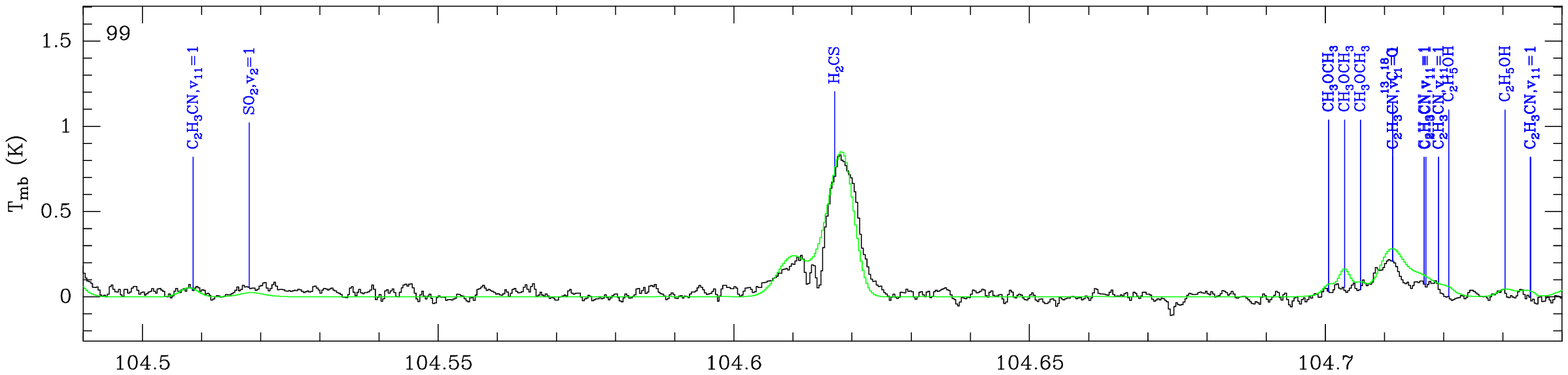}}}
\vspace*{1ex}\centerline{\resizebox{1.0\hsize}{!}{\includegraphics[angle=270]{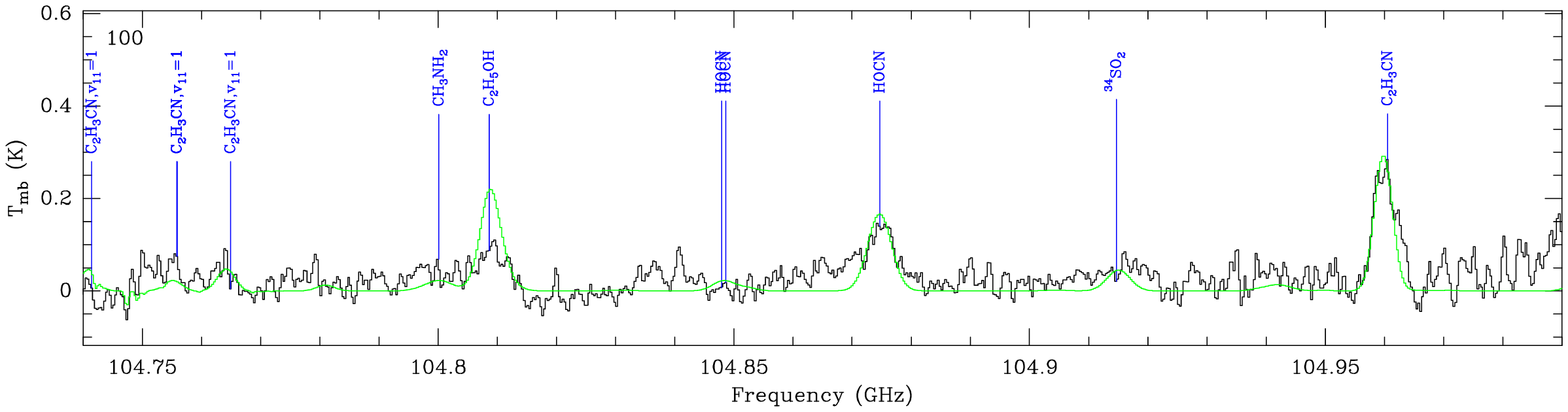}}}
\caption{
continued.
}
\end{figure*}
 \clearpage
\begin{figure*}
\addtocounter{figure}{-1}
\centerline{\resizebox{1.0\hsize}{!}{\includegraphics[angle=270]{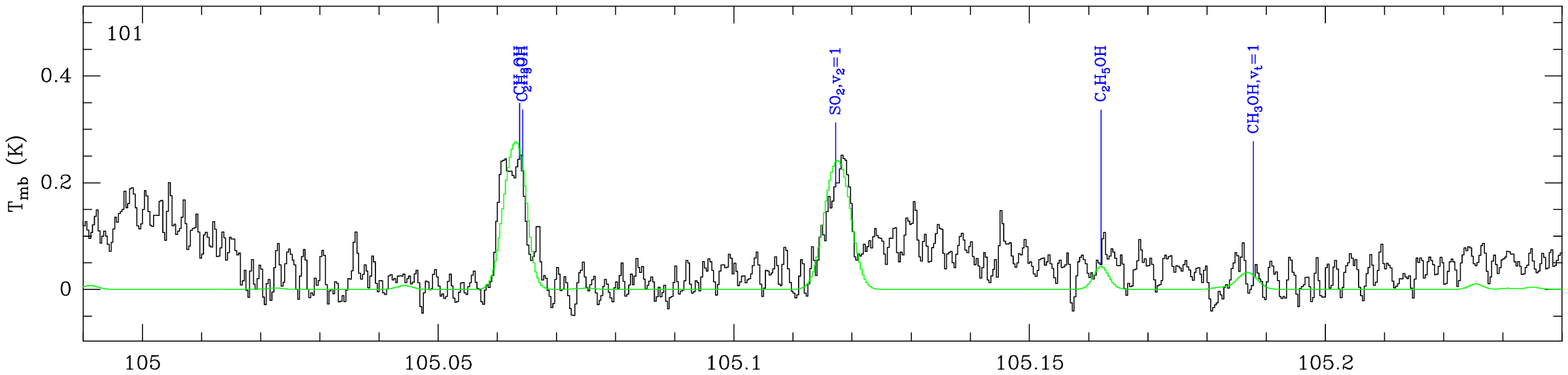}}}
\vspace*{1ex}\centerline{\resizebox{1.0\hsize}{!}{\includegraphics[angle=270]{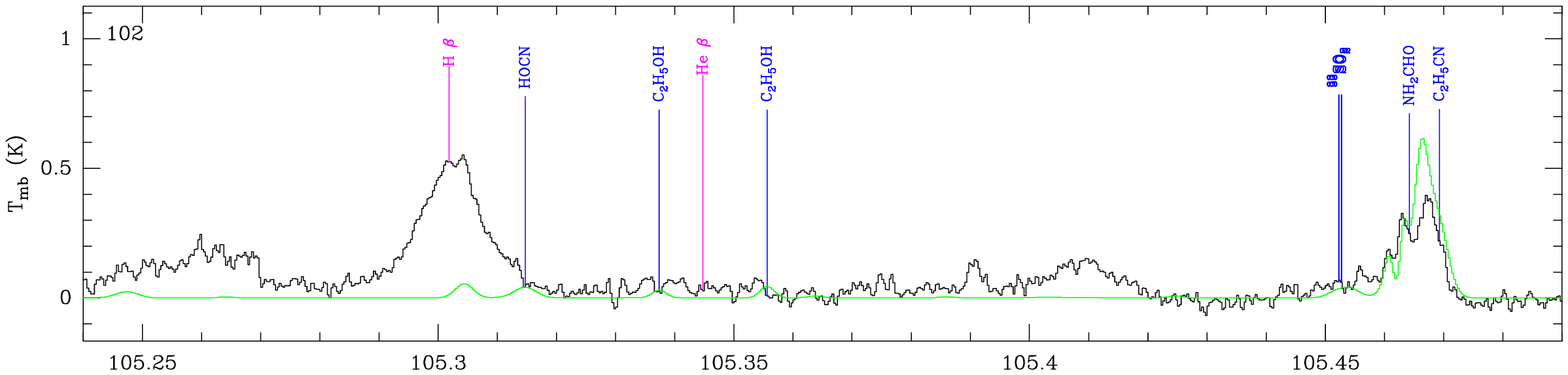}}}
\vspace*{1ex}\centerline{\resizebox{1.0\hsize}{!}{\includegraphics[angle=270]{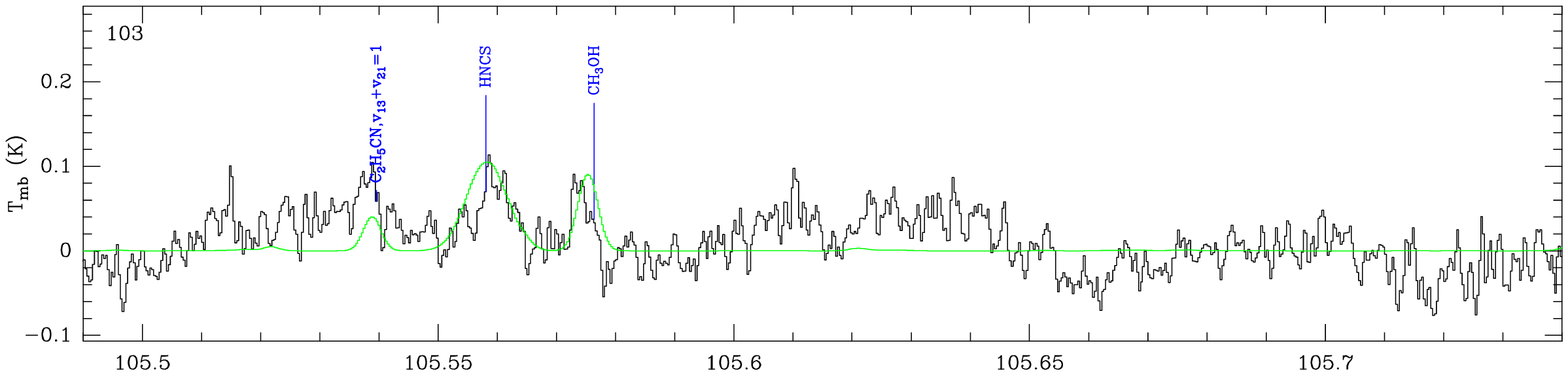}}}
\vspace*{1ex}\centerline{\resizebox{1.0\hsize}{!}{\includegraphics[angle=270]{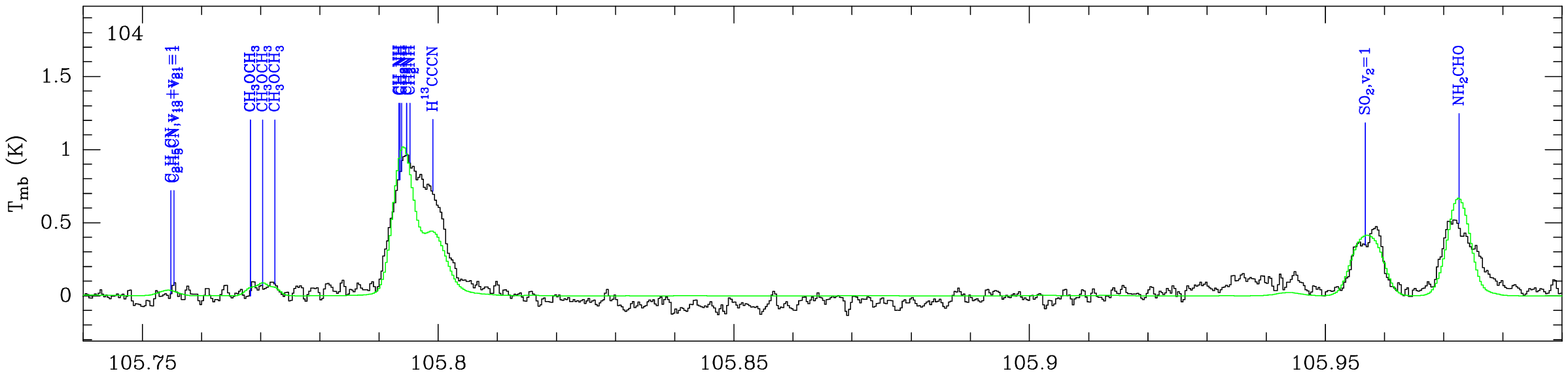}}}
\vspace*{1ex}\centerline{\resizebox{1.0\hsize}{!}{\includegraphics[angle=270]{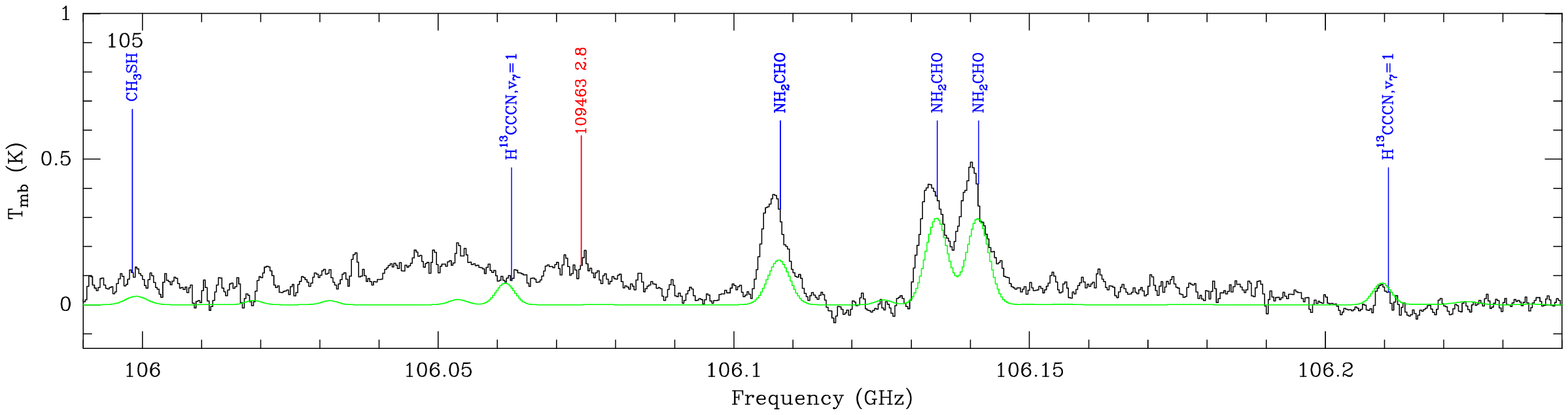}}}
\caption{
continued.
}
\end{figure*}
 \clearpage
\begin{figure*}
\addtocounter{figure}{-1}
\centerline{\resizebox{1.0\hsize}{!}{\includegraphics[angle=270]{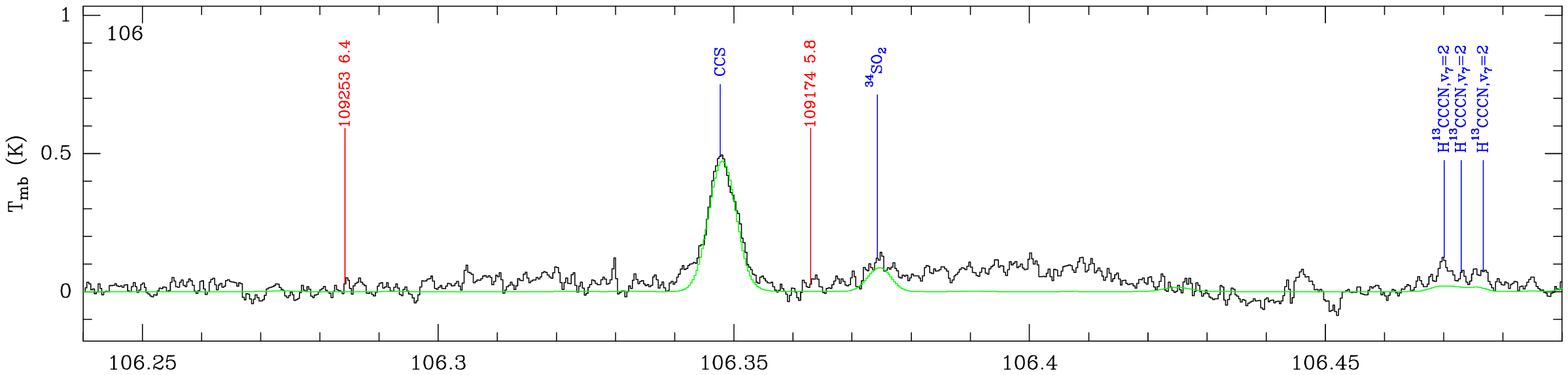}}}
\vspace*{1ex}\centerline{\resizebox{1.0\hsize}{!}{\includegraphics[angle=270]{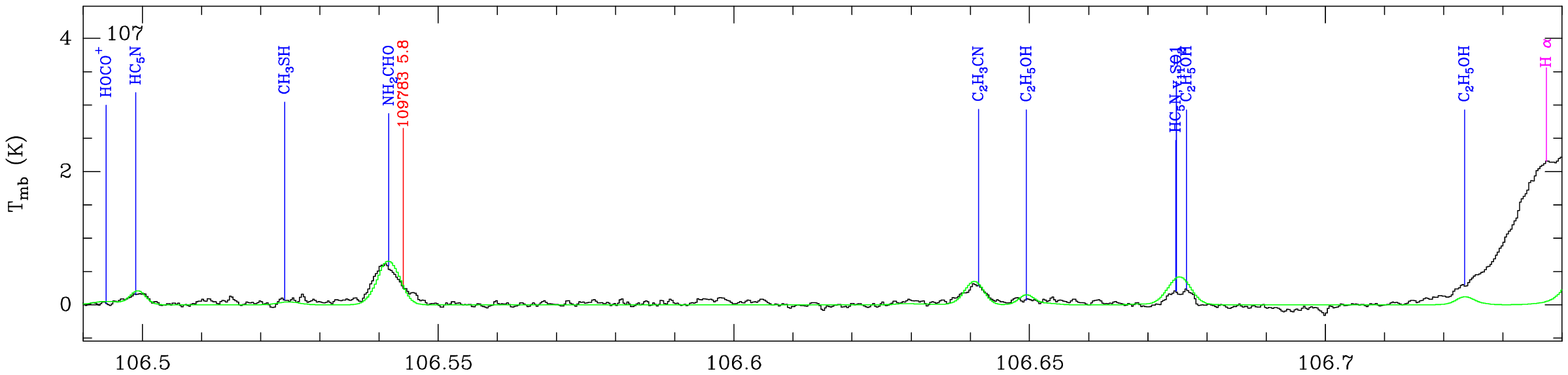}}}
\vspace*{1ex}\centerline{\resizebox{1.0\hsize}{!}{\includegraphics[angle=270]{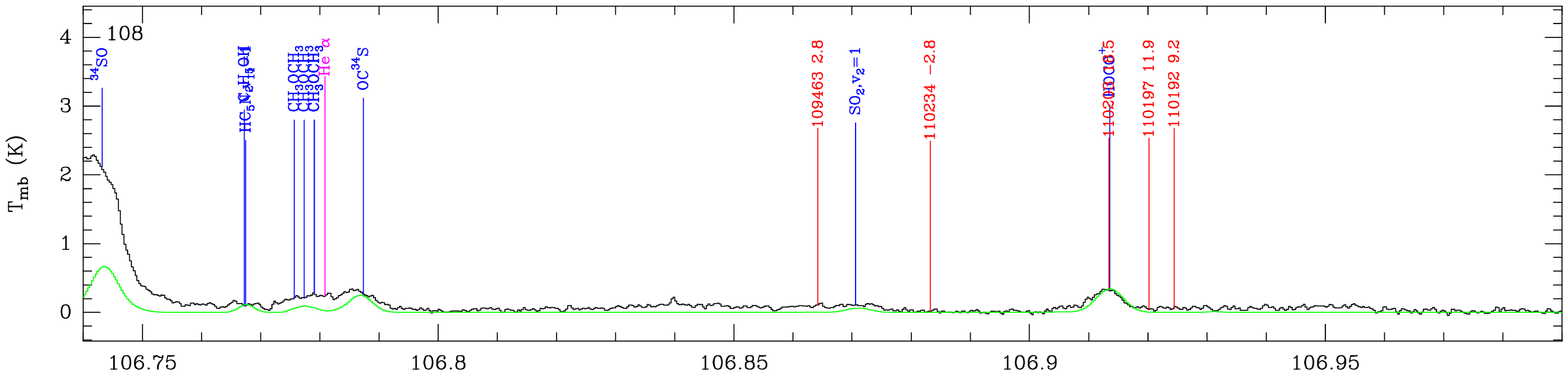}}}
\vspace*{1ex}\centerline{\resizebox{1.0\hsize}{!}{\includegraphics[angle=270]{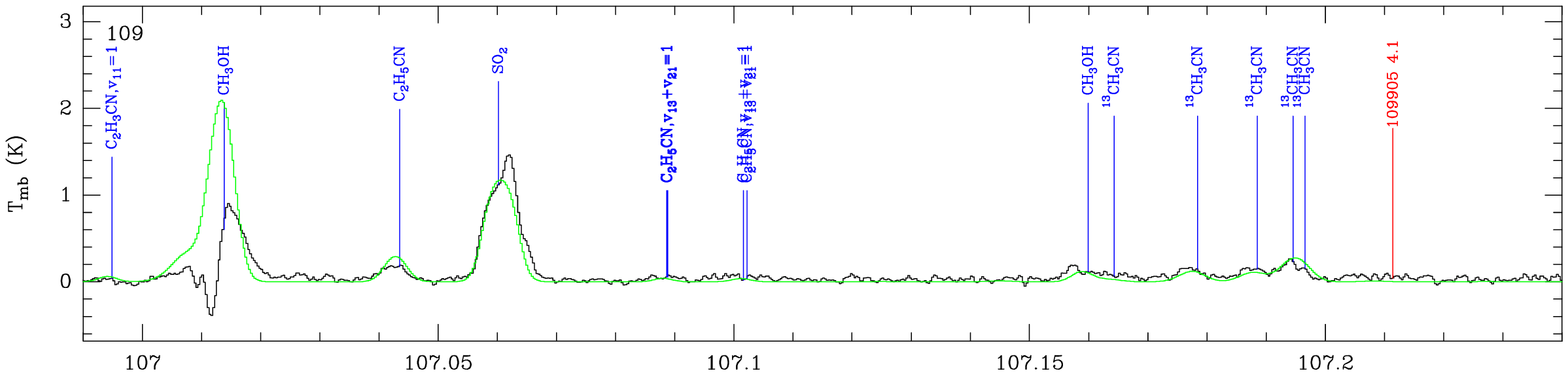}}}
\vspace*{1ex}\centerline{\resizebox{1.0\hsize}{!}{\includegraphics[angle=270]{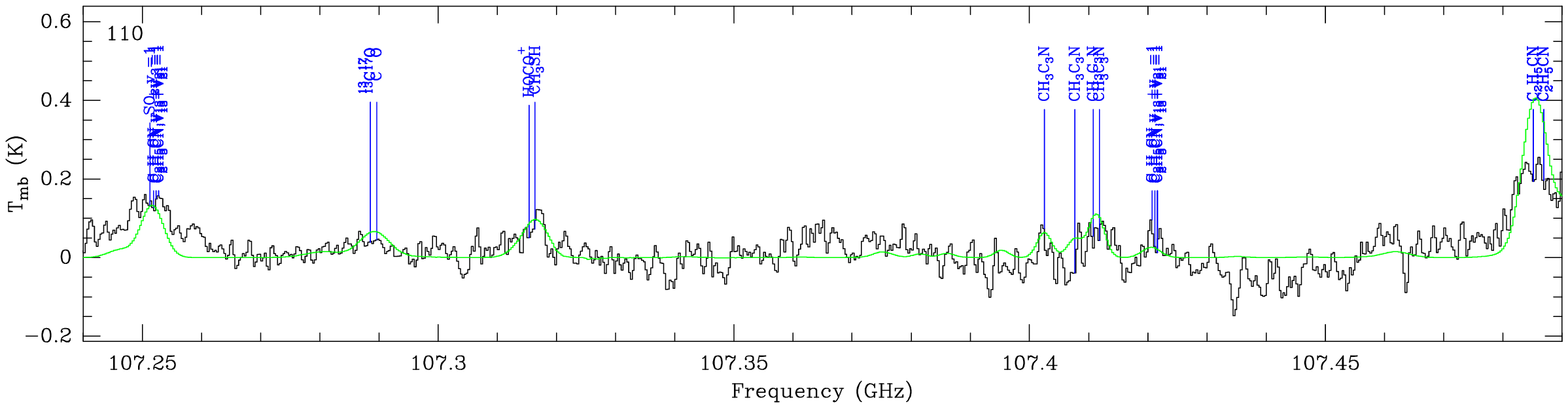}}}
\caption{
continued.
}
\end{figure*}
 \clearpage
\begin{figure*}
\addtocounter{figure}{-1}
\centerline{\resizebox{1.0\hsize}{!}{\includegraphics[angle=270]{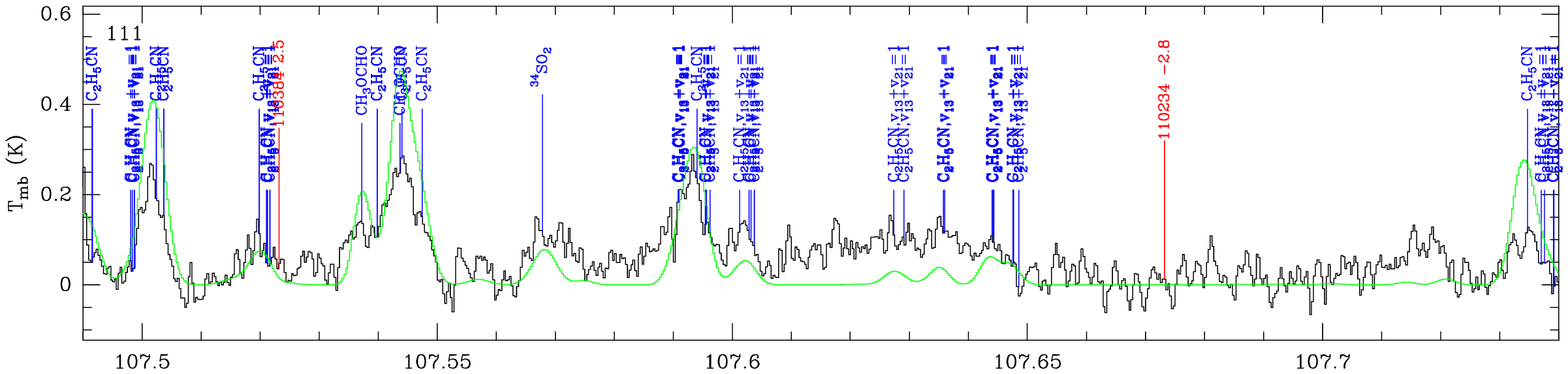}}}
\vspace*{1ex}\centerline{\resizebox{1.0\hsize}{!}{\includegraphics[angle=270]{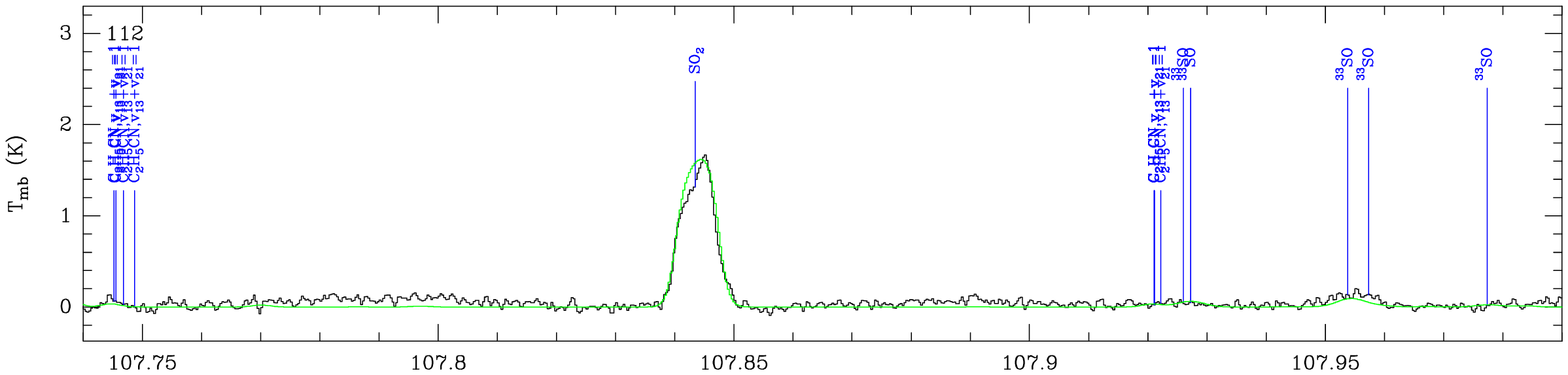}}}
\vspace*{1ex}\centerline{\resizebox{1.0\hsize}{!}{\includegraphics[angle=270]{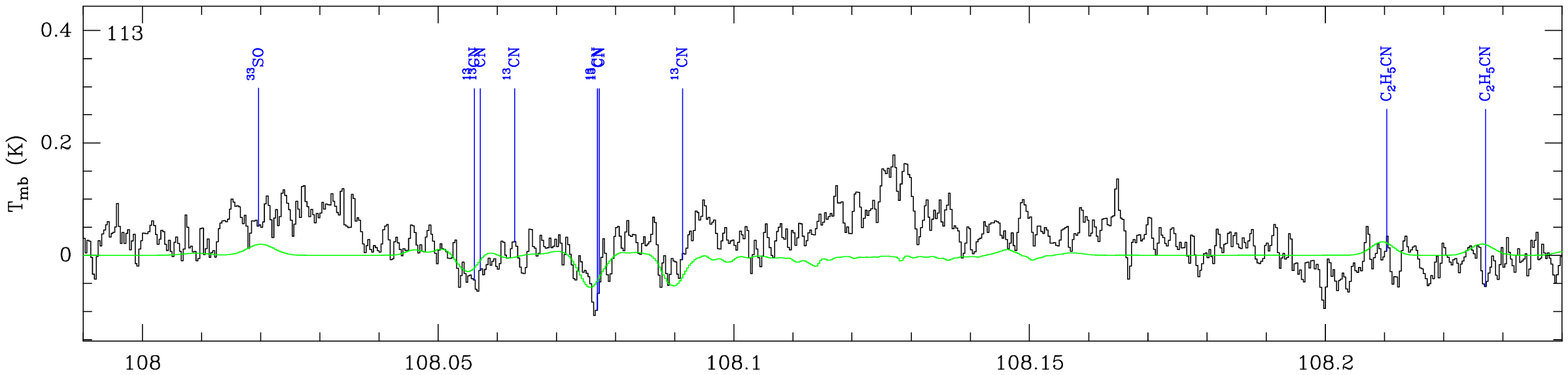}}}
\vspace*{1ex}\centerline{\resizebox{1.0\hsize}{!}{\includegraphics[angle=270]{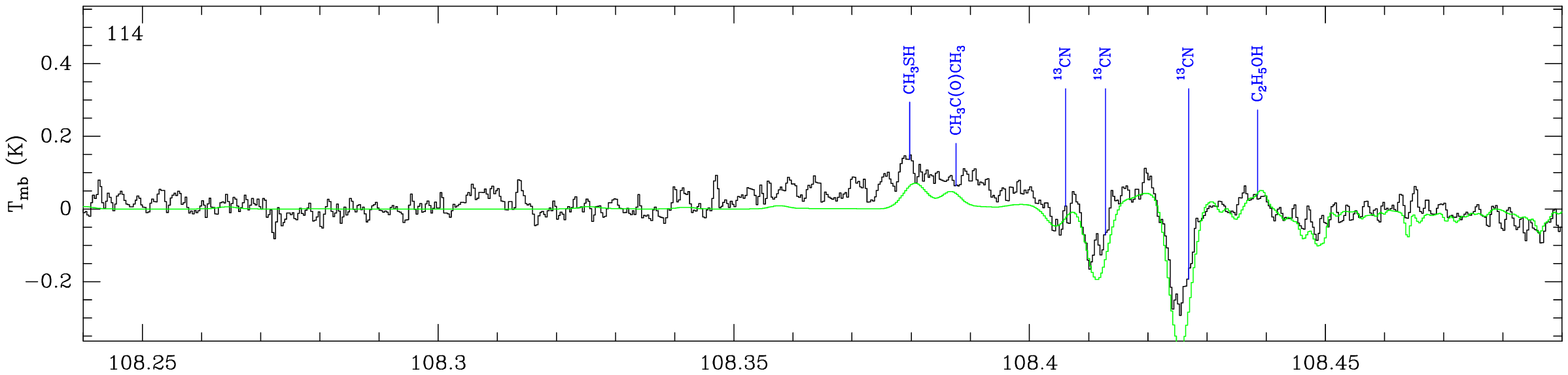}}}
\vspace*{1ex}\centerline{\resizebox{1.0\hsize}{!}{\includegraphics[angle=270]{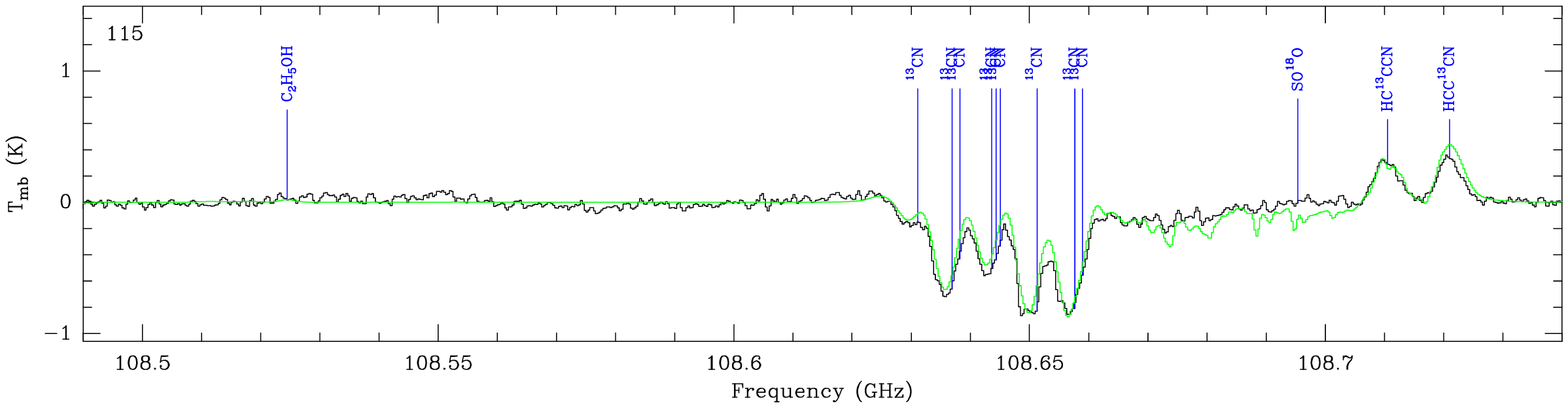}}}
\caption{
continued.
}
\end{figure*}
 \clearpage
\begin{figure*}
\addtocounter{figure}{-1}
\centerline{\resizebox{1.0\hsize}{!}{\includegraphics[angle=270]{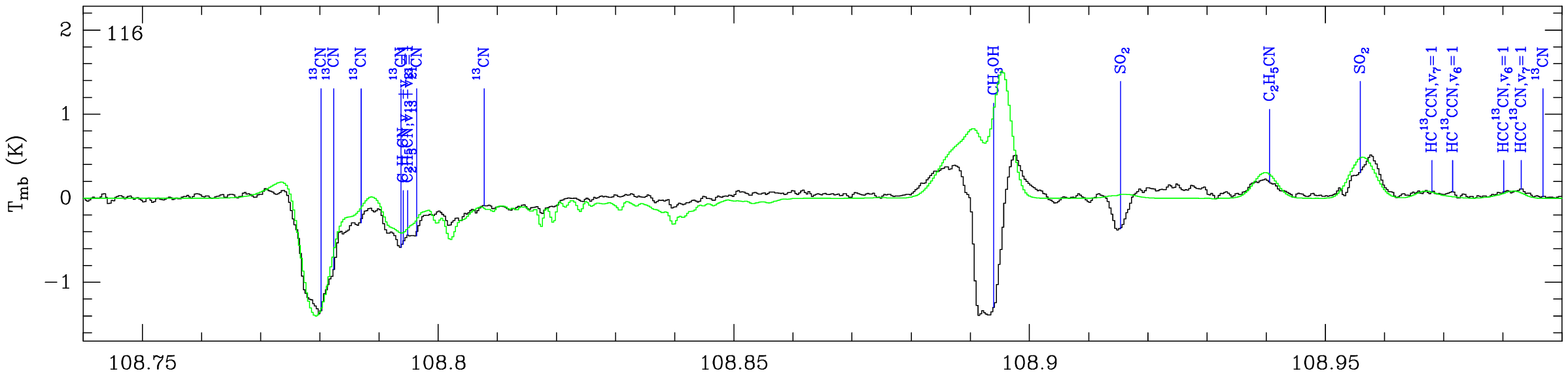}}}
\vspace*{1ex}\centerline{\resizebox{1.0\hsize}{!}{\includegraphics[angle=270]{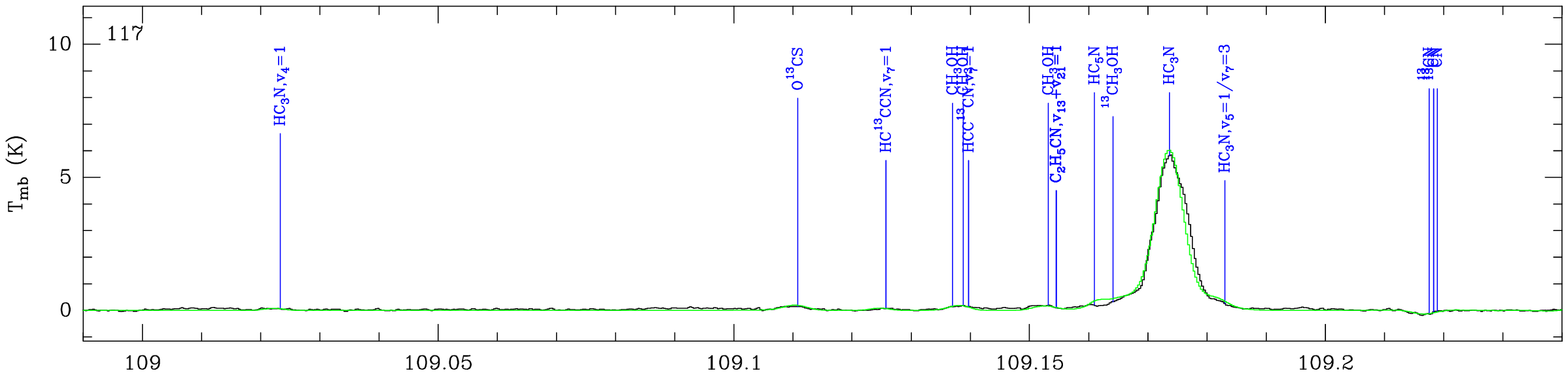}}}
\vspace*{1ex}\centerline{\resizebox{1.0\hsize}{!}{\includegraphics[angle=270]{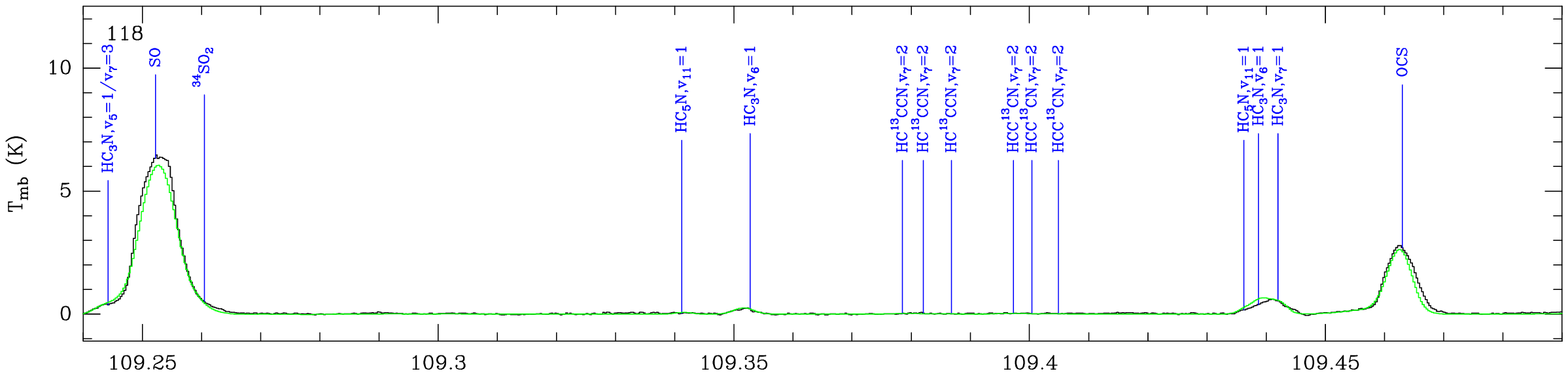}}}
\vspace*{1ex}\centerline{\resizebox{1.0\hsize}{!}{\includegraphics[angle=270]{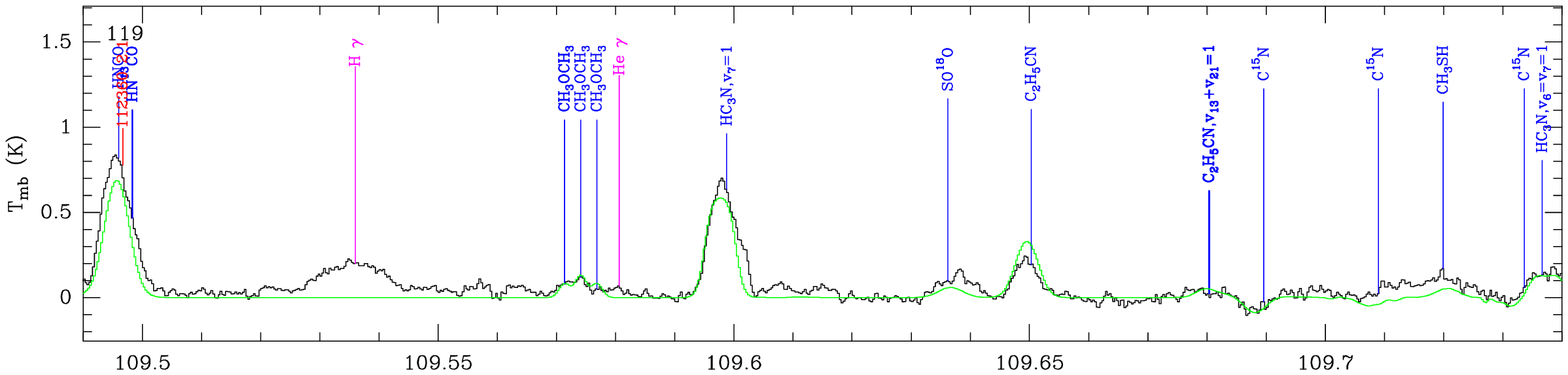}}}
\vspace*{1ex}\centerline{\resizebox{1.0\hsize}{!}{\includegraphics[angle=270]{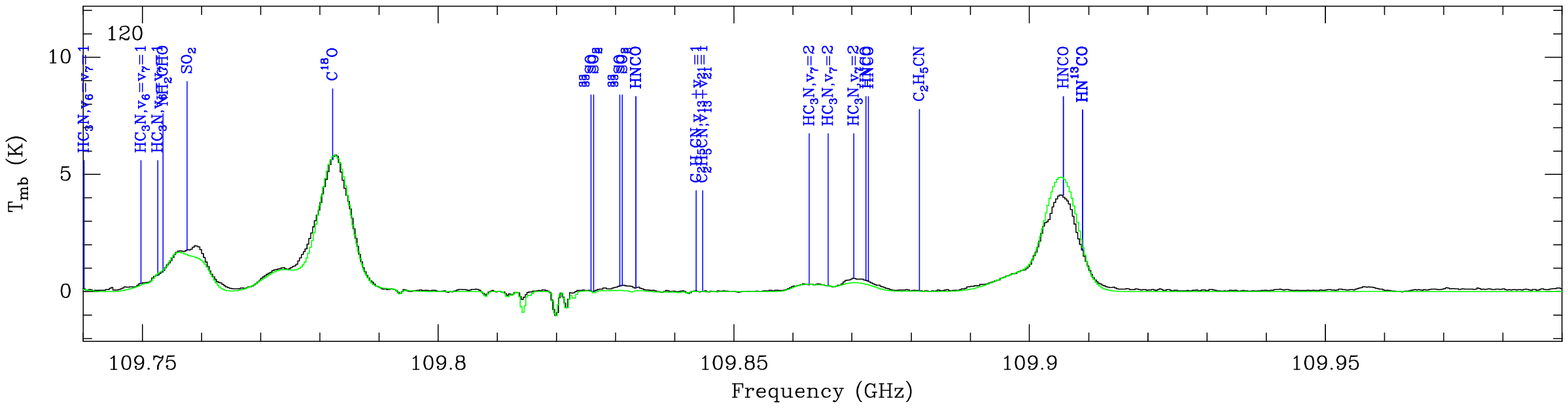}}}
\caption{
continued.
}
\end{figure*}
 \clearpage
\begin{figure*}
\addtocounter{figure}{-1}
\centerline{\resizebox{1.0\hsize}{!}{\includegraphics[angle=270]{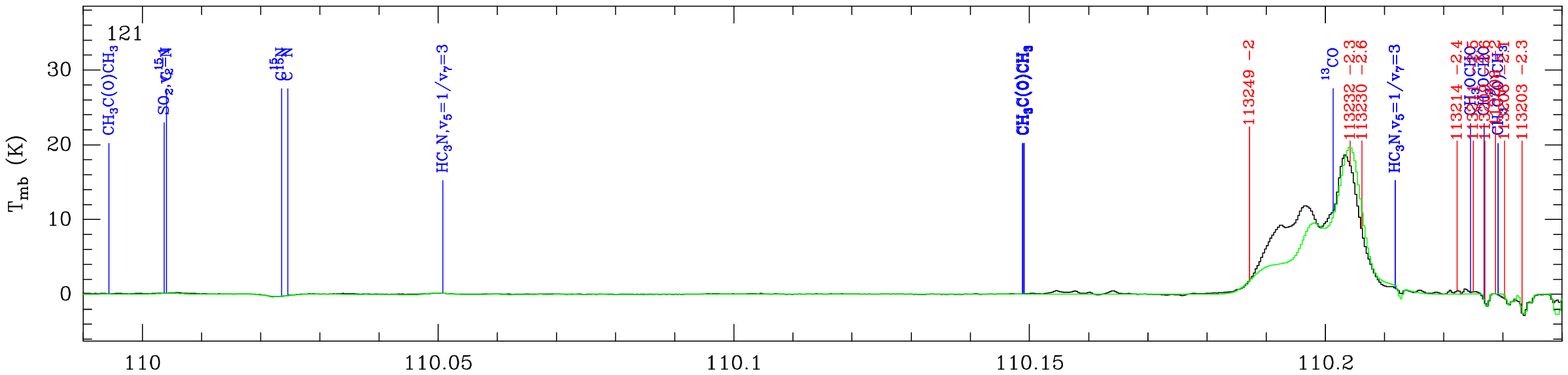}}}
\vspace*{1ex}\centerline{\resizebox{1.0\hsize}{!}{\includegraphics[angle=270]{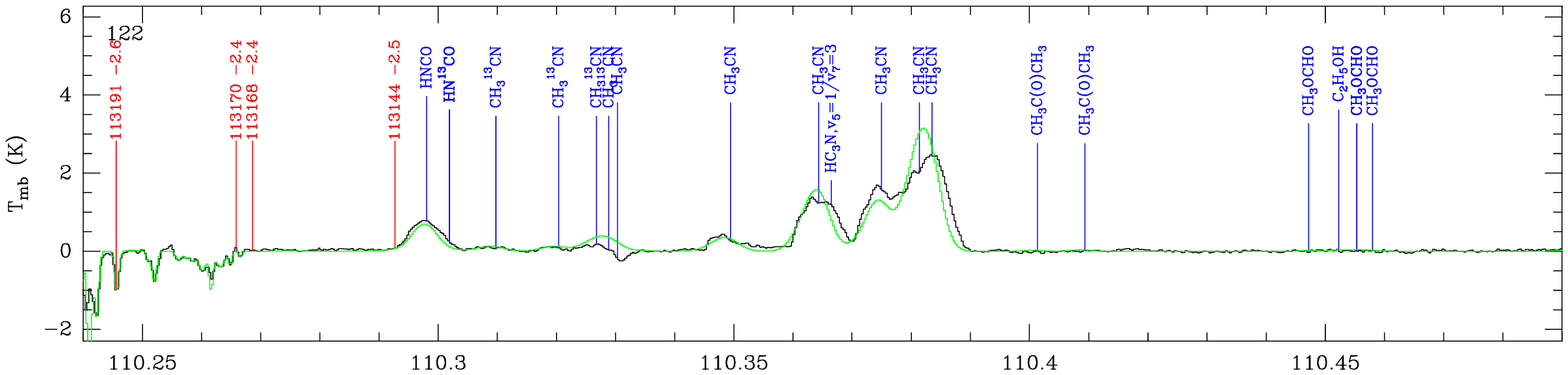}}}
\vspace*{1ex}\centerline{\resizebox{1.0\hsize}{!}{\includegraphics[angle=270]{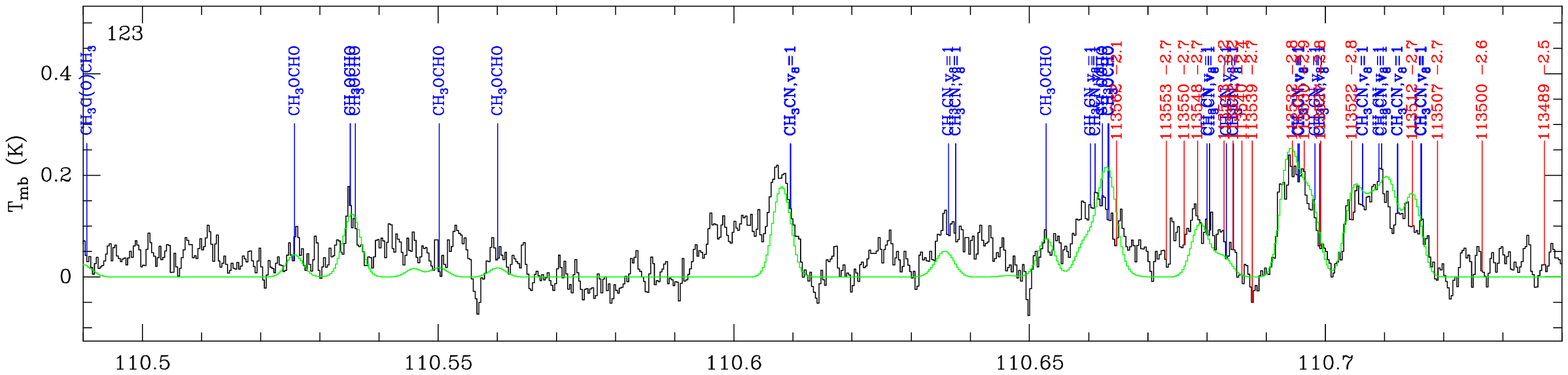}}}
\vspace*{1ex}\centerline{\resizebox{1.0\hsize}{!}{\includegraphics[angle=270]{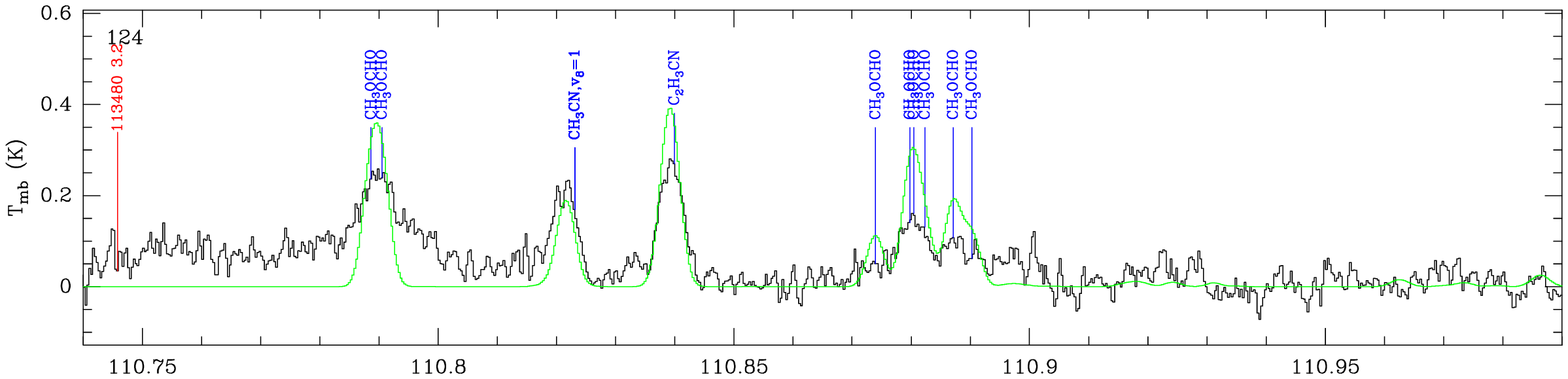}}}
\vspace*{1ex}\centerline{\resizebox{1.0\hsize}{!}{\includegraphics[angle=270]{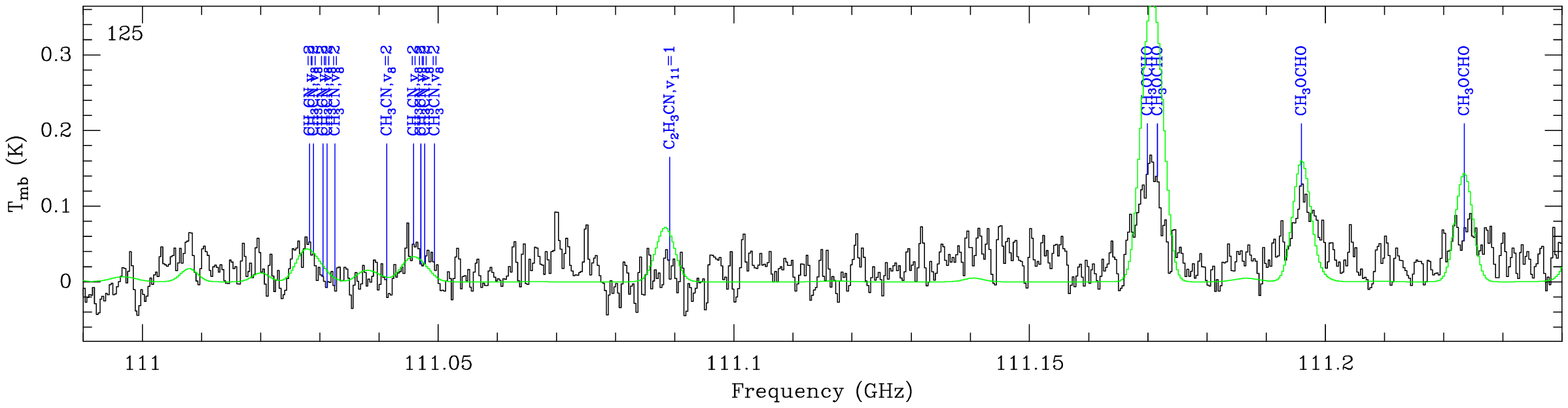}}}
\caption{
continued.
}
\end{figure*}
 \clearpage
\begin{figure*}
\addtocounter{figure}{-1}
\centerline{\resizebox{1.0\hsize}{!}{\includegraphics[angle=270]{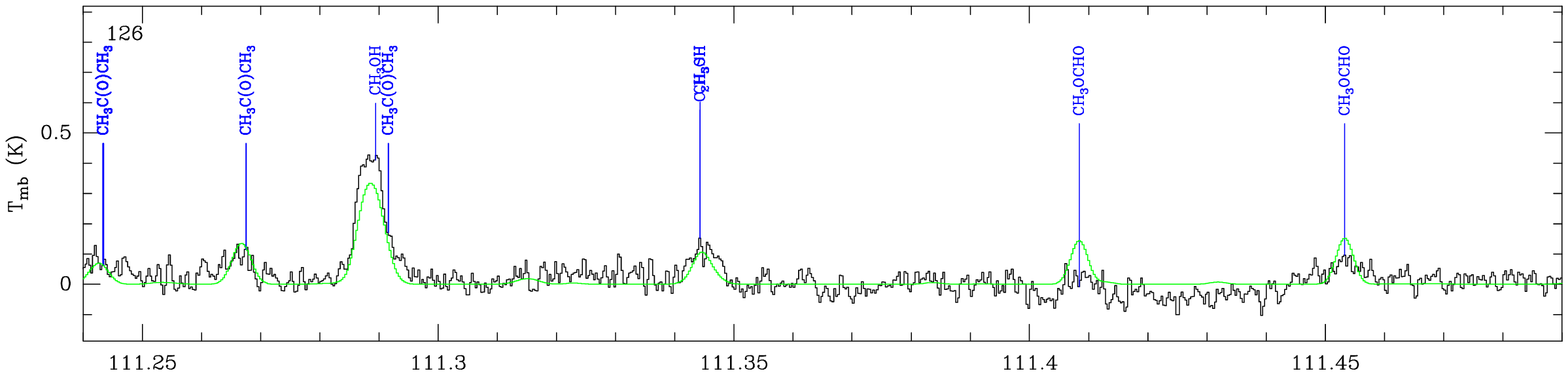}}}
\vspace*{1ex}\centerline{\resizebox{1.0\hsize}{!}{\includegraphics[angle=270]{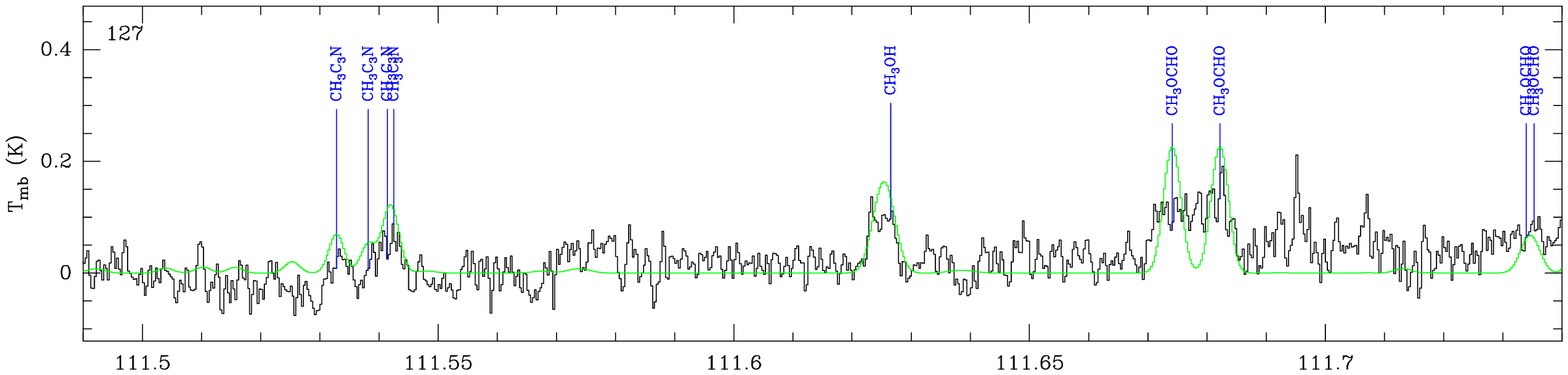}}}
\vspace*{1ex}\centerline{\resizebox{1.0\hsize}{!}{\includegraphics[angle=270]{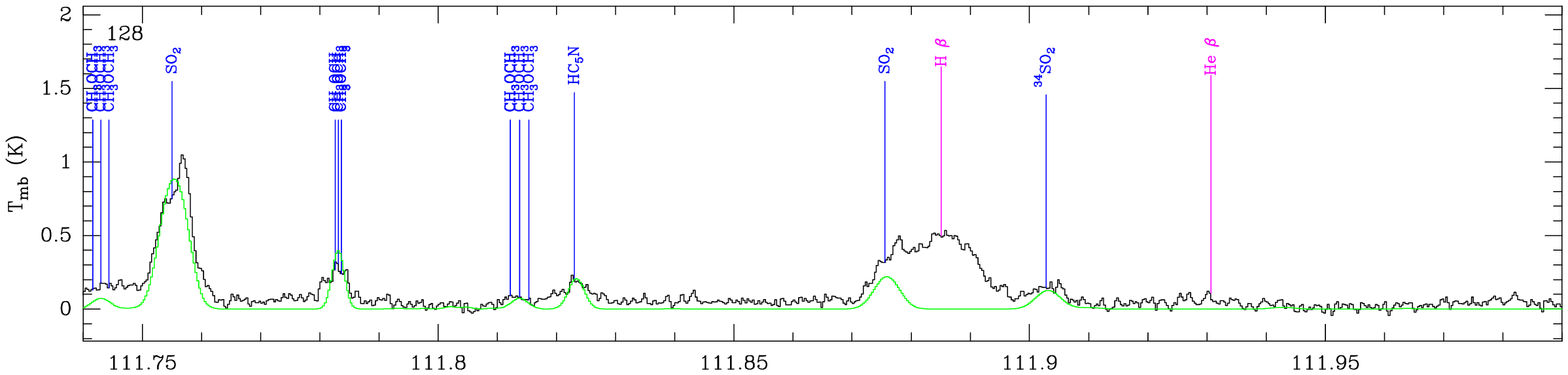}}}
\vspace*{1ex}\centerline{\resizebox{1.0\hsize}{!}{\includegraphics[angle=270]{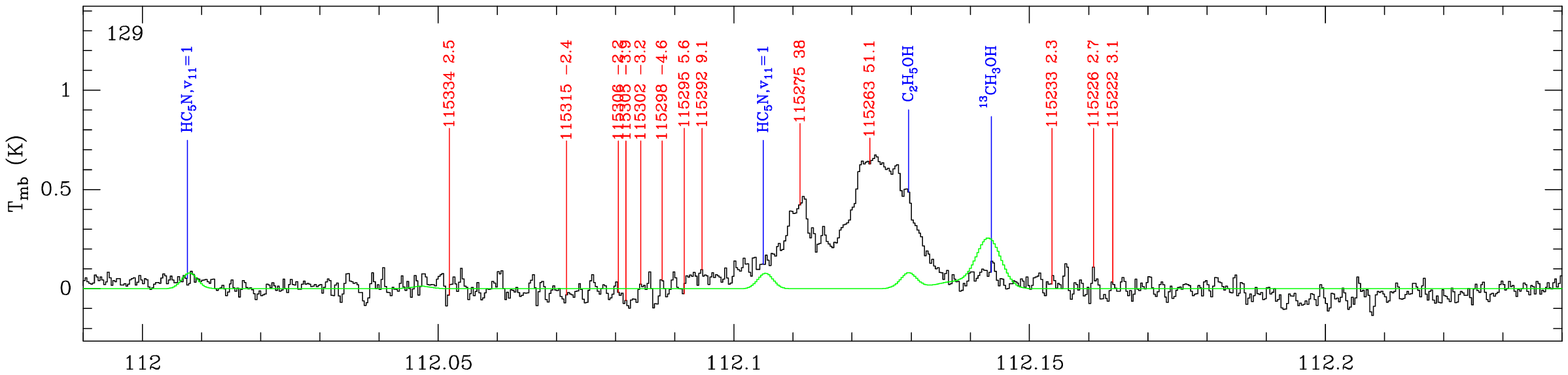}}}
\vspace*{1ex}\centerline{\resizebox{1.0\hsize}{!}{\includegraphics[angle=270]{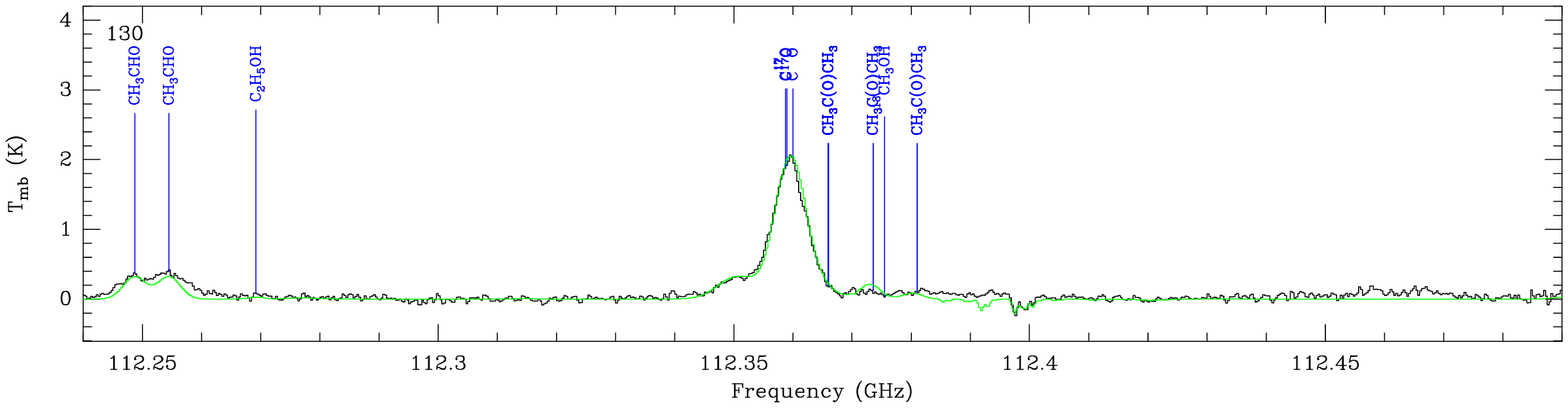}}}
\caption{
continued.
}
\end{figure*}
 \clearpage
\begin{figure*}
\addtocounter{figure}{-1}
\centerline{\resizebox{1.0\hsize}{!}{\includegraphics[angle=270]{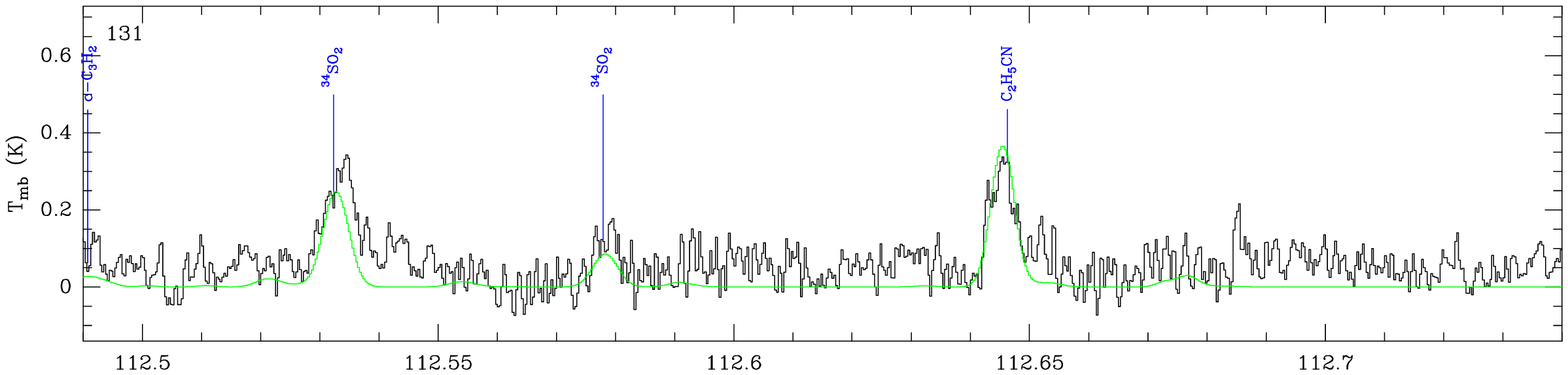}}}
\vspace*{1ex}\centerline{\resizebox{1.0\hsize}{!}{\includegraphics[angle=270]{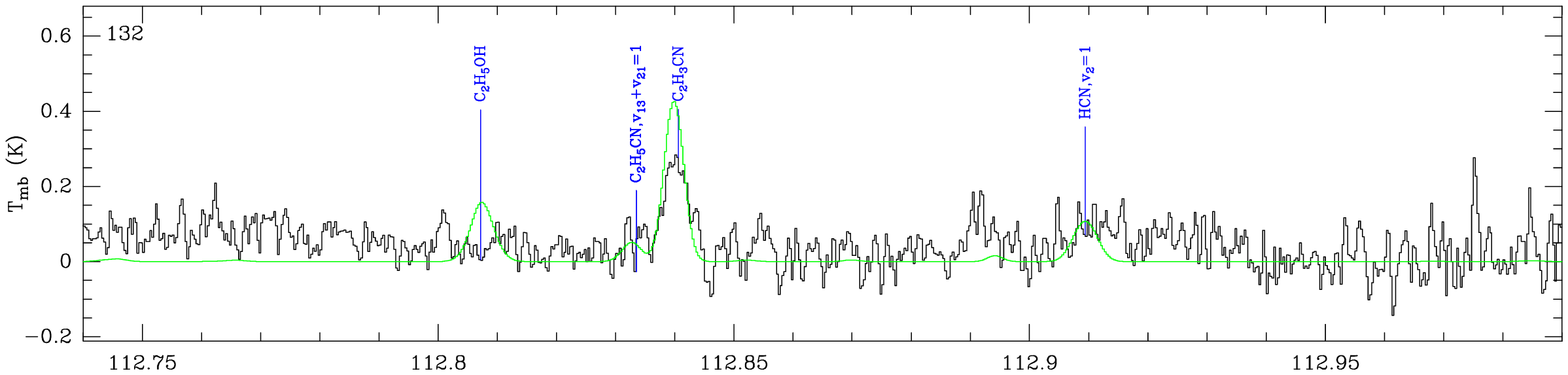}}}
\vspace*{1ex}\centerline{\resizebox{1.0\hsize}{!}{\includegraphics[angle=270]{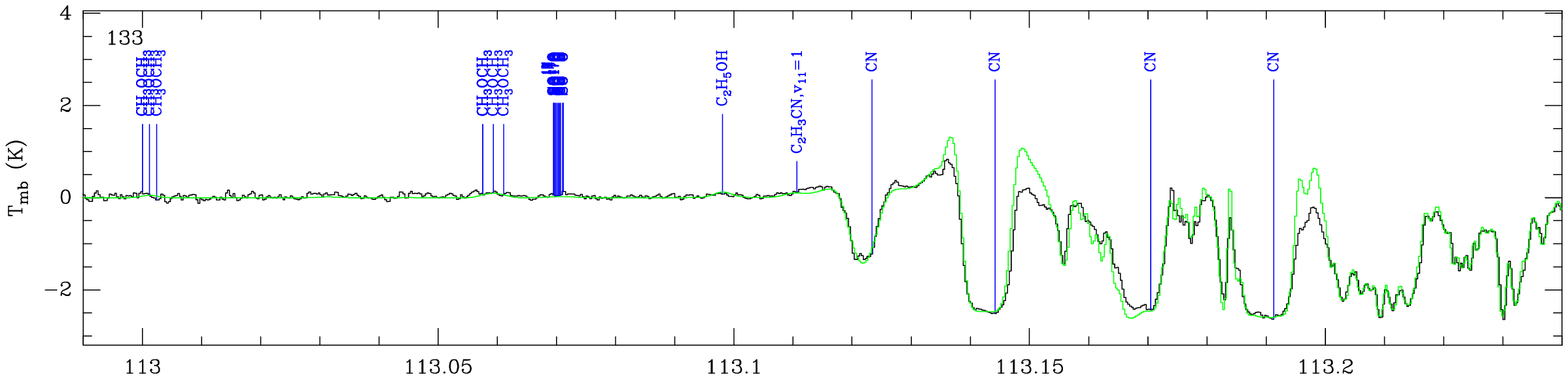}}}
\vspace*{1ex}\centerline{\resizebox{1.0\hsize}{!}{\includegraphics[angle=270]{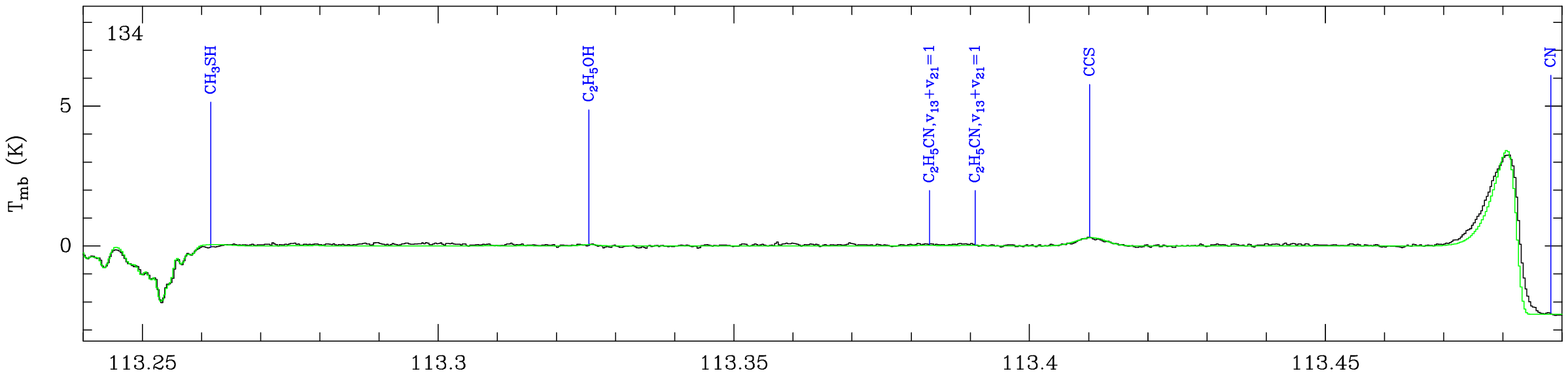}}}
\vspace*{1ex}\centerline{\resizebox{1.0\hsize}{!}{\includegraphics[angle=270]{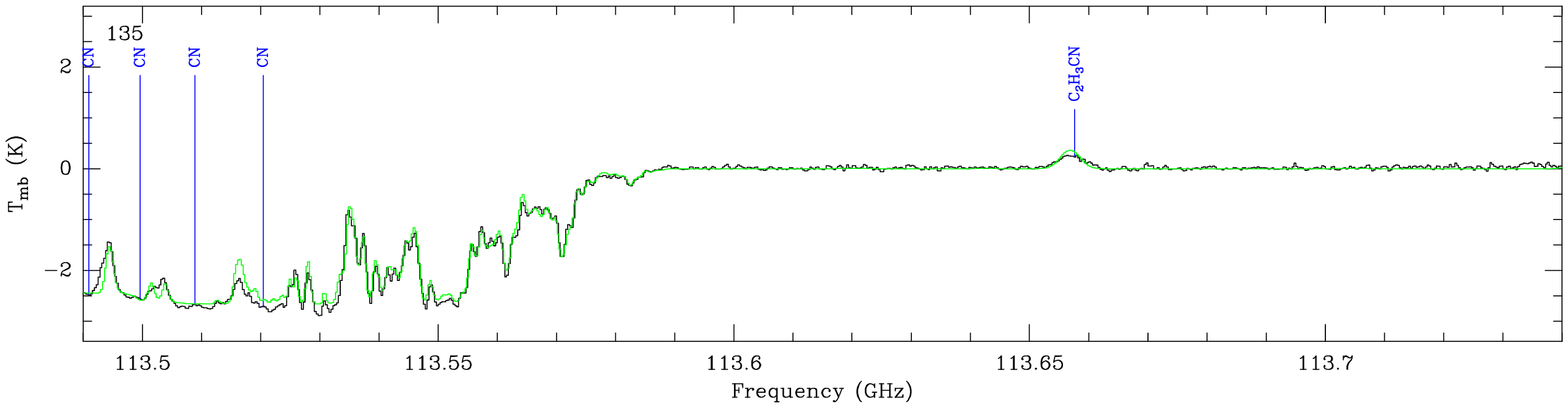}}}
\caption{
continued.
}
\end{figure*}
 \clearpage
\begin{figure*}
\addtocounter{figure}{-1}
\centerline{\resizebox{1.0\hsize}{!}{\includegraphics[angle=270]{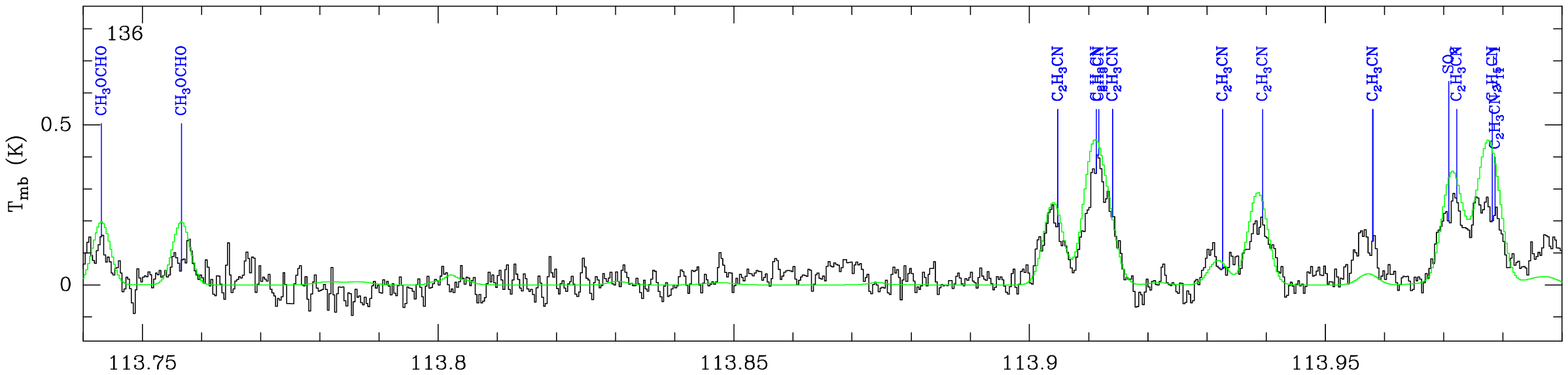}}}
\vspace*{1ex}\centerline{\resizebox{1.0\hsize}{!}{\includegraphics[angle=270]{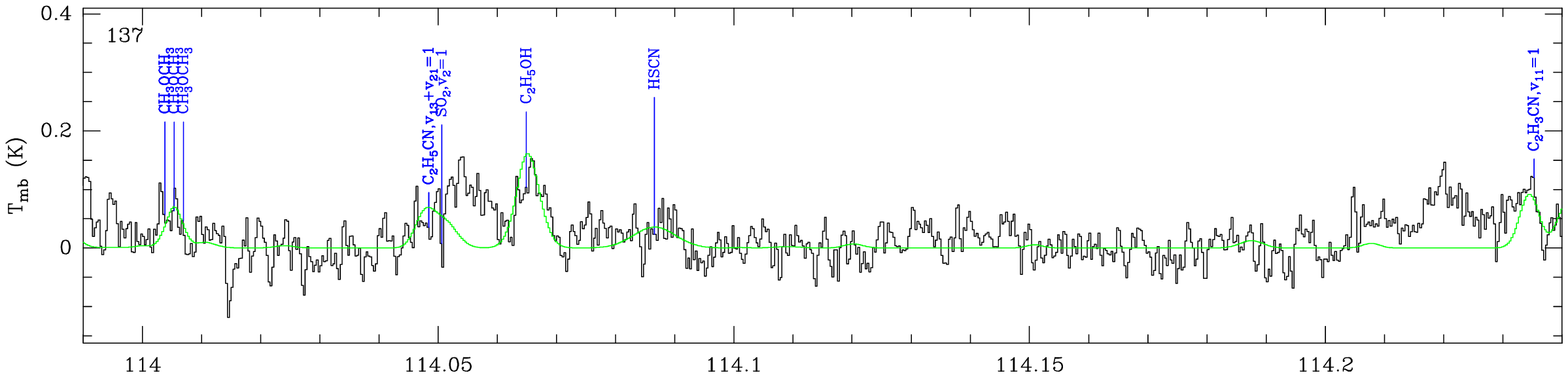}}}
\vspace*{1ex}\centerline{\resizebox{1.0\hsize}{!}{\includegraphics[angle=270]{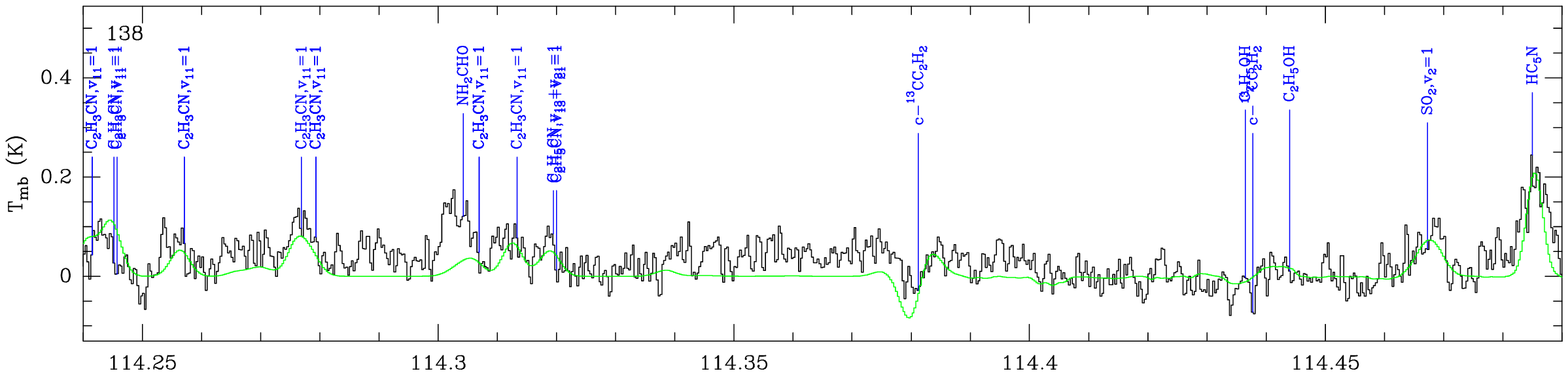}}}
\vspace*{1ex}\centerline{\resizebox{1.0\hsize}{!}{\includegraphics[angle=270]{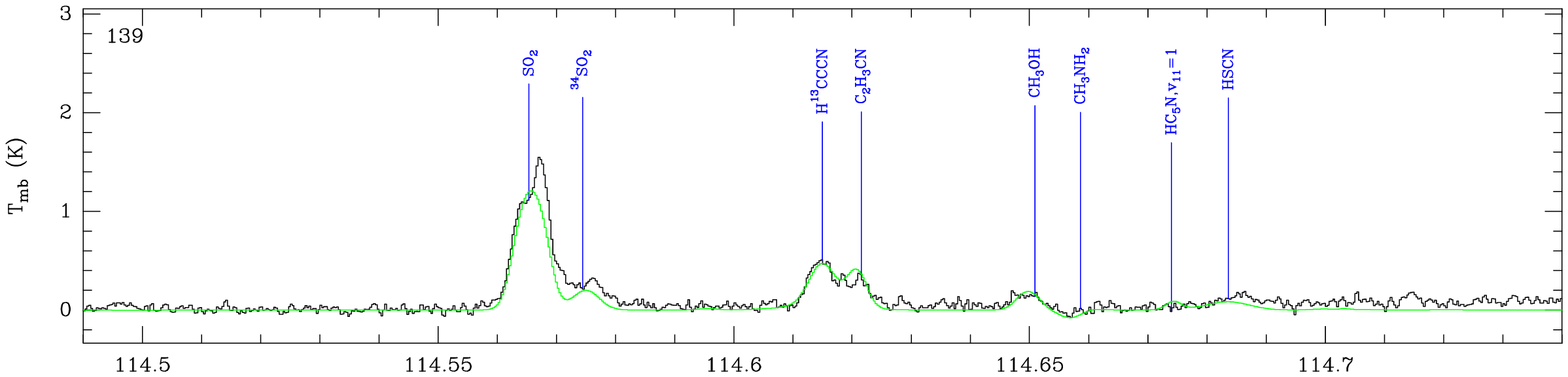}}}
\vspace*{1ex}\centerline{\resizebox{1.0\hsize}{!}{\includegraphics[angle=270]{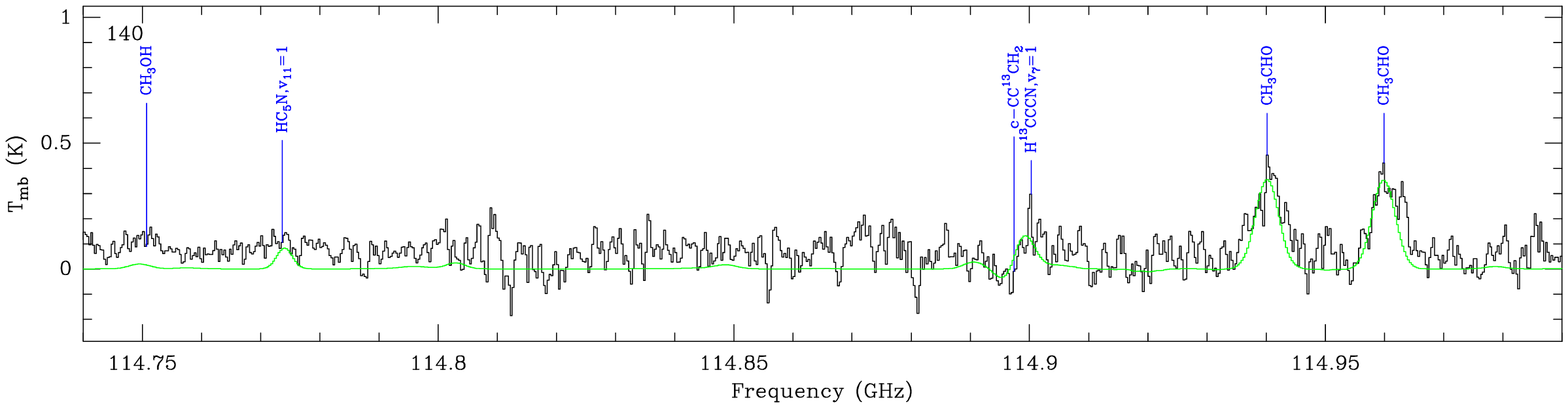}}}
\caption{
continued.
}
\end{figure*}
 \clearpage
\begin{figure*}
\addtocounter{figure}{-1}
\centerline{\resizebox{1.0\hsize}{!}{\includegraphics[angle=270]{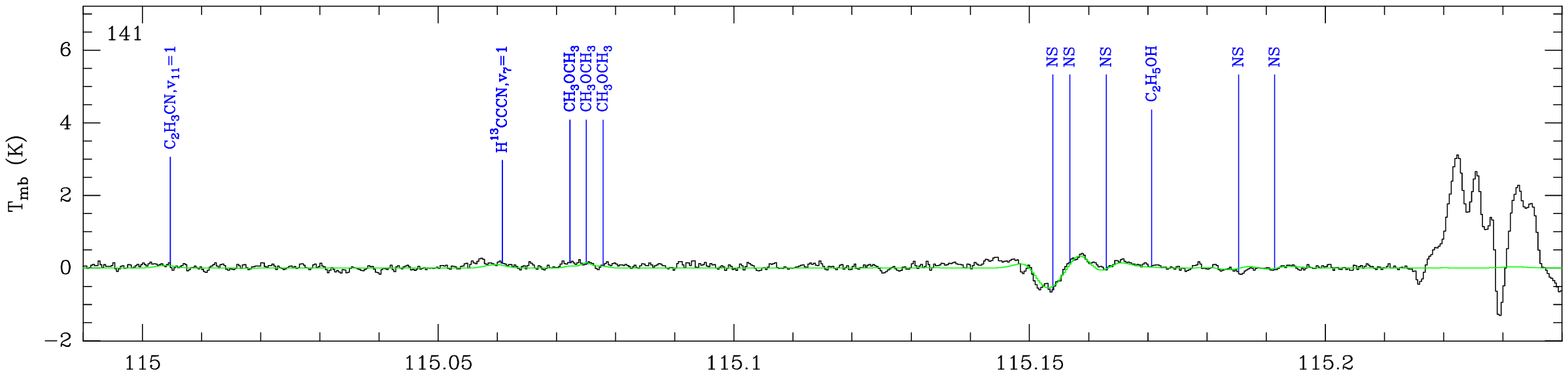}}}
\vspace*{1ex}\centerline{\resizebox{1.0\hsize}{!}{\includegraphics[angle=270]{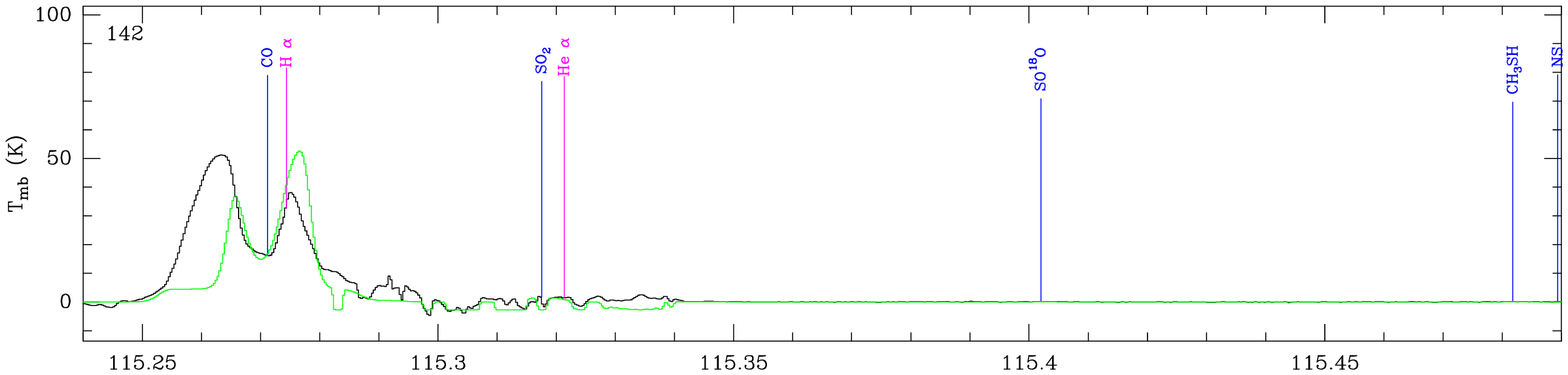}}}
\vspace*{1ex}\centerline{\resizebox{1.0\hsize}{!}{\includegraphics[angle=270]{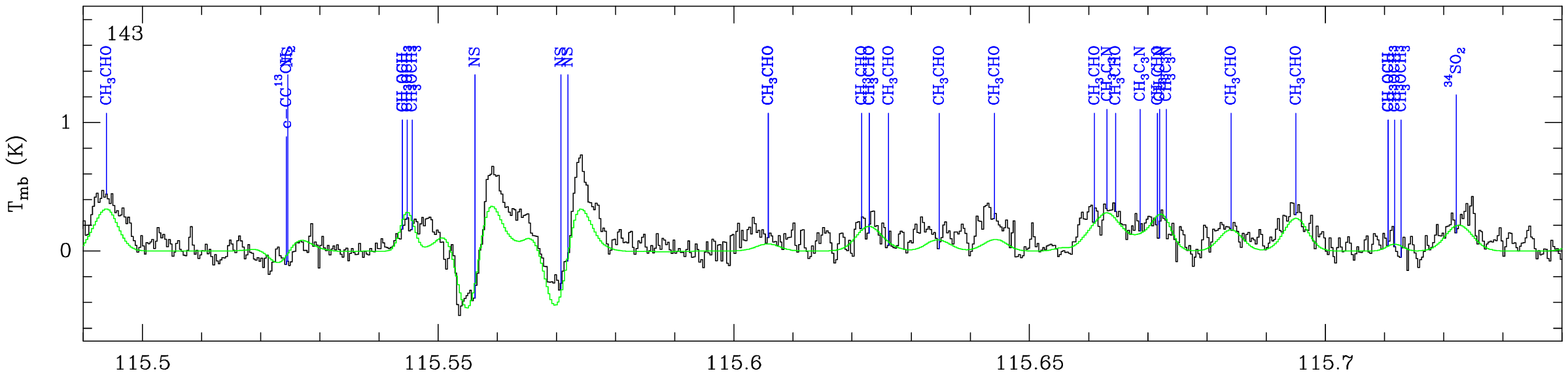}}}
\vspace*{1ex}\centerline{\resizebox{1.0\hsize}{!}{\includegraphics[angle=270]{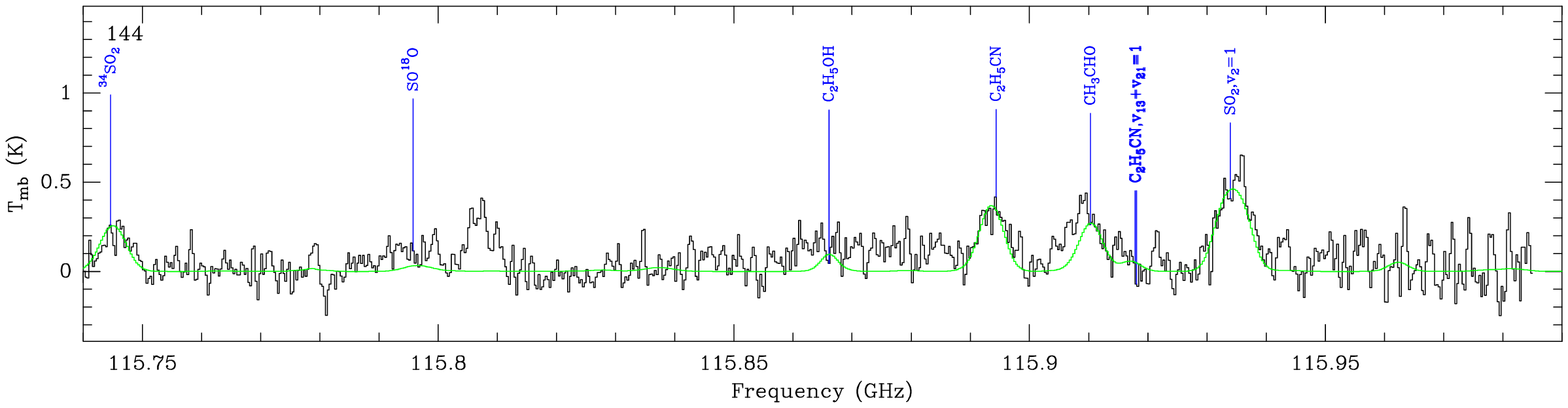}}}
\caption{
continued.
}
\end{figure*}
 \clearpage

}
\onlfig{\clearpage
\begin{figure*}
\centerline{\resizebox{1.0\hsize}{!}{\includegraphics[angle=270]{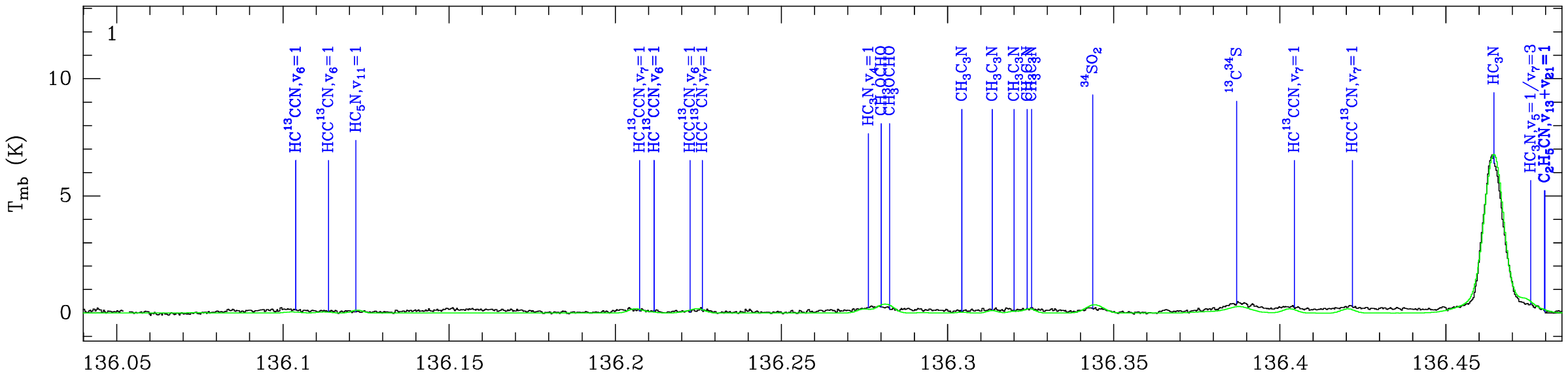}}}
\vspace*{1ex}\centerline{\resizebox{1.0\hsize}{!}{\includegraphics[angle=270]{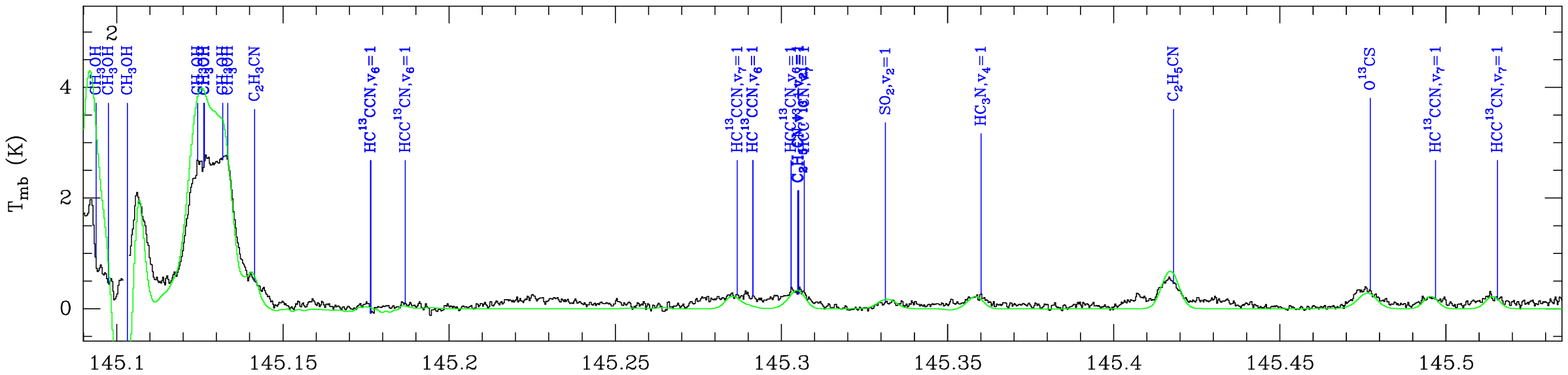}}}
\vspace*{1ex}\centerline{\resizebox{1.0\hsize}{!}{\includegraphics[angle=270]{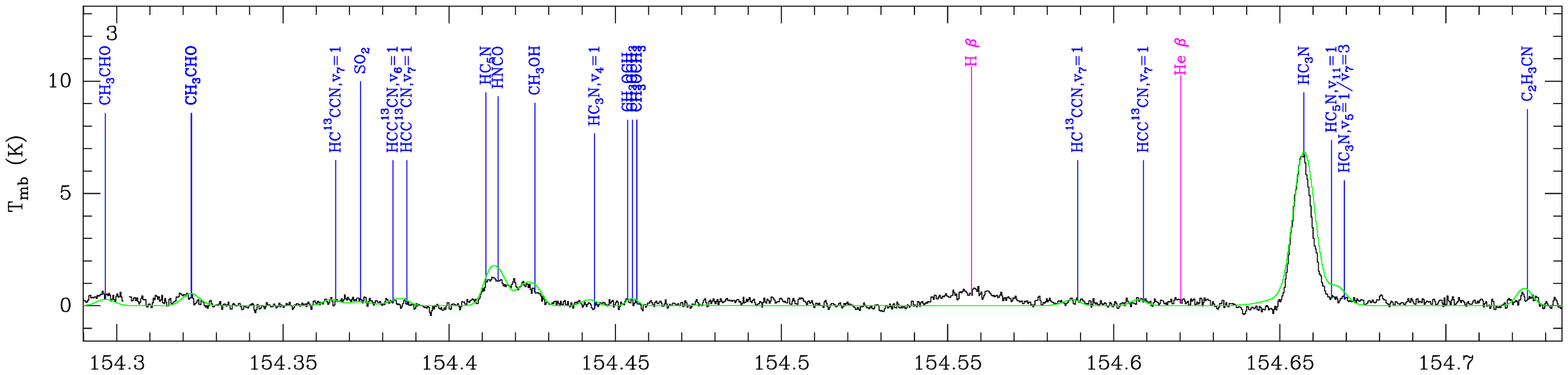}}}
\vspace*{1ex}\centerline{\resizebox{1.0\hsize}{!}{\includegraphics[angle=270]{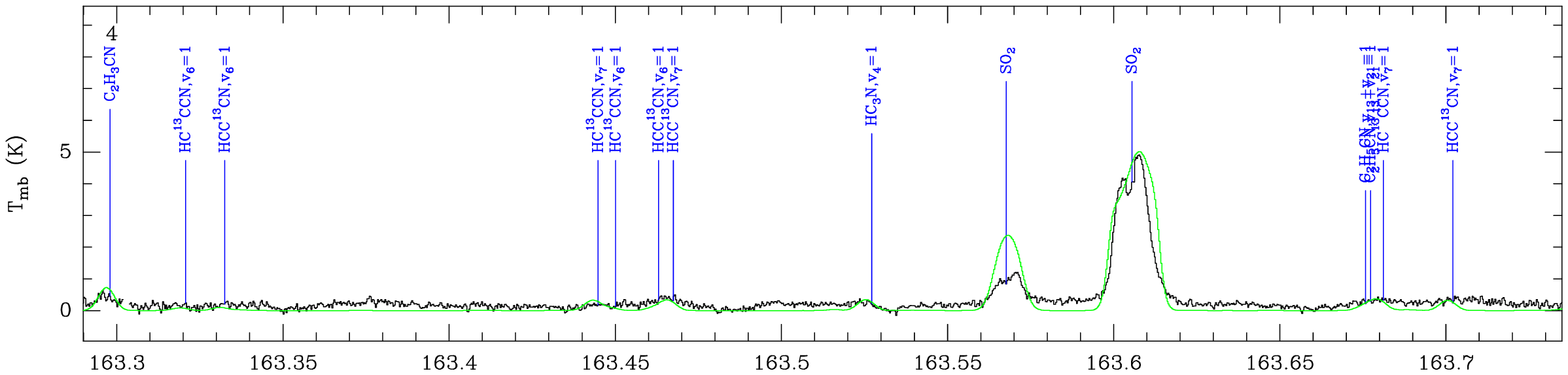}}}
\vspace*{1ex}\centerline{\resizebox{1.0\hsize}{!}{\includegraphics[angle=270]{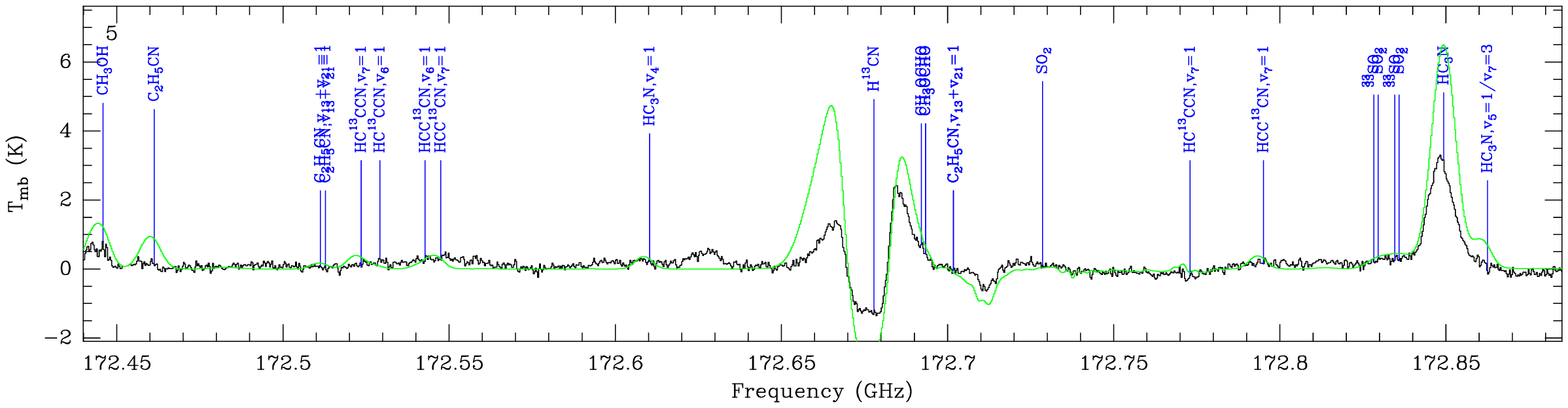}}}
\caption{
Spectrum obtained toward Sgr~B2(M)
in the 2~mm window
with the IRAM~30\,m telescope in main-beam temperature scale. The synthetic model is overlaid in green and its relevant lines are labeled in blue.
The frequencies of the hydrogen and helium recombination lines are indicated with a pink label.
}
\label{f:survey_b2m_2mm}
\end{figure*}
 \clearpage
\begin{figure*}
\addtocounter{figure}{-1}
\centerline{\resizebox{1.0\hsize}{!}{\includegraphics[angle=270]{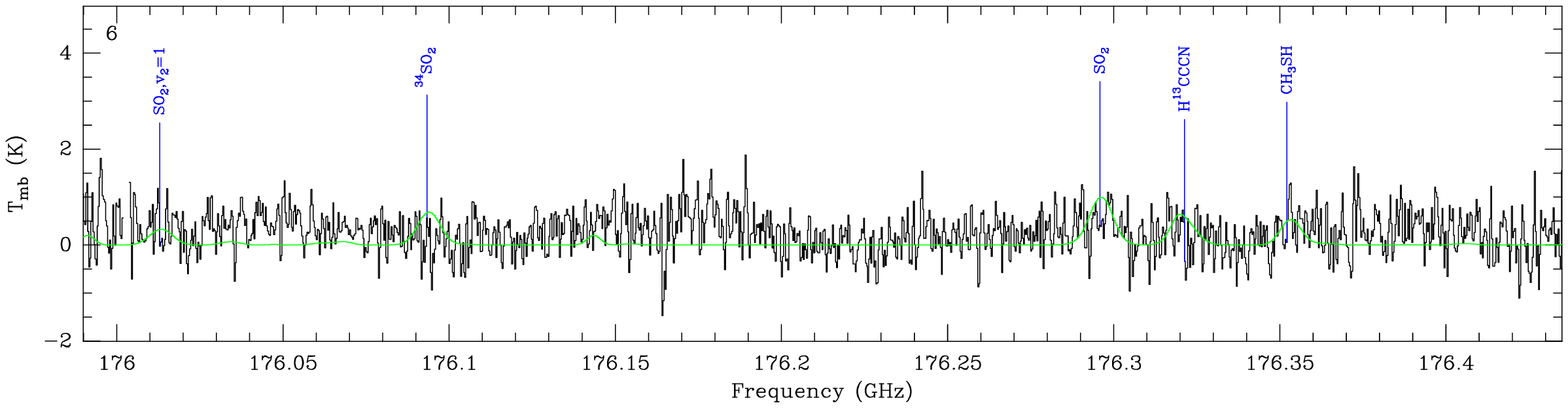}}}
\caption{
continued.
}
\end{figure*}
 \clearpage

}
\onlfig{\clearpage
\begin{figure*}
\centerline{\resizebox{1.0\hsize}{!}{\includegraphics[angle=270]{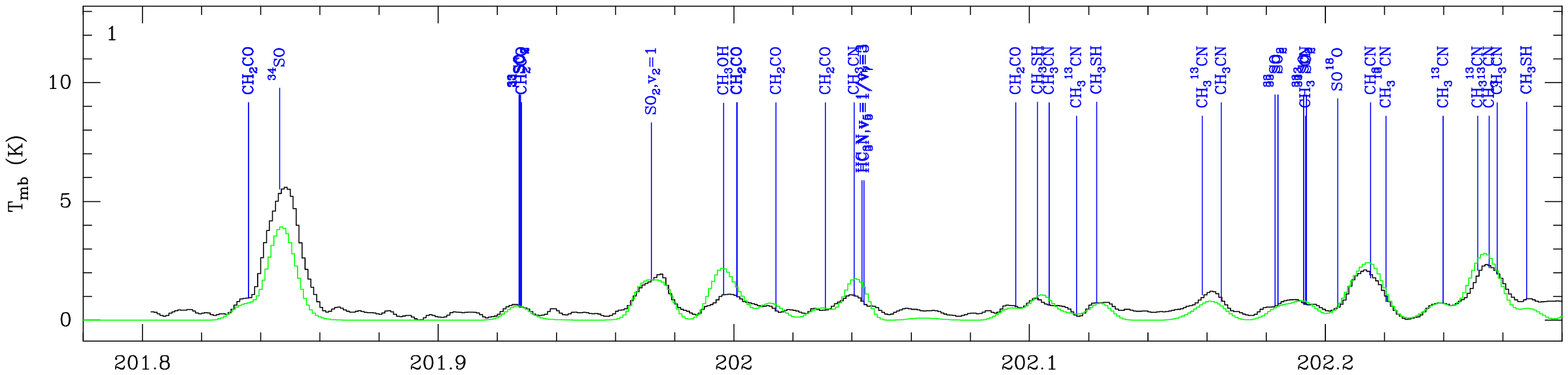}}}
\vspace*{1ex}\centerline{\resizebox{1.0\hsize}{!}{\includegraphics[angle=270]{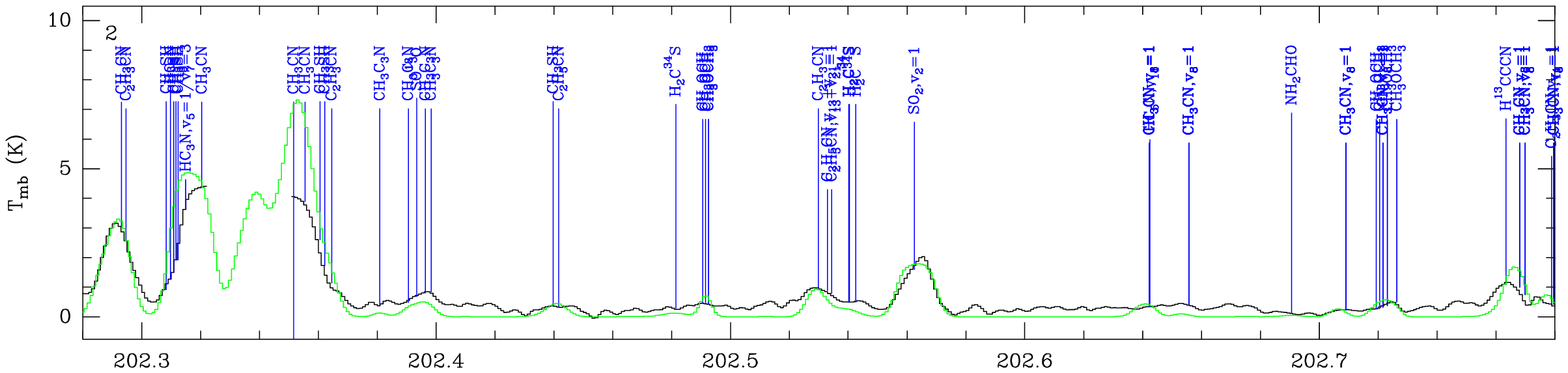}}}
\vspace*{1ex}\centerline{\resizebox{1.0\hsize}{!}{\includegraphics[angle=270]{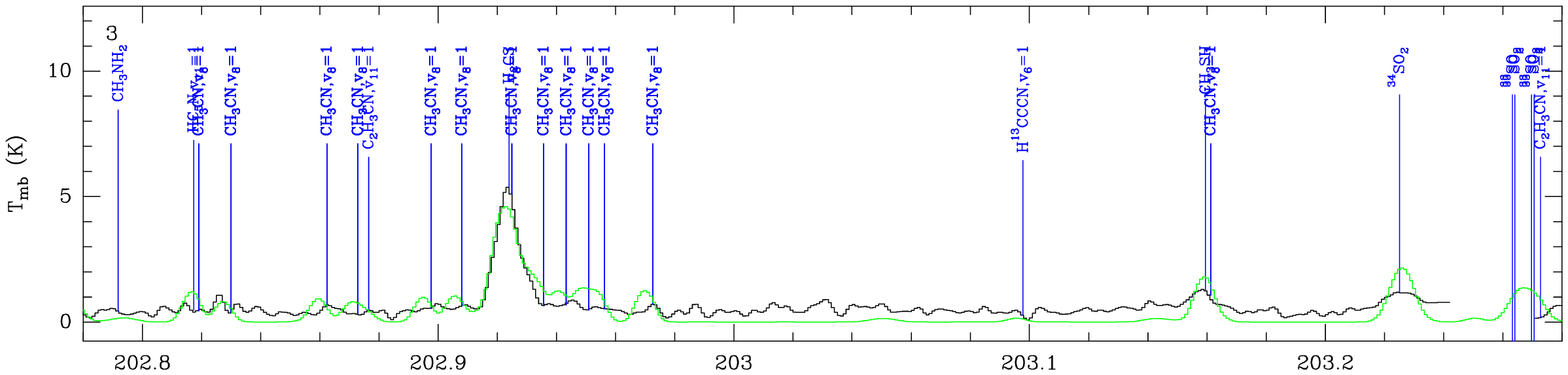}}}
\vspace*{1ex}\centerline{\resizebox{1.0\hsize}{!}{\includegraphics[angle=270]{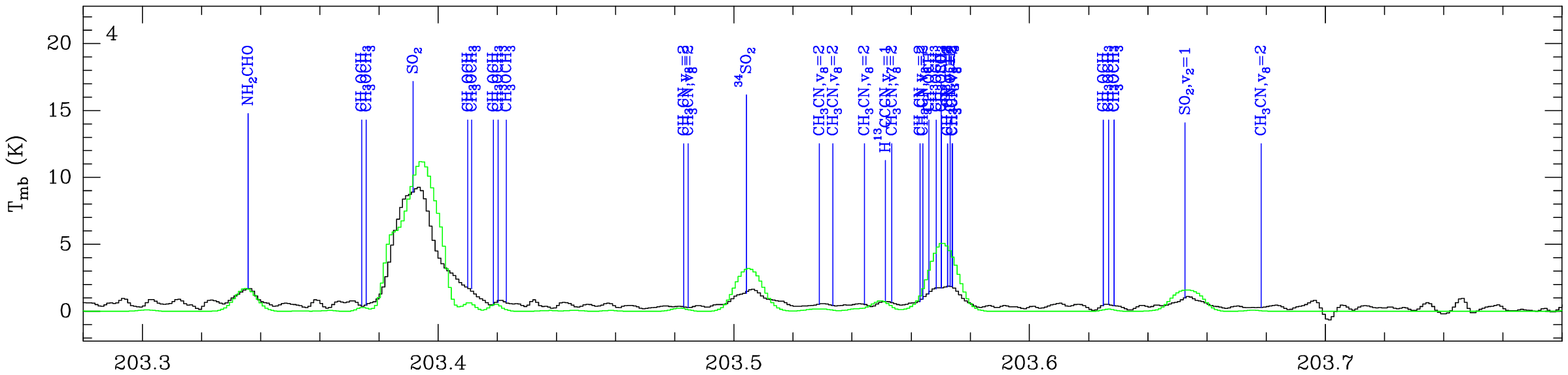}}}
\vspace*{1ex}\centerline{\resizebox{1.0\hsize}{!}{\includegraphics[angle=270]{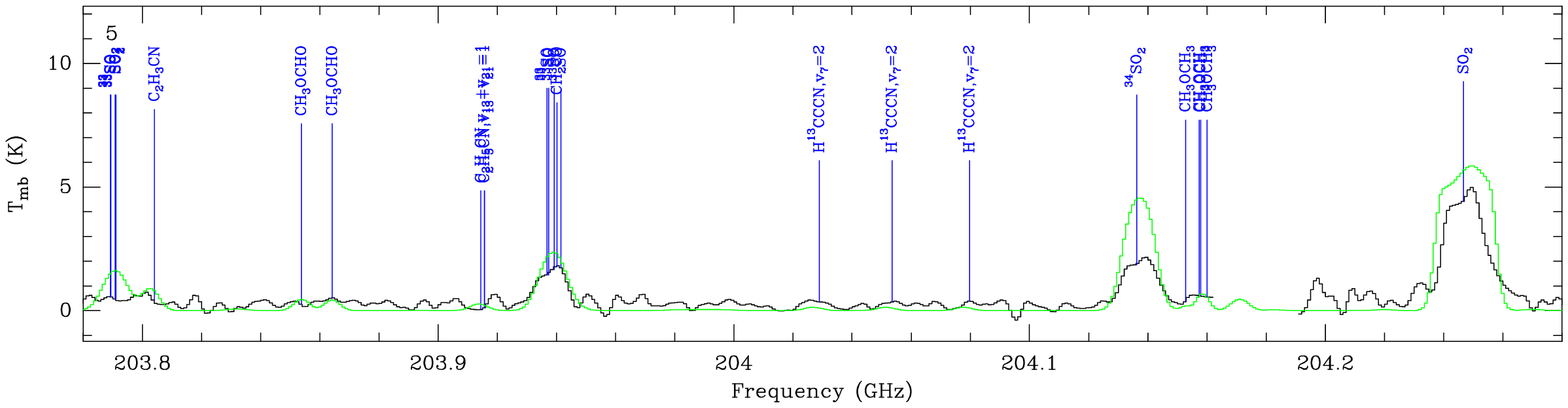}}}
\caption{
Spectrum obtained toward Sgr~B2(M)
in the 1~mm window
with the IRAM~30\,m telescope in main-beam temperature scale. The synthetic model is overlaid in green and its relevant lines are labeled in blue.
The frequencies of the hydrogen and helium recombination lines are indicated with a pink label.
}
\label{f:survey_b2m_1mm}
\end{figure*}
 \clearpage
\begin{figure*}
\addtocounter{figure}{-1}
\centerline{\resizebox{1.0\hsize}{!}{\includegraphics[angle=270]{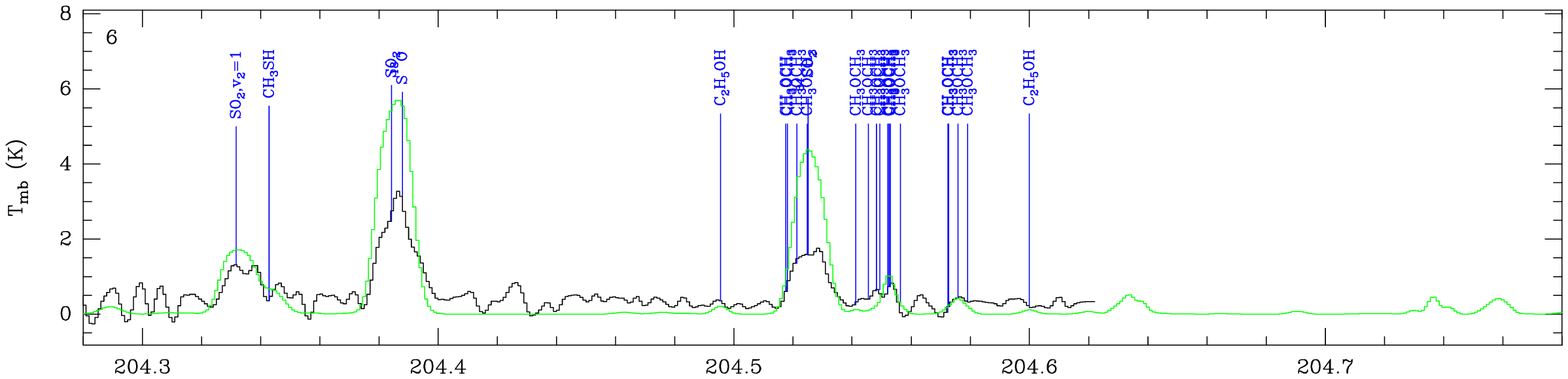}}}
\vspace*{1ex}\centerline{\resizebox{1.0\hsize}{!}{\includegraphics[angle=270]{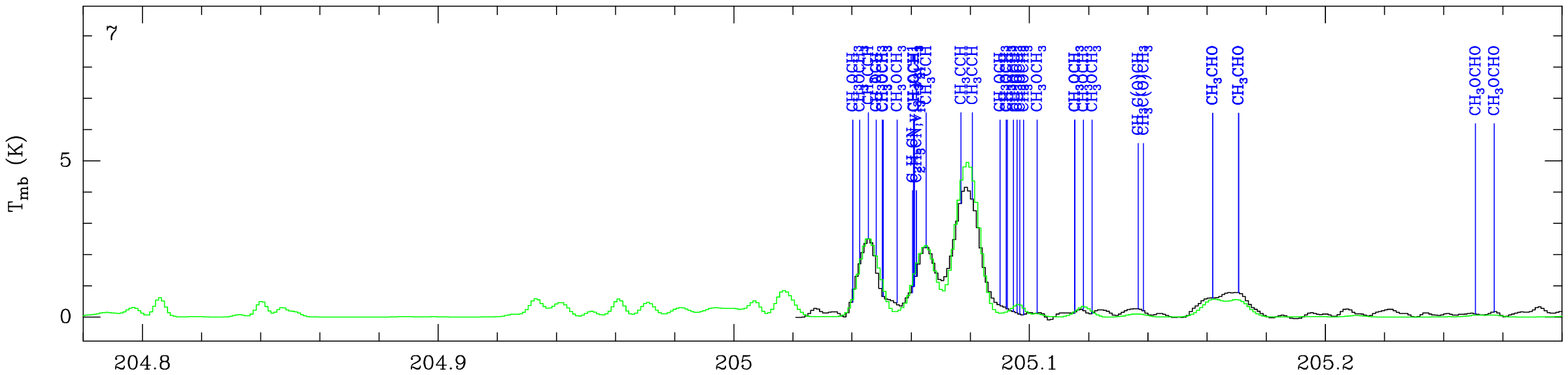}}}
\vspace*{1ex}\centerline{\resizebox{1.0\hsize}{!}{\includegraphics[angle=270]{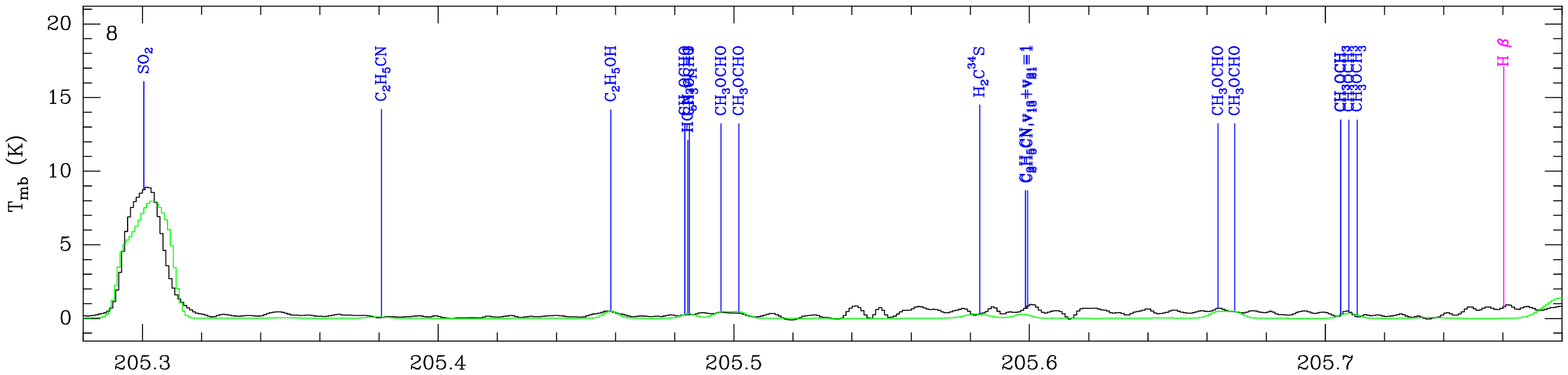}}}
\vspace*{1ex}\centerline{\resizebox{1.0\hsize}{!}{\includegraphics[angle=270]{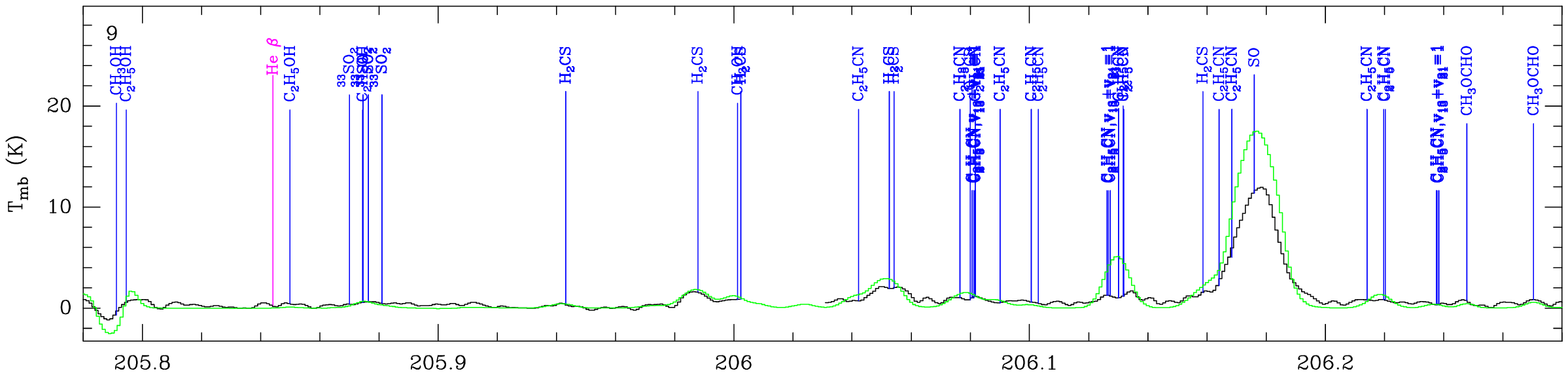}}}
\vspace*{1ex}\centerline{\resizebox{1.0\hsize}{!}{\includegraphics[angle=270]{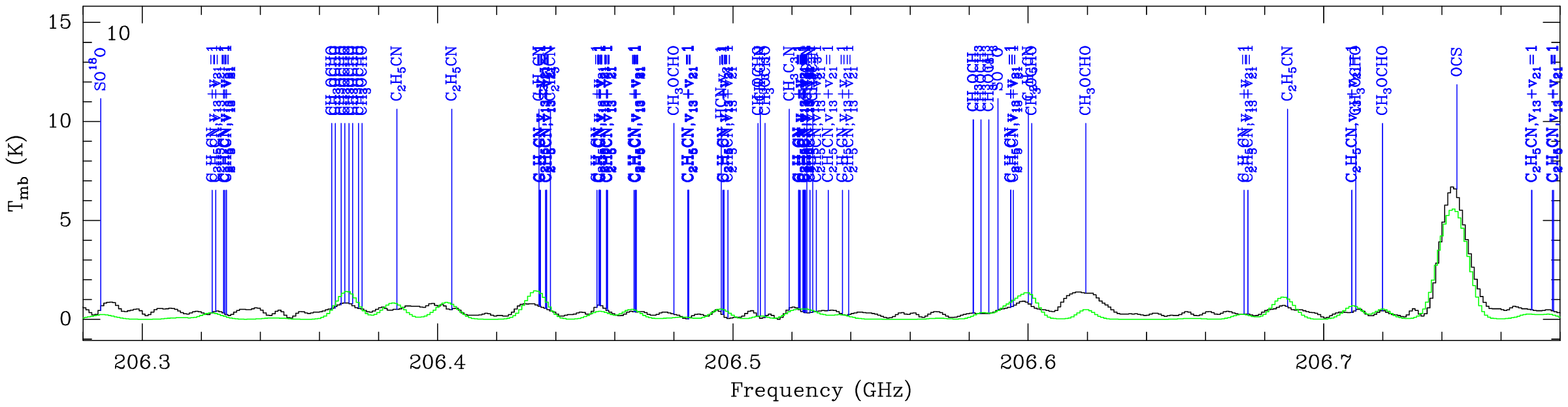}}}
\caption{
continued.
}
\end{figure*}
 \clearpage
\begin{figure*}
\addtocounter{figure}{-1}
\centerline{\resizebox{1.0\hsize}{!}{\includegraphics[angle=270]{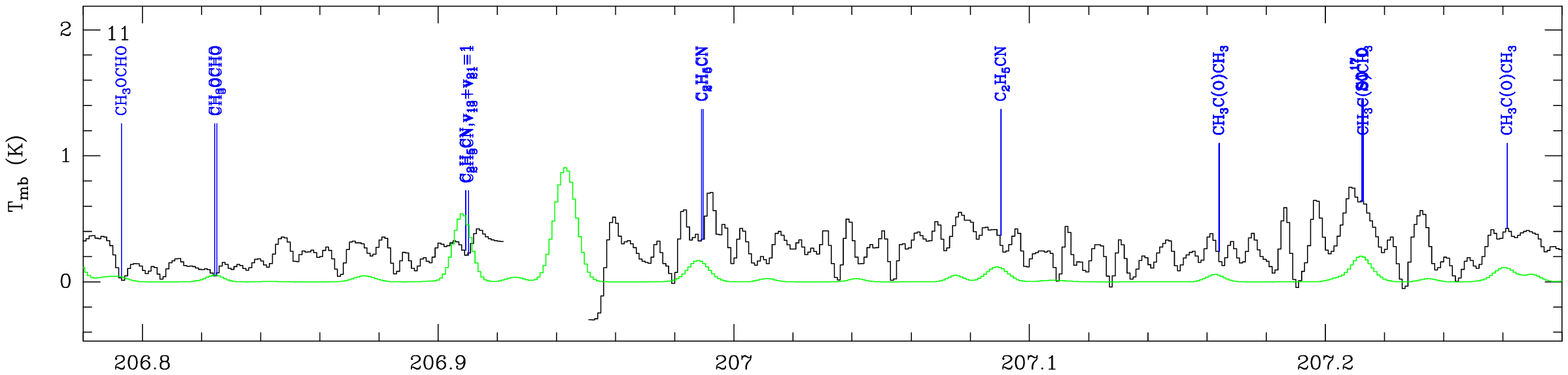}}}
\vspace*{1ex}\centerline{\resizebox{1.0\hsize}{!}{\includegraphics[angle=270]{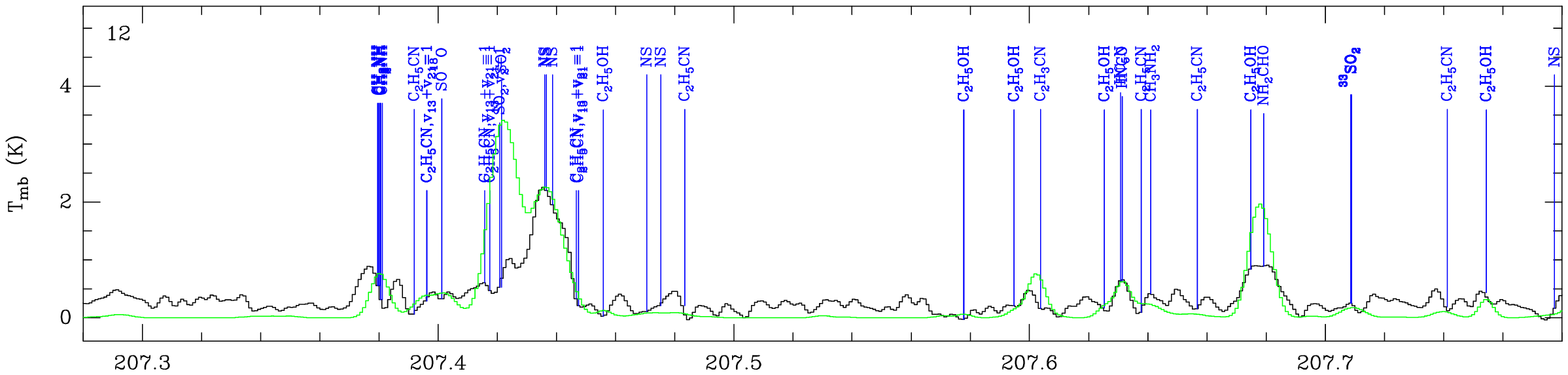}}}
\vspace*{1ex}\centerline{\resizebox{1.0\hsize}{!}{\includegraphics[angle=270]{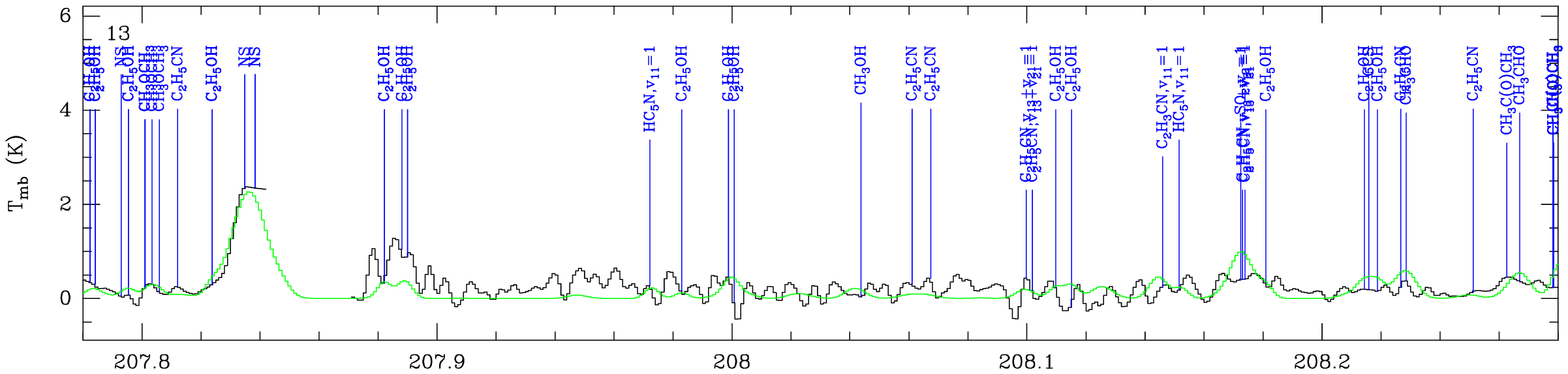}}}
\vspace*{1ex}\centerline{\resizebox{1.0\hsize}{!}{\includegraphics[angle=270]{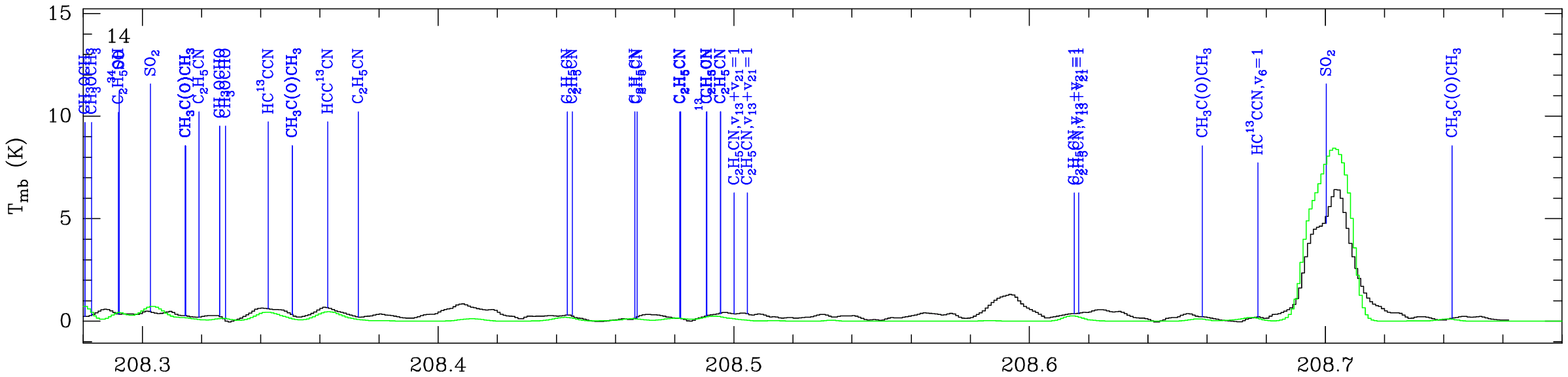}}}
\vspace*{1ex}\centerline{\resizebox{1.0\hsize}{!}{\includegraphics[angle=270]{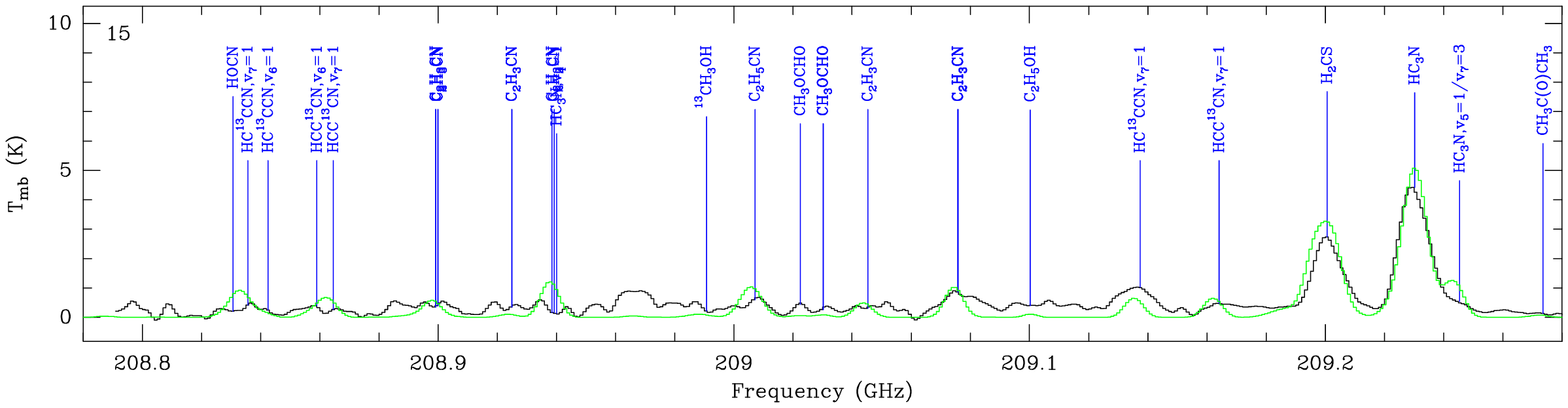}}}
\caption{
continued.
}
\end{figure*}
 \clearpage
\begin{figure*}
\addtocounter{figure}{-1}
\centerline{\resizebox{1.0\hsize}{!}{\includegraphics[angle=270]{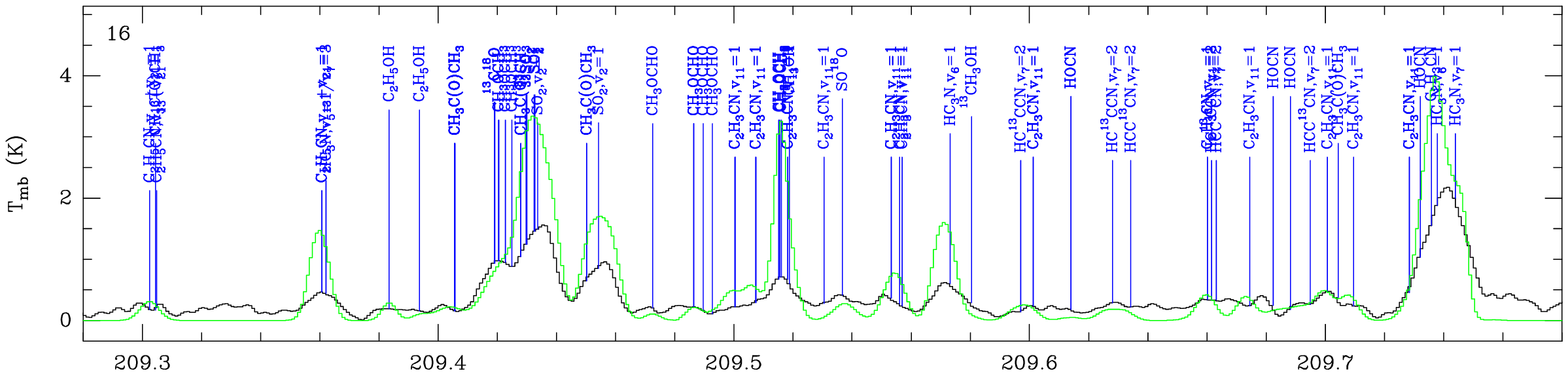}}}
\vspace*{1ex}\centerline{\resizebox{1.0\hsize}{!}{\includegraphics[angle=270]{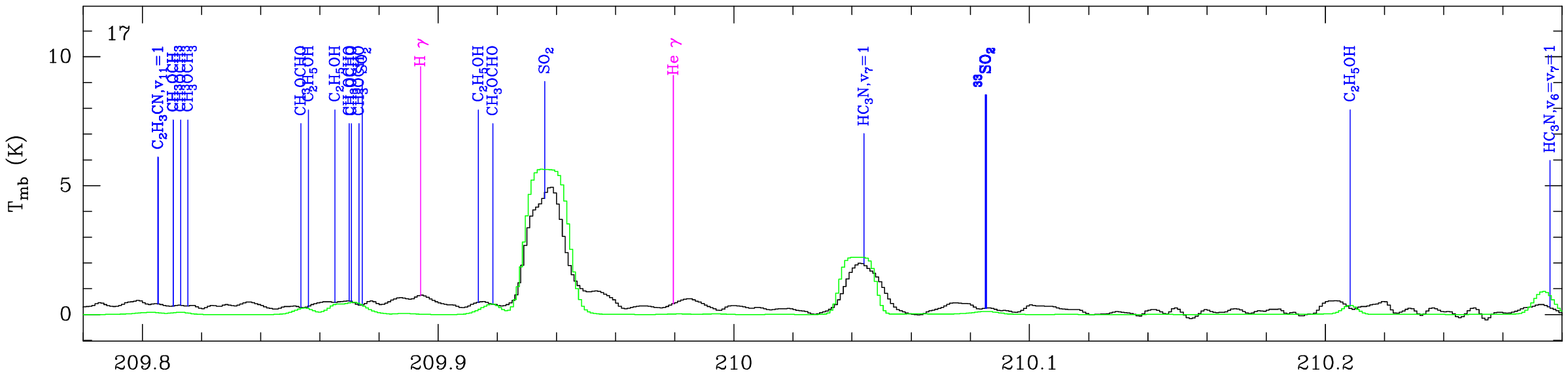}}}
\vspace*{1ex}\centerline{\resizebox{1.0\hsize}{!}{\includegraphics[angle=270]{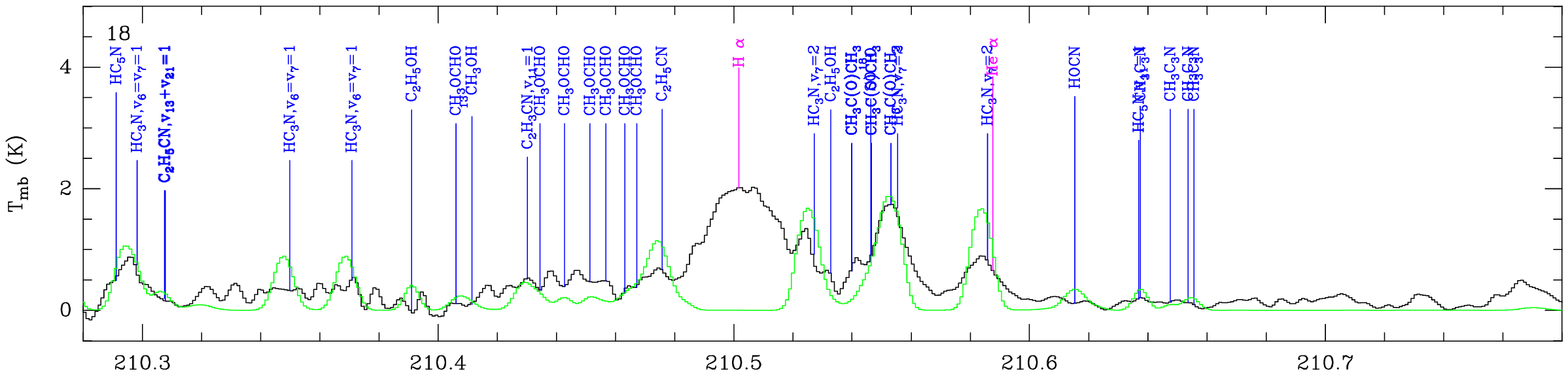}}}
\vspace*{1ex}\centerline{\resizebox{1.0\hsize}{!}{\includegraphics[angle=270]{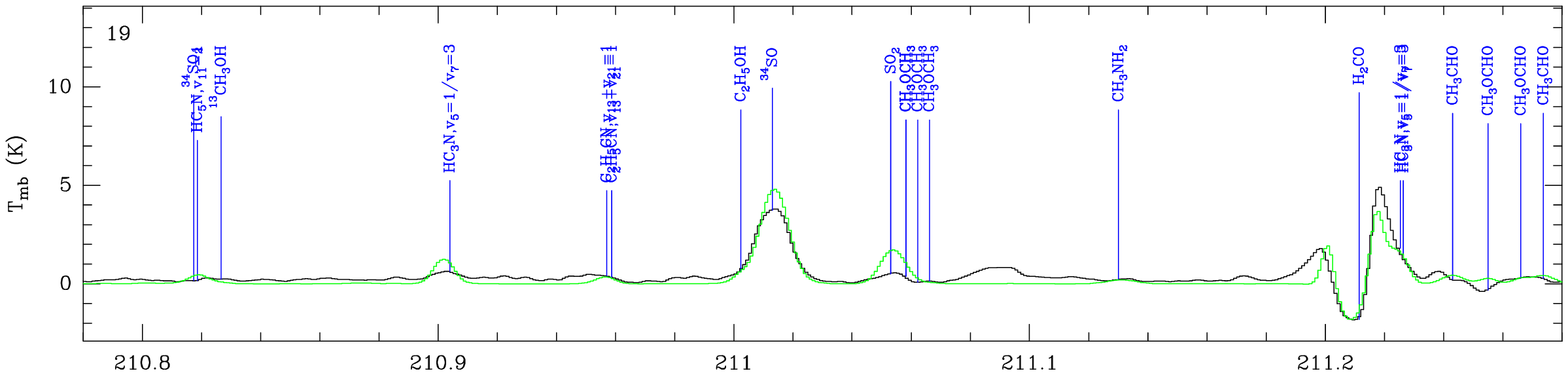}}}
\vspace*{1ex}\centerline{\resizebox{1.0\hsize}{!}{\includegraphics[angle=270]{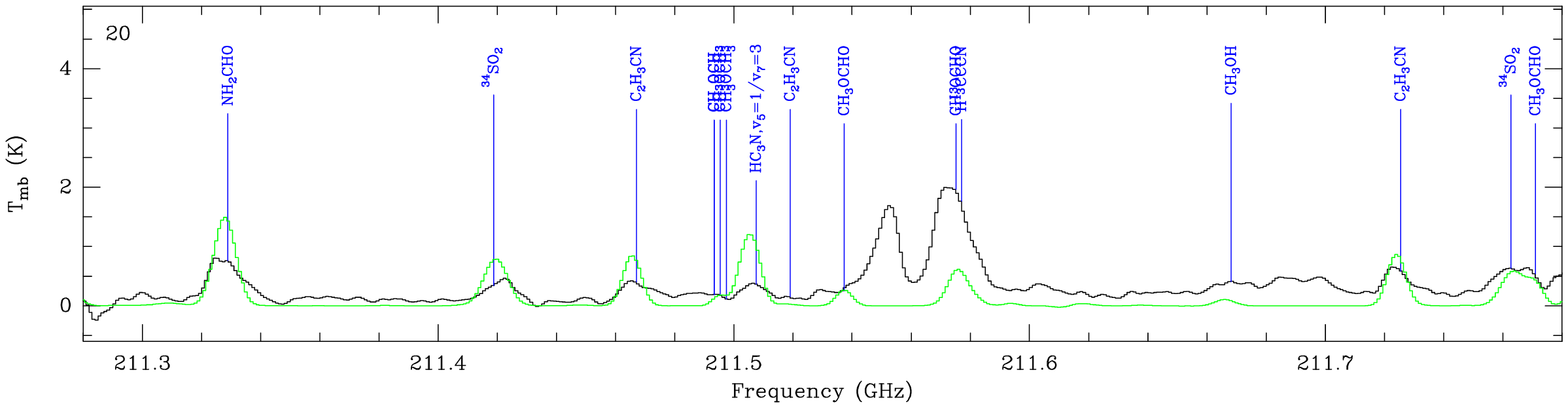}}}
\caption{
continued.
}
\end{figure*}
 \clearpage
\begin{figure*}
\addtocounter{figure}{-1}
\centerline{\resizebox{1.0\hsize}{!}{\includegraphics[angle=270]{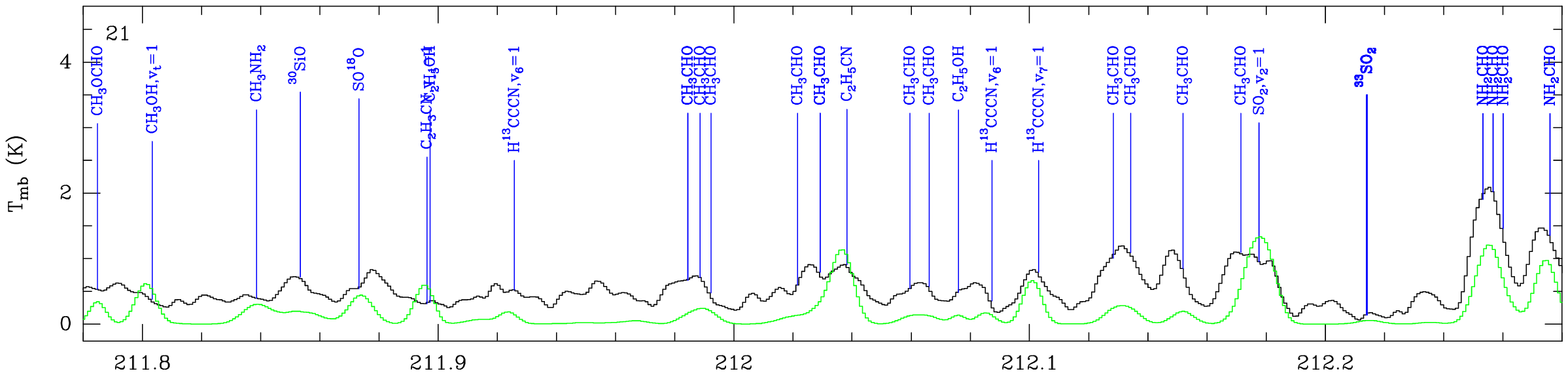}}}
\vspace*{1ex}\centerline{\resizebox{1.0\hsize}{!}{\includegraphics[angle=270]{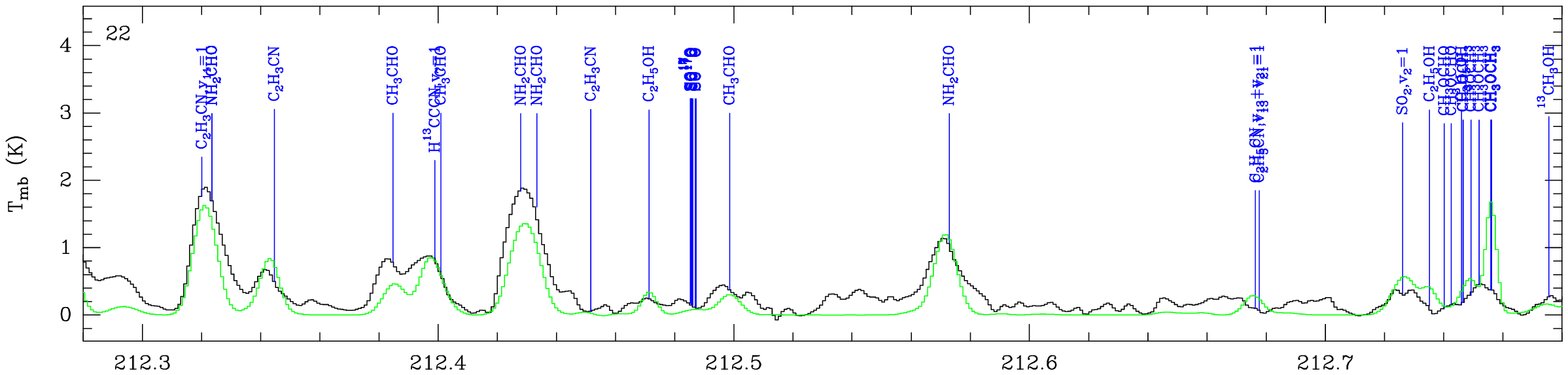}}}
\vspace*{1ex}\centerline{\resizebox{1.0\hsize}{!}{\includegraphics[angle=270]{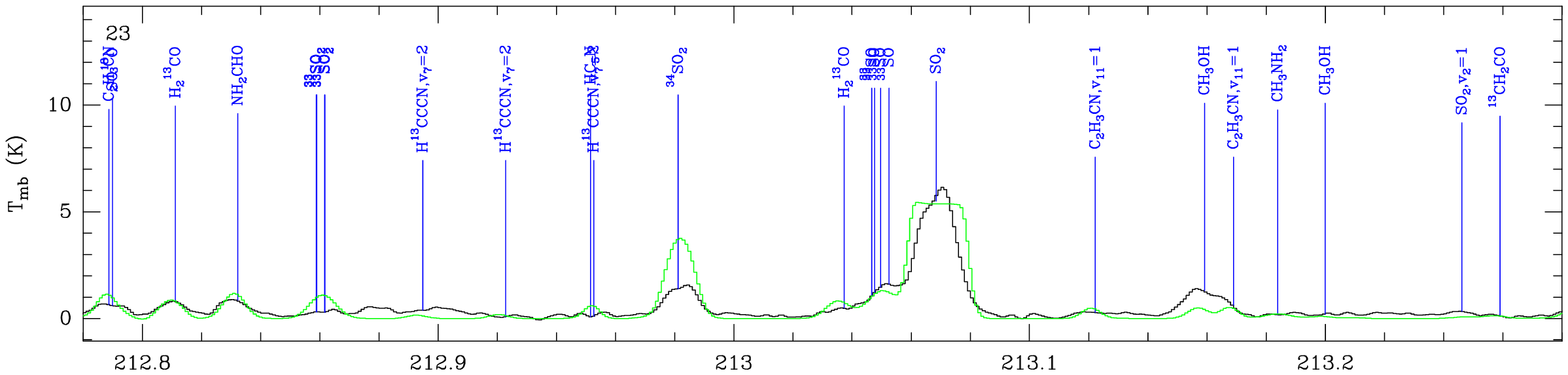}}}
\vspace*{1ex}\centerline{\resizebox{1.0\hsize}{!}{\includegraphics[angle=270]{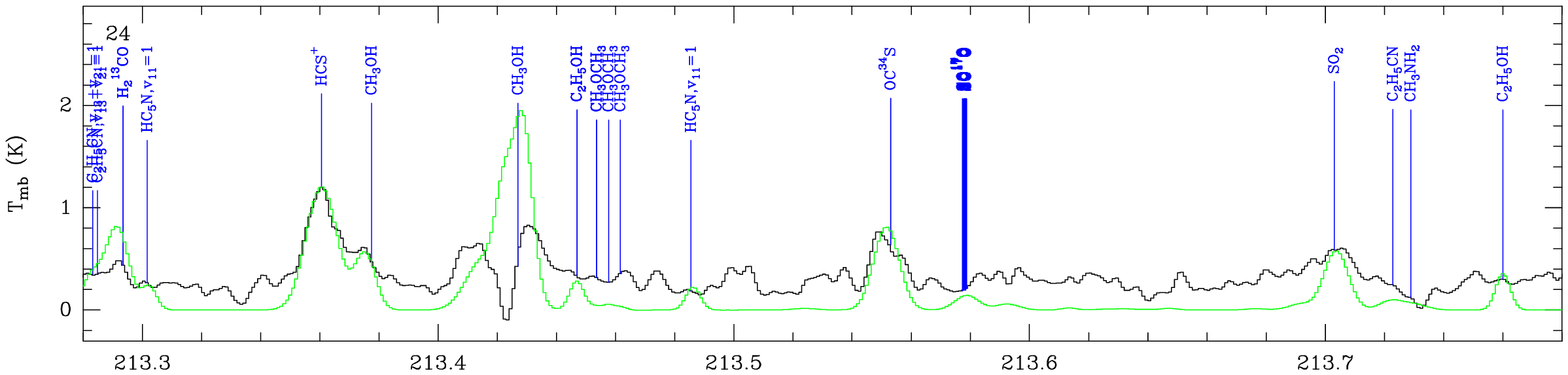}}}
\vspace*{1ex}\centerline{\resizebox{1.0\hsize}{!}{\includegraphics[angle=270]{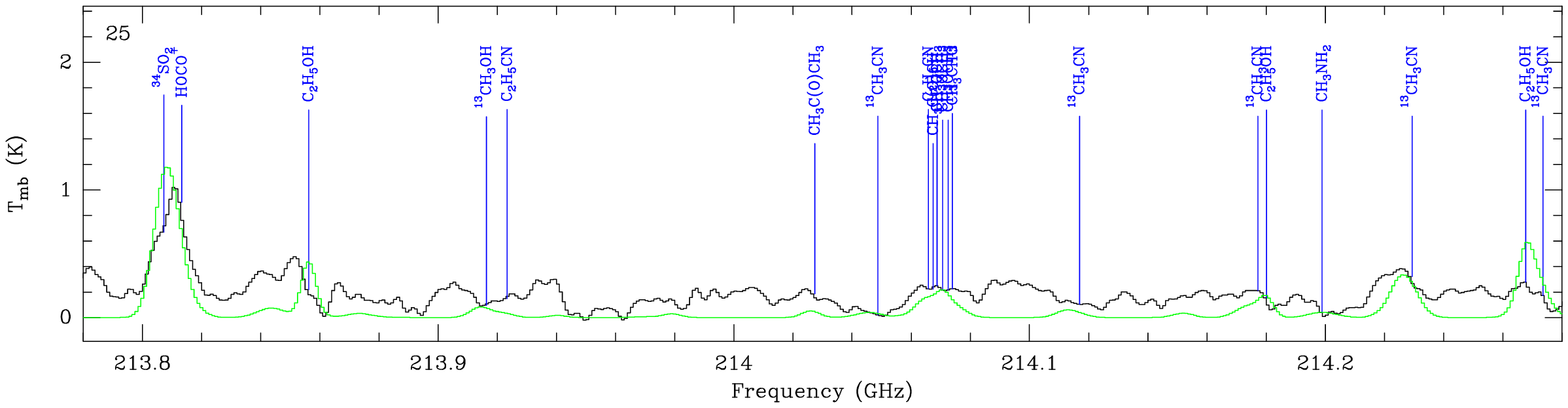}}}
\caption{
continued.
}
\end{figure*}
 \clearpage
\begin{figure*}
\addtocounter{figure}{-1}
\centerline{\resizebox{1.0\hsize}{!}{\includegraphics[angle=270]{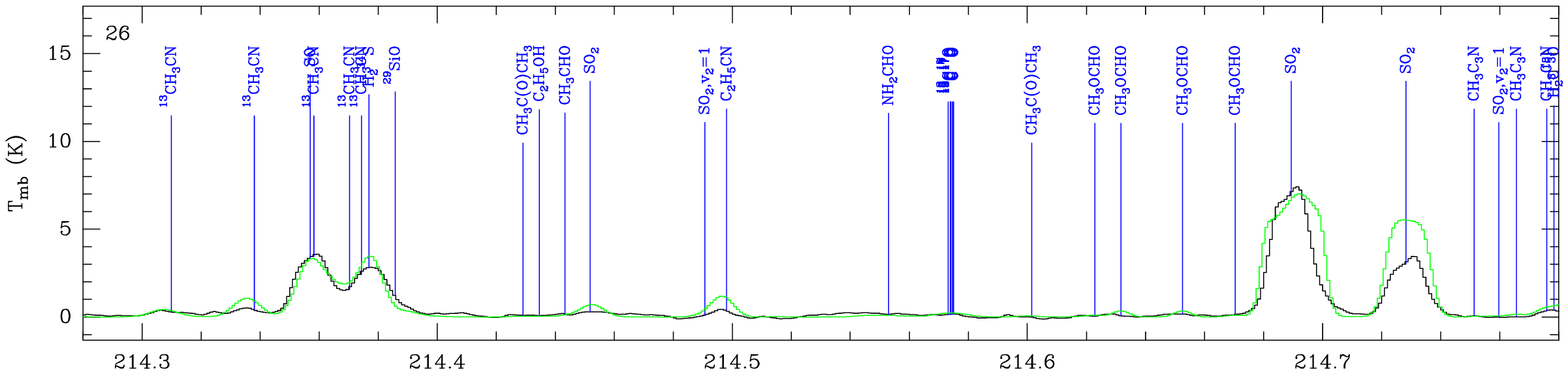}}}
\vspace*{1ex}\centerline{\resizebox{1.0\hsize}{!}{\includegraphics[angle=270]{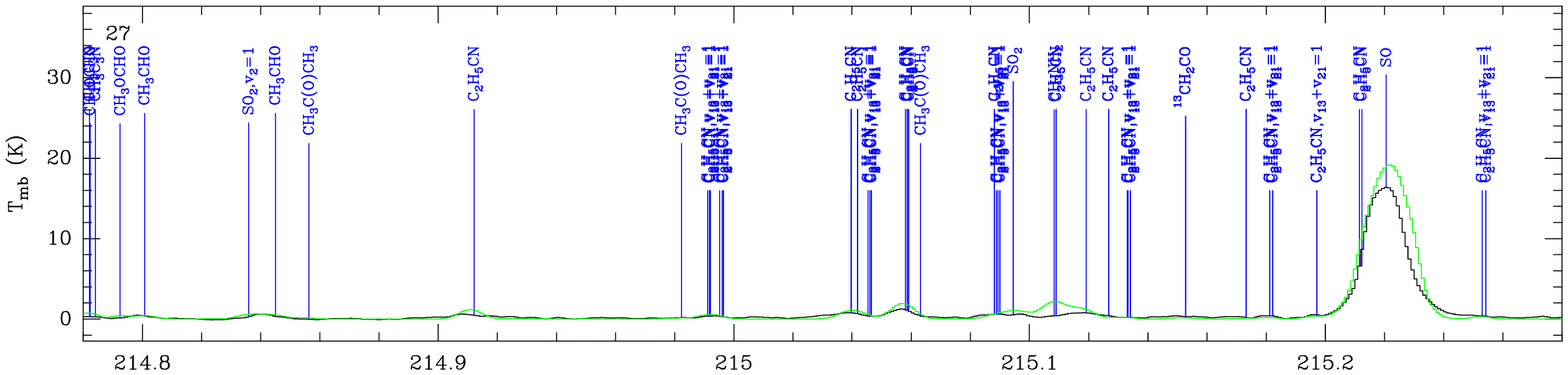}}}
\vspace*{1ex}\centerline{\resizebox{1.0\hsize}{!}{\includegraphics[angle=270]{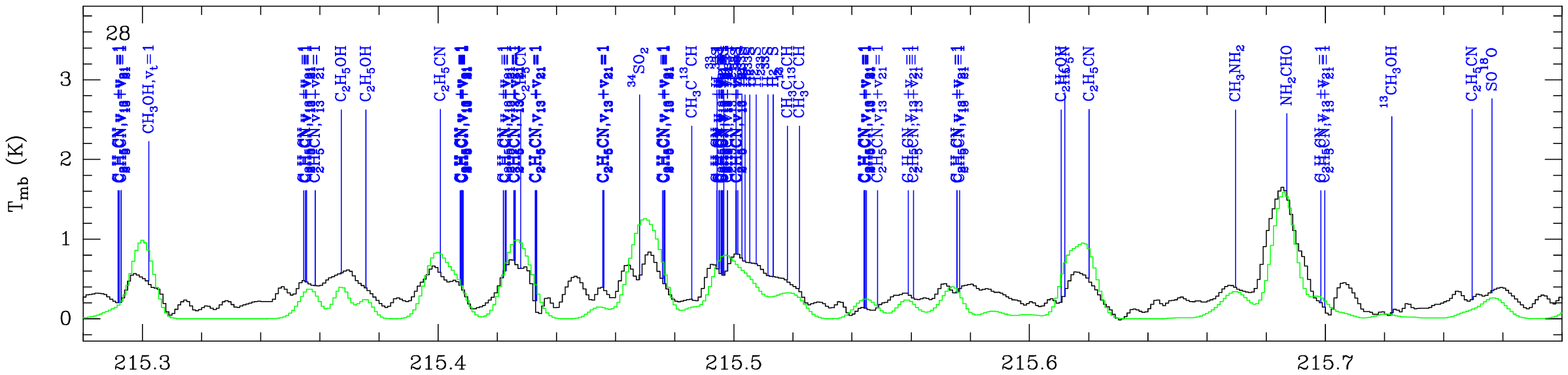}}}
\vspace*{1ex}\centerline{\resizebox{1.0\hsize}{!}{\includegraphics[angle=270]{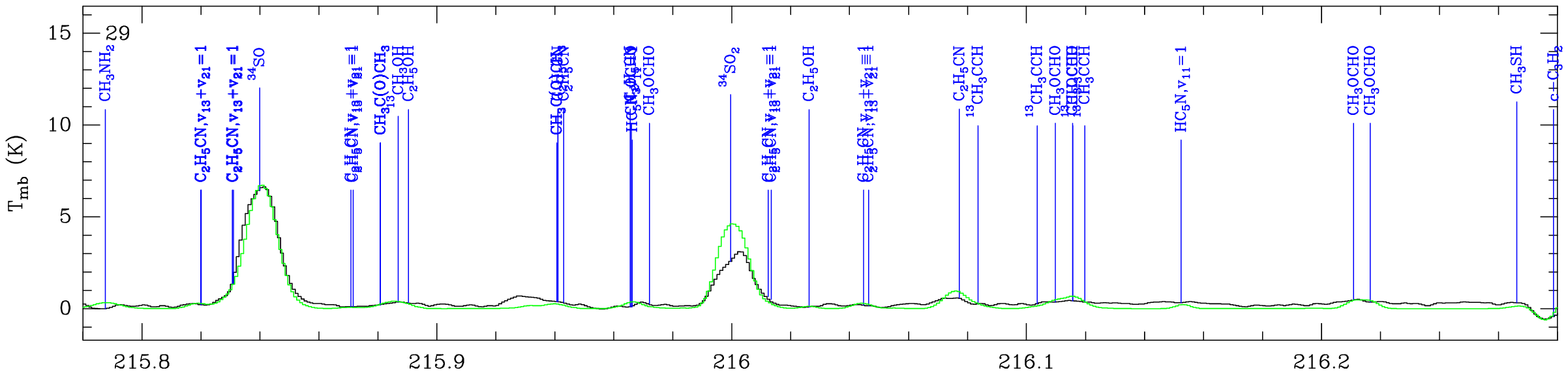}}}
\vspace*{1ex}\centerline{\resizebox{1.0\hsize}{!}{\includegraphics[angle=270]{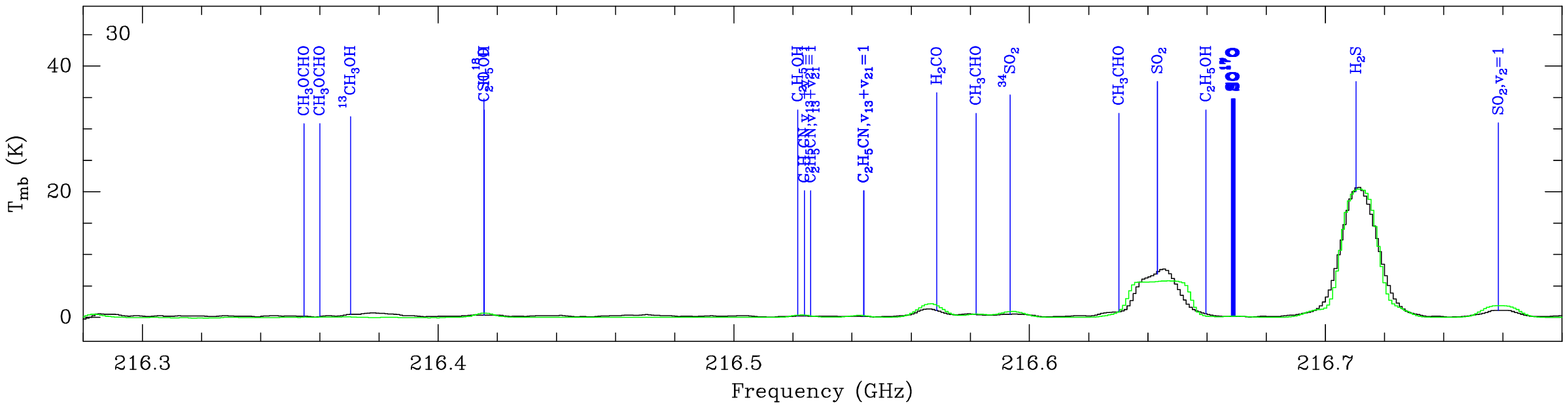}}}
\caption{
continued.
}
\end{figure*}
 \clearpage
\begin{figure*}
\addtocounter{figure}{-1}
\centerline{\resizebox{1.0\hsize}{!}{\includegraphics[angle=270]{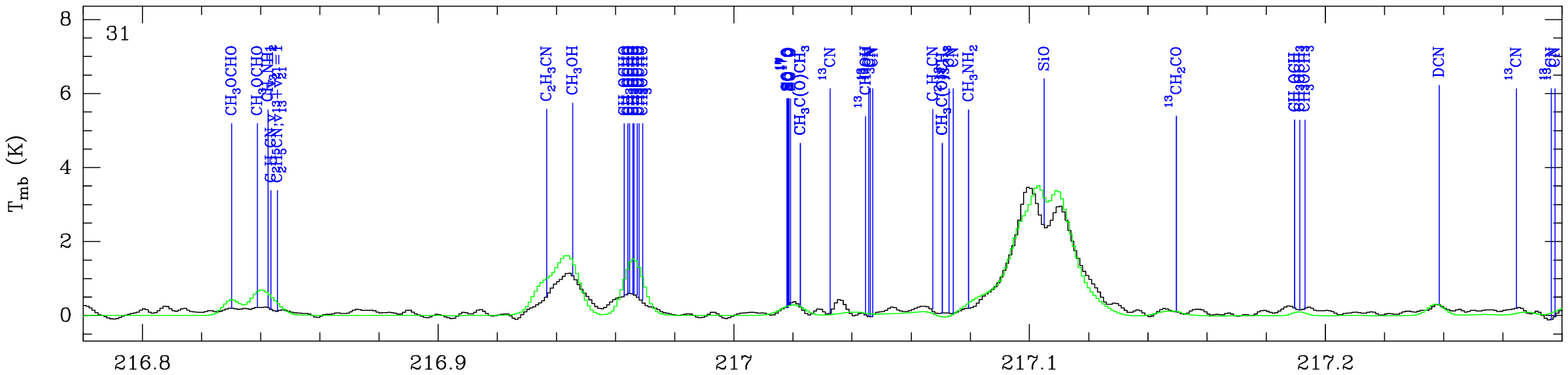}}}
\vspace*{1ex}\centerline{\resizebox{1.0\hsize}{!}{\includegraphics[angle=270]{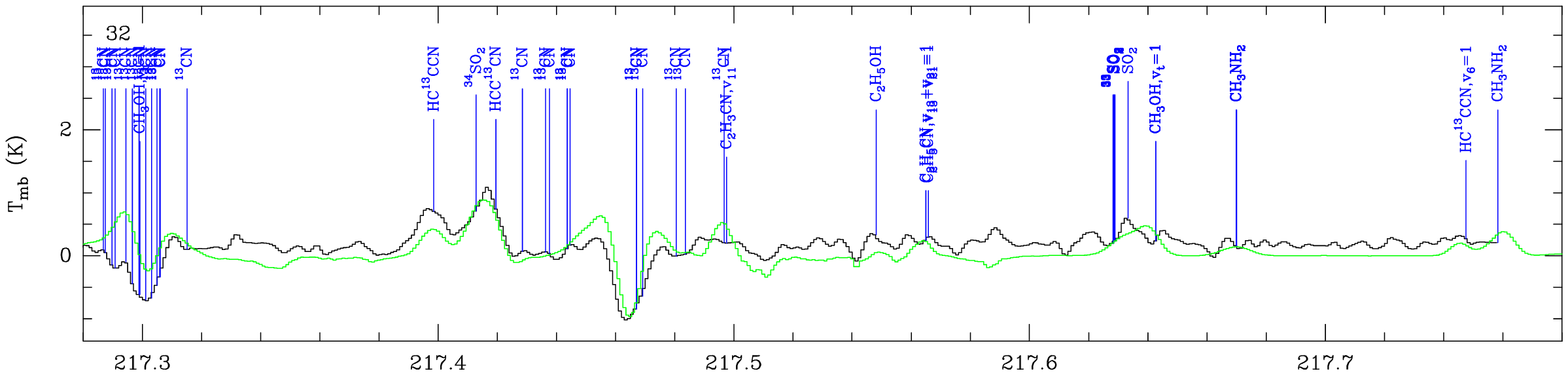}}}
\vspace*{1ex}\centerline{\resizebox{1.0\hsize}{!}{\includegraphics[angle=270]{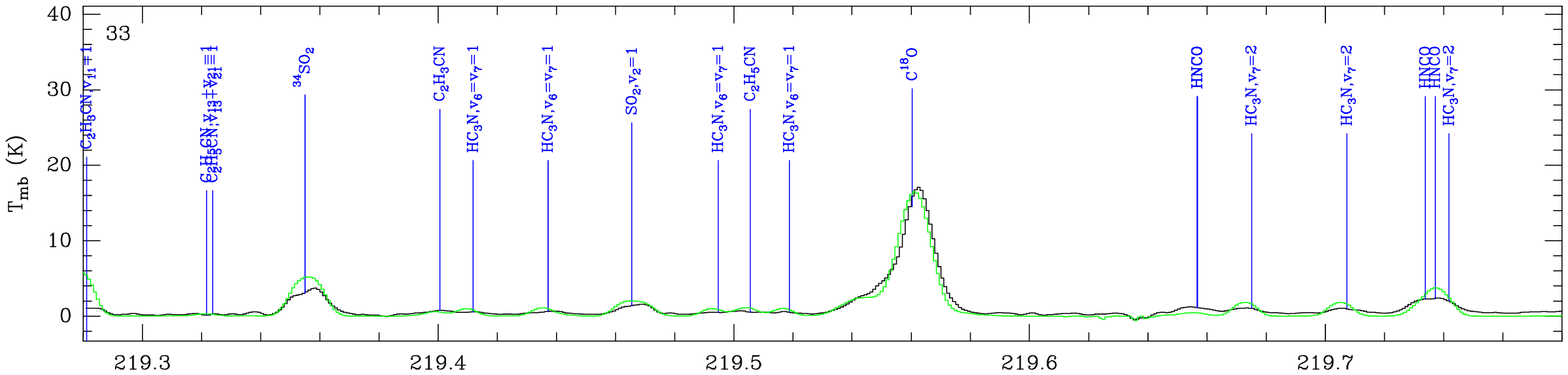}}}
\vspace*{1ex}\centerline{\resizebox{1.0\hsize}{!}{\includegraphics[angle=270]{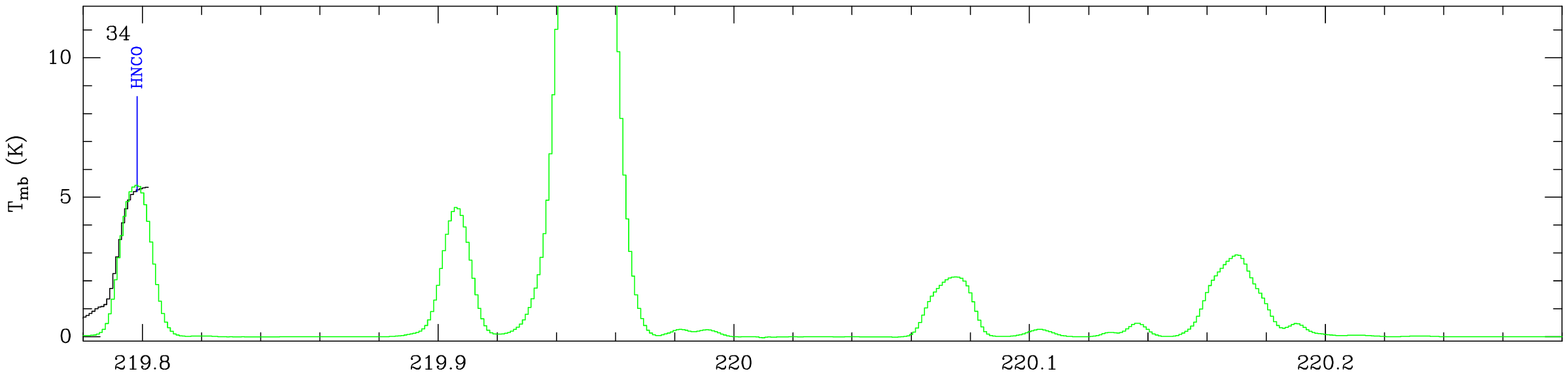}}}
\vspace*{1ex}\centerline{\resizebox{1.0\hsize}{!}{\includegraphics[angle=270]{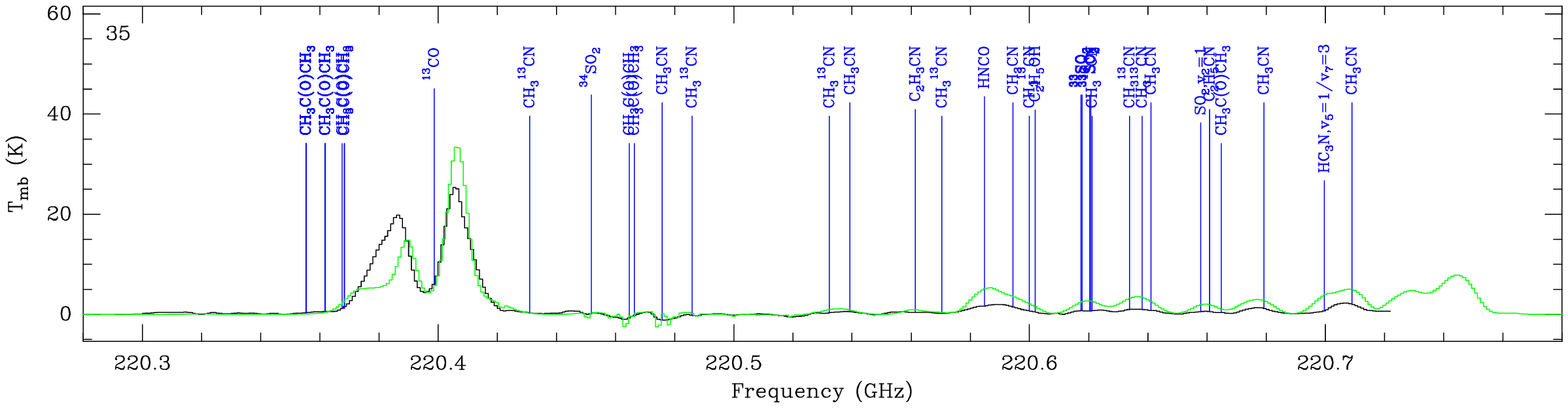}}}
\caption{
continued.
}
\end{figure*}
 \clearpage
\begin{figure*}
\addtocounter{figure}{-1}
\centerline{\resizebox{1.0\hsize}{!}{\includegraphics[angle=270]{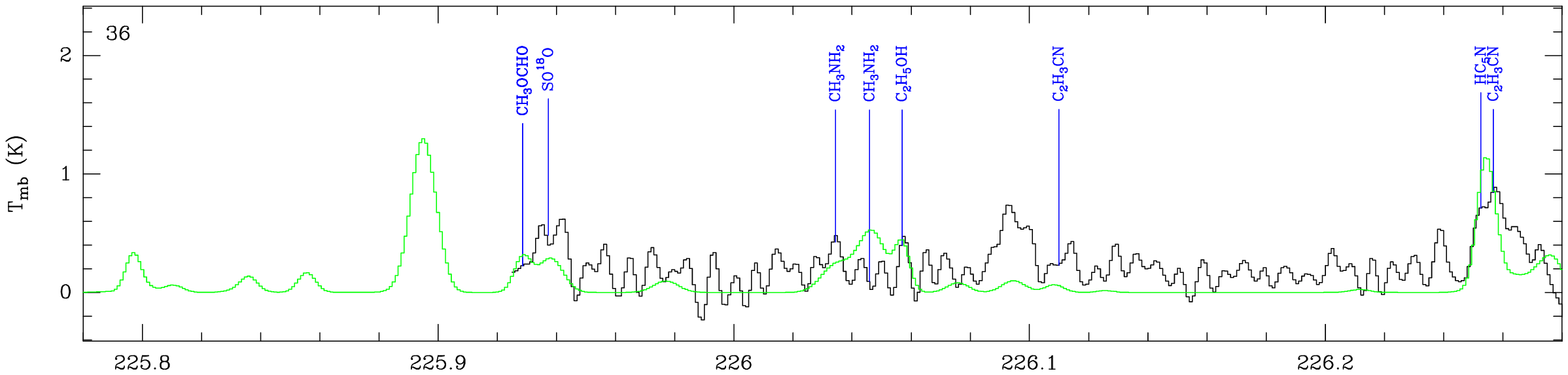}}}
\vspace*{1ex}\centerline{\resizebox{1.0\hsize}{!}{\includegraphics[angle=270]{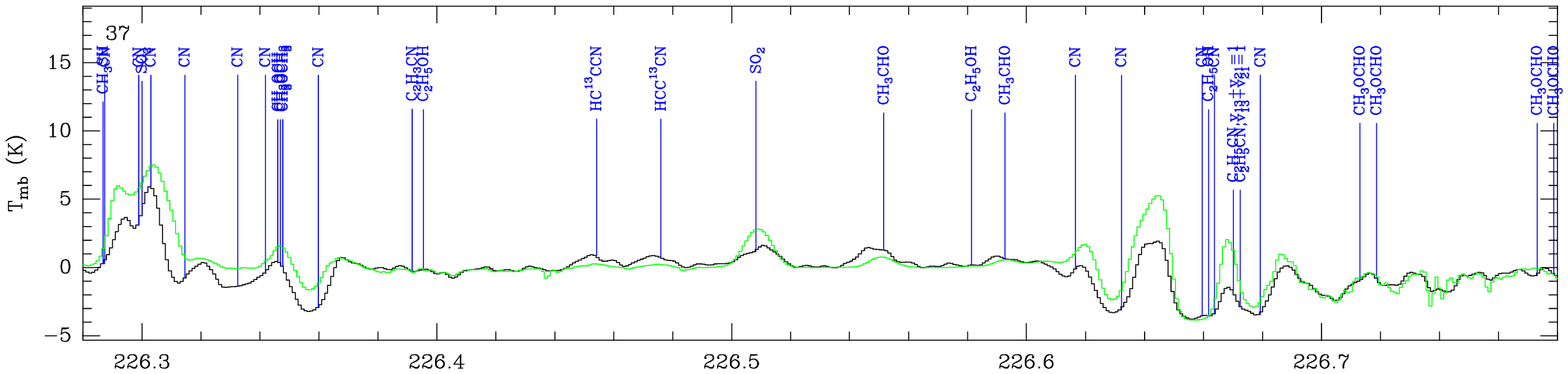}}}
\vspace*{1ex}\centerline{\resizebox{1.0\hsize}{!}{\includegraphics[angle=270]{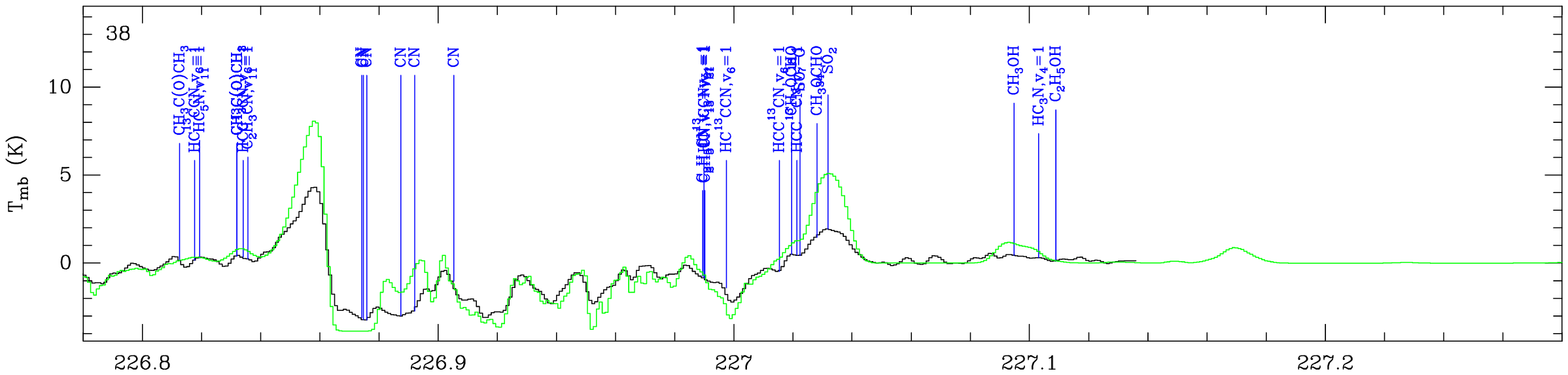}}}
\vspace*{1ex}\centerline{\resizebox{1.0\hsize}{!}{\includegraphics[angle=270]{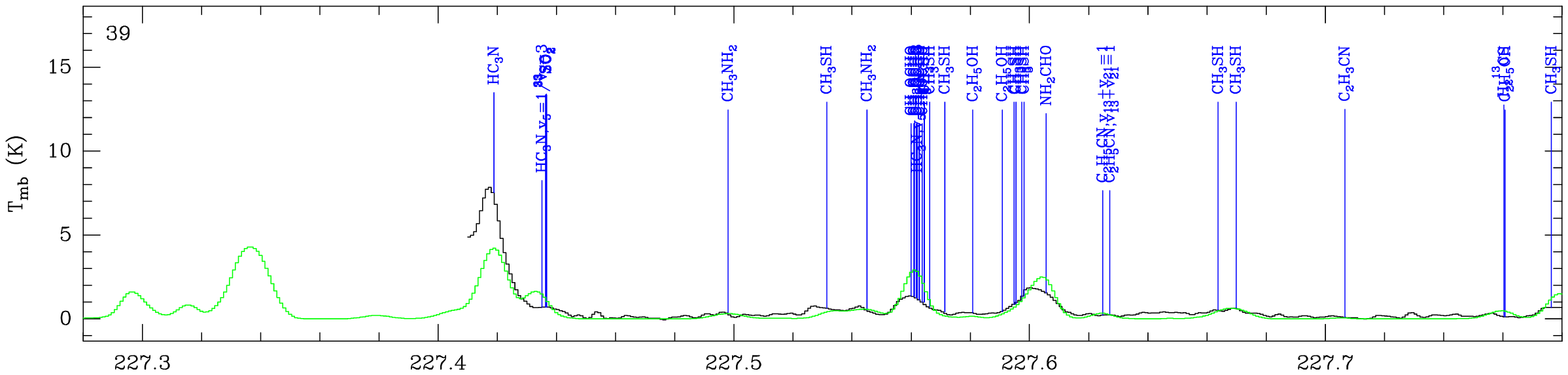}}}
\vspace*{1ex}\centerline{\resizebox{1.0\hsize}{!}{\includegraphics[angle=270]{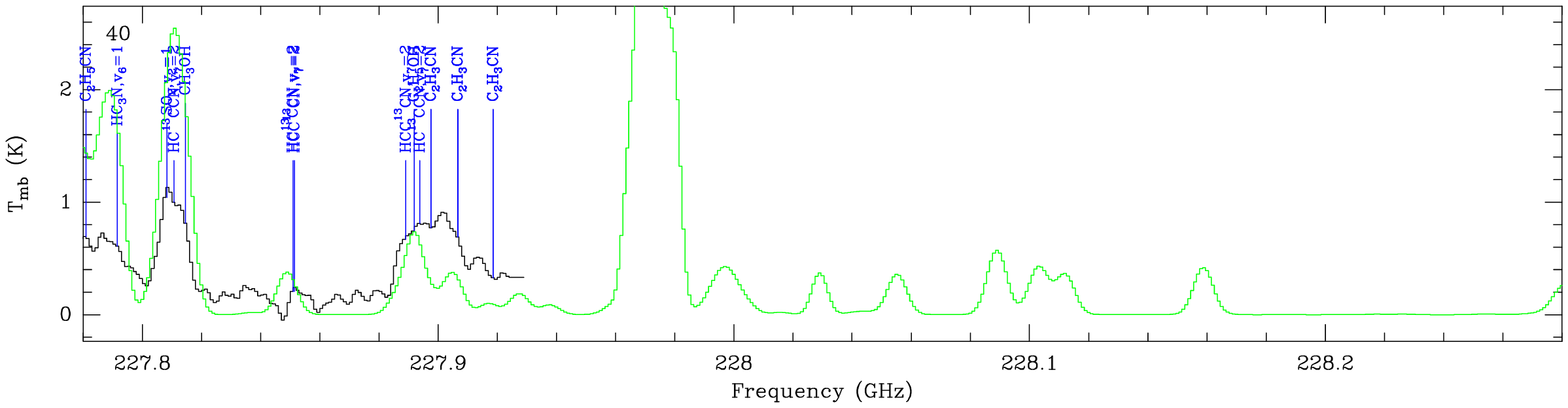}}}
\caption{
continued.
}
\end{figure*}
 \clearpage
\begin{figure*}
\addtocounter{figure}{-1}
\centerline{\resizebox{1.0\hsize}{!}{\includegraphics[angle=270]{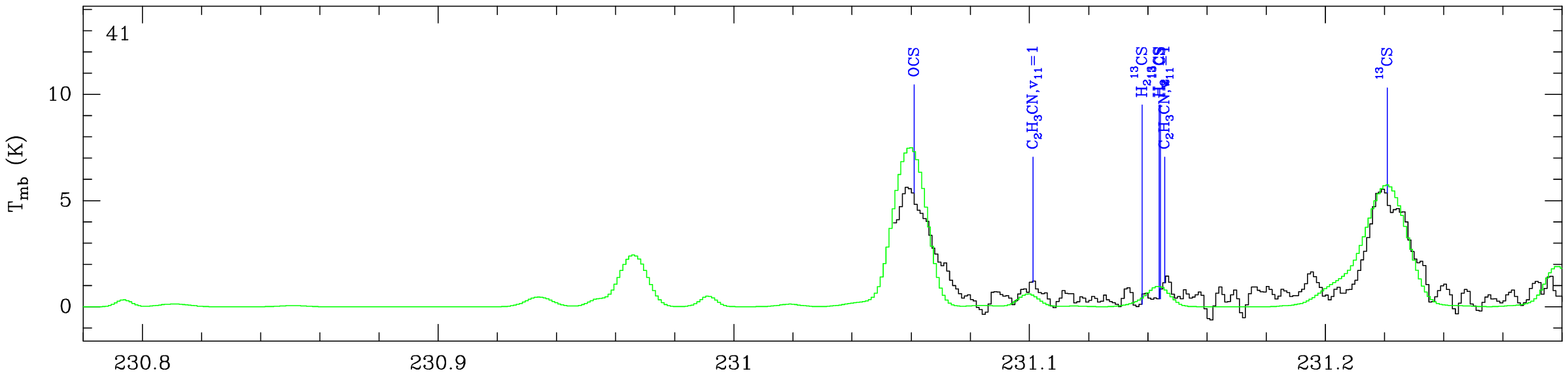}}}
\vspace*{1ex}\centerline{\resizebox{1.0\hsize}{!}{\includegraphics[angle=270]{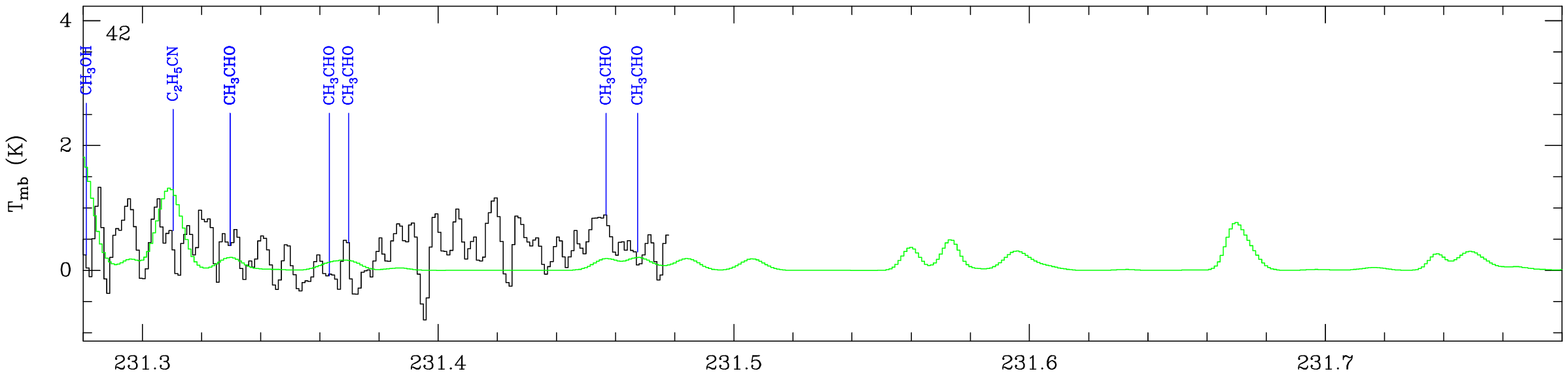}}}
\vspace*{1ex}\centerline{\resizebox{1.0\hsize}{!}{\includegraphics[angle=270]{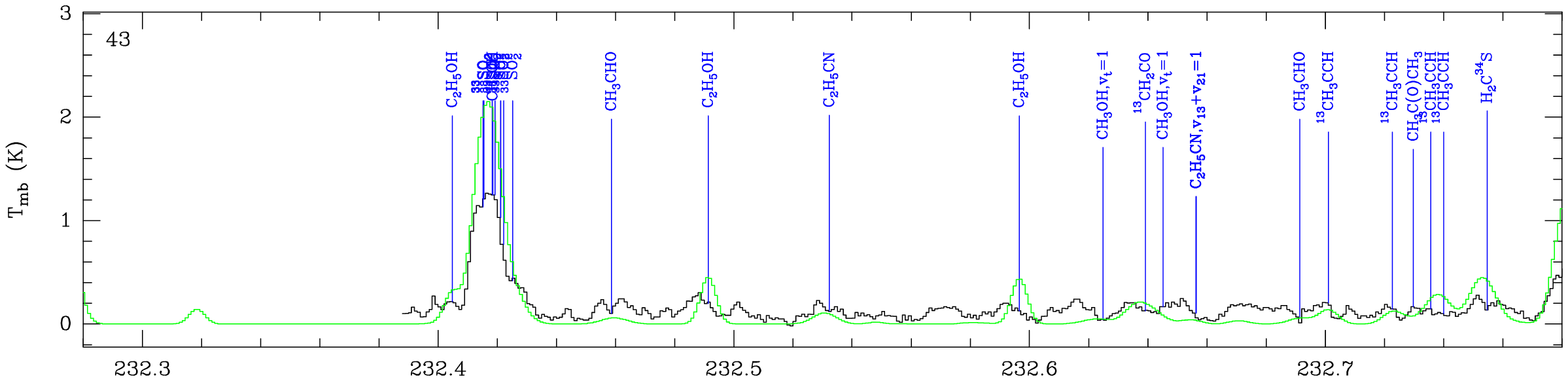}}}
\vspace*{1ex}\centerline{\resizebox{1.0\hsize}{!}{\includegraphics[angle=270]{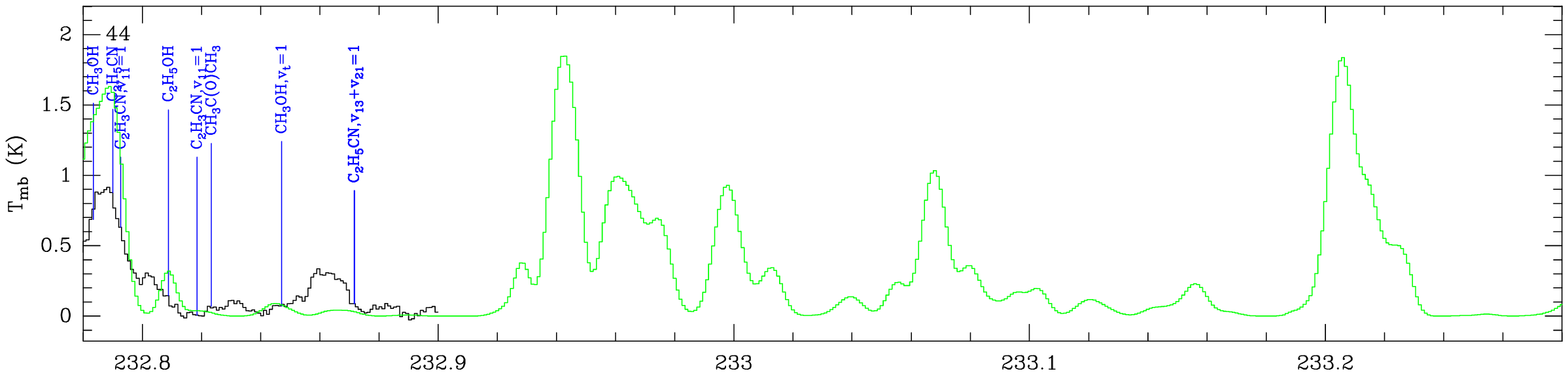}}}
\vspace*{1ex}\centerline{\resizebox{1.0\hsize}{!}{\includegraphics[angle=270]{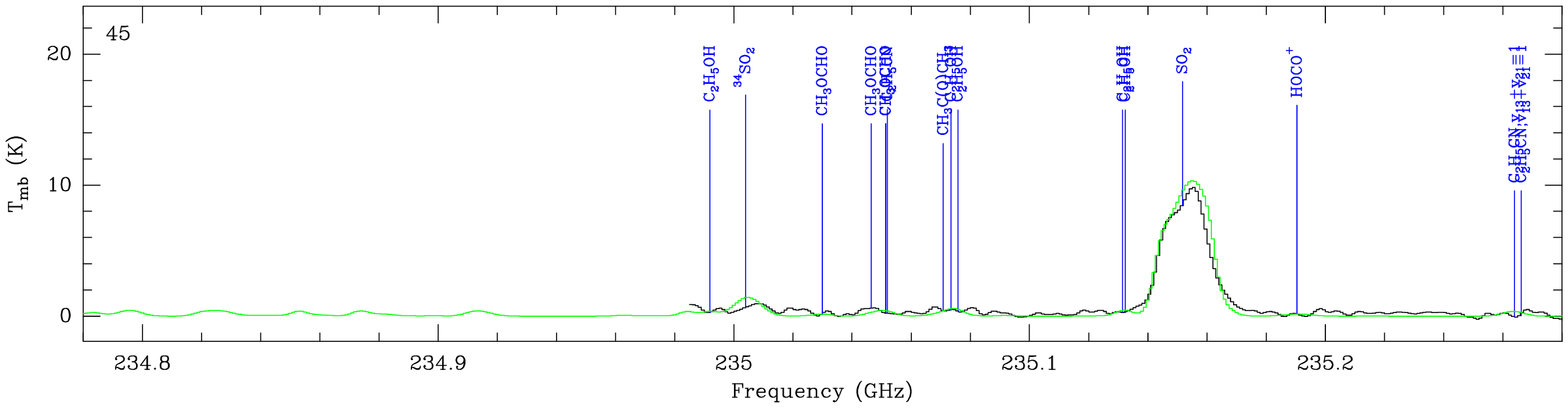}}}
\caption{
continued.
}
\end{figure*}
 \clearpage
\begin{figure*}
\addtocounter{figure}{-1}
\centerline{\resizebox{1.0\hsize}{!}{\includegraphics[angle=270]{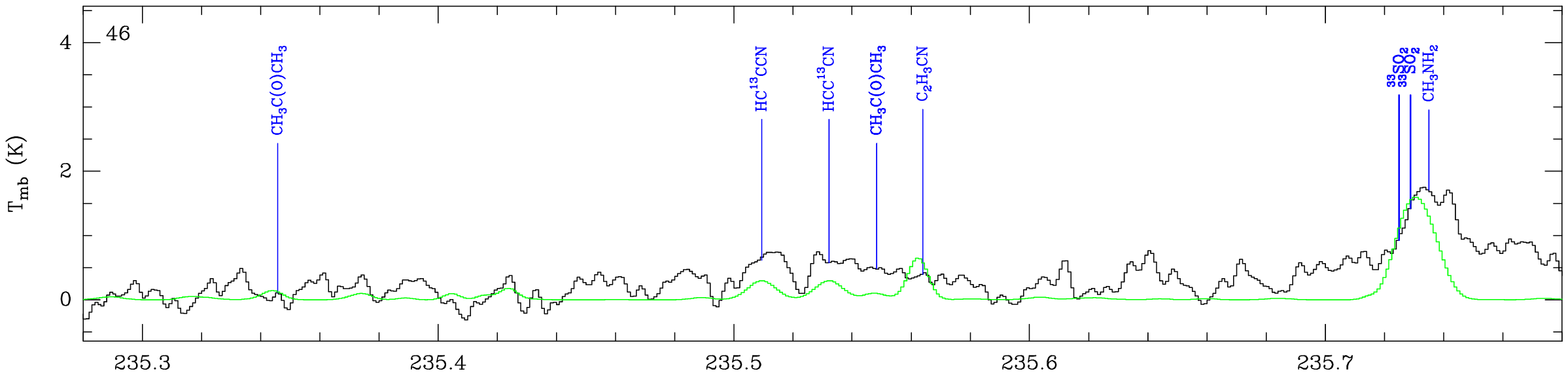}}}
\vspace*{1ex}\centerline{\resizebox{1.0\hsize}{!}{\includegraphics[angle=270]{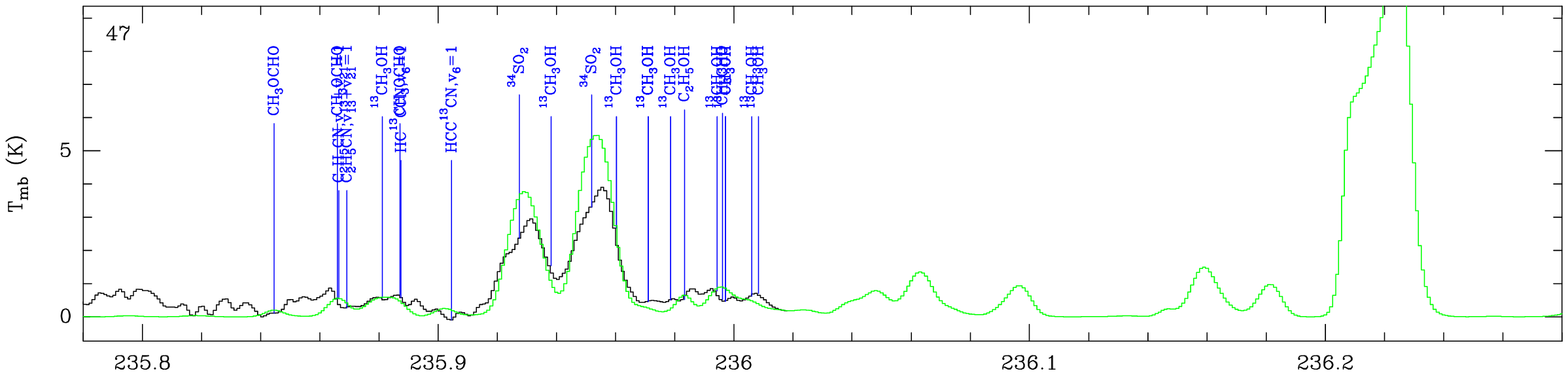}}}
\vspace*{1ex}\centerline{\resizebox{1.0\hsize}{!}{\includegraphics[angle=270]{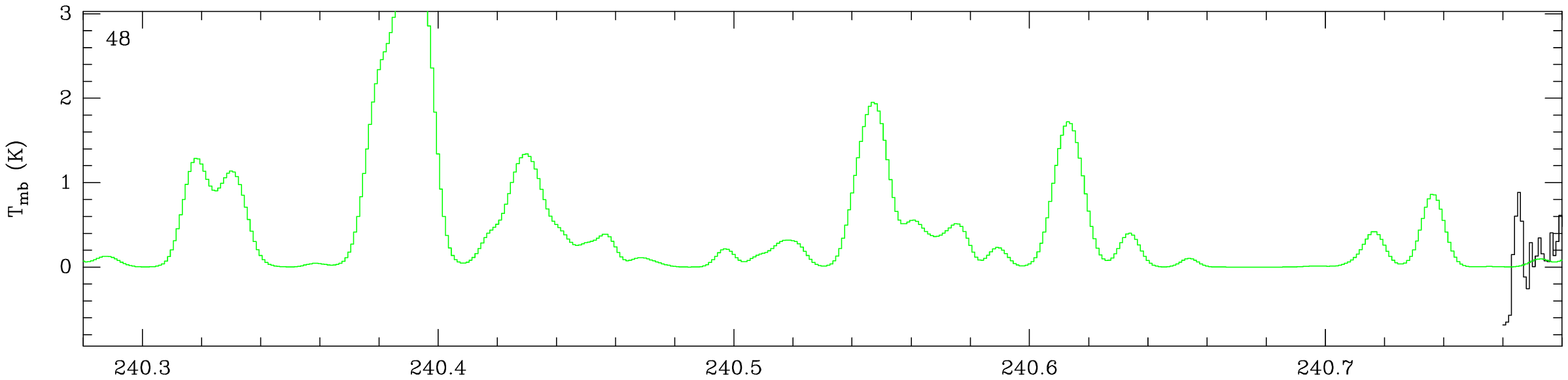}}}
\vspace*{1ex}\centerline{\resizebox{1.0\hsize}{!}{\includegraphics[angle=270]{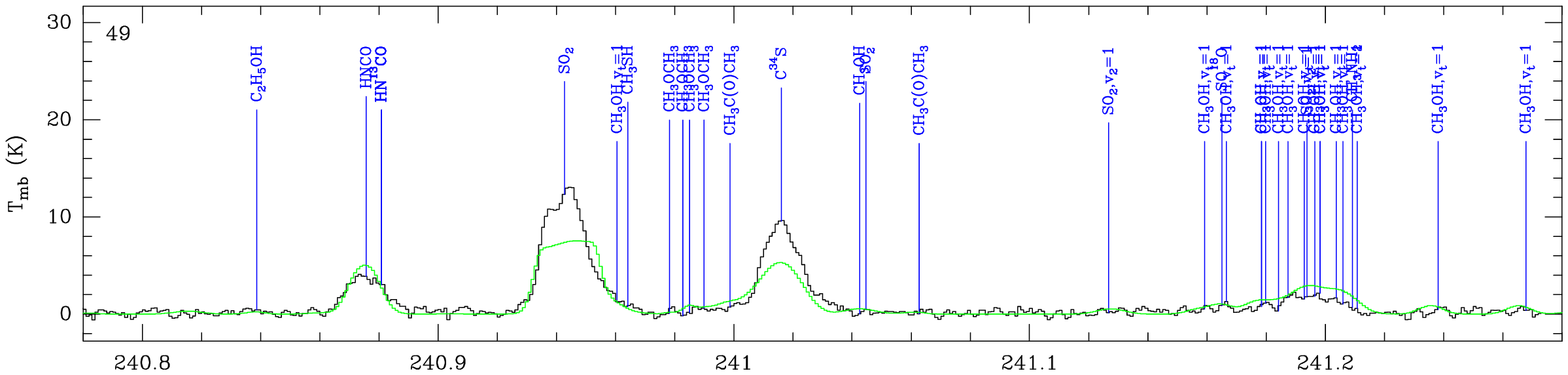}}}
\vspace*{1ex}\centerline{\resizebox{1.0\hsize}{!}{\includegraphics[angle=270]{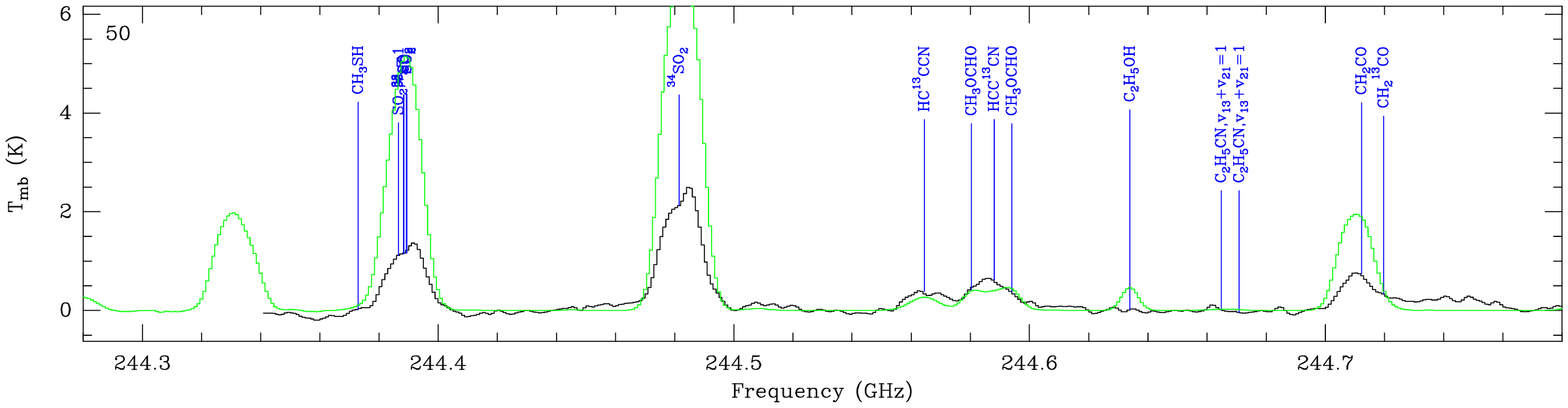}}}
\caption{
continued.
}
\end{figure*}
 \clearpage
\begin{figure*}
\addtocounter{figure}{-1}
\centerline{\resizebox{1.0\hsize}{!}{\includegraphics[angle=270]{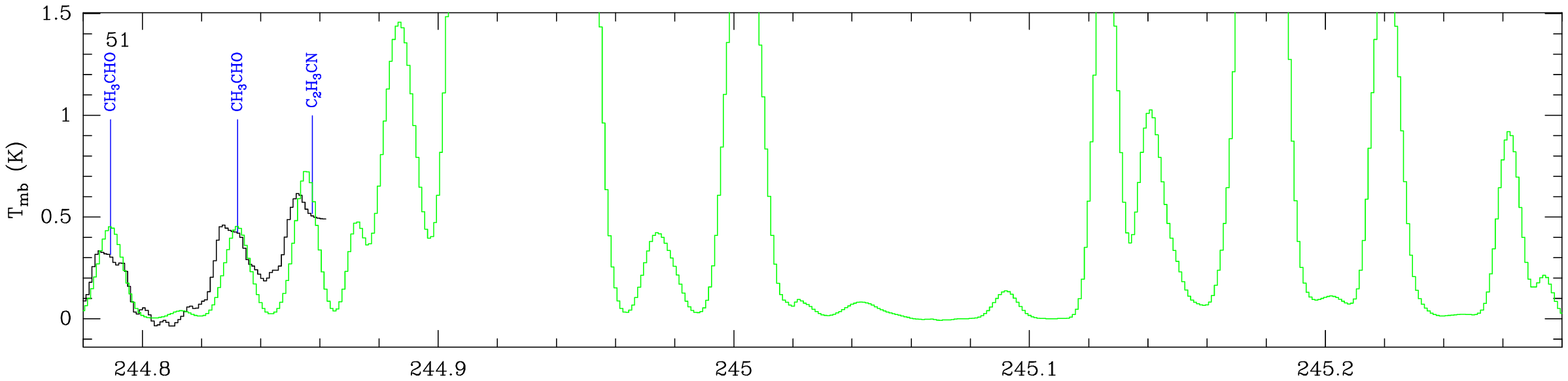}}}
\vspace*{1ex}\centerline{\resizebox{1.0\hsize}{!}{\includegraphics[angle=270]{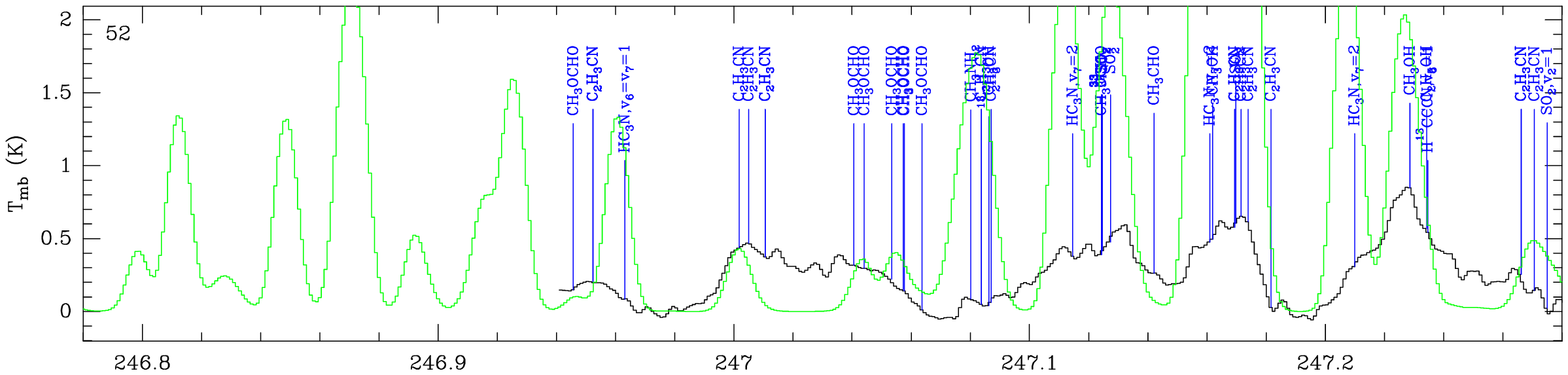}}}
\vspace*{1ex}\centerline{\resizebox{1.0\hsize}{!}{\includegraphics[angle=270]{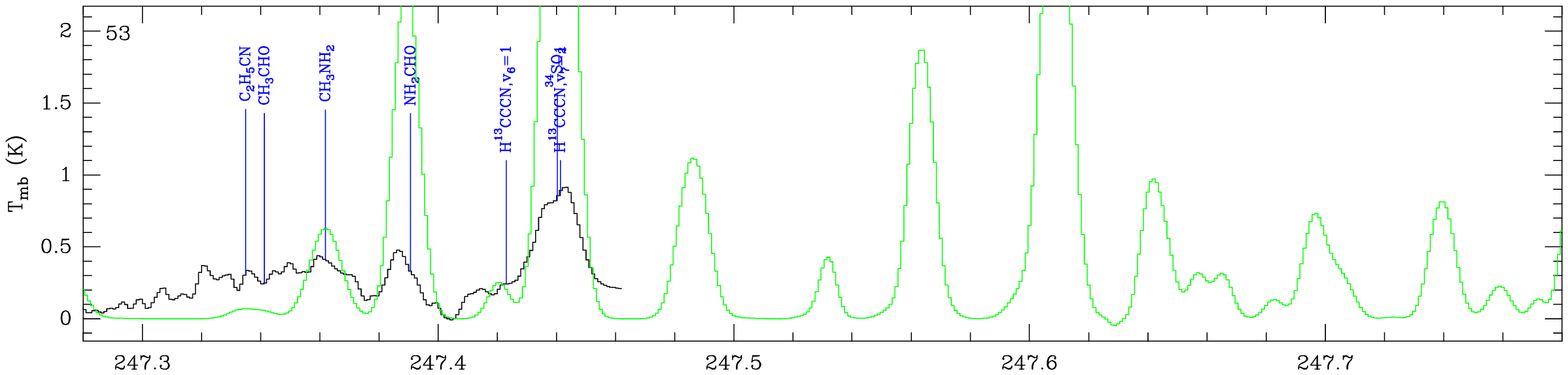}}}
\vspace*{1ex}\centerline{\resizebox{1.0\hsize}{!}{\includegraphics[angle=270]{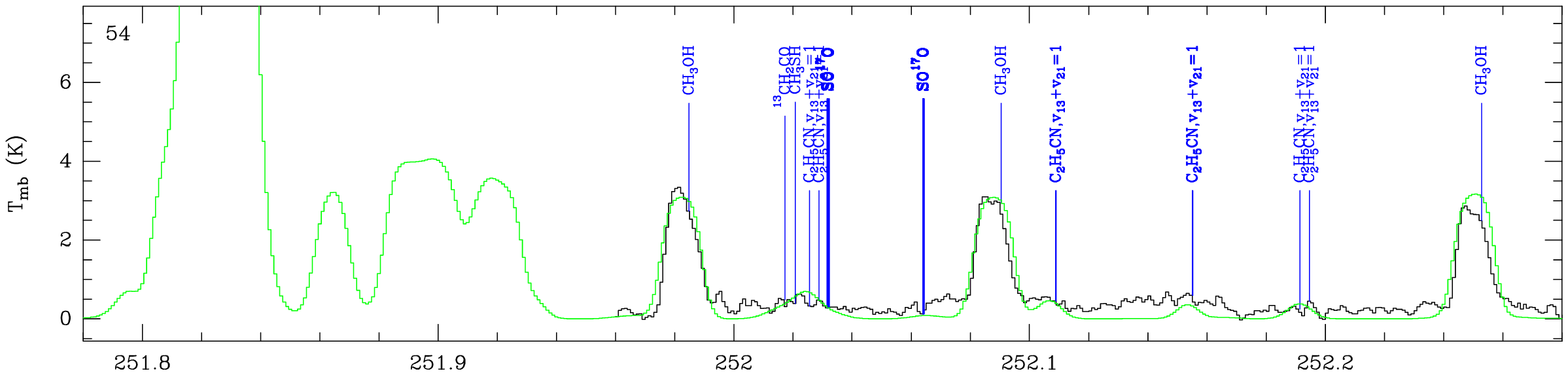}}}
\vspace*{1ex}\centerline{\resizebox{1.0\hsize}{!}{\includegraphics[angle=270]{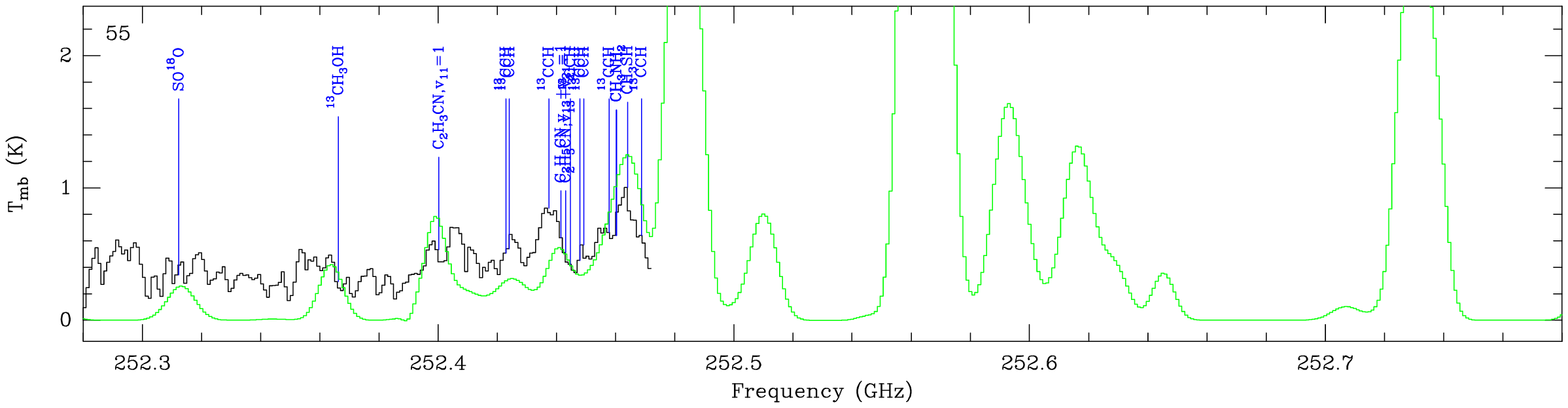}}}
\caption{
continued.
}
\end{figure*}
 \clearpage
\begin{figure*}
\addtocounter{figure}{-1}
\centerline{\resizebox{1.0\hsize}{!}{\includegraphics[angle=270]{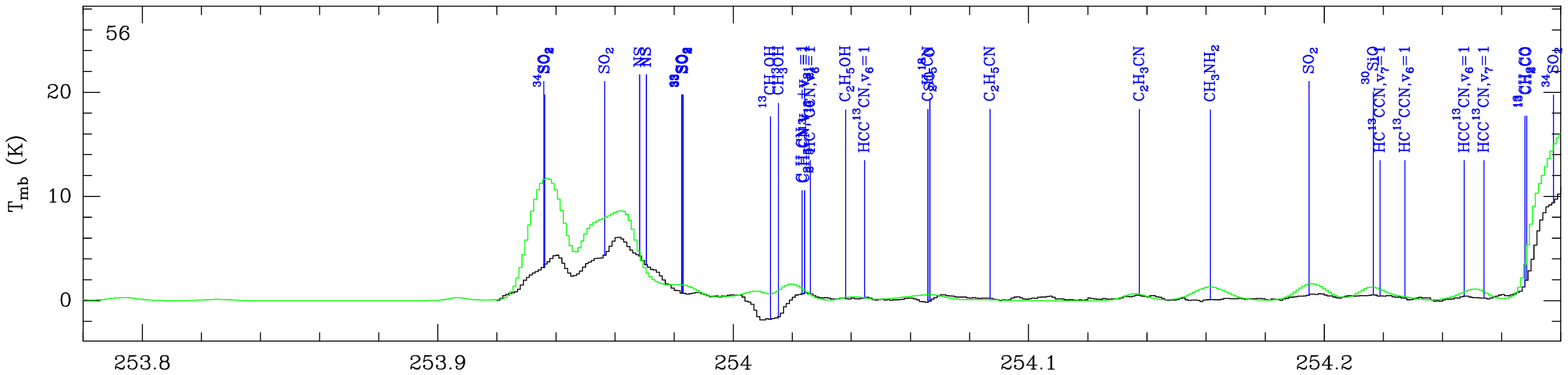}}}
\vspace*{1ex}\centerline{\resizebox{1.0\hsize}{!}{\includegraphics[angle=270]{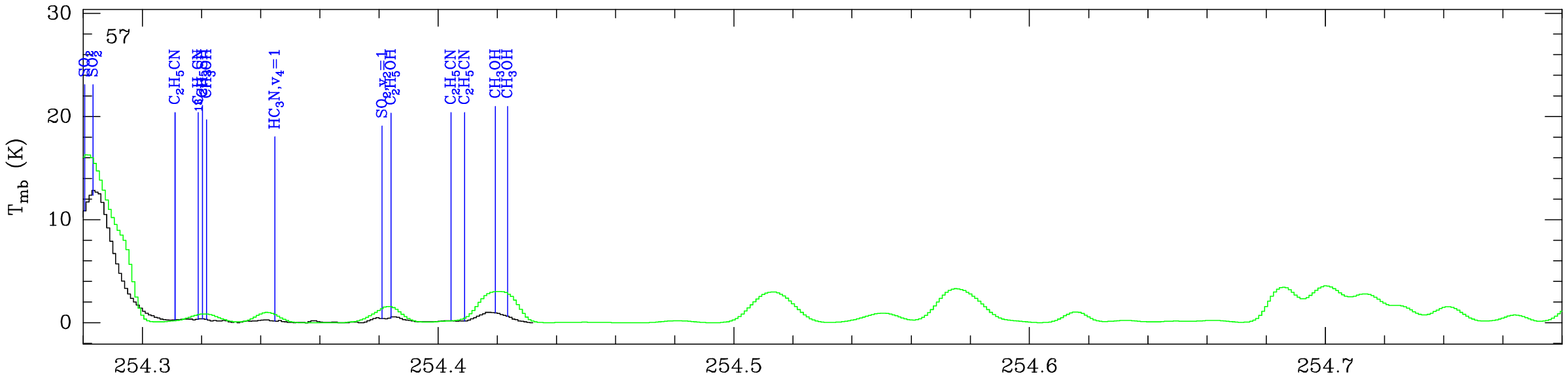}}}
\vspace*{1ex}\centerline{\resizebox{1.0\hsize}{!}{\includegraphics[angle=270]{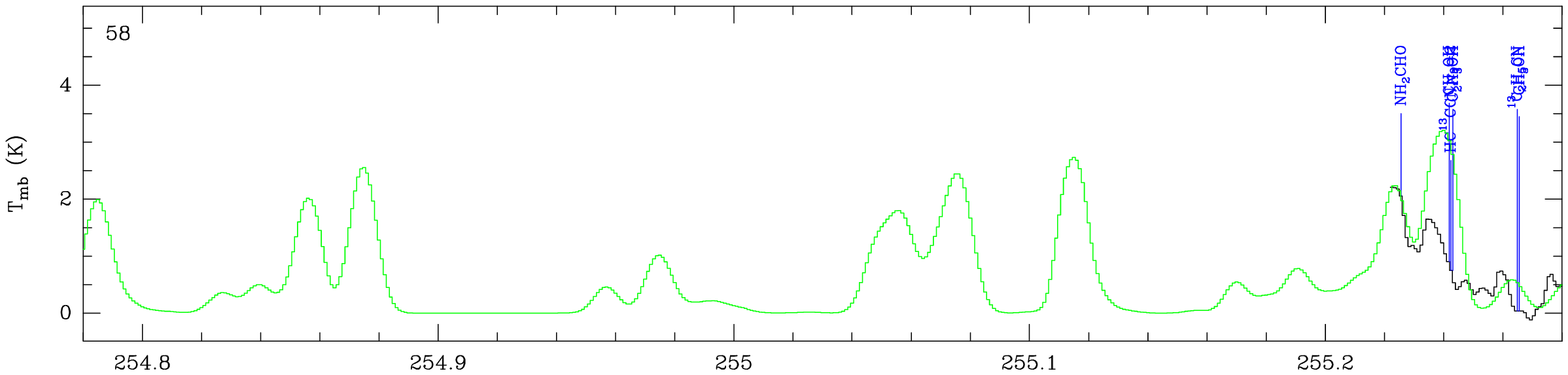}}}
\vspace*{1ex}\centerline{\resizebox{1.0\hsize}{!}{\includegraphics[angle=270]{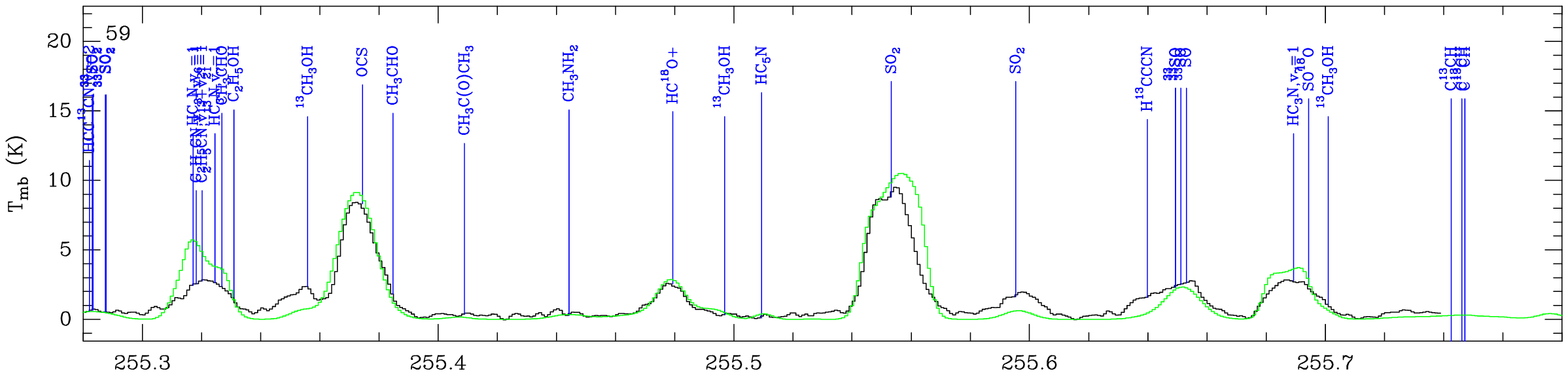}}}
\vspace*{1ex}\centerline{\resizebox{1.0\hsize}{!}{\includegraphics[angle=270]{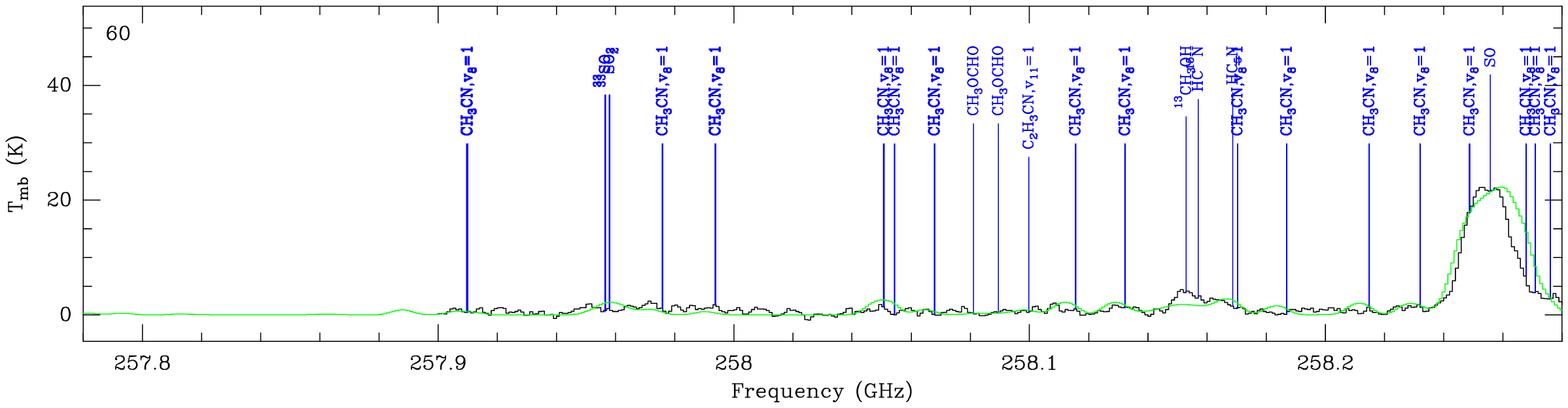}}}
\caption{
continued.
}
\end{figure*}
 \clearpage
\begin{figure*}
\addtocounter{figure}{-1}
\centerline{\resizebox{1.0\hsize}{!}{\includegraphics[angle=270]{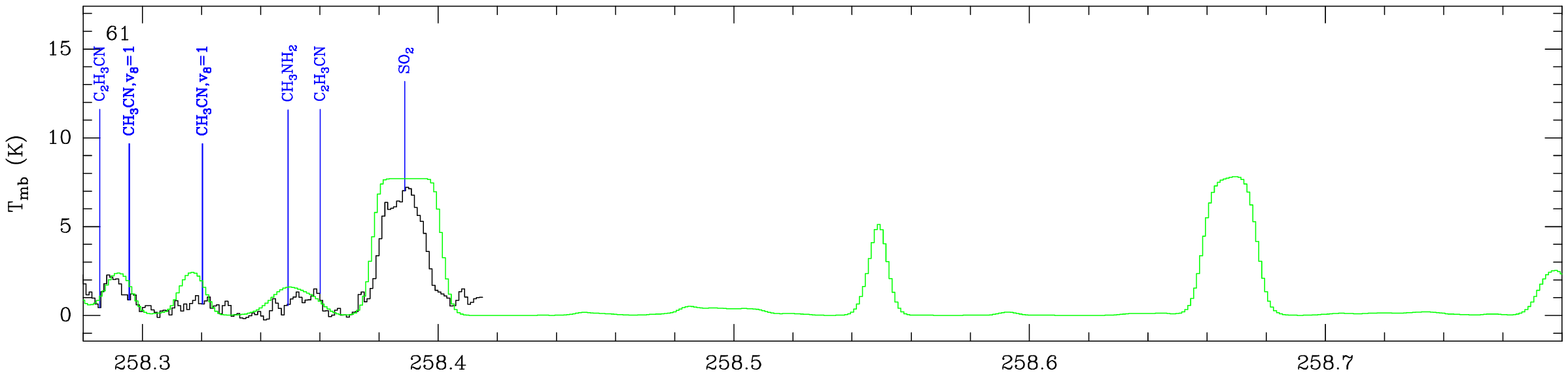}}}
\vspace*{1ex}\centerline{\resizebox{1.0\hsize}{!}{\includegraphics[angle=270]{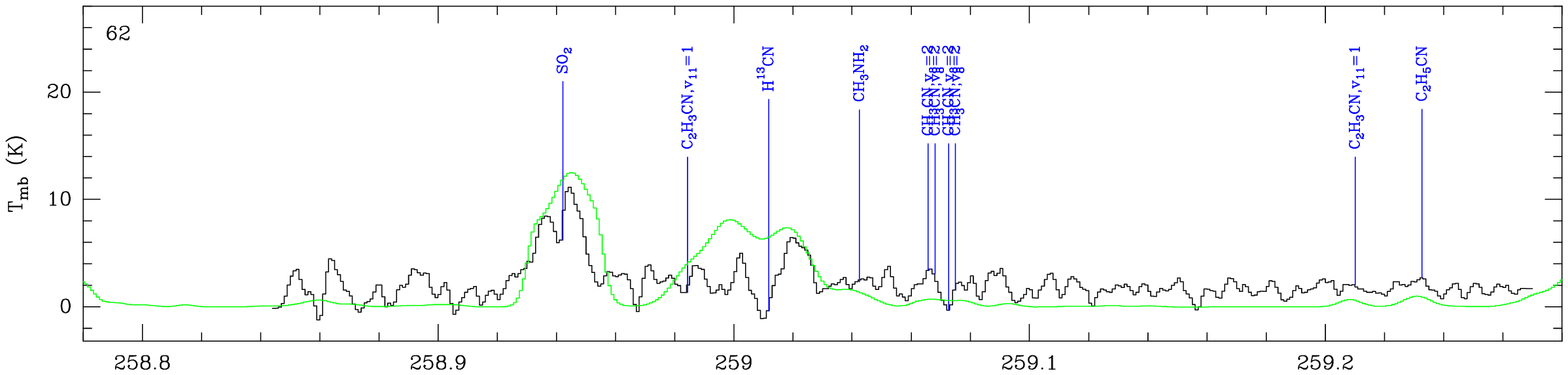}}}
\vspace*{1ex}\centerline{\resizebox{1.0\hsize}{!}{\includegraphics[angle=270]{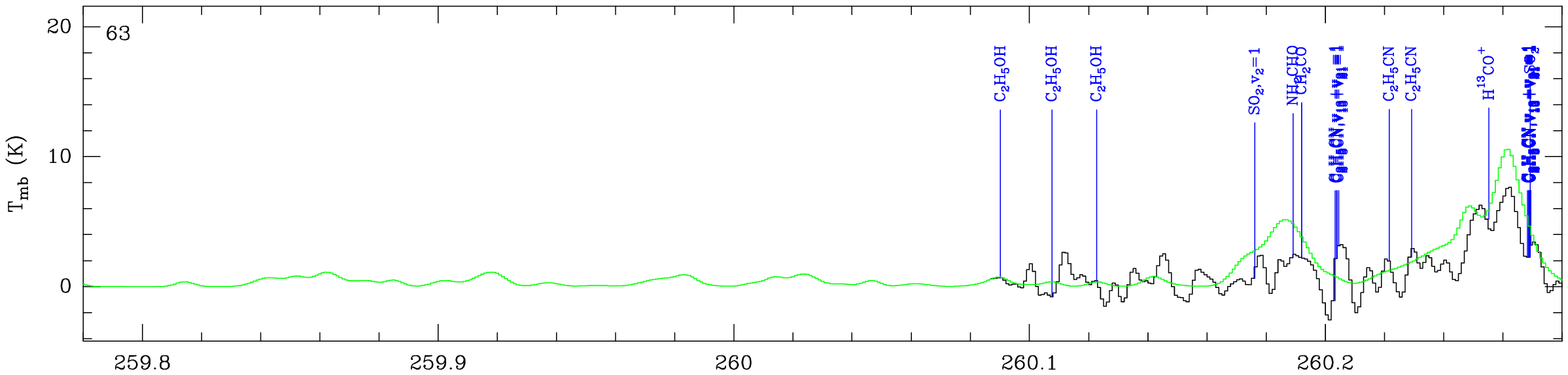}}}
\vspace*{1ex}\centerline{\resizebox{1.0\hsize}{!}{\includegraphics[angle=270]{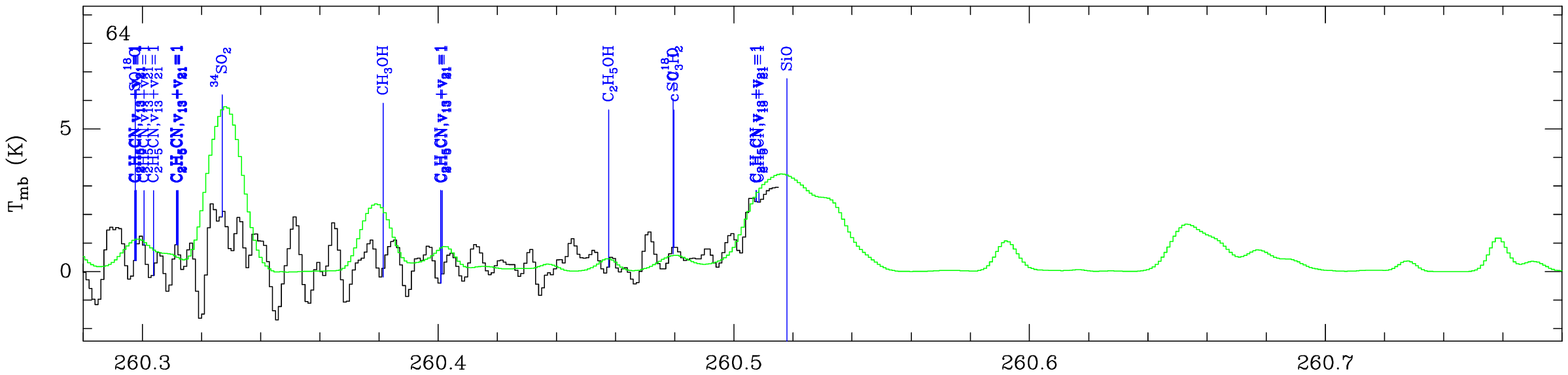}}}
\vspace*{1ex}\centerline{\resizebox{1.0\hsize}{!}{\includegraphics[angle=270]{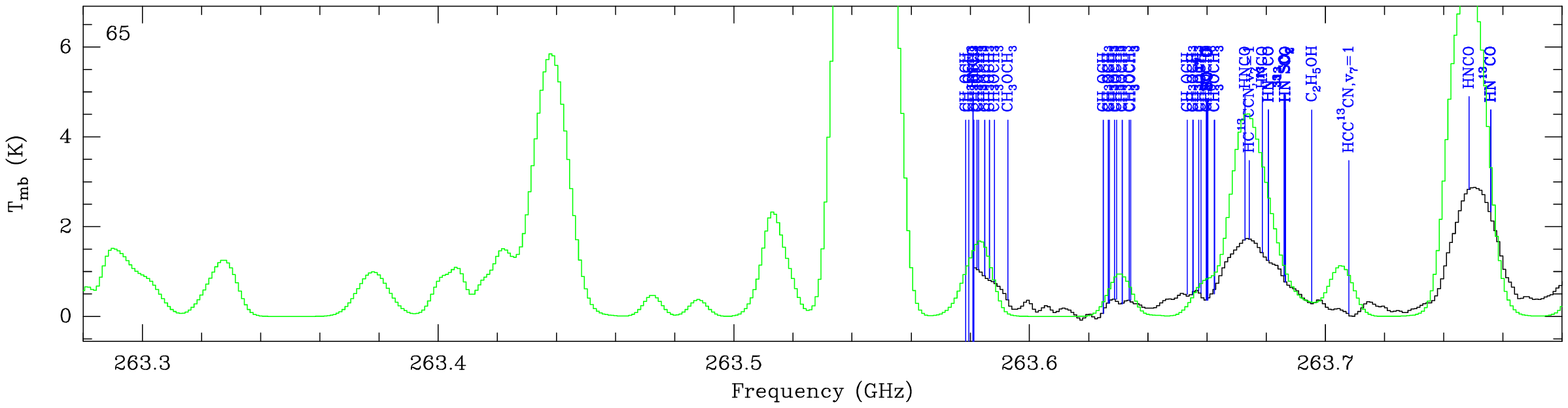}}}
\caption{
continued.
}
\end{figure*}
 \clearpage
\begin{figure*}
\addtocounter{figure}{-1}
\centerline{\resizebox{1.0\hsize}{!}{\includegraphics[angle=270]{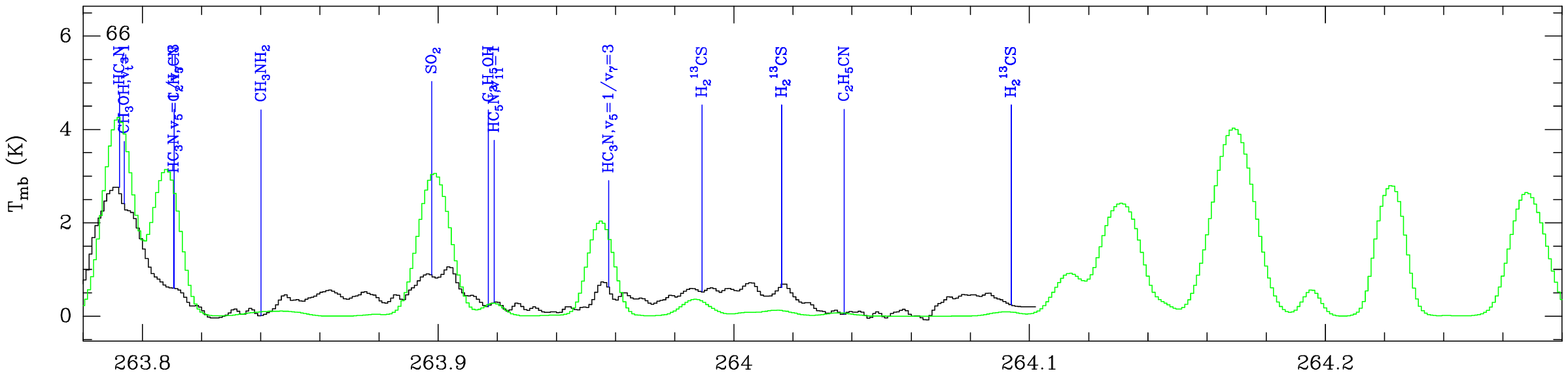}}}
\vspace*{1ex}\centerline{\resizebox{1.0\hsize}{!}{\includegraphics[angle=270]{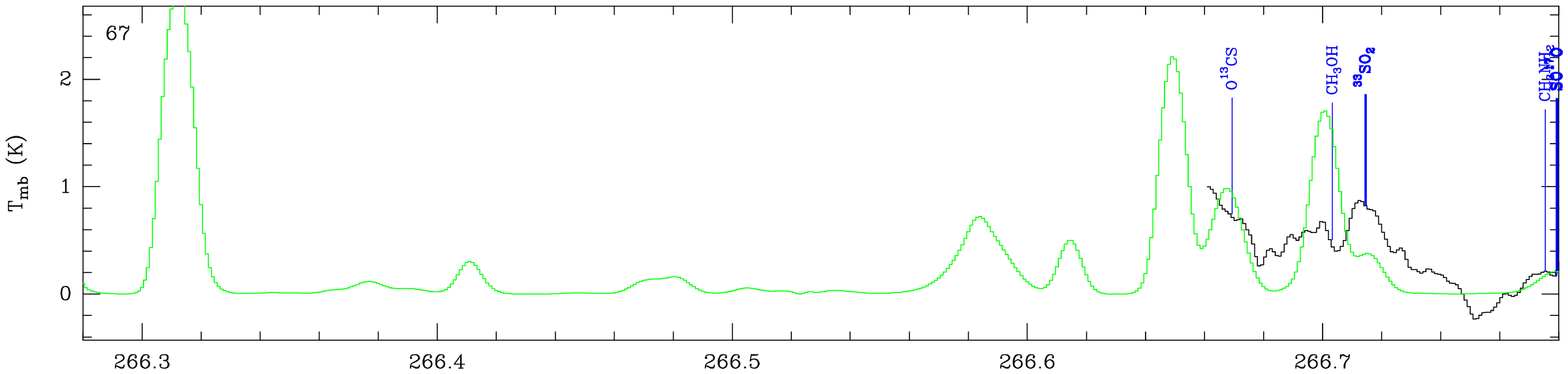}}}
\vspace*{1ex}\centerline{\resizebox{1.0\hsize}{!}{\includegraphics[angle=270]{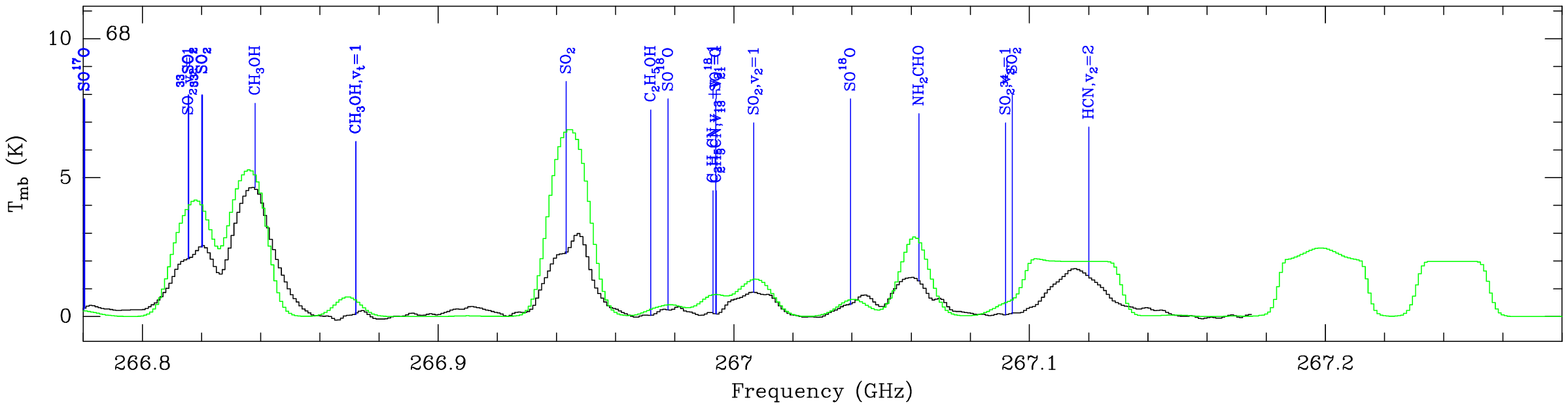}}}
\caption{
continued.
}
\end{figure*}
 \clearpage

}

\subsection{Statistics}
\label{ss:stats}

\begin{table*}
 \caption{
 Statistics about the identification of the lines detected above a threshold of 90~mK in the 3~mm spectra of Sgr~B2(N) and (M).
}
\label{t:stats}
 \centering
 \begin{tabular}{@{\extracolsep{-2.7pt}}rrrrrrrrrrcccccc}
 \hline\hline
\multicolumn{2}{c}{Frequency range} & \multicolumn{1}{c}{\hspace*{-0.9ex}$N_{\mathrm{tot}}$\tablefootmark{a}} & \multicolumn{1}{c}{\hspace*{-0.9ex}$N_{\mathrm{sel}}$\tablefootmark{b}} & \multicolumn{1}{c}{\hspace*{-0.9ex}$N_{\mathrm{det}}$\tablefootmark{c}} & \multicolumn{1}{c}{\hspace*{-0.9ex}$N_{\mathrm{id}}$\tablefootmark{d}} & \multicolumn{1}{c}{\hspace*{-0.9ex}$L_{\mathrm{det}}$\tablefootmark{e}} & \multicolumn{1}{c}{\hspace*{-0.9ex}$I_{\mathrm{obs}}$\tablefootmark{f}} & \multicolumn{1}{c}{\hspace*{-0.9ex}$I_{\mathrm{mod}}$\tablefootmark{g}} & \multicolumn{1}{c}{\hspace*{-0.9ex}$I_{\mathrm{diff}}$\tablefootmark{h}} & \multicolumn{1}{c}{\hspace*{-0.9ex}$I_{\mathrm{diffPos}}$\tablefootmark{i}} & \multicolumn{1}{c}{\hspace*{-0.9ex}$f_{N,\mathrm{sel}}$\tablefootmark{j}} & \multicolumn{1}{c}{\hspace*{-0.9ex}$f_{N,\mathrm{det}}$\tablefootmark{k}} & \multicolumn{1}{c}{\hspace*{-0.9ex}$f_{I,\mathrm{rec}}$\tablefootmark{l}} & \multicolumn{1}{c}{\hspace*{-0.9ex}$f_{N,\mathrm{rec}}$\tablefootmark{m}} & \multicolumn{1}{c}{\hspace*{-0.9ex}$f_{N,\mathrm{id}}$\tablefootmark{n}} \\ 
 \multicolumn{2}{c}{\scriptsize (GHz)} & \multicolumn{1}{c}{} & \multicolumn{1}{c}{} & \multicolumn{1}{c}{} & \multicolumn{1}{c}{} & \multicolumn{1}{c}{} & \multicolumn{2}{c}{\hspace*{-1.8ex}\scriptsize (K~km~s$^{-1}$)} & \multicolumn{2}{c}{\hspace*{-1.8ex}\scriptsize (K~km~s$^{-1}$)} & \multicolumn{1}{c}{\hspace*{-0.9ex}\scriptsize (\%)} & \multicolumn{1}{c}{\hspace*{-0.9ex}\scriptsize (\%)} & \multicolumn{1}{c}{\hspace*{-0.9ex}\scriptsize (\%)} & \multicolumn{1}{c}{\hspace*{-0.9ex}\scriptsize (\%)} & \multicolumn{1}{c}{\hspace*{-0.9ex}\scriptsize (\%)}  \\ 
 \multicolumn{2}{c}{(1)} & \multicolumn{1}{c}{\hspace*{-0.9ex}(2)} & \multicolumn{1}{c}{\hspace*{-0.9ex}(3)} & \multicolumn{1}{c}{\hspace*{-0.9ex}(4)} & \multicolumn{1}{c}{\hspace*{-0.9ex}(5)} & \multicolumn{1}{c}{\hspace*{-0.9ex}(6)} & \multicolumn{1}{c}{\hspace*{-0.9ex}(7)} & \multicolumn{1}{c}{\hspace*{-0.9ex}(8)} & \multicolumn{1}{c}{\hspace*{-0.9ex}(9)} & \multicolumn{1}{c}{\hspace*{-0.9ex}(10)} & \multicolumn{1}{c}{\hspace*{-0.9ex}(11)} & \multicolumn{1}{c}{\hspace*{-0.9ex}(12)} & \multicolumn{1}{c}{\hspace*{-0.9ex}(13)} & \multicolumn{1}{c}{\hspace*{-0.9ex}(14)} & \multicolumn{1}{c}{\hspace*{-0.9ex}(15)}  \\ 
\hline
\noalign{\smallskip}\multicolumn{5}{l}{Sgr~B2(N) } & & & & & & & & & \\ 
\noalign{\smallskip}  79.98 -- & \hspace*{-3ex}   90.00 & \hspace*{-0.9ex} 32033 & \hspace*{-0.9ex} 32033 & \hspace*{-0.9ex} 11444 & \hspace*{-0.9ex} 8393 & \multicolumn{1}{c}{\hspace*{-0.9ex} --} & \hspace*{-0.9ex}    4820 & \hspace*{-0.9ex}    4467 & \hspace*{-0.9ex}     353 & \hspace*{-0.9ex}     895 & \hspace*{-0.9ex} 100.0 & \hspace*{-0.9ex}  35.7 & \hspace*{-0.9ex}  81.4 & \hspace*{-0.9ex}  70.5 & \hspace*{-0.9ex}  73.3 \\ 
  90.00 -- & \hspace*{-3ex}  100.00 & \hspace*{-0.9ex} 32000 & \hspace*{-0.9ex} 32000 & \hspace*{-0.9ex} 14408 & \hspace*{-0.9ex} 10444 & \multicolumn{1}{c}{\hspace*{-0.9ex} --} & \hspace*{-0.9ex}    5048 & \hspace*{-0.9ex}    4914 & \hspace*{-0.9ex}     134 & \hspace*{-0.9ex}     880 & \hspace*{-0.9ex} 100.0 & \hspace*{-0.9ex}  45.0 & \hspace*{-0.9ex}  82.6 & \hspace*{-0.9ex}  70.4 & \hspace*{-0.9ex}  72.5 \\ 
 100.00 -- & \hspace*{-3ex}  110.00 & \hspace*{-0.9ex} 32000 & \hspace*{-0.9ex} 31990 & \hspace*{-0.9ex} 17978 & \hspace*{-0.9ex} 12612 & \multicolumn{1}{c}{\hspace*{-0.9ex} --} & \hspace*{-0.9ex}    6575 & \hspace*{-0.9ex}    7141 & \hspace*{-0.9ex}    -566 & \hspace*{-0.9ex}     958 & \hspace*{-0.9ex} 100.0 & \hspace*{-0.9ex}  56.2 & \hspace*{-0.9ex}  85.4 & \hspace*{-0.9ex}  69.6 & \hspace*{-0.9ex}  70.2 \\ 
 110.00 -- & \hspace*{-3ex}  116.00 & \hspace*{-0.9ex} 19152 & \hspace*{-0.9ex} 19152 & \hspace*{-0.9ex} 13465 & \hspace*{-0.9ex} 9454 & \multicolumn{1}{c}{\hspace*{-0.9ex} --} & \hspace*{-0.9ex}    7318 & \hspace*{-0.9ex}    7340 & \hspace*{-0.9ex}     -22 & \hspace*{-0.9ex}    1158 & \hspace*{-0.9ex} 100.0 & \hspace*{-0.9ex}  70.3 & \hspace*{-0.9ex}  84.2 & \hspace*{-0.9ex}  69.9 & \hspace*{-0.9ex}  70.2 \\ 
  79.98 -- & \hspace*{-3ex}  116.00 & \hspace*{-0.9ex} 115185 & \hspace*{-0.9ex} 115175 & \hspace*{-0.9ex} 57295 & \hspace*{-0.9ex} 40903 & \multicolumn{1}{c}{\hspace*{-0.9ex} --} & \hspace*{-0.9ex}   23761 & \hspace*{-0.9ex}   23861 & \hspace*{-0.9ex}    -101 & \hspace*{-0.9ex}    3890 & \hspace*{-0.9ex} 100.0 & \hspace*{-0.9ex}  49.7 & \hspace*{-0.9ex}  83.6 & \hspace*{-0.9ex}  70.0 & \hspace*{-0.9ex}  71.4 \\ 
\noalign{\smallskip}\multicolumn{5}{l}{Sgr~B2(N) without absorption parts} & & & & & & & & & \\ 
\noalign{\smallskip}  79.98 -- & \hspace*{-3ex}   90.00 & \hspace*{-0.9ex} 32033 & \hspace*{-0.9ex} 30733 & \hspace*{-0.9ex} 10144 & \hspace*{-0.9ex} 7160 &\hspace*{-0.9ex} 803 & \hspace*{-0.9ex}    4027 & \hspace*{-0.9ex}    3574 & \hspace*{-0.9ex}     453 & \hspace*{-0.9ex}     858 & \hspace*{-0.9ex}  95.9 & \hspace*{-0.9ex}  33.0 & \hspace*{-0.9ex}  78.7 & \hspace*{-0.9ex}  67.8 & \hspace*{-0.9ex}  70.6 \\ 
  90.00 -- & \hspace*{-3ex}  100.00 & \hspace*{-0.9ex} 32000 & \hspace*{-0.9ex} 31634 & \hspace*{-0.9ex} 14042 & \hspace*{-0.9ex} 10109 &\hspace*{-0.9ex} 993 & \hspace*{-0.9ex}    4874 & \hspace*{-0.9ex}    4712 & \hspace*{-0.9ex}     163 & \hspace*{-0.9ex}     864 & \hspace*{-0.9ex}  98.9 & \hspace*{-0.9ex}  44.4 & \hspace*{-0.9ex}  82.3 & \hspace*{-0.9ex}  70.0 & \hspace*{-0.9ex}  72.0 \\ 
 100.00 -- & \hspace*{-3ex}  110.00 & \hspace*{-0.9ex} 32000 & \hspace*{-0.9ex} 31645 & \hspace*{-0.9ex} 17633 & \hspace*{-0.9ex} 12337 &\hspace*{-0.9ex} 1131 & \hspace*{-0.9ex}    6465 & \hspace*{-0.9ex}    6978 & \hspace*{-0.9ex}    -514 & \hspace*{-0.9ex}     947 & \hspace*{-0.9ex}  98.9 & \hspace*{-0.9ex}  55.7 & \hspace*{-0.9ex}  85.4 & \hspace*{-0.9ex}  69.4 & \hspace*{-0.9ex}  70.0 \\ 
 110.00 -- & \hspace*{-3ex}  116.00 & \hspace*{-0.9ex} 19152 & \hspace*{-0.9ex} 18296 & \hspace*{-0.9ex} 12609 & \hspace*{-0.9ex} 8688 &\hspace*{-0.9ex} 752 & \hspace*{-0.9ex}    6618 & \hspace*{-0.9ex}    6555 & \hspace*{-0.9ex}      62 & \hspace*{-0.9ex}    1159 & \hspace*{-0.9ex}  95.5 & \hspace*{-0.9ex}  68.9 & \hspace*{-0.9ex}  82.5 & \hspace*{-0.9ex}  68.2 & \hspace*{-0.9ex}  68.9 \\ 
  79.98 -- & \hspace*{-3ex}  116.00 & \hspace*{-0.9ex} 115185 & \hspace*{-0.9ex} 112308 & \hspace*{-0.9ex} 54428 & \hspace*{-0.9ex} 38294 &\hspace*{-0.9ex} 3680 & \hspace*{-0.9ex}   21984 & \hspace*{-0.9ex}   21820 & \hspace*{-0.9ex}     164 & \hspace*{-0.9ex}    3828 & \hspace*{-0.9ex}  97.5 & \hspace*{-0.9ex}  48.5 & \hspace*{-0.9ex}  82.6 & \hspace*{-0.9ex}  69.0 & \hspace*{-0.9ex}  70.4 \\ 
\hline 
\noalign{\smallskip}\multicolumn{5}{l}{Sgr~B2(M) } & & & & & & & & & \\ 
\noalign{\smallskip}  79.98 -- & \hspace*{-3ex}   90.00 & \hspace*{-0.9ex} 32033 & \hspace*{-0.9ex} 32033 & \hspace*{-0.9ex} 4619 & \hspace*{-0.9ex} 3221 & \multicolumn{1}{c}{\hspace*{-0.9ex} --} & \hspace*{-0.9ex}    2928 & \hspace*{-0.9ex}    2702 & \hspace*{-0.9ex}     225 & \hspace*{-0.9ex}     464 & \hspace*{-0.9ex} 100.0 & \hspace*{-0.9ex}  14.4 & \hspace*{-0.9ex}  84.1 & \hspace*{-0.9ex}  66.3 & \hspace*{-0.9ex}  69.7 \\ 
  90.00 -- & \hspace*{-3ex}  100.00 & \hspace*{-0.9ex} 32000 & \hspace*{-0.9ex} 31984 & \hspace*{-0.9ex} 5050 & \hspace*{-0.9ex} 2523 & \multicolumn{1}{c}{\hspace*{-0.9ex} --} & \hspace*{-0.9ex}    2115 & \hspace*{-0.9ex}    1725 & \hspace*{-0.9ex}     389 & \hspace*{-0.9ex}     649 & \hspace*{-0.9ex} 100.0 & \hspace*{-0.9ex}  15.8 & \hspace*{-0.9ex}  69.3 & \hspace*{-0.9ex}  49.4 & \hspace*{-0.9ex}  50.0 \\ 
 100.00 -- & \hspace*{-3ex}  110.00 & \hspace*{-0.9ex} 32000 & \hspace*{-0.9ex} 31997 & \hspace*{-0.9ex} 6538 & \hspace*{-0.9ex} 3144 & \multicolumn{1}{c}{\hspace*{-0.9ex} --} & \hspace*{-0.9ex}    2145 & \hspace*{-0.9ex}    1602 & \hspace*{-0.9ex}     543 & \hspace*{-0.9ex}     661 & \hspace*{-0.9ex} 100.0 & \hspace*{-0.9ex}  20.4 & \hspace*{-0.9ex}  69.2 & \hspace*{-0.9ex}  47.7 & \hspace*{-0.9ex}  48.1 \\ 
 110.00 -- & \hspace*{-3ex}  116.00 & \hspace*{-0.9ex} 19152 & \hspace*{-0.9ex} 19152 & \hspace*{-0.9ex} 5329 & \hspace*{-0.9ex} 2686 & \multicolumn{1}{c}{\hspace*{-0.9ex} --} & \hspace*{-0.9ex}    4862 & \hspace*{-0.9ex}    3683 & \hspace*{-0.9ex}    1180 & \hspace*{-0.9ex}    1630 & \hspace*{-0.9ex} 100.0 & \hspace*{-0.9ex}  27.8 & \hspace*{-0.9ex}  66.5 & \hspace*{-0.9ex}  50.1 & \hspace*{-0.9ex}  50.4 \\ 
  79.98 -- & \hspace*{-3ex}  116.00 & \hspace*{-0.9ex} 115185 & \hspace*{-0.9ex} 115166 & \hspace*{-0.9ex} 21536 & \hspace*{-0.9ex} 11574 & \multicolumn{1}{c}{\hspace*{-0.9ex} --} & \hspace*{-0.9ex}   12050 & \hspace*{-0.9ex}    9713 & \hspace*{-0.9ex}    2337 & \hspace*{-0.9ex}    3405 & \hspace*{-0.9ex} 100.0 & \hspace*{-0.9ex}  18.7 & \hspace*{-0.9ex}  71.7 & \hspace*{-0.9ex}  52.7 & \hspace*{-0.9ex}  53.7 \\ 
\noalign{\smallskip}\multicolumn{5}{l}{Sgr~B2(M) without absorption parts} & & & & & & & & & \\ 
\noalign{\smallskip}  79.98 -- & \hspace*{-3ex}   90.00 & \hspace*{-0.9ex} 32033 & \hspace*{-0.9ex} 30332 & \hspace*{-0.9ex} 2918 & \hspace*{-0.9ex} 1670 &\hspace*{-0.9ex} 186 & \hspace*{-0.9ex}    1468 & \hspace*{-0.9ex}    1156 & \hspace*{-0.9ex}     312 & \hspace*{-0.9ex}     426 & \hspace*{-0.9ex}  94.7 & \hspace*{-0.9ex}   9.6 & \hspace*{-0.9ex}  71.0 & \hspace*{-0.9ex}  51.7 & \hspace*{-0.9ex}  57.2 \\ 
  90.00 -- & \hspace*{-3ex}  100.00 & \hspace*{-0.9ex} 32000 & \hspace*{-0.9ex} 31124 & \hspace*{-0.9ex} 4190 & \hspace*{-0.9ex} 1948 &\hspace*{-0.9ex} 238 & \hspace*{-0.9ex}    1670 & \hspace*{-0.9ex}    1206 & \hspace*{-0.9ex}     464 & \hspace*{-0.9ex}     627 & \hspace*{-0.9ex}  97.3 & \hspace*{-0.9ex}  13.5 & \hspace*{-0.9ex}  62.4 & \hspace*{-0.9ex}  44.8 & \hspace*{-0.9ex}  46.5 \\ 
 100.00 -- & \hspace*{-3ex}  110.00 & \hspace*{-0.9ex} 32000 & \hspace*{-0.9ex} 31510 & \hspace*{-0.9ex} 6051 & \hspace*{-0.9ex} 2777 &\hspace*{-0.9ex} 314 & \hspace*{-0.9ex}    2011 & \hspace*{-0.9ex}    1479 & \hspace*{-0.9ex}     532 & \hspace*{-0.9ex}     632 & \hspace*{-0.9ex}  98.5 & \hspace*{-0.9ex}  19.2 & \hspace*{-0.9ex}  68.6 & \hspace*{-0.9ex}  46.0 & \hspace*{-0.9ex}  45.9 \\ 
 110.00 -- & \hspace*{-3ex}  116.00 & \hspace*{-0.9ex} 19152 & \hspace*{-0.9ex} 18056 & \hspace*{-0.9ex} 4233 & \hspace*{-0.9ex} 1734 &\hspace*{-0.9ex} 203 & \hspace*{-0.9ex}    3762 & \hspace*{-0.9ex}    2389 & \hspace*{-0.9ex}    1373 & \hspace*{-0.9ex}    1699 & \hspace*{-0.9ex}  94.3 & \hspace*{-0.9ex}  23.4 & \hspace*{-0.9ex}  54.8 & \hspace*{-0.9ex}  37.0 & \hspace*{-0.9ex}  41.0 \\ 
  79.98 -- & \hspace*{-3ex}  116.00 & \hspace*{-0.9ex} 115185 & \hspace*{-0.9ex} 111022 & \hspace*{-0.9ex} 17392 & \hspace*{-0.9ex} 8129 &\hspace*{-0.9ex} 942 & \hspace*{-0.9ex}    8911 & \hspace*{-0.9ex}    6229 & \hspace*{-0.9ex}    2682 & \hspace*{-0.9ex}    3384 & \hspace*{-0.9ex}  96.4 & \hspace*{-0.9ex}  15.7 & \hspace*{-0.9ex}  62.0 & \hspace*{-0.9ex}  44.5 & \hspace*{-0.9ex}  46.7 \\ 
\hline
 \end{tabular}
 \tablefoot{
\tablefoottext{a}{Total number of channels in the selected frequency range.}
 \tablefoottext{b}{Number of selected channels, excluding the blanked channels, and those in the absorption parts of the spectrum when explicitely mentioned.}
 \tablefoottext{c}{Number of channels above the detection threshold (90~mK in main-beam temperature scale).}
\tablefoottext{d}{Number of identified channels, i.e. channels for which the temperature of the synthetic spectrum reaches at least 50\% of the observed temperature.}
\tablefoottext{e}{Estimated number of lines detected in emission.}
 \tablefoottext{f}{Integrated intensity of the absolute value of the observed spectrum over the range of detected channels.}
 \tablefoottext{g}{Integrated intensity of the synthetic spectrum over the range of detected channels.}
 \tablefoottext{h}{Difference between Cols.~7 and 8 ($I_{\mathrm{obs}} - I_{\mathrm{mod}}$).}
\tablefoottext{i}{Integrated intensity of the difference between the absolute value of the observed spectrum and the absolute value of the clipped synthetic spectrum over the range of detected channels. The synthetic spectrum is clipped to the observed spectrum in the channels for which the synthetic temperature is higher than the observed temperature in absolute value.}
 \tablefoottext{j}{Fraction of selected channels ($N_{\mathrm{sel}}/N_{\mathrm{tot}}$).}
 \tablefoottext{k}{Fraction of detected channels ($N_{\mathrm{det}}/N_{\mathrm{sel}}$).}
 \tablefoottext{l}{Fraction of recovered integrated intensity ($1-I_{\mathrm{diffPos}}/I_{\mathrm{obs}}$).}
\tablefoottext{m}{Average fraction of recovered temperature per channel, after clipping the synthetic spectrum like in Col.~10.}
 \tablefoottext{n}{Fraction of identified channels ($N_{\mathrm{id}}/N_{\mathrm{det}}$).}
}
 \end{table*}

The synthetic spectra of Sgr~B2(N) and (M) containing all the detected
species are used to estimate the fraction of lines that remain unidentified in 
the 3~mm range. The statistics about the identification are summarized in 
Table~\ref{t:stats}. The analysis is performed over four frequency ranges, 
as well as over the full frequency coverage from 80 to 116~GHz. The goal is to 
estimate the fraction of channels for which the detected emission (or 
absorption) can be considered as identified based on our model. For each 
source, the calculation is done twice, once including both the parts in 
emission and in absorption, and the second time including only the emission 
part. The analysis is performed on the channels that have a main-beam 
temperature higher than 90~mK ($4\sigma$) in absolute value. The number of such 
``detected'' channels is given in Col.~4. About 50\% of all channels are 
``detected'' toward Sgr~B2(N), and 19\% toward Sgr~B2(M) (Col.~12). Given that
the detected lines have wings extending below the $4\sigma$ threshold and other
lines exist between 3 and $4\sigma$, we can conclude that less than 50\% of 
all channels are free of emission or absorption toward Sgr~B2(N), meaning that
its 3~mm spectrum is close to the confusion limit.

We measure the performance of our model in two ways. First we compute the
average fraction of the observed spectrum that is accounted for by our model
based on the integrated intensities. The integrated intensities of the observed
and synthetic spectra are listed in Cols.~7 and 8. Their difference is given in
Col.~9, but, given that the model sometimes overestimates the observed 
signal\footnote{As described in Sect.~\ref{ss:xclass}, XCLASS
adds the contributions of different emission components linearly. This can 
lead the model to overestimate the observed emission when spectrally blended
transitions of two different molecules or two components of the same molecule 
with similar velocities (e.g. a cold one and a warm one) overlap along the 
line of sight and the one in the front is partially optically thick. The model 
can also overestimate the observed emission if the level of the baseline of
the observed spectrum is overestimated due to line confusion, or if there
is a gradient of kinetic temperature along the line of sight.}, 
a more relevant quantity is the integrated intensity of the difference spectrum
after clipping the synthetic spectrum to the observed one in the channels 
where the former overestimates the latter. The resulting integrated intensity
is given in Col.~10. It corresponds to the integrated intensity of the signal
that is not accounted for by the model. Using the values of Cols.~7 and 10, we 
find that 83\% of the part of the spectrum of Sgr~B2(N) detected in emission 
is well reproduced by our model, and 62\% for Sgr~B2(M) (Col.~13). This 
estimate is biased in the sense that channels that have strong emission or 
absorption have more weight than channels that have weak emission or absorption.

The second method, which does not suffer from the previous bias, is performed 
channel-wise. For each ``detected'' channel, we compute the fraction of its 
emission that is recovered by the model (clipping this fraction to 1 when the 
model overestimates the observed signal). We find that, on average, 69\% of 
the emission of the ``detected'' channels toward Sgr~B2(N) is recovered by 
the model, and 45\% toward Sgr~B2(M) (Col. 14).

We set the identification threshold to 50\%, i.e. we consider the 
emission/absorption in a channel to be identified when the synthetic spectrum 
accounts for at least 50\% of the detected level of emission/absorption. The
number of such ``identified'' channels is given in Col.~5. They represent 70\%
of the channels ``detected'' in emission toward Sgr~B2(N), and 47\% toward
Sgr~B2(M) (Col.~15). These fractions are very similar to the ones obtained in
the previous paragraph because we assumed an identification threshold of 50\%.
They would be lower if we would increase this threshold.

Finally, Col.~6 gives an \textit{estimate} of the number of detected lines. 
This estimate is done by assuming a typical line width in velocity (a 
priori \textit{not} equal to the typical $FWHM$) such that the derived number 
of lines detected in the full 3~mm window corresponds to the one obtained by
eye in Sect.~\ref{ss:overview}. The typical line widths we had to use to have a
good match are 13.9 and 17.2 km~s$^{-1}$ for Sgr~B2(N) and (M), respectively.
Column~6 shows that the density of detected lines increases with frequency for 
both sources, by a factor of 1.6 for Sgr~B2(N) and 1.7 for Sgr~B2(M) from 80
to 116~GHz.

\begin{figure*}
\centerline{\resizebox{1.0\hsize}{!}{\includegraphics[angle=0]{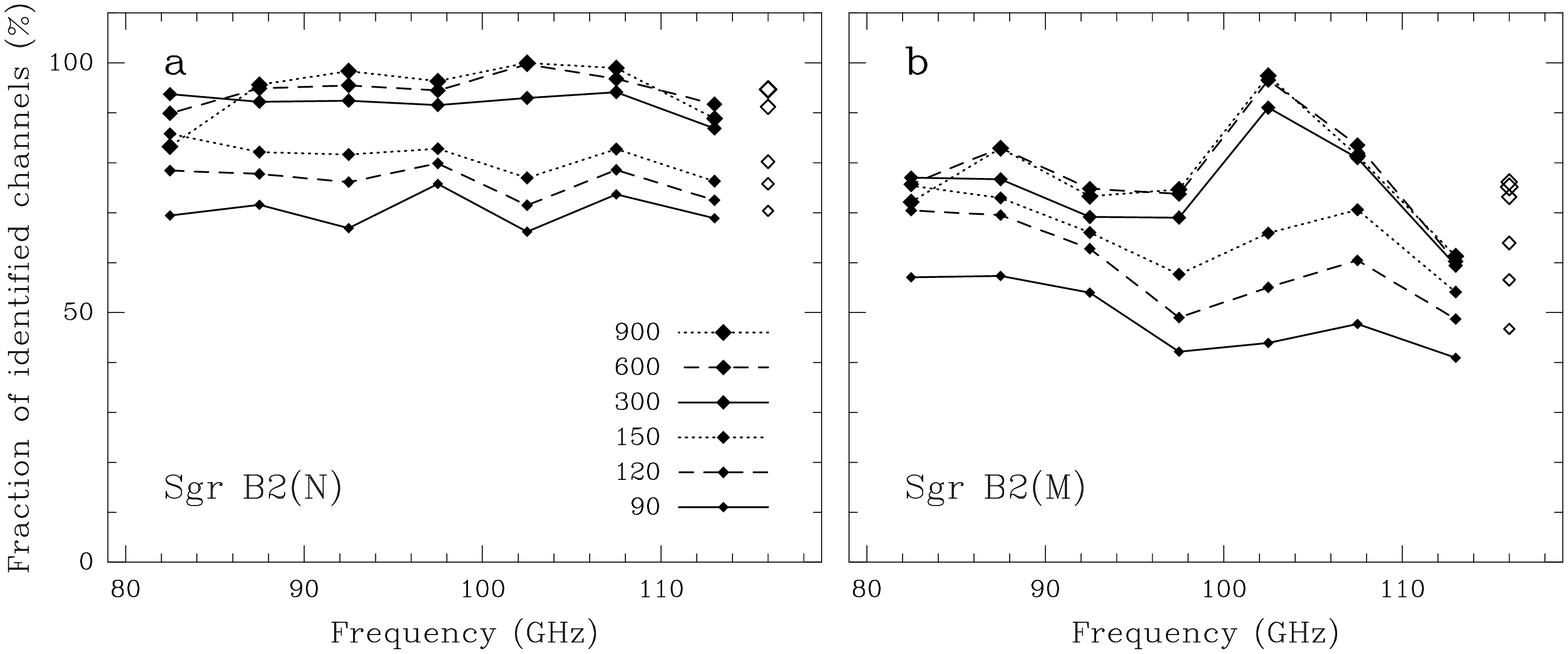}}}
\caption{Fraction of identified channels in the spectra of \textbf{a} Sgr~B2(N) 
and \textbf{b} Sgr~B2(M) as a function of frequency. Each curve
corresponds to a different temperature threshold, which is given in mK in the
lower right corner of panel \textbf{a}. Each data point (filled diamond) 
corresponds to the average fraction of identified channels over a frequency 
range of 5~GHz, except for the point at highest frequency which is an 
average over 6~GHz. The unfilled diamonds to the right of each panel are the 
average fractions over the full frequency range 80 to 116 GHz, for the same
temperature thresholds. The size of the filled and unfilled diamonds increases 
with the temperature threshold.}
\label{f:stats}
\end{figure*}

Figure~\ref{f:stats} plots the fraction of identified channels as a function
of frequency for each source. This calculation was done for different 
thresholds in main-beam temperature, from 90 to 900~mK (4 to $40\sigma$). There
is no obvious trend in frequency, except maybe for Sgr~B2(M) with a slight
decrease in the fraction of identified channels with increasing frequency.
But there is a clear increase in the fraction of identified channels as a 
function of temperature threshold, from 70\% to 95\% for 
Sgr~B2(N) and from 47\% to 75\% for Sgr~B2(M).

The fraction of identified lines above a threshold of 0.9~K estimated in the 
previous paragraph is affected by the shortcomings of the model. We counted by 
eye 388 identified lines out of 411 lines detected above this threshold toward 
Sgr~B2(N), and 80 out of 103 toward Sgr~B2(M). However, the observed CO and 
$^{13}$CO 1--0 lines are very strong and consist of multiple components both 
in emission and absorption. The model was not fine-tuned to fit all the 
emission components because the LTE assumption is certainly wrong for these 
components. If we ignore the frequency ranges  where the CO and 
\hbox{$^{13}$CO~1--0}
transitions are detected, then we find 367 identified lines out of 371 
detected toward Sgr~B2(N) and 68 out of 72 toward Sgr~B2(M). The four lines 
missing in the model of each source correspond to H$\alpha$ recombination 
lines that cannot be modeled with XCLASS, at 85.7, 92.0, 99.0, and 106.7~GHz. 
All the lines (100\%) stronger than 0.9~K are therefore identified in the 3~mm 
spectra of Sgr~B2(N) and (M). The somewhat lower fractions obtained in the 
previous paragraph (95\% and 75\%, respectively) are due to the unmodeled 
H$\alpha$ emission, the complicated CO and $^{13}$CO 1--0 emission, and a 
significant underestimate of the strength of the methanol transitions 
at 84.5 and 95.2~GHz (see Sects.~\ref{sss:ch3oh} and 
\ref{ss:discussion_ch3oh}).

\subsection{Overview of the source properties}
\label{ss:source}

\paragraph{Sgr\,\,B2(N):} About 75\% of the molecules (main 
isotopologues) tracing 
warm/hot gas ($> 30$~K) toward Sgr~B2(N) emit in two velocity components at LSR 
velocities of $\sim$ 63--64 and 73--74~km~s$^{-1}$ (e.g., CH$_3$C$_3$N, 
CH$_3$CHO). They correspond to the two hot cores embedded in Sgr~B2(N) and 
separated by 5$\arcsec$ in the north-south direction 
(see Sect.~\ref{s:intro}). The linewidth of each velocity component is 
typically on the order of 7~km~s$^{-1}$. Six complex organic molecules emit
in only one component (C$_2$H$_5$OCHO, \textit{c}-C$_2$H$_4$O, 
\textit{aGg'}-(CH$_2$OH)$_2$, CH$_2$(OH)CHO, CH$_3$C(O)NH$_2$, NH$_2$CH$_2$CN).
This is most likely due to a lack of sensitivity to the less prominent, 
northern hot core.

A few species with strong emission show the 
presence of additional wing emission (e.g., HC$_3$N, C$_2$H$_5$CN). At 
higher angular resolution, this wing emission is resolved in the east-west 
direction 
\citep[see, e.g., HC$_3$N $\varv_7 = 1$ in Figs.~5k, l, and m of][, or 
\citeauthor{Lis93} \citeyear{Lis93}, \citeauthor{Liu99} \citeyear{Liu99},
\citeauthor{Hollis03} \citeyear{Hollis03}]{Belloche08}. Given that this wing 
emission is also traced by SO, SO$_2$, and SiO, it may be due to an outflow 
\citep[see, e.g.,][, which show that these molecules are good outflow 
tracers]{RodriguezFernandez10,Tafalla10}. Higher angular resolution with, 
e.g., ALMA will certainly help in understanding the physical origin of this 
wing emission. Here, we use one or several additional (Gaussian) component(s) 
at blue- and/or red-shifted velocities to account for the extra wing emission 
when necessary. The Gaussian approximation is a priori not optimum to model 
wing emission if it is produced by an outflow. This can prevent us from 
perfectly matching the model to the observed emission in the cases where wing 
emission is present.

Most simple molecules are less sensitive to the two hot cores embedded 
in Sgr~B2(N) and rather trace colder, more extended emission in the envelope.
They are often self-absorbed, and also suffer from absorption by multiple 
diffuse clouds along the line-of-sight.

The rotation temperatures derived for the complex molecules detected 
toward Sgr~B2(N) range from 50~K (e.g., \textit{aGg'-}(CH$_2$OH)$_2$) to 200~K
(e.g., CH$_3$CN, CH$_3$OH). A few vibrationally or torsionally excited states
seem to require higher rotation temperatures (230~K for HC$_3$N, $\varv_7=1$,
500~K for HCN, $\varv_2=1$).

\paragraph{Sgr\,\,B2(M):} The emission of most complex molecules 
detected toward Sgr~B2(M) is
relatively well reproduced with only one velocity component, with linewidths 
ranging from $\sim$ 6 to 15~km~s$^{-1}$ and LSR velocities from $\sim$ 60 to 
66~km~s$^{-1}$. The simpler molecules tracing colder, more extended emission 
have linewidths up to $\sim$ 22~km~s$^{-1}$.

The rotation temperatures derived for the complex molecules detected 
toward Sgr~B2(M) range from 30--40~K (e.g., CH$_3$C(O)CH$_3$, CH$_3$CHO) to 
200~K (e.g., CH$_3$CN, CH$_3$OH).

\subsection{Detected molecules}
\label{ss:molecules}

The model parameters for each detected molecule are listed in 
Tables~\ref{t:modparam_lmh_c2h3cn} to \ref{t:modparam_b2m_so} (online 
material). In this 
section we briefly describe the fit to the emission or absorption lines of 
each detected molecule. When the frequency range is not mentioned 
(e.g., when the maximum opacity of the modeled transitions is 
mentioned), it is by default the 3~mm range.

For a catalog entry containing a vibrationally or torsionally excited
state (with the level energies including the vibrational or torsional energy), 
each column density given in Tables~\ref{t:modparam_lmh_c2h3cn} to 
\ref{t:modparam_b2m_so} represents the total column density of the molecule
(assuming LTE), not the partial column density of the excited state only. If 
all levels of a molecule are populated following LTE, then our modeling should 
yield the same column density parameter for the entry of the ground state and 
the entries of all vibrationally or torsionally excited states. When this is 
not the case, we estimate a vibration (or torsion) temperature for each 
excited state with the following equation:
\begin{equation}
\frac{N_j}{N_i} = \frac{e^{-\Delta E_{\mathrm{vib}}/k T_{\mathrm{vib}}}}{e^{-\Delta E_{\mathrm{vib}}/k T_{\mathrm{LTE}}}} \, ,
\end{equation}
which yields
\begin{equation}
\label{e:tvib}
T_{\mathrm{vib}} = \frac{1}{\frac{1}{T_{\mathrm{LTE}}}-\frac{k}{\Delta E_{\mathrm{vib}}} \ln{\frac{N_j}{N_i}}} \, ,
\end{equation}
with $T_{\mathrm{vib}}$ the vibration (or torsion) temperature, 
$T_{\mathrm{LTE}}$ the 
LTE temperature (equal to the rotation temperature), $E_j$ the vibration (or 
torsion) energy of the excited state, $E_i$ the vibration (or torsion) energy 
of the reference state, $\Delta E_{\mathrm{vib}} = E_j -E_i$, $N_j$ and $N_i$ 
the column density parameters (i.e. total column density of the molecule, see 
above) derived from our LTE modeling and corresponding to the entries of the 
excited and reference states, respectively, and $k$ the Boltzmann constant. 
When the LTE models for the ground state and all vibrationally or torsionally 
excited states have the same source size and rotation temperature, we use the 
ground state as reference state to compute the vibration (or torsion) 
temperature of each excited state. When it is not the case, we indicate which
excited state is used as reference to 
compute the vibration (or torsion) temperature of each higher excited 
state. Assuming that the relative uncertainty $U_{\mathrm{vib}}$ on 
$T_{\mathrm{vib}}$ only depends on the relative uncertainty $U_N$ on the 
column density ratio, we derive
\begin{equation}
U_{\mathrm{vib}} = \frac{\Delta T_{\mathrm{vib}}}{T_{\mathrm{vib}}} = 
\frac{k T_{\mathrm{vib}}}{\Delta E_{\mathrm{vib}}}\,U_N \, .
\end{equation}
With a typical relative uncertainty ($\sim 3\sigma$) of 10\% on each 
column density $N_i$ and $N_j$ (for source sizes and rotation temperatures 
kept fixed), $U_N$ is about 14\%.

Uncertainties are not given in Tables~\ref{t:modparam_lmh_c2h3cn} to 
\ref{t:modparam_b2m_so}. Because of the blendings of multiple components and 
species, degeneracies of the parameters in certain cases (e.g., between source 
size and column density for optically thin emission), small number or limited 
energy range of detected transitions for some molecules (preventing a reliable 
estimate of the temperature), limitations due to the simplicity of the 
model (e.g., uniform temperature, non-interacting components, same 
background
temperature for compact and extended components, dust attenuation not accounted
for), uncertainties are very hard to quantify in a satisfactory way. Cases 
when some or even all parameters are poorly or not even constrained are 
explicitely mentioned in the following subsections and we invite the reader to 
carefully read them before using the values listed in 
Tables~\ref{t:modparam_lmh_c2h3cn} to \ref{t:modparam_b2m_so}. In the general 
case, the uncertainties on source size, temperature, and column density 
are probably not smaller than 20--30\%. Linewidth and velocity are certainly
better constrained.

Population (or ``Boltzmann'' or ``rotation'') diagrams are not shown 
in this article. Examples of such diagrams can be found in 
\citet{Belloche08,Belloche09}, where the optical depths effects were properly 
taken into account. \citet{Belloche09} even showed an improved version of 
such diagrams by removing the contribution of contaminating species, based on 
the complete model presented in the present article. These diagrams illustrate
the reliability of our modeling approach.

We provide information on the source of the rest frequencies, e.g. 
the CDMS or JPL catalogs, as well as information on the laboratory data 
on which these rest frequencies are based\footnote{It is worthwhile mentioning 
that the CDMS and JPL catalogs are intended to be available together in a 
database environment in the near future. Recommendations, information on 
laboratory rest frequencies, as well as additional information will eventually 
be automatically provided with each request. The implementation of the CDMS 
into a database environment within the framework of the Virtual Atomic and 
Molecular Data Centre \citep[VAMDC;][]{Dubernet10} is approaching the official 
release of a test version.}. 
The number of detected species reported in this section is 
large. Therefore we limit the information on laboratory data somewhat because 
the number of references is rather large for some species. We provide in all 
instances a primary reference that usually summarizes the (mostly laboratory) 
rest frequencies used to generate the catalog entry or that provides 
spectroscopic parameters most closely resembling those employed for the 
catalog entry. The primary reference may or may not contain measurements in the 
frequency ranges of our line survey. We also mention references for rest 
frequencies that cover the ranges of this survey. Dipole moment information is 
given in special cases only. Not all entries from the CDMS or JPL catalogs or 
from private sources include all of the presently available data. This 
shortcoming should have a small, probably negligible effect on our 
results. Several of the very old entries as well as a small number of recent 
entries provide incomplete lists of references in the corresponding 
documentation files. We have inspected the references given in the 
documentation files for previous original rest frequency determinations in the 
range of our survey.

As described in Sect.~\ref{ss:xclass}, the partition function at any temperature
$T$ is computed based on a global linear fit performed by XCLASS in log-log
space to the values tabulated between 9.375 and 300~K. This method is not very
accurate when the partition function contains contributions of vibrationally 
or torsionally excited states in addition to the ground state. The extrapolation
at temperatures below 9.375~K (for the absorption components) or above 300~K 
based on this fit is sometimes also inaccurate. Therefore,  in order to 
compensate for these inaccuracies, we applied a correction to the derived 
column densities \textit{a posteriori}. The corrected partition functions were 
computed with a piecewise, linear interpolation in the range 9.375 to 300~K.
Most species with components below 9.375~K have partition function values at 
2.725 and 5~K in the CDMS catalog which allowed for a similar piecewise 
interpolation. For the species lacking these low-temperature values, we 
extrapolated the partition function from a linear fit (in log-log space) to the 
values at 9.375 and 18.75~K. We also corrected the partition functions for 
the contribution of excited states, for as many species as we could. The
global correction factor used to rescale the column density is listed in 
Col.~7 of the table listing the model parameters for each species.

\subsubsection{Vinyl cyanide C$_2$H$_3$CN}
\label{sss:c2h3cn}

We use the CDMS entries for the ground state of the $^{12}$C and $^{13}$C 
isotopologues (tags 53515, 54506, 54507, 54508, all version 1). 
The entries were based on data described in \citet{Mueller08}. This work 
includes in particular laboratory data from \citet{VyCN_1973},
\citet{VyCN_1996}, and \citet{VyCN-vib_1988} providing rest frequencies for the 
main isotopic species in the range of our line survey. In the case of the 
$^{13}$C isotopologues, \citet{isos_VyCN_1997} cover additional frequency 
ranges similar to the ones of our survey.

Recently, \citet{Krasnicki11} redetermined the dipole
moment components of vinyl cyanide and ethyl cyanide. While most values
agreed well with previous measurements, the $b$-component of vinyl cyanide
turned out to be 23\% smaller than the value currently used in the CDMS.
This has not been taken into account in our model yet.

The catalog entries for the vibrationally excited states of the main 
isotopologue were prepared by one of us (HSPM).
The predictions were based on assignments by \citet{VyCN-vib_1988} for 
$\varv_{11} = 1$, $\varv_{15} = 1$, $\varv_{11} = 2$, and $\varv_{11} = 3$ 
obtained in the frequency range 20--40~GHz as well as parts of the 
100--184~GHz range. Additional $\varv_{11} = 1$ data (350--630~GHz) 
were taken from \citet{VyCN_1994}. Further measurements of selected 
transition frequencies pertaining to a variety of vibrational states 
were made between 64 and 119~GHz as well as between 239 and 353~GHz 
as described for the ground state of the main species and the singly $^{13}$C 
and $^{15}$N isotopic species \citep{Mueller08}. Broader scans were 
also taken in the lower frequency range.

Pseudoequilibrium spectroscopic parameters were determined 
in that unpublished study. Vibrational corrections for $\varv_{11}$ 
up to third order ensured a good reproduction for $\varv_{11} = 1$, 
a reasonably good reproduction for $\varv_{11} = 2$ and a somewhat 
satisfactory reproduction for $\varv_{11} = 3$. 
Systematic deviations between measured and calculated frequencies 
suggested $\varv_{11} = 3$ to be severly perturbed by one or more 
vibrational states. Some of these deviations were already observed 
by \citet{VyCN-vib_1988}, but were not discussed, presumably because 
they appeared to be rather random. Assignments in the 3~mm range and 
below were secure at least in most cases even if the frequencies were 
not reproduced well. Assignments in the 239--353~GHz range are 
less certain for $\varv_{11} = 3$.

First order $\varv _{15}$ corrections for some spectroscopic parameters not 
only accounted quite well for the $\varv _{15} = 1$ data, but also permitted 
assignments for $\varv _{15} = 2$ and for $\varv _{11} = \varv _{15} = 1$ 
to be made for the first time. These transition frequencies were better
reproduced by including a small number of second order vibrational corrections. 
Some of the assignments in the 3~mm region and below appear to be safe, 
those at higher frequencies are probably less certain as both states 
are perturbed, the former possibly by $\varv _{11} = 3$ and very likely 
by $\varv _{14} = 1$, the latter by $\varv _{10} = 1$.

\citet{VyCN_2009} recently extended the analyses of vinyl cyanide 
in its ground and first excited $\varv _{11} = 1$ states considerably in 
frequency. Even though $\varv _{11} = 1$ is much higher than the ground 
vibrational state, these authors observed perturbations between these states 
which permitted the $\varv _{11} = 1$ origin to be determined very accurately 
to 228.300~cm$^{-1}$ or 328.47~K. The effective matrix elements of such 
interactions usually decrease rapidly with increasing $\Delta K_a$, but the 
effect for a given $\Delta K_a$ increases with $J$ and $K_a$. Very local 
effects were observed for rather high values of $\Delta K_a$ (6 and 5) with 
resonant interactions beyond $J = 100$. There were also more pronounced 
near-resonant $\Delta K_a = 4$ interactions, however, these occured at rather 
high values of $K_a$ (21/17 and 22/18), and the effects became noticeable 
in the spectrum only above $J \approx 40$; their strong $a$-type $R$-branch 
transitions are at almost 400~GHz or higher. Transition within 
$\varv _{11} = 1$ observed in our line survey are thus unaffected by these 
perturbations.

More recently, \citet{VyCN_isos_2011} investigated perturbations between 
the ground state and $\varv_{11} = 1$ for four singly substituted isotopic 
species and published also data for the ground vibrational states of some 
doubly substituted species. The investigations of the main isotopologue 
were extended to $\varv_{15} = 1$ and $\varv_{11} = 2$ which perturb 
each other as well as $\varv_{11} = 1$ \citep{VyCN_2012}. The band origins 
were determined very accurately to 332.678 and 457.175~cm$^{-1}$, respectively.
Transitions within $\varv_{15} = 1$ observed in our line survey are 
probably unaffected by perturbations. The situation is less clear for 
$\varv_{11} = 2$. Transitions pertaining to even higher lying vibrational 
states may well be affected by perturbations, and identifications in our 
survey should be viewed with some caution.

\paragraph{Sgr\,\,B2(N):} The parameters of all entries related to vinyl 
cyanide are listed in Table~\ref{t:modparam_lmh_c2h3cn}. The first detection 
in space of the $^{13}$C isotopologues based on this survey was already 
reported in \citet{Mueller08}. A $^{12}$C to $^{13}$C isotopic ratio of 21 was 
derived in that work, in excellent agreement with the value of 20 determined by 
\citet{Wilson94}. The lines of these isotopologues are optically 
thin ($\tau_{\rm max} \sim 0.3$), so their relative intensities give a direct
constraint on the temperature, which is relatively high due to the detection 
of transitions with $E_l/k$ up to $\sim 200$~K (e.g., 168~K at 
113.45~GHz, 235~K at 103.96~GHz). Many transitions of the main isotopologue 
are optically thick ($\tau_{\rm max} \sim 7$). With the temperature derived 
from the $^{13}$C isotopologues, the optically thick $^{12}$C transitions in 
turn constrain the size of the emitting region. The column density is then 
constrained by the optically thin lines. We see no obvious sign of column 
density differences, i.e. differential fractionation, between the three 
$^{13}$C isotopologues. 

The northern hot core is also clearly detected at, e.g., 
84.95 and 85.30~GHz for the main isotopologue, and 85.00 and 85.28~GHz for 
the $^{13}$C isotopologues. In addition, a third velocity component is 
necessary to reproduce the blue wing of the optically thick $^{12}$C 
transitions (e.g., 84.95 and 85.30~GHz). Its temperature and source size
are not well constrained. We use the same values as for the two main 
components.

We also detect emission lines from within six vibrationally excited states of 
the main isotopologue: $\varv_{11} = 1$, $\varv_{15} = 1$, $\varv_{11} = 2$, 
$\varv_{11} = \varv_{15} = 1$, $\varv_{15} = 2$, $\varv_{11} = 3$ 
\citep[229.4, 333.3, 458.8, 562.7, 666.6, 688.2~cm$^{-1}$, respectively,
derived from the apparent band centers of $\nu _{11}$ and $\nu _{15}$ in the 
far-infrared spectrum,][]{VyCN_FIR_1973}.  We fixed the rotation
temperature to the value obtained from the ground 
state transitions but we had to reduce the source size to fit the optically 
thick lines of $\varv_{11} = 1$ and $\varv_{15} = 1$ ($\tau_{\rm max} \sim 4$ 
and 3, respectively). The northern hot core is also clearly detected for 
$\varv_{11} = 1$, $\varv_{15} = 1$, and $\varv_{11} = 2$  (e.g., 85.99, 106.75, 
and 105.64~GHz, respectively) and it is likely detected for 
$\varv_{11} = \varv_{15} = 1$ (e.g., 114.38~GHz). There may be signs of an 
additional blueshifted component for $\varv_{11} = 1$ at certain frequencies 
(e.g., 114.25 and 114.28~GHz), but it is less obvious at other frequencies so 
we did not include it in the model. The transitions in the higher vibrational 
states ($\varv_{11} = 2$ and higher) are optically thin 
($\tau_{\rm max} =$ 0.4--0.9), so their source size is not constrained.
We set it to the size derived for $\varv_{11} = 1$ and $\varv_{15} = 1$.
Finally, we note that transitions from within $\varv_{10} = 1$ (560 cm$^{-1}$) 
should be easily detected, but no spectroscopic data are available yet. They
may be as strong and numerous as the transitions from within 
$\varv_{11} = \varv_{15} = 1$ detected here (a dozen at 3~mm with peak 
temperatures ranging from 0.1 to 0.4~K).

While the rotation temperature derived from the ground state fits the 
relative intensities of the lines from within the vibrationally excited states 
well, the column density parameter required to fit these excited 
states increases with 
their energy, which means that the ``vibration'' temperature is 
actually higher than the rotation temperature (170~K). This 
could be due to radiative pumping. Using equation~\ref{e:tvib} with 
$\varv_{11} = 1$ as reference state and only for the first velocity component, 
we derive vibration temperatures of $230 \pm 50$, 
$210 \pm 20$, $200 \pm 10$, $230 \pm 10$, and $240 \pm 10$~K for the other 
excited states in the same order as in Table~\ref{t:modparam_lmh_c2h3cn}.
If a direct pumping mechanism from the ground state exists for each of 
the six excited states listed above, then it has to occur at $\sim$~44, 30, 
22, 18, 15, and 14.5~$\mu$m, respectively. However, the vibrational band 
$\nu _{11}$ is fairly weak, and $\nu _{15}$ is weak. Therefore, direct infrared 
pumping is unlikely to explain the apparent higher vibration temperatures of
$\varv_{11} = 1$ and $\varv_{15} = 1$. The combination and overtone bands seem 
similarly too weak to permit direct pumping of the associated states.

\citet{Schilke97} and \citet{Nummelin99} reported the detection of transitions 
pertaining to the $\varv_{11} = 1$ and $\varv_{15} = 1$ excited states in 
\object{Orion~KL} and Sgr~B2(N), respectively. The higher-lying vibrational 
states have been detected here for the first time as far as we know. 

\paragraph{Sgr\,\,B2(M):} The parameters of all entries related to vinyl 
cyanide are listed in Table~\ref{t:modparam_b2m_c2h3cn}. Many transitions
of vinyl cyanide are detected at 3~mm with lower-level energies ranging 
from 16~K (84.95~GHz) to 136~K (113.93~GHz). We note that a few transitions
with even higher energies ($E_l/k =$ 159, 163, and 205~K at 94.96, 104.46, and 
113.99~GHz, respectively) coincide with detected lines but are significantly 
underestimated by the current low-temperature model. Increasing the 
temperature to, e.g., 150~K improves the fit to these high-energy transitions, 
but degrades it for some of the lower-energy ones and produces too strong 
lines at 1.3~mm. The level of the baseline at 1.3~mm is uncertain and the 
spectrum may be affected by dust attenuation, but both effects may not be 
sufficient to make the 1.3~mm spectrum of vinyl cyanide consistent with a 
higher temperature. In any case, the model parameters should be viewed with 
caution. In addition, the lines of the current model being optically thin 
($\tau_{\rm max} \sim 0.25$), the source size is not constrained. The
$^{13}$C isotopologues are not detected.

A few emission lines from within the vibrationally excited state 
$\varv_{11} = 1$ are detected (e.g., 85.67, 104.51, 114.24, and 114.28~GHz). 
Because the lines have low signal-to-noise ratios, the detection and the model 
parameters should be viewed with caution. A temperature of 150~K fits the 
ratios of the five transitions detected around 114.3~GHz a bit better than the 
temperature of 50~K used for the ground state. In addition, the latter would 
require a column density parameter 30 times higher than the former to 
fit roughly the detected transitions.

\subsubsection{Ethyl cyanide C$_2$H$_5$CN}
\label{sss:c2h5cn}

We use the CDMS entries for the ground state of the $^{12}$C, $^{13}$C, and 
$^{15}$N isotopologues (tags 55502, 56504, 56505, 56506, and 56508, all 
version 2). The entry of the main isotopologue is based on \citet{EtCN_2009} 
which contained measurements in the frequency range of our survey from 
\citet{EtCN_8-200_1996}. The entries for the $^{13}$C and $^{15}$N 
isotopologues are based on \citet{13C-EtCN_2012} and \citet{15N-D-EtCN_2009},
respectively.

Entries for the vibrationally excited states $\varv_{13} + \varv_{21} = 1$, 
$\varv _{20}=1$, and $\varv _{12}=1$ of the main isotopologue were provided 
by J. Pearson. They are based on unpublished, preliminary, though rather 
extensive analyses involving these vibrational 
states\footnote{\citet{Daly13} published preliminary analyses of
$\varv _{20}=1$ and $\varv _{12}=1$ during the reviewing process of this
article.}. A limited and 
preliminary account on the first three vibrational states was given by 
\citet{vib-EtCN_1999}. A more extensive analysis of $\varv_{13}+\varv_{21}=1$ 
was described by \citet{Mehringer04}. The low-lying vibrational states of 
ethyl cyanide form polyads, groups of vibrational states which are fairly 
close in energy. The first group is a dyad described by 
$\varv_{13}+\varv_{21}=1$. Both states are rather close, at 206.5 and 
212.7~cm$^{-1}$, respectively, and they are interacting very strongly 
\citep{vib-EtCN_1999,Mehringer04}. The next group is a tetrad described by 
$\varv_{13}+2\varv_{20}+\varv_{21}=2$. $\varv_{20}=1$ at 378~cm$^{-1}$ is 
somewhat lower than the three states described by $\varv_{13}+\varv_{21} = 2$ 
near 400~cm$^{-1}$. Even though some transitions show torsional splitting, 
transition frequencies pertaining to this state can be fitted quite well as 
isolated state up to fairly high quantum numbers \citep{EtCN_v20_2009_WH13}. 
No data have been published for the remaining three states. The next higher 
group of states is already a heptad described by 
$3\varv_{12}+\varv_{13}+2\varv_{20}+\varv_{21}=3$. We have only data for the 
lowest, apparently fairly isolated state $\varv_{12}=1$ at 523~cm$^{-1}$.

\paragraph{Sgr\,\,B2(N):} The parameters of all entries related to ethyl 
cyanide are listed in Table~\ref{t:modparam_lmh_c2h5cn}. The model for the 
ground state of ethyl cyanide ($\tau_{\rm max} \sim 10$) was already 
reported in Sect.~4.4 of \citet{Belloche09} and for the $^{13}$C isotopologues 
($\tau_{\rm max} \sim 0.35$) in Sect.~6 of \citet{Mueller08}. The source 
sizes derived for the main isotopologue from our Plateau de Bure 
interferometric observations \citep[][]{Belloche08} are $2.2\arcsec \pm 
0.2\arcsec$ and $1.5\arcsec \pm 0.2\arcsec$ for the southern and northern hot
cores, respectively. Our model uses slightly larger sizes in order to fit better
both the optically thick and thin transitions of the main isotopologue. 
Some optically thick transitions are too strong by 
about 20--30\% (e.g. 80.60, 89.57, 96.92, and 107.49~GHz), but other optically 
thick lines with similar lower-level energies and similar opacities
are well reproduced, so the 
reason for this discrepancy is unclear. Many optically thin lines of the main 
isotopologue with a broad range of lower-level energies from 1.6 to 
$\sim 500$~K are also detected and well fitted, implying that the temperature 
is well constrained. Three velocity components (both hot cores and an additional
blue wing) are clearly detected with the $^{12}$C and $^{13}$C isotopologues
\citep[see discussion in Sect.~4.4 of][]{Belloche09}. There is no significant
difference in column density between the three $^{13}$C isotopologues.
The column density ratio of the $^{12}$C to $^{13}$C isotopologues is about 30
for the hot core components, which is puzzling given the isotopic ratio of 
$\sim 20$ usually found toward the Galactic center.
We note that a model with source sizes closer to the sizes derived 
interferometrically requires a temperature of 280~K to reproduce the 
intensities of the optically thick transitions of the main isotopologue and 
yields a $^{12}$C to $^{13}$C isotopic ratio of $\sim$~24. However, this 
alternative model overestimates the (optically thin) high-energy 
($E_l/k >$~300~K) transitions of the main isotopologue. We favored the (simple) 
model with the slightly larger sizes in order to better fit \textit{all} the 
detected transitions. A more complex model would probably help in deriving a 
physically more consistent description of the structure of the hot core.
The $^{15}$N isotopologue is not detected. With a 
$\frac{^{14}\rm N}{^{15}\rm N}$ isotopic ratio of 300, we
expect emission above the $3\sigma$ level, but it is always blended with
emission of other species (e.g., 87.01, 95.72, and 95.75~GHz), which prevents a 
secure identification. It is for that reason not included in the full model.

We detect emission lines from within four vibrationally excited states of 
the main isotopologue: $\varv_{13} + \varv_{21} = 1$, $\varv_{20} = 1$, and 
$\varv_{12} = 1$ (206.5/212.7, 
378, and 523~cm$^{-1}$). The peak opacities of the synthetic model are 
6.5, 1.2, and 0.36, respectively.  We fixed the temperature to the value 
derived for the ground state. The source size had to be reduced for the 
excited states to simultaneously match their optically thick and thin lines. 
Both hot-core components are clearly detected in all four states (e.g., 
89.45~GHz for $\varv_{12} = 1$). The third, blueshifted component is well 
detected for $\varv_{13} + \varv_{21} = 1$ and reasonably well detected 
for $\varv_{20} = 1$. It is less secure for  $\varv_{12} = 1$ but the model 
including this component gives a good match (e.g., 98.38~GHz). Like for vinyl 
cyanide (see Sect.~\ref{sss:c2h3cn}), the column density parameter 
required to fit the vibrationally excited states is higher than for the ground 
state, which suggests the presence of radiative pumping\footnote{If a 
direct pumping mechanism from the ground state exists for each of the four 
excited states
listed above, then it has to occur at $\sim$~48, 47, 26, and 19~$\mu$m, 
respectively. The intensities of the infrared bands $\nu _{13}$, $\nu _{20}$, 
and $\nu _{12}$ are sufficiently high to facilitate direct infrared pumping. 
The weak torsional mode $\nu _{21}$ may be pumped through the Coriolis 
interaction with $\nu _{13}$.}. However, surprisingly, it does not 
increase monotonically with energy: $\varv_{20} = 1$ has a lower column 
density parameter than the lower-energy states $\varv_{13} = 1$ and 
$\varv_{21} = 1$. 
We note that our catalog file for $\varv_{13} + \varv_{21} = 1$ contains
intensities at 200~K instead of 300~K, but correcting for this does not
solve the discrepancy and we have not identified the origin of the problem so 
far. Given this issue, we do not attempt to estimate any vibration 
temperature.

Given the detections above, we expect all vibrationally excited states with
$\varv_{13} + \varv_{21} = 2$ to be easy to detect in our 3~mm survey but no 
predictions are available yet. The transitions from within each of these three 
states may be as strong and numerous 
as the transitions from within $\varv_{20}=1$ detected here (about two dozens 
with peak temperatures ranging from 0.2 to 0.9~K). Transitions from within 
$\varv_{13} + \varv_{21} = 3$ (4 states) or from within 
$\varv_{20}=1/\varv_{13} + \varv_{21} = 1$ (2 states) may even be detected once 
predictions become available, each of them maybe as strong and numerous as the 
transitions from within 
$\varv_{12}=1$ detected here (about 30 with peak temperatures ranging from 0.1 
to 0.3~K). 
We also expect a few transitions of the $^{13}$C isotopologues from 
within the vibrationally excited states with $\varv_{13} + \varv_{21} = 1$ to 
be detectable in our survey at 3~mm. 

\citet{Gibb00} and \citet{Mehringer04} report the detection of transitions 
in the $\varv_{13} = 1$ and $\varv_{21} = 1$ excited states toward G327.3-0.6 
and Sgr~B2(N), respectively. \citet{Daly13} report the detection of 
transitions in the $\varv_{20} = 1$ and $\varv_{12} = 1$ excited states toward 
\object{Orion~KL} during the reviewing process of this article. Both states 
are detected here toward Sgr~B2(N) for the first time as far as we know.

\paragraph{Sgr\,\,B2(M):} The parameters of all entries related to ethyl 
cyanide are listed in Table~\ref{t:modparam_b2m_c2h5cn}. The lower-level 
energies of the detected transitions range from 19~K (88.32~GHz) to 74~K 
(80.60~GHz). The optical depths are low ($\tau_{\rm max} \sim 0.25$), which 
lets the source size unconstrained. The temperature is somewhat constrained 
around, e.g., 80.6, 89.6, and 98.6~GHz, where group of transitions with 
different lower-level energies are detected. The $^{13}$C isotopologues are not
unambiguously detected.

Emission lines from within the  $\varv_{13} = 1$ and $\varv_{21} = 1$ 
vibrationally excited states are detected at, e.g., 89.65, 97.15, 98.62, and 
98.74~GHz. The temperature is not constrained at 3~mm, but detections at 1.3~mm 
favor a model with a temperature higher than the one derived for the ground
state. The source size is not constrained. One transition from within the 
$\varv_{20} = 1$ vibrationally excited state may be detected at 98.60~GHz,
but this is not sufficient to claim a detection and we do not include this
state in the complete model.

\subsubsection{Ethyl formate C$_2$H$_5$OCHO}
\label{sss:c2h5ocho}

We use the CDMS entry (tag 74514 version 1) which is based on \citet{EtFo_2009}.

\paragraph{Sgr\,\,B2(N):} The parameters of ethyl formate are listed in 
Table~\ref{t:modparam_lmh_c2h5ocho}. The detection of this complex organic 
molecule was reported for the first time in space in \citet{Belloche09}. 
\citet{Kalenskii10} claim a (somewhat indirect) detection of this molecule 
toward the \object{W51~e1/e2} star-forming region based on a stacking analysis 
of their 3~mm spectral survey obtained with the Onsala 20~m telescope.

\paragraph{Sgr\,\,B2(M):} Ethyl formate is not detected.

\subsubsection{Ethanol C$_2$H$_5$OH}
\label{sss:c2h5oh}

We use the JPL entry for the main isotopologue (tag 46004 version 4) 
based on the very extensive analysis in \citet{EtOH_2008}. 
Additional data in the range of our survey were taken from 
\citet{t-EtOH_1995,g-EtOH_1996,g-EtOH_1997}. We use the CDMS entries
for the $^{13}$C isotopologues (tags 47511 and 47512, both version 1) which 
are based on \citet{13C-EtOH_2012}.

\paragraph{Sgr\,\,B2(N):} The parameters of ethanol are 
listed in Table~\ref{t:modparam_lmh_c2h5oh}. The synthetic spectrum is 
optically thin ($\tau_{\rm max} \sim 0.6$). The source size is therefore 
not constrained and was fixed to 3$\arcsec$. The temperature is well 
constrained thanks to many transitions detected over a broad range of 
lower-level energies (from 2~K at 112.81~GHz to $\sim 430$~K at 97.86~GHz).
The other parameters were slightly readjusted compared to the model reported 
in \citet{Belloche09}. The northern hot core is clearly detected in addition to 
the main one. Two low-energy transitions (at 85.27 and 90.12~GHz with 
$E_l/k = 13$ and 5~K, respectively) are underestimated by the model by a factor 
of about 2. However, other transitions with similar energies are well 
reproduced (at 84.60, 87.72, and 112.81~GHz with $E_l/k = 9$, 13, and 2~K, 
respectively), so we exclude the additional presence of a significant, 
low-excitation (and extended?) component. The model overestimates by a factor 
$\gtrsim 3$ the observed spectrum for two transitions at 99.54 and 101.36~GHz 
with $E_l/k = 223$ and 158~K. Many other transitions with similar energies 
(e.g., the nearby transition at 99.52~GHz with $E_l/k = 137$~K) are well 
fitted however. The two 
problematic transitions are close to large gauche-gauche perturbations and 
their line strengths could be wrong because of that (J. Pearson, 
\textit{priv. comm.}).

With a $\frac{^{12}\rm C}{^{13}\rm C}$ isotopic ratio of 20, the 
emission of the $^{13}$C isotopologues of ethanol is expected to be weak. It 
is consistent with the observed spectrum within the noise, with possible 
detection at 100.70 and 101.40~GHz (CH$_3$$^{13}$CH$_2$OH and 
$^{13}$CH$_3$CH$_2$OH, respectively). Even if one transition per species is not 
sufficient to claim a secure detection, we include the $^{13}$C isotopologues 
in the complete model.

\paragraph{Sgr\,\,B2(M):} The parameters of ethanol are listed in 
Table~\ref{t:modparam_b2m_c2h5oh}. Most lines detected at 3~mm are reasonably
well fitted with a low-excitation component. Several higher-energy 
transitions, however, seem to be detected too and suggest the presence of an 
additional, warm component (e.g., $E_l/k = 94$~K at 87.96~GHz, 154~K at 
98.59~GHz). All lines are optically thin ($\tau_{\rm max} = 0.018$ for the
low-excitation component, 0.030 for the warm component), 
implying that the source size of each component is not constrained. There are
two issues with the current model at 84.60~GHz ($E_l/k = 9$~K) and 112.81~GHz 
(2~K). The reason for this discrepancy is unclear because the transition at 
114.06~GHz ($E_l/k = 2$~K), for instance, is well reproduced. The $^{13}$C 
isotopologues are not detected.

\subsubsection{$n$-Propyl cyanide $n$-C$_3$H$_7$CN}
\label{sss:c3h7cn}

We use the CDMS entry (tag 69505 version 1) which is based on the analysis in 
\citet{Belloche09}. Transition frequencies related to the range of our survey 
were published by \citet{n-PrCN_1988}.

\paragraph{Sgr\,\,B2(N):} The parameters of $n$-propyl cyanide are listed in 
Table~\ref{t:modparam_lmh_c3h7cn}. The detection of this complex organic 
molecule was reported in \citet{Belloche09}. It has only been detected in 
Sgr~B2(N) so far.

\paragraph{Sgr\,\,B2(M):} $n$-Propyl cyanide is not detected.

\subsubsection{Ethylene oxide $c$-C$_2$H$_4$O}
\label{sss:c-c2h4o}

We use the CDMS entry (tag 44504 version 2). The entry is based on data 
summarized in \citet{c-C2H4O_1998}. The spectroscopic parameters are very 
different from those in that work and have been mentioned in a very recent 
far-infrared study \citep{c-C2H4O_2012} which improves predictions somewhat 
at very high frequencies. \citet{c-C2H4O_1974} published 
transition frequencies that in part fall into the frequency 
ranges of our survey.

\paragraph{Sgr\,\,B2(N):} The parameters of ethylene oxide are listed in 
Table~\ref{t:modparam_lmh_c-c2h4o}. The synthetic spectrum is optically thin
($\tau_{\rm max} \sim 0.15$), so the source size is not constrained and was
set to 3$\arcsec$. The model is based on eight clearly detected transitions,
plus a number of other transitions that are partly blended with emission from
other species. The detected transitions span a range of lower-level energies 
from 5~K (at 94.66~GHz) to 147~K (at 110.81~GHz). At this low level of 
detection, there is no evidence for emission from the northern hot core 
(see, e.g., 102.42~GHz).

\paragraph{Sgr\,\,B2(M):} Ethylene oxide is not detected.

\subsubsection{Cyclopropenylidene $c$-C$_3$H$_2$}
\label{sss:c-c3h2}

We use the CDMS entries for the $^{12}$C and $^{13}$C isotopologues 
(tags 38508, version 2, as well as 39509 and 39510, both version 1).
The entries are based roughly on the spectroscopic parameters reported 
by \citet{isos-c-C3H2_1987}. Most of the rest frequencies for the 
$^{13}$C isotopologues were also taken from that work. Transition 
frequencies for the main isotopic species were largely taken from 
\citet{c-C3H2_1986}. Spectroscopic information for a number of 
isotopic species, though mostly outside the frequency range of our survey, 
were published by \citet{c-C3H2_div_2012}. A small number of transition 
frequencies for the main isotopologue in the range of our survey were 
reported by \citet{c-C3H2_1987}.

\paragraph{Sgr\,\,B2(N):} The parameters of all entries related to 
cyclopropenylidene are listed in Table~\ref{t:modparam_lmh_c-c3h2}. The 
molecule is detected both in emission (at 82.96, 84.73, and 85.65~GHz) and in 
absorption (at 82.09 and 85.34 GHz). The absorption consists of contributions
from the envelope of Sgr~B2(N) itself and several diffuse clouds along the
line of sight \citep[see][ for more details]{Menten11}. The temperature of the
emission component is not very well constrained because of the narrow range of
lower-level energies of the few detected transitions.

One of the $^{13}$C isotopologue ($c$-CC$^{13}$CH$_2$) is clearly detected in 
absorption (at 80.05 and 84.18~GHz). Only the self-absorption components are
detected, but the diffuse-cloud components are included in the model with 
the isotopic ratios listed in Table~\ref{t:iso}. The situation for the other 
isotopologue ($c$-$^{13}$CC$_2$H$_2$) is less clear: its predicted absorption 
($\sim -0.2$~K) is blended with emission from C$_2$H$_5$OCHO, HCC$^{13}$CN, 
$\varv_5=1$/$\varv_7=3$, and HC$_3$N, $\varv_6=\varv_7=1$ at 82.30~GHz. The 
synthetic spectrum including all species is not consistent with the observed 
spectrum at this frequency: either the contribution in absorption of 
$c$-$^{13}$CC$_2$H$_2$ is overestimated or the contribution in emission of the 
contaminating species is underestimated. The situation at 81.15~GHz is similar, 
the absorption due to $c$-$^{13}$CC$_2$H$_2$ being likely blended with emission 
from other (still unidentified) species. Despite these uncertainties, we kept 
both isotopologues in the complete model, with the same column densities.

\paragraph{Sgr\,\,B2(M):} The parameters of all entries related to 
cyclopropenylidene are listed in Table~\ref{t:modparam_b2m_c-c3h2}. The model
for the main isotopologue was already reported in \citet{Menten11}.
The self-absorption component of the $^{13}$C isotopologues is clearly 
detected at 80.05, 84.19 ($c$-CC$^{13}$CH$_2$), and 82.30~GHz 
($c$-$^{13}$CC$_2$H$_2$). The latter is somewhat overestimated and the 
transition of the same isotopologue expected at 81.15~GHz seems to be blended 
with an unidentified line in emission (see similar issues for Sgr~B2(N) above).
The model for the $^{13}$C isotopologues also contains the diffuse-cloud 
components derived for the main isotopologue and scaled with the isotopic 
ratios listed in Table~\ref{t:iso}, but these components are too weak to be 
detected, except for a hint of weak detection at 80.06~GHz.

\subsubsection{Ethynyl CCH}
\label{sss:cch}

We use the CDMS entries for the $^{12}$C and $^{13}$C isotopologues (tags 25501
version 3, 26502 version 1, and 26503 version 2) which are based on 
\citet{C2H_2009}, \citet{C-13-CH_1995}, and \citet{CC-13-H_2010}, respectively.

\paragraph{Sgr\,\,B2(N):} The parameters of all entries related to ethynyl are 
listed in Table~\ref{t:modparam_lmh_cch}. The molecule is clearly detected in
absorption, both from the envelope of Sgr~B2(N) itself and from the diffuse
clouds along the line of sight. Including a component in emission turns out
to be necessary to consistently reproduce the self-absorption features
of all three transitions at 87.28, 87.32, and 87.33~GHz, but the parameters 
of this emission component are very poorly constrained. Such a very extended
emission component is seen in the large-scale maps of \citet{Jones08} taken 
with the Mopra telescope. Like for $c$-C$_3$H$_2$, only one of the $^{13}$C 
isotopologues ($^{13}$CCH) is detected, in absorption at the 
$3\sigma$ level at 84.12~GHz, with a further hint of detection at 84.21~GHz. 
The other isotopologue (C$^{13}$CH) is not detected at 85.23 GHz, probably 
because of blends with emission from other (unidentified) species. It is
still included in the complete model.

\paragraph{Sgr\,\,B2(M):} The parameters of all entries related to ethynyl 
are listed in Table~\ref{t:modparam_b2m_cch}. Ethynyl is detected both in 
emission and in absorption. The emission part is modeled with two components,
one for the core of the line and one for the redshifted wing. Their parameters
are poorly constrained. The presence of a blueshifted wing cannot be seen in 
our data because, if it exists, it is masked by the deep absorption components 
of spiral-arm diffuse clouds. The diffuse-cloud components were fitted on the 
first group of hyperfine transitions (at 87.28, 87.32, and 87.33~GHz). The 
second group of hyperfine transitions (at 87.40, 87.41, and 87.45~GHz) is well 
fitted with the same parameters.

There are hints of detection of the $^{13}$C isotopologues at 84.1--84.2~GHz 
($^{13}$CCH) and 85.2--85.3~GHz (C$^{13}$CH). Both are included in the complete 
model, with the same parameters as the main isotopologue after rescaling the
column densities using the isotopic ratios listed in Table~\ref{t:iso}. The 
first one is partly contaminated by absorption lines of $c$-H$^{13}$CCCH.

\subsubsection{Thioethenyl CCS}
\label{sss:ccs}

We use the CDMS entry for the main isotopologue (tag 56502 version 1) which is 
based mainly on \citet{CCS_1987}.

\paragraph{Sgr\,\,B2(N):} The parameters of thioethenyl are listed in 
Table~\ref{t:modparam_lmh_ccs}. It is detected with no or little contamination 
at 81.51, 86.18, 93.87, 103.64, and 106.35~GHz (with $E_l/k = 11$, 
19, 15, 26, and 20~K, respectively). The energy range of the detected 
transitions is too narrow to fully constrain the temperature. A comparison
of the 30~m detection to the peak temperatures obtained by \citet{Friedel04} 
with the NRAO 12~m telescope suggests that the emission is more extended than 
the 30~m beam but less extended than the 12~m beam, which led us to assume a 
source size of $60\arcsec$. The emission is optically thin 
($\tau_{\rm max} = 0.03$). The $^{13}$C and $^{34}$S isotopologues are 
not detected.

\paragraph{Sgr\,\,B2(M):} The parameters of thioethynyl are listed in 
Table~\ref{t:modparam_b2m_ccs}. It is clearly detected at the same frequencies
as toward Sgr~B2(N). We assume the same source size as for the latter source. 
The emission is optically thin ($\tau_{\rm max} = 0.03$). The $^{13}$C and 
$^{34}$S isotopologues are not detected.

\subsubsection{Methylenimine CH$_2$NH}
\label{sss:ch2nh}

We use the CDMS entry (tag 29518, version 1), which is based largely on 
\citet{H2CNH_2012}. Transition frequencies in the range of our survey 
were published by \citet{H2CNH_2010}. These transitions display hyperfine 
splitting and we use the CDMS entry that accounts for it.

\paragraph{Sgr\,\,B2(N):} The parameters of methylenimine are listed in 
Table~\ref{t:modparam_lmh_ch2nh}. The detection relies, at 3~mm, on only four 
lines with no or little contamination from other species (at 87.53, 95.31, 
95.51, and 114.55~GHz with $E_l/k = 270$, 114, 597, and 492~K). The transition 
at 85.21~GHz is blended with diffuse-cloud absorption components of 
HC$^{18}$O$^+$ 1--0. There are five additional lines with little contamination 
detected in the 2~mm and 1.3~mm windows. The emission is optically thin at 
3~mm ($\tau_{\rm max} = 0.85$), which allows for a reasonable estimate of the 
rotation temperature. The lines 
detected at 1.3~mm are optically thick ($\tau_{\rm max} = 5.5$) and in turn
constrain the source size (but see Sect.~\ref{ss:optimization} for the caveats 
about the modeling at 1.3~mm). The $^{13}$C and $^{15}$N isotopologues are 
not detected.

\paragraph{Sgr\,\,B2(M):} The parameters of methylenimine are listed in 
Table~\ref{t:modparam_b2m_ch2nh}. The observed spectrum is not consistent with
a warm component like in Sgr~B2(N), for instance at 87.53~GHz. Only one 
transition is detected at 3~mm (at 105.79~GHz with $E_l/k = 26$~K), with a high 
signal-to-noise ratio. This transition is severely blended with the
second velocity component of H$^{13}$CCCN toward Sgr~B2(N) but only its 
blueshifted wing is partially blended toward Sgr~B2(M) because this source
does not harbor such a second velocity component.

\subsubsection{Ethylene glycol $aGg'$-(CH$_2$OH)$_2$}
\label{sss:ch2oh-2}

We use the CDMS entry for the two lowest-state conformers (tags 62503 and 
62504, both version 1) which are based on \citet{aGg_eglyc_2003}
and \citet{gGg_eglyc_2004}.

\paragraph{Sgr\,\,B2(N):} The parameters of ethylene glycol are listed in 
Table~\ref{t:modparam_lmh_ch2oh-2-a}. The 3~mm emission of the lowest state 
conformer, \textit{aGg'}, is weak and optically thin 
($\tau_{\rm max} = 0.024$), so the source size is not constrained.
A comparison of our 30~m spectrum around 93.0~GHz to the NRAO~12~m spectrum 
of \citet{Hollis02} suggests that the emission of ethylene glycol and 
glycolaldehyde are in the same proportion more extended than the emission of 
ethyl cyanide. As a result, we arbitrarily set the source size to $10\arcsec$. 
The detection relies on about 13 weak lines that are not or only weakly blended
with emission from other species. The lower-level energies of the detected 
transitions range from 13~K (83.62~GHz) to 40~K (95.53~GHz), which implies 
that the rotation temperature is not very well constrained. There is 
no evidence for the presence of several velocity components (e.g. 87.56, 
99.51~GHz). The second lowest state conformer, \textit{gGg'}, is not detected.

\paragraph{Sgr\,\,B2(M):} Ethylene glycol is not detected.

\subsubsection{Glycolaldehyde CH$_2$(OH)CHO}
\label{sss:ch2ohcho}

We use the CDMS entry for the vibrational ground state (tag 60501 version 2)
for technical reasons, the quality of the CDMS and JPL entries being about 
the same in the frequency range of our survey.
We use a subset of the JPL entry for the first vibrationally excited state 
(tag 60006 version 2). The ground state data are based on 
\citet{Butler01}, those of the vibrationally excited states on 
\citet{glycolaldehyde_2005}. 

\paragraph{Sgr\,\,B2(N):} The parameters of glycolaldehyde are listed in 
Table~\ref{t:modparam_lmh_ch2ohcho}. The 3~mm emission is weak and optically 
thin ($\tau_{\rm max} = 0.017$). The source size is not constrained and was
set to the same value as ethylene glycol (Sect.~\ref{sss:ch2oh-2}). The 
detection relies on only five weak lines that are not or only weakly blended
with emission from other species. The lower-level energies of the detected 
transitions range from 4~K (97.92~GHz) to 78~K (96.87~GHz). The temperature
is not well constrained (toward lower values). There is no evidence for the 
presence of several velocity components, but this may be due to the limited 
signal-to-noise ratios.

No transition from within the vibrationally excited state $\varv=1$ is detected.

\paragraph{Sgr\,\,B2(M):} Glycolaldehyde is not detected.

\subsubsection{Methylcyanoacetylene CH$_3$C$_3$N}
\label{sss:ch3c3n}

We use the CDMS entry (tag 65503 version 1) which is based on 
\citet{MeC3N_1983}.

\paragraph{Sgr\,\,B2(N):} The parameters of methylcyanoacetylene are listed in 
Table~\ref{t:modparam_lmh_ch3c3n}. The 3~mm emission is optically thin
($\tau_{\rm max} = 0.017$). The source size was arbitrarily set to 
$3\arcsec$. The observed spectrum is well fitted with two velocity components,
one emitted by each hot core. A temperature of 100~K gives a good fit. A
somewhat higher temperature (150~K) would still give a reasonable fit but tends
to produce too strong emission at 107.395 and 111.533~GHz.

\paragraph{Sgr\,\,B2(M):} The parameters of methylcyanoacetylene are listed 
in Table~\ref{t:modparam_b2m_ch3c3n}. The molecule is detected in emission at
several frequencies, e.g., at 82.63, 90.89, 99.15, 107.41, and 111.54~GHz, but 
the signal-to-noise ratio for each transition is low.  The spectrum is 
somewhat better fitted with a temperature of 100~K than with 50 or 150~K, but 
this is uncertain. The emission is optically thin
($\tau_{\rm max} = 0.028$) and the source size is not constrained.

\subsubsection{Propyne CH$_3$CCH}
\label{sss:ch3cch}

We use the CDMS entry for the vibrational ground state \citep{propyne_2008}
and first excited state $\varv_{10} = 1$ \citep{propyne_2002} of the main 
isotopologue (tag 40502 version 3, 40504 version 1), and the JPL entries for 
the $^{13}$C isotopologues (tags 41002, 
41003, and 41004, all version 1) which are based on
\citet{13C-propyne_1978}\footnote{The documentation in the JPL catalog 
quotes a wrong reference.}.

\paragraph{Sgr\,\,B2(N):} The parameters of all entries related to propyne are 
listed in Table~\ref{t:modparam_lmh_ch3cch}. The 3~mm emission is optically 
thin ($\tau_{\rm max} = 0.23$). Two groups of transitions with lower-level
energies $E_l/k$ ranging from 8 to 128~K are detected, which constrains the 
temperature relatively well. A source size of 10$\arcsec$ was arbitrarily 
chosen but we note that a large-scale map obtained with Mopra shows emission 
extended over several arc minutes \citep[see Fig.~7 of][]{Jones08}. No 
transition from within the vibrationally excited state $\varv_{10} = 1$ is 
detected.

The $^{13}$C isotopologues were modeled with the same parameters as the main
isotopologue, assuming an isotopic ratio of 20. The synthetic model is 
consistent with the observations, with one line of $^{13}$CH$_3$CCH 
(83.13~GHz), one weak line of CH$_3$$^{13}$CCH (85.41~GHz), and one line of 
CH$_3$C$^{13}$CH (99.48~GHz) detected with no or little contamination. 

\paragraph{Sgr\,\,B2(M):} The parameters of all entries related to propyne 
are listed in Table~\ref{t:modparam_b2m_ch3cch}. Two groups of transitions
(around 85.45 and 102.54~GHz) are detected.  Their lower-level energies 
$E_l/k$ range from 8 to 128~K and constrain the temperature to about 70~K. The 
source size has in turn to be larger than the beam at 1.3~mm to fit the 
transitions detected around 205.05~GHz. For a smaller size, the synthetic 
model becomes too strong (but remember the caveats about the modeling at 
1.3~mm). With the current parameters, the transitions are all optically thin
at 3~mm ($\tau_{\rm max} = 0.045$). The $^{13}$C isotopologues are detected
with little contamination at 82.90, 99.48 (CH$_3$C$^{13}$CH), 99.75 
($^{13}$CH$_3$CCH), and 102.51~GHz (CH$_3$$^{13}$CCH).

\subsubsection{Acetone CH$_3$C(O)CH$_3$}
\label{sss:ch3coch3}

We use the JPL entry for the vibrational ground state (tag 58003 version 3)
which is based on \citet{acetone_2002} with measurements in the range of our 
survey from \citet{acetone_v0_1986}. An entry for the lowest $\varv _{12} = 1$ 
excited torsional state, based on an analysis presented by 
\citet{acetone_2006}, was provided by P. Groner. For both states, we use the 
partition function provided by the JPL catalog for entry 58003. We note that
this partition function has a strong kink at 75~K which looks suspicious
even if it includes the contribution of low-lying vibrationally excited states. 
We suspect that some of the tabulated values are erroneous, which could then
affect the derived column densities.

\paragraph{Sgr\,\,B2(N):} The parameters of acetone are listed in 
Table~\ref{t:modparam_lmh_ch3ch3co}. Most transitions at 3~mm are optically 
thin with $\tau < 0.3$, and only a few are marginally optically thin 
($\tau_{\rm max} \sim 0.8$). The source size was set arbitrarily. The detected 
transitions have lower-level energies $E_l/k$ ranging from 8~K (at 98.30~GHz) 
to 194~K (at 92.93~GHz), which constrains the temperature relatively well.
Several transitions are consistent with the presence of a second velocity 
component (see, e.g., at 90.29, 92.74, 99.42, and 101.45~GHz). Overall, the 
synthetic spectrum fits better the observed one with two velocity components. 

We also detect emission from within the first torsionally excited
state $\varv _{12} = 1$ ($\sim$~78~cm$^{-1}$). Many lines are clearly 
detected at 3~mm, at, e.g., 
92.65 ($E_l/k = 130$~K), 92.71 (130 K), 94.95 (123 K), 101.38 (133 K), 102.47 
(135 K), and 106.39~GHz (227 K). All 3~mm transitions are optically thin, with 
$\tau_{\rm max} \sim 0.35$. Two issues occur at 81.77 (124~K) and 90.50~GHz 
(155~K) but the discrepancies are certainly due to an overestimate of the 
baseline level. This is, to the best of our knowledge, the first report on the 
detection of torsionally excited acetone in space.

\paragraph{Sgr\,\,B2(M):} The parameters of acetone are listed in 
Table~\ref{t:modparam_b2m_ch3ch3co}. Acetone is clearly detected at several
frequencies in the 3~mm window (e.g., groups of three transitions around 82.92, 
92.74, and 111.25~GHz). The 1.3~mm range seems to constrain the temperature to 
low values. The source size is not constrained because all transitions detected
at 3~mm are optically thin ($\tau_{\rm max} \sim 0.16$).

\subsubsection{Acetaldehyde CH$_3$CHO}
\label{sss:ch3cho}

We use the JPL entry (tag 44003 version 3) and split it to model the ground 
state and the first vibrationally excited state separately. The entry is based 
on \citet{acetaldehyde_1996} with laboratory rest frequencies in the range of 
our survey taken from \citet{Bau76}, \citet{Lia86}, \citet{Mae87}, 
\citet{Kle91,Kle92}, \citet{Bar93}, and \citet{Bel93}.

\paragraph{Sgr\,\,B2(N):} The parameters of all entries related to 
acetaldehyde are listed in Table~\ref{t:modparam_lmh_ch3cho}. The 3~mm emission
is optically thin ($\tau_{\rm max} \sim 0.7$). The source size was 
arbitrarily set. The detected transitions have lower-level energies ranging 
from  9~K (at 95.96~GHz) to 70~K (at 115.61~GHz). The temperature is
not very well constrained. A temperature of 150~K yields a somewhat poorer fit 
than 100~K for the $J=$~6--5 multiplet around 115.6~GHz. A temperature of 60~K 
also fits the detected transitions relatively well. Two velocity components 
are clearly detected.

Transitions from within the torsionally excited state 
$\varv_{\rm t}=1$ (142~cm$^{-1}$) are clearly detected at, e.g., 
95.65, 95.76, 96.80, 97.34, and 114.72 GHz. The column density 
parameter required to fit the torsionally excited state is 
a factor of two higher than for the ground state. With a temperature
of 60~K, the discrepancy would be a factor of ten, which is why we favor a
temperature of 100~K for both states. The higher column density 
parameter required to fit $\varv_{\rm t}=1$ suggests 
the presence of radiative pumping. Using equation~\ref{e:tvib} with the 
ground state as reference state, we derive a torsion temperature of 
$150 \pm 20$~K for $\varv_{\rm t}=1$. If a direct pumping mechanism 
exists, then it has to occur at $\sim$~70~$\mu$m. The torsional mode is 
sufficiently strong to facilitate direct infrared pumping. The 
presence of a second velocity component looks plausible. The only issue is at 
95.76~GHz where the second component is not seen, but this could be due to an 
overstimate of the baseline level. This is, to the best of our knowledge, the 
first report on the detection of torsionally excited acetaldehyde.

\paragraph{Sgr\,\,B2(M):} The parameters of acetaldehyde are listed in 
Table~\ref{t:modparam_b2m_ch3cho}. Many transitions of acetaldehyde are 
detected in the 3~mm window, with lower-level energies ranging from 
9~K (95.95~GHz) to 34~K (115.66~GHz). Higher energy transitions (up to 
$\sim 70~K$) are detected in the 2 and 1.3~mm ranges and constrain the 
temperature below 80~K and the source size above $20\arcsec$. The line ratios
around 96.35~GHz and around 115.65~GHz, and the detection at 205.15~GHz 
suggest a temperature higher than 20~K. All transitions detected at 3~mm are
optically thin ($\tau_{\rm max} \sim 0.016$). No transition from within the 
vibrationally excited state $\varv_{\rm t}=1$ is detected.

\subsubsection{Methyl cyanide CH$_3$CN}
\label{sss:ch3cn}

We use the CDMS entries for the vibrational ground state of the $^{12}$C, 
$^{13}$C, and $^{15}$N isotopologues (tags 41505, 42508, 42509, 42510, and 
43513, all version 1), and the JPL entry for the vibrationally excited state 
$\varv_8 = 1$ of the main isotopologue (tag 41010 version 1). The entries for 
the vibrationally excited state $\varv_8 = 1$ of the $^{13}$C isotopologues and 
$\varv_8 = 2$ and $\varv_4 = 1$ of the main isotopologue were prepared by one 
of us (HSPM). The ground-state entries were all based on \citet{MeCN_2009}, 
the $\varv_8 = 1$ entry was based on \citet{MeCN-vibs_1988}, and the 
$\varv_8 = 2$ and $\varv_4 = 1$ on \citet{MeCN-conf_2010}. The excited states 
of the $^{13}$C isotopologues are based on unpublished work. Ground-state data 
in the range of our survey were taken from \citet{MeCN_2006}, 
\citet{13C-MeCN_1979}, and \citet{MeCN-15_div-v_1975} for the main, the 
$^{13}$C, and the $^{15}$N isotopologues, respectively. \citet{MeCN_v8_1969}, 
\citet{MeCN_2v8_1971}, and \citet{MeCN_v4_1991} provided corresponding data 
for the $\varv_8 = 1$, $\varv_8 = 2$, and $\varv_4 = 1$ states, 
respectively, of the main isotopic species.

The cyano C of methyl cyanide is rather 
close to the center of mass of the molecule. Therefore, transitions of 
CH$_3$$^{13}$CN and $^{13}$CH$_3$$^{13}$CN occur close to those of the more 
abundant CH$_3$CN and $^{13}$CH$_3$CN, respectively, in particular at 
wavelengths as long as 3~mm.

\paragraph{Sgr\,\,B2(N):} The parameters of all entries related to methyl 
cyanide are listed in Table~\ref{t:modparam_lmh_ch3cn}. The model for the 
ground state of methyl cyanide ($\tau_{\rm max} \sim 28$) was already 
reported in Sect.~4.4 of \citet{Belloche09}. The $^{13}$C isotopologues are 
clearly detected ($\tau_{\rm max} \sim 1.7$). The temperature is relatively
well constrained by their $K$-ladders, and 
their few marginally optically thick lines also constrain the source size to 
some extent. Two velocity components are clearly detected. The derived 
parameters are used for the main isotopologue assuming an isotopic ratio of 
20. The synthetic spectrum of the latter does not fit the shape of the 
observed spectrum very well because our radiative transfer method is too 
simplistic for such transitions with high optical depths. Like for ethyl 
cyanide (see Sect.~\ref{sss:c2h5cn}), a third, redshifted, component is added
to better fill the linewing. The $^{15}$N isotopologue is not directly detected.
Its transitions around 89.2 and 107.05~GHz are blended with deep absorption 
components of HCO$^+$ and emission lines of C$_2$H$_5$$^{13}$CN and SO$_2$, 
respectively. We still include this isotopologue in the complete model because 
it significantly improves the fit between 107.05 and 107.06~GHz, range over 
which it contributes to about 30\% of the detected emission. Even if it is
not clearly detected as such, the doubly-substituted $^{13}$C isotopologue is 
also included in the complete model because it significantly contributes to 
the detected emission at 89.27 and 107.12~GHz.

We also detect emission lines from within three vibrationally excited states of 
the main isotopologue: $\varv_{8} = 1$, $\varv_{8} = 2$, and $\varv_{4} = 1$, 
(365, 717, and 920~cm$^{-1}$, respectively), with
$\tau_{\rm max} \sim 5.5$, 0.65, and 0.15, respectively. The optically thick
lines of $\varv_{8} = 1$ require a source size a bit smaller than for the ground
state\footnote{A model with the same source size as the ground state 
but a lower rotation temperature of 140~K still fits the optically thick 
lines reasonably well but not anymore the optically thin lines at 92.22 and 
92.23~GHz.}. The column density parameter needed to fit the emission 
of the excited 
states increases with their energy, suggesting that the vibration
temperature is higher than the rotation temperature, probably because 
of radiative pumping.
Using equation~\ref{e:tvib} with $\varv_{8} = 1$ as reference state
and only for the first velocity component, we derive vibration temperatures 
of $290 \pm 20$ and $340 \pm 20$~K for $\varv_{8} = 2$ and $\varv_{4} = 1$, 
respectively. The intensities of the infrared bands $\nu _8$, 2$\nu _8$, and 
$\nu _4$ are possibly sufficient to permit direct infrared pumping.
If such a direct pumping mechanism from the ground state exists for each
of the three excited states $\varv_{8} = 1$, $\varv_{8} = 2$, and 
$\varv_{4} = 1$, then it has to occur at $\sim$~27, 14, and 11~$\mu$m, 
respectively.
The second velocity component is clearly detected for $\varv_{8} = 1$ and 
$\varv_{8} = 2$, and consistent with the observed spectrum for $\varv_{4} = 1$. 
The third, redshifted, component, is significantly detected for 
$\varv_{8} = 1$ only. Emission lines from within the excited state 
$\varv_{8} = 1$ of the $^{13}$C isotopologues are also detected, e.g., at 
89.52, 92.19, and 110.63~GHz. The column density ratio between the $^{12}$C and 
$^{13}$C isotopologues is only 14 for $\varv_{8} = 1$. A few detected 
transitions of the $^{12}$C isotopologue are optically thin (e.g., at
92.22~GHz with $\tau = 0.4$) so the deviation from the usual 
$\frac{^{12}{\rm C}}{^{13}{\rm C}}$ isotopic ratio of 20 does not seem to be 
related to an optical depth effect. The reason for this discrepancy is unclear.

\citet{Goldsmith83} have detected transitions of the main isotopologue
from within $\varv_{8} = 1$ and \citet{Fortman12} have very recently reported 
the detection of $\varv_{8} = 2$ in \object{Orion~KL}. 
Lines from within higher excited vibrational 
states as well as the $\varv_{8} = 1$ lines of the $^{13}$C isotopologues have 
been detected here for the first time.

\paragraph{Sgr\,\,B2(M):} The parameters of all entries related to methyl 
cyanide are listed in Table~\ref{t:modparam_b2m_ch3cn}. Following the results 
of \citet{deVicente97}, we used two temperature components to model the 
vibrational ground state of the $^{12}$C and $^{13}$C isotopologues. However,
the parameters are not well constrained, and the fit is not very satisfactory,
especially for the main isotopologue. Like \citet{deVicente97}, we detect the
$J = K$ components of the $J = 5-4$ and $6-5$ $K$-ladders in absorption.
These authors accounted for these absorption features by adding a third
component corresponding to a diffuse ($10^3$~cm$^{-3}$), hot (300~K), outer 
layer and computed their synthetic spectrum with a non-LTE model. This is 
beyond the scope of our simple, LTE modeling, which cannot reproduce these 
absorption features. The doubly-substituted $^{13}$C isotopologue is not 
detected. 

We detect emission lines from within two vibrationally excited states of the 
main isotopologue: $\varv_{8} = 1$ and $\varv_{8} = 2$. A model with the same 
source size and temperature as for the ground state ($1.5\arcsec$ and 200~K) 
fits the $\varv_{8} = 1$ transitions at 3~mm as well as our model with 
$1\arcsec$ and 300~K, but the former would require a column density twice as 
high to also fit the $\varv_{8} = 2$ transitions while the latter fits both 
vibrationally excited states well with the same column density. If the 
presence of radiative pumping can explain the former model, then it is unclear 
which model is more realistic.

\subsubsection{Acetamide CH$_3$C(O)NH$_2$} 
\label{sss:ch3conh2}

The entry for acetamide was provided by V. Ilyushin and is based on 
\citet{Ilyushin04}. We used partition functions for the \textit{A} and 
\textit{E} species considered as independent. They were computed and 
provided by V. Ilyushin (2012, \textit{priv. comm.}). The ground state and the 
first two torsionally excited states of each species were splitted into 
different entries in our catalog.

The first astronomical detection of acetamide was reported by \citet{Hollis06} 
toward Sgr~B2(N). Eight low-energy transitions (four from the \textit{A} 
species and four from the \textit{E} species) were detected with the GBT 
between 9 and 48~GHz, in absorption or in emission. The detected signals were 
interpreted as 
coming from a low-excitation ($\sim 8$~K) halo surrounding both hot cores. 
\citet{Halfen11} published a detailed analysis of the emission of acetamide
based on their spectral survey of Sgr~B2(N) performed with the ARO 12~m 
telescope ($HPBW \sim 60\arcsec$ at 100~GHz) and the SMT at 3, 2, and 1~mm. 
They claim the detection of 132 transitions of acetamide, the majority of them 
being ``partially blended'', and they identify two temperature components 
based on a rotation-diagram analysis ($17 \pm 4$ and $171 \pm 4$~K, 
respectively). 

\paragraph{Sgr\,\,B2(N):} The parameters of all entries related to acetamide 
are listed in Table~\ref{t:modparam_lmh_ch3conh2-a}. In our spectrum, most of 
these transitions are significantly blended with emission from other species 
and only a few transitions can be reliably identified. The ground state of the 
\textit{A} species is reasonably well detected at 87.63 ($E_l/k = 15$~K), 92.38 
(109~K), 97.905 (19~K), 97.911 (19~K), and 97.94~GHz (20~K), and there are 
hints of detection at 87.72 (159~K), 88.08 (162~K), 90.91 (126~K), 101.28 
(138~K), and 114.50~GHz (115~K), pointing to a high temperature. In comparison 
to the 12~m observations, our 30~m observations are relatively less sensitive 
to an extended, low-excitation component. The temperature is however not very 
well constrained in our data and we set it to the value derived by 
\citet{Halfen11} for their warm component, which is very close to the 
temperature we derive for formamide (see Sect.~\ref{sss:nh2cho}). We 
arbitrarily set the source size to the one that we derive for formamide (see 
Sect.~\ref{sss:nh2cho}). Two transitions at 108.21 (23~K) and 108.26~GHz 
(24~K) are overestimated by the model, by a factor of two for the latter. We do 
not think that it invalidates the detection because nearby transitions of 
ethyl cyanide (at 108.21 and 108.23~GHz) suffer from the same issue. Since the 
detection of ethyl cyanide is robust and its emission is very well reproduced 
at other frequencies, the issue around 108.2~GHz points to a calibration or 
pointing problem for that frequency range. The ground state of the 
\textit{E} species is also reasonably well detected at 86.46 (279~K), 92.28 
(174~K), 97.89 (19~K), 99.70 (152~K), 107.99 (23~K), and 108.19~GHz (24~K). We 
note however that the best fit yields a column density lower than for the 
\textit{A} species by 25\%. 

We also include the contributions of the first two torsionally excited states 
of each species in the model. For the \textit{A} species, transitions from
within $\varv=1$ are weakly detected at 92.34 (81~K) and 100.37~GHz (85~K) and 
from within $\varv=2$ at 99.63~GHz (216~K), with the same parameters as for 
the ground state. The model for $\varv=1$ overestimates the transition at 
98.96~GHz (82.5~K) but we believe that it is due to the level of the baseline 
being overestimated. For the \textit{E} species, transitions from within 
$\varv=1$ are weakly detected at 97.40 (141~K), 99.19 (47~K), 101.61 (110~K), 
and 111.92~GHz (51~K) and from within $\varv=2$ at 114.50 GHz (133~K), again 
with the same parameters as for the ground state. These detections are only 
very tentative \textit{per se}, but we keep these entries in the complete model
because their parameters are the same as for the ground states.

\paragraph{Sgr\,\,B2(M):} Acetamide is not detected.

\subsubsection{Acetic acid CH$_3$COOH}
\label{sss:ch3cooh}

We use entries provided by I. Kleiner for the ground state and the 
torsionally excited states $\varv_{\rm t} = 1$ and $\varv_{\rm t} = 2$. 
The entry was generated from \citet{HAc_2008}. Ground state transition 
frequencies at 3~mm and also to a large extent in the higher frequency windows 
of our survey were published by \citet{HAc_2001}. The partition function was
computed including levels up to $J=130$ and the first 11 torsionally excited
states (V. Ilyushin, 2012, \textit{priv. comm.}).

\paragraph{Sgr\,\,B2(N):} The parameters of acetic acid are listed in 
Table~\ref{t:modparam_lmh_ch3cooh}. The 3~mm emission is optically 
thin ($\tau_{\rm max} \sim 0.05$), so the source size is not constrained.
The range of lower-energy levels of the detected transitions is very narrow,
from 16~K (at 90.25~GHz) to 25~K (at 111.55~GHz), which is not sufficient to
constrain the temperature. There is a hint of second velocity component at 
100.86~GHz. The vibrationally excited states $\varv_{\rm t} = 1$ and 
$\varv_{\rm t} = 2$ are not detected.

\paragraph{Sgr\,\,B2(M):} Acetic acid is not detected.

\subsubsection{Methylamine CH$_3$NH$_2$}
\label{sss:ch3nh2}

We use the JPL entry (tag 31008 version 1) which employed data summarized in 
\citet{MeNH2_2005}. Additional data in the range of our survey were published 
by \citet{MeNH2_1992}.

\paragraph{Sgr\,\,B2(N):} The parameters of methylamine are listed in 
Table~\ref{t:modparam_lmh_ch3nh2}. The 3~mm emission is optically thin 
($\tau_{\rm max} \sim 0.32$), so the source size is not constrained. The 
transitions detected in emission have lower-level energies $E_l/k$ ranging from
6.4~K (at 89.08~GHz) to 286~K (at 97.48~GHz), which constrains the temperature
relatively well. The presence of a second velocity component in emission is
clearly seen. A few transitions are detected in absorption with two velocity
components (e.g., at 84.31~GHz). One issue is that the transition at 89.96~GHz
with $E_l/k = 2.1$~K is also expected in absorption but is not seen in the 
observed spectrum. This may be due to contamination by emission from another 
species that has not been identified so far. The emission being weak 
($T_{\rm mb} < 0.35$~K), no other isotopologue is expected to be detectable 
in our survey.

\paragraph{Sgr\,\,B2(M):} The parameters of methylamine are listed in 
Table~\ref{t:modparam_b2m_ch3nh2}. A few transitions of methylamine are 
detected at 3~mm, both in emission (at, e.g., 83.98 and 84.31~GHz) and in 
absorption (at, e.g., 82.23 and 89.96~GHz). The lines detected in emission at 
3~mm have lower-level energies ranging from 13~K (87.78~GHz) to 45~K 
(81.52~GHz), which does not constrain much the temperature. A model with 100~K 
fits the 3~mm spectrum reasonably well, but it produces too strong lines at 
1.3~mm. 50~K is more consistent with the upper limits at 1.3~mm, but this is 
quite uncertain because of the dust-absorption and baseline-level issues 
mentioned earlier for this atmospheric window. In particular, there are still
three low-energy transitions significantly overestimated by the 50~K synthetic
spectrum in the 1.3~mm window, at 215.79, 247.08, and 254.61~GHz 
($E_l/k = 7$, 26, and 17~K, respectively). The lines in emission at 3~mm are 
optically thin ($\tau_{\rm max} \sim 0.07$), implying that the source size 
is not constrained. The absorption component is shifted by $\sim 4$~km~s$^{-1}$ 
in velocity compared to the emission component.

\subsubsection{Dimethyl ether CH$_3$OCH$_3$}
\label{sss:ch3och3}

We use the CDMS entry for the ground state of the main isotopologue (tag 46514 
version 1). It is based on the very extensive analysis of \citet{DME_2009} and
employed additional rest frequencies in the range of our survey from 
\citet{DME_1979} and \citet{DME_1990}. 
The entries for the vibrationally excited states $\varv_{11}= 1$ and
$\varv_{15} = 1$ were provided by C. Endres. The entry
for the ground state of the $^{13}$C isotopologue was provided by M. Koerber. 

\paragraph{Sgr\,\,B2(N):} The parameters of all entries related to dimethyl 
ether are listed in Table~\ref{t:modparam_lmh_ch3och3}. The model for the 
ground state of dimethyl ether was already reported in Sect.~3.4 of 
\citet{Belloche09}. Some transitions detected in emission at 3~mm are 
marginally optically thick, others are optically thin 
($\tau_{\rm max} \sim 1.8$). The temperature is well constrained by the 
broad range of lower-energy levels of the optically thin transitions, from 
5~K (at 99.32~GHz) to $\sim 430$~K (at 97.24~GHz). The source size is in turn
constrained by the optically thick lines. The presence of a second velocity 
component is clearly seen. Assuming an isotopic ratio of 20, there are several 
hints of emission from the $^{13}$C isotopologue at 81.81, 93.28, 108.20, and 
114.14~GHz. There may be an issue at 86.87~GHz, but it is probably due to 
blending with velocity components of SiO in absorption that have not been 
included in the complete model.

We also detect emission lines from within two vibrationally excited states of 
the main isotopologue: $\varv_{11} = 1$ and $\varv_{15} = 1$ (199 and 
241~cm$^{-1}$, respectively), with $\tau_{\rm max} \sim 0.12$ and 0.08, 
respectively. We used the same model parameters as for the ground state. About 
ten transitions of $\varv_{11} = 1$ are weakly detected (e.g., 82.93, 111.93, 
and 113.02~GHz), and five for $\varv_{15} = 1$ (e.g., 81.83, 90.24, and 
93.33~GHz).
We note that the detection of numerous transitions from within both 
vibrationally excited states were recently reported toward G327.3-0.6  
\citep[][]{Bisschop13}.

\paragraph{Sgr\,\,B2(M):} The parameters of dimethyl ether are listed in 
Table~\ref{t:modparam_b2m_ch3och3}. The transitions detected at 3~mm have
lower-level energies ranging from 7~K (91.47~GHz) to 164~K (111.81~GHz),
which yields a relatively well constrained temperature. The transitions at 3~mm
are optically thin ($\tau_{\rm max} \sim 0.045$) and do not constrain the 
source size. However the lines detected at 1.3~mm suggest a beam filling factor
at 1.3~mm close to 1. A smaller source size, with a higher column density to 
still fit the 3~mm spectrum, tends to produce too strong lines at 1.3~mm. The
$^{13}$C isotopologue  is not detected. The vibrationally excited states 
$\varv_{11} = 1$ and $\varv_{15} = 1$ of the main isotopologue are not detected 
either.

\subsubsection{Methyl formate CH$_3$OCHO}
\label{sss:ch3ocho}

We use the JPL entry (tag 60003 version 2) which we have split into two
entries to model the ground state and the first vibrationally excited state
separately. The entry is based on \citet{MeFo_2009} with additional data from 
\citet{MeFo_7-200_2001}.

\paragraph{Sgr\,\,B2(N):} The parameters of all entries related to methyl 
formate are listed in Table~\ref{t:modparam_lmh_ch3ocho}. The model for the 
ground state of methyl formate was already reported in Sect.~3.4 of 
\citet{Belloche09} but was slightly readjusted. Most detected transitions are
(marginally) optically thin ($\tau_{\rm max} \sim 1.3$). We refer to 
\citet{Belloche09} for a comparison of our fit results to those obtained by
\citet{Nummelin00}. The temperature is relatively well constrained while the
source size is not. With a peak temperature of $\sim 1.7$~K at 3~mm for the 
main isotopologue, we expect the $^{13}$C isotopologues to be detectable in 
our survey but we do not have a catalog file at our disposal yet. 

We also detect emission lines from within the torsionally excited 
state $\varv_{\rm t} = 1$ (132~cm$^{-1}$), with $\tau_{\rm max} \sim 0.2$. Like 
for other species, the fit yields a column density parameter 
much higher than the one derived for the ground state, suggesting
the presence of radiative pumping. Using equation~\ref{e:tvib} with the 
ground state as reference state and only for the first velocity component, we 
derive a torsional temperature of $120 \pm 10$~K for $\varv_{\rm t}=1$. 
If a direct pumping mechanism exists, then it has to occur at 
$\sim$~76~$\mu$m. However, the intensity of the methyl torsion is very low 
such that direct infrared pumping is very unlikely. The presence of a second 
velocity component is clearly seen at, e.g., 89.73, 95.24, and 100.22~GHz.

\paragraph{Sgr\,\,B2(M):} The parameters of methyl formate are listed in 
Table~\ref{t:modparam_b2m_ch3ocho}. The transitions detected at 3~mm have
lower-level energies ranging from 14~K (88.85~GHz) to 54~K (110.54~GHz) and are
optically thin ($\tau_{\rm max} \sim 0.26$). The temperature is somewhat
constrained, but the source size is not. The first vibrationally excited state 
$\varv_{\rm t} = 1$ is not unambiguously detected.

\subsubsection{Methanol CH$_3$OH}
\label{sss:ch3oh}

The entry for the main isotopologue was taken from the JPL catalog (tag 32003, 
version 3) and the one for the $^{13}$C isotopologue from the CDMS (tag 
33502, version 1). These entries are from \citet{MeOH_2008} and 
\citet{12_13C-MeOH_1997}. The entry for the $^{18}$O species was prepared by 
one of us (HSPM). Transitions in the range of our survey were taken 
mainly from \citet{MeOH_1968}, \citet{MeOH_1981}, \citet{HER84}, \citet{SAS84},
\citet{MeOH_1990}, \citet{MeOH_2004}, and \citet{MeOH_2008} for the 
main isotopologue, from \citet{13C-MeOH_1974}, \citet{13C-MeOH_1984}, 
\citet{13C-MeOH_1986}, \citet{13C-MeOH_1987}, and \citet{13C-MeOH_1990} 
for $^{13}$CH$_3$OH, and from \citet{18O-MeOH_1998} for CH$_3$$^{18}$OH.

\paragraph{Sgr\,\,B2(N):} The parameters of all entries related to methanol are 
listed in Table~\ref{t:modparam_lmh_ch3oh}. Many transitions are detected at
3~mm, up to at least 720~K. However, methanol is very difficult to
model. We model its emission and absorption with six temperature/velocity 
components (with $\tau_{\rm max} \sim 7.5$ for the main hot component and 
1.7 for the 15~K component), but several issues remain (see discussion 
in Sect.~\ref{ss:discussion_ch3oh}): at 97.20 and 
97.21~GHz, the synthetic lines (with $E_l/k = 947$~K) have no counterpart in 
the observed spectrum; at 107.01 and 108.89~GHz ($E_l/k = 23$ and 
8~K, respectively), the model predicts lines in emission while the observed 
spectrum is mostly in absorption; the model largely underestimates the emission 
of the transition detected at 84.52~GHz, which is known to be a maser line 
\citep[see Sect.~\ref{ss:discussion_ch3oh} and][]{MeOH_2004}.
The starting parameters for the cold component (15~K) were taken from our 
model of Sgr~B2(M). This component is mostly (but poorly) constrained
by the transitions at 95.91, 96.74, and 97.58 GHz ($E_l/k$ in the range 2.3 to 
17~K). The two absorption components at $\varv_{\rm off} \sim -170$~km~s$^{-1}$
were taken from our model toward Sgr~B2(M). They are (weakly) detected close to 
96.8~GHz only. The 3~mm emission of the $^{13}$C isotopologue is clearly 
detected and is optically thin ($\tau_{\rm max} \sim 0.34$ and 0.08 for the hot 
and cold components, respectively) and is dominated by the hot component for 
the transitions with $E_l/k > 20$~K. The lower-level energies of the detected 
transitions range from 2.3~K (at 94.41~GHz) to 470~K (at 90.86~GHz), which 
constrains the temperature of the hot component relatively well. The source 
size is in turn constrained by the optically thick lines of the main 
isotopologue.The 3~mm emission of the $^{18}$O isotopologue is (barely) 
detected at 112.95 and 114.18~GHz ($\tau_{\rm max} \sim 0.02$). The cold 
component is not constrained for this isotopologue.

We also detect emission lines from within two torsionally excited states of 
the main isotopologue: $\varv_{\rm t} = 1$ and $\varv_{\rm t} = 2$ 
(204.2~cm$^{-1}$ for $J = K = 1$ of $E$ symmetry and 
353.2~cm$^{-1}$ for $J = K = 0$ of $A$ symmetry, respectively), with 
$\tau_{\rm max} \sim 0.9$ and 0.2, respectively. About 15 transitions are 
detected at 3~mm for $\varv_{\rm t} = 1$, but only three transitions are 
detected for $\varv_{\rm t} = 2$, at 96.38, 96.55, and 102.42~GHz. The first 
one is partially blended with emission from CH$_3$CHO, the second one from 
CH$_3$OCHO and C$_2$H$_5$CN, $\varv_{13}+\varv_{21} = 1$, and the third 
one is weak. We had to reduce the source size compared to the model of the 
ground state in order to fit both the optically thin lines of 
$\varv_{\rm t} = 1$ and those with an opacity close to 1. The presence of a 
second velocity component is clearly seen for $\varv_{\rm t} = 1$ (e.g. 
96.49~GHz). Finally, we also detect the $^{13}$C isotopologue in 
its torsionally excited state $\varv_{\rm t} = 1$ (e.g., 93.44, 102.71, and 
110.73~GHz).

\paragraph{Sgr\,\,B2(M):} The parameters of all entries related to methanol 
are listed in Table~\ref{t:modparam_b2m_ch3oh}. Transitions with lower-level
energies ranging from 2.3~K (96.74~GHz) to 489~K (81.65~GHz) are detected at
3~mm. There is a clear indication of at least two temperature components
\citep[see also][]{Sutton91}. We use a temperature of 15~K for the cold 
component, similar to the values found for methanol by \citet{Sutton91} toward 
Sgr~B2(M) at higher frequency and by \citet{Requena06} toward Galactic-center 
clouds without hot cores. Our model largely underestimates the transitions at
84.52~GHz ($E_l/k = 36$~K) and 95.17~GHz (79~K), which are both known to be 
maser lines 
\citep[see Sect.~\ref{ss:discussion_ch3oh} and][]{MeOH_2004}. One 
component in absorption is added to fit the 
self-absorption seen at 96.74~GHz ($E_l/k = 2.3$ and 8~K) and 205.79~GHz (7~K), 
but it does not account well for the self-absorption seen at 107.01~GHz 
(23.2~K) and 108.89~GHz (8~K), where the model is much too strong (see 
discussion in Sect.~\ref{ss:discussion_ch3oh}). An 
additional cold component is added to account for the redshifted wing emission 
seen at, e.g., 84.52 (36~K), 95.91 (17~K), and 97.58~GHz (17~K). Several 
diffuse-cloud components are detected in absorption at 96.8~GHz. A hot 
component is necessary to account for the detected transitions with higher 
energies. Its temperature is relatively well constrained. The source size was 
determined by the optically thick transitions at 1.3~mm. At 3~mm, the hot 
component has a maximum opacity $\tau_{\rm max} \sim 1.5$, similar to the 
maximum opacity of the main cold component (1.6). The $^{13}$C isotopologue
is clearly detected, especially at 94.41~GHz. Only the cold components in 
emission are detected at 3~mm. The hot component is tentatively detected in
the 1.3~mm window at 235.9 and 236.0~GHz. The $^{18}$O isotopologue is not 
detected.

We also detect emission lines from within the torsionally excited state
$\varv_{\rm t}=1$. Only two transitions at 99.73 and 102.96~GHz are clearly 
seen at 3~mm, with $\tau_{\rm max} = 0.19$.

\subsubsection{Methyl mercaptan CH$_3$SH}
\label{sss:ch3sh}

The entries for the ground state, first and second torsionally excited states 
of methyl mercaptan were provided by L.-H. Xu prior to the publication of the 
recent extensive analysis \citep{MeSH_2012}. Transition frequencies in the 
range of our survey were taken mainly from \citet{MeSH_1980} and 
\citet{MeSH_1986}. 

\paragraph{Sgr\,\,B2(N):} The parameters of methyl mercaptan are listed in 
Table~\ref{t:modparam_lmh_ch3sh}. Many lines of methyl mercaptan are blended
with emission of other species, but four transitions are clearly detected 
at 101.14, 101.16, 101.96, and 102.20~GHz ($E_l/k$ from 7 to 48~K) with two 
velocity components. The emission is weak ($T_{\rm mb} < 0.6$~K) and 
optically thin ($\tau_{\rm max} \sim 0.3$). The source size is not 
constrained and the temperature not very well either. The $^{13}$C and $^{34}$S 
isotopologues are not expected to be detectable in our survey. No transition 
from within the torsionally excited state $\varv_{\rm t} = 1$ is unambiguously 
detected.

\paragraph{Sgr\,\,B2(M):} The parameters of methyl mercaptan are listed in 
Table~\ref{t:modparam_b2m_ch3sh}. A group of transitions with $E_l/k$ ranging 
from 7 to 48~K is clearly detected around 101.2~GHz, which somewhat constrains
the temperature. A temperature of 30~K yields a better fit to the 
$E_l/k = 60$~K transitions at 1.3~mm than 40~K. The lines are optically thin 
at 3~mm ($\tau_{\rm max} = 0.33$), so the source size is not constrained. No 
transition from within the torsionally excited state $\varv_{\rm t} = 1$ is 
detected.

\subsubsection{Cyanide radical CN}
\label{sss:cn}

We use the CDMS entries for the $^{12}$C, $^{13}$C, and $^{15}$N isotopologues
(tags 26504, 27505, and 27506, all version 1). The predictions are based on a 
combined fit involving several isotopologues which is similar to the one in 
\citet{CN-15_combined_1994}. Rest frequencies for C$^{15}$N from astronomical 
observations were also published in that work. Ground state data for $^{13}$CN 
were published by \citet{C-13-N_v0_1984}. Unpublished data from C. Gottlieb 
were employed for the main isotopologue.

\paragraph{Sgr\,\,B2(N):} The parameters of all entries related to the cyanide 
radical are listed in Table~\ref{t:modparam_lmh_cn}. The cyanide radical is
essentially detected in absorption around 113.2 and 113.5~GHz and is modeled 
with 38 velocity components. The emission component on top of which the 
self-absorption and part of the diffuse-cloud absorption occur is very poorly 
constrained. The spectrum is not easy to model because of the hyperfine 
structure. The $^{13}$C isotopologue is also detected in 
absorption around 108.40, 108.65, 108.78, and 109.22~GHz, mainly the 
self-absorption component but also a few diffuse-cloud components (e.g., at 
108.84~GHz). We nevertheless keep the contribution of all diffuse-cloud
components assuming 
the isotopic ratios listed in Table~\ref{t:iso}. There is a hint of detection
of the $^{15}$N isotopologue in absorption at 110.02~GHz, where the level of 
the baseline may have been underestimated. We only include the emission and 
self-absorption components in the model of this isotopologue.

\paragraph{Sgr\,\,B2(M):} The parameters of all entries related to the 
cyanide radical are listed in Table~\ref{t:modparam_b2m_cn}. The model is 
constructed with 31 velocity components in a similar way as for Sgr~B2(N).
The emission component is poorly constrained. The $^{13}$C isotopologue is
clearly detected both at 3 and 1.3~mm, but may be affected by blends with 
emission lines. The $^{15}$N isotopologue is weakly detected at 109.7~GHz,
but clearly at 110.02~GHz (mainly the self-absorption component).

\subsubsection{Carbon monoxide CO}
\label{sss:co}

We use the CDMS entries for CO, $^{13}$CO, C$^{18}$O, C$^{17}$O, $^{13}$C$^{18}$O,
and $^{13}$C$^{17}$O (tags 28503, 29501, 30502, 29503, 31502, and 30503, all
version 1 except version 2 for $^{13}$CO). The entries are based on 
\citet{12C16O}, \citet{13C16O}, \citet{12C18O}, \citet{12_13C17O}, 
\citet{13C18O}, and \citet{12_13C17O}, respectively.

\paragraph{Sgr\,\,B2(N):} The parameters of all entries related to carbon 
monoxide are listed in Table~\ref{t:modparam_lmh_co}. Three isotopologues in
addition to the main one are clearly detected ($^{13}$CO, C$^{18}$O, and 
C$^{17}$O). The emission of $^{13}$C$^{17}$O is partially blended with 
emission from aminoacetonitrile and possibly another unidentified species. 
The emission of $^{13}$C$^{18}$O is completely blended with emission from 
C$_2$H$_3$CN, $\varv_{11}=1$, but it is still included in the full model because
$^{13}$C$^{17}$O is partially detected. The width and centroid velocity of the 
main emission component is constrained by the C$^{18}$O and C$^{17}$O spectra.
The diffuse-cloud absorption components were first adjusted on $^{13}$CO, then 
transfered to the other isotopologues assuming the isotopic ratios listed in 
Table~\ref{t:iso}. Two components were then readjusted on C$^{18}$O. They are 
in turn too deep in $^{13}$CO. This may point to deviations from the assumed 
isotopic ratios or contamination of the $^{13}$CO absorption components by 
emission from a species not included in the complete model. Apart for the 
self-absorption, absorption components with $\varv_{\rm off} > -72$~km~s$^{-1}$ 
were not included in the model because they are blended with broad $^{13}$CO 
emission, which implies too many degeneracies. The model for CO was derived 
from the $^{13}$CO one, but it was not optimized further. Here again, the 
mixture of broad, multiple, emission and absorption components cannot be 
modeled in a simple way. Finally, we note that the observed CO emission shows 
blueshifted components up to $V_{\rm lsr} \sim 200$~km~s$^{-1}$ 
\citep[see Fig.~6 of][ for Sgr~B2(M)]{Menten11}, but we did not include these 
components in the model.

\paragraph{Sgr\,\,B2(M):} The parameters of all entries related to carbon 
monoxide are listed in Table~\ref{t:modparam_b2m_co}. All isotopologues but 
$^{13}$C$^{17}$O are clearly detected. $^{13}$C$^{17}$O has a low signal-to-noise 
ratio at 3~mm but is well detected at 1.3~mm. The optically thick CO 
transition requires a temperature of at least 58~K. This transition is 
complex, with many emission and absorption components, and it is not well
fitted. The linewidth of the main emission component was fitted on C$^{18}$O. 
The diffuse-cloud absorption components were first fitted on C$^{18}$O, then
transfered to $^{13}$CO for which additional components were added. All 
absorption components were then propagated to the other isotopologues
assuming the isotopic ratios listed in Table~\ref{t:iso}. The diffuse-cloud
absorption components are not detected in $^{13}$C$^{18}$O and $^{13}$C$^{17}$O.
Like for Sgr~B2(N), the observed CO (and likely $^{13}$CO too) emission shows 
blueshifted components up to $V_{\rm lsr} \sim 200$~km~s$^{-1}$ , but we
did not include these components in the model.

\subsubsection{Carbon monosulfide CS}
\label{sss:cs}

We use the CDMS entries for CS, $^{13}$CS, C$^{34}$S, C$^{33}$S, and 
$^{13}$C$^{34}$S (tags 44501, 45501, 46501, 45502, and 47501, all version 2).
The spectroscopic parameters were taken from \citet{Mueller05}. Ground state 
transition frequencies in the range of our survey were taken from 
\citet{CS_CS-34_2003} for CS and C$^{34}$S, \citet{isos-CS_1982} for $^{13}$CS, 
and \citet{rare-isos-CS_1981} for C$^{33}$S and $^{13}$C$^{34}$S.

\paragraph{Sgr\,\,B2(N):} The parameters of all entries related to carbon 
monosulfide are listed in Table~\ref{t:modparam_lmh_cs}. Three isotopologues
are clearly detected in addition to the main one ($^{13}$CS, C$^{34}$S, and
C$^{33}$S). $^{13}$C$^{34}$S is not directly detected but it contributes to
about 30\% of the peak temperature at 90.92~GHz so we include it in the 
complete model. The emission of $^{13}$CS and C$^{34}$S at 1.3~mm and C$^{33}$S 
at 3~mm cannot be simultaneously reproduced with a cold component only. A hot 
component is needed in addition to the cold component in emission to match the
three lines. Its parameters are however not very well constrained. The 
diffuse-cloud absorption components were first adjusted on CS, then transfered 
to $^{13}$CS and C$^{34}$S and readjusted, and then transfered back to CS and 
C$^{33}$S, assuming the isotopic ratios listed in Table~\ref{t:iso}.
Most absorption components of $^{13}$CS and some of C$^{34}$S are blended with 
emission of identified species. The complete model takes this into account but,
since it is not perfectly matching the emission lines, some uncertainty remains
for some of the absorption components. Several absorption components are too
deep for CS. The reason may be blending with emission of still unidentified
species or blueshifted wing emission of CS not included in the model, or an
underestimate of the excitation temperature, or erroneous isotopic ratios.

\paragraph{Sgr\,\,B2(M):} The parameters of all entries related to carbon 
monosulfide are listed in Table~\ref{t:modparam_b2m_cs}. Four isotopologues
are clearly detected in addition to the main one ($^{13}$CS, C$^{34}$S, 
C$^{33}$S, and $^{13}$C$^{34}$S). C$^{34}$S~5--4 is twice as strong as 
$^{13}$CS~5--4, which is not expected given that both isotopologues should 
have roughly the same column density. This may be an isotopic anomaly, or 
contamination by an unknown species. Contrary to Sgr~B2(N), no warm component 
is needed to fit both the 3~mm and 1.3~mm transitions. The diffuse-cloud
absorption components were modeled in the same way as for Sgr~B2(N). None of 
them is detected in $^{13}$C$^{34}$S.

\subsubsection{Ethenone CH$_2$CO}
\label{sss:ch2co}

We use the CDMS entries for the $^{12}$C and $^{13}$C isotopologues (tags 
42501 version 2, 43505, and 43506, both version 1). The entry for the main 
isotopic species is based on \citet{H2C2O-param_1992} with additional 
measurements in the range of our survey from \citet{isos-H2C2O_2003}. The 
$^{13}$C isotopologue entries are based on \citet{isos-H2C2O_2003}. Predictions 
for excited states were generated by one of us (HSPM) from an unpublished fit 
based on the infrared analysis from \citet{H2C2O_vibs_2000} with rotational 
transitions in the range of our survey from \citet{H2C2O_rot-vibs_1996}. 

The carbonyl C of ethenone is very close 
to the center of mass of the molecule. Therefore, the transitions of 
CH$_2$$^{13}$CO are frequently overlapped by those of the main species, in 
particular at wavelengths as long as 3~mm.

\paragraph{Sgr\,\,B2(N):} The parameters of all entries related to ethenone are 
listed in Table~\ref{t:modparam_lmh_h2cco}. The 3~mm emission is marginally 
optically thick for most transitions ($\tau_{\rm max} \sim 2.2$), but some
are optically thin (e.g., at 80.82~GHz). Altogether, the source size and
temperature are both partially constrained. The presence of a second velocity
component is clearly seen. Most transitions of the $^{13}$C isotopologues are
close to transitions of the main isotopologue and partly hidden in their 
blueshifted wings, which prevents a firm identification. Two transitions of
$^{13}$CH$_2$CO (at 96.95 and 97.80~GHz) are not in this situation. Both look 
shifted by about 3~MHz compared to the features detected in the observed 
spectrum. While the discrepancy could be related to the uncertain level of the
baseline in the first case, this seems to be unlikely for the second case. We
do not have any good explanation for this discrepancy.
The model is not very well constrained for the $^{13}$C isotopologues and is 
simply a scaled version of the model of the main isotopologue. No transition
from within the vibrationally excited states $\varv_5=1$, $\varv_6=1$, and 
$\varv_9=1$ is detected.

\paragraph{Sgr\,\,B2(M):} The parameters of all entries related to ethenone 
are listed in Table~\ref{t:modparam_b2m_h2cco}. Two temperature components
are needed to fit both the 3 and 1.3~mm ranges. At 3~mm, the detected
transitions have lower-level energies ranging from 6~K (80.83~GHz) to 127~K 
(101.00~GHz) and are all optically thin ($\tau_{\rm max} = 0.4$ for the hot 
component, 0.02 for the cold component). The transitions detected at 1.3~mm 
are dominated by the warm component and are slightly optically thick 
($\tau_{\rm max} = 2.4$). They partially constrain the source size and 
temperature, especially around 202.0~GHz where several transitions with 
different lower-level energies (from 43 to 252~K) are detected. The $^{13}$C 
isotopologues are not detected at 3~mm but there are hints of detection at 
213.26, 217.15, and 232.64~GHz for $^{13}$CH$_2$CO.

\subsubsection{Formaldehyde H$_2$CO}
\label{sss:h2co}

We use the CDMS entries for the $^{12}$C, $^{13}$C, and $^{18}$O isotopologues 
(tags 30501 version 2, 31503, and 32503, both version 1). The entries are 
based on \citet{H2CO_2003}, \citet{H2C-13-O_2000}, and \citet{H2CO-18_2000}, 
respectively, with additional measurements in the range of our survey 
from \citet{div-H2CO_1980} for all three isotopologues and from 
\citet{H2CO_H2CS_FA_1972} and \citet{div-H2CO_1978} for the main and $^{13}$C 
species.

\paragraph{Sgr\,\,B2(N):} The parameters of all entries related to 
formaldehyde are listed in Table~\ref{t:modparam_lmh_h2co}. They are not
very well constrained. The absorption components were introduced to fit the
transition at 211.21~GHz. Only one transition is detected in the 3~mm window 
(at 101.33~GHz). The $^{13}$C isotopologue is also detected with only one 
transition at 3~mm (at 96.38~GHz), but several at 1.3~mm. The $^{18}$O 
isotopologue is not detected.

\paragraph{Sgr\,\,B2(M):} The parameters of all entries related to 
formaldehyde are listed in Table~\ref{t:modparam_b2m_h2co}. Only one transition
of the main isotopologue is detected at 3~mm (at 101.33~GHz), and two at 
1.3~mm. The model parameters are not well constrained. The $^{13}$C isotopologue
is detected at 1.3~mm only, but the model is a bit too strong (or the level of 
the baselines overestimated). The $^{18}$O isotopologue is detected at 
214.78~GHz.

\subsubsection{Thioformaldehyde H$_2$CS}
\label{sss:h2cs}

We use the CDMS entries for the $^{12}$C, $^{13}$C, and $^{34}$S isotopologues
(tags 46509 version 2, 47505, and 48508, both version 1). The entry for the
main species is based on the extensive account by \citet{H2CS_2008}. A small 
number of transition frequencies in the 3~mm range were taken from 
\citet{H2CS_1972}. The entry for the $^{34}$S isotopologue was based on 
\citet{H2CS_1993} with some higher order parameters derived from those of the 
main species in that study. The parameters for the $^{13}$C isotopologue were 
also derived from those of the main species. The very limited data from 
\citet{H2C-13-S_1987} and references therein permitted only a very small 
number of spectroscopic parameters to be determined.

\paragraph{Sgr\,\,B2(N):} The parameters of all entries related to 
thioformaldehyde are listed in Table~\ref{t:modparam_lmh_h2cs}. Two velocity
components are clearly detected. A combination of hot and cold components is
required to fit the transitions both in the 3 and 1.3~mm windows, but their
parameters are not very well constrained. The $^{34}$S isotopologue is detected
at 102.81~GHz, and significantly contributes to the ``wing'' of the emission 
detected at 99.77~GHz. The $^{13}$C isotopologue is not unambiguously detected,
but it is included in the complete model because it contributes to the 
emission detected at 97.63 and 100.53~GHz significantly ($>50\%$). The $^{33}$S 
isotopologue is not detected.

\paragraph{Sgr\,\,B2(M):} The parameters of all entries related to 
thioformaldehyde are listed in Table~\ref{t:modparam_b2m_h2cs}. The main 
isotopologue is detected both at 3 and 1.3~mm. The higher-energy transitions 
in the 1.3~mm window are shifted in velocity compared to the lower-energy 
ones in the 3~mm window. This is a clear indication of two components. At 
3~mm, both components are optically thin, with $\tau_{\rm max} = 0.08$ for 
the cold one and 0.4 for the warm one. Some transitions of the latter are 
slightly optically thick at 1.3~mm ($\tau_{\rm max} \sim 2$), but others
are optically thin. As a result, both the temperature and the source size are
somewhat constrained, but may be affected by the uncertainties of the baseline
level. The third component accounts for the wing emission seen mainly for the
low-energy transitions at 3~mm. Its source size is not constrained. Only one
transition of the $^{34}$S isotopologue is detected, at 232.75~GHz. The
$^{13}$C isotopologue is tentatively detected at 97.63 and 227.76~GHz. It is
included in the complete model.

\subsubsection{Aminoacetonitrile NH$_2$CH$_2$CN}
\label{sss:nh2ch2cn}

We use the CDMS entry (tag 56507 version 1) which is based on spectroscopic 
parameters from \citet{Belloche08}. Transition frequencies in the range of 
our survey were published by \citet{AAN_1990}.

\paragraph{Sgr\,\,B2(N):} The parameters of all entries related to 
aminoacetonitrile are listed in Table~\ref{t:modparam_lmh_h2nch2cn}. The
detection of this complex organic molecule was reported in \citet{Belloche08}. 
It has only been detected in Sgr~B2(N) so far, and a correction of the column
density for the contribution of vibrational modes to the partition function
was reported by \citet{Braakman10}.

\paragraph{Sgr\,\,B2(M):} Aminoacetonitrile is not detected.

\subsubsection{Hydrogen sulfide H$_2$S}
\label{sss:h2s}

We use the CDMS entries for the $^{32}$S, $^{34}$S, and $^{33}$S isotopologues
(tags 34502, 36504, and 35503, all version 1). The entry for the main species 
is based on a modification of results in \citet{H2S_1995}. The entries for the 
$^{34}$S and $^{33}$S isotopologues are based on \citet{isos-H2S_1995}. The 
frequencies for the only observed transition were published by \citet{H2S_1971} 
for the $^{32}$S and $^{34}$S isotopologues.

\paragraph{Sgr\,\,B2(N):} The parameters of hydrogen sulfide 
are listed in Table~\ref{t:modparam_lmh_h2s}. Only one transition is detected
in the 1.3~mm window. It is very strong (23~K). The model parameters are not
constrained. The transitions of the $^{34}$S and $^{33}$S isotopologues are too
heavily blended with emission from other species to be included in the complete
model.

\paragraph{Sgr\,\,B2(M):} The parameters of all entries related to hydrogen 
sulfide are listed in Table~\ref{t:modparam_b2m_h2s}. Like toward Sgr~B2(N),
only one transition of the main isotopologue is detected at 216.71~GHz 
($E_l/k = 74$~K). It is very strong (20.6~K), and thus requires a high 
temperature. The corresponding $^{34}$S transition is detected at 214.38~GHz 
with a peak temperature of 2.8~K, indicating that the transition of the main 
isotopologue is optically thick. Even with this indication, we do not have 
enough constraints to derive the temperature and source size independently.
Two additional components are added to the model to account for the wing
emission. We arbitrarily use the same temperature and source size as for the
main component. The $^{33}$S isotopologue is blended with other species but it 
contributes significantly (about 50\%) to the observed emission at 215.51~GHz, 
which led us to include it in the complete model.

\subsubsection{Cyanoacetylene HC$_3$N}
\label{sss:hc3n}

We use the CDMS entries for all isotopologues and vibrationally excited states,
all version 1 except version 2 for the states $\varv_6=\varv_7=1$ and 
$\varv_4=1$ of the main isotopologue. Most entries for the main 
isotopologue are based on \citet{HC3N_Koeln_2000}. Entries of the interacting 
$\varv_4=1$/$\varv_7=4$/$\varv_5 = \varv_7=1$/$\varv_6=2$ states were 
generated by extension and modification of \citet{HC3N_1986}. The $\varv_3=1$ 
and $\varv_2=1$ entries are based on \citet{excited-HC3N_etc_2005}. The 
$^{13}$C entries, including the doubly substituted ones, are based on
\citet{isos-HC3N_2001}, and those of HC$_3$$^{15}$N on \citet{HC3N-15_2004}. 
Additional measurements in the range of our survey were taken from 
\citet{div-HC3N_v0_1977} for the ground vibrational states of the main as well 
as singly substituted $^{13}$C and $^{15}$N entries, from \citet{HC3N_v0_1995} 
for $\varv = 0$ of the main species, \citet{HC3N_1986} and 
\citet{HC3N_Lille_2000} for several excited states of the main isotopologue, 
and \citet{isos-HC3N_2001} for the HC$_3$$^{15}$N entries.

The resonant interactions between $\varv_7=4$ and \hbox{$\varv_5 = \varv_7=1$} 
have been approached by the measurements of \citet{HC3N_1986} and 
\citet{HC3N_Lille_2000}, but not reached. Therefore, predicted transition 
frequencies involving these resonances may differ from the actual frequencies 
by amounts exceeding three times the predicted uncertainties. The interacting 
levels in the frequency range of our survey are $J = 10$ of $\varv_7=4$, 
$l = 2^-$, and $\varv_5 = \varv_7=1$, $l = 0^-$ and possibly $J = 26$  of 
$\varv_7=4$, $l = 2^+$, and $\varv_5 = \varv_7=1$, $l = 2^+$ as well as 
$J = 27$  of $\varv_7=4$, $l = 2^-$, and $\varv_5 = \varv_7=1$, $l = 2^-$.

\paragraph{Sgr\,\,B2(N):} The parameters of all entries related to 
cyanoacetylene are listed in Table~\ref{t:modparam_lmh_hc3n}. 
Cyanoacetylene is difficult to model because it is widespread in the Sgr~B2
molecular cloud \citep[see, e.g., Fig.~4 of][]{Jones08} and the emission at
3~mm is optically thick ($\tau_{\rm max} \sim 4.5$). We model the emission
with four velocity components: one component for each hot core, and one 
blueshifted and one redshifted components to account for the line wings.
The components associated with the hot cores have to be moderately compact to
match the opacities directly derived from the ratio of the $^{12}$C and 
$^{13}$C detected transitions. The temperature has in turn to be higher than 
50~K otherwise the optically thick lines saturate too much and do not match the 
observed spectrum. The $^{13}$C isotopologues are clearly detected at 3~mm, 
with optically thin transitions ($\tau_{\rm max} \sim 0.24$), but the range 
of lower-level energies is too narrow (15 to 33~K) to constrain the temperature
well. The $^{15}$N isotopologue is clearly detected at 106.00~GHz (with little
contamination from CH$_3$SH and C$_2$H$_5$CN, $\varv_{13}+\varv_{21}=1$), 
and it significantly contributes to the emission detected at other 
frequencies. One doubly-substituted $^{13}$C isotopologue is detected at 
105.33~GHz (H$^{13}$CC$^{13}$CN) and it, as well as the two other 
doubly-substituted $^{13}$C isotopologues, significantly contributes to the 
emission detected at other frequencies. This justifies their inclusion in the 
complete model. The model for the wing components of the $^{13}$C 
isotopologues has a $\frac{^{12}\rm C}{^{13}\rm C}$ isotopic ratio 
lower than 20, but this is rather a shortcoming of our simple radiative 
transfer method than a true indication that the isotopic ratio is lower than 
20. The lineshape of the main isotopologue is indeed not very well reproduced. 
HCC$^{13}$C$^{15}$N is not detected.

We also detect emission lines from within ten vibrationally excited states of 
the main isotopologue: $\varv_7 = 1$, $\varv_7 = 2$, $\varv_6 = 1$, 
$\varv_7 = 3$, $\varv_5 = 1$, $\varv_6 = \varv_7 =1$, $\varv_4 = 1$, 
$\varv_7 = 4$, $\varv_5 = \varv_7 = 1$, and $\varv_6 = 2$ (223, 446, 
499, 663, 663, 721, 864, 883, 886, and 1013~cm$^{-1}$, respectively), with 
$\tau_{\rm max} \sim 5$, 4, 3, 4, 4, 1.7, 0.6, 0.8, 0.8, and 0.6, 
respectively. Transitions of some of the $^{13}$C isotopologues from within the 
first six states are also detected (for one, two or three of them, depending 
on the state) and 
are all optically thin. This is, to the best of our knowledge, the first 
detection in space of the $^{13}$C isotopologues in their $\varv_7 = 3$ 
(HCC$^{13}$CN), $\varv_5 = 1$ (H$^{13}$CCCN), and $\varv_6=\varv_7=1$ 
(HC$^{13}$CCN and HCC$^{13}$CN, and H$^{13}$CCCN tentatively) excited 
states\footnote{\citet{Pardo07} report the detection of one transition of 
H$^{13}$CCCN in its $\varv_7 = 3$ state toward the protoplanetary nebula 
CRL~618. Transitions from within the $\varv_7 = 1$, $\varv_7 = 2$, and 
$\varv_6 = 1$ states were reported by 
\citet{Wyrowski99} toward the hot core \object{G10.47+0.03}.}. Our previous 
interferometric detection of transitions of HC$_3$N, $\varv_7=1$, HC$_3$N, 
$\varv_4=1$, and HC$^{13}$CCN, $\varv_7=1$ indicates that most of the emission 
detected with the 30~m telescope is very compact 
\citep[see Fig.~5 of][]{Belloche08}. Source sizes were estimated from the 
interferometric maps: 
$\sim 2.5\arcsec$ for the northern hot core for HC$_3$N, $\varv_7=1$;
$\sim 1.5\arcsec$ for the components detected in each wing for HC$_3$N, 
$\varv_7=1$;
$\sim 1.6\arcsec$ for the main (southern) hot core for HC$^{13}$CCN, 
$\varv_7=1$;
$\sim 1.4\arcsec$ for the main hot core for HC$_3$N, $\varv_4=1$. These sizes
were used as guidelines to model the 30~m emission of all states. A temperature
of $\sim 200$~K was  derived from the optically thick lines of 
$\varv_6 = \varv_7 =1$
assuming the source size measured for $\varv_4=1$. This temperature was used for
(nearly) all other states and the column densities were adjusted with the 
$^{13}$C isotopologues if detected, or the optically thin lines of the main 
isotopologue if not. Like for methyl cyanide (Sect.~\ref{sss:ch3cn}), the 
column density parameter needed to fit the emission of the excited 
states increases with 
their energy, suggesting that the vibration temperature is 
higher than the rotation temperature, probably because of radiative 
pumping\footnote{If a direct pumping mechanism from the ground state 
exists for each of the ten excited states listed above, then it has to occur at 
$\sim$~45, 22, 20, 15, 15, 14, 12, 11, 11, and 10~$\mu$m, respectively.
Direct infrared pumping should be possible for $\varv_6 = 1$, $\varv_5 = 1$, 
$\varv_6 = \varv_7 = 1$, and maybe for $\varv_7 = 1$. The intensities of 
2$\nu_7$ and $\nu _4$ are probably too low to support direct infrared pumping
for $\varv_7 = 2$ and  $\varv_4 = 1$.}.
Given that the LTE modeling yields different sizes and rotation
temperatures for the different states, we do not attempt to estimate any
vibration temperature.
There is no clear indication for wing emission for 
$\varv_7 = 3$ and the states further above, hence these states are modeled with 
only two components. No transition from  $\varv_4 = \varv_7 = 1$, 
$(\varv_4=1,\varv_7=2)$, $\varv_5=2$, $\varv_3=1$ (2079~cm$^{-1}$), and 
$\varv_2=1$ (2274~cm$^{-1}$) of the main 
isotopologue is detected, neither from  $\varv_4 = 1$ of the $^{13}$C 
isotopologues and from $\varv_7=1$ of the $^{15}$N isotopologue.

We note one minor issue: the transition of the entry 
$\varv_7=4/\varv_5=\varv_7=1$ of 
HC$_3$N at 100.431~GHz would fit better the observed spectrum if its frequency 
was 2 or 3~MHz higher. Its uncertainty (0.3~MHz) is much higher than for the 
other nearby transitions which are well fitted. This discrepancy may be related
to resonant interactions between $\varv_7=4$ and $\varv_5 = \varv_7=1$ as 
mentioned above.

\paragraph{Sgr\,\,B2(M):} The parameters of all entries related to 
cyanoacetylene are listed in Table~\ref{t:modparam_b2m_hc3n}. The emission
is modeled with three velocity components: one main component and two 
components to account for the linewing emission. The line ratios between
the $^{12}$C and $^{13}$C isotopologues indicate that all transitions are 
optically thin. The source size is not constrained. A temperature of 60~K
fits relatively well all detected transitions. The transitions detected in the
2~mm window seem to indicate a lower temperature for the wing components. The 
$^{15}$N isotopologue is not detected.

We also detect emission lines from within seven vibrationally excited states of 
the main isotopologue: $\varv_7 = 1$, $\varv_7 = 2$, $\varv_6 = 1$, 
$\varv_7 = 3$, $\varv_5 = 1$, $\varv_6 = \varv_7 =1$, and $\varv_4 = 1$. 
Transitions from within the first three vibrationally excited states of the 
$^{13}$C isotopologues are also detected. The interferometric observations of 
\citet{deVicente00} indicate that the source size of these excited states is 
smaller than $3\arcsec \times 5\arcsec$. We find that the transitions from 
within $\varv_7 = 1$ of the main isotopologue are optically thick, while all 
transitions from within the higher excited states are optically thin. A source 
size of 1.3$\arcsec$ and a temperature of 200~K fit the optically thick 
transitions of $\varv_7 = 1$ both at 3 and 1.3~mm relatively well, as well as 
the $^{13}$C optically thin transitions of $\varv_7 = 1$. We keep these 
parameters for all other states. Like for Sgr~B2(N), the column density
parameter needed to fit the emission
of the excited states increases with their energy. We did not find a good
fit to the observed spectra when we added to the model of the ground state
a hot component similar to the one used for the vibrationally excited states.
Such a component produces in particular too strong $^{13}$C lines at 1.3~mm, 
although this may be partly due to an overestimate of the level of the 
baseline in some places. A more detailed analysis would be necessary to
resolve this inconsistency.

\subsubsection{Cyanodiacetylene HC$_5$N}
\label{sss:hc5n}

We use the CDMS entries for the ground state and the first vibrationally
excited state $\varv_{11}=1$ 
\citep[tags 75503 and 75504, both version 1,][]{HC5N_2004}.

\paragraph{Sgr\,\,B2(N):} Cyanodiacetylene is not detected.

\paragraph{Sgr\,\,B2(M):} The parameters of all entries related to 
cyanodiacetylene are listed in Table~\ref{t:modparam_b2m_hc5n}. Simultaneously
fitting the transitions detected at 3 and 1.3~mm requires two temperature 
components. All transitions are optically thin ($\tau_{\rm max} = 0.11$).
The $^{13}$C isotopologues are not detected.

We also detect emission lines from within the vibrationally excited state 
$\varv_{11}=1$ (106.8~cm$^{-1}$), but with a low signal-to-noise ratio. These 
transitions are well fitted with the hot component used for the ground state.

\subsubsection{Ethynyl isocyanide HCCNC}
\label{sss:hccnc}

We use the JPL entry 51004 \citep[version 1,][]{HCCNC_1992}.

\paragraph{Sgr\,\,B2(N):} The parameters of ethynyl isocyanide are listed in 
Table~\ref{t:modparam_lmh_hccnc}. The detection relies on only two transitions,
one uncontaminated at 99.35~GHz and one blended with the second component of 
CH$_3$OCHO at 109.29~GHz where, depending on the true level of the baseline, 
they may be in conflict. This detection should therefore be viewed as very
tentative. If confirmed, it would be the first detection of HCCNC in a hot 
core \citep[see][ for detections toward \object{TMC1} and \object{IRC+10216}, 
respectively]{Kawaguchi92,Gensheimer97}.
We assumed the same source size and temperature as for the main 
component of HC$_3$N. The $^{13}$C and $^{15}$N isotopologues are not detected.
No transition from within the vibrationally excited states $\varv_7=1$,
$\varv_6=1$, and $\varv_4=1$ of the main isotopologue is detected.

\paragraph{Sgr\,\,B2(M):} Ethynyl isocyanide is not detected.

\subsubsection{Hydrogen cyanide HCN}
\label{sss:hcn}

We use the CDMS entries for all isotopologues and vibrationally excited states
(version 4 for HCN, version 2 for HCN, $\varv_2=1$ and H$^{13}$CN, and 
version 1 for all other entries). The HCN, $\varv_2=0$ and $\varv_2=1$ entries 
are based mainly on \citet{HCN_2003}, those for higher excited states mainly 
on \citet{HCN_hohe-v_2003}. Among the transitions observed in the current 
survey, $\varv_2=0$ data were taken from \citet{HCN_v0_1969}, and for several 
of the higher excited transitions from \citet{HCN_hohe-v_1977}. The H$^{13}$CN 
entries are mostly based on \citet{HC-13-N_v2_2004} and \citet{HC-13-N_v0_2005} 
for $\varv_2=1$ and $\varv_2=0$, respectively. The entries for the ground 
vibrational states of HC$^{15}$N, H$^{13}$C$^{15}$N, and DCN were based largely 
on \citet{HCN-15_2005}, \citet{HC-13-N_v2_2004}, and \citet{DCN_2004}, 
respectively.

\paragraph{Sgr\,\,B2(N):} The parameters of all entries related to hydrogen 
cyanide are listed in Table~\ref{t:modparam_lmh_hcn}. The molecule is
detected at 3~mm both in emission and absorption. Its $^{13}$C isotopologue
is detected at 3, 2, and 1.3~mm and its $^{15}$N isotopologue at 3 and 1.3~mm. 
H$^{13}$C$^{15}$N is detected at 83.73~GHz, to our knowledge for the first time
in space. The parameters of the emission components are not well constrained. 
The diffuse-cloud absorption components were 
modeled first on the $^{13}$C isotopologue, then transfered to the main 
isotopologue and readjusted when they were too weak, and finally transfered 
back to the $^{13}$C and $^{15}$N isotopologues. Only a few diffuse-cloud 
components are detected for the latter, but the two self-absorption components
are prominent. Some absorption components are too deep for the main 
isotopologue. The possible reasons for this discrepancy are the same as for CS 
(see Sect.~\ref{sss:cs}).

Following the detection and modeling of several transitions from within 
vibrationally excited states of HCN toward Sgr~B2 by \citet{Rolffs11}, we 
include $\varv_2=1$ (712.0~cm$^{-1}$)  and $\varv_2=2$ 
(1411.4~cm$^{-1}$) in our model, with the source size and 
rotation temperature derived by these authors. Four transitions from within 
$\varv_2=1$ are detected at 3~mm, only one from within $\varv_2=2$. There is 
no detection for $\varv_2=3$, $\varv_1=1$, and $\varv_3=1$. One transition of 
the $^{13}$C isotopologue from within $\varv_2=1$ is detected at 2~mm.

\paragraph{Sgr\,\,B2(M):} The parameters of all entries related to hydrogen 
cyanide are listed in Table~\ref{t:modparam_b2m_hcn}. Like for Sgr~B2(N),
the parameters of the emission components are not well constrained. We
followed the same strategy as for Sgr~B2(N) to model the diffuse-cloud 
absorption 
components. Only a few of them are detected in HC$^{15}$N. The H$^{13}$C$^{15}$N
isotopologue is tentatively detected in absorption at 83.73~GHz. We also
detect DCN at 1.3~mm, but this detection is based on only one transition,
so it is uncertain.

We included $\varv_2=1$ and $\varv_2=2$ in our model with the same temperature 
as for Sgr~B2(N). We reduced the source size to limit the strength of the 
transition of $\varv_2=2$ in the 1.3~mm window, but its lineshape is not good 
(too much saturation). The vibrational states $\varv_2 = 3$, $\varv_3 = 1$, and 
$\varv_1 = 1$ are likely detected at 3~mm too, but the LTE model that fits them
predicts much too strong lines at 1.3~mm, so we did not include them in the
complete model. The vibrational state $\varv_2 = 1$ of the $^{13}$C isotopologue
may be detected at 172.63~GHz, but the assignment is uncertain and we do not
include it in the complete model.

\subsubsection{Formamide NH$_2$CHO}
\label{sss:nh2cho}

We use the CDMS entries for the vibrational ground state of the $^{12}$C and
$^{13}$C isotopologues (tags 45512 and 46512, both version 2)
as well as for $\varv_{12} = 1$ of the main species (tag 45516 
version 1). These entries have been updated or created in the review 
process of this article. They are based largely on \citet{Motiyenko12}, 
but also contain additional data. Laboratory data in the range of our 
survey were published by \citet{HCONH2_2009}. The entry for
$\varv_{12} = 1$ of the $^{13}$C isotopologue was prepared by one of us (HSPM) 
based on data from \citet{HCONH2_1978}.

\paragraph{Sgr\,\,B2(N):} The parameters of all entries related to formamide 
are listed in Table~\ref{t:modparam_lmh_hconh2}. The strongest transitions
detected at 3~mm are optically thick ($\tau_{\rm max} \sim 7.5$).
Transitions with $E_l/k < 10$~K are seen both in emission and in absorption 
(with a velocity offset of 19~km~s$^{-1}$, see, e.g., at 81.69, 84.54, and 
87.85~GHz). The $^{13}$C isotopologue is detected at 3~mm with 
optically thin ($\tau_{\rm max} \sim 0.38$) transitions of
lower-level energy up to 57~K (at 105.95~GHz). The temperature is somewhat 
better constrained with the weak transitions of the main isotopologue that are 
optically thin like, e.g., at 101.30, 104.18, and 108.33~GHz ($E_l/k = 185$, 
385, and 150~K, respectively). The source size is in turn constrained by the
optically thick lines. The presence of a second velocity component from the
northern hot core is clearly seen. The $^{15}$N and $^{18}$O isotopologues are
not detected at 3~mm because of blends with emission from other species.

We also detect emission lines from within the vibrationally excited state 
$\varv_{12}=1$ of the main isotopologue (289~cm$^{-1}$). They are optically thin
($\tau_{\rm max} \sim$~0.75) and two velocity components are clearly detected.
Transitions of the $^{13}$C isotopologue from within $\varv_{12}=1$ are 
predicted to be weak. A few of them contribute significantly ($>30\%$) to the 
detected emission, and one is potentially detected at 84.33~GHz, which justifies
including this entry in the complete model. Transitions pertaining to  
$\varv_{12}=1$ of the main isotopologue have been reported in the ISM very 
recently only \citep[][]{Motiyenko12}. In addition, it is
worthwhile mentioning that ten unidentified lines observed toward Sgr~B2(N) by 
\citet{Nummelin00} can be assigned, at least in part, to formamide in its 
$\varv_{12} = 1$ excited vibrational state. 

\paragraph{Sgr\,\,B2(M):} The parameters of all entries related to formamide 
are listed in Table~\ref{t:modparam_b2m_hconh2}. Transitions of formamide are 
detected at 3~mm with lower-level energies ranging from 6~K (84.54~GHz) to 58~K 
(106.11~GHz). At 1.3~mm, they range up to 256~K (212.25~GHz). Two components, 
a cold one and a warm one, are necessary to fit the 3 and 1.3~mm transitions
simultaneously. All transitions are optically thin ($\tau_{\rm max} = 0.12$ 
and 0.035 for the cold and warm components at 3~mm, respectively, and 0.07
and 0.17 at 1.3~mm). The source sizes are thus not constrained. Two additional
components in absorption are added to the model to account for the
self-absorption seen at, e.g., 84.54 and 87.85~GHz.

\subsubsection{Formyl cation HCO$^+$}
\label{sss:hcop}

We use the CDMS entries for the $^{12}$C, $^{13}$C, $^{18}$O, and $^{17}$O 
isotopologues (tags 29507, 30504, both version 2, and 31506, 30505, both 
version 1). The entries are based on \citet{HCO+_2007}, \citet{HC-13-O+_2007},
\citet{HC-13-O+_2004}, and \citet{HCO-17+_2001a}, respectively. The 
$J =$~1--0 transitions were taken from \citet{HCO+_1994} for HCO$^+$, 
\citet{HC-13-O+_2004} for the $^{13}$C and $^{18}$O isotopologues, and from 
\citet{HCO-17+_2001b} for the $^{17}$O isotopologue.

\paragraph{Sgr\,\,B2(N):} The parameters of all entries related to the formyl 
cation are listed in Table~\ref{t:modparam_lmh_hcop}. The model for this ion
and its isotopologues is based on the model constructed for Sgr~B2(M) and 
already reported in \citet[][]{Menten11}. The components in emission are
very poorly constrained. The parameters for the diffuse-cloud absorption 
components were essentially optimized for the $^{13}$C isotopologue and then 
transfered to the other isotopologues assuming the isotopic ratios listed
in Table~\ref{t:iso}. Both self-absorption components (at 
$\varv_{\rm off} \sim 0$ and 16~km~s$^{-1}$) are prominent for the $^{18}$O and
$^{17}$O isotopologues. A few diffuse-cloud components appear to be too weak 
for the former, but this could be due to an improper estimate of the level of 
the baseline. The main isotopologue suffers from the opposite effect: many 
diffuse-cloud components appear to be too deep (see Sect.~\ref{sss:cs} for 
possible explanations).

\paragraph{Sgr\,\,B2(M):} The parameters of all entries related to the 
formyl cation are listed in Table~\ref{t:modparam_b2m_hcop}. The model was 
already reported in \citet[][]{Menten11}. The emission components are poorly
constrained. Some diffuse-cloud absorption components seem to be detected in
the $^{17}$O isotopologue, but the uncertain level of the baseline prevents 
an accurate analysis.

\subsubsection{Thiomethylium HCS$^+$}
\label{sss:hcsp}

We use the CDMS entry (tag 45506) which is based on frequencies summarized
by \citet{Margules03}. Measurements in the range of our survey were reported by
\citet{Gudeman81} and \citet{Tang95}.

\paragraph{Sgr\,\,B2(N):} The parameters of thiomethylium are listed in 
Table~\ref{t:modparam_lmh_hcsp}. The 5--4 transition of HCS$^+$ is clearly 
detected at 213.36~GHz. However the 2--1 transition at 85.35~GHz is blended 
with a diffuse-cloud absorption component of \textit{c}-C$_3$H$_2$ and we see 
no evidence for emission at that frequency. Our LTE modeling yields an 
acceptable fit at both frequencies only if we assume a compact source 
($FWHM < 10 \arcsec$) and a temperature higher than $\sim 50$~K. A more 
extended source would produce emission at 85.35~GHz that would not be 
consistent with our spectrum. Given that HCS$^+$ was also claimed to
be detected toward Sgr~B2(N) at 256.03~GHz by \citet{Nummelin98}, we consider 
the detection as reliable, even if we detect only one line in our survey. 
However, the parameters of our model are very uncertain.

\paragraph{Sgr\,\,B2(M):} The parameters of thiomethylium are listed in 
Table~\ref{t:modparam_b2m_hcsp}. The situation is very similar to the case of 
Sgr~B2(N). In addition, we exclude that the column density of the overlapping 
absorption component of \textit{c}-C$_3$H$_2$ is underestimated (which would
allow for a more extended HCS$^+$ emission) because it fits the absorption 
feature at 82.10~GHz reasonably well. \citet{Nummelin98} and \citet{Sutton91} 
report the detection of HCS$^+$ at 256.03 and 341.35~GHz, respectively, toward 
Sgr~B2(M), which again gives us confidence in the identification of this 
molecule in our survey.

\subsubsection{Singly deuterated water HDO}
\label{sss:hdo}

We use the JPL entry (tag 19002 version 3). The entry is based on 
\citet{HDO_1985}, and rest frequencies of transitions at 3~mm were taken in 
particular from \citet{HDO_D2O_1956} and \citet{HDO_1971}.

\paragraph{Sgr\,\,B2(N):} The parameters of singly deuterated water are listed 
in Table~\ref{t:modparam_lmh_hdo}. Only one line is detected at 80.58~GHz 
($E_l/k = 43$~K) with two velocity components. The temperature has to be lower 
than 300~K to be consistent with the upper limit derived for another 
transition with $E_l/k = 826$~K at 87.96~GHz. The detection of other, 
higher-frequency transitions of HDO toward Sgr~B2(N) and (M) gives us 
confidence that the detection at 3~mm can be trusted 
\citep[see, e.g.,][]{Comito03,Comito10}. However, our model parameters based
on the 3~mm window only are not constrained, except for the upper limit on the 
temperature mentioned above.

\paragraph{Sgr\,\,B2(M):} The parameters of singly deuterated water are 
listed in Table~\ref{t:modparam_b2m_hdo}. The situation is the same as for
Sgr~B2(N).

\subsubsection{Hydrogen isocyanide HNC}
\label{sss:hnc}

We use the CDMS entries for the $^{12}$C and $^{13}$C isotopologues (tags 27502 
and 28515, versions 2 and 1, respectively), and the JPL entry for the 
$^{15}$N isotopologue (tag 28006 version 1). The entries are based on results 
from \citet{HNC_2000}, \citet{HNC-13_etc_2009}, and \citet{HN-15-C_1976}, 
respectively. The $J =$~1--0 transition of the main isotopologue was taken from 
\citet{HNC_1976}.

\paragraph{Sgr\,\,B2(N):} The parameters of all entries related to hydrogen 
isocyanide are listed in Table~\ref{t:modparam_lmh_hnc}. The emission and
absorption components of hydrogen isocyanide were modeled following the same 
strategy as for hydrogen cyanide (Sect.~\ref{sss:hcn}). Both the $^{13}$C and 
$^{15}$N isotopologues are detected, mainly the two self-absorption 
components. A few diffuse-cloud components are also detected for the $^{13}$C 
isotopologue. The other ones are too weak or blended with emission from other 
species.

\paragraph{Sgr\,\,B2(M):} The parameters of all entries related to hydrogen 
isocyanide are listed in Table~\ref{t:modparam_b2m_hnc}. The model was prepared
following the same strategy as for Sgr~B2(N). The poor baseline quality 
around 88.9~GHz prevents the secure detection of diffuse-cloud absorption 
components for the $^{15}$N isotopologue.

\subsubsection{Isocyanic acid HNCO}
\label{sss:hnco}

We use the CDMS entries for the vibrational ground state of the $^{12}$C 
isotopologue (tag 43511 version 1) by \citet{HNCO_2007} with additional 
measurements in the range of our survey by \citet{isos-HNCO_1975}, the JPL 
entries for the $^{13}$C and $^{18}$O isotopologues 
\citep[tags 44008 and 45006, both version 1,][]{isos-HNCO_1975}, and private
entries for the vibrationally excited states $\varv_5 = 1$, $\varv_6 = 1$, and
$\varv_4 = 1$ of the main isotopologue, prepared by one of us (HSPM) based on 
a preliminary, unpublished analysis of the ground and the four lowest excited 
states.  The excited states data were summarized in \citet{vib-HNCO_1996}.
Transition frequencies in the range of our survey were published by 
\citet{vib-HNCO-b_1977} and \citet{vib-HNCO-a_1977}.

The C atom of HNCO is close to the center of mass of the molecule. 
Therefore, the transitions of HN$^{13}$CO are frequently overlapped by those 
of the main species, especially at low frequencies.

\paragraph{Sgr\,\,B2(N):} The parameters of all entries related to isocyanic 
acid are listed in Table~\ref{t:modparam_lmh_hnco}. Our model for the 
vibrational ground state of isocyanic acid was already reported in 
\citet{Bruenken10}. The model for the $^{13}$C isotopologue is difficult to
constrain because its transitions are always blended with the emission of the 
blueshifted wing of the main isotopologue. We simply used a scaled version of
the model of the latter. The $^{18}$O isotopologue is detected at only one
frequency (83.19~GHz).

We also detect emission lines from within three vibrationally excited states of 
the main isotopologue: $\varv_5 = 1$, $\varv_6 = 1$, and $\varv_4 = 1$ (577, 
656, and 777~cm$^{-1}$, respectively). The detected 
transitions are all optically thin ($\tau_{\rm max} \sim 0.17$, 0.18, and 0.26, 
respectively), and the presence of a second velocity component is plausible
for $\varv_5 = 1$. The column density parameters required to fit the 
vibrationally excited states are higher than for the ground state. This may 
be due to radiative pumping. Using equation~\ref{e:tvib} with the 
ground state as reference state and only for the velocity component with a
source size of 2.4$\arcsec$, we derive vibration temperatures of 
$310 \pm 20$, $290 \pm 10$, and $390 \pm 20$~K for $\varv_5 = 1$, 
$\varv_6 = 1$, and $\varv_4 = 1$, respectively. If a direct pumping 
mechanism from the ground state exists for each of the three excited states
listed above, then it has to occur at $\sim$~17, 15, and 13~$\mu$m, 
respectively. Direct infrared pumping is likely for $\varv_5 = 1$ and 
$\varv_6 = 1$. The larger effect seen for $\varv_4 = 1$ is compatible with the 
higher intensity of the infrared band $\nu_4$.

\paragraph{Sgr\,\,B2(M):} The parameters of all entries related to 
isocyanic acid are listed in Table~\ref{t:modparam_b2m_hnco}. The model was
already reported in \citet{Bruenken10}, except that we added a third component
to account for the wing emission, which is prominent for the low-energy
transitions only, indicating that it has a low excitation temperature. Like 
for Sgr~B2(N), 
the emission of the $^{13}$C isotopologue is always blended with the main 
isotopologue. The $^{18}$O isotopologue is not detected.

\subsubsection{Isothiocyanic acid HNCS}
\label{sss:hncs}

We use the CDMS entry (tag 59503 version 1). The rest frequencies were 
summarized in \citet{HNCS_1995}, those in the range of our survey were taken 
from \citet{HNCS_1979}.

\paragraph{Sgr\,\,B2(N):} The parameters of isothiocyanic acid are listed in 
Table~\ref{t:modparam_lmh_hncs-a}. Our survey alone does not provide a
secure detection of this molecule: the three lines potentially detected at 
3~mm are all partially blended (with absorption components of $c$-C$_3$H$_2$, 
unidentified lines, and CH$_3$OCH$_3$, respectively). However, the 
detection by 
\citet{Halfen09} at a position about 10$\arcsec$ offset from Sgr~B2(N) looks 
secure enough to include this molecule in our complete model. We assumed a 
rotation temperature of 20~K, consistent with the temperature derived 
by these authors ($18 \pm 3$~K), and a beam filling factor of unity, consistent 
with the extent of the emission (at least $3\arcmin \times 6\arcmin$) seen in 
maps of \citet{Adande10}. All three lines are optically thin 
($\tau_{\rm max} \sim 0.014$). The column density derived from our modeling is
about three times higher than the value obtained by \citet{Halfen09} from an
excitation diagram analysis. The position offset between both datasets is 
small compared to the large beam of the ARO 12~m telescope. If the emission is 
extended and relatively uniform, this should not affect the column density 
estimates. We note that we assumed a linewidth smaller than the one measured 
by \citet{Halfen09} (20 versus 25~km~s$^{-1}$), but using a higher value would 
increase the discrepancy even further. The reason for this discrepancy is 
still unclear.

\paragraph{Sgr\,\,B2(M):} The parameters of isothiocyanic acid are listed 
in Table~\ref{t:modparam_b2m_hncs-a}. The emission of this molecule is less
blended than toward Sgr~B2(N). It is detected at 93.83 and 105.56~GHz. We model
the emission with the same parameters as HSCN, apart for the column density 
(see Sect.~\ref{sss:hscn}).

\subsubsection{Cyanic acid HOCN}
\label{sss:hocn}

We use the CDMS entry \citep[tag 43510 version 1,][]{HOCN_2009}.

\paragraph{Sgr\,\,B2(N):} The parameters of cyanic acid are listed in 
Table~\ref{t:modparam_lmh_hocn}. Our model for the vibrational ground state of 
cyanic acid was already reported in \citet{Bruenken10}. It has been slightly 
updated and we removed the two high-temperature components that were not 
really constrained. One line is clearly detected at 83.90~GHz but the other 
main transition in the 3~mm window (at 104.87~GHz) is blended with emission 
from C$_2$H$_3$CN, $\varv_{11}=1$/$\varv_{15}=1$ and $^{13}$CH$_3$CH$_2$CN. Our 
complete model overestimates the observed intensity of this blended feature, 
but this is likely due to an overestimate of the baseline level. The 
transitions are optically thin ($\tau_{\rm max} \sim 0.028$).

\paragraph{Sgr\,\,B2(M):} The parameters of cyanic acid are listed in 
Table~\ref{t:modparam_b2m_hocn}. The model was already reported in 
\citet{Bruenken10}.

\subsubsection{Protonated carbon dioxide HOCO$^+$}
\label{sss:hocop}

We use the JPL entry \citep[tag 45010 version 1,][]{HOCO+_1988}.

\paragraph{Sgr\,\,B2(N):} The parameters of protonated carbon dioxide are 
listed in Table~\ref{t:modparam_lmh_hocop}. Two transitions are detected at 
3~mm, one at 85.53~GHz ($E_l/k = 6.2$~K) blended with C$_2$H$_3$CN, 
$\varv_{15}=1$, and one at 106.91~GHz ($E_l/k = 10$~K) partially blended with 
an unidentified line. The emission at 85.53~GHz looks extended in the 
large-scale maps of \citet{Jones08} so we assumed a beam-filling factor of 
unity. The peak-temperature ratio of both lines plus the upper limit for a 
transition at 85.85~GHz ($E_l/k = 44$~K) constrain the temperature to be lower 
than 20~K, provided the LTE assumption is valid. The transitions are all 
optically thin ($\tau_{\rm max} \sim 0.017$).

\paragraph{Sgr\,\,B2(M):} The parameters of protonated carbon dioxide are 
listed in Table~\ref{t:modparam_b2m_hocop}. The same transitions as toward
Sgr~B2(N) are detected. They are also optically thin ($\tau_{\rm max} \sim 
0.026$).

\subsubsection{Hydroxymethylidynium HOC$^+$}
\label{sss:hocp}

We use the CDMS entry (tag 29504 version 1). Rest frequencies for the ground 
and first excited bending mode were published by \citet{HOC+_2000}, 3~mm data 
were taken from \citet{HOC+_3mm_1982}.

\paragraph{Sgr\,\,B2(N):} The parameters of hydroxymethylidynium are listed in 
Table~\ref{t:modparam_lmh_hocp}. Only one transition is detected at 3~mm, both 
in emission and absorption. We assumed a beam-filling factor of unity and a 
temperature of 20~K for the emission component. The $^{13}$C isotopologue is 
not detected. We do not detect any transition from within the vibrationally
excited state $\varv_2=1$ of the main isotopologue either.

\paragraph{Sgr\,\,B2(M):} The parameters of hydroxymethylidynium are listed 
in Table~\ref{t:modparam_b2m_hocp}. Only one line is detected, mostly in 
absorption. The parameters are very poorly constrained.

\subsubsection{Thiocyanic acid HSCN}
\label{sss:hscn}

We use the CDMS entry \citep[tag 59505 version 1,][]{HSCN_2009}.

\paragraph{Sgr\,\,B2(N):} The parameters of thiocyanic acid 
are listed in Table~\ref{t:modparam_lmh_hscn}. Our survey alone is not 
sufficient to secure the detection of this molecule but, like for 
isothiocyanic acid (see Sect.~\ref{sss:hncs}), we include it in our complete 
model because it was rather convincingly detected by \citet{Halfen09}. We 
assumed the same size and temperature as for HNCS, which is consistent with
the findings of these authors (extended emission and 
$T_{\rm rot} = 19 \pm 2$~K). All transitions in the 3~mm window are at 
least partially blended with emission from other species. At 91.75~GHz, the
synthetic spectrum accounts for only $\sim 50\%$ of the intensity of the 
detected feature, suggesting that the HSCN emission is blended with a still 
unidentified line. All transitions are optically thin 
($\tau_{\rm max} \sim 0.014$).

\paragraph{Sgr\,\,B2(M):} The parameters of thiocyanic acid are listed in 
Table~\ref{t:modparam_b2m_hscn}. The molecule is clearly detected at 
91.75 and 103.22~GHz. The other transitions at 3~mm are too weak or blended 
with other species. We model the emission with a beam filling factor of unity
because the map of \citet{Adande10} shows that it is extended over several 
arcminutes. We use a temperature close to the one measured by \citet{Halfen09} 
toward Sgr~B2(N). A higher temperature produces too strong $K = 1$ transitions.
All transitions are optically thin  ($\tau_{\rm max} \sim 0.009$).

\subsubsection{Diazenylium N$_2$H$^+$}
\label{sss:n2hp}

We use the CDMS entries with hyperfine structure for the $^{14}$N and $^{15}$N 
isotopologues (tags 29506 version 3, 30507 and 30508, both version 2).
Predictions for the main isotopologue were generated from an 
unpublished analysis using hyperfine-free center frequencies mainly from 
\citet{N2H+_2005} and \citet{N2H+_2009}. The $^{14}$N hyperfine splitting 
values were taken from \citet{N2H+_1995}. Entries for the $^{15}$N 
isotopologues were generated from results published by 
\citet{15N-N2H+2009}. The $J =$~1--0 transitions were taken from 
\citet{15N-N2H+_1982}. 

\paragraph{Sgr\,\,B2(N):} The parameters of diazenylium are listed in 
Table~\ref{t:modparam_lmh_n2hp}. The 1--0 transition of diazenylium is detected
both in emission and in multiple absorption components produced by spiral-arm 
diffuse clouds along the line of sight. The hyperfine structure of the 
emission component is not resolved but it is for some of the absorption 
components. The shape of the emission and self-absorption components is 
relatively well reproduced, but the model parameters are certainly not unique. 
Fifteen diffuse-cloud components were needed to fit the absorption part. Some 
of them are blended with emission from other species, which implies that their 
parameters are model-dependent and thus more uncertain. The $^{15}$N 
isotopologues are not (and cannot be) detected because their transitions are 
heavily blended with emission from $^{13}$CH$_3$OH and HC$_3$N, $\varv_7=1$.

\paragraph{Sgr\,\,B2(M):} The parameters of all entries related to 
diazenylium are listed in Table~\ref{t:modparam_b2m_n2hp}. Diazenylium is
detected in emission and in absorption. We fit the diffuse-cloud absorption with
9 components. $^{15}$NNH$^+$ is detected in absorption at 90.26~GHz, but 
N$^{15}$NH$^+$ is blended at 91.20~GHz with a transition from within the 
vibrationally excited state $\varv_7=1$ of HC$_3$N, which prevents a firm 
identification.

\subsubsection{Cyanamide NH$_2$CN}
\label{sss:nh2cn}

We use the JPL entry for the $^{12}$C isotopologue (tag 42003 version 1),
based on \citet{H2NCN_1993} with measurements in the frequency range 
of our survey from \citet{H2NCN_1976} and \citet{H2NCN_1986}, and the CDMS 
entry for the $^{13}$C isotopologue (tag 43515 version 1) based on 
\citet{isos-H2NCN_2011}.

The C atom of cyanamide is very close to the center of mass of the 
molecule. Therefore, the transitions of NH$_2$$^{13}$CN are frequently 
overlapped by those of the main species, in particular at long wavelengths.

\paragraph{Sgr\,\,B2(N):} The parameters of all entries related to cyanamide 
are listed in Table~\ref{t:modparam_lmh_nh2cn}. About ten transitions of
cyanamide are detected at 3~mm, with lower-level energies ranging from 9.6~K 
(at 99.97~GHz) to 136~K (80.06~GHz). A fit with two temperature components 
yields a better result than with only one. The presence of a low-excitation, 
extended component is strongly suggested by the large-scale map of 
\citet[][, see their Fig.~9]{Jones08}. Although not fully certain, the 
presence of a second (warm) velocity component is plausible. The emission
corresponding to the warm component is optically thin for a source size of 
2$\arcsec$ ($\tau_{\rm max} \sim 0.9$), but this size is not constrained. 
The $^{13}$C isotopologues were included in the complete model assuming a 
$^{12}$C/$^{13}$C isotopic ratio of 20 because there is a hint of weak 
detection at 100.59~GHz and their emission is consistent with the observed
spectrum at other frequencies, but the detection is not secure in itself.

\paragraph{Sgr\,\,B2(M):} The parameters of cyanamide are listed in 
Table~\ref{t:modparam_b2m_nh2cn}. We detect four transitions with lower-level
energies ranging from 10~K (99.97~GHz) to 24~K (99.31~GHz). We see no evidence
at 3 and 1.3~mm for the presence of a warm component (e.g., upper limit of a 
transition with $E_l/k = 81$~K at 99.89~GHz). The $^{13}$C isotopologue is
not detected.

\subsubsection{Singly deuterated ammonia NH$_2$D}
\label{sss:nh2d}

We use the CDMS entry (tag 18501 version 1). The spectroscopic parameters were 
taken from \citet{H2DO+_w-NH2D_2010}. That work employed transition 
frequencies in the range of our survey from \citet{NH2D_NHD2_1975} and 
\citet{NH2D_1982}.

\paragraph{Sgr\,\,B2(N):} The parameters of singly deuterated ammonia are 
listed in Table~\ref{t:modparam_lmh_nh2d}. Only one transition is clearly 
detected at 99.12~GHz ($E_l/k = 257$~K), with two velocity components. Another 
transition (with $E_l/k = 29$~K) significantly contributes ($\sim 30\%$) to a 
feature detected at 85.93~GHz, the rest of the emission coming from 
C$_2$H$_3$CN, $\varv_{11}=2$ and CH$_3$OCHO. The temperature is not well 
constrained. The emission is optically thin ($\tau_{\rm max} \sim 0.17$).

\paragraph{Sgr\,\,B2(M):} Singly deuterated ammonia is not detected.

\subsubsection{Nitrogen sulfide radical NS}
\label{sss:ns}

We use the CDMS entry (tag 46515 version 1). Predictions were generated from a 
combined fit of several isotopic species, similar to the one from 
\citet{NS_combined_1995}. The data for the main isotopologue were 
contributed by \citet{NS_main_1995}.

\paragraph{Sgr\,\,B2(N):} The parameters of the nitrogen sulfide radical are 
listed in Table~\ref{t:modparam_lmh_ns}. All transitions detected at 3~mm 
have the same lower-level energy (4.5~K). However, several higher-energy 
transitions are detected in our 1.3~mm survey, implying a temperature of 20~K
at least. An additional self-absorption component is needed to reproduce the 
shape of the lines detected at 3~mm. No minor isotopologue is detected.

\paragraph{Sgr\,\,B2(M):} The parameters of nitrogen sulfide radical are 
listed in Table~\ref{t:modparam_b2m_ns}. The model is constructed in a similar
way as for Sgr~B2(N). The emission component underestimates the transitions 
around 115.57~GHz while those around 115.16~GHz, with the same lower-level 
energy, are well fitted.

\subsubsection{Carbonyl sulfide OCS}
\label{sss:ocs}

We use the CDMS entries for the vibrational ground state of the $^{12}$C
\citep{OCS_main_2005}, $^{13}$C \citep{OCS_1980}, $^{34}$S 
\citep{OCS-34_1989}, $^{33}$S, and $^{18}$O \citep{OCS_w18_33_1981} 
isotopologues and for the vibrationally excited state $\varv_2 = 1$ of the
 main isotopologue \citep{OCS_excited_2000} (tags 60503 
version 2, 61502, 62505, 61503, 62506, and 60504, all version 1).
(Additional) transition frequencies in the range of our survey were 
contributed by \citet{OCS_1980} for all species, by 
\citet{OCS_main_1970} and \citet{DBr_OCS_97GHz_1974} for the main isotopologue 
in its ground vibrational state, and by \citet{OC-13-S_1974} for the $^{13}$C 
isotopologue and for $\varv_2 = 1$ of the main species.

\paragraph{Sgr\,\,B2(N):} The parameters of all entries related to carbonyl 
sulfide are listed in Table~\ref{t:modparam_lmh_ocs}. The $^{13}$C, $^{34}$S,
and $^{33}$S isotopologues are clearly detected at 3~mm with two velocity 
components. The flux of the $^{34}$S isotopologue detected by 
\citet{Friedel04} at 106.79~GHz with BIMA with a beam of 
$23\arcsec \times 5\arcsec$ toward Sgr~B2(N) agrees within 25\% with the flux 
that they detected with the NRAO 12~m telescope, which means that the emission 
is dominated by a compact component. However, the large-scale Mopra maps of 
\citet{Jones08} also show some extended emission from the main isotopologue 
(see their Fig.~6). The range of lower-level energies of the transitions 
detected at 3~mm is limited, so the temperature is not well constrained by 
the minor isotopologues. However, the transitions of the main isotopologue are 
optically thick ($\tau_{\rm max} \sim 4$) and thus constrain the product of 
the source size and temperature. We arbitrarily assumed a source size of 
8$\arcsec$, then implying a temperature of $\sim 60$~K. Two additional 
components are needed to account for the line wings of the main isotopologue. 
Their parameters are poorly constrained. This wing emission is barely detected 
in the $^{13}$C isotopologue, and is too weak (or blended with emission from 
other species) in the other isotopologues. The shape of the core of the lines 
detected in the main isotopologue is not well reproduced with the two velocity
components. This may be due to the high optical depths and the presence of a 
colder, optically thick (?), foreground component as suggested by the Mopra 
maps but not accounted for by our simple model. The $^{18}$O isotopologue may
be detected at 102.68~GHz. It is included in the complete model assuming
the isotopic ratio given in Table~\ref{t:iso}, but a $^{16}$O/$^{18}$O ratio 
twice lower would still be consistent with the data. The emission of the other 
isotopologues is too weak to be detected.

We also detect several transitions from within the vibrationally excited state
$\varv_2=1$ of the main isotopologue (e.g., 97.42, 97.52, and 109.71~GHz), but 
the model parameters are not constrained. The wings detected in the ground-state
transitions are not seen for these transitions.

Transitions belonging to the $\varv _2=1$ and $\varv _3=1$ 
vibrationally excited 
states of the main isotopologue were reported toward \object{Orion~KL} by 
\citet{Orion_Pepe_S-species}.

\paragraph{Sgr\,\,B2(M):} The parameters of all entries related to carbonyl 
sulfide are listed in Table~\ref{t:modparam_b2m_ocs}. We detect transitions
from the main isotopologue and the $^{34}$S isotopologue at 3 and 1.3~mm.
The $^{13}$C isotopologue is detected at 3 and 2~mm, and requires two 
temperature components to be well fitted. A third component is detected in the 
redshifted wing of the low-energy transitions of the main isotopologue at 
3~mm, but not for the higher-energy transitions at 1.3~mm, indicating a low 
temperature. The other isotopologues are not detected.

\subsubsection{Phosphorus nitride PN}
\label{sss:pn}

We use the CDMS entry with hyperfine structure (tag 45511 version 1) based on 
\citet{PN_2006}. The hyperfine parameters were taken from \citet{PN_HFS_1971}.

\paragraph{Sgr\,\,B2(N):} The parameters of phosphorus nitride are listed in 
Table~\ref{t:modparam_lmh_pn}. The detection is uncertain because it relies
on only one feature seen in absorption with two velocity components at 
93.98~GHz. Nevertheless, we include it in the complete model because 
phosphorus nitride was already detected (in emission) toward \object{Orion-KL} 
and \object{Sgr~B2(OH)} \citep[][]{Ziurys87}, and more recently toward two 
shocked regions related to the outflow of the protostar 
\object{IRAS~20386+6751} in \object{L~1157} \citep[][]{Yamaguchi11}

\paragraph{Sgr\,\,B2(M):} The parameters of phosphorus nitride are listed 
in Table~\ref{t:modparam_b2m_pn}. The case is the same as for Sgr~B2(N).

\subsubsection{Silicon monoxide SiO}
\label{sss:sio}

We use the CDMS entries for SiO and its $^{29}$Si, $^{30}$Si, and $^{18}$O 
isotopologues (tags 44505, 45504, 46502, and 46503, all version 1). The 
predictions are based on an unpublished isotope-invariant fit employing 
rest frequencies in the range of our survey from \citet{SiO_1977} and 
\citet{SiO_1991} for the main isotopic species and from \citet{SiO-18_1998}. 
Isotopic information on the $^{29}$SiO and $^{30}$SiO isotopologues come 
mainly from \citet{isos-SiO_1968}.

\paragraph{Sgr\,\,B2(N):} The parameters of all entries related to silicon 
monoxide are listed in Table~\ref{t:modparam_lmh_sio}. Silicon monoxide is
detected both in emission and absorption in the 3~mm window. The absorption
consists of self-absorption, which we decompose into four components based on
the detection in the $^{29}$Si and $^{30}$Si isotopologues, and spiral-arm
diffuse-cloud components, which are detected only in the main isotopologue.
Two blueshifted and redshifted components are also added in emission to account
for wing emission in the main isotopologue. The parameters of the three 
components in emission are poorly constrained. Part of the self-absorption 
components of the $^{30}$Si isotopologue is blended with emission features of 
NH$_2$CHO, $\varv_{12}=1$ and CH$_3$OH. The $^{18}$O isotopologue is not 
detected because of a blend with C$_2$H$_5$CN, $\varv_{13}+\varv_{21}=1$.

\paragraph{Sgr\,\,B2(M):} The parameters of all entries related to silicon 
monoxide are listed in Table~\ref{t:modparam_b2m_sio}. The main isotopologue
is detected at 3 and 1.3~mm, while the $^{29}$Si, $^{30}$Si, and $^{18}$O are 
detected at 3~mm only, the latter only in absorption. Three components in 
emission are used to reproduce the core of the lines of the main isotopologue 
and their wings. Their temperature has to be lower than 20~K to fit both the
3 and 1.3~mm transitions. Self-absorption is detected for all isotopologues, 
with two velocity components. Diffuse-cloud absorption components are detected 
for the main isotopologue, but the baseline level is apparently overestimated, 
which in turn leads to an overestimate of the column densities of these 
components and makes the identification of weak absorption components 
unreliable. The parameters of the diffuse-cloud absorption components are
therefore very uncertain. The $^{29}$Si isotopologue may also show the 
signature of a few diffuse-cloud components, but it suffers from the same 
baseline issue as the main isotopologue. The $^{30}$Si and $^{18}$O 
isotopologues are too weak for the diffuse-cloud components to be detected.

\subsubsection{Sulfur dioxide SO$_2$}
\label{sss:so2}

We use the CDMS entries for the vibrational ground state of all isotopologues
and for the vibrationally excited state $\varv_2 = 1$ of the main isotopologue
(tags 64502 version 2, 66501, 65501, 66502, and 65502, all version 1, and 
64503 version 2). The entries for the ground and first vibrationally
excited states of the main isotopologue are based on \citet{SO2_2005}, those 
for $^{34}$SO$_2$ and SO$^{18}$O on \citet{34SO2_etc_1998}, and those for 
$^{33}$SO$_2$ and SO$^{17}$O on \citet{17O-SO2_33SO3_2000}. Important additional 
data for the ground and first vibrationally excited states of the main 
isotopologue as well as for $^{34}$SO$_2$ in the range of our survey were 
taken from \citet{SO2_etc_2-3mm_1996}.

\paragraph{Sgr\,\,B2(N):} The parameters of all entries related to sulfur 
dioxide are listed in Table~\ref{t:modparam_lmh_so2}. Many transitions of 
sulfur dioxide are detected in the 3~mm window, with lower-level energies 
ranging from 2.8~K (at 104.03~GHz) to 545~K (at 84.32~GHz). A complex mixture 
of velocity and temperature components is needed to reproduce the features 
detected in emission and absorption. Three hot velocity components are required
by the high-energy transitions (e.g., 82.95~GHz). An additional redshifted 
component is necessary to reproduce the wing emission of the low-energy
transitions (e.g., 83.69 and 104.03~GHz). The low-excitation component in 
emission close
to the systemic velocity is needed to fit the transition at 83.69~GHz. The 
absorption components account for the self-absorption detected at 104.03~GHz. 
The $^{34}$S isotopologue is detected at 83.04, 102.03, and 104.39~GHz. The 
$^{33}$S, $^{18}$O, and $^{17}$O isotopologues are not detected. No transition 
from within the vibrationally excited state $\varv_2=1$ is detected.

\paragraph{Sgr\,\,B2(M):} The parameters of all entries related to sulfur 
dioxide are listed in Table~\ref{t:modparam_b2m_so2}. Numerous transitions
of the main isotopologue are detected in all bands. At 3~mm, the lower-level
energies range from 2.8~K (104.03~GHz) to 1151~K (90.00~GHz), and we need
two temperature components to fit all the detected transitions. The colder
component is in particular necessary for the transitions at 83.69, 104.03, and 
104.24~GHz ($E_l/k = 33$, 2.8, and 50~K). The detected transitions are a mixture
of optically thin and thick transitions ($\tau_{\rm max} = 19$ for the hot 
component, and 0.08 for the colder component), which simultaneously allows an 
estimate of the temperature and the source size. The $^{34}$S isotopologue is
also detected with many transitions ($\tau_{\rm max} = 0.8 $ for the hot 
component, and 0.0033 for the colder one). The $^{33}$S isotopologue is weak 
but clearly detected, e.g., at 83.35 and 103.00~GHz, but the model tends to
underestimate the emission. The $^{18}$O isotopologue is weak. It is clearly
detected at 267.04~GHz and maybe at 109.64~GHz (but underestimated by the 
model?). The $^{17}$O isotopologue is not detected at 3~mm, but it is likely 
detected at 216.67 and 217.02~GHz and was for this reason included in the
complete model.

We also detect many transitions from within the vibrationally excited state
$\varv_2 = 1$ of the main isotopologue. A few lines are optically thick at 3~mm 
($\tau_{\rm max} = 1.8$), but the majority is optically thin. A very small 
size is needed to limit the strength of the (optically thick) lines at 1.3~mm.

\subsubsection{Sulfur monoxide SO}
\label{sss:so}

We use the CDMS entries for the main isotopologue and its $^{34}$S, $^{33}$S, 
and $^{18}$O isotopologues (tags 48501, 50501, 49501, and 50502, all version 1).
The entry for the main isotopologue is based on \citet{SO_1997}, the remaining 
entries on \citet{isos-SO_1996}. Data in the range of our survey were 
taken mainly from \citet{div-SO_1982} for the main, $^{34}$S, and 
$^{18}$O isotopologues and from \citet{SO_34_18_1982} for the $^{34}$S and 
$^{18}$O species.

\paragraph{Sgr\,\,B2(N):} The parameters of all entries related to sulfur 
monoxide are listed in Table~\ref{t:modparam_lmh_so}. Four transitions of 
sulfur monoxide are detected at 3~mm, one of them both in emission and 
absorption, with two self-absorption components (at 99.30~GHz). The flux 
density of the transition at 86.09~GHz ($E_l/k = 15$~K) represents about 75\% 
of the flux density detected with the NRAO 12~m telescope 
\citep[][]{Friedel04}. Since the 12~m flux density is partly contaminated by 
Sgr~B2(M) which emits more than Sgr~B2(N) in this transition, the contribution 
of Sgr~B2(N) to the emission detected with the 12~m telescope is mainly 
confined to the 30~m beam. In addition, the 30~m flux density at that 
frequency is about twice the flux density detected with BIMA with an angular 
resolution of $27.2\arcsec \times 6.6\arcsec$ \citep[][]{Friedel04}, which 
implies that at least half of the 30~m flux density is emitted by a region 
more extended than the minor size of the BIMA beam, but also that there is at 
least one additional component more compact than the BIMA beam. Since both SO 
and SO$_2$ are shock tracers, we expect that they should trace similar regions 
and we constructed a model for SO similar to the one for SO$_2$, the latter 
being better constrained thanks to the wider range of lower-level energies of 
its detected transitions. Two transitions of the $^{34}$S isotopologue are 
clearly detected at 3~mm, the other two being blended with emission from other 
species, in particular from a recombination line of hydrogen close to 
106.74~GHz. The $^{33}$S isotopologue is not unambigusously detected, but its 
transitions around 98.49~GHz contribute to nearly 50\% of the observed 
features so we include it in the complete model. The other isotopologues 
($^{18}$O, $^{17}$O, $^{36}$S) are not detected. No transition from within the 
vibrationally excited state $\varv=1$ is detected.

\paragraph{Sgr\,\,B2(M):} The parameters of all entries related to sulfur 
monoxide are listed in Table~\ref{t:modparam_b2m_so}. Sulfur monoxide is 
detected with four lines at 3~mm and four lines at 1~mm. The transition at
99.30~GHz ($E_l/k = 4$~K) is detected both in emission and absorption, even
from spiral-arm diffuse clouds. The emission is modeled with three components,
one for the core of the lines and two for the wings. The Mopra maps of 
\citet[][]{Jones08} show that there is a component extended over several 
arcminutes (mainly seen at 99.30~GHz), but also a more compact component
peaking toward Sgr~B2(M). The $^{34}$S isotopologue is clearly detected in 
emission at 3 and 1.3~mm (e.g., 97.72 and 211.01~GHz), but the diffuse-cloud 
absorption components are not detected. The $^{33}$S and $^{18}$O isotopologues 
are clearly detected in emission at 3~mm (e.g., 98.49 and 93.27~GHz). 
Assuming the isotopic ratios listed in Table~\ref{t:iso}, a size of the
main cold emission component similar to the beam at 3~mm gives a reasonable 
fit to all isotopologues. The 3~mm lines are dominated by the main cold 
component of the model (with $\tau_{\rm max} = 0.9$), while the warm 
component ($\tau_{\rm max} = 6$ at 3~mm) contributes mainly at 214.36~GHz 
($E_l/k = 71$~K) and less than 50\% for the other 1.3~mm transitions. We note
however that the transition at 100.03~GHz ($E_l/k = 34$~K) is underestimated
by the model. Finally, one transition of the vibrationally excited state
$\varv=1$ of the main isotopologue may be detected at 213.50~GHz, but it 
requires a column density ten times higher than the one of the warm component 
of the ground state, so the detection is highly uncertain and this state is not
included in the complete model.

\subsubsection{Formic acid t-HCOOH}
\label{sss:hcooh}

We use the CDMS entry (tag 46506 version 1) which is based on 
\citet{HCOOH_2002}, with additional data in the range of our survey mainly 
from \citet{HCOOH_1971}.

\paragraph{Sgr\,\,B2(N):} The parameters of formic acid in the \textit{trans}
form are listed in Table~\ref{t:modparam_lmh_t-hcooh}. This model was already 
reported in Sect.~3.4 of \citet{Belloche09}. The detected transitions are
all optically thin ($\tau_{\rm max} \sim 0.14$). The $^{13}$C isotopologue 
is not detected.

\paragraph{Sgr\,\,B2(M):} Formic acid is not detected.

\onltab{\clearpage
\begin{table*}
 \caption{Parameters of our best-fit model of C$_2$H$_3$CN toward Sgr~B2(N).
 }
 \label{t:modparam_lmh_c2h3cn}
 \centering
 \begin{tabular}{lccccccccc}
 \hline\hline
 \multicolumn{1}{c}{Molecule\tablefootmark{a}} & \multicolumn{1}{c}{Size\tablefootmark{b}} & \multicolumn{1}{c}{$T_{\mathrm{rot}}$\tablefootmark{c}} & \multicolumn{1}{c}{$N$\tablefootmark{d}} & \multicolumn{1}{c}{$C_{\rm int}$\tablefootmark{e}} & \multicolumn{1}{c}{$C_{\rm vib}$\tablefootmark{f}} & \multicolumn{1}{c}{$C_N$\tablefootmark{g}} & \multicolumn{1}{c}{$\Delta V$\tablefootmark{h}} & \multicolumn{1}{c}{$V_{\mathrm{off}}$\tablefootmark{i}} & \multicolumn{1}{c}{$F$\tablefootmark{j}} \\ 
  & \multicolumn{1}{c}{($''$)} & \multicolumn{1}{c}{(K)} & \multicolumn{1}{c}{(cm$^{-2}$)} & & & & \multicolumn{2}{c}{(km~s$^{-1}$)} & \\ 
 \multicolumn{1}{c}{(1)} & \multicolumn{1}{c}{(2)} & \multicolumn{1}{c}{(3)} & \multicolumn{1}{c}{(4)} & \multicolumn{1}{c}{(5)} & \multicolumn{1}{c}{(6)} & \multicolumn{1}{c}{(7)} & \multicolumn{1}{c}{(8)} & \multicolumn{1}{c}{(9)} & \multicolumn{1}{c}{(10)} \\ 
 \hline
\noalign{\smallskip} 
 C$_2$H$_3$CN & 2.3 &  170 & $ 7.38 \times 10^{17}$ & 0.92 & 1.00 & 0.92 & 7.0 & -1.0 & e \\ 
 & 2.3 &  170 & $ 2.21 \times 10^{17}$ & 0.92 & 1.00 & 0.92 & 7.0 & 9.0 & e \\ 
 & 2.3 &  170 & $ 9.22 \times 10^{16}$ & 0.92 & 1.00 & 0.92 & 10.0 & -9.0 & e \\ 
\noalign{\smallskip} 
 C$_2$H$_3$CN, $\varv_{11}$=1 & 1.6 &  170 & $ 3.69 \times 10^{18}$ & 0.92 & 1.00 & 0.92 & 7.0 & -1.0 & e \\ 
 & 1.6 &  170 & $ 9.22 \times 10^{17}$ & 0.92 & 1.00 & 0.92 & 7.0 & 9.0 & e \\ 
\noalign{\smallskip} 
 C$_2$H$_3$CN, $\varv_{15}$=1 & 1.6 &  170 & $ 4.61 \times 10^{18}$ & 0.92 & 1.00 & 0.92 & 7.0 & -1.0 & e \\ 
 & 1.6 &  170 & $ 1.11 \times 10^{18}$ & 0.92 & 1.00 & 0.92 & 7.0 & 9.0 & e \\ 
\noalign{\smallskip} 
 C$_2$H$_3$CN, $\varv_{11}$=2 & 1.6 &  170 & $ 5.53 \times 10^{18}$ & 0.92 & 1.00 & 0.92 & 7.0 & -1.0 & e \\ 
 & 1.6 &  170 & $ 1.38 \times 10^{18}$ & 0.92 & 1.00 & 0.92 & 7.0 & 9.0 & e \\ 
\noalign{\smallskip} 
 C$_2$H$_3$CN, $\varv_{11}$=1/$\varv_{15}$=1 & 1.6 &  170 & $ 5.53 \times 10^{18}$ & 0.92 & 1.00 & 0.92 & 7.0 & -1.0 & e \\ 
 & 1.6 &  170 & $ 1.38 \times 10^{18}$ & 0.92 & 1.00 & 0.92 & 7.0 & 9.0 & e \\ 
\noalign{\smallskip} 
 C$_2$H$_3$CN, $\varv_{15}$=2 & 1.6 &  170 & $ 1.01 \times 10^{19}$ & 0.92 & 1.00 & 0.92 & 7.0 & -1.0 & e \\ 
\noalign{\smallskip} 
 C$_2$H$_3$CN, $\varv_{11}$=3 & 1.6 &  170 & $ 1.20 \times 10^{19}$ & 0.92 & 1.00 & 0.92 & 7.0 & -1.0 & e \\ 
\noalign{\smallskip} 
 CH$_2$CH$^{13}$CN & 2.3 &  170 & $ 3.77 \times 10^{16}$ & 1.00 & 1.26 & 1.26 & 7.0 & -1.0 & e \\ 
 & 2.3 &  170 & $ 1.13 \times 10^{16}$ & 1.00 & 1.26 & 1.26 & 7.0 & 9.0 & e \\ 
 & 2.3 &  170 & $ 4.72 \times 10^{15}$ & 1.00 & 1.26 & 1.26 & 10.0 & -9.0 & e \\ 
\noalign{\smallskip} 
 CH$_2$$^{13}$CHCN & 2.3 &  170 & $ 3.77 \times 10^{16}$ & 1.00 & 1.26 & 1.26 & 7.0 & -1.0 & e \\ 
 & 2.3 &  170 & $ 1.13 \times 10^{16}$ & 1.00 & 1.26 & 1.26 & 7.0 & 9.0 & e \\ 
 & 2.3 &  170 & $ 4.72 \times 10^{15}$ & 1.00 & 1.26 & 1.26 & 10.0 & -9.0 & e \\ 
\noalign{\smallskip} 
 $^{13}$CH$_2$CHCN & 2.3 &  170 & $ 3.77 \times 10^{16}$ & 1.00 & 1.26 & 1.26 & 7.0 & -1.0 & e \\ 
 & 2.3 &  170 & $ 1.13 \times 10^{16}$ & 1.00 & 1.26 & 1.26 & 7.0 & 9.0 & e \\ 
 & 2.3 &  170 & $ 4.72 \times 10^{15}$ & 1.00 & 1.26 & 1.26 & 10.0 & -9.0 & e \\ 
 \hline
 \end{tabular}
 \tablefoot{
 \tablefoottext{a}{Torsionally or vibrationally excited states are indicated after the name of the molecule and are modeled as separate entries. Entries with only the name of the molecule correspond to the ground state.}
 \tablefoottext{b}{Source diameter (\textit{FWHM}).}
 \tablefoottext{c}{Rotational temperature.}
 \tablefoottext{d}{Column density.}
 \tablefoottext{e}{Correction factor that was applied to the column density to account for the correction to the approximate interpolation performed by XCLASS to compute the partition function.}
 \tablefoottext{f}{Correction factor that was applied to the column density to account for the contribution of vibrationally or torsionally excited states or other conformers to the partition function.}
 \tablefoottext{g}{Global correction factor that was applied to the column density ($C_{\rm int} \times C_{\rm vib}$).}
 \tablefoottext{h}{Linewidth (\textit{FWHM}).}
 \tablefoottext{i}{Velocity offset with respect to the systemic velocity of Sgr~B2(N) $V_{\mathrm{lsr}} = 64$ km~s$^{-1}$.}
 \tablefoottext{j}{Flag indicating if the component belongs to the "emission" (e) or "absorption" (a) group.}
 }
\end{table*}
} 
\onltab{\clearpage
\begin{table*}
 \caption{Parameters of our best-fit model of C$_2$H$_5$CN toward Sgr~B2(N).
 }
 \label{t:modparam_lmh_c2h5cn}
 \centering
 \begin{tabular}{lccccccccc}
 \hline\hline
 \multicolumn{1}{c}{Molecule\tablefootmark{a}} & \multicolumn{1}{c}{Size\tablefootmark{b}} & \multicolumn{1}{c}{$T_{\mathrm{rot}}$\tablefootmark{c}} & \multicolumn{1}{c}{$N$\tablefootmark{d}} & \multicolumn{1}{c}{$C_{\rm int}$\tablefootmark{e}} & \multicolumn{1}{c}{$C_{\rm vib}$\tablefootmark{f}} & \multicolumn{1}{c}{$C_N$\tablefootmark{g}} & \multicolumn{1}{c}{$\Delta V$\tablefootmark{h}} & \multicolumn{1}{c}{$V_{\mathrm{off}}$\tablefootmark{i}} & \multicolumn{1}{c}{$F$\tablefootmark{j}} \\ 
  & \multicolumn{1}{c}{($''$)} & \multicolumn{1}{c}{(K)} & \multicolumn{1}{c}{(cm$^{-2}$)} & & & & \multicolumn{2}{c}{(km~s$^{-1}$)} & \\ 
 \multicolumn{1}{c}{(1)} & \multicolumn{1}{c}{(2)} & \multicolumn{1}{c}{(3)} & \multicolumn{1}{c}{(4)} & \multicolumn{1}{c}{(5)} & \multicolumn{1}{c}{(6)} & \multicolumn{1}{c}{(7)} & \multicolumn{1}{c}{(8)} & \multicolumn{1}{c}{(9)} & \multicolumn{1}{c}{(10)} \\ 
 \hline
\noalign{\smallskip} 
 C$_2$H$_5$CN & 3.0 &  170 & $ 1.84 \times 10^{18}$ & 1.00 & 1.53 & 1.53 & 6.5 & -1.0 & e \\ 
 & 2.3 &  170 & $ 2.14 \times 10^{18}$ & 1.00 & 1.53 & 1.53 & 6.5 & 9.0 & e \\ 
 & 1.7 &  170 & $ 1.38 \times 10^{18}$ & 1.00 & 1.53 & 1.53 & 8.0 & -11.0 & e \\ 
\noalign{\smallskip} 
 C$_2$H$_5$CN, $\varv_{13}$$+$$\varv_{21}$=1 & 2.0 &  170 & $ 1.07 \times 10^{19}$ & 1.00 & 1.53 & 1.53 & 6.5 & -1.0 & e \\ 
 & 1.5 &  170 & $ 9.19 \times 10^{18}$ & 1.00 & 1.53 & 1.53 & 6.5 & 9.0 & e \\
\hline
 \end{tabular}
 \tablefoot{
 See notes of Table~\ref{t:modparam_lmh_c2h3cn}.
 }
\end{table*}
} 
\onltab{\clearpage
\begin{table*}
 \caption{Parameters of our best-fit model of C$_2$H$_5$OCHO toward Sgr~B2(N).
 }
 \label{t:modparam_lmh_c2h5ocho}
 \centering
 \begin{tabular}{lccccccccc}
 \hline\hline
 \multicolumn{1}{c}{Molecule\tablefootmark{a}} & \multicolumn{1}{c}{Size\tablefootmark{b}} & \multicolumn{1}{c}{$T_{\mathrm{rot}}$\tablefootmark{c}} & \multicolumn{1}{c}{$N$\tablefootmark{d}} & \multicolumn{1}{c}{$C_{\rm int}$\tablefootmark{e}} & \multicolumn{1}{c}{$C_{\rm vib}$\tablefootmark{f}} & \multicolumn{1}{c}{$C_N$\tablefootmark{g}} & \multicolumn{1}{c}{$\Delta V$\tablefootmark{h}} & \multicolumn{1}{c}{$V_{\mathrm{off}}$\tablefootmark{i}} & \multicolumn{1}{c}{$F$\tablefootmark{j}} \\ 
  & \multicolumn{1}{c}{($''$)} & \multicolumn{1}{c}{(K)} & \multicolumn{1}{c}{(cm$^{-2}$)} & & & & \multicolumn{2}{c}{(km~s$^{-1}$)} & \\ 
 \multicolumn{1}{c}{(1)} & \multicolumn{1}{c}{(2)} & \multicolumn{1}{c}{(3)} & \multicolumn{1}{c}{(4)} & \multicolumn{1}{c}{(5)} & \multicolumn{1}{c}{(6)} & \multicolumn{1}{c}{(7)} & \multicolumn{1}{c}{(8)} & \multicolumn{1}{c}{(9)} & \multicolumn{1}{c}{(10)} \\ 
 \hline
\noalign{\smallskip} 
 C$_2$H$_5$OCHO & 3.0 &  100 & $ 5.41 \times 10^{16}$ & 1.00 & 1.00 & 1.00 & 7.0 & 0.0 & e \\ 
 \hline
 \end{tabular}
 \tablefoot{
 See notes of Table~\ref{t:modparam_lmh_c2h3cn}.
 }
\end{table*}
} 
\onltab{\clearpage
\begin{table*}
 \caption{Parameters of our best-fit model of C$_2$H$_5$OH toward Sgr~B2(N).
 }
 \label{t:modparam_lmh_c2h5oh}
 \centering
 \begin{tabular}{lccccccccc}
 \hline\hline
 \multicolumn{1}{c}{Molecule\tablefootmark{a}} & \multicolumn{1}{c}{Size\tablefootmark{b}} & \multicolumn{1}{c}{$T_{\mathrm{rot}}$\tablefootmark{c}} & \multicolumn{1}{c}{$N$\tablefootmark{d}} & \multicolumn{1}{c}{$C_{\rm int}$\tablefootmark{e}} & \multicolumn{1}{c}{$C_{\rm vib}$\tablefootmark{f}} & \multicolumn{1}{c}{$C_N$\tablefootmark{g}} & \multicolumn{1}{c}{$\Delta V$\tablefootmark{h}} & \multicolumn{1}{c}{$V_{\mathrm{off}}$\tablefootmark{i}} & \multicolumn{1}{c}{$F$\tablefootmark{j}} \\ 
  & \multicolumn{1}{c}{($''$)} & \multicolumn{1}{c}{(K)} & \multicolumn{1}{c}{(cm$^{-2}$)} & & & & \multicolumn{2}{c}{(km~s$^{-1}$)} & \\ 
 \multicolumn{1}{c}{(1)} & \multicolumn{1}{c}{(2)} & \multicolumn{1}{c}{(3)} & \multicolumn{1}{c}{(4)} & \multicolumn{1}{c}{(5)} & \multicolumn{1}{c}{(6)} & \multicolumn{1}{c}{(7)} & \multicolumn{1}{c}{(8)} & \multicolumn{1}{c}{(9)} & \multicolumn{1}{c}{(10)} \\ 
 \hline
\noalign{\smallskip} 
 C$_2$H$_5$OH & 3.0 &  150 & $ 9.14 \times 10^{17}$ & 1.05 & 1.00 & 1.05 & 7.0 & -0.5 & e \\ 
 & 3.0 &  150 & $ 3.99 \times 10^{17}$ & 1.05 & 1.00 & 1.05 & 7.0 & 9.5 & e \\ 
\noalign{\smallskip} 
 \textit{a}-CH$_3$$^{13}$CH$_2$OH & 3.0 &  150 & $ 4.34 \times 10^{16}$ & 1.00 & 2.39 & 2.39 & 7.0 & -0.5 & e \\ 
 & 3.0 &  150 & $ 1.90 \times 10^{16}$ & 1.00 & 2.39 & 2.39 & 7.0 & 9.5 & e \\ 
\noalign{\smallskip} 
 \textit{a}-$^{13}$CH$_3$CH$_2$OH & 3.0 &  150 & $ 4.34 \times 10^{16}$ & 1.00 & 2.39 & 2.39 & 7.0 & -0.5 & e \\ 
 & 3.0 &  150 & $ 1.90 \times 10^{16}$ & 1.00 & 2.39 & 2.39 & 7.0 & 9.5 & e \\ 
 \hline
 \end{tabular}
 \tablefoot{
 See notes of Table~\ref{t:modparam_lmh_c2h3cn}.
 }
\end{table*}
} 
\onltab{\clearpage
\begin{table*}
 \caption{Parameters of our best-fit model of \textit{n}-C$_3$H$_7$CN toward Sgr~B2(N).
 }
 \label{t:modparam_lmh_c3h7cn}
 \centering
 \begin{tabular}{lccccccccc}
 \hline\hline
 \multicolumn{1}{c}{Molecule\tablefootmark{a}} & \multicolumn{1}{c}{Size\tablefootmark{b}} & \multicolumn{1}{c}{$T_{\mathrm{rot}}$\tablefootmark{c}} & \multicolumn{1}{c}{$N$\tablefootmark{d}} & \multicolumn{1}{c}{$C_{\rm int}$\tablefootmark{e}} & \multicolumn{1}{c}{$C_{\rm vib}$\tablefootmark{f}} & \multicolumn{1}{c}{$C_N$\tablefootmark{g}} & \multicolumn{1}{c}{$\Delta V$\tablefootmark{h}} & \multicolumn{1}{c}{$V_{\mathrm{off}}$\tablefootmark{i}} & \multicolumn{1}{c}{$F$\tablefootmark{j}} \\ 
  & \multicolumn{1}{c}{($''$)} & \multicolumn{1}{c}{(K)} & \multicolumn{1}{c}{(cm$^{-2}$)} & & & & \multicolumn{2}{c}{(km~s$^{-1}$)} & \\ 
 \multicolumn{1}{c}{(1)} & \multicolumn{1}{c}{(2)} & \multicolumn{1}{c}{(3)} & \multicolumn{1}{c}{(4)} & \multicolumn{1}{c}{(5)} & \multicolumn{1}{c}{(6)} & \multicolumn{1}{c}{(7)} & \multicolumn{1}{c}{(8)} & \multicolumn{1}{c}{(9)} & \multicolumn{1}{c}{(10)} \\ 
 \hline
\noalign{\smallskip} 
 \textit{n}-C$_3$H$_7$CN & 3.0 &  150 & $ 1.54 \times 10^{16}$ & 1.02 & 1.00 & 1.02 & 7.0 & -1.0 & e \\ 
 & 3.0 &  150 & $ 6.76 \times 10^{15}$ & 1.02 & 1.00 & 1.02 & 7.0 & 9.0 & e \\ 
 \hline
 \end{tabular}
 \tablefoot{
 See notes of Table~\ref{t:modparam_lmh_c2h3cn}.
 }
\end{table*}
} 
\onltab{\clearpage
\begin{table*}
 \caption{Parameters of our best-fit model of \textit{c}-C$_2$H$_4$O toward Sgr~B2(N).
 }
 \label{t:modparam_lmh_c-c2h4o}
 \centering
 \begin{tabular}{lccccccccc}
 \hline\hline
 \multicolumn{1}{c}{Molecule\tablefootmark{a}} & \multicolumn{1}{c}{Size\tablefootmark{b}} & \multicolumn{1}{c}{$T_{\mathrm{rot}}$\tablefootmark{c}} & \multicolumn{1}{c}{$N$\tablefootmark{d}} & \multicolumn{1}{c}{$C_{\rm int}$\tablefootmark{e}} & \multicolumn{1}{c}{$C_{\rm vib}$\tablefootmark{f}} & \multicolumn{1}{c}{$C_N$\tablefootmark{g}} & \multicolumn{1}{c}{$\Delta V$\tablefootmark{h}} & \multicolumn{1}{c}{$V_{\mathrm{off}}$\tablefootmark{i}} & \multicolumn{1}{c}{$F$\tablefootmark{j}} \\ 
  & \multicolumn{1}{c}{($''$)} & \multicolumn{1}{c}{(K)} & \multicolumn{1}{c}{(cm$^{-2}$)} & & & & \multicolumn{2}{c}{(km~s$^{-1}$)} & \\ 
 \multicolumn{1}{c}{(1)} & \multicolumn{1}{c}{(2)} & \multicolumn{1}{c}{(3)} & \multicolumn{1}{c}{(4)} & \multicolumn{1}{c}{(5)} & \multicolumn{1}{c}{(6)} & \multicolumn{1}{c}{(7)} & \multicolumn{1}{c}{(8)} & \multicolumn{1}{c}{(9)} & \multicolumn{1}{c}{(10)} \\ 
 \hline
\noalign{\smallskip} 
 \textit{c}-C$_2$H$_4$O & 3.0 &  100 & $ 3.89 \times 10^{16}$ & 1.00 & 1.00 & 1.00 & 7.0 & -1.0 & e \\ 
 \hline
 \end{tabular}
 \tablefoot{
 See notes of Table~\ref{t:modparam_lmh_c2h3cn}.
 }
\end{table*}
} 
\onltab{\clearpage
\begin{table*}
 \caption{Parameters of our best-fit model of \textit{c}-C$_3$H$_2$ toward Sgr~B2(N).
 }
 \label{t:modparam_lmh_c-c3h2}
 \centering
 \begin{tabular}{lccccccccc}
 \hline\hline
 \multicolumn{1}{c}{Molecule\tablefootmark{a}} & \multicolumn{1}{c}{Size\tablefootmark{b}} & \multicolumn{1}{c}{$T_{\mathrm{rot}}$\tablefootmark{c}} & \multicolumn{1}{c}{$N$\tablefootmark{d}} & \multicolumn{1}{c}{$C_{\rm int}$\tablefootmark{e}} & \multicolumn{1}{c}{$C_{\rm vib}$\tablefootmark{f}} & \multicolumn{1}{c}{$C_N$\tablefootmark{g}} & \multicolumn{1}{c}{$\Delta V$\tablefootmark{h}} & \multicolumn{1}{c}{$V_{\mathrm{off}}$\tablefootmark{i}} & \multicolumn{1}{c}{$F$\tablefootmark{j}} \\ 
  & \multicolumn{1}{c}{($''$)} & \multicolumn{1}{c}{(K)} & \multicolumn{1}{c}{(cm$^{-2}$)} & & & & \multicolumn{2}{c}{(km~s$^{-1}$)} & \\ 
 \multicolumn{1}{c}{(1)} & \multicolumn{1}{c}{(2)} & \multicolumn{1}{c}{(3)} & \multicolumn{1}{c}{(4)} & \multicolumn{1}{c}{(5)} & \multicolumn{1}{c}{(6)} & \multicolumn{1}{c}{(7)} & \multicolumn{1}{c}{(8)} & \multicolumn{1}{c}{(9)} & \multicolumn{1}{c}{(10)} \\ 
 \hline
\noalign{\smallskip} 
 \textit{c}-C$_3$H$_2$ &  200 &   14 & $ 2.50 \times 10^{14}$ & 1.00 & 1.00 & 1.00 & 15.0 & 3.0 & e \\ 
 &  120 &  2.7 & $ 3.39 \times 10^{13}$ & 1.06 & 1.00 & 1.06 & 9.0 & 17.0 & a \\ 
 &  120 &  2.7 & $ 1.27 \times 10^{14}$ & 1.06 & 1.00 & 1.06 & 10.5 & 0.5 & a \\ 
 &  120 &  2.7 & $ 2.33 \times 10^{12}$ & 1.06 & 1.00 & 1.06 & 5.0 & -32.0 & a \\ 
 &  120 &  2.7 & $ 9.42 \times 10^{11}$ & 1.06 & 1.00 & 1.06 & 5.0 & -42.0 & a \\
\hline
 \end{tabular}
 \end{table*}
\begin{table*}
 \addtocounter{table}{-1}
 \caption{
 continued.
 }
 \centering
 \begin{tabular}{lccccccccc}
 \hline\hline
 \multicolumn{1}{c}{Molecule\tablefootmark{a}} & \multicolumn{1}{c}{Size\tablefootmark{b}} & \multicolumn{1}{c}{$T_{\mathrm{rot}}$\tablefootmark{c}} & \multicolumn{1}{c}{$N$\tablefootmark{d}} & \multicolumn{1}{c}{$C_{\rm int}$\tablefootmark{e}} & \multicolumn{1}{c}{$C_{\rm vib}$\tablefootmark{f}} & \multicolumn{1}{c}{$C_N$\tablefootmark{g}} & \multicolumn{1}{c}{$\Delta V$\tablefootmark{h}} & \multicolumn{1}{c}{$V_{\mathrm{off}}$\tablefootmark{i}} & \multicolumn{1}{c}{$F$\tablefootmark{j}} \\ 
  & \multicolumn{1}{c}{($''$)} & \multicolumn{1}{c}{(K)} & \multicolumn{1}{c}{(cm$^{-2}$)} & & & & \multicolumn{2}{c}{(km~s$^{-1}$)} & \\ 
 \multicolumn{1}{c}{(1)} & \multicolumn{1}{c}{(2)} & \multicolumn{1}{c}{(3)} & \multicolumn{1}{c}{(4)} & \multicolumn{1}{c}{(5)} & \multicolumn{1}{c}{(6)} & \multicolumn{1}{c}{(7)} & \multicolumn{1}{c}{(8)} & \multicolumn{1}{c}{(9)} & \multicolumn{1}{c}{(10)} \\ 
 \hline
\noalign{\smallskip} 
 \textit{c}-$^{13}$CC$_2$H$_2$ &  200 &   14 & $ 1.25 \times 10^{13}$ & 1.00 & 1.00 & 1.00 & 15.0 & 3.0 & e \\ 
 &  120 &  2.7 & $ 2.96 \times 10^{12}$ & 1.06 & 1.00 & 1.06 & 9.0 & 17.0 & a \\ 
 &  120 &  2.7 & $ 9.19 \times 10^{12}$ & 1.06 & 1.00 & 1.06 & 10.5 & 0.5 & a \\ 
 &  120 &  2.7 & $ 5.81 \times 10^{10}$ & 1.06 & 1.00 & 1.06 & 5.0 & -32.0 & a \\ 
 &  120 &  2.7 & $ 1.56 \times 10^{10}$ & 1.06 & 1.00 & 1.06 & 5.0 & -42.0 & a \\
\hline
 \end{tabular}
 \tablefoot{
 See notes of Table~\ref{t:modparam_lmh_c2h3cn}.
 }
\end{table*}
} 
\onltab{\clearpage
\begin{table*}
 \caption{Parameters of our best-fit model of CCH toward Sgr~B2(N).
 }
 \label{t:modparam_lmh_cch}
 \centering
 \begin{tabular}{lccccccccc}
 \hline\hline
 \multicolumn{1}{c}{Molecule\tablefootmark{a}} & \multicolumn{1}{c}{Size\tablefootmark{b}} & \multicolumn{1}{c}{$T_{\mathrm{rot}}$\tablefootmark{c}} & \multicolumn{1}{c}{$N$\tablefootmark{d}} & \multicolumn{1}{c}{$C_{\rm int}$\tablefootmark{e}} & \multicolumn{1}{c}{$C_{\rm vib}$\tablefootmark{f}} & \multicolumn{1}{c}{$C_N$\tablefootmark{g}} & \multicolumn{1}{c}{$\Delta V$\tablefootmark{h}} & \multicolumn{1}{c}{$V_{\mathrm{off}}$\tablefootmark{i}} & \multicolumn{1}{c}{$F$\tablefootmark{j}} \\ 
  & \multicolumn{1}{c}{($''$)} & \multicolumn{1}{c}{(K)} & \multicolumn{1}{c}{(cm$^{-2}$)} & & & & \multicolumn{2}{c}{(km~s$^{-1}$)} & \\ 
 \multicolumn{1}{c}{(1)} & \multicolumn{1}{c}{(2)} & \multicolumn{1}{c}{(3)} & \multicolumn{1}{c}{(4)} & \multicolumn{1}{c}{(5)} & \multicolumn{1}{c}{(6)} & \multicolumn{1}{c}{(7)} & \multicolumn{1}{c}{(8)} & \multicolumn{1}{c}{(9)} & \multicolumn{1}{c}{(10)} \\ 
 \hline
\noalign{\smallskip} 
 CCH &  200 &   20 & $ 2.00 \times 10^{15}$ & 1.00 & 1.00 & 1.00 & 15.0 & 3.0 & e \\ 
 &   20 &  2.7 & $ 2.43 \times 10^{15}$ & 1.43 & 1.00 & 1.43 & 8.0 & 0.0 & a \\ 
 &   20 &  2.7 & $ 2.85 \times 10^{14}$ & 1.43 & 1.00 & 1.43 & 6.0 & 19.0 & a \\ 
 &  120 &  2.7 & $ 2.57 \times 10^{13}$ & 1.43 & 1.00 & 1.43 & 5.0 & -29.5 & a \\ 
 &  120 &  2.7 & $ 4.99 \times 10^{13}$ & 1.43 & 1.00 & 1.43 & 5.0 & -33.5 & a \\
\hline
 \end{tabular}
 \end{table*}
\begin{table*}
 \addtocounter{table}{-1}
 \caption{
 continued.
 }
 \centering
 \begin{tabular}{lccccccccc}
 \hline\hline
 \multicolumn{1}{c}{Molecule\tablefootmark{a}} & \multicolumn{1}{c}{Size\tablefootmark{b}} & \multicolumn{1}{c}{$T_{\mathrm{rot}}$\tablefootmark{c}} & \multicolumn{1}{c}{$N$\tablefootmark{d}} & \multicolumn{1}{c}{$C_{\rm int}$\tablefootmark{e}} & \multicolumn{1}{c}{$C_{\rm vib}$\tablefootmark{f}} & \multicolumn{1}{c}{$C_N$\tablefootmark{g}} & \multicolumn{1}{c}{$\Delta V$\tablefootmark{h}} & \multicolumn{1}{c}{$V_{\mathrm{off}}$\tablefootmark{i}} & \multicolumn{1}{c}{$F$\tablefootmark{j}} \\ 
  & \multicolumn{1}{c}{($''$)} & \multicolumn{1}{c}{(K)} & \multicolumn{1}{c}{(cm$^{-2}$)} & & & & \multicolumn{2}{c}{(km~s$^{-1}$)} & \\ 
 \multicolumn{1}{c}{(1)} & \multicolumn{1}{c}{(2)} & \multicolumn{1}{c}{(3)} & \multicolumn{1}{c}{(4)} & \multicolumn{1}{c}{(5)} & \multicolumn{1}{c}{(6)} & \multicolumn{1}{c}{(7)} & \multicolumn{1}{c}{(8)} & \multicolumn{1}{c}{(9)} & \multicolumn{1}{c}{(10)} \\ 
 \hline
\noalign{\smallskip} 
 $^{13}$CCH &  200 &   20 & $ 9.93 \times 10^{13}$ & 0.99 & 1.00 & 0.99 & 15.0 & 3.0 & e \\ 
 &   20 &  2.7 & $ 1.01 \times 10^{14}$ & 1.19 & 1.00 & 1.19 & 8.0 & 0.0 & a \\ 
 &   20 &  2.7 & $ 1.19 \times 10^{13}$ & 1.19 & 1.00 & 1.19 & 6.0 & 19.0 & a \\ 
 &  120 &  2.7 & $ 5.34 \times 10^{11}$ & 1.19 & 1.00 & 1.19 & 5.0 & -29.5 & a \\ 
 &  120 &  2.7 & $ 1.04 \times 10^{12}$ & 1.19 & 1.00 & 1.19 & 5.0 & -33.5 & a \\
\hline
 \end{tabular}
 \end{table*}
\begin{table*}
 \addtocounter{table}{-1}
 \caption{
 continued.
 }
 \centering
 \begin{tabular}{lccccccccc}
 \hline\hline
 \multicolumn{1}{c}{Molecule\tablefootmark{a}} & \multicolumn{1}{c}{Size\tablefootmark{b}} & \multicolumn{1}{c}{$T_{\mathrm{rot}}$\tablefootmark{c}} & \multicolumn{1}{c}{$N$\tablefootmark{d}} & \multicolumn{1}{c}{$C_{\rm int}$\tablefootmark{e}} & \multicolumn{1}{c}{$C_{\rm vib}$\tablefootmark{f}} & \multicolumn{1}{c}{$C_N$\tablefootmark{g}} & \multicolumn{1}{c}{$\Delta V$\tablefootmark{h}} & \multicolumn{1}{c}{$V_{\mathrm{off}}$\tablefootmark{i}} & \multicolumn{1}{c}{$F$\tablefootmark{j}} \\ 
  & \multicolumn{1}{c}{($''$)} & \multicolumn{1}{c}{(K)} & \multicolumn{1}{c}{(cm$^{-2}$)} & & & & \multicolumn{2}{c}{(km~s$^{-1}$)} & \\ 
 \multicolumn{1}{c}{(1)} & \multicolumn{1}{c}{(2)} & \multicolumn{1}{c}{(3)} & \multicolumn{1}{c}{(4)} & \multicolumn{1}{c}{(5)} & \multicolumn{1}{c}{(6)} & \multicolumn{1}{c}{(7)} & \multicolumn{1}{c}{(8)} & \multicolumn{1}{c}{(9)} & \multicolumn{1}{c}{(10)} \\ 
 \hline
\noalign{\smallskip} 
 C$^{13}$CH &  200 &   20 & $ 9.92 \times 10^{13}$ & 0.99 & 1.00 & 0.99 & 15.0 & 3.0 & e \\ 
 &   20 &  2.7 & $ 1.02 \times 10^{14}$ & 1.20 & 1.00 & 1.20 & 8.0 & 0.0 & a \\ 
 &   20 &  2.7 & $ 1.20 \times 10^{13}$ & 1.20 & 1.00 & 1.20 & 6.0 & 19.0 & a \\ 
 &  120 &  2.7 & $ 5.39 \times 10^{11}$ & 1.20 & 1.00 & 1.20 & 5.0 & -29.5 & a \\ 
 &  120 &  2.7 & $ 1.05 \times 10^{12}$ & 1.20 & 1.00 & 1.20 & 5.0 & -33.5 & a \\
\hline
 \end{tabular}
 \tablefoot{
 See notes of Table~\ref{t:modparam_lmh_c2h3cn}.
 }
\end{table*}
} 
\onltab{\clearpage
\begin{table*}
 \caption{Parameters of our best-fit model of CCS toward Sgr~B2(N).
 }
 \label{t:modparam_lmh_ccs}
 \centering
 \begin{tabular}{lccccccccc}
 \hline\hline
 \multicolumn{1}{c}{Molecule\tablefootmark{a}} & \multicolumn{1}{c}{Size\tablefootmark{b}} & \multicolumn{1}{c}{$T_{\mathrm{rot}}$\tablefootmark{c}} & \multicolumn{1}{c}{$N$\tablefootmark{d}} & \multicolumn{1}{c}{$C_{\rm int}$\tablefootmark{e}} & \multicolumn{1}{c}{$C_{\rm vib}$\tablefootmark{f}} & \multicolumn{1}{c}{$C_N$\tablefootmark{g}} & \multicolumn{1}{c}{$\Delta V$\tablefootmark{h}} & \multicolumn{1}{c}{$V_{\mathrm{off}}$\tablefootmark{i}} & \multicolumn{1}{c}{$F$\tablefootmark{j}} \\ 
  & \multicolumn{1}{c}{($''$)} & \multicolumn{1}{c}{(K)} & \multicolumn{1}{c}{(cm$^{-2}$)} & & & & \multicolumn{2}{c}{(km~s$^{-1}$)} & \\ 
 \multicolumn{1}{c}{(1)} & \multicolumn{1}{c}{(2)} & \multicolumn{1}{c}{(3)} & \multicolumn{1}{c}{(4)} & \multicolumn{1}{c}{(5)} & \multicolumn{1}{c}{(6)} & \multicolumn{1}{c}{(7)} & \multicolumn{1}{c}{(8)} & \multicolumn{1}{c}{(9)} & \multicolumn{1}{c}{(10)} \\ 
 \hline
\noalign{\smallskip} 
 CCS &   60 &   20 & $ 8.29 \times 10^{13}$ & 1.04 & 1.00 & 1.04 & 13.0 & 1.0 & e \\ 
 \hline
 \end{tabular}
 \tablefoot{
 See notes of Table~\ref{t:modparam_lmh_c2h3cn}.
 }
\end{table*}
} 
\onltab{\clearpage
\begin{table*}
 \caption{Parameters of our best-fit model of CH$_2$NH toward Sgr~B2(N).
 }
 \label{t:modparam_lmh_ch2nh}
 \centering
 \begin{tabular}{lccccccccc}
 \hline\hline
 \multicolumn{1}{c}{Molecule\tablefootmark{a}} & \multicolumn{1}{c}{Size\tablefootmark{b}} & \multicolumn{1}{c}{$T_{\mathrm{rot}}$\tablefootmark{c}} & \multicolumn{1}{c}{$N$\tablefootmark{d}} & \multicolumn{1}{c}{$C_{\rm int}$\tablefootmark{e}} & \multicolumn{1}{c}{$C_{\rm vib}$\tablefootmark{f}} & \multicolumn{1}{c}{$C_N$\tablefootmark{g}} & \multicolumn{1}{c}{$\Delta V$\tablefootmark{h}} & \multicolumn{1}{c}{$V_{\mathrm{off}}$\tablefootmark{i}} & \multicolumn{1}{c}{$F$\tablefootmark{j}} \\ 
  & \multicolumn{1}{c}{($''$)} & \multicolumn{1}{c}{(K)} & \multicolumn{1}{c}{(cm$^{-2}$)} & & & & \multicolumn{2}{c}{(km~s$^{-1}$)} & \\ 
 \multicolumn{1}{c}{(1)} & \multicolumn{1}{c}{(2)} & \multicolumn{1}{c}{(3)} & \multicolumn{1}{c}{(4)} & \multicolumn{1}{c}{(5)} & \multicolumn{1}{c}{(6)} & \multicolumn{1}{c}{(7)} & \multicolumn{1}{c}{(8)} & \multicolumn{1}{c}{(9)} & \multicolumn{1}{c}{(10)} \\ 
 \hline
\noalign{\smallskip} 
 CH$_2$NH & 2.0 &  200 & $ 8.02 \times 10^{17}$ & 1.00 & 1.00 & 1.00 & 7.0 & -1.0 & e \\ 
 & 2.0 &  200 & $ 3.01 \times 10^{17}$ & 1.00 & 1.00 & 1.00 & 7.0 & 9.0 & e \\ 
 \hline
 \end{tabular}
 \tablefoot{
 See notes of Table~\ref{t:modparam_lmh_c2h3cn}.
 }
\end{table*}
} 
\onltab{\clearpage
\begin{table*}
 \caption{Parameters of our best-fit model of \textit{aGg'}-(CH$_2$OH)$_2$ toward Sgr~B2(N).
 }
 \label{t:modparam_lmh_ch2oh-2-a}
 \centering
 \begin{tabular}{lccccccccc}
 \hline\hline
 \multicolumn{1}{c}{Molecule\tablefootmark{a}} & \multicolumn{1}{c}{Size\tablefootmark{b}} & \multicolumn{1}{c}{$T_{\mathrm{rot}}$\tablefootmark{c}} & \multicolumn{1}{c}{$N$\tablefootmark{d}} & \multicolumn{1}{c}{$C_{\rm int}$\tablefootmark{e}} & \multicolumn{1}{c}{$C_{\rm vib}$\tablefootmark{f}} & \multicolumn{1}{c}{$C_N$\tablefootmark{g}} & \multicolumn{1}{c}{$\Delta V$\tablefootmark{h}} & \multicolumn{1}{c}{$V_{\mathrm{off}}$\tablefootmark{i}} & \multicolumn{1}{c}{$F$\tablefootmark{j}} \\ 
  & \multicolumn{1}{c}{($''$)} & \multicolumn{1}{c}{(K)} & \multicolumn{1}{c}{(cm$^{-2}$)} & & & & \multicolumn{2}{c}{(km~s$^{-1}$)} & \\ 
 \multicolumn{1}{c}{(1)} & \multicolumn{1}{c}{(2)} & \multicolumn{1}{c}{(3)} & \multicolumn{1}{c}{(4)} & \multicolumn{1}{c}{(5)} & \multicolumn{1}{c}{(6)} & \multicolumn{1}{c}{(7)} & \multicolumn{1}{c}{(8)} & \multicolumn{1}{c}{(9)} & \multicolumn{1}{c}{(10)} \\ 
 \hline
\noalign{\smallskip} 
 \textit{aGg'}-(CH$_2$OH)$_2$ &   10 &   50 & $ 2.30 \times 10^{15}$ & 1.00 & 1.00 & 1.00 & 6.0 & -1.0 & e \\ 
 \hline
 \end{tabular}
 \tablefoot{
 See notes of Table~\ref{t:modparam_lmh_c2h3cn}.
 }
\end{table*}
} 
\onltab{\clearpage
\begin{table*}
 \caption{Parameters of our best-fit model of CH$_2$(OH)CHO toward Sgr~B2(N).
 }
 \label{t:modparam_lmh_ch2ohcho}
 \centering
 \begin{tabular}{lccccccccc}
 \hline\hline
 \multicolumn{1}{c}{Molecule\tablefootmark{a}} & \multicolumn{1}{c}{Size\tablefootmark{b}} & \multicolumn{1}{c}{$T_{\mathrm{rot}}$\tablefootmark{c}} & \multicolumn{1}{c}{$N$\tablefootmark{d}} & \multicolumn{1}{c}{$C_{\rm int}$\tablefootmark{e}} & \multicolumn{1}{c}{$C_{\rm vib}$\tablefootmark{f}} & \multicolumn{1}{c}{$C_N$\tablefootmark{g}} & \multicolumn{1}{c}{$\Delta V$\tablefootmark{h}} & \multicolumn{1}{c}{$V_{\mathrm{off}}$\tablefootmark{i}} & \multicolumn{1}{c}{$F$\tablefootmark{j}} \\ 
  & \multicolumn{1}{c}{($''$)} & \multicolumn{1}{c}{(K)} & \multicolumn{1}{c}{(cm$^{-2}$)} & & & & \multicolumn{2}{c}{(km~s$^{-1}$)} & \\ 
 \multicolumn{1}{c}{(1)} & \multicolumn{1}{c}{(2)} & \multicolumn{1}{c}{(3)} & \multicolumn{1}{c}{(4)} & \multicolumn{1}{c}{(5)} & \multicolumn{1}{c}{(6)} & \multicolumn{1}{c}{(7)} & \multicolumn{1}{c}{(8)} & \multicolumn{1}{c}{(9)} & \multicolumn{1}{c}{(10)} \\ 
 \hline
\noalign{\smallskip} 
 CH$_2$(OH)CHO &   10 &   80 & $ 1.80 \times 10^{15}$ & 1.00 & 1.00 & 1.00 & 6.0 & -1.0 & e \\ 
 \hline
 \end{tabular}
 \tablefoot{
 See notes of Table~\ref{t:modparam_lmh_c2h3cn}.
 }
\end{table*}
} 
\onltab{\clearpage
\begin{table*}
 \caption{Parameters of our best-fit model of CH$_3$C$_3$N toward Sgr~B2(N).
 }
 \label{t:modparam_lmh_ch3c3n}
 \centering
 \begin{tabular}{lccccccccc}
 \hline\hline
 \multicolumn{1}{c}{Molecule\tablefootmark{a}} & \multicolumn{1}{c}{Size\tablefootmark{b}} & \multicolumn{1}{c}{$T_{\mathrm{rot}}$\tablefootmark{c}} & \multicolumn{1}{c}{$N$\tablefootmark{d}} & \multicolumn{1}{c}{$C_{\rm int}$\tablefootmark{e}} & \multicolumn{1}{c}{$C_{\rm vib}$\tablefootmark{f}} & \multicolumn{1}{c}{$C_N$\tablefootmark{g}} & \multicolumn{1}{c}{$\Delta V$\tablefootmark{h}} & \multicolumn{1}{c}{$V_{\mathrm{off}}$\tablefootmark{i}} & \multicolumn{1}{c}{$F$\tablefootmark{j}} \\ 
  & \multicolumn{1}{c}{($''$)} & \multicolumn{1}{c}{(K)} & \multicolumn{1}{c}{(cm$^{-2}$)} & & & & \multicolumn{2}{c}{(km~s$^{-1}$)} & \\ 
 \multicolumn{1}{c}{(1)} & \multicolumn{1}{c}{(2)} & \multicolumn{1}{c}{(3)} & \multicolumn{1}{c}{(4)} & \multicolumn{1}{c}{(5)} & \multicolumn{1}{c}{(6)} & \multicolumn{1}{c}{(7)} & \multicolumn{1}{c}{(8)} & \multicolumn{1}{c}{(9)} & \multicolumn{1}{c}{(10)} \\ 
 \hline
\noalign{\smallskip} 
 CH$_3$C$_3$N & 3.0 &  100 & $ 1.09 \times 10^{15}$ & 0.99 & 1.00 & 0.99 & 7.0 & 0.0 & e \\ 
 & 3.0 &  100 & $ 5.54 \times 10^{14}$ & 0.99 & 1.00 & 0.99 & 7.0 & 10.0 & e \\ 
 \hline
 \end{tabular}
 \tablefoot{
 See notes of Table~\ref{t:modparam_lmh_c2h3cn}.
 }
\end{table*}
} 
\onltab{\clearpage
\begin{table*}
 \caption{Parameters of our best-fit model of CH$_3$CCH toward Sgr~B2(N).
 }
 \label{t:modparam_lmh_ch3cch}
 \centering
 \begin{tabular}{lccccccccc}
 \hline\hline
 \multicolumn{1}{c}{Molecule\tablefootmark{a}} & \multicolumn{1}{c}{Size\tablefootmark{b}} & \multicolumn{1}{c}{$T_{\mathrm{rot}}$\tablefootmark{c}} & \multicolumn{1}{c}{$N$\tablefootmark{d}} & \multicolumn{1}{c}{$C_{\rm int}$\tablefootmark{e}} & \multicolumn{1}{c}{$C_{\rm vib}$\tablefootmark{f}} & \multicolumn{1}{c}{$C_N$\tablefootmark{g}} & \multicolumn{1}{c}{$\Delta V$\tablefootmark{h}} & \multicolumn{1}{c}{$V_{\mathrm{off}}$\tablefootmark{i}} & \multicolumn{1}{c}{$F$\tablefootmark{j}} \\ 
  & \multicolumn{1}{c}{($''$)} & \multicolumn{1}{c}{(K)} & \multicolumn{1}{c}{(cm$^{-2}$)} & & & & \multicolumn{2}{c}{(km~s$^{-1}$)} & \\ 
 \multicolumn{1}{c}{(1)} & \multicolumn{1}{c}{(2)} & \multicolumn{1}{c}{(3)} & \multicolumn{1}{c}{(4)} & \multicolumn{1}{c}{(5)} & \multicolumn{1}{c}{(6)} & \multicolumn{1}{c}{(7)} & \multicolumn{1}{c}{(8)} & \multicolumn{1}{c}{(9)} & \multicolumn{1}{c}{(10)} \\ 
 \hline
\noalign{\smallskip} 
 CH$_3$CCH &   10 &   75 & $ 3.84 \times 10^{16}$ & 0.85 & 1.00 & 0.85 & 7.0 & 0.0 & e \\ 
 &   10 &   75 & $ 2.39 \times 10^{16}$ & 0.85 & 1.00 & 0.85 & 7.0 & 8.5 & e \\ 
\noalign{\smallskip} 
 CH$_3$C$^{13}$CH &   10 &   75 & $ 2.21 \times 10^{15}$ & 0.98 & 1.00 & 0.98 & 7.0 & 0.0 & e \\ 
\noalign{\smallskip} 
 CH$_3$$^{13}$CCH &   10 &   75 & $ 2.21 \times 10^{15}$ & 0.98 & 1.00 & 0.98 & 7.0 & 0.0 & e \\ 
\noalign{\smallskip} 
 $^{13}$CH$_3$CCH &   10 &   75 & $ 2.21 \times 10^{15}$ & 0.98 & 1.00 & 0.98 & 7.0 & 0.0 & e \\ 
 \hline
 \end{tabular}
 \tablefoot{
 See notes of Table~\ref{t:modparam_lmh_c2h3cn}.
 }
\end{table*}
} 
\onltab{\clearpage
\begin{table*}
 \caption{Parameters of our best-fit model of CH$_3$C(O)CH$_3$ toward Sgr~B2(N).
 }
 \label{t:modparam_lmh_ch3ch3co}
 \centering
 \begin{tabular}{lccccccccc}
 \hline\hline
 \multicolumn{1}{c}{Molecule\tablefootmark{a}} & \multicolumn{1}{c}{Size\tablefootmark{b}} & \multicolumn{1}{c}{$T_{\mathrm{rot}}$\tablefootmark{c}} & \multicolumn{1}{c}{$N$\tablefootmark{d}} & \multicolumn{1}{c}{$C_{\rm int}$\tablefootmark{e}} & \multicolumn{1}{c}{$C_{\rm vib}$\tablefootmark{f}} & \multicolumn{1}{c}{$C_N$\tablefootmark{g}} & \multicolumn{1}{c}{$\Delta V$\tablefootmark{h}} & \multicolumn{1}{c}{$V_{\mathrm{off}}$\tablefootmark{i}} & \multicolumn{1}{c}{$F$\tablefootmark{j}} \\ 
  & \multicolumn{1}{c}{($''$)} & \multicolumn{1}{c}{(K)} & \multicolumn{1}{c}{(cm$^{-2}$)} & & & & \multicolumn{2}{c}{(km~s$^{-1}$)} & \\ 
 \multicolumn{1}{c}{(1)} & \multicolumn{1}{c}{(2)} & \multicolumn{1}{c}{(3)} & \multicolumn{1}{c}{(4)} & \multicolumn{1}{c}{(5)} & \multicolumn{1}{c}{(6)} & \multicolumn{1}{c}{(7)} & \multicolumn{1}{c}{(8)} & \multicolumn{1}{c}{(9)} & \multicolumn{1}{c}{(10)} \\ 
 \hline
\noalign{\smallskip} 
 CH$_3$C(O)CH$_3$ & 3.0 &  100 & $ 1.49 \times 10^{17}$ & 0.50 & 1.00 & 0.50 & 7.0 & 0.0 & e \\ 
 & 3.0 &  100 & $ 3.73 \times 10^{16}$ & 0.50 & 1.00 & 0.50 & 7.0 & 10.0 & e \\ 
\noalign{\smallskip} 
 CH$_3$C(O)CH$_3$, $\varv_{12}$=1 & 3.0 &  100 & $ 1.79 \times 10^{17}$ & 0.50 & 1.00 & 0.50 & 7.0 & 0.0 & e \\ 
 & 3.0 &  100 & $ 4.47 \times 10^{16}$ & 0.50 & 1.00 & 0.50 & 7.0 & 10.0 & e \\ 
 \hline
 \end{tabular}
 \tablefoot{
 See notes of Table~\ref{t:modparam_lmh_c2h3cn}.
 }
\end{table*}
} 
\onltab{\clearpage
\begin{table*}
 \caption{Parameters of our best-fit model of CH$_3$CHO toward Sgr~B2(N).
 }
 \label{t:modparam_lmh_ch3cho}
 \centering
 \begin{tabular}{lccccccccc}
 \hline\hline
 \multicolumn{1}{c}{Molecule\tablefootmark{a}} & \multicolumn{1}{c}{Size\tablefootmark{b}} & \multicolumn{1}{c}{$T_{\mathrm{rot}}$\tablefootmark{c}} & \multicolumn{1}{c}{$N$\tablefootmark{d}} & \multicolumn{1}{c}{$C_{\rm int}$\tablefootmark{e}} & \multicolumn{1}{c}{$C_{\rm vib}$\tablefootmark{f}} & \multicolumn{1}{c}{$C_N$\tablefootmark{g}} & \multicolumn{1}{c}{$\Delta V$\tablefootmark{h}} & \multicolumn{1}{c}{$V_{\mathrm{off}}$\tablefootmark{i}} & \multicolumn{1}{c}{$F$\tablefootmark{j}} \\ 
  & \multicolumn{1}{c}{($''$)} & \multicolumn{1}{c}{(K)} & \multicolumn{1}{c}{(cm$^{-2}$)} & & & & \multicolumn{2}{c}{(km~s$^{-1}$)} & \\ 
 \multicolumn{1}{c}{(1)} & \multicolumn{1}{c}{(2)} & \multicolumn{1}{c}{(3)} & \multicolumn{1}{c}{(4)} & \multicolumn{1}{c}{(5)} & \multicolumn{1}{c}{(6)} & \multicolumn{1}{c}{(7)} & \multicolumn{1}{c}{(8)} & \multicolumn{1}{c}{(9)} & \multicolumn{1}{c}{(10)} \\ 
 \hline
\noalign{\smallskip} 
 CH$_3$CHO & 3.0 &  100 & $ 1.43 \times 10^{17}$ & 0.89 & 1.00 & 0.89 & 8.0 & -1.0 & e \\ 
 & 3.0 &  100 & $ 8.04 \times 10^{16}$ & 0.89 & 1.00 & 0.89 & 8.0 & 9.0 & e \\ 
\noalign{\smallskip} 
 CH$_3$CHO, $\varv_{\rm t}$=1 & 3.0 &  100 & $ 2.86 \times 10^{17}$ & 0.89 & 1.00 & 0.89 & 8.0 & -1.0 & e \\ 
 & 3.0 &  100 & $ 1.61 \times 10^{17}$ & 0.89 & 1.00 & 0.89 & 8.0 & 9.0 & e \\ 
 \hline
 \end{tabular}
 \tablefoot{
 See notes of Table~\ref{t:modparam_lmh_c2h3cn}.
 }
\end{table*}
} 
\input{tab_modparam_survey30m_lmh_17.tex} 
\input{tab_modparam_survey30m_lmh_18.tex} 
\input{tab_modparam_survey30m_lmh_19.tex} 
\input{tab_modparam_survey30m_lmh_20.tex} 
\input{tab_modparam_survey30m_lmh_21.tex} 
\input{tab_modparam_survey30m_lmh_22.tex} 
\input{tab_modparam_survey30m_lmh_23.tex} 
\input{tab_modparam_survey30m_lmh_24.tex} 
\input{tab_modparam_survey30m_lmh_25.tex} 
\input{tab_modparam_survey30m_lmh_26.tex} 
\input{tab_modparam_survey30m_lmh_27.tex} 
\input{tab_modparam_survey30m_lmh_28.tex} 
\input{tab_modparam_survey30m_lmh_29.tex} 
\input{tab_modparam_survey30m_lmh_30.tex} 
\input{tab_modparam_survey30m_lmh_31.tex} 
\input{tab_modparam_survey30m_lmh_32.tex} 
\input{tab_modparam_survey30m_lmh_33.tex} 
\input{tab_modparam_survey30m_lmh_34.tex} 
\input{tab_modparam_survey30m_lmh_35.tex} 
\input{tab_modparam_survey30m_lmh_36.tex} 
\input{tab_modparam_survey30m_lmh_37.tex} 
\input{tab_modparam_survey30m_lmh_38.tex} 
\input{tab_modparam_survey30m_lmh_39.tex} 
\input{tab_modparam_survey30m_lmh_40.tex} 
\input{tab_modparam_survey30m_lmh_41.tex} 
\input{tab_modparam_survey30m_lmh_42.tex} 
\input{tab_modparam_survey30m_lmh_43.tex} 
\input{tab_modparam_survey30m_lmh_44.tex} 
\input{tab_modparam_survey30m_lmh_45.tex} 
\input{tab_modparam_survey30m_lmh_46.tex} 
\input{tab_modparam_survey30m_lmh_47.tex} 
\input{tab_modparam_survey30m_lmh_48.tex} 
\input{tab_modparam_survey30m_lmh_49.tex} 
\input{tab_modparam_survey30m_lmh_50.tex} 
\input{tab_modparam_survey30m_lmh_51.tex} 
\input{tab_modparam_survey30m_lmh_52.tex} 
\input{tab_modparam_survey30m_lmh_53.tex} 
\input{tab_modparam_survey30m_lmh_54.tex} 
\input{tab_modparam_survey30m_lmh_55.tex} 
\input{tab_modparam_survey30m_lmh_56.tex} 
\onltab{\clearpage
\begin{table*}
 \caption{Parameters of our best-fit model of C$_2$H$_3$CN toward Sgr~B2(M).
 }
 \label{t:modparam_b2m_c2h3cn}
 \centering
 \begin{tabular}{lccccccccc}
 \hline\hline
 \multicolumn{1}{c}{Molecule\tablefootmark{a}} & \multicolumn{1}{c}{Size\tablefootmark{b}} & \multicolumn{1}{c}{$T_{\mathrm{rot}}$\tablefootmark{c}} & \multicolumn{1}{c}{$N$\tablefootmark{d}} & \multicolumn{1}{c}{$C_{\rm int}$\tablefootmark{e}} & \multicolumn{1}{c}{$C_{\rm vib}$\tablefootmark{f}} & \multicolumn{1}{c}{$C_N$\tablefootmark{g}} & \multicolumn{1}{c}{$\Delta V$\tablefootmark{h}} & \multicolumn{1}{c}{$V_{\mathrm{off}}$\tablefootmark{i}} & \multicolumn{1}{c}{$F$\tablefootmark{j}} \\ 
  & \multicolumn{1}{c}{($''$)} & \multicolumn{1}{c}{(K)} & \multicolumn{1}{c}{(cm$^{-2}$)} & & & & \multicolumn{2}{c}{(km~s$^{-1}$)} & \\ 
 \multicolumn{1}{c}{(1)} & \multicolumn{1}{c}{(2)} & \multicolumn{1}{c}{(3)} & \multicolumn{1}{c}{(4)} & \multicolumn{1}{c}{(5)} & \multicolumn{1}{c}{(6)} & \multicolumn{1}{c}{(7)} & \multicolumn{1}{c}{(8)} & \multicolumn{1}{c}{(9)} & \multicolumn{1}{c}{(10)} \\ 
 \hline
\noalign{\smallskip} 
 C$_2$H$_3$CN & 5.0 &   50 & $ 3.54 \times 10^{15}$ & 0.89 & 1.00 & 0.89 & 10.0 & 2.0 & e \\ 
\noalign{\smallskip} 
 C$_2$H$_3$CN, $\varv_{11}$=1 & 2.0 &  150 & $ 1.14 \times 10^{17}$ & 0.88 & 1.00 & 0.88 & 10.0 & 2.0 & e \\ 
 \hline
 \end{tabular}
 \tablefoot{
 \tablefoottext{a}{Torsionally or vibrationally excited states are indicated after the name of the molecule and are modeled as separate entries. Entries with only the name of the molecule correspond to the ground state.}
 \tablefoottext{b}{Source diameter (\textit{FWHM}).}
 \tablefoottext{c}{Rotational temperature.}
 \tablefoottext{d}{Column density.}
 \tablefoottext{e}{Correction factor that was applied to the column density to account for the correction to the approximate interpolation performed by XCLASS to compute the partition function.}
 \tablefoottext{f}{Correction factor that was applied to the column density to account for the contribution of vibrationally or torsionally excited states or other conformers to the partition function.}
 \tablefoottext{g}{Global correction factor that was applied to the column density ($C_{\rm int} \times C_{\rm vib}$).}
 \tablefoottext{h}{Linewidth (\textit{FWHM}).}
 \tablefoottext{i}{Velocity offset with respect to the systemic velocity of Sgr~B2(M) $V_{\mathrm{lsr}} = 62$ km~s$^{-1}$.}
 \tablefoottext{j}{Flag indicating if the component belongs to the "emission" (e) or "absorption" (a) group.}
 }
\end{table*}
}
 
\input{tab_modparam_survey30m_b2m_2.tex} 
\input{tab_modparam_survey30m_b2m_3.tex} 
\input{tab_modparam_survey30m_b2m_4.tex} 
\input{tab_modparam_survey30m_b2m_5.tex} 
\input{tab_modparam_survey30m_b2m_6.tex} 
\input{tab_modparam_survey30m_b2m_7.tex} 
\input{tab_modparam_survey30m_b2m_8.tex} 
\input{tab_modparam_survey30m_b2m_9.tex} 
\input{tab_modparam_survey30m_b2m_10.tex} 
\input{tab_modparam_survey30m_b2m_11.tex} 
\input{tab_modparam_survey30m_b2m_12.tex} 
\input{tab_modparam_survey30m_b2m_13.tex} 
\input{tab_modparam_survey30m_b2m_14.tex} 
\input{tab_modparam_survey30m_b2m_15.tex} 
\input{tab_modparam_survey30m_b2m_16.tex} 
\input{tab_modparam_survey30m_b2m_17.tex} 
\input{tab_modparam_survey30m_b2m_18.tex} 
\input{tab_modparam_survey30m_b2m_19.tex} 
\input{tab_modparam_survey30m_b2m_20.tex} 
\input{tab_modparam_survey30m_b2m_21.tex} 
\input{tab_modparam_survey30m_b2m_22.tex} 
\input{tab_modparam_survey30m_b2m_23.tex} 
\input{tab_modparam_survey30m_b2m_24.tex} 
\input{tab_modparam_survey30m_b2m_25.tex} 
\input{tab_modparam_survey30m_b2m_26.tex} 
\input{tab_modparam_survey30m_b2m_27.tex} 
\input{tab_modparam_survey30m_b2m_28.tex} 
\input{tab_modparam_survey30m_b2m_29.tex} 
\input{tab_modparam_survey30m_b2m_30.tex} 
\input{tab_modparam_survey30m_b2m_31.tex} 
\input{tab_modparam_survey30m_b2m_32.tex} 
\input{tab_modparam_survey30m_b2m_33.tex} 
\input{tab_modparam_survey30m_b2m_34.tex} 
\input{tab_modparam_survey30m_b2m_35.tex} 
\input{tab_modparam_survey30m_b2m_36.tex} 
\input{tab_modparam_survey30m_b2m_37.tex} 
\input{tab_modparam_survey30m_b2m_38.tex} 
\input{tab_modparam_survey30m_b2m_39.tex} 
\input{tab_modparam_survey30m_b2m_40.tex} 
\input{tab_modparam_survey30m_b2m_41.tex} 
\input{tab_modparam_survey30m_b2m_42.tex} 
\input{tab_modparam_survey30m_b2m_43.tex} 
\input{tab_modparam_survey30m_b2m_44.tex} 
\input{tab_modparam_survey30m_b2m_45.tex} 
\input{tab_modparam_survey30m_b2m_46.tex}

\subsection{Non-detections}
\label{ss:nondetections}

Upper limits to the column densities of glycine (NH$_2$CH$_2$COOH), 
methoxyacetonitrile (CH$_3$OCH$_2$CN), 2-cyanoethanol (HOCH$_2$CH$_2$CN),
2-aminopropionitrile (CH$_3$CH(NH$_2$)CN), and \textit{n}-butyl cyanide 
(\textit{n}-C$_4$H$_9$CN) based on our spectral survey of Sgr~B2(N) were 
reported elsewhere \citep[][]{Belloche08,Braakman10,Mollendal12,Ordu12}. In 
this section, we list a few additional upper limits derived from our survey
for species that were either claimed to be detected toward Sgr~B2(N) by other 
authors, or that are of some interest for astrochemistry.

\subsubsection{Vinyl alcohol CH$_2$CHOH}
\label{sss:ch2choh}

We use the CDMS entries (tags 44506 and 44507, both version 1) for the 
\textit{syn} and \textit{anti} conformers of vinyl alcohol. They are based on 
the measurements performed by \citet{Saito76} and \citet{Rodler83} for the 
former, and \citet{Rodler85} for the latter.

\citet{Turner01} report the detection of vinyl alcohol toward Sgr~B2(N) with
the Kitt Peak 12~m telescope at 3 and 2~mm. They claim the detection of five 
transitions of the \textit{anti} conformer and two transitions of the 
\textit{syn}
conformer, with linewidths ($FWHM$) in the range 8--16~km~s$^{-1}$. With a beam 
filling factor of unity, they derive an excitation temperature of 
$11.6^{+13.5}_{-4.7}$~K and a column density of
$(2.4 \pm 0.6) \times 10^{13}$~cm$^{-2}$ for the former and a column density 
of $2.0 \times 10^{14}$~cm$^{-2}$ for the latter, assuming the same temperature.
The observation of the high-lying \textit{anti} conformer together with the 
low rotation temperature suggests, assuming the detection of vinyl alcohol 
is correct, that the molecule resides in the extended part of Sgr~B2, which 
often shows large deviations from LTE. Since the beam in our survey is much 
smaller, our survey is relatively less sensitive to such an extended, 
low-excitation component. 

With a beam filling factor of unity, a linewidth of 9~km~s$^{-1}$, and a 
temperature of 11.6~K, we derive an upper limit of $2 \times 10^{14}$~cm$^{-2}$
for the \textit{anti} conformer, which is less constraining than the detection
reported by \citet{Turner01}. Our upper limit is mainly set by the non-detection
of the transition at 84.83~GHz. On the other hand, if we assume a source size 
($3\arcsec$), temperature (100~K), and linewidth (7~km~s$^{-1}$) similar to
the values used for its isomer ethylene oxide ($c$-C$_2$H$_4$O, see 
Sect.~\ref{sss:c-c2h4o}), 
then we obtain an upper limit of $1.5 \times 10^{16}$~cm$^{-2}$, i.e. a ratio of
\textit{anti}-vinyl alcohol to ethylene oxide column densities lower than 0.4.
Assuming the same parameters as for acetaldehyde (CH$_3$CHO, 5$\arcsec$, 60~K, 
and 8~km~s$^{-1}$, see Sect.~\ref{sss:ch3cho}), we obtain an upper limit of
$3 \times 10^{15}$~cm$^{-2}$, i.e. a ratio of \textit{anti}-vinyl alcohol to 
acetaldehyde column densities lower than 0.1. For the \textit{syn} form, we 
find an upper limit of $2 \times 10^{14}$~cm$^{-2}$ for a low-excitation, 
extended component, marginally consistent with the detection of
\citet{Turner01}. For a 100~K, compact component, the upper limit is 
$3 \times 10^{16}$~cm$^{-2}$, implying a \textit{syn}-vinyl alcohol to ethylene 
oxide column densities lower than 0.8. For a 60~K, compact component, the 
upper limit is $8 \times 10^{15}$~cm$^{-2}$, i.e. a ratio of \textit{syn}-vinyl 
alcohol to acetaldehyde column densities lower than 0.3.

\subsubsection{Ketenimine CH$_2$CNH}
\label{sss:ch2nh}

We use the CDMS entry (tag 41503 version 1) which is based on measurements 
performed by \citet{Rodler84,Rodler86}.

The first detection of ketenimine was claimed by \citet{Lovas06} toward 
Sgr~B2(N) based on GBT observations of three lines in absorption, with 
lower-level energies ranging from 33 to 50~K. Assuming a source size of 
$5\arcsec$, they derive an excitation temperature of 65~K and a column density 
of $\sim 1.5 \times 10^{16}$~cm$^{-2}$, with an uncertainty of a factor of two.
Assuming the same source size and temperature and a linewidth ($FWHM$) of 
20~km~s$^{-1}$, our spectral survey at 3~mm yields a column-density upper 
limit of $2 \times 10^{16}$~cm$^{-2}$ toward Sgr~B2(N), which is less 
constraining than the detection reported by \citet{Lovas06}.

\subsubsection{\textit{trans} methyl formate CH$_3$OCHO}
\label{sss:ch3ocho-t}

The entry for the \textit{trans} conformer of methyl formate (only the 
\textit{A} species) was provided by B.~H. Pate in March 2010, prior to
publication \citep[][]{Neill12}.

A tentative detection of the \textit{trans} conformer of methyl formate,
both the \textit{A} and \textit{E} species, was reported in absorption 
toward Sgr~B2(N) by \citet{Neill12} based on low-frequency observations with 
the GBT. They derive an excitation temperature of $7.6 \pm 1.5$~K and estimate
the column density to be $1.2^{+1,2}_{-0.6} \times 10^{13}$~cm$^{-2}$.
These authors also report that the \textit{cis} conformer is detected 
in emission in the GBT survey and conclude that the two conformers have a 
different spatial distribution, the \textit{trans} conformer coming from a 
colder, more extended region.

The \textit{trans} conformer of methyl formate is not unambiguously detected
in our 3~mm survey toward Sgr~B2(N). Using the same source size and 
temperature  as we derived for the \textit{cis} conformer ($4\arcsec$ and 80~K, 
see Sect.~\ref{sss:ch3ocho}), we obtain an upper limit of 
$8 \times 10^{14}$~cm$^{-2}$ 
for the column density of the \textit{trans} conformer for the main hot core 
with $V_{\rm lsr} = 64$~km~s$^{-1}$. Hints of emission may be seen at 82.15 
($E_l/k \sim 41$--61~K), 91.28 (45--65~K), and 109.54~GHz (74~K), but they are 
not sufficient to claim a detection. As a result, we obtain an upper limit to 
the ratio of the \textit{trans} to \textit{cis} column densities of 0.2\%. For
comparison, the thermal-equilibrium value would be $\sim 10^{-16}$ at 80~K
\citep[see][]{Neill12}. If we assume an emission more extended than the beam 
and an excitation temperature of 7.6~K, we derive an upper limit of 
$2 \times 10^{13}$~cm$^{-2}$ for a linewidth of 7~km~s$^{-1}$, which is less 
constraining than the tentative detection of the \textit{A} species by 
\citet{Neill12}. 

\subsubsection{Propanal C$_2$H$_5$CHO, propenal C$_2$H$_3$CHO, and propynal 
CHCCHO}
\label{sss:c2h5cho}

We use the CDMS entry (tag 58505 version 1) for the \textit{syn} form of
propanal, the JPL entry (tag 56008 version 1, with the partition function 
values updated in October 2012) for propenal, and the CDMS entry (tag 54510
version 1) for propynal. The entry for propanal is based on measurements 
reported by \citet{Hardy82} and \citet{Demaison87}. The part of the entry 
dealing with the \textit{trans} conformer of propenal is mostly based on 
measurements reported by \citet{Winnewisser75}. The \textit{cis} 
conformer is much higher in energy ($\sim 594$~cm$^{-1}$). The entry for 
propynal is based on measurements performed mainly
by \citet{Costain59} and \citet{Winnewisser73}.

\citet{Hollis04} report the detection of six transitions of propanal, two
transitions of propenal, and one transition of propynal toward Sgr~B2(N) with 
the GBT ($HPBW = 28$--40$\arcsec$), with a mixture of emission and absorption 
in several velocity components that prevented them to derive reliable column 
densities. Propynal was reported for the first time in space toward 
\object{TMC1} by \citet{Irvine88}.
Assuming a beam filling factor of unity, a linewidth of 8~km~s$^{-1}$, and a 
temperature of 50~K, our 3~mm survey of Sgr~B2(N) yields column-density upper 
limits of $1.5 \times 10^{14}$~cm$^{-2}$ for the \textit{syn} form of propanal, 
$3 \times 10^{13}$~cm$^{-2}$ for propenal, and $3 \times 10^{13}$~cm$^{-2}$ for 
propynal. For a temperature of 15~K, the upper limits become
$1.0 \times 10^{14}$, $2 \times 10^{13}$, and $2 \times 10^{13}$~cm$^{-2}$, 
respectively.

\subsubsection{Silicon monosulfide SiS}
\label{sss:sis}

We use the CDMS entry \citep[tag 60506 version 1, see][]{Mueller07}.

Silicon monosulfide was reported toward \object{Sgr~B2(OH)} at 3~mm by 
\citet{Cummins86} and \citet{Turner89,Turner91}. Both detections are somewhat
uncertain as they rely on only one transition at 90.77~GHz with a low 
signal-to-noise ratio. The 5--4 and 6--5 transitions fall into the range of 
our survey at 3~mm. At both frequencies (90.77 and 108.92~GHz), there is 
emission detected toward Sgr~B2(N) and (M) with a peak temperature of 
$\sim 0.2$ and 0.1~K but this emission is broad and likely results from the 
overlap of several lines. Therefore we are not able to firmly identify SiS in 
our spectra. With an excitation temperature of 25~K, a linewidth of 
18~km~s$^{-1}$, and a beam filling factor of unity, as assumed by 
\citet{Turner91}, we derive column density upper limits of $4.5 \times 10^{13}$
and $3 \times 10^{13}$~cm$^{-2}$ toward Sgr~B2(N) and (M), respectively.
The line tentatively reported toward Sgr~B2(OH), which also includes Sgr~B2(M)
given the large beam of the NRAO 12~m telescope, yields a column density of 
$1.2 \times 10^{13}$~cm$^{-2}$ \citep[][]{Turner91}. We note that 
\citet{Dickinson81} report the detection of both transitions at 3~mm toward
Sgr~B2(OH) at a velocity of 56--58~km~s$^{-1}$. They derive a column density of
1--$2 \times 10^{13}$~cm$^{-2}$ assuming a temperature of 20~K.

\subsubsection{Glycine NH$_2$CH$_2$COOH}
\label{sss:nh2ch2cooh}

Upper limits to the column density of glycine based on our survey toward 
Sgr~B2(N) were already published in \citet{Belloche08}. More stringent 
constraints were reported by \citet{Jones07} and \citet{Cunningham07}.

\subsubsection{Deuterated methanol CH$_2$DOH}
\label{sss:ch2doh}

We use the JPL entry (tag 33004, version 1).

Using the same parameters as for $^{13}$CH$_3$OH (see Sect.~\ref{sss:ch3oh}),
we derive an upper limit of $1.0 \times 10^{17}$ or
$1.0 \times 10^{14}$~cm$^{-2}$ for the column density of a warm or cold 
component, respectively, toward Sgr~B2(N). This yields an upper limit to the
deuterium fractionation of methanol of 0.7\% or 0.3\%, respectively.
Following the same procedure, we find an upper limit to the deuterium 
fractionation of 7\% or 0.4\% for a warm or cold component toward Sgr~B2(M),
respectively. These upper limits are consistent with the low levels of 
deuterium fractionation measured toward Sgr~B2 for, e.g., water, ammonia, 
and HCO$^+$ \citep[$1.6 \times 10^{-3}$, $1.1 \times 10^{-3}$, and 
$2 \times 10^{-4}$ , see][, respectively]{Jacq90,Peng93,Jacq99}.

\subsubsection{Deuterated formamide}
\label{sss:nh2cho_d}

We use the CDMS entries for the three isotopologues NH$_2$CDO, 
\textit{c-}NHDCHO, and \textit{t-}NHDCHO (tags 46520, 46521, and 46522, all 
version 1). They are based on \citet{Kutsenko13} with additional data from 
\citet{1973_CJ_Nielsen}.

Using the same parameters as for NH$_2$CHO (see Sect.~\ref{sss:nh2cho}),
we derive upper limits of $8.0 \times 10^{15}$, $8.0 \times 10^{15}$, and 
$1.2 \times 10^{16}$~cm$^{-2}$ for the column densities of NH$_2$CDO, 
\textit{c-}NHDCHO, and \textit{t-}NHDCHO, respectively, toward Sgr~B2(N). This 
yields upper limits to the deuterium fractionation of formamide of 0.6\%,
0.6\%, and 0.9\%, respectively. These upper limits are consistent with the low 
levels of deuterium fractionation mentioned in Sect.~\ref{sss:ch2doh}.

\subsubsection{Vibrationally excited CN and SiO}

Data for both excited vibrational states are contained in the entries 
dealing with the ground states. Laboratory data for CN were taken from
\citet{Dixon77}, those for SiO from \citet{SiO_1991}.

The only CN transitions from within $\varv = 1$ covered by our survey 
are at $\sim$~112.1 and 112.4 GHz. Assuming the same source size (0.9$\arcsec$) 
and temperature (500~K) as for HCN, $\varv_2 =1$ (see Sect.~\ref{sss:hcn}), we 
obtain an upper limit on the column density parameter of 
$\sim 5 \times 10^{19}$~cm$^{-2}$ for a linewidth of 7~km~s$^{-1}$.

No transition of SiO from within $\varv = 1$ is detected. The most
stringent constraint is obtained at 86.243370~GHz. Assuming again the same 
source size, temperature, and linewidth (12~km~s$^{-1}$) as for HCN, 
$\varv_2 =1$, we obtain an upper limit on the column density parameter of 
$\sim 5 \times 10^{17}$~cm$^{-2}$.

\subsection{Strong unidentified lines}
\label{ss:strongUlines}

\input{tab_ulines_lmh_survey30m}

\begin{table}
 \caption{
 Frequencies of unidentified lines in the 3~mm line survey of Sgr~B2(M) with a peak temperature higher than 0.3~K in main-beam temperature scale (about $13\,\sigma$).
 }
 \label{t:ulines_b2m}
 \centering
 \begin{tabular}{cccc}
 \hline\hline
  \multicolumn{1}{c}{Frequency} & \multicolumn{1}{c}{Frequency} & \multicolumn{1}{c}{Frequency} & \multicolumn{1}{c}{Frequency}\\ 
  \multicolumn{1}{c}{(GHz)} & \multicolumn{1}{c}{(GHz)} & \multicolumn{1}{c}{(GHz)} & \multicolumn{1}{c}{(GHz)} \\ 
 \hline
        95145.3 &        110154.5 &        112110.9$^\star$ &        112124.3$^\star$ \\ 
        95948.9 &        110157.7 &        112115.2 &        115806.4 \\ 
        95964.6 &        110164.1 &  & \\ 
  \hline
 \end{tabular}
 \tablefoot{
 The typical uncertainty is on the order of 1~MHz but could also be higher in some cases.
 The rest frequencies assume a systemic velocity of 62~km~s$^{-1}$.
 Frequencies marked with a star indicate lines which are likely contamination from the image band (CO 1--0).
 }
 \end{table}

Tables~\ref{t:ulines_lmh} and \ref{t:ulines_b2m} list the frequencies of 
the 83 and 10 unidentified lines that have a peak temperature higher than 
0.3~K (i.e. about $13\,\sigma$) in the 3~mm surveys of Sgr~B2(N) and (M), 
respectively. In some cases, these frequencies are uncertain because of 
blends with other lines. For Sgr~B2(N), some of them may also correspond to 
the second velocity component of a line at higher (by $\sim 2.7$--3.9~MHz) 
frequency. For both sources, a few unidentified lines may also correspond to
contamination by CO 1--0 from the image band.

\section{Discussion}
\label{s:discussion}

\subsection{Comparison to previous surveys}
\label{ss:comp2prevsurv}

\input{tab_comp2prevsurv}

The molecules and their isotopologues detected in the present survey of 
Sgr~B2(N) and (M) are listed in Table~\ref{t:previous}. Their detection
or non-detection in previous spectral surveys of Sgr~B2 (either N, M, or OH) 
is also indicated. The statistics of the present survey are compared to these
previous surveys in Table~\ref{t:prevstats}. With respect to the 3~mm survey of
\citet{Turner89}, the density of detected lines has increased by 70\% for 
Sgr~B2(M). Surprisingly enough, the density of \textit{identified} lines has 
increased by about 20\% only, while we could have expected a larger improvement
given the growth of the CDMS and JPL spectroscopic databases over the years. 
While we set our identification threshold to 50\% of the detected level of 
emission/absorption (see Sect.~\ref{ss:stats}), the criterion used by 
\citet{Turner89,Turner91} to mark a line as identified may have been less 
stringent and their number of identified lines was in turn perhaps 
overestimated. Another explanation may be that Turner's survey is a priori less
sensitive to hot-core species: its angular resolution is poorer by a factor 3
and its target, \object{Sgr~B2(OH)}, is located $\sim\,$25$\arcsec$ south of 
Sgr~B2(M).

In contrast, our survey toward Sgr~B2(N) shows a much clearer improvement 
(factor 2) both in terms of line detection and identification compared to 
previous surveys.
Further progress in reducing the fraction of unidentified lines in our survey
of Sgr~B2(N) is expected when spectroscopic predictions for vibrationally or 
torsionally excited states of known species will become available. For 
instance, we expect about 300 additional lines of ethyl cyanide to be 
detectable in our survey, and about ten for vinyl cyanide (see 
Sects.~\ref{sss:c2h5cn} and \ref{sss:c2h3cn}, respectively). The $^{13}$C 
isotopologues of methyl formate in their vibrational ground state may also 
contribute two dozens of lines (see Sect.~\ref{sss:ch3ocho}). Provided these 
estimates are correct, these entries missing in our catalog would already 
reduce the number of unidentified lines to $\sim\,$750 and increase the 
fraction of identified lines to $\sim\,$80\%.

Finally, we emphasize that, although the fraction of identified lines in our
survey of Sgr~B(M) may appear surprisingly low compared to Sgr~B2(N) (see 
Col.~10 of Table~\ref{t:prevstats}), the density of unidentified lines per 
unit frequency is actually a factor of two higher for the latter (see Col.~9 of 
Table~\ref{t:prevstats}).

\subsection{First detections in space}
\label{ss:newdetections}

The high sensitivity of our survey toward Sgr~B2(N) and/or the recent 
availability of spectroscopic predictions for new molecules, rare 
isotopologues, and vibrationally or torsionally excited states of already 
known molecules, allowed us to identify a number of such species/entries for 
the first time in space. Our detections of three new molecules -- 
aminoacetonitrile, ethyl formate, and \textit{n-}propyl cyanide -- were already 
reported in \citet{Belloche08} and \citet{Belloche09}, and the detection of 
the $^{13}$C isotopologues of vinyl cyanide in \citet{Mueller08}. Here, we 
report the first detection of two other rare isotopologues, H$^{13}$C$^{15}$N 
and H$^{13}$CC$^{13}$CN (see Sects.~\ref{sss:hcn} and \ref{sss:hc3n}, 
respectively). We also report the first detection of transitions from within 
twelve new vibrationally or torsionally excited states of known
molecules: four states of vinyl cyanide, one
state of acetone, one state of acetaldehyde, one state of methyl cyanide,
one state of the $^{13}$C isotopologues of methyl cyanide, and four states of
the $^{13}$C isotopologues of cyanoacetylene
(see Sects.~\ref{sss:c2h3cn}, \ref{sss:ch3coch3}, \ref{sss:ch3cho}, 
\ref{sss:ch3cn}, \ref{sss:ch3cn}, and \ref{sss:hc3n} respectively). 

While the line density and the line intensity of typical hot-core molecules 
increase with frequency from the 3~mm atmospheric window to the 1~mm window, 
we would like to stress the fact that the 3~mm range appears to be best suited
in our survey
for the detection of new molecules toward Sgr~B2(N). The line confusion at 2 
and 1~mm is such that the weak emission of rare molecules is most of the time 
hidden below the stronger emission of more abundant species. In addition, the 
frequent overlap between transitions from different species and the 
significant opacity of the continuum emission makes the modeling more 
difficult and the line identification in turn less reliable at 2 and 1~mm.
We note that quite a number of interstellar complex organic molecules were 
first detected toward Sgr~B2 at the beginnings of molecular radio astronomy, 
surprizingly, at long centimeter wavelengths, namely methanol, formic acid, 
formamide, acetaldehyde, methanimine, methylamine, vinyl cyanide, and
methyl formate 
\citep[for a historical perspective and references, see][]{Menten04}. More 
recently, propanal, acetamide, and cyanoformaldehyde were added 
\citep{Hollis04,Hollis06,Remijan08}. These frequency ranges have an even 
lower level of line confusion than the 3~mm range, which in some cases 
facilitates the line identification. They seem, however, to trace extended 
emission with low excitation rather than the high-temperature/high-density 
emission of the embedded hot cores that is traced at 3~mm and at higher 
frequencies. Moreover, the cm observations indicate that the populations of 
the low-lying energy levels giving rise to lines at these wavelengths are far 
from LTE. This causes some lines to become inverted, causing low gain 
amplification of the strong extended radio continuum emission and absorption 
against that emission in others \citep[see][]{Menten04}.

The large number of unidentified lines in our survey of Sgr~B2(N) may still 
allow future identifications of new molecules. We expect however a high 
fraction of these unidentified lines to be emitted by transitions from within 
vibrationally or torsionally excited states of known molecules for which 
spectroscopic predictions are currently missing. In particular, ethyl cyanide 
is expected to be a major contributor (see Sect.~\ref{ss:comp2prevsurv}).

\input{tab_prevsurvstats}

\subsection{Implications for interstellar chemistry}
\label{ss:chemistry}

The implications for the interstellar chemistry of the detection of the three 
molecules found for the first time in space in the course of this project 
(aminoacetronitrile, propyl cyanide, and ethyl formate) were already discussed 
by \citet{Belloche09} based on a comparison to predictions of a numerical, 
gas-grain, chemical model. The main conclusion is that the abundance ratios in 
the chemical classes of interstellar esters and alkyl cyanides 
(HCOOH/CH$_3$OCHO/C$_2$H$_5$OCHO and CH$_3$CN/C$_2$H$_5$CN/C$_3$H$_7$CN) can be 
relatively well reproduced if the main mechanism leading to the formation of 
these complex organic molecules occurs on the grain surfaces as a piecewise 
construction from their constituent functional-group radicals. In particular, 
the hydrogenation of less saturated species appears to be much less 
significant for the formation of alkyl cyanides than this piecewise formation 
mode.

The detection of aminoacetonitrile (NH$_2$CH$_2$CN) allowed us to predict a 
very low column density for glycine (NH$_2$CH$_2$COOH), based on the measured 
ratio of acetic acid to methyl cyanide column densities 
([CH$_3$CN]/[CH$_3$COOH] $\sim 200$). If both pairs of molecules (carboxylic 
acid/cyanide) share similar formation routes, then we expect glycine to be 
about 200 times less abundant than aminoacetronitrile \citep[][]{Belloche08}. 
The detection of glycine in Sgr~B2(N), and by extension in the interstellar 
medium, will therefore likely have to wait for the full completion of ALMA.

Our first detection of the $^{13}$C isotopologues of vinyl cyanide led 
us to conclude that the northern hot core in Sgr~B2(N) is in a different 
evolutionary stage than the main hot core, based on a comparison of the measured
ratio of vinyl cyanide to ethyl cyanide column densities 
([C$_2$H$_3$CN]/[C$_2$H$_5$CN]) to predictions of older chemical models 
\citep[see][, and references therein]{Mueller08}. We were, however, not able to 
determine which source is the youngest one. A more detailed 
comparison based on several sets of molecules should help braking the 
degeneracy.

Although the results reported in the present article could allow for a more 
detailed comparison to predictions of chemical models, the lack of angular 
resolution implies ambiguities that cannot be solved easily with the current
dataset and may lead to erroneous conclusions. Even if we do constrain the 
size of the emitting region for some species that have both optically thick 
and thin lines in our single-dish
survey, nothing guarantees that two species with the same source size are 
emitted from the same region. A very good counter-example can be seen in the 
interferometric maps shown in Fig.~5 of \citet{Belloche08}: the emission peak 
of methyl formate (panel n) is displaced by about 1.7$\arcsec$ compared to 
the emission peaks of aminoacetonitrile (panels a to e) and ethyl cyanide 
(panel h). Since we currently have an ongoing project with ALMA to complete a 
line survey of Sgr~B2(N) in the 3~mm window with an improvement by 1.5 orders
of magnitude in both signal-to-noise ratio and angular resolution (P.I.: A. 
Belloche), we plan to discuss the implications in terms of interstellar 
chemistry in more details based on this coming new dataset, taking advantage
of the line identification already performed for this article.

\subsection{Peculiar case of methanol}
\label{ss:discussion_ch3oh}

That our LTE model significantly underpredicts the emission in the 
84.52 and 95.17~GHz transitions \textit{and} underpredicts or does not predict
absorption in  the transitions at 107.01 and 108.89~GHz can both be understood 
from excitation peculiarities of the methanol molecule. As described by 
\citet{Menten91a,Menten91b}, in the absence of a dominating far infrared 
field, the combination of (1) a propensity for $\Delta k =0$ collisions over 
$|\Delta k| =1$ collisions and (2) the fact that spontaneous transitions out 
of the $k=-1$ ladder are much slower than $J_{k=-1} - (J-1)_{k=-1}$ 
transitions, the molecules funnel down the $k=-1$ ladder relative to levels of 
the neighboring $k$ ladders, overpopulating the levels in that ladder. For 
$J_{-1} - J_0$ transitions, the upper energy levels are in the $k=-1$ ladder 
for $J\gtrsim4$, while for $J=3,$ 2, and 1 this is true for the lower energy
levels. This leads to inversion (and so called \textit{class I}) methanol 
maser emission in the $4_{-1}-3_0~E$ and $5_{-1}-4_0~E$ lines at 36.17 and 
84.52~GHz, respectively, and to anti-inversion (over-cooling) in the 
$2_0-3_{-1}~E$, $1_0-2_{-1}~E$, and $0_0-1_{-1}~E$ lines at 12.18, 60.53, and 
108.89 GHz, respectively.

For $A$-type methanol, analogously, the levels in the $K=0$ ladder are 
overpopulated relative to levels in the $K=1$ ladder. This causes inversion in 
the $7_0-6_1~A^+$ and $8_0-7_1~A^+$ transitions at 44.07 and 95.17 GHz, and 
anti-inversion in, e.g., the $5_1-6_0~A^+$, $4_1-5_0~A^+$, and $3_1-4_0~A^+$ 
transitions at 6.67, 57.03, and 107.01~GHz, respectively, and also lower 
$J$ $A$-type lines.

The above explains the excitation anomalies we observe in the 84.52 and 
95.17~GHz lines (both inverted) and the 107.01 and 108.89~GHz lines (both 
anti-inverted). In addition deep absorption has been observed (with large beam 
widths) in the 6.68 and 12.18~GHz lines \citep{Whiteoak88,Menten91b}. We note 
that the above scenario only holds  for a range of densities and temperatures 
that are more moderate than encountered in hot cores proper. Under hot core 
proper conditions, $\gtrsim 150$~K, $n \gtrsim 10^7$~cm$^{-3}$, the levels in 
the transitions discussed above are thermalized and the maser action and over-
cooling is quenched. In fact the FIR emission in many, but not all, hot cores 
leads to inversion in the 6.67~GHz $5_1-6_0~A^+$ and the 12.18~GHz 
$2_0-3_{-1}~E$ transitions, which are the strongest \textit{class II} methanol 
maser 
transitions. Therefore, the observed emission and absorption discussed above 
must arise, in both Sgr~B2(M) and (N),  from material surrounding the hot
cores. In contrast, class II methanol maser emission has been found from 12 
sites scattered over the greater Sgr B2 region \citep[][]{Caswell96}. None of 
these is closer than several arc seconds to any of the compact submillimeter 
sources/hot cores imaged by \citet{Qin11}.

\section{Conclusions}
\label{s:conclusions}

We performed a complete and sensitive spectral line survey toward Sgr~B2(N) 
and (M) with the IRAM~30~m telescope in the 3~mm atmospheric window as well as 
partial surveys at 2 and 1.3~mm in order to search for new complex organic 
molecules in the interstellar medium. The spectra were analyzed under the 
assumption of local thermodynamic equilibrium. Optical depth effects and the
finite angular resolution of the telescope were taken into account. We made 
extensive use of the CDMS and JPL spectroscopic databases to identify the 
detected lines. The main results of this project are the following:

\begin{enumerate}

\item About 3675 and 945 spectral lines are detected at 3~mm with a peak 
signal-to-noise ratio higher than 4 toward Sgr~B2(N) and (M), 
respectively. This yields densities of 102 and 26 lines per GHz, respectively,
i.e. an increase by about a factor of two over previous surveys of Sgr~B2.

\item About 70\% and 47\% of the lines detected toward Sgr~B2(N) and (M) at 
3~mm are identified and assigned to 56 and 46 distinct molecules as well as 66 
and 54 less abundant isotopologues of these molecules, respectively. In 
addition, 
transitions from 59 and 24 catalog entries corresponding to vibrationally or 
torsionally excited states of some of these molecules were detected toward 
Sgr~B2(N) and (M), respectively, up to a vibration energy of 
1400~cm$^{-1}$ (2000~K). The density of unidentified lines per 
unit frequency is a factor of two higher for Sgr~B2(N) than for (M).

\item The line confusion and the significant opacity of the continuum emission
at 2 and 1.3~mm make the modeling more difficult and the line identification 
process less reliable. The 3~mm window turns out to be the best suited to 
search for weak emission of rare hot-core species in our survey.

\item Absorption features produced by diffuse clouds along the line of 
sight are detected in transitions with low rotation quantum numbers of many 
simple molecules and are modeled with $\sim$~30--40 velocity components with 
typical linewidths of $\sim$~3--5~km~s$^{-1}$.

\item Three complex organic molecules -- aminoacetonitrile, ethyl formate, and 
\textit{n}-propyl cyanide -- and five rare isotopologues -- the $^{13}$C 
isotopologues of vinyl cyanide, H$^{13}$C$^{15}$N, and H$^{13}$CC$^{13}$CN --
were detected for the first time in space. The detection of transitions from 
within twelve new vibrationally or torsionally excited states of known 
molecules is also reported for the first time.

\item Although the large number of unidentified lines may still allow
future identifications of new molecules, we expect most of these lines to 
belong to 
vibrationally or torsionally excited states or to rare isotopologues of known 
molecules for which spectroscopic predictions are currently missing. In 
particular, ethyl cyanide (mainly), vinyl cyanide, and the $^{13}$C 
isotopologues of methyl formate could account for about 30\% of the 
unidentified lines toward Sgr~B2(N).

\item Excitation temperatures, column densities, systemic velocities, and 
linewidths were derived for each detected species under the LTE approximation,
sometimes with large uncertainties however. They should be used with 
caution. Most complex species detected toward Sgr~B2(N) have two velocity 
components corresponding to the two unresolved hot cores embedded in this 
source. 

\item The 3~mm observed and synthetic spectra of each source are made 
publicly available to allow potentially interested readers to search for their 
favorite molecules, either on their own or on a collaborative basis.
\end{enumerate}

The size of the often unresolved emitting region(s) could be to some extent 
derived for the species emitting both optically thick and optically 
thin lines. They are, however, somewhat uncertain and, as was shown 
interferometrically elsewhere, two species with the same source size are not 
necessarily located in the very same region. Therefore, we did not attempt to 
compare the derived column densities of all detected species to predictions of 
state-of-the-art chemical models in a systematic way. Interferometric 
observations will be necessary to break the remaining degeneracies. 
This is 
one of the main goals of our ongoing unbiased spectral line survey toward 
Sgr~B2(N) at 3~mm with ALMA in its cycles 0 and 1 (P.I.: A. Belloche). With an 
improvement in angular resolution and sensitivity by more than one order of 
magnitude compared to our single-dish survey, we expect to set tighter 
constraints on the location, origin, and abundance of complex organic 
molecules and to detect a few, or maybe even a dozen of, new species. We are 
therefore confident that the field of astrochemistry will, in the near future,
largely benefit from the tremendous opportunities offered by this new, 
powerful interferometer.

\begin{acknowledgements}
We thank the IRAM staff in Granada for service observing in January 2005 and 
Sergio Mart\'in for providing the coordinates of the reference (off) position
for our observations. We are indebted to Claus J. Nielsen for providing the 
frequency  part of the PhD thesis of M. Stubgaard and to Aur\'elia Bouchez, 
Brian J. Drouin, Christian P. Endres, Peter Groner, Vadim Ilyushin, Isabelle 
Kleiner, Monika Koerber, Laurent Margul{\`e}s, John C. Pearson, and Li-Hong Xu 
for providing entries in suitable formats. Some of these entries were made 
available to us considerably prior to 
publication. This work has been supported in part by the Deutsche 
Forschungsgemeinschaft (DFG) initially through the collaborative research 
grant SFB~494, more recently through SFB~956 ''Conditions and Impact of Star 
Formation'', project area B3. HSPM is very grateful to the Bundesministerium 
f\"ur Bildung und Forschung (BMBF) for recent support through project 
FKZ 50OF0901 (ICC HIFI \textit{Herschel}) aimed at maintaining the Cologne 
Database for Molecular Spectroscopy, CDMS.
\end{acknowledgements}

\begin{appendix} 
\section{Star formation rate of Sgr~B2}
\label{a:sfr}

Combining their results with those of \citet{Mehringer93}, 
\citet{Gaume95} report that there are at least 57 \ion{H}{ii} regions in the 
whole Sgr~B2 complex, 41 of which being ultracompact (UC) with diameters less 
than 0.1~pc. These 41 UC\ion{H}{ii} regions correspond 
to OB stars with spectral types ranging from B0.5 to O5 
\citep[see Table 3 of][]{Gaume95}. We use Table~1 of \citet{Panagia73} to 
convert the spectral types into stellar luminosities. We then compute the 
stellar masses with the equation 
$\frac{L}{L_\odot} = 2.3 \times \left(\frac{M}{M_\odot}\right)^{3.51}$, valid
for $M > 5$~$M_\odot$ \citep[][]{Griffiths88}. We obtain a total stellar 
mass of 675~$M_\odot$ for the 41 UC\ion{H}{ii} regions detected in Sgr~B2.

The spectral type B0.5 corresponds to a stellar mass of 
$\sim 11$~$M_\odot$ with the equation above. Given an initial mass function 
(IMF), we call $f_{11}$ the ratio of the mass of all stars with a mass between 
11 and 120~$M_\odot$ to the mass of all stars with a mass between 0.01 and 
120~$M_\odot$. We find $f_{11} = 0.179$ for the IMF of \citet{Chabrier05}, 0.174
for the IMF of \citet{Kroupa01}, and 0.242 for the Orion IMF of \citet{Muench02}
modified like in \citet{Murray10}. The total stellar mass of Sgr~B2 based
on the detected UC\ion{H}{ii} regions and corrected for the IMF with the 
extrapolation factor computed above is therefore $\sim$ 3800, 3900, and 
2800~$M_\odot$ for the three IMFs listed above, respectively.

As a consistency check, we also compute the total luminosity of the
49 \ion{H}{ii} regions detected by \citet{Gaume95}. We find 
$4.7 \times 10^6$~$L_\odot$. We use equation 4 of \citet{Griffiths88} to
evaluate the contribution $f_{L,11}$ of all stars with a mass between 11 and 
120~$M_\odot$ to the total luminosity of all stars with a mass between 0.01 and
120~$M_\odot$. We find $f_{L,11} = 99.4$\%, 99.5\%, and 99.6\% for the three 
IMFs listed above, respectively. We deduce a total stellar luminosity of 
$4.7 \times 10^6$~$L_\odot$ for Sgr~B2. This is within a factor of two 
consistent with the luminosity of $\sim 8.5 \times 10^6$~$L_\odot$ derived for 
Sgr~B2(N) and (M) by \citet{Goldsmith92} from the modeling of the far-infrared 
continuum emission\footnote{\citet{Goldsmith92} assumed a distance of 
8.5~kpc, while we use 7.9 kpc in this article. However, we do not correct the 
luminosity of \citet{Goldsmith92} for the distance in this appendix because we 
compare it with the luminosity derived from the emission of UC\ion{H}{ii} 
regions that was analyzed by \citet{Gaume95} who assumed the same distance as 
\citet{Goldsmith92}.}.

The lifetime of the UC\ion{H}{ii} regions detected in Sgr~B2 is 
uncertain. With a lifetime of $10^5$~yr \citep[see][]{Peters10}, we 
obtain a star formation rate of 
0.028--0.039~$M_\odot$~yr$^{-1}$ averaged over $10^5$~yr for the Sgr~B2 complex. 
This rate represents about 2--3\% of the global star formation rate of the 
Galaxy \citep[$1.2\pm 0.2$~$M_\odot$~yr$^{-1}$, see][]{Lee12}.

If the current star formation rate of Sgr~B2 could be 
sustained over a longer period, it would take 130--360 Myr to consume the total 
mass of the complex \citep[5--$10 \times 10^6$~$M_\odot$,][]{Lis90}. This 
timescale is about two orders of magnitude shorter than the Hubble time and
2.5 orders of magnitude shorter than the timescale of 
69~Gyr obtained for the whole Galaxy with the star formation rate mentioned 
above and a mass of $8.3 \times 10^{10}$~$M_\odot$ 
\citep[without the dark halo,][]{Sofue09}. This qualifies Sgr~B2 as a 
ministarburst region.

\end{appendix}

\listofobjects


\begin{thebibliography}{}

\bibitem[Adande et al.(2010)]{Adande10} Adande, G.~R., Halfen, D.~T., Ziurys, 
L.~M., Quan, D., \& Herbst, E.\ 2010, \apj, 725, 561 

\bibitem[Alekseev et al.(1996)]{SO2_etc_2-3mm_1996} 
Alekseev, E., Dyubko, S.~F., Ilyushin, V.~V. \& Podnos, S.~V.\ 
1996, J. Mol. Spectrosc., 176, 316 

\bibitem[Amano \& Maeda(2000)]{HOC+_2000}
Amano, T., \& Maeda, A.\ 
2000, J. Mol. Spectrosc., 203, 140 

\bibitem[Amano et al.(2005)]{N2H+_2005}
Amano, T., Hirao, T., \& Takano, J.\ 
2005, J. Mol. Spectrosc., 234, 170 

\bibitem[Anderson et al.(1987)]{13C-MeOH_1987} 
Anderson, T., Herbst, E., \& De Lucia, F.~C.\ 
1987, \apjs, 64, 703 

\bibitem[Anderson et al.(1990a)]{MeOH_1990} 
Anderson, T., De Lucia, F., \& Herbst, E.\ 
1990a, \apjs, 72, 797 

\bibitem[Anderson et al.(1990b)]{13C-MeOH_1990} 
Anderson, T., Herbst, E., \& De Lucia, F.~C.\ 
1990b, \apjs, 74, 647 

\bibitem[Barclay et al.(1993)]{Bar93}
Barclay, W.~L., Jr., Anderson, M.~A., Ziurys, L.~M., Kleiner, I., \& Hougen, 
J.~T.\ 
1993, \apjs, 89, 221 

\bibitem[Baskakov(1996)]{VyCN_1996}
Baskakov, O., Dyubko, S.~F. Ilyushin, V.~V., et al.\
1996, J. Mol. Spectrosc., 179, 94 

\bibitem[Bauder et al.(1976)]{Bau76}
Bauder, A., Lovas, F.~J., \& Johnson, D.~R.\ 
1976, J. Phys. Chem. Ref. Data, 5, 53 

\bibitem[Bauer \& Maes(1969)]{MeCN_v8_1969} 
Bauer, A. \& Maes, S.\ 
1969, J. Phys. France, 30, 169 

\bibitem[Bauer(1971)]{MeCN_2v8_1971} 
Bauer, A.\ 
1971, J. Mol. Spectrosc., 40, 183 

\bibitem[Bauer et al.(1975)]{MeCN-15_div-v_1975} 
Bauer, A., Tarrago, G., \& Remy, A.\ 
1975, J. Mol. Spectrosc., 58, 111 

\bibitem[Beers et al.(1972)]{H2CS_1972} 
Beers, Y., Klein, G.~P., Kirchhoff, W.~H., \& Johnson, D.~R.\ 
1972, J. Mol. Spectrosc., 44, 553 

\bibitem[Bellet et al.(1971)]{HCOOH_1971} 
Bellet, J., Samson, C., Steenbeckliers, G., \& Wertheimer, R.\ 
1971, J. Mol. Struct., 9, 49 

\bibitem[Belloche et al.(2008)]{Belloche08} Belloche, A., Menten, 
K.~M., Comito, C., et al.\ 2008, \aap, 482, 179 + Erratum \aap, 492, 769

\bibitem[Belloche et al.(2009)]{Belloche09} Belloche, A., Garrod, R.~T., 
M{\"u}ller, H.~S.~P., et al.\ 2009, \aap, 499, 215

\bibitem[Belov et al.(1993)]{Bel93}
Belov, S.~P., Tretyakov, M.~Y., Kleiner, I., \& Hougen, J.~T.\ 
1993, J. Mol. Spectrosc., 160, 61 

\bibitem[Belov et al.(1995)]{H2S_1995}
Belov, S., Yamada, K.~M.~T., Winnewisser, G., et al.\ 
1995, J. Mol. Spectrosc., 173, 380 

\bibitem[Belov et al.(1998)]{34SO2_etc_1998} 
Belov, S.~P., Tretyakov, M.~Y., Kozin, I.~N., et al.\ 
1998, J. Mol. Spectrosc., 191, 17 

\bibitem[Bernstein et al.(2002)]{Bernstein02} 
Bernstein, M.~P., Dworkin, J.~P., Sandford, S.~A., Cooper, G.~W., \& 
Allamandola, L.~J.\
2002, \nat, 416, 401

\bibitem[Bester et al.(1983)]{MeC3N_1983}
Bester, M., Tanimoto, M., Vowinkel, B., Winnewisser, G., \& Yamada, K.\ 
1983, Z. Naturforsch. A, 38, 64 

\bibitem[Birk et al.(1993)]{H2NCN_1993}
Birk, M., Winnewisser, M., \& Cohen, E.~A.\ 
1993, J. Mol. Spectrosc., 159, 69 

\bibitem[Bisschop et al.(2013)]{Bisschop13} Bisschop, S.~E., 
Schilke, P., Wyrowski, F., et al.\ 2013, \aap, 552, A122

\bibitem[Blake et al.(1984)]{13C-MeOH_1984} 
Blake, G.~A., Sutton, E.~C., Masson, C.~R., et al.\ 
1984, \apj, 286, 586 

\bibitem[Bocquet et al.(1988)]{MeCN-vibs_1988} 
Bocquet, R., Wlodarczak, G., Bauer, A., \& Demaison, J.\ 
1988, J. Mol. Spectrosc., 127, 382 

\bibitem[Bogey et al.(1981)]{rare-isos-CS_1981}
Bogey, M., Demuynck, C., \& Destombes, J.~L.\ 
1981, Chem. Phys. Lett., 81, 256 

\bibitem[Bogey et al.(1982a)]{isos-CS_1982}
Bogey, M., Demuynck, C., \& Destombes, J.~L.\ 
1982a, J. Mol. Spectrosc., 95, 35 

\bibitem[Bogey et al.(1982b)]{SO_34_18_1982} 
Bogey, M., Demuynck, C., \& Destombes, J.~L.\  
1982b, Chem. Phys., 66, 99 

\bibitem[Bogey et al.(1984)]{C-13-N_v0_1984} 
Bogey, M., Demuynck, C., \& Destombes, J.~L.\ 
1984, Can. J. Phys., 62, 1248 

\bibitem[Bogey et al.(1986)]{c-C3H2_1986} 
Bogey, M., Demuynck, C., \& Destombes, J.~L.\ 
1986, Chem. Phys. Lett., 125, 383 

\bibitem[Bogey et al.(1987)]{isos-c-C3H2_1987} 
Bogey, M., Demuynck, C., Destombes, J.~L., \& Dubus, H.\ 
1987, J. Mol. Spectrosc., 122, 313 

\bibitem[Bogey et al.(1988)]{HOCO+_1988}
Bogey, M., Demuynck, C., Destombes, J., \& Krupnov, A.\ 
1988, J. Mol. Struct., 190, 465 

\bibitem[Bogey et al.(1990)]{AAN_1990} 
Bogey, M., Dubus, H., \& Guillemin, J.~C.\ 
1990, J. Mol. Spectrosc., 143, 180 

\bibitem[Bogey et al.(1997)]{SO_1997}
Bogey, M., Civi{\v s}, S., Delcroix, B., et al.\ 
1997, J. Mol. Spectrosc., 182, 85 

\bibitem[Bouchez et al.(2012)]{13C-EtOH_2012} 
Bouchez, A., Walters, A., M{\"u}ller, H.~S.~P., et al.\ 
2012, \jqsrt, 113, 1148 

\bibitem[Braakman et al.(2010)]{Braakman10} Braakman, R., Belloche, A., Blake, 
G.~A., \& Menten, K.~M.\ 2010, \apj, 724, 994

\bibitem[Brauer et al.(2009)]{EtCN_2009}
Brauer, C.~S., Pearson, J.~C., Drouin, B.~J., \& Yu, S.\ 
2009, \apjs, 184, 133 

\bibitem[Brown et al.(1987)]{H2C-13-S_1987} 
Brown, R.~D., Godfrey, P.~D., McNaughton, D., \& Yamanouchi, K.\ 
1987, Mol. Phys., 62, 1429 

\bibitem[Br{\"u}nken et al.(2003)]{H2CO_2003}
Br{\"u}nken, S., M{\"u}ller, H.~S.~P., Lewen, F., \& Winnewisser, G.\ 
2003, Phys. Chem. Chem. Phys., 5, 1515 

\bibitem[Br{\"u}nken et al.(2004)]{DCN_2004}
Br{\"u}nken, S., Fuchs, U., Lewen, F., et al.\ 
2004, J. Mol. Spectrosc., 225, 152 

\bibitem[Br{\"u}nken et al.(2009a)]{HOCN_2009} 
Br{\"u}nken, S., Gottlieb, C.~A., McCarthy, M.~C., \& Thaddeus, P.\ 
2009a, \apj, 697, 880 

\bibitem[Br{\"u}nken et al.(2009b)]{HSCN_2009} 
Br{\"u}nken, S., Yu, Z., Gottlieb, C.~A., McCarthy, M.~C., \& Thaddeus, P.\ 
2009b, \apj, 706, 1588 

\bibitem[Br{\"u}nken et al.(2010)]{Bruenken10} Br{\"u}nken, S., Belloche, A., 
Mart{\'{\i}}n, S., Verheyen, L., \& Menten, K.~M.\ 2010, \aap, 516, A109

\bibitem[Buffa et al.(1994)]{HCO+_1994} 
Buffa, G., Tarrini, O., Cazzoli, G., \& Dore, L.\ 
1994, \pra, 49, 3557 

\bibitem[Burenin et al.(1981)]{OCS_w18_33_1981}
Burenin, A.~V., Val'Dov, A.~N., Karyakin, E.~N., Krupnov, A.~F., \& Shapin,
S.~M.\ 
1981, J. Mol. Spectrosc., 87, 312 

\bibitem[Butler et al.(2001)]{Butler01} Butler, R.~A.~H., De 
Lucia, F.~C., Petkie, D.~T., et al.\ 2001, \apjs, 134, 319 

\bibitem[Caselli et al.(1995)]{N2H+_1995}
Caselli, P., Myers, P.~C., \& Thaddeus, P.\ 
1995, \apjl, 455, L77 

\bibitem[Caswell(1996)]{Caswell96} Caswell, J.~L.\ 1996, \mnras, 283, 
606

\bibitem[Cazzoli \& Kisiel(1988)]{VyCN-vib_1988}
Cazzoli, G., \& Kisiel, Z.\ 
1988, J. Mol Spectrosc., 130, 303 

\bibitem[Cazzoli et al.(2004)]{13C16O} 
Cazzoli, G., Puzzarini, C., \& Lapinov, A.~V.\ 
2004, \apj, 611, 615 

\bibitem[Cazzoli \& Puzzarini(2005)]{HC-13-N_v0_2005}
Cazzoli, G., \& Puzzarini, C.\ 
2005, J. Mol. Spectrosc., 233, 280 

\bibitem[Cazzoli et al.(2005)]{HCN-15_2005}
Cazzoli, G., Puzzarini, C., \& Gauss, J.\ 
2005, \apjs, 159, 181 

\bibitem[Cazzoli \& Puzzarini(2006)]{MeCN_2006} 
Cazzoli, G., \& Puzzarini, C.\ 
2006, J. Mol. Spectrosc., 240, 153 

\bibitem[Cazzoli et al.(2006)]{PN_2006}
Cazzoli, G., Cludi, L., \& Puzzarini, C.\ 
2006, J. Mol. Struct., 780, 260 

\bibitem[Cazzoli \& Puzzarini(2008)]{propyne_2008}
Cazzoli, G., \& Puzzarini, C.\ 
2008, \aap, 487, 1197 

\bibitem[Chabrier(2005)]{Chabrier05} Chabrier, G.\ 2005, in The 
Initial Mass Function 50 Years Later, ed. E. Corbelli, F. Palla, \& H.
Zinnecker (Astrophysics and Space Science Library; Dordrecht: Springer), 
327, 41

\bibitem[Cho \& Saito(1998)]{SiO-18_1998} 
Cho, S.-H., \& Saito, S.\ 
1998, \apjl, 496, L51 

\bibitem[Christen \& M\"uller(2003)]{aGg_eglyc_2003} 
Christen, D. \& M\"uller, H.~S.~P.\ 
2003, Phys. Chem. Chem. Phys., 5, 3600 

\bibitem[Cohen \& Pickett(1982)]{NH2D_1982} 
Cohen, E.~A., \& Pickett, H.~M.\ 
1982, J. Mol. Spectrosc., 93, 83 

\bibitem[Cole \& Green(1973)]{VyCN_FIR_1973}
Cole, A.~R.~H., \& Green, A.~A.\ 
1973, J. Mol. Spectrosc., 48, 246 

\bibitem[Colmont et al.(1997)]{isos_VyCN_1997}
Colmont, J.~M., Wlodarczak, G., Priem, D., et al.\ 
1997, J. Mol. Spectrosc., 181, 330 

\bibitem[Comito et al.(2003)]{Comito03} Comito, C., Schilke, P., Gerin, M., et 
al.\ 2003, \aap, 402, 635

\bibitem[Comito et al.(2005)]{Comito05} Comito, C., Schilke, P., Phillips, 
T.~G., Lis, D.~C., Motte, F., \& Mehringer, D.\ 2005, \apjs, 156, 127

\bibitem[Comito et al.(2010)]{Comito10} Comito, C., Schilke, P., Rolffs, R., 
et al.\ 2010, \aap, 521, L38

\bibitem[Cornet \& Winnewisser(1980)]{div-H2CO_1980} 
Cornet, R.~A., \& Winnewisser, G.\ 
1980, J. Mol. Spectrosc, 80, 438 

\bibitem[Cosleou et al.(1991)]{MeCN_v4_1991} 
Cosleou, J., Wlodarczak, G., Boucher, D., \& Demaison, J.\ 
1991, J. Mol. Spectrosc., 146, 49 

\bibitem[Costain \& Morton(1959)]{Costain59} Costain, C.~C., \& Morton, J.~R.\ 
1959, \jcp, 31, 389 

\bibitem[Creswell et al.(1977)]{div-HC3N_v0_1977} 
Creswell, R.~A., Winnewisser, G., \& Gerry, M.~C.~L.\ 
1977, J. Mol. Spectrosc., 65, 420 

\bibitem[Cummins et al.(1986)]{Cummins86} Cummins, S.~E., Linke, R.~A., \& 
Thaddeus, P.\ 1986, \apjs, 60, 819 

\bibitem[Cunningham et al.(2007)]{Cunningham07} Cunningham, M.~R., 
et al.\ 2007, \mnras, 376, 1201

\bibitem[Dahmen et al.(1995)]{Dahmen95} Dahmen, G., Wilson, T.~L., \& 
Matteucci, F.\ 1995, \aap, 295, 194

\bibitem[Daly et al.(2013)]{Daly13} Daly, A.~M., Berm{\'u}dez, 
C., L{\'o}pez, A., et al.\ 2013, \apj, 768, 81

\bibitem[Dangoisse et al.(1978)]{div-H2CO_1978} 
Dangoisse, D., Willemot, E., \& Bellet, J.\ 
1978, J. Mol. Spectrosc., 71, 414 

\bibitem[de Lucia \& Gordy(1969)]{HCN_v0_1969} 
de Lucia, F., \& Gordy, W.\ 
1969, Phys. Rev., 187, 58 

\bibitem[de Lucia et al.(1971)]{HDO_1971} 
de Lucia, F.~C., Cook, R.~L., Helminger, P., \& Gordy, W.\ 
1971, \jcp, 55, 5334 

\bibitem[de Lucia \& Helminger(1975)]{NH2D_NHD2_1975} 
de Lucia, F.~C., \& Helminger, P.\ 
1975, J. Mol. Spectrosc., 54, 200 

\bibitem[de Lucia \& Helminger(1977)]{HCN_hohe-v_1977} 
de Lucia, F.~C., \& Helminger, P.~A.\ 
1977, \jcp, 67, 4262 

\bibitem[Demaison et al.(1979)]{13C-MeCN_1979} 
Demaison, J., Dubrulle, A., Boucher, D., Burie, J., \& Typke, V.\ 
1979, J. Mol. Spectrosc., 76, 1 

\bibitem[Demaison et al.(1987)]{Demaison87} Demaison, J., Maes, H., 
van Eijck, B.~P., Wlodarczak, G., \& Lasne, M.~C.\ 1987, J. Mol. Spectrosc., 
125, 214 

\bibitem[Demaison et al.(1994)]{VyCN_1994}
Demaison, J., Cosl\'eou, J., Bocquet, R., \& Lesarri, A.~G.\ 
1994, J. Mol. Spectrosc., 167, 400 

\bibitem[de Pree et al.(1998)]{dePree98} de Pree, C.~G., Goss, W.~M., \& 
Gaume, R.~A.\ 1998, \apj, 500, 847

\bibitem[de Vicente et al.(1997)]{deVicente97} de Vicente, P., Martin-Pintado, 
J., \& Wilson, T.~L.\ 1997, \aap, 320, 957

\bibitem[de Vicente et al.(2000)]{deVicente00} de Vicente, P., 
Mart{\'{\i}}n-Pintado, J., Neri, R., \& Colom, P.\ 2000, \aap, 361, 1058

\bibitem[Dickinson \& Kuiper(1981)]{Dickinson81} Dickinson, D.~F., \& 
Kuiper, E.~N.~R.\ 1981, \apj, 247, 112 

\bibitem[Dixon \& Woods(1977)]{Dixon77} Dixon, T.~A., \& Woods, R.~C.\
1977, \jcp, 67, 3956

\bibitem[Dore et al.(2001a)]{HCO-17+_2001a}
Dore, L., Puzzarini, C., \& Cazzoli, G.\ 
2001a, Can. J. Phys., 79, 359 

\bibitem[Dore et al.(2001b)]{HCO-17+_2001b} 
Dore, L., Cazzoli, G., \& Caselli, P.\ 
2001b, \aap, 368, 712 

\bibitem[Dore et al.(2009)]{15N-N2H+2009}
Dore, L., Bizzocchi, L., Degli Esposti, C., \& Tinti, F.\ 
2009, \aap, 496, 275 

\bibitem[Dore et al.(2010)]{H2CNH_2010} 
Dore, L., Bizzocchi, L., Degli Esposti, C., \& Gauss, J.\ 
2010, J. Mol. Spectrosc., 263, 44 

\bibitem[Dore et al.(2012)]{H2CNH_2012} 
Dore, L., Bizzocchi, L., \& Degli Esposti, C.\ 
2012, \aap, 544, A19 

\bibitem[Dubernet et al.(2010)]{Dubernet10}
Dubernet, M.~L., Boudon, V., Culhane, J.~L., et al.\
2010, \jqsrt, 111, 2151

\bibitem[Dubrulle et al.(1978)]{13C-propyne_1978}
Dubrulle, A., Boucher, D., Burie, J., \& Demaison, J.\ 
1978, J. Mol. Spectrosc., 72, 158 

\bibitem[Dubrulle et al.(1980)]{OCS_1980}
Dubrulle, A., Demaison, J., Burie, J., \& Boucher, D.\ 
1980, Z. Naturforsch. A, 35, 471 

\bibitem[Ehrenfreund et al.(2001)]{Ehrenfreund01} 
Ehrenfreund, P., Glavin, D.~P., Botta, O., Cooper, G., \& Bada, J.~L.\ 
2001, Proc. Nat. Acad. Sci., 98, 2138

\bibitem[Elsila et al.(2007)]{Elsila07} 
Elsila, J.~E., Dworkin, J.~P., Bernstein, M.~P., Martin, M.~P., \& Sandford, 
S.~A.\
2007, \apj, 660, 911

\bibitem[Elsila et al.(2009)]{Elsila09} Elsila, J.~E., Glavin, D.~P., \& 
Dworkin, J.~P.\ 2009, Meteorit. Planet. Sci., 44, 1323

\bibitem[Endres et al.(2009)]{DME_2009}
Endres, C.~P., Drouin, B.~J., Pearson, J.~C., et al.\ 
2009, \aap, 504, 635 

\bibitem[Erlandsson \& Cox(1956)]{HDO_D2O_1956} 
Erlandsson, G. \& Cox, J.\
1956, \jcp, 25, 778 

\bibitem[Fayt et al.(2004)]{HC3N-15_2004} 
Fayt, A., Vigouroux, C., Willaert, F., et al.\ 
2004, J. Mol. Struct., 695, 295 

\bibitem[Fortman et al.(2012)]{Fortman12} Fortman, S.~M., 
McMillan, J.~P., Neese, C.~F., et al.\ 2012, Journal of Molecular 
Spectroscopy, 280, 11 

\bibitem[Friedel et al.(2004)]{Friedel04} Friedel, D.~N., Snyder, L.~E., 
Turner, B.~E., \& Remijan, A.\ 2004, \apj, 600, 234

\bibitem[Fuchs et al.(2004)]{HC-13-N_v2_2004}
Fuchs, U., Br\"unken, S., Fuchs, G.~W., et al.\ 
2004, Z. Naturforsch. A, 59, 861 

\bibitem[Fukuyama et al.(1996)]{EtCN_8-200_1996} 
Fukuyama, Y., Odashima, H., Takagi, K., \& Tsunekawa, S.\ 
1996, \apjs, 104, 329 

\bibitem[Fukuyama et al.(1999)]{vib-EtCN_1999} 
Fukuyama, Y., Omori, K., Odashima, H., Takagi, K., \& Tsunekawa, S.\ 
1999, J. Mol. Spectrosc., 193, 72 

\bibitem[Furuya et al.(2003)]{Furuya03} Furuya, R.~S., Walmsley, C.~M., 
Nakanishi, K., Schilke, P., \& Bachiller, R.\ 2003, \aap, 409, L21 

\bibitem[Gaume \& Claussen(1990)]{Gaume90} Gaume, R.~A., \& Claussen, 
M.~J.\ 1990, \apj, 351, 538

\bibitem[Gaume et al.(1995)]{Gaume95} Gaume, R.~A., Claussen, 
M.~J., de Pree, C.~G., Goss, W.~M., \& Mehringer, D.~M.\ 1995, \apj, 449, 663

\bibitem[Gensheimer(1997)]{Gensheimer97} Gensheimer, P.~D.\ 1997, 
\apjl, 479, L75

\bibitem[Gerry \& Winnewisser(1973)]{VyCN_1973} 
Gerry, M.~C.~L., \& Winnewisser, G.\ 1973, 
J. Mol. Spectrosc., 48, 1 

\bibitem[Gibb et al.(2000)]{Gibb00}
Gibb, E., Nummelin, A., Irvine, W.~M., Whittet, D.~C.~B., \& Bergman, P.\ 
2000, \apj, 545, 309 

\bibitem[Goldsmith et al.(1983)]{Goldsmith83}
Goldsmith, P.~F., Krotkov, R., Snell, R.~L., Brown, R.~D., 
\& Godfrey, P.\ 
1983, \apj, 274, 184 

\bibitem[Goldsmith et al.(1992)]{Goldsmith92} Goldsmith, P.~F., 
Lis, D.~C., Lester, D.~F., \& Harvey, P.~M.\ 1992, \apj, 389, 338

\bibitem[Golubiatnikov et al.(2005)]{OCS_main_2005}
Golubiatnikov, G.~Y., Lapinov, A.~V., Guarnieri, A., \& Kn{\"o}chel, R.\ 
2005, J. Mol. Spectrosc., 234, 190 

\bibitem[Gottlieb et al.(2003)]{CS_CS-34_2003} 
Gottlieb, C.~A., Myers, P.~C., \& Thaddeus, P.\ 
2003, \apj, 588, 655 

\bibitem[Griffiths et al.(1988)]{Griffiths88} Griffiths, S.~C., 
Hicks, R.~B., \& Milone, E.~F.\ 1988, \jrasc, 82, 1

\bibitem[Groner et al.(2002)]{acetone_2002}
Groner, P., Albert, S., Herbst, E., et al.\ 
2002, \apjs, 142, 145 

\bibitem[Groner et al.(2006)]{acetone_2006}
Groner, P., Herbst, E., De Lucia, F.~C., Drouin, B.~J. \& M\"ader, H.\ 
2006, J. Mol. Struct., 795, 173 

\bibitem[Guarnieri et al.(1992)]{HCCNC_1992} 
Guarnieri, A., Hinze, R., Kr{\"u}ger, M., et al.\ 
1992, J. Mol. Spectrosc., 156, 39 

\bibitem[Guarnieri \& Huckauf(2003)]{isos-H2C2O_2003}
Guarnieri, A., \& Huckauf, A.\
2003, Z. Naturforsch. A, 58, 272

\bibitem[Gudeman et al.(1981)]{Gudeman81}
Gudeman, C.~S., Haese, N.~N., Piltch, N.~D., \& Woods, R.~C.\
1981, \apjl, 246, L47

\bibitem[Gudeman(1982)]{15N-N2H+_1982} 
Gudeman, C.~S.\ 
1982, Ph.D.~Thesis, University of Wisconsin, Madison, WI, USA

\bibitem[Gudeman \& Woods(1982)]{HOC+_3mm_1982} 
Gudeman, C.~S., \& Woods, R.~C.\ 
1982, Phys. Rev. Lett., 48, 1344 

\bibitem[Halfen et al.(2009)]{Halfen09} Halfen, D.~T., Ziurys, L.~M., 
Br{\"u}nken, S., et al.\ 2009, \apjl, 702, L124

\bibitem[Halfen et al.(2011)]{Halfen11} Halfen, D.~T., Ilyushin, V., \& 
Ziurys, L.~M.\ 2011, \apj, 743, 60 

\bibitem[Haque et al.(1974)]{13C-MeOH_1974} 
Haque, S.~S., Lees, R.~M., Saint Clair, J.~M., Beers, Y., \& Johnson, D.~R.\ 
1974, \apjl, 187, L15 

\bibitem[Hardy et al.(1982)]{Hardy82} Hardy, J.~A., Cox, A.~P., Fliege, E., 
\& Dreizler, H.\ 1982, Z. Naturforsch., 37a, 1035 

\bibitem[Herbst et al.(1984)]{HER84} 
Herbst, E., Messer, J.~K., de Lucia, F.~C., \& Helminger, P.\ 
1984, J. Mol. Spectrosc., 108, 42 

\bibitem[Herbst \& van Dishoeck(2009)]{Herbst09} Herbst, E., \& 
van Dishoeck, E.~F.\ 2009, \araa, 47, 427

\bibitem[Hieret(2005)]{Hieret05} Hieret, C.~O.\ 2005, Absorption studies along
the line of sight towards Sgr~B2(M), Diplomarbeit in Physik, 
Mathematisch-Naturwissenschaftlichen Fakult\"at der Rheinischen 
Friedrich-Wilhelms-Universit\"at Bonn, Germany

\bibitem[Hinze et al.(1996)]{H2C2O_rot-vibs_1996} 
Hinze, R., Zerbe-Foese, H. Doose, J. \& Guarnieri, A.\ 
1996, J. Mol. Spectrosc., 176, 133 

\bibitem[Hirose(1974)]{c-C2H4O_1974} 
Hirose, C.\ 
1974, \apjl, 189, L145 

\bibitem[Hocking et al.(1975)]{isos-HNCO_1975}
Hocking, W.~H., Gerry, M.~C.~L., \& Winnewisser, G.\ 
1975, Can. J. Phys., 53, 1869 

\bibitem[Hollis et al.(2000)]{Hollis00} Hollis, J.~M., Lovas, F.~J., \& 
Jewell, P.~R.\ 2000, \apjl, 540, L107

\bibitem[Hollis et al.(2002)]{Hollis02} Hollis, J.~M., Lovas, F.~J., Jewell, 
P.~R., \& Coudert, L.~H.\ 2002, \apjl, 571, L59

\bibitem[Hollis et al.(2003)]{Hollis03} Hollis, J.~M., Pedelty, J.~A., 
Boboltz, D.~A., et al.\ 2003, \apjl, 596, L235 

\bibitem[Hollis et al.(2004)]{Hollis04} Hollis, J.~M., Jewell, P.~R., 
Lovas, F.~J., Remijan, A., \& M{\o}llendal, H.\ 2004, \apjl, 610, L21 

\bibitem[Hollis et al.(2006)]{Hollis06} Hollis, J.~M., Lovas, F.~J., Remijan, 
A.~J., et al.\ 2006, \apjl, 643, L25 

\bibitem[Huiszoon(1971)]{H2S_1971} Huiszoon, C.\ 
1971, Rev. Sci. Instrum., 42, 477 

\bibitem[Ikeda et al.(1998)]{18O-MeOH_1998} 
Ikeda, M., Duan, Y.-B., Tsunekawa, S., \& Takagi, K.\ 
1998, \apjs, 117, 249 

\bibitem[Ilyushin et al.(2001)]{HAc_2001}
Ilyushin, V.~V., Alekseev, E.~A., Dyubko, S.~F., et al.\ 
2001, J. Mol. Spectrosc., 205, 286 

\bibitem[Ilyushin et al.(2004)]{Ilyushin04} Ilyushin, V.~V., Alekseev, E.~A., 
Dyubko, S.~F., Kleiner, I., \& Hougen, J.~T.\ 2004, J. Molec. Spectrosc., 227, 
115 

\bibitem[Ilyushin et al.(2005)]{MeNH2_2005}
Ilyushin, V.~V., Alekseev, E.~A., Dyubko, S.~F., Motiyenko, R.~A., \& 
Hougen, J.~T.\ 
2005, J. Mol. Spectrosc., 229, 170 

\bibitem[Ilyushin et al.(2008)]{HAc_2008}
Ilyushin, V., Kleiner, I., \& Lovas, F.~J.\ 
2008, J. Phys. Chem. Ref. Data, 37, 97 

\bibitem[Ilyushin et al.(2009)]{MeFo_2009}
Ilyushin, V., Kryvda, A., \& Alekseev, E.\ 
2009, J. Mol. Spectrosc., 255, 32 

\bibitem[Irvine et al.(1988)]{Irvine88} Irvine, W.~M., Brown, R.~D., 
Cragg, D.~M., et al.\ 1988, \apjl, 335, L89

\bibitem[Jacq et al.(1990)]{Jacq90} Jacq, T., Walmsley, C.~M., Henkel, C., et 
al.\ 1990, \aap, 228, 447

\bibitem[Jacq et al.(1999)]{Jacq99} Jacq, T., Baudry, A., Walmsley, C.~M., \& 
Caselli, P.\ 1999, \aap, 347, 957 

\bibitem[Johns(1985)]{HDO_1985}
Johns, J.~W.~C.\ 
1985, J. Opt. Soc. Am. B, 2, 1340 

\bibitem[Johns et al.(1992)]{H2C2O-param_1992}
Johns, J.~W.~C., Nemes, L., Yamada, K.~M.~T., et al.\ 
1992, J. Mol. Spectrosc., 156, 501 

\bibitem[Johnson et al.(1972)]{H2CO_H2CS_FA_1972} 
Johnson, D.~R., Lovas, F.~J., \& Kirchhoff, W.~H.\ 
1972, J. Phys. Chem. Ref. Data, 1, 1011 

\bibitem[Johnson et al.(1976)]{H2NCN_1976} 
Johnson, D.~R., Suenram, R.~D., \& Lafferty, W.~J.\ 
1976, \apj, 208, 245 

\bibitem[Jones et al.(2007)]{Jones07} Jones, P.~A., Cunningham, M.~R., 
Godfrey, P.~D., \& Cragg, D.~M.\ 2007, \mnras, 374, 579

\bibitem[Jones et al.(2008)]{Jones08} Jones, P.~A., Burton, M.~G., Cunningham, 
M.~R., et al.\ 2008, \mnras, 386, 117

\bibitem[Kalenskii \& Johansson(2010)]{Kalenskii10} Kalenskii, S.~V., \& 
Johansson, L.~E.~B.\ 2010, Astronomy Reports, 54, 1084 

\bibitem[Karakawa et al.(2001)]{MeFo_7-200_2001} 
Karakawa, Y., Oka, K., Odashima, H., Takagi, K., \& Tsunekawa, S.\ 
2001, J. Mol. Spectrosc., 210, 196 

\bibitem[Kawaguchi et al.(1992)]{Kawaguchi92} Kawaguchi, K., Ohishi, M., 
Ishikawa, S.-I., \& Kaifu, N.\ 1992, \apjl, 386, L51 

\bibitem[Kisiel et al.(2009)]{VyCN_2009} 
Kisiel, Z., Pszcz{\'o}{\l}kowski, L., Drouin, B.~J., et al.\ 
2009, J. Mol. Spectrosc., 258, 26 

\bibitem[Kisiel et al.(2012)]{VyCN_2012} 
Kisiel, Z., Pszcz{\'o}{\l}kowski, L., Drouin, B.~J., et al.\ 
2012, J. Mol. Spectrosc., 280, 134 

\bibitem[Klapper et al.(2000)]{13C18O} 
Klapper, G., Lewen, F., Belov, S.~P., \& Winnewisser, G.\
2000, Z. Naturforsch. A, 55, 441

\bibitem[Klapper et al.(2001)]{12C18O} 
Klapper, G., Lewen, F., Gendriesch, R., Belov, S.~P., \& Winnewisser, G.\
2001, Z. Naturforsch. A, 56, 329

\bibitem[Klapper et al.(2003)]{12_13C17O} 
Klapper, G., Surin, L., Lewen, F., et al.\ 
2003, \apj, 582, 262 

\bibitem[Klaus et al.(1996)]{isos-SO_1996}
Klaus, T., Saleck, A.~H., Belov, S.~P., et al.\ 
1996, J. Mol. Spectrosc., 180, 197 

\bibitem[Kleiner et al.(1991)]{Kle91}
Kleiner, I., Hougen, J.~T., Suenram, R.~D., Lovas, F.~J., \& Godefroid, M.\ 
1991, J. Mol. Spectrosc., 148, 38 

\bibitem[Kleiner et al.(1992)]{Kle92}
Kleiner, I., Hougen, J.~T., Suenram, R.~D., Lovas, F.~J., \& Godefroid, M.\ 
1992, J. Mol. Spectrosc., 153, 578 

\bibitem[Kleiner et al.(1996)]{acetaldehyde_1996}
Kleiner, I., Lovas, F.~J., \& Godefroid, M.\ 
1996, J. Phys. Chem. Ref. Data, 25, 1113 

\bibitem[Kra{\'s}nicki \& Kisiel(2011)]{Krasnicki11} Kra{\'s}nicki, A., \& 
Kisiel, Z.\ 2011, J. Mol. Spectrosc., 270, 83 

\bibitem[Kra{\'s}nicki et al.(2011a)]{VyCN_isos_2011} 
Kra{\'s}nicki, A., Kisiel, Z., Drouin, B.~J., \& Pearson, J.~C.\ 
2011a, J. Mol. Struct., 1006, 20 

\bibitem[Kra{\'s}nicki et al.(2011b)]{isos-H2NCN_2011}
Kra{\'s}nicki, A., Kisiel, Z., Jabs, W., Winnewisser, B.~P., \& 
Winnewisser, M.\ 
2011b, J. Mol. Spectrosc., 267, 144 

\bibitem[Kr{\'e}glewski \& Wlodarczak(1992)]{MeNH2_1992} 
Kr{\'e}glewski, M., \& Wlodarczak, G.\ 
1992, J. Mol. Spectrosc., 156, 383 

\bibitem[Kroupa(2001)]{Kroupa01} Kroupa, P.\ 2001, \mnras, 322, 231

\bibitem[Kryvda et al.(2009)]{HCONH2_2009}
Kryvda, A.~V., Gerasimov, V.~G., Dyubko, S.~F., Alekseev, E.~A., \& Motiyenko, 
R.~A.\ 
2009, J. Mol. Spectrosc., 254, 28 

\bibitem[Kuan et al.(1996)]{Kuan96} Kuan, Y.-J., Mehringer, D.~M., \& Snyder, 
L.~E.\ 1996, \apj, 459, 619

\bibitem[Kuriyama et al.(1986)]{13C-MeOH_1986} 
Kuriyama, H., Takagi, K., Takeo, H., \& Matsumura, C.\ 
1986, \apj, 311, 1073 

\bibitem[Kutsenko et al.(2013)]{Kutsenko13}
Kutsenko, A.~S., Motiyenko, R.~A., Margul{\`e}s, L., \& 
Guillemin, J.-C.\ 2013, \aap, 549, A128

\bibitem[Lapinov et al.(2007)]{HNCO_2007}
Lapinov, A.~V., Golubiatnikov, G.~Y., Markov, V.~N., \& Guarnieri, A.\ 
2007, Astron. Lett., 33, 121 

\bibitem[Larsen \& Winnewisser(1974)]{OC-13-S_1974} 
Larsen, N.~W., \& Winnewisser, B.~P.\ 
1974, Z. Naturforsch. A, 29, 1213 

\bibitem[Lattanzi et al.(2007)]{HC-13-O+_2007}
Lattanzi, V., Walters, A., Drouin, B.~J., \& Pearson, J.~C.\ 
2007, \apj, 662, 771 

\bibitem[Lee et al.(2012)]{Lee12} Lee, E.~J., Murray, N., \& Rahman, 
M.\ 2012, \apj, 752, 146

\bibitem[Lee et al.(1995)]{NS_main_1995} 
Lee, S.~K., Ozeki, H., \& Saito, S.\ 
1995, \apjs, 98, 351 

\bibitem[Lees \& Baker(1968)]{MeOH_1968} 
Lees, R.~M., \& Baker, J.~G.\ 
1968, \jcp, 48, 5299 

\bibitem[Lees \& Mohammadi(1980)]{MeSH_1980} 
Lees, R.~M., \& Mohammadi, M.~A.\ 
1980, Can. J. Phys., 58, 1640 

\bibitem[Liang et al.(1986)]{Lia86}
Liang, W., Baker, J.~G., Herbst, E., Booker, R.~A., \& de Lucia, F.~C.\ 
1986, J. Mol. Spectrosc., 120, 298 

\bibitem[Lis \& Goldsmith(1990)]{Lis90} Lis, D.~C., \& Goldsmith, 
P.~F.\ 1990, \apj, 356, 195

\bibitem[Lis et al.(1993)]{Lis93} Lis, D.~C., Goldsmith, P.~F., Carlstrom, 
J.~E., \& Scoville, N.~Z.\ 1993, \apj, 402, 238

\bibitem[Liu \& Snyder(1999)]{Liu99} Liu, S., \& Snyder, L.~E.\ 1999, \apj, 
523, 683

\bibitem[Lovas et al.(1979)]{DME_1979} 
Lovas, F.~J., Lutz, H., \& Dreizler, H.\ 
1979, J. Phys. Chem. Ref. Data, 8, 1051 

\bibitem[Lovas et al.(2006)]{Lovas06} Lovas, F.~J., Hollis, J.~M., Remijan, 
A.~J., \& Jewell, P.~R.\ 2006, \apjl, 645, L137 

\bibitem[Lowry Manson et al.(1977)]{SiO_1977} 
Lowry Manson, E., Jr., Clark, W.~W., De Lucia, F.~C., \& Gordy, W.\ 
1977, \pra, 15, 223 

\bibitem[Maeda et al.(2008)]{H2CS_2008}
Maeda, A., Medvedev, I.~R., Winnewisser, M., et al.\ 
2008, \apjs, 176, 543 

\bibitem[Maes et al.(1987)]{Mae87}
Maes, H., Wlodarczak, G., Boucher, D., \& Demaison, J.\ 
1987, Z. Naturforsch. A, 42, 97 

\bibitem[Margul{\`e}s et al.(2003)]{Margules03} Margul{\`e}s, L., Lewen, F., 
Winnewisser, G., Botschwina, P., M{\"u}ller, H.~S.~P.\ 2003, Phys. Chem. Chem. 
Phys., 5, 2770 

\bibitem[Margul{\`e}s et al.(2009)]{15N-D-EtCN_2009} 
Margul{\`e}s, L., Motiyenko, R., Demyk, K., et al.\ 
2009, \aap, 493, 565 

\bibitem[Mbosei et al.(2000)]{HC3N_Lille_2000} 
Mbosei, L., Fayt, A., Dr\'ean, P. \& Cosl\'eou, J.\ 
2000, J. Mol. Struct., 517, 271 

\bibitem[McCarthy et al.(1995)]{C-13-CH_1995} 
McCarthy, M.~C., Gottlieb, C.~A. \& Thaddeus, P.\ 
1995, J. Mol. Spectrosc., 173, 303 

\bibitem[McNaughton \& Bruget(1993)]{H2CS_1993}
McNaughton, D., \& Bruget, D.~N.\ 1993, 
J. Mol. Spectrosc., 159, 340 

\bibitem[Medcraft et al.(2012)]{c-C2H4O_2012}
Medcraft, C., Thompson, C.~D., Robertson, E.~G., Appadoo, D.~R.~T., \& 
McNaughton, D.\ 
2012, \apj, 753, 18 

\bibitem[Medvedev et al.(2009)]{EtFo_2009}
Medvedev, I.~R., De Lucia, F.~C., \& Herbst, E.\ 
2009, \apjs, 181, 433 

\bibitem[Mehringer et al.(1993)]{Mehringer93} Mehringer, D.~M., 
Palmer, P., Goss, W.~M., \& Yusef-Zadeh, F.\ 1993, \apj, 412, 684

\bibitem[Mehringer et al.(1997)]{Mehringer97} Mehringer, D.~M., 
Snyder, L.~E., Miao, Y., \& Lovas, F.~J.\ 1997, \apjl, 480, L71

\bibitem[Mehringer et al.(2004)]{Mehringer04}
Mehringer, D.~M., Pearson, J.~C., Keene, J., \& Phillips, T.~G.\ 
2004, \apj, 608, 306 

\bibitem[Menten(1991a)]{Menten91a} Menten, K.~M.\ 1991a, in Atoms, 
Ions and Molecules: New Results in Spectral Line Astrophysics, ed. A.~D. 
Haschick \& P.~T.~P. Ho, ASP Conf. Ser. (San Francisco, CA: ASP), 16, 119

\bibitem[Menten(1991b)]{Menten91b} Menten, K.~M.\ 1991b, \apjl, 380, L75

\bibitem[Menten(2004)]{Menten04} Menten, K.~M.\ 2004, in The Dense 
Interstellar Medium in Galaxies, ed. Pfalzner, Kramer, Staubmeier \& 
Heithausen, Springer proceedings in physics (Berlin, Heidelberg: Springer), 
91, 69

\bibitem[Menten et al.(2011)]{Menten11} Menten, K.~M., Wyrowski, F., Belloche, 
A., et al.\ 2011, \aap, 525, A77

\bibitem[Milam et al.(2005)]{Milam05} Milam, S.~N., Savage, C., Brewster, 
M.~A., Ziurys, L.~M., \& Wyckoff, S.\ 2005, \apj, 634, 1126

\bibitem[Mollaaghababa et al.(1991)]{SiO_1991} 
Mollaaghababa, R., Gottlieb, C.~A., Vrtilek, J.~M., \& Thaddeus, P.\ 
1991, \apjl, 368, L19 

\bibitem[M{\o}llendal et al.(2012)]{Mollendal12} M{\o}llendal, H., 
Margul{\`e}s, L., Belloche, A., et al.\ 2012, \aap, 538, A51 

\bibitem[Morino et al.(2000)]{OCS_excited_2000}
Morino, I., Yamada, K.~M.~T., \& Maki, A.~G.\ 
2000, J. Mol. Spectrosc., 200, 145 

\bibitem[Motiyenko et al.(2012)]{Motiyenko12} Motiyenko, R.~A., Tercero, B., 
Cernicharo, J., \& Margul{\`e}s, L.\ 2012, \aap, 548, A71 

\bibitem[Muench et al.(2002)]{Muench02} Muench, A.~A., Lada, E.~A., 
Lada, C.~J., \& Alves, J.\ 2002, \apj, 573, 366

\bibitem[M{\"u}ller et al.(2000a)]{17O-SO2_33SO3_2000}
M{\"u}ller, H.~S.~P., Farhoomand, J., Cohen, E.~A., et al.\ 
2000a, J. Mol. Spectrosc., 201, 1

\bibitem[M{\"u}ller et al.(2000b)]{H2CO-18_2000}
M{\"u}ller, H.~S.~P., Gendriesch, R., Lewen, F., \& Winnewisser, G.\ 
2000b, Z. Naturforsch. A, 55, 486 

\bibitem[M{\"u}ller et al.(2000c)]{H2C-13-O_2000}
M{\"u}ller, H.~S.~P., Gendriesch, R., Margul{\`e}s, L., et al.\ 
2000c, Phys. Chem. Chem. Phys., 2, 3401 

\bibitem[M{\"u}ller et al.(2001)]{Mueller01} M{\"u}ller, H.~S.~P., Thorwirth, 
S., Roth, D.~A., \& Winnewisser, G.\ 2001, \aap, 370, L49

\bibitem[M{\"u}ller et al.(2002)]{propyne_2002}
M{\"u}ller, H.~S.~P., Pracna, P., \& Horneman, V.-M.\ 
2002, J. Mol. Spectrosc., 216, 397 

\bibitem[M\"uller \& Christen(2004)]{gGg_eglyc_2004} 
M\"uller, H.~S.~P. \& Christen, D.\ 
2004, J. Mol. Spectrosc., 228, 298 

\bibitem[M{\"u}ller et al.(2004)]{MeOH_2004} 
M{\"u}ller, H.~S.~P., Menten, K.~M., \& M\"ader, H.\ 
2004, \aap, 428, 1019 

\bibitem[M{\"u}ller \& Br{\"u}nken(2005)]{SO2_2005} 
M{\"u}ller, H.~S.~P., \& Br{\"u}nken, S.\ 
2005, J. Mol. Spectrosc., 232, 213 

\bibitem[M{\"u}ller et al.(2005)]{Mueller05} M{\" u}ller, H.~S.~P.,
Schl{\" o}der, F., Stutzki, J., \& Winnewisser, G.\ 2005, J. Mol. Struct.,
742, 215

\bibitem[M{\"u}ller et al.(2007)]{Mueller07} M{\"u}ller, H.~S.~P., 
McCarthy, M.~C., Bizzocchi, L., et al.\ 2007, Phys. Chem. Chem. Phys., 9, 1579 

\bibitem[M{\"u}ller et al.(2008)]{Mueller08} M{\"u}ller, H.~S.~P., Belloche, 
A., Menten, K.~M., Comito, C., \& Schilke, P.\ 2008, J. Mol. Spectrosc., 251, 
319

\bibitem[M{\"u}ller et al.(2009)]{MeCN_2009} 
M{\"u}ller, H.~S.~P., Drouin, B.~J., \& Pearson, J.~C.\ 
2009, \aap, 506, 1487 

\bibitem[M{\"u}ller et al.(2010a)]{MeCN-conf_2010} 
M{\"u}ller, H.~S.~P., Drouin, B.~J., Pearson, J.~C., et al.\ 
2010a, contribution RC12, 65th International Symposium On Molecular 
Spectroscopy,  Columbus, OH, USA

\bibitem[M{\"u}ller et al.(2010b)]{H2DO+_w-NH2D_2010} 
M{\"u}ller, H.~S.~P., Dong, F., Nesbitt, D.~J., Furuya, T., \& Saito, S.\ 
2010b, Phys. Chem. Chem. Phys., 12, 8362 

\bibitem[Murray \& Rahman(2010)]{Murray10} Murray, N., \& Rahman, M.\ 
2010, \apj, 709, 424

\bibitem[Neill et al.(2012)]{Neill12} Neill, J.~L., Muckle, M.~T., Zaleski, 
D.~P., et al.\ 2012, \apj, 755, 153 

\bibitem[Nemes et al.(2000)]{H2C2O_vibs_2000}
Nemes, L., Luckhaus, D., Quack, M. \& Johns, J.~W.~C.\ 
2000, J. Mol. Struct., 517, 217 

\bibitem[Neustock et al.(1990)]{DME_1990} 
Neustock, W., Guarnieri, A., Demaison, J. \& Wlodarczak, G.\ 
1990, Z. Naturforsch. A, 45, 702

\bibitem[Niedenhoff et al.(1995)]{HNCS_1995}
Niedenhoff, M., Winnewisser, G., Yamada, K.~M.~T. \& Belov, S.~P.\ 
1995, J. Mol. Spectrosc., 169, 224 

\bibitem[Niedenhoff et al.(1996)]{vib-HNCO_1996}
Niedenhoff, M., Yamada, K.~M.~T. \& Winnewisser, G.\ 
1996, J. Mol. Spectrosc., 176, 342 

\bibitem[Nielsen(1973)]{1973_CJ_Nielsen} Nielsen, C.~J.,
1973, Ph.D. Thesis, K{\o}benhavns Universitet, Denmark

\bibitem[Nummelin et al.(1998)]{Nummelin98} Nummelin, A., Bergman, P., 
Hjalmarson, A., et al.\ 1998, \apjs, 117, 427 

\bibitem[Nummelin \& Bergman(1999)]{Nummelin99}
Nummelin, A., \& Bergman, P.\ 
1999, \aap, 341, L59 

\bibitem[Nummelin et al.(2000)]{Nummelin00} Nummelin, A., Bergman, P.,
Hjalmarson, {\AA}., et al.\ 2000, \apjs, 128, 213

\bibitem[Ordu et al.(2012)]{Ordu12} Ordu, M.~H., M{\"u}ller, H.~S.~P., 
Walters, A., et al.\ 2012, \aap, 541, A121 

\bibitem[Padovani et al.(2009)]{C2H_2009}
Padovani, M., Walmsley, C.~M., Tafalla, M., Galli, D. \& M\"uller, H.~S.~P.\ 
2009, \aap, 505, 1199 

\bibitem[Pagani et al.(2009)]{N2H+_2009}
Pagani, L., Daniel, F., \& Dubernet, M.-L.\ 
2009, \aap, 494, 719 

\bibitem[Pan et al.(1998)]{c-C2H4O_1998} 
Pan, J., Albert, S., Sastry, K.~V.~L.~N., Herbst, E., \& De Lucia, F.~C.\ 
1998, \apj, 499, 517 

\bibitem[Panagia(1973)]{Panagia73} Panagia, N.\ 1973, \aj, 78, 929

\bibitem[Pardo et al.(2007)]{Pardo07} Pardo, J.~R., Cernicharo, J., 
Goicoechea, J.~R., Gu{\'e}lin, M., \& Asensio Ramos, A.\ 2007, \apj, 661, 250 

\bibitem[Pearson et al.(1976)]{HN-15-C_1976}
Pearson, E.~F., Creswell, R.~A., Winnewisser, M., \& Winnewisser, G.\ 
1976, Z. Naturforsch. A, 31, 1394 

\bibitem[Pearson et al.(1995)]{t-EtOH_1995} 
Pearson, J.~C., Sastry, K.~V.~L.~N., Winnewisser, M., Herbst, E., \& de Lucia, 
F.~C.\ 
1995, J. Phys. Chem. Ref. Data, 24, 1 

\bibitem[Pearson et al.(1996)]{g-EtOH_1996} 
Pearson, J.~C., Sastry, K.~V.~L.~N., Herbst, E., \& De Lucia, F.~C.\ 
1996, J. Mol. Spectrosc., 175, 246 

\bibitem[Pearson et al.(1997)]{g-EtOH_1997} 
Pearson, J.~C., Sastry, K.~V.~L.~N., Herbst, E., \& De Lucia, F.~C.\ 
1997, \apj, 480, 420 

\bibitem[Pearson et al.(2008)]{EtOH_2008} 
Pearson, J.~C., Brauer, C.~S., \& Drouin, B.~J.\ 
2008, J. Mol. Spectrosc., 251, 394

\bibitem[Pearson et al.(2009)]{EtCN_v20_2009_WH13} 
Pearson, J.~C., Brauer, C.~S., Yu, S., \& Drouin, B.~J.\ 
2009, contribution WH13, 64th International Symposium On Molecular 
Spectroscopy,  Columbus, OH, USA

\bibitem[Peng et al.(1993)]{Peng93} Peng, Y., Vogel, S.~N., \& 
Carlstrom, J.~E.\ 1993, \apj, 418, 255 

\bibitem[Penzias(1981)]{Penzias81} Penzias, A.~A.\ 1981, \apj, 249, 513 

\bibitem[Peters et al.(2010)]{Peters10} Peters, T., Mac Low, M.-M., 
Banerjee, R., Klessen, R.~S., \& Dullemond, C.~P.\ 2010, \apj, 719, 831

\bibitem[Pickett et al.(1981)]{MeOH_1981} 
Pickett, H.~M., Cohen, E.~A., Brinza, D.~E., \& Schaefer, M.~M.\ 
1981, J. Mol. Spectrosc., 89, 542 

\bibitem[Pickett et al.(1998)]{Pickett98} Pickett, H.~M., Poynter, R.~L., 
Cohen, E.~A., Delitsky, M.~L., Pearson, J.~C., \& M{\"u}ller, H.~S.~P.\ 1998
, \jqsrt, 60, 883

\bibitem[Qin et al.(2011)]{Qin11} Qin, S.-L., Schilke, P., Rolffs, R., et al.\ 
2011, \aap, 530, L9 

\bibitem[Raymonda \& Klemperer(1971)]{PN_HFS_1971} 
Raymonda, J., \& Klemperer, W.\ 
1971, \jcp, 55, 232 

\bibitem[Read et al.(1986)]{H2NCN_1986} 
Read, W.~G., Cohen, E.~A., \& Pickett, H.~M.\ 
1986, J. Mol. Spectrosc., 115, 316 

\bibitem[Reid et al.(2009)]{Reid09} Reid, M.~J., Menten, K.~M., Zheng, X.~W., 
Brunthaler, A., \& Xu, Y.\ 2009, \apj, 705, 1548 

\bibitem[Remijan et al.(2008)]{Remijan08} Remijan, A.~J., Hollis, J.~M., 
Lovas, F.~J., et al.\ 2008, \apjl, 675, L85

\bibitem[Requena-Torres et al.(2006)]{Requena06} Requena-Torres, M.~A., 
Mart{\'{\i}}n-Pintado, J., Rodr{\'{\i}}guez-Franco, A., et al.\ 2006, \aap, 
455, 971 

\bibitem[Richard et al.(2012)]{13C-EtCN_2012} 
Richard, C., Margul{\`e}s, L., Motiyenko, R.~A., \& Guillemin, J.-C.\ 
2012, \aap, 543, A135 

\bibitem[Rodler(1983)]{Rodler83} Rodler, M., PhD thesis, 1983, 
Eidgen{\"o}ssische Technische Hochschule, Z{\"u}rich, Switzerland

\bibitem[Rodler et al.(1984)]{Rodler84} Rodler, M., Brown, R.~D., Godfrey, 
P.~D., \& Tack, L.~M.\ 1984, Chem. Phys. Lett., 110, 447 

\bibitem[Rodler(1985)]{Rodler85} Rodler, M.\ 1985, J. Mol. Spectrosc., 114, 23 

\bibitem[Rodler et al.(1986)]{Rodler86} Rodler, M., Brown, R.~D., Godfrey, 
P.~D., \& Kleib{\"o}mer, B.\ 1986, J. Mol. Spectrosc., 118, 267

\bibitem[Rodr{\'{\i}}guez-Fern{\'a}ndez et al.(2010)]{RodriguezFernandez10} 
Rodr{\'{\i}}guez-Fern{\'a}ndez, N.~J., Tafalla, M., Gueth, F., \& 
Bachiller, R.\ 2010, \aap, 516, A98

\bibitem[Rolffs et al.(2011)]{Rolffs11} Rolffs, R., Schilke, P., Wyrowski, F., 
et al.\ 2011, \aap, 529, A76 

\bibitem[Saito(1976)]{Saito76} Saito, S.\ 1976, Chem. Phys. Lett., 42, 399 

\bibitem[Saito et al.(1987)]{CCS_1987}
Saito, S., Kawaguchi, K., Yamamoto, S., et al.\ 
1987, \apjl, 317, L115 

\bibitem[Sakai et al.(2010)]{CC-13-H_2010} 
Sakai, N., Saruwatari, O., Sakai, T., Takano, S., \& Yamamoto, S.\ 
2010, \aap, 512, A31 

\bibitem[Saleck et al.(1994)]{CN-15_combined_1994} 
Saleck, A.~H., Simon, R., \& Winnewisser, G.\ 
1994, \apj, 436, 176 

\bibitem[Saleck et al.(1995a)]{isos-H2S_1995}
Saleck, A., Tanimoto, M., Belov, S.~P., Klaus, T., Winnewisser, G.\ 
1995a, J. Mol. Spectrosc., 171, 481 

\bibitem[Saleck et al.(1995b)]{NS_combined_1995}
Saleck, A.~H., Ozeki, H., \& Saito, S.\ 
1995b, Chem. Phys. Lett., 244, 199 

\bibitem[Sanz et al.(2005)]{excited-HC3N_etc_2005} 
Sanz, M.~E., McCarthy, M.~C., \& Thaddeus, P.\ 
2005, \jcp, 122, 194319 

\bibitem[Sastry et al.(1984)]{SAS84} 
Sastry, K.~V.~L.~N., Lees, R.~M., \& de Lucia, F.~C.\ 
1984, J. Mol. Spectrosc., 103, 486 

\bibitem[Sastry et al.(1986)]{MeSH_1986} 
Sastry, K.~V.~L.~N., Herbst, E., Booker, R.~A., \& de Lucia, F.~C.\ 
1986, J. Mol. Spectrosc., 116, 120 

\bibitem[Saykally et al.(1976)]{HNC_1976} 
Saykally, R.~J., Szanto, P.~G., Anderson, T.~G., \& Woods, R.~C.\ 
1976, \apjl, 204, L143 

\bibitem[Schilke et al.(1997)]{Schilke97}
Schilke, P., Groesbeck, T.~D., Blake, G.~A., \& Phillips, T.~G.\ 
1997, \apjs, 108, 301

\bibitem[Schilke et al.(1999)]{Schilke99} Schilke, P., Phillips, T.~G., 
\& Mehringer, D.~M.\ 1999, in The Physics and Chemistry of the Interstellar 
Medium, ed. V. Ossenkopf, J. Stutzki \& G. Winnewisser (GCA-Verlag Herdecke), 
330 

\bibitem[Schmid-Burgk et al.(2004)]{HC-13-O+_2004}
Schmid-Burgk, J., Muders, D., M{\"u}ller, H.~S.~P., \& Brupbacher-Gatehouse, 
B.\ 
2004, \aap, 419, 949 

\bibitem[Scoville et al.(1975)]{Scoville75} Scoville, N.~Z., 
Solomon, P.~M., \& Penzias, A.~A.\ 1975, \apj, 201, 352

\bibitem[Snyder et al.(2005)]{Snyder05} Snyder, L.~E., Lovas, F.~J., 
Hollis, J.~M., Friedel, D.~N., et al.\ 2005, \apj, 619, 914

\bibitem[Sofue et al.(2009)]{Sofue09} Sofue, Y., Honma, M., 
\& Omodaka, T.\ 2009, \pasj, 61, 227

\bibitem[Spezzano et al.(2012)]{c-C3H2_div_2012}
Spezzano, S., Tamassia, F., Thorwirth, S., et al.\ 
2012, \apjs, 200, 1 

\bibitem[Stubgaard(1978)]{HCONH2_1978} 
Stubgaard, M.\ 
1978, Ph.D.~Thesis, K{\o}benhavns Universitet, Denmark

\bibitem[Sutton et al.(1991)]{Sutton91} Sutton, E.~C., Jaminet, P.~A., Danchi, 
W.~C., \& Blake, G.~A.\ 1991, \apjs, 77, 255 

\bibitem[Tafalla et al.(2010)]{Tafalla10} Tafalla, M., 
Santiago-Garc{\'{\i}}a, J., Hacar, A., \& Bachiller, R.\ 2010, \aap, 522, A91

\bibitem[Tang \& Saito(1995)]{Tang95} Tang, J., \& Saito, S.\ 1995, \apjl, 
451, L93

\bibitem[Tercero et al.(2010)]{Orion_Pepe_S-species} 
Tercero, B., Cernicharo, J., Pardo, J.~R., \& Goicoechea, J.~R.\ 
2010, \aap, 517, A96 

\bibitem[Thorwirth et al.(2000a)]{HC3N_Koeln_2000} 
Thorwirth, S., M{\"u}ller, H.~S.~P., \& Winnewisser, G.\ 
2000a, J. Mol. Spectrosc., 204, 133 

\bibitem[Thorwirth et al.(2000b)]{HNC_2000}
Thorwirth, S., M{\"u}ller, H.~S.~P., Lewen, F., Gendriesch, R., \& 
Winnewisser, G.\ 
2000b, \aap, 363, L37 

\bibitem[Thorwirth et al.(2001)]{isos-HC3N_2001} 
Thorwirth, S., M\"uller, H.~S.~P., \& Winnewisser, G.\ 
2001, Phys. Chem. Chem. Phys., 3, 1236 

\bibitem[Thorwirth et al.(2003)]{HCN_2003}
Thorwirth, S., M{\"u}ller, H.~S.~P., Lewen, F., et al.\ 
2003, \apjl, 585, L163 

\bibitem[Tiemann(1982)]{div-SO_1982} 
Tiemann, E.\ 
1982, J. Mol. Spectrosc., 91, 60 

\bibitem[Tinti et al.(2007)]{HCO+_2007}
Tinti, F., Bizzocchi, L., Degli Esposti, C., \& Dore, L.\ 
2007, \apjl, 669, L113 

\bibitem[T\"orring(1968)]{isos-SiO_1968}
T\"orring, T.\
1968, Z. Naturforsch. A, 23, 777

\bibitem[Turner(1989)]{Turner89} Turner, B.~E.\ 1989, \apjs, 70, 539 

\bibitem[Turner(1991)]{Turner91} Turner, B.~E.\ 1991, \apjs, 76, 617 

\bibitem[Turner \& Apponi(2001)]{Turner01} Turner, B.~E., \& Apponi, A.~J.\ 
2001, \apjl, 561, L207 

\bibitem[Vacherand et al.(1986)]{acetone_v0_1986} 
Vacherand, J.~M., van Eijck, B.~P., Burie, J., \& Demaison, J.\ 
1986, J. Mol. Spectrosc., 118, 355 

\bibitem[van der Tak et al.(2009)]{HNC-13_etc_2009} 
van der Tak, F.~F.~S., M{\"u}ller, H.~S.~P., Harding, M.~E., \& Gauss, J.\ 
2009, \aap, 507, 347 

\bibitem[van Dijk \& Dymanus(1974)]{DBr_OCS_97GHz_1974} 
van Dijk, F. \& Dymanus, A.\ 
1974, Chem. Phys., 6, 474 

\bibitem[Vanek et al.(1989)]{OCS-34_1989}
Vanek, M.~D., Jennings, D.~A., Wells, J.~S., \& Maki, A.~G.\ 
1989, J. Mol. Spectrosc., 138, 79 

\bibitem[Vrt\'ilek et al.(1987)]{c-C3H2_1987} 
Vrt\'ilek, J.~M., Gottlieb, C.~A., \& Thaddeus, P.\ 
1987, \apj, 314, 716 

\bibitem[Whiteoak et al.(1988)]{Whiteoak88} Whiteoak, J.~B., 
Gardner, F.~F., Caswell, J.~L., et al.\ 1988, \mnras, 235, 655

\bibitem[Widicus Weaver et al.(2005)]{glycolaldehyde_2005}
Widicus Weaver, S.~L., Butler, R.~A.~H., Drouin, B.~J., et al.\ 
2005, \apjs, 158, 188 

\bibitem[Wilson \& Rood(1994)]{Wilson94} Wilson, T.~L., \& Rood, R.\ 1994, 
\araa, 32, 191

\bibitem[Winnewisser(1973)]{Winnewisser73} Winnewisser, G.\ 1973, 
J. Mol. Spectrosc., 46, 16 

\bibitem[Winnewisser et al.(1997)]{12C16O} 
Winnewisser, G., Belov, S.~P., Klaus, T., \& Schieder, R.\ 
1997, J. Mol. Spectrosc., 184, 468 

\bibitem[Winnewisser et al.(1975)]{Winnewisser75} Winnewisser, M., 
Winnewisser, G., Honda, T., \& Hirota, E.\ 1975, Z. Naturforsch. A, 30, 1001

\bibitem[Winnewisser et al.(2002)]{HCOOH_2002}
Winnewisser, M., Winnewisser, B.~P., Stein, M., et al.\ 
2002, J. Mol. Spectrosc., 216, 259 

\bibitem[Winton \& Gordy(1970)]{OCS_main_1970} 
Winton, R.~S., \& Gordy, W.\ 
1970, Phys. Lett. A, 32, 219 

\bibitem[Wlodarczak et al.(1988)]{n-PrCN_1988} 
Wlodarczak, G., Martinache, L., Demaison, J., Marstokk, K.-M., \& 
M{\o}llendal, H.\ 
1988, J. Mol. Spectrosc., 127, 178 

\bibitem[Wouterloot et al.(2008)]{Wouterloot08} Wouterloot, J.~G.~A., Henkel, 
C., Brand, J., \& Davis, G.~R.\ 2008, \aap, 487, 237

\bibitem[Wyrowski et al.(1999)]{Wyrowski99} Wyrowski, F., Schilke, P., \& 
Walmsley, C.~M.\ 1999, \aap, 341, 882 

\bibitem[Xu \& Lovas(1997)]{12_13C-MeOH_1997} 
Xu, L.-H., \& Lovas, F.~J.\ 
1997, J. Phys. Chem. Ref. Data, 26, 17 

\bibitem[Xu et al.(2008)]{MeOH_2008} 
Xu, L.-H., Fisher, J., Lees, R.~M., et al.\ 
2008, J. Mol. Spectrosc., 251, 305 

\bibitem[Xu et al.(2012)]{MeSH_2012}
Xu, L.-H., Lees,R.~M., Crabbe, G.~T., et al.\ 
2012, \jcp, 137, 104313

\bibitem[Yamada \& Winnewisser(1977)]{vib-HNCO-b_1977} 
Yamada, K., \& Winnewisser, M.\ 
1977, J. Mol. Spectrosc., 68, 307 

\bibitem[Yamada(1977)]{vib-HNCO-a_1977} 
Yamada, K.\ 
1977, J. Mol. Spectrosc., 68, 423 

\bibitem[Yamada et al.(1979)]{HNCS_1979} 
Yamada, K., Winnewisser, M., Winnewisser, G., Szalanski, L.~B., \& Gerry, 
M.~C.~L.\ 
1979, J. Mol. Spectrosc., 78, 189 

\bibitem[Yamada \& Creswell(1986)]{HC3N_1986} 
Yamada, K.~M.~T., \& Creswell, R.~A.\ 
1986, J. Mol. Spectrosc., 116, 384 

\bibitem[Yamada et al.(1995)]{HC3N_v0_1995}
Yamada, K.~M.~T., Moravec, A. \& Winnewisser, G.\
1995, Z. Naturforsch. A, 50, 1179

\bibitem[Yamada et al.(2004)]{HC5N_2004}
Yamada, K.~M.~T., Degli Esposti, C., Botschwina, P., et al.\ 
2004, \aap, 425, 767 

\bibitem[Yamaguchi et al.(2011)]{Yamaguchi11} Yamaguchi, T., Takano, S., 
Sakai, N., et al.\ 2011, \pasj, 63, L37 

\bibitem[Zelinger et al.(2003)]{HCN_hohe-v_2003}
Zelinger, Z., Amano, T., Ahrens, V., et al.\ 
2003, J. Mol. Spectrosc., 220, 223 

\bibitem[Zernickel et al.(2012)]{Zernickel12} Zernickel, A., Schilke, P., 
Schmiedeke, A., et al.\ 2012, \aap, 546, A87 

\bibitem[Ziurys(1987)]{Ziurys87} Ziurys, L.~M.\ 1987, \apjl, 321, L81 

\end{thebibliography}
\end{document}